\documentclass[review,3p,times]{elsarticle}

\usepackage{lineno,hyperref}
\usepackage{floatrow}

\modulolinenumbers[5]

\journal{Physics Reports}

\biboptions{sort&compress}

\usepackage{amssymb,amsmath,textcomp}
\usepackage{float}
\usepackage{bm}

\usepackage{graphicx}
\usepackage{epsfig,epsf}
\usepackage{epstopdf}

\usepackage{makeidx}

\usepackage{color}
\usepackage{caption}
\usepackage{subcaption}

\usepackage{setspace}

\makeindex

\def\bk{{\bf k}}
\newcommand{\beq}{\begin{eqnarray}}
\newcommand{\eeq}{\end{eqnarray}}

\def\be{\begin{equation}}
\def\ee{\end{equation}}
\def\la{\langle}
\def\ra{\rangle}
\def\DS{\displaystyle}
\def\sign{\mathrm{sign}}

\def\Cas{\mathrm{Cas}}

\def\ex{\mathrm{ex}}
\def\kB{k_{\mathrm{B}}}

\def\dmb{\Delta \bar{\mu}}

\def\xt{\xi_\tau}
\def\xop{\xi_0^+}
\def\xom{\xi_0^-}
\def\xm{\xi_\mu}
\def\Capp{\Delta_\Cas^{(+,+)}}

\def\Caoo{\Delta_\Cas^{(O,O)}}
\def\bc{(\zeta)}
\def\BC{\rm boundary conditions}

\def\rme{{\rm e}}

\newcommand\reff[1]{(\ref{#1})}

\newcommand\eq[1]{Eq.~(\ref{#1})}

\newcommand{\eg}{\textit{e.g.}}

\newcommand{\tcr}[1]{\textcolor{red}{#1}}

\newcommand{\tb}{\mathring{\tau}}

\newcommand{\mbf}[1]{\mathbf #1}

\numberwithin{equation}{section}

\begin{document}

\begin{frontmatter}

\title{\centering{Critical Casimir Effect: Exact Results}}
\author[DMD,SD,DMDSD]{D. M. Dantchev\corref{CorrespondingAuthor}}
\ead{daniel@imbm.bas.bg}
\author[SD,DMDSD]{S. Dietrich}
\ead{dietrich@is.mpg.de}

\address[DMD]{Institute of
	Mechanics, Bulgarian Academy of Sciences, Academic Georgy Bonchev St. building 4,
	1113 Sofia, Bulgaria}
\address[SD]{Max-Planck-Institut f\"{u}r Intelligente Systeme, Heisenbergstrasse 3, D-70569 Stuttgart, Germany} 
\address[DMDSD]{
	IV. Institut f\"{u}r Theoretische Physik, Universit\"{a}t Stuttgart, Pfaffenwaldring 57, D-70569 Stuttgart, Germany}

\cortext[CorrespondingAuthor]{Corresponding author}


\begin{abstract}
In any medium there are fluctuations due to temperature or due to the quantum nature of its constituents. If a material body is immersed into such a medium, its shape and the properties of its constituents modify the properties of the surrounding medium and its fluctuations. If in the same medium there is a second body then --- in addition to all direct interactions between them --- the modifications  due to the first body influence the modifications due to the second body. This mutual influence results in a force between these bodies. If the excitations of the medium, which mediate the effective interaction between the bodies, are massless,   this force is long-ranged and nowadays known as a Casimir force. If the fluctuating medium consists of the confined electromagnetic field in vacuum, one speaks of the quantum mechanical Casimir effect. In the case that the order parameter of material fields fluctuates - such as differences of number densities or concentrations - and that the corresponding fluctuations of the order parameter are long-ranged, one speaks of the critical Casimir effect. This holds, e.g., in the case of systems which undergo a second-order phase transition and which are thermodynamically located near the corresponding critical point, or for systems with a broken continuous symmetry exhibiting Goldstone mode excitations. Here we review the currently available exact results concerning the critical Casimir effect in systems encompassing the one-dimensional Ising, XY,  and Heisenberg models, the two-dimensional Ising model, the Gaussian and the spherical models, as well as the mean field results for the Ising and the XY model. Special attention is paid to the influence of the boundary conditions on the behavior of the Casimir force. We present results both for the case of classical critical fluctuations if the system possesses  a critical point at a non-zero temperature, as well as the case of  quantum systems undergoing a continuous phase transition at zero temperature as function of certain parameters.  As confinements we consider the film, the sphere - plane, and the sphere - sphere geometries. We discuss systems governed by short-ranged, by subleading long-ranged (i.e., of the van der Waals type), and by leading long-ranged interactions. In order to put the critical Casimir effect into the proper context and in order to make the review as self-contained as possible,  basic facts about the theory of phase transitions, the theory of critical phenomena in classical and quantum systems, and  finite-size scaling theory are recalled. Whenever possible, a discussion of the relevance of the exact results towards an understanding of available experiments is presented.  The eventual applicability of the present results for certain devices is pointed out, too. 
\end{abstract}

\date{\today}

\end{frontmatter}

\newpage

\tableofcontents


\newpage
\section{Introduction}

\subsection{Preface}

If the mean value of a nonzero random variable vanishes, its variance nonetheless differs from zero. This simple mathematical fact leads to nontrivial physical consequences such as the occurrence of the so-called Casimir force \cite{C48}. In 1948 \cite{C48}, after a discussion with Niels Bohr, the Dutch physicist H. B. G.  Casimir realized that the zero-point fluctuations of the quantum electromagnetic field in vacuum lead to the remarkable mechanical effect of the appearance of a  long-ranged attractive \textit{force} between two perfectly conducting, uncharged, parallel plates at a distance $L$ from each other, and he calculated this force. In the absence of charges on
the plates the vacuum expectation value of both the electric field $\bf E$ and the magnetic field $\bf B$ between the plates vanishes, i.e.,
$$
\la {\bf E}\ra = 0 \qquad \mbox{and} \qquad \la {\bf B} \ra =0,\quad
\mbox{but} \quad 
\la {\bf E}^2 \ra \ne 0 \qquad
\mbox{and} \qquad \la {\bf B}^2 \ra \ne 0,
$$
so that the expectation value of the energy due to the electromagnetic field in the volume between the plates, i.e., $\langle{\cal H}\rangle$ with 
$$
{\cal H}=\int  \left[\frac{1}{2}\varepsilon_0 \textbf{E}^2(\textbf{r})+\frac{1}{2\mu_0} \textbf{B}^2(\textbf{r})\right]\;d^3 r,
$$
is nonzero.
Here $\varepsilon_0$ and $\mu_0$ are the dielectric permittivity and the magnetic permeability of the vacuum, respectively. Since the uncharged plates are supposed to be ideal conductors (i.e., $\textbf{E}=0$ inside the plates), the electric field component parallel to the plates has to be zero, i.e., in the lateral direction one has Dirichlet boundary conditions for the $\textbf{E}$ field. The latter cases lead to a $L$-dependence of the energy of the field in between the plates.  Thus, upon changing the distance $L$ between them, a \textit{force} emerges which acts on the plates in normal direction. Nowadays this phenomenon is known as the quantum-mechanical Casimir effect~\cite{C48} and currently there is a vast number of publications devoted to this effect (see the few corresponding reviews
\cite{MT97,M94,KG99,M2004,CP2011}). 

Naturally, the question arises whether the phenomenon described above holds for the electromagnetic field only. In that case the interaction between the bodies is mediated by photons which are the massless excitations of the field. However, there are also other fields with massless excitations.  Accordingly, thirty years after Casimir, in 1978 Fisher and De Gennes \cite{FG78} pointed out that a very similar effect occurs in fluids\footnote{Actually, in Ref. \cite{FG78} no connection to the Casimir effect has been made. The insight that the authors actually observed a similar effect, in which the role of massless excitations is played by the critical fluctuations of the order parameter, has been made only later.} with the fluctuating field given by its order parameter. In this case the interactions in the system are mediated not by photons but by various types of massless excitations such as critical fluctuations or Goldstone bosons (i.e., spin waves). Nowadays one usually calls the corresponding Casimir
effect the critical or the thermodynamic Casimir effect \cite{K94, BDT2000}. So far the
critical Casimir effect has enjoyed only two general reviews \cite{K94,BDT2000}. 

Guided by the general idea outlined above the reader should not be surprised that currently the Casimir effect and the Casimir-like effects are studied in
quantum electrodynamics, quantum chromodynamics, cosmology, condensed matter physics, biology, and some elements of it in nano-science. 

The present review is focused on exact results concerning the thermodynamic Casimir effect, with an emphasis on the film, the sphere-plane, and the sphere-sphere  geometry.

We shall present results for various models of statistical physics. Any such model corresponds to a certain  approximation of physical reality. However, once the model is defined, the so-called exact results follow from using the methods of contemporary mathematics in order to derive expressions for quantities of interest without making any further assumptions or approximations. The above notion can be clarified by referring to the standard example of mean field models. As it is well known, for $d\le 4$ they represent only an approximation to the behavior of the actual physical systems. Once the corresponding mean field model is formulated, one can study it, however, in a way that does not involve any further simplification. In this sense the corresponding analytical result is "exact", although the model  itself is an obvious approximation. Furthermore, if numerical results of more advanced models, elucidating the considered problems, are available, we shall mention them in order to make a comparison, despite they are not "exact". Finally, let us note that we have chosen the phrase "exact results" in the same spirit as in the book of Baxter \cite{B82}:
 "exact" is not necessarily the same as "rigorous" results.

\subsection{Background}
\label{subsec:Background}

Immersing bodies of given shapes and materials into a fluid medium changes its fluctuation
spectrum, which has to be in accord with the geometry, the relative positions and
orientations, and the materials properties of the bodies. If these
fluctuations are correlated in space, the dependence of their spectrum on the  relative
positions and orientations of the bodies generates an effective force and torque, respectively, acting between them. If the
excitations of the fluctuations lack an energy gap, as it is the case, e.g., for photons,
Goldstone bosons, and the fluctuations of an order parameter at criticality, the
fluctuation induced force acquires an algebraic decay and, thus, becomes
long-ranged. Concerning the occurrence of such a force the notion of
\textit{Casimir effect} has emerged,  while the force itself is called  \textit{Casimir force} \cite{C48}. As already mentioned, it has been predicted first in
1948 by the Dutch physicist Hendrik Casimir for the zero-point
quantum fluctuations of the electromagnetic field in the space between two
perfectly conducting, parallel metallic plates\footnote{The first contribution
	by H. B. G. Casimir ($^*$July 15, 1909 -- $^\dagger$May 4, 2000) concerning this effect, which has been named  after him later on, carries the title "On the attraction between two perfectly conducting plates" and was presented as a talk at the Royal Netherlands Academy of Arts and Sciences on May 29, 1948. On that occasion Casimir showed that the boundary conditions imposed by two perfectly conducting plates onto the spectrum of the quantum-mechanical zero-point fluctuations of the electromagnetic field lead to a long-ranged, attractive force between the plates. This result was published the same year in Dutch in the Proceedings of the Koninklijke Nederlandse Akademie van Wetenschappen, Amsterdam, Vol. 51(7), 1948, p. 793-796.  At that time Casimir worked at the Philips Research Laboratories in the Netherlands and was studying the
	properties of colloidal suspensions. }.  Nowadays one broadly uses the notion 
"Casimir effect" in order to refer to the emergence of the effective  interaction between meso- and macroscopic pieces of materials, which is mediated by a fluctuating
field exhibiting massless excitations. 

The general mechanism leading to the occurrence of such a force can be understood readily. The surfaces of the bodies involved impose
boundary conditions onto the allowed spectrum of the  fluctuations. This
leads to a dependence of the ground state or of the
thermodynamic potential of the system, such as the grand canonical free energy, on the
geometry of the system (bodies plus the fluid), on the distances between its
parts and their mutual orientations. Changing, say,  these distances requires a force which depends on the induced
change of the allowed fluctuations.  Considered in this very general context the Casimir effect has naturally become a subject of studies in many diverse fields, already outlined in the Preface: condensed matter physics,
quantum electrodynamics, quantum chromodynamics, cosmology, nano-mechanics, and biophysics.

Fluctuations are ubiquitous: they unavoidably appear in any matter either due to its quantum nature or due to nonzero temperature of the material bodies and of the confined medium. In general,  any of these bodies and the medium can be even at different temperatures thus creating a set of non-equilibrium phenomena.  The bodies can also be in motion with respect to the medium or each other.

As expected, the fluctuation-induced  forces have strengths proportional to the driving energy of the fluctuations, and thus to Planck's constant $h$ in quantum systems and temperature $T$ in classical systems. In general, the QED Casimir effect, similar to the thermodynamic one, depends on both the temperature and applied external fields. 

If the fluctuating field is the electromagnetic one, one speaks of
the so-called {\it quantum mechanical Casimir effect}, while in the case of the
fluctuating field of an order parameter describing a continuous phase
transition one deals with the so-called
{\it critical Casimir effect}. In this latter form, the effect was
first discussed by Fisher and de Gennes\footnote{See also Ref. \cite{deG2003}, pp. 237-241, where the original article is reprinted and certain  additional comments are provided.} in 1978 \cite{FG78} for binary demixing of liquid mixtures and unrelated three years later by K. Symanzik \cite{S81}
in the context of the Schr\"{o}dinger representation of the
interaction of world lines. If the systems exhibit a non-zero
critical temperature, the thermal fluctuations
are dominant. The quantum effects in such systems are usually negligible. There are, however, systems in which
the critical point has a quantum origin and instead of temperature
certain quantum parameters govern the fluctuations in the system.
In that case one speaks of the {\it quantum critical Casimir effect} \cite{CDT2000,BDT2000}. There are also systems with  massless excitations of fluctuations at nonzero temperatures which are, however, thermodynamically apart from any critical point. These are, e.g., systems with a continuous symmetry, which undergo spontaneous symmetry breaking and thus  exhibit Goldstone modes within a given temperature interval, as well as certain systems with a discrete symmetry in which an interface is present due to opposing  boundary conditions at the confining  surfaces of the system. These cases can be considered as extensions of the ones discussed above, which correspond to the occurrence of a given critical point nearby. In order to encompass both versions we shall use the slightly more general notion of the {\it thermodynamic Casimir effect}. 

Being negligible at macroscopic distances, the Casimir force can, however, become quite strong at micro- and nano-scales and thus affects the design and the functioning of devices at these scales. For example, if two perfectly conducting parallel
metal plates are facing each other at a distance of the order of $10$ nm in
vacuum and at zero temperature,  the attractive Casimir  force per area,
i.e., the Casimir pressure, can be as large as one atmosphere. Obviously, such a
large force strongly influences the performance of micro- and nano-machines by
causing stiction, i.e., their moving parts stick together (de facto irreversibly) and stop working. These observations, inter alia, give rise to the  following questions:
\begin{itemize}
	\item Can one explain  and predict  this force quantitatively? 
	\item Can this force also be repulsive?
	\item What kind of changes occur if the fluctuating medium is not the vacuum with the  quantum electromagnetic field present, but a fluid?
	\item How is this effect related to the gross features of the medium and of the
	immersed bodies, as well as to their materials properties?
	\item How does this force depend on the shape of the bodies?
	\item Can one tune the force by using geometrically or chemically patterned
	surfaces?
	\item Are there potential applications of the Casimir effect?
\end{itemize}

The current review aims at presenting answers to these and further questions,
reporting theoretical achievements in terms of exact results derived from studies of specific models. 

While the quantum
mechanical Casimir effect has been covered by a series of reviews
\cite{PMG86,MT88,LM93,MT97,M94,KG99,B99,BMM2001,M2001,M2004,L2005,KM2006,GLR2008,BKMM2009,KMM2009,FPPRJLACCGKKKLLLLWWWMHLLAOCZ2010,OGS2011,KMM2011,RCJ2011,MAPPBE2012,B2012,Bo2012,Ba2012,C2012,DGT2014,RHWJLC2015,BW2007,KM2015c,SL2015,ZLP2015,WDTRRP2016,BEKK2017,WKD2021}
(see also the set of reviews contained in Ref. \cite{CP2011}), so far the
critical Casimir effect has enjoyed only two general reviews \cite{K94,BDT2000}
and a few concerning specific aspects of it \cite{K99,G2009,TD2010,GD2011,D2012,V2015,NDNS2016,MD2018}.
Since these general reviews do not cover the substantial progress made during the last decade
concerning the thermodynamic Casimir effect, here we aim at partially 
closing this gap with respect to exact theoretical results which have been obtained in this research field. 

Exact results in any research field, including the ones associated with the Casimir effect, are useful  in at least  two respects:

\textit{(i)} They render a detailed understanding of the phenomena of interest within a given model.

\textit{(ii)}  They provide benchmarks for approximate and numerical methods which facilitate to study  more sophisticated and complex models.  

Unfortunately, such results are available only for a rather limited set of models which themselves  have to be considered as serious approximations to actual materials. However, certain quantities are universal (see below); the Casimir force scaling function in critical systems belongs to them.  Their properties are identical to those which can be measured for materials belonging to the same universality class. The meaning of the notions "universality",  "universality classes", and "scaling functions" will be explained in the remainder of this review. 

\subsection{Basic examples of fluctuation-induced effective forces}
\label{sec:FIEF}

As explained above, the confinement of a fluctuating field generates effective forces on the confining surfaces which run under the general notion of fluctuation-induced forces.

Below we list some of the presently known fluctuation-induced forces, without striving for completeness. This list is supposed to indicate how broad and diverse the scientific field associated with such forces is. 

\begin{enumerate}
	
	\item It might come as no surprise, that one example of such a type of force  appears among Einstein's publications.  Voltage fluctuations in capacitor systems, due to nonzero temperature, have been considered by Einstein as early as in 1907  \cite{E07}. Similar effects are also known to occur in  
	wires\footnote{For example, the famous Johnson-Nyquist formula concerning the noise current through a resistor tells that the mean-square noise current $\langle I^2\rangle$ depends on temperature and the resistivity $R$ of a resistor according to  $\langle I^2\rangle=4k_bT\Delta f/R$, where $\Delta f$ is the measurement bandwidth.} \cite{J28,N28}.  Such fluctuations lead to forces which are contemporary of particular interest  for the operation of electromechanical devices \cite{RRJ2013}.  
	
	\item The currently most prominent example of a fluctuation-induced force is the  one which we already alluded to and which is due to quantum or thermal fluctuations of the electromagnetic field.
	 There is a vast amount of literature concerning this very active field of research. We just mention the review articles in Refs. 
	\cite{PMG86,MT88,LM93,MT97,M94,KG99,B99,BMM2001,M2001,M2004,L2005,KM2006,GLR2008,BKMM2009,KMM2009,FPPRJLACCGKKKLLLLWWWMHLLAOCZ2010,CP2011,OGS2011,KMM2011,RCJ2011,MAPPBE2012,B2012,Bo2012,Ba2012,C2012,DGT2014,RHWJLC2015,BW2007,KM2015c,SL2015,ZLP2015,WDTRRP2016,BEKK2017,WKD2021,GCMSM2021,Bimonte2022a}, recent studies of the dynamical Casimir effect (in which actual
	photons can be created if a single mechanical mirror undergoes accelerated
	motion in vacuum) \cite{M70,GK98,JJWF2009,FC2011,WJPJDND2011,NJBN2012,LPHH2013}, and studies of the effects which emerge in systems out of thermodynamic equilibrium (in which the material bodies are characterized by different temperatures) \cite{APSS2008,B2009,KEK2011,KEBK2011,MA2011,LBBAM2017,IF2021}.

	\item The fluctuations of the order parameter describing a continuous phase
	transition of a many-body system leads, as explained above, to the so-called critical Casimir effect \cite{FG78}. In the case that the critical point has a quantum origin, and instead of temperature
	certain quantum parameters govern the fluctuations in the system, one speaks of a quantum critical Casimir effect \cite{CDT2000,BDT2000}. In addition, systems like liquid $^4$He and liquid crystals, i.e., so-called correlated fluids, exhibit gapless excitations called Goldstone modes \cite{LK91,APP91,LK92,KG99}. These fluctuations lead also to long-ranged forces between the boundaries of the systems,  although such systems are thermodynamically positioned below their respective critical points.  For these cases one speaks of the noncritical or, more generally, the thermodynamic Casimir effect. We shall use the latter notion as a general one that encompasses all cases in which the Casimir effect is due to the fluctuations of a certain order parameter.

	\item Several fluctuation-induced forces are related to charge fluctuations: 
	\begin{itemize}
		\item In Ref. \cite{PS1998} it has been shown that charge fluctuations can lead to strong attraction between membranes. In the case that the counterions
			are strongly localized in the membrane planes, the attraction scales as $L^{-3}$ at large
			distances $L$ between the membranes, and as $L^{-1}$ if they are closer than a typical screening length. Recently, in Ref. \cite{AAEH2022} a prediction of enhanced attraction between drops carrying fluctuating charge distributions has been reported.
		
		\item Thermal charge fluctuations in ionic solutions can generate an attractive long-ranged dispersive force even between like-charged molecules \cite{KS52,P89}. Counterion-mediated attraction between two like-charged rods has been predicted in Refs. \cite{HL97,HP2005}. The proposed mechanism  suggests that condensed counterions introduce charge fluctuations along the axes of the rods, which give rise to attractive interactions similar to the van der Waals interaction. 
		\item Charge disorder effects between neutral dielectric slabs \cite{NDSHP2010}:
		\\
		{\it (i)}  Quenched bulk charge disorder gives rise to an additive contribution to the net force acting on the slabs,  which decays as the inverse distance between them and which may completely mask the standard Casimir–van der Waals force at large separations. 
		\\
		{\it (ii)} Annealed (bulk or surface) charge disorder leads to a net force the large-distance behavior of which agrees with the universal Casimir force between ideal conductors, which scales as the inverse cubic distance; the dielectric properties enter only the subleading contributions.
		\item Charge fluctuations in nano-circuits with capacitor components give rise to a long-ranged interaction which competes with the regular Casimir - van der Waals force \cite{DBWPW2016}.
	\end{itemize}

	\item Fluctuation induced forces between objects on a fluctuating membrane or on fluid interfaces have been theoretically predicted and some of their properties have been studied in Refs. \cite{GBP93,BDF2010,LOD2006,OD2008,BRF2011,MVS2012,NWZ2013}. In Ref. \cite{GBP93} the authors report a long-ranged interaction between inclusions, such as proteins, on a fluid membrane with the interaction being mediated by the membrane itself. One finds that the interaction falls off with the distance $R$ between the inclusions as $1/R^4$; it can be attractive or repulsive depending on the temperature and the elastic properties of the inclusion and of the membrane. Using the membrane stress tensor, in Ref. \cite{BDF2010} one has studied the fluctuations of the membrane-mediated Casimir-like force. In Ref. \cite{BRF2011} the fluctuation-induced, Casimir-like interaction between four types of pairs of parallel rods adsorbed on a fluid membrane have been calculated within Gaussian approximation. In Refs. \cite{LOD2006,OD2008} the force between two colloids positioned at interfaces has been studied. This Casimir type interaction depends on the boundary conditions imposed at the three-phase contact line. In Ref. \cite{MVS2012} the authors state that experiments \cite{BHSHHBW2007,VCSHHB2008} suggest that membranes of living cells are tuned close to a miscibility critical point in the two-dimensional Ising universality class. This leads to the proposition that membrane bound proteins experience weak yet long-ranged forces mediated by critical composition fluctuations in the plasma membranes of living cells. In Ref. \cite{NWZ2013} a scattering formalism has been developed in order to calculate the interaction between colloidal particles trapped at a fluid interface.
	
	\item Fluctuations of fluid velocities \cite{DW2006} as well as fluctuations in electric fields may both give rise to forces acting on the solute particles in colloidal suspensions. 
	
	\item Pseudo-Casimir\footnote{Here we used the notion "pseudo" in order to indicate that the degrees of freedom to enter and to leave the region between the interacting objects are missing.} stresses and elasticity of a confined elastomer film. In Ref. \cite{LNP2016} the authors study the impact, which thermal fluctuations of the cross-links in elastomers have
	on the free energy of the elastic deformations of the system, subject to the requirement that the fluctuating
	elastomer cannot detach from two large, hard, and co-planar substrates. They find an  attractive fluctuation-induced  pseudo-Casimir stress with a power law
	decay away from the substrates, if a rigid pinning boundary condition is applied.
	
\item In condensed matter systems one studies the phonon Casimir effect, which describes the phonon-mediated interaction between defects \cite{Ro2019,LR2020}. For pairs of impurities, this interaction has been shown \cite{Ro2019} to
follow a quasi-power-law\footnote{i.e.,  fitted best by a power law} at zero-temperature and evolve into an exponentially decaying form as the temperature is increased.
	
\item Non-equilibrium thermodynamic (hydrodynamic) Casimir-like effect.
	
	In a  fluid out of equilibrium, correlations are generally of longer range than the ones in equilibrium, including  even those near an equilibrium critical point.

	\begin{itemize}
	\item Correlations in fluids in nonequilibrium steady states are long ranged.  A giant Casimir-like effect for fluids in nonequilibrium steady states has been  reported in Ref. \cite{KOS2013}; fluctuation-induced pressures in fluids have been  investigated in Ref. \cite{KOS2014}; a binary liquid mixture, confined between two thermally conducting walls and subjected to a stationary temperature gradient, is studied in Refs. \cite{KOS2015,KOS2016}. From that it follows that fluctuations in a binary liquid mixture out of equilibrium induce a Casimir-like force on the walls. In Ref. \cite{KOS2016b} the physical origin of nonequilibrium fluctuation-induced forces in fluids is elucidated. In Ref.  \cite{KDS2016} the notion of Casimir work is introduced and an alternative way to measure the nonequilibrium Casimir force related to heat flux is proposed, which differs from the approach in equilibrium where a volume derivative of the free energy is taken. Further examples of fluctuation-induced forces in nonequilibrium systems include the case of a driven diffusive system consisting of a slab of thickness L \cite{AKK2015}. For this it is shown that the force between the parallel surfaces decays as $k_BT/L$ (in all spatial dimensions). It can be attractive or repulsive, exhibiting  a nonuniversal amplitude which  explicitly depends on the details of the dynamics.
		
	\item Systems out of equilibrium exhibit long-ranged correlations caused by the  conservation of specific global  quantities like the number of particles or momentum. This, in turn, leads to long-ranged, fluctuation-induced Casimir-like forces.  Such forces, for classical systems after a temperature quench, are studied in Refs. \cite{RKK2016,RSKK2017}.
	\end{itemize}

\item Fluctuation-induced Casimir forces in granular fluids have been reported in Ref. \cite{CBMNS2006}. The authors studied  driven, noncohesive granular media and found that two fixed, large intruder particles immersed in a sea of small particles experience, in addition to a short-ranged depletion force, a long-ranged, fluctuation-induced repulsive force because the hydrodynamic fluctuations are geometrically confined in the space between the intruders.
	
\item The fluctuations of the director in nematic liquid crystals have been investigated early on in Refs. \cite{APP91,ADHPP92,LKS93}. In this system the fluctuations of the
	nematic director are responsible for the
	long-ranged nature of the corresponding Casimir force. Close to the phase transition to the
	isotropic phase (which ultimately is of first order so that no bona fide critical fluctuations can emerge), fluctuations of the degree of nematic order and the degree
	of biaxiality generate short-ranged corrections to the leading Casimir force. Further information about this aspect is given in Refs.  \cite{ZZ96,ZPZ99} and references therein. Concerning the influence of patterned substrates on the liquid crystalline Casimir effect see Refs. \cite{HSD2004,HSD2005,HD2006}.  
	
\item Recently, in Ref. \cite{DPKXSF2022} the experimental observation has been made that there are long-ranged attractive, fluctuation-induced forces in laboratory hydrodynamic turbulent flow between two walls which locally confine the hydrodynamic fluctuations. We mention here also studies of the Casimir effect in active matter  systems \cite{RRR2014,RSKK2017,Kjeldbjerg2021,Tayar2022,BACV2022,Fava2022}. In Ref. \cite{RRR2014} the authors study numerically run-and-tumble active matter particles which move in a two-dimensional slit formed by two finite, 
	parallel, one-dimensional walls. They observe an attractive force between the walls of a magnitude that increases upon
	increasing the run length. One finds that the attraction is robust as long as the wall length
	is comparable to or smaller than the run length of the swimmer. In Ref. \cite{RSKK2017}   transient Casimir forces are studied which arise from the presence of parallel plates or compact
	inclusions in a gas of active particles, following a change (“quench”) of temperature or activity of the medium. In Ref. \cite{RSKK2017} it has been observed in simulations that two parallel plates
	experience a ‘‘Casimir effect’’ and attract each other when placed in a dilute bath of active Brownian particles \cite{Kjeldbjerg2021}. The liquid-liquid "phase transition" considered in Ref. \cite{Tayar2022} resembles
	the one in a water-lutidine mixture which is commonly used to carry out experiments concerned with critical Casimir forces in thermal equilibrium as well as out of equilibrium, such as heated Janus
	particles suspended in lutidine-water mixtures. In Ref. \cite{BACV2022}
	the authors report on experiments with toy robots, the so-called "Hexbugs", considered as active particles. They demonstrate the emergence of Casimir-like activity-induced attraction between planar objects in the presence of active particles in the environment. A two-dimensional system  of aligning, self-propelled particles  is considered in the bulk flocking phase in Ref.  \cite{Fava2022}. The particles are transversally confined by reflecting or partly repelling walls. The authors show that the finite-size contribution leads to the emergence of a Casimir-like pressure on the walls, which decays slowly and algebraically upon increasing the distance between the walls, with a certain degree of universality.

\end{enumerate}
The present review covers the available \textit{exact} results pertinent to  item 3, i.e., the thermodynamic Casimir effect. Since this effect depends on the geometry of the system, there is a plethora of phenomena. In order to keep the volume of the review within reasonable limits, we focus on the  film, the sphere-wall, and the sphere-sphere geometry. 



\newpage
\section{Thermodynamic Casimir effect}
\label{sec:TCE}

We start by considering the behavior of a medium in the film geometry $L\times \infty^2$, which can be described in terms of standard statistical mechanics and thermodynamics, and which exhibits massless excitations. As already explained in Sec. \ref{subsec:Background}, in such a system the pressure acting on its bounding surfaces in general differs from the pressure in the same system, characterized by the same thermodynamic parameters, but being macroscopically large, i.e., $L\to\infty$. If the medium is fluid, in distinct scientific communities and for various  ranges of parameters one uses the notions of disjoining pressure, solvation force, critical Casimir force, or the thermodynamic Casimir force \cite{E90book, K94,  BDT2000,KD91,KD91b,KD92a,KD92b}.    

It is important to note that the critical Casimir force manifests itself only for fluid media, which allow the ordering degrees of freedom to leave or enter the confined space - in contrast to, say, magnetostriction where the spins are fixed at lattice sites. This offers the possibility to realize the critical Casimir force within the grand canonical ensemble. 

If the system is at temperature $T$ and is exposed to an external ordering field $h$, which couples linearly to its order parameter - such as the number density, the concentration difference, the magnetization etc, the thermodynamic Casimir force $F_{\rm Cas}(T,h,L)$ per area is the excess pressure over the bulk one due to the finite size $(L<\infty)$ of that system:
\begin{equation} \label{eq:Casimir}
F_{\rm Cas}(T,h,L)= P_L(T,h)-P_b(T,h).
\end{equation} 
$P_L$ is the pressure in the finite system, while $P_b$ is the one in the infinite, i.e., macroscopically large, system. The above definition is actually equivalent to another one, which is also commonly used \cite{E90book,K94,BDT2000}:
\begin{equation} 
\label{eq:grand_can}
F_{\rm Cas}(T,h,L)=-\frac{\partial\omega_{\rm ex}(T,h,L)}{\partial L},
\end{equation}
where $\omega_{\rm ex}=\omega_L-L\,\omega_b$ is the excess grand potential per area, $\omega_L$ is the grand canonical potential per area of the finite system,  and $\omega_b$ is the grand potential per volume $V$ for the macroscopically large system. The equivalence between the definitions in Eqs. \reff{eq:Casimir} and \reff{eq:grand_can} stems from the observation that for the finite system with surface area $A$ and thickness $L$ (so that $V=AL$) one has $\omega_L=\lim_{A\to \infty}\Omega_L/A$ with  $P_L=-\partial\omega_L/\partial{L}$  and $\omega_b=-P_b$, where $\Omega_L$ and $\Omega_b$ are the grand canonical free energy of the finite and macroscopic systems, respectively. 

One can view the above discussion also from a somewhat more general perspective. To this end we consider a medium which is characterized by its bulk grand potential $\Omega_b$ with $\Omega_{\rm ex}^{\mathcal{A}\mathcal{,B}}(L)\equiv \Omega^{\mathcal{A},\mathcal{B}}(L)-\Omega_b$ as the change in this potential when two macroscopic material bodies $\mathcal{A}$ and $\mathcal{B}$ are immersed in it such that $L$ is the distance of closest approach between them. One can consider a generalized force conjugate to the distance $L$ acting between these bodies which is defined as 
\begin{equation}
\label{eq:grand_can_A_B}
F_{\rm Cas}^{\mathcal{A,B}}(L)=-\frac{\partial\Omega_{\rm ex}^{\mathcal{A,B}}(L)}{\partial L}.
\end{equation}
We note that while $F_{\rm Cas}$ in \eq{eq:grand_can} is the force per area, in  \eq{eq:grand_can_A_B}, however, $F_{\rm Cas}^{\mathcal{A,B}}(L)$ is the total force between the bodies $\mathcal{A}$ and $\mathcal{B}$. Obviously, taking this different normalization into account, in the case of plates immersed in a fluid one will obtain the same result for the force from both equations.

We recall that if at least one of the spatial extensions of the system is finite, one denotes the corresponding system to be a finite one. If a finite system is thermodynamically located close to its corresponding bulk critical point  $(T=T_c, h=0)$, its correlation length $\xi$ becomes comparable to  $L$. As a consequence, the thermodynamic functions describing its behavior depend  on $L$ and $\xi$ via the dimensionless ratio $L/\xi$ and attain a scaling form as provided by the finite-size scaling theory \cite{Ba83,C88,Ped90,PE90,BDT2000}. 

In a system, which is finite in the above sense and undergoes a phase transition, its phase behavior can differ significantly from its bulk counterpart  \cite{FB72,Ba83,C88,Bb83,D86,Di88,P90,E90book,BDT2000}. One observes effects like shifts of the critical points, both with respect to $T$ and $h$. Or transitions of their own appear at some new critical points $(T_{c,L}, h_L)$, such as capillary condensation, provided the dimensionality of the finite system is high enough. Near the bulk critical point, the behavior of the bulk system is characterized by universal critical exponents and scaling functions, which depend only on gross features of the system such as its spatial dimension  $d$, the $O(n)$ symmetry of the ordered state, and the range of the interaction involved, which together form the so-called bulk universality class. In addition, the behavior of a finite system depends on the so-called surface universality class, which is determined  by the boundary conditions on the surfaces of the finite system, as well as on its geometry. In a system with film geometry, if the finite system exhibits a phase transition of its own, it belongs to the universality class of the $(d-1)$-dimensional infinite system. One of the quantities of particular interest concerning a finite critical system  is the thermodynamic Casimir force and, more specifically, the one which is observed near the critical point of the bulk system, often called {\it critical Casimir force}.

\subsection{Critical Casimir effect}

The critical Casimir effect has been observed directly, via light scattering from a spherical colloid interacting  with a planar substrate \cite{HHGDB2008}, both of which are immersed in a critical binary liquid mixture. Very recently the nonadditivity of critical Casimir forces (CCF) has been experimentally demonstrated \cite{PCTBDGV2016}. Indirectly, as  a balancing force which determines the thickness of a complete wetting film in the vicinity of its bulk critical point, the Casimir force has been also studied in $^4$He \cite{GC99,GSGC2006}, as well as in $^3$He--$^4$He mixtures \cite{GC2002}. In  Refs. \cite{FYP2005,RBM2007} measurements of the Casimir force acting on thin wetting films of a binary liquid mixture have been also performed. The dynamics of colloidal aggregation in microgravity induced by CCF has been considered in Ref. \cite{PMVWMSW2013}.  The synchronization of the motion of two colloid beads has been studied by using optical tweezers in a binary liquid mixture close to the critical point of its demixing transition \cite{MDPC2017}. The control of colloidal phase transitions via  CCF  was investigated in Ref. \cite{NFHVS2013}. In Ref. \cite{Schmidt2022} the authors demonstrate experimentally that repulsive critical Casimir forces can counteract the Casimir-Lifshitz attraction.  Further  experiments related to CCF are reported in Refs.  \cite{BeE1985,SZHHB2007,NHC2009,BCPP2010,TZGVHBD2011,ZAB2011,HKTAGB2021}. The CCF in soft matter systems attract considerable attention because they can be
precisely and fully reversibly tuned by small changes of  
temperature. 

The theoretical studies in this research field have triggered considerable attention on the experimental side and vice verse.  Reviews on the corresponding theoretical results can be found in Refs. \cite{K99,G2009,TD2010,GD2011,D2012,V2015}. 

Up to now, the critical Casimir effect has been studied theoretically for the following models: 
\begin{itemize}
	\item one-dimensional Ising model \cite{RZSA2010}
	\item one-dimensional XY model \cite{F64,DaR2017}
	\item one-dimensional Heisenberg model \cite{DaR2017}
	\item two-dimensional Ising model  \cite{ES94,MCD99,INW86,DME2000,DMC2000,BDT2000,DK2004,MDB2004,AEM2007,NN2008,NN2009,AM2010,RZSA2010,Iz2011,HGS2011,DrzMaBa2011,Z2012,WIG2012,DM2013,ZMD2013,VED2013,HH2014,WI2015,HH2015,JRT2015,NMD2016,NN2016,H2016,H2017,HH2017,MP2018,MP2019,MZR2019,BE2020,SME2020,DW2021,DWKS2021}
	
	\item three-dimensional Ising model \cite{BU98,DK2004,FYP2005,VGMD2007,BU2008,VGMD2009,GMHNHBD2009,TD2010,MMD2010,H10b,H2011,VMD2011,H2012,OO2012,H2013,UB2013,TTD2013,VD2013,V2014,H2015a,H2015b,MVDD2015,V2015,B2015,THD2015,T2015,B2016,V2016,PLCH2016,SRP2022}
	
	\item three-dimensional XY model \cite{DK2004,DKD2005,VGMD2007,VGMD2009,V2015,BDR2011,DaR2017}
	
	\item three-dimensional Heisenberg model
	\cite{DK2004}
	
	\item three- and $d$-dimensional $O(n\to\infty)$ (spherical) model \cite{D96,D98,BDT2000,CDT2000,DDG2006,DK2004,CD2004,ChD2004,DG2009,DGHHRS2012,DR2014,DBR2014,DGHHRS2014,D2020,DWKS2021}
	
	\item Bose gas \cite{Da72,FB72,MZ2006,GD2006,Bi2007a,Bi2007b,BBMSC2010,BBSBH2010,NJN2013,BBA2014,MRC2016,TT2017,DiRu2017,DVZ2018,RPR2019,FB2018,RRP2019a,D2020,LJ2020,Thu2020,TTH2020,TS2020,NPT2020,SoVa2020,LJ2020b,Song2020,A2020,Bhu2021,NK2021,Song2022,NP2022,Pruszczyk2022}

	\item $d$-dimensional Gaussian model \cite{KD92a,BDT2000,DK2004,KD2010,DiRu2017,DaR2017,Gross2021,GGD2021}
	
	\item {mean field and effectively mean-field models:}  \cite{PE92,K97,GaD2006,SSD2006,DSD2007,CM2010,BDR2011,VMD2011,Bier2011,PCM2012,VDK2012,DV2012,LTHD2014,Pousaneh2014,VD2015,LD2016,DVD2016,GVGD2016,DaR2017,DDV2017,VD2017,BMD2017,DVD2018,DVD2020b,Dan2021,Zhang2021,AS2021,CSS2021}
	
	\item $\Psi$ model for $^4$He \cite{DRVD2019}
	
	\item general $O(n\ge 1)$ models \cite{KD91,KD91b,KD92a,KD92b,BDT2000,D2009,Do2011}.
	
\end{itemize}

The theoretical techniques, which have been used, can be summarized as follows: 
\begin{itemize}
\item 
studies based on exact calculations 
 \begin{itemize}
 \item 
 for the two-dimensional Ising model \cite{INW86,ES94,MDB2004,AEM2007,NN2008,NN2009,AM2010,RZSA2010,Iz2011,HGS2011,WIG2012,ZMD2013,DM2013,NMD2016,NN2016,H2016,H2017,HH2017,BPK2017};

\item for the $d$-dimensional Gaussian model \cite{KD92a,DK2004,G2008,DaR2017,DiRu2017};
 
 \item for the three- and $d$-dimensional $O(n\to\infty)$ (spherical) model
 \cite{D96,D98,BDT2000,CDT2000,DDG2006,DK2004,CD2004,DG2009,DGHHRS2012,DR2014,DBR2014,DGHHRS2014,RD2015,DR2017b,D2020,DWKS2021};
 
\item via conformal field-theoretical methods \cite{A86,BCN86,BE95,ER95,HSED98,BE2020,VED2013,BEK2013,BEK2015,DSE2015,JRT2015,EB2016,Ra2016,DSE2016,SME2020,DW2021};

\item for the	$\Psi$ model of $^4$He \cite{DRVD2019};
 
\item  within mean field type calculations 
\begin{itemize}
	\item for Ising-like models  \cite{K97,PE92,GaD2006,DSD2007,ZSRKC2007,VMD2011,MKMD2014,DVD2016,GVGD2016,DaR2017,DDV2017,VD2017,DVD2018,VDD2019b,DVD2020b};
	
\item for $XY$-like models \cite{BDR2011,DaR2017};
\end{itemize}

\item Bose gas models 
\begin{itemize}
	\item perfect Bose gas \cite{MZ2006,GD2006,DiRu2017};
	\item imperfect Bose gas \cite{NP2011,DiRu2017,NJN2013,JN2013};
	\item relativistic Bose gas \cite{D2020};
\end{itemize}
\end{itemize}

\item exact variational approach due to Mikheev and Fisher applied to the Ising model \cite{BU98,BU2008,B2015,B2016,Z2012,UB2013}

\item studies using  numerical density-matrix renormalization-group techniques \cite{MCD99,DMC2000,DME2000,MDB2004,DrzMaBa2011,ZMD2013}

\item 
 studies based on renormalization-group theory using the  $\varepsilon$-expansion technique  \cite{KD92a,KD92b,DKD2003,DGS2006,GD2008,SD2008,D2009,DS2011,BDS2011a,D2013}
 
 \item studies using fixed dimension $d$ techniques \cite{D2009,Do2011,D2013,D2014,D2018} of $O(n)$-symmetric  models
 
 \item  Monte-Carlo simulations  \cite{KL96,DK2004,DKD2005,H2007,VGMD2007,VGMD2009,H2009,H2010,H10b,TD2010,TD2010rev,H2011,HGS2011,VMD2011,H2012,H2013,TTD2013,VED2013,VD2013,ETBERD2014,V2014,HH2014,T2015,MVDD2015,TTD2015,V2016,HH2015,V2015,H2015a,H2015b,GVGD2016,DWKS2021}
 
 \item studies partially using mean-field approximation\footnote{In these articles the set-up of the investigation is in line with the general renormalization group argumentation. However, the numerical part of these studies is carried out based on a certain mean-field approximation. In some other studies the mean-field type theory is combined with the Debye–H\"{u}ckel theory, like in the case of electrolytes  (Ref. \cite{Bier2011} and the references therein).} \cite{MMD2010,CM2010,Bier2011,PCM2012,MKMD2014,Pousaneh2014,THD2015,VD2015,DSD2007,VD2017,DVD2016,DVD2018}.
 
 \item renormalized local functional theory \cite{OO2012}
 
 \item vortex-loop renormalization group techniques \cite{W2004}
 
 \item fluid particle dynamics method \cite{FGDT2013} 
 
 \item studies based on density functional theory \cite{MZR2019,MR2020}.

\end{itemize}

\subsection{Casimir-like effects off criticality}
\label{sec:Cas-offcrit}

This subsection provides a summary of the investigations concerning Casimir-like effects off criticality:

\begin{itemize}
\item two-dimensional Ising model \cite{NN2008,NN2009,AM2010,DM2013}
\item fluctuations confined by membranes \cite{YRD2011,NMD2016,VCSHHB2008,MVS2012,GBP93,BDF2010,LZMP2011,SISS2009,AEM2007}
\item liquid crystals \cite{LK92,ZPZ98,APP91,BLF2000,U2001,DzK2004,GAF2001,LKS93,HNSP2014,RBS2011,Ha2016,KNSP2013}
\item $O(n \ge 2)$ vector models below $T_c$ \cite{D96,D98,D2013,D2014} 
\item three-dimensional Ising model \cite{UB2013,BU2008}

\item non-equilibrium fluctuation-induced forces in fluids \cite{KOS2016b,KOS2013,KOS2014,KOS2015,OKS2015,CZS2016,KDS2016,RKK2016}.
\end{itemize}

\subsection{Quantum critical Casimir effect} 

In all of the above mentioned models the important fluctuations are of thermal origin because all these models possess  non-zero critical temperatures. In some systems, however, certain quantum parameters govern the
fluctuations near their critical point, which is usually close to or at zero temperature \cite{SK2011,S2008,S2000,Sa2011}.  As explained in Section \ref{subsec:Background}, in this particular case one speaks of a \textit{quantum critical} Casimir (QCC) effect. Results for this case have been presented in Refs. \cite{CDT2000,BDT2000,PCC2009,RHRLDHR2016,GC2018}. Exact results for the QCC effect are reviewed in Sec. \ref{QuantumCritCasEffect}.

\section{Theoretical background: finite-size scaling and critical Casimir effect}
\label{FSS_theory}

As discussed in Chapter \ref{sec:TCE}, we consider two materials bodies $\mathcal{A}$ and $\mathcal{B}$ immersed in a fluid. They exert an effective force $F^{\rm \mathcal{A,B}}$ onto each other which is mediated by the fluid. This includes, inter alia, the direct interaction between the material bodies $\mathcal{A}$ and $\mathcal{B}$. If the thermodynamic state of the fluid is far away from a bulk phase transition at $T_c$, this force varies slowly and smoothly as function of temperature. Upon approaching $T_c$ of a continuous phase transition, $F^{\rm \mathcal{A,B}}$  acquires in addition a contribution  $F_\Cas^{\rm\mathcal{ A,B}}$ due to the critical fluctuations of the confined fluid. This additive, singular contribution encompasses both the distortion of the local, nonzero order parameter (if this is the case), due to the finite distance between $\mathcal{A}$ and $\mathcal{B}$, and the fluctuations of the vanishing mean order parameter. The singular contribution  $F_\Cas^{\rm \mathcal{A,B}}$ follows by subtracting the aforementioned smooth background contribution (after extrapolating it to the neighborhood of $T_c$) from $F^{\rm \mathcal{A,B}}$. This corresponds to the standard procedure for obtaining the singular behavior of thermodynamic quantities such as, e.g., the specific heat (see Ref.  \cite{KD92a}). Upon construction, for the disordered phase $F^{\rm\mathcal{ A,B}}$ and  $F_\Cas^{\rm \mathcal{A,B}}$ vanish in the limit of increasing separation between $\mathcal{A}$ and $\mathcal{B}$. 

It will turn out that the effective force $F_\Cas^{\rm \mathcal{A,B}}$ between $\mathcal{A}$ and $\mathcal{B}$ can be attractive or repulsive. As expected on general grounds and as already partially outlined                                                                   above, the critical Casimir effect depends on the parameters describing the thermodynamic state of the critical medium, say the temperature and an externally applied field (e.g., pressure, excess chemical potential, magnetic field), as well as on the distance $L$ between $\mathcal{A}$ and $\mathcal{B}$ [see Eqs. \reff{eq:Casimir} and \reff{eq:grand_can}], i.e., the observed phenomenon is a finite size effect: if $L$ increases the effect and therefore the magnitude of the associated force decreases and eventually vanishes. 

Any thermodynamic system, which is of finite extent at least in one spatial direction, is called a finite-size system.  The corresponding  modification of its phase behavior, compared with its bulk one, is described by finite-size scaling theory \cite{Ba83,P90,BDT2000}. Because of this profound interconnection between the theory of the thermodynamic Casimir effect and finite-size scaling theory, in the present chapter we shall summarize some basic knowledge concerning  finite-size scaling theory which  will be relevant for studying the thermodynamic Casimir effect. We start by recalling some basic properties of critical phenomena in bulk systems as far as needed for finite-size scaling.

\subsection{Scaling and universality}
\label{sec:scaling} 

According to the universality hypothesis, as formulated by Kadanoff (see page 103 in Ref. \cite{K71}), ``all\footnote{continuous} phase transition 
 problems can be divided into a small number of different classes depending upon the dimensionality of the system and the
symmetries of the ordered state. Within each class, all phase transitions have identical behavior in the critical region, only the names of thermodynamic variables are changed." In other words, if two model systems, embedded in a space with the same dimensionality $d$, undergo a continuous phase transition which corresponds to the same breaking of a common symmetry of the disordered state (shorthand labeled as $n$),  the scaling functions of the thermodynamic potentials in terms of suitably normalized pertinent thermodynamic variables will be the same upon approaching the phase transition, together with the associated critical exponents\footnote{In order to be more accurate, we recall that for certain systems, with the interactions decaying sufficiently slowly,  an additional exponent $\sigma$, which quantifies how fast the interactions decay as function of distance, is needed in order to precisely specify the corresponding universality class. Since in the current review we do not discuss in detail such systems,  we do not  elaborate on that aspect further. However, we shall present the corresponding specific details upon discussing the behavior of the Casimir force in such systems. }.
All such systems are part of the same universality class. This general property of the critical behavior, which has been extensively checked and verified (see Refs.~\cite{B82,F83,A84,Do96,C96,BDT2000,PV2002,ZJ2002,K2007} and references therein), is a very powerful instrument in constructing model systems which describe the behavior of the corresponding actual physical system. In fact, in order to determine the critical behavior of any actual system, one can study the simplest possible model,  which shares the above-mentioned gross features with that system. 
Based on these theoretical grounds, for instance, carbon dioxide, xenon,
and the three-dimensional Ising model are expected to have the same critical exponents. This has been checked experimentally  \cite{HM76}. These properties hold for the asymptotically leading critical behavior\footnote{i.e., for correlation lengths much larger than any specific microscopic length.}. The strength of the subdominant terms, and thus the size of the critical region, is nonuniversal.

However, there are exceptional systems for which the hypothesis of universality as  formulated above does not 
hold. Perhaps the most notable example of that is the so-called eight-vertex model, the critical exponents of which vary continuously with the parameters entering the partition function \cite{B82}. This feature can be understood  within the renormalization-group framework in terms of the presence of a so-called marginal operator in the Hamiltonian of the system, which is responsible for the emergence of such "violations" of universality (for details see Refs. \cite{K71,KW71,B82,F98}).

The universality hypothesis is commonly stated together with the hypothesis of
scaling for the thermodynamic functions close to the critical
point. We emphasize that these are, in fact, two independent assumptions which might not hold simultaneously,  as in the case of the eight-vertex model for which universality fails in its stricter sense, while scaling is still observed \cite{B82}.
Historically, the scaling forms of the free energy \cite{W65a} and of the correlation function \cite{Ka66} were formulated as a hypothesis before the renormalization group was invented.
We briefly recall the essence of the scaling hypothesis  mentioned above. 

In 1965 Widom \cite{W65a,W65b} and Domb and Hunter \cite{DH65} suggested that around the bulk critical temperature $T_c$ the singular parts of the thermodynamic functions of a macroscopically large system, which undergoes a second-order phase transition, are homogeneous functions of their variables. This is the essence of the so-called  scaling hypothesis. For example, Griffiths  suggested \cite{G67} that the equation of state $h=h(m,t)$ for a
ferromagnet  takes on the scaling form
\begin{equation}
h=m|m|^{\delta -1}h_s(\tau|m|^{-1/\beta }),  \label{Griffiths}
\end{equation}
i.e., it is a generalized homogeneous function\footnote{\label{footnote_on_scaling}A function $f(x_1,x_2,\cdots,x_n)$ 
is called to be 
a generalized homogeneous function with respect to its variables $x_1, x_2,\cdots, x_n$ if, for any $\lambda> 0$, there exist numbers (exponents) $a_1,a_2,\cdots, a_n,a_f$ such that 
$f(\lambda^{a_1} x_1,\lambda^{a_2} x_2,\cdots,\lambda^{a_n} x_n)=\lambda^{a_f} f(x_1,x_2,\cdots,x_n)$,  
where the exponent $a_i$ describes the degree of homogeneity with respect to the $i$-th argument. 
By replacing,  in the above equation $\lambda$ with $\lambda^{1/a_f}$, one finds that each homogeneous function 
also satisfies the equation  $f(\lambda^{a_1/a_f} x_1,\lambda^{a_2/a_f} x_2,\cdots,\lambda^{a_n/a_f} x_n)=\lambda f(x_1,x_2,\cdots,x_n)$. Hence, the number of independent exponents characterizing a generalized 
homogeneous function of $n$ variables is
$n$. In the context of
critical phenomena one usually 
specializes 
the homogeneity property with, say,  $\lambda=|x_1|^{-1/a_1}$ so that $f(x_1,x_2,\cdots,x_n)=|x_1|^{a_f/a_1}f({\rm sign} (x_1), x_ 2 |x_1|^{-a_2/a_1},\cdots,x_n |x_1|^{-a_n/a_1})$. This means that homogeneity reduces the function of $n$ variables to one of $n-1$ variables.} of $\tau=(T-T_c)/T_c$ and
$m$, where $m$ is the order parameter of the system (i.e., the magnetization) conjugate to the external magnetic field $h$; $\delta$ and $\beta$ are the corresponding critical exponents, while $h_s(x)$ is the scaling function. According to the universality hypothesis $\delta$, $\beta$, and the dimensionless
scaling function $h_s(x)$, after a suitable normalization of its argument, are the same for all
systems within a certain universality class.

The bulk free energy density $f_b(T,h)$ splits into a singular part
$f^{\rm (s)}_b$ and into a nonsingular, i.e., analytic one  $f^{\rm (ns)}_b$: 
\begin{equation} \label{singreg}
f_b(T,h)=f^{\rm (s)}_b(T,h)+f^{\rm (ns)}_b(T,h).
\end{equation}
The singular part is responsible for the critical behavior and encompasses those terms which have a non-analytic dependence on the (reduced) thermodynamic variables.
The scaling
hypothesis concerns the singular part of the free energy density $f^{\rm (s)}_b(T,h)$. As confirmed by renormalization group theory (see, e.g., Refs. \cite{WK74,Fi1974,F83,F98,A84,P90,C96,BDT2000,ZJ2002,PV2002}), in the case of a simple isotropic\footnote{The corresponding statements can be also extended  to the case of anisotropic systems, but they become more complicated. For the case of weakly anisotropic systems, the reader can consult, e.g., Refs. \cite{DW2021,DWKS2021} and the literature cited therein. Here we refrain from discussing such systems. Some additional information about them is given in Sect. \ref{more_on_Casimir}. } system (fluid, or ferromagnet) the asymptotic
behavior of $f^{\rm (s)}_b(T,h)$ in the limit $(T,h) \rightarrow (T_c,0)$ has the following scaling form:
\begin{equation}
\beta f^{\rm (s)}_b(T,h)\simeq A_1|\tau|^{2-\alpha}W^{\pm}(A_2h|\tau|^{-\Delta}) ,
\label{BFEScaling}
\end{equation}
where $\pm$ refers to $\tau>0$ and $\tau<0$, respectively, and $\beta=1/(k_B T)$ (distinct from the critical exponent $\beta$). The scaling functions
$W^{\pm}$ and the exponents $\alpha $ and $\Delta =\beta +\gamma $ are the
same for all members of a given universality class. The latter implies that, within a given universality
class, lattice structures, coupling constants,
etc., may vary, but all these variations are subsumed under the values of only two
nonuniversal metric factors $A_1$ and $A_2$. Equation (\ref{BFEScaling}) is consistent with the tacit assumption that in the system under study there are only  {\it two relevant} scaling fields - the ones related to $\tau$ and $h$. In the general case, according to renormalization group (RG) theory \cite{W76,PV2002}, one can write $f^{\rm (s)}_b$ in
terms of the nonlinear scaling fields $\{u_i\}$ associated with the RG eigenoperators at the fixed point of the theory\footnote{Compare this with the corresponding hypothesis that any thermodynamic function near the critical point of the system is a generalized homogeneous function of its parameters (see footnote \ref{footnote_on_scaling}). Equation (\ref{generlaRG}) represents  the generic scaling form. For completeness, we mention that in certain specific cases logarithmic terms can appear which may be due to a degeneracy of the RG critical exponents, due to the presence of marginal operators,
etc. (see Refs.~\cite{W72,W76}).}:
\begin{equation}
\label{generlaRG}
f^{\rm (s)}_b(u_1,u_2,\cdots,u_n, \cdots)=b^{-d}f^{\rm (s)}_b(b^{y_1}u_1,b^{y_2}u_2,\cdots,b^{y_n}u_n, \cdots),
\end{equation}
where $b>0$ is any positive number (which sets the spatial scale of the coarse-graining),  and $y_n \in \mathbb{R}$  are the RG dimensions of the scaling fields which are {\it analytic} functions of $\tau$, $h$, and of any other parameter appearing in the Hamiltonian. In \eq{generlaRG}, $d$ indicates the spatial dimension of the system. The scaling fields characterized by $y_i>0$ are called {\it relevant}, those with $y_i<0$ {\it irrelevant}, while {\it marginal} ones have  $y_i=0$. The number of relevant fields is finite. In the systems mentioned above there are two relevant fields $u_1$ and $u_2$, with $y_1>0$ and $y_2>0$ related to $\tau$ and $h$, respectively, and an infinite set of irrelevant fields (indicated by the ellipses in \eq{generlaRG}). Since the scaling fields $u_i$ are analytic functions of $\tau$ and $h$, one has 
\begin{equation}
\label{defscalingfields}
u_1=a_\tau \tau +o(\tau,h)\qquad \mbox{and} \qquad u_2=a_h h +o(\tau,h).
\end{equation}
 If in \eq{generlaRG} one chooses $b$ such that $b^{y_1}|u_1|=1$, one obtains 
\begin{eqnarray}
\label{generlaRGfinal}
\lefteqn{f^{\rm (s)}_b(u_1,u_2,\cdots,u_n, \cdots)=} \\
&& |u_1|^{d/y_1}f^{\rm (s)}_b({\rm sign}(u_1), |u_1|^{-y_2/y_1} u_2,\cdots, |u_1|^{-y_n/y_1}u_n, \cdots). \nonumber
\end{eqnarray}
In the case of two relevant fields, one has $y_i<0$ for $i>2$ so that $|u_1|^{-y_n/y_1}u_i\to 0$ for $|\tau| \to 0$. If the  corresponding limit of $f^{\rm (s)}_b$ exists, and after making the identifications $y_1=1/\nu$ and $y_2=\Delta/\nu$, one arrives at \eq{BFEScaling}, provided that the hyperscaling relation\footnote{This is the case only below the upper critical dimension $d_u$ of the system \cite{PV2002,P90}; for further details see the end of the present section.} 
\begin{equation}
2-\alpha =d\nu 
\label{hypsc}
\end{equation} 
holds. 

In the remainder, if not stated otherwise, we shall assume that the systems we discuss possess only two  relevant bulk scaling fields. For such systems another basic hypothesis --- again supported by renormalization group theory and by available exact results --- concerns the scaling form of the connected two-point correlation function:
\begin{equation}\label{eq:corrfunction}
G({\bf r},{\bf 0};T,H) \equiv \langle S({\bf r}) S({\bf 0})\rangle -
\langle S({\bf r})\rangle \langle S({\bf 0})\rangle,
\end{equation}
where $S({\bf r})$ is the local field of the ordering degrees of freedom (e.g., spin density)  embedded at point ${\bf r}$ of the system.  The mean value $\la \cdots \ra$ follows from the Hamiltonian ${\cal H}$ which defines the model. The corresponding scaling hypothesis states that asymptotically\footnote{Here and in the remainder the symbol $\simeq$ means that the corresponding quantities are equal up to the order of terms kept on the right-hand side of the expression.}
\begin{equation}
G(\mathbf{r};T,h) \simeq D_1r^{-(d-2+\eta)}X_G^{\pm}(\mathbf{r}/\xi ;
D_2h|\tau|^{-\Delta}) ,  \label{BCLScaling}
\end{equation}
where $X_G^{\pm }$ are universal scaling functions for $\tau \gtrless 0$; $D_1$ and $D_2$ are nonuniversal metric factors which can be linked to $A_1$ and $A_2$ (see below).  One can take this postulate
as the basic one to find the scaling form of the free energy density (Eq. (\ref{BFEScaling}))
with
\begin{equation}
\label{eq:scalrelation}
\alpha =2\left( 1-\Delta +\nu \right) -\eta \nu,
\end{equation}
where $\Delta$ is the so-called ``gap exponent" \cite{H87}. This follows from the fluctuation-dissipation
relationship\footnote{In \eq{flucdiss} the symbol $\sum$ is to be understood in a general way --- as the $d$-dimensional summation over the discrete volume of the system if it is defined on a lattice, or as a $d$-dimensional integral in the case of a continuum model. We note that the sums and the integrals carry different physical units.} for the susceptibility
\begin{equation} \label{flucdiss}
	\chi (T,h=0) \equiv \chi_0(\tau) =(k_B T)^{-1} \sum_{\mathbf r}G(\mathbf{r};T,h=0),
\end{equation}
and from the standard relation
\begin{equation}
	\label{chi_def}
	\chi (T,h)\equiv -\frac{\partial^2}{\partial h^2} f_b(T,h).
\end{equation}
This form of scaling is called three-exponent scaling, because  \textit{all} bulk critical exponents can be expressed in terms of the basic triplet $\eta ,\nu$, and $\Delta $. For
example, from Eqs.~(\ref{BFEScaling}) and (\ref{BCLScaling}) it follows that
\begin{equation}\label{eq:scalrelationcritexp}
\alpha +2\beta +\gamma =2, \quad \beta +\gamma =\Delta ,
\quad (2-\eta) \nu =\gamma,
\end{equation}
where $\gamma$ is the critical exponent characterizing the temperature behavior of the zero-field susceptibility  (compressibility) $\chi(T,h=0)$ near the bulk critical point. 
Such relations are called scaling relations between the
critical exponents. If in addition the hyperscaling relation in \eq{hypsc} holds, the thermodynamic exponents $\alpha, \beta, \gamma$, and $\delta$\footnote{At $T=T_c$ one has $h\propto m |m|^{\delta-1}$.} follow from the exponents $\nu$ and $\eta$, which are related to the
scaling of the correlation function, and from scaling relations. Accordingly, in this case there are only two independent critical exponents, i.e., all critical exponents can be expressed in terms of a suitably chosen \textit{pair} of critical exponents. 

Upon approaching the bulk critical point $\xi$ diverges. In this limit (i.e., $\xi \gg $ lattice constant) 
one can replace the summation over ${\mathbf r}$
on the right-hand side of Eq. (\ref{flucdiss}) by an integration over 
${\mathbf x} = {\mathbf r}/\xi$, which yields a scaling
expression for the susceptibility:
\begin{equation} \label{susscf1}
k_B T\; \chi(T,h) \simeq D_1\xi^{2-\eta}X_{\chi}^{\pm}\left(
D_2h|\tau|^{-\Delta }\right),
\end{equation}
where $X_{\chi}^{\pm}$ are universal functions, and the amplitudes $D_1$ and $D_2$ are those appearing in \eq{BCLScaling}. On the other hand, by differentiating the scaling form of the free energy
density \eq{BFEScaling} with respect to the magnetic field variable $h$,
one obtains successively
\begin{equation} \label{magscf2}
m(T,h) \simeq A_1A_2|\tau|^\beta W_1^{\pm }\left(A_2h|\tau|^{-\Delta}\right)
\end{equation}
with $\beta =2-\alpha -\Delta$, and
\begin{equation} \label{susscf2}
k_B T\; \chi(T,h) \simeq A_1A_2^2|\tau|^{-\gamma }W_2^{\pm}
\left(A_2h|\tau|^{-\Delta }\right)
\end{equation}
with $\gamma =2\Delta +\alpha -2$; $W_{1,2}^{\pm}$ are universal
functions. The scaling expression for $\chi(T,h)$, obtained above in two
different ways, can be used to derive relations between the metric factors entering into Eqs. (\ref{susscf1}) and (\ref{susscf2}), i.e., to link the constants $D_1$ and $D_2$ to the constants $A_1$ and $A_2$ (see, e.g., Ref. \cite{PF84}). In addition, one again obtains the scaling relation $\gamma =(2-\eta )\nu$.

Under some plausible assumptions 
it is also possible to show \cite{SFW72,BDT2000} that
\begin{equation} \label{HScalingH}
\lim_{T\rightarrow T_c, h\rightarrow 0}[\beta f_{\rm sing}(T,h)\,\xi^d(T,h)] =Q=
\mbox{universal constant}\: ,
\end{equation}
which represents the hyperuniversality
hypothesis \cite{SFW72} in the form of two-scale
factor universality. Obviously, Eq. (\ref{HScalingH}) yields the
hyperscaling relation in \eq{hypsc}.

The finite-size scaling theory, based on the
two-scale factor universality, predicts that the singular part of the free
energy density of a system in a finite volume $L^d$ {\it at the bulk critical
point} is a universal quantity. With respect to the Casimir effect, this quantity is directly related to the so-called {\it Casimir amplitude} (see Sec. \ref{sec:Cas_definition_simplest} below). Formally, within the justification of the finite-size scaling theory in terms of renormalization-group framework, the variable $1/L$ effectively plays the role of an additional relevant scaling field which does not require a new scaling pre-factor; it combines with the other length scales --- typically the correlation length $\xi$ --- in order to give rise to universal scaling functions \cite{Ba83,C88}.

The universality of critical exponents and of  certain
critical amplitude ratios is a central concept in the theory of critical phenomena. The universality of scaling functions naturally leads to a variety of universal critical amplitudes and amplitude combinations, which characterize
a given universality class to the same extent as the critical
exponents do. Two reviews concerning the available  theoretical and experimental data corresponding to 
this topic can be found in Refs. \cite{PHA91,PV2002}.

Finally, it should be emphasized that the three-exponent scaling relations
do not involve explicitly the spatial dimension $d$ and thus are  generally valid. On the other
hand, it is well known \cite{P90,BDT2000}, that above the upper critical dimension
$d_u$ hyperscaling does not hold. For $d>d_u$ the critical exponents of the system become the ones predicted by mean-field theory. 
For Ising-type and $O(n\geq 2)$ models with short-ranged interactions (see the definitions of these models in Sect. \ref{sec:UC}), one has $d_u=4$. 
For systems in which the interactions decay algebraically  $\propto r^{-d-\sigma}$, with $0<\sigma<2$, one has $d_u=2\sigma$ \cite{FMN72}. It turns out that if $d>d_u$,  in \eq{generlaRGfinal} one cannot simply carry out the limit  $|u_1|^{-y_n/y_1}u_i\to 0$ for $|\tau| \to 0$  and for {\it all} $i>2$. In this context the notion of "dangerous" irrelevant variables has been coined. This mechanism renders the mean-field exponents for all dimensions $d>d_u$ \cite{P90,BDT2000}.

The  main features of the critical Casimir force
can be predicted on the basis of finite-size scaling theory \cite{F71,P90,Ped90,BDT2000,C88,Ba83}. The corresponding universal scaling functions follow from studying specific model
systems. One has to model the fluctuating  medium, the confining geometries, and finally, the possible interactions
of the medium  with the confining objects.
Because the critical Casimir effect is a genuine critical phenomenon, for modeling the critical medium one can use 
the simplest representatives of the corresponding  bulk universality class.
These representative models can be discrete or continuous.

\subsection{Short-ranged, subleading, and long-ranged interactions}
\label{sec:UC}
\label{sec:interactions}

In order to set the stage for the discussion which will follow,
we start by presenting some definitions concerning the types of interactions which we shall mainly deal with throughout the present review. 

In the remainder we shall deal with short-ranged and algebraically decaying long-ranged interactions. To this end we introduce the following definitions:
\begin{itemize}
\item An interaction $J({\bf r})$
will be called {\it short-ranged} if for any finite $m$ its
$m$-th moment is finite, i.e., if $\sum_{\bf r} r^m |J({\bf r})|<\infty$ for all $m$.

\item An interaction is called {\it long-ranged} if there exists a finite $m$ such that
the corresponding $m$-th moment diverges.
\begin{itemize}
\item If $m = 2$ this is
a leading-order long-ranged interaction.

\item If $m> 2$ this is
a subleading long-ranged interaction.
\end{itemize}
\end{itemize}

Thus nearest-neighbor, next-nearest neighbor, etc. interactions, i.e., all interactions, which have a finite support, are short-ranged interactions, while van der Waals and the retarded van der Waals interactions provide two important examples of physically relevant subleading long-ranged interactions\footnote{Obviously, the above classification can be extended, and in the present review we shall do so, to the terminology used for \textit{forces}, such as  short- or long-ranged forces.}.

For interactions $J({\bf r})$, which decay algebraically as a function of the distance $r=|{\bf r}|$, we shall use the notation $J({\bf r})\simeq J/r^{\,d+\sigma}$, where $J>0$, $d$ is the spatial dimension, and $\sigma$ is a parameter which controls how fast $J({\bf r})$ decays. Obviously, $\sigma\le 2$ corresponds to leading-order long-ranged interactions\footnote{Formally, also $\sigma<0$ can be  considered. This case leads to a class of systems which are non-extensive. As an example, one can consider the case $d=3$ and $\sigma=-2$, which can be thought of as being brought about by gravity. However, such systems are beyond the scope of the present review.}, while $\sigma>2$ corresponds to subleading long-ranged interactions. Among the latter, the case $d=3$ and $\sigma=3$ mirrors the long-ranged attractive part of the standard van der Waals interaction, while the case $d=3$ and $\sigma=4$ represents the corresponding retarded van der Waals interaction.

Below we introduce the lattice and the continuum models which will be surveyed in the present review. 

Often, lattice models  are more amenable to analytic treatment.
Furthermore, extensive Monte Carlo simulations can be performed for large system sizes, 
which is necessary to reach the scaling limit where the universal behavior is expected to hold asymptotically. 
For continuous models, large system sizes are difficult to reach. 

We start the discussion with defining of a set of lattice models.

\subsection{Lattice models}
\label{sec:models}

In this section we present the definitions and the notations used for presenting the results for the basic lattice models discussed  in this review. In the spirit of this review, we focus on models for which exact results are available.

\subsubsection{The one-dimensional Ising model}

This model deals with a chain array of $N$ spins with the Hamiltonian 
\begin{equation}\label{HIsingdef_one}
{\cal H}\left(\left\{s_i\right\}\right)=-J\sum_{ \la i,j \ra } s_{i}s_{j},
\end{equation}
where $s_{i}=\pm 1$, $J>0$ for the ferromagnetic and $J<0$ for the anti-ferromagnetic case, and the sum is over the nearest neighbors.  One can fix the  spin state at the left boundary of the $N$-spin array to one of the values $\pm 1$ ("fixed" boundary conditions), or one considers the case that the leftmost spin has no neighbor ("free" boundary conditions). This can be combined with conditions on the corresponding state of the rightmost spin.  

The infinite one-dimensional Ising chain with short-ranged interactions exhibits an essential critical point at $T=0$. 

\subsubsection{The two-dimensional Ising model}
\label{eq:two-dimensional-Ising-model-definition}

\begin{figure}[!htb]
	\includegraphics[width=0.95\columnwidth]{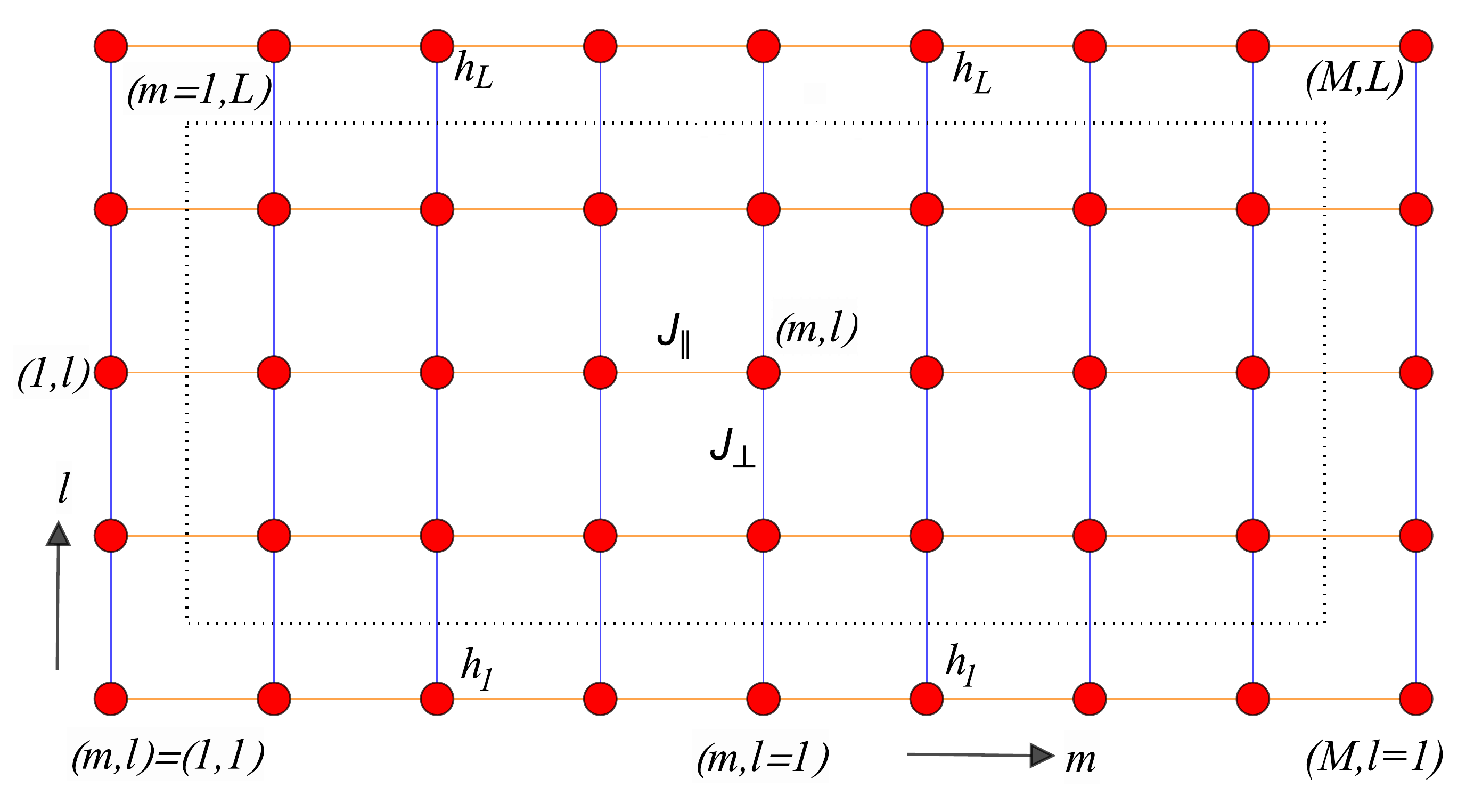}
	\caption{Example of a nearest neighbor  Ising model on a square lattice with the Hamiltonian given by \eq{HIsingdef}. This  example consists of $M=9$ columns and $L=5$ rows. The red dots represent the spins $\pm 1$. $J_\parallel$ and $J_\perp$ denote the horizontal (orange) and vertical (blue) bond strengths, respectively. The spins inside the dotted rectangle have the full number of nearest neighbors (four) and can be called "bulk spins". The "surface spins" have only three nearest neighbors, and the corner spins have only two nearest neighbors. At the "surfaces" $m=1, M$ there are open boundary conditions, whereas the "surfaces" $l=1, L$ are exposed to surface fields $h_1$ and $h_L$, respectively. The quantities $J_\perp, J_\|, h_1$, and $h_L$ are spatially homogeneous. The bulk field $h$ acts on all spins.   }
	\label{Fig:square}
\end{figure}

We consider an Ising model on a square lattice consisting of $L$ rows and $M$ columns, i.e., $L, M\in \mathbb{N}^+$, (see Fig. \ref{Fig:square})
with the Hamiltonian 
\begin{eqnarray}\label{HIsingdef}
	{\cal H}\left(\left\{s_i\right\}\right)&=&-J_\perp\sum_{m=1}^M \sum_{l=1}^{L-1} s_{m,\,l}\,s_{m,\,l+1}-J_\parallel\sum_{m=1}^{M-1} \sum_{l=1}^L s_{m,\,l}\,s_{m+1,\,l}-h_1\sum_{m=1}^M s_{m,1}-h_L \sum_{m=1}^M s_{m,L}-h \sum_{m=1}^M \sum_{l=1}^{L} s_{m,\,l}\,.
\end{eqnarray}
Here $s_{m,\,l}=\pm 1$ denotes the spin located in the $m$-th column and $l$-th row. 
$J_\parallel$ is the coupling constant along all rows, while $J_\perp$ is the corresponding constant  along all columns.  
In the following equation (for $h=0$ )
	\begin{eqnarray}\label{HIsingdef2} 
	{\cal H}\left(\left\{s_i\right\}\right)	&=& -J_\perp \sum_{m=2}^{M-1} \sum_{l=2}^{L-2} s_{m,\,l}\,s_{m,\,l+1}-J_\parallel \sum_{m=2}^{M-2} \sum_{l=2}^{L-1} s_{m,\,l}\,s_{m+1,\,l}  \\
	&& -J_\perp \left[ \sum_{l=2}^{L-2}\left( s_{1,\,l}\,s_{1,\,l+1} + s_{M,\,l}\,s_{M,\,l+1} \right)+\sum_{m=2}^{M-1}\left( s_{m,\,1}\,s_{m,\,2} + s_{m,\,L-1}\,s_{m,\,L} \right)\right]\nonumber\\
	&& -J_\parallel \left[\sum_{m=2}^{M-2} \left( s_{m,\,1}\,s_{m+1,\,1}+s_{m,\,L}\,s_{m+1,\,L}\right) +\sum_{l=2}^{L-1}\left( s_{1,\,l}\,s_{2,\,l} + s_{M-1,\,l}\,s_{M,\,l} \right)\right] -h_1\sum_{m=1}^{M} s_{m,1}-h_L \sum_{m=1}^M s_{m,L}\nonumber  \\ 
	&&-s_{1,1}\left(J_\parallel s_{2,1}+J_\perp s_{1,2}\right)-s_{1,L}\left(J_\parallel s_{2,L}+J_\perp s_{1,L-1}\right)-s_{M,1}\left(J_\parallel s_{M-1,1}+J_\perp s_{M,2}\right)-s_{M,L}\left(J_\parallel s_{M-1,L}+J_\perp s_{M,L-1}\right)\nonumber
\end{eqnarray}
 the Hamiltonian is written in a form in which \textit{all} spins contributing to the first line in \eq{HIsingdef2} have the full number (four) of nearest neighbors (they lie inside the dotted rectangular in Fig. 	\ref{Fig:square} and represent the "bulk" spins), the second and third  line contain the spins which miss one neighbor ("surface" spins), and the last line contains the spins at the corners of the system with two neighbors missing.
The surface field $h_1$  acts on the first row, while the surface field $h_L$ acts on the $L$-th row. In the ground state, say, $s_{l,m}=1 \;\forall\; l, m$, one has 
\begin{equation}\label{HIsingdef3} 
	{\cal H}\left(\left\{s_i=1\right\}\right)	= - (M-2)(L-2) (J_\perp+J_\parallel)-\left[2 (L-2)+2(M-2)\right] (J_\perp +J_\parallel)-M (h_1+h_L)-4\left(J_\parallel +J_\perp \right)
\end{equation}
these groups of terms give rise to bulk $\left[- (M-2)(L-2) (J_\perp+J_\parallel)\right]$, surface $\left[-\left(2 (L-2)+2(M-2)\right) (J_\perp +J_\parallel)\right]$, and corner $\left[-4\left(J_\parallel +J_\perp \right)\right]$ contributions to the energy, where the bulk  and surface ones scale $\propto L\times M, \propto L$, or $\propto M$,  respectively\footnote{As \eq{HIsingdef3} shows, this decomposition is not unique and, thus, definition dependent.}.  The variation of the parameters $h_1, h_l, h, J_\|$, and $J_\perp$ leads to particularly rich wetting and adsorption phenomena (see below). 
 
For this model the bulk critical temperature $T_c=1/(k_B\beta_c)$ is implicitly given by the equation \cite{O44,MW73,B82}
\begin{equation}\label{IsingTcani}
\sinh(2\beta_cJ_\perp)\sinh(2\beta_cJ_\parallel)=1.
\end{equation}
The critical exponents of the model are known exactly, e.g.,
\begin{equation}
\label{eq:Ising2d_crit_exp}
\alpha=0,\quad \nu=1, \quad \eta=1/4, \quad \beta=1/8, \quad \gamma =7/4, \quad \delta=15.
\end{equation}
One often uses the notations $K_\parallel=J_\parallel/(k_B T)$, $K_\perp=J_\perp/(k_B T)$, and $\tilde h_1=h_1/(k_B T)$. In the isotropic case $J_\perp=J_\parallel=J$, i.e., $K_\perp=K_\parallel=K=\beta J$, Eq. (\ref{IsingTcani}) leads to 
\begin{equation}
\label{eq:2d_Ising_crit_point}
K_c=\left[\ln (1+\sqrt{2})\right]/2,
\end{equation}
where $K_c=\beta_c J$. For the amplitudes of the correlation lengths  $\xi_0^\pm=\lim \xi^{\pm}(\tau\to 0^\pm)|\tau|^\nu$ one  finds\footnote{Here $\xop$ is a nonuniversal quantity which depends on its definition. Commonly, $\xop$ and $\xom$ are inferred from the exponential decay of the two-point correlation function. The ratio $\xop/\xom$ is, however, universal.}
\begin{equation}
\label{IsingInfo_prop}
\xop=\frac{1}{4K_c}, \quad \mbox{and} \quad \frac{\xop}{\xom}=2.
\end{equation}

In  a semi--infinite system with the surface field $0<h_{1}/J<1$
acting on the first row, there occurs a surface transition at the temperature
$T=T_w=1/(k_B\beta_w)$, which is the analogue of the  critical wetting transition in fluids \cite{Di88}.
At this transition an  interface, separating a thin layer of positive spins ("liquid") located near
the surface from the negatively magnetized
bulk phase ("gas"), unbinds continuously.
The so called wetting temperature $T_w$ is given by the equation \cite{A80}
\begin{equation}
\label{wetT2dI}
\cosh 2 \beta_w h_1=\cosh 2J_\parallel \beta_w-\rme^{-2J_\perp\beta_w}\sinh 2J_\perp \beta_w.
\end{equation}
In the isotropic case for small surface fields one has \cite{NN2008}. 
\begin{equation}
\label{eq:Tw_small_fields}
\frac{T_c-T_w}{T_c}=\frac{1+\sqrt{2}}{2}K_c \left(\frac{h_1}{J}\right)^2, \quad \left(\frac{h_1}{J}\right)\to 0.
\end{equation}
We recall that if $h_1 \ge J$ one has $T_w=0$, while for $h_1=0$ one has $T_w=T_c$.

The two-dimensional bulk Ising model is exactly solvable \cite{O44}
also in confined geometries for which there are 
analytical results.  For $d=2$ Ising strips, a fruitful approach uses the transfer matrix, which  in many cases can be diagonalized exactly.  On ones disposal is also the quasi-exact numerical  density matrix renormalization-group method
\cite{MCD99}. As in general for $d=2$, the conformal invariance of critical systems \cite{A86,BCN86,C86,C87,C89,BEK2013,BE2020} can be exploited, too (see Sec. \ref{conformal_invariance}). 

\subsubsection{The $d$-dimensional Ising model}

The extension of the above definitions to $d$ dimensions is obvious: at each lattice site ${ \mathbf{r}}$ on the $d$-dimensional lattice, there is a spin variable $s_{{ \mathbf{r}}}$
which can take the values $+1$ and $-1$. The spins interact via a nearest neighbor
exchange interaction $J$.
In the presence of the external bulk field $h$ the Hamiltonian of this model   is
\begin{equation}
\label{eq:Ham_I}
{\cal H} = - J \sum_{\langle{\mathbf{r}, \mathbf{r'}}\rangle} s_{{\mathbf{r}}}s_{{\mathbf{r'}}} - h \sum_{\mathbf{r}} s_{\mathbf{r}},
\end{equation}
where the sum runs over nearest neighbor pairs $\langle{\mathbf{r}, \mathbf{r'}}\rangle$ on the lattice. The Ising model possesses a global $Z_2$ symmetry. 
Unfortunately, for $d>2$ there are no exact results for the Ising model. Therefore we do not go into any further details concerning the definition of this model in $d>2$. We only mention that the inverse critical temperature for the macroscopically large Ising model on a cubic three-dimensional lattice is  
$K_c\equiv J/ (k_B T_c)=0.2216544(3)$~\cite{TB96}.

The models defined above exhibit a {\it discrete} symmetry of the order parameter. Here,  we introduce a few models with a {\it continuous} symmetry for which exact results for the Casimir force are available. We start with the simplest case, i.e.,  $d=1$.  

We consider two one-dimensional models ($d=1$) with a continuous $O(n)$ spin symmetry: the XY ($n=2$) and the Heisenberg ($n=3$) chains. 
All chains considered are taken to consist of $N$ spins interacting with each other via a ferromagnetic interaction $J$ between nearest-neighbor spins at a temperature $T$.  We suppose that boundary fields ${\mathbf H}_1$ and ${\mathbf H}_2$ act on the end spins of the chain. Accordingly, for $n=2$ all spins lie in the plane spanned by ${\mathbf H}_1$ and ${\mathbf H}_2$, which form an angle $0\le\psi\le \pi$; for $n=3$ see below.  It follows rigorously from the Mermin-Wagner theorem \cite{MW66} that these systems do not exhibit spontaneous symmetry breaking at non-zero temperatures given  their low dimension and the short range of the interactions between the spins. Nevertheless, they posses an essential singular point at $T=0$; in that limit, long-ranged order is supported. 

\subsubsection{The one-dimensional $XY$ model}

The Hamiltonian of this system is given by 
\begin{equation}
\label{eq:def_1d_Ham}
{\cal H} = -J \sum _{i=1}^{N-1} {\mathbf S}_i \cdot{\mathbf S}_{i+1}-{\mathbf H}_1\cdot{\mathbf S}_1-{\mathbf H}_N\cdot{\mathbf S}_N 
\end{equation}
where ${\mathbf S}_i$, with ${\mathbf S}_i^2=1$ and ${\mathbf S}_i \in \mathbb{R}^2$, $i=1,\cdots,N$,  are $N$ spins arranged along a straight line. The Hamiltonian can be written in  the form 
\begin{eqnarray}
\label{eq:system_angles}
{\cal H} &=& -J \sum _{i=1}^{N-1} \cos \left(\varphi _{i+1}-\varphi _i\right)
-H_1 \cos \left(\psi _1-\varphi _1\right)-H_N
\cos \left(\psi _N-\varphi _N\right),
\end{eqnarray}
where the angles $\psi_1, \psi_N$ characterize the orientation of the magnetic field vectors, and $\varphi _1,\cdots,\varphi _N$ characterize the corresponding spins,  are measured with respect to the, say, $x$ axis of the plane spanned by ${\mathbf H}_1$ and ${\mathbf H}_2$. The model possesses only an essential singular point at $T=0$, in accordance with the Mermin-Wagner theorem \cite{MW66}.

\subsubsection{The one-dimensional Heisenberg model}

The Hamiltonian of this system is again given by \eq{eq:def_1d_Ham} with the conditions that here the $N$ spins  ${\mathbf S}_i$, $i=1,\cdots N$, arranged along a straight line, are three-component vectors ${\mathbf S}_i \in \mathbb{R}^3$, $i=1,\cdots,N$,  ${\mathbf S}_i^2=1$. The last one means that the vectors in \eq{eq:def_1d_Ham} in spherical coordinates can be written as
\begin{eqnarray}
\label{def}
{\mathbf H}_1 &=& H_1\left \{\sin\varphi_1^H\cos\theta_1^H,\sin\varphi_1^H\sin\theta_1^H,\cos\varphi_1^H\right \}, \\
{\mathbf H}_N &=& H_N\left \{\sin\varphi_N^H\cos\theta_N^H,\sin\varphi_N^H\sin\theta_N^H,\cos \varphi_N^H\right \},
\nonumber \\
{\mathbf S}_i &=& \left \{\sin\varphi_i\cos\theta_i,\sin \varphi_i\sin \theta_i,\cos\varphi_i\right \}, i= 1,\cdots,N. \nonumber
\end{eqnarray}
As the  1d $XY$ model, this model possesses only an essential singular point at $T=0$, in accordance with the Mermin - Wagner theorem \cite{MW66}.

\subsubsection{Gaussian model}

The Gaussian model \cite{PB2011,BDT2000}, or Gaussian approximation \cite{Ma_S_K}, considers spins with continuously variable amplitude and a polynomial free energy which is at most bilinear  in the amplitude of the spins. Here, we focus on such a system with scalar spins. As an example \cite{DaR2017} we consider a planar discrete system containing $L$ two-dimensional layers with a Gaussian type Hamiltonian:
	\begin{eqnarray}
	\label{eq:def_Ham_GM_def}
	-\beta \mathcal{H}&=&\sum _{x=1}^M \sum _{y=1}^N \Bigg\{K^\| \sum _{z=1}^L S_{x,y,z} \left(S_{x+1,y,z}+S_{x,y+1,z}\right)+K^\perp\sum _{z=1}^{L-1} 
	S_{x,y,z} S_{x,y,z+1}+h_1 S_{x,y,1} \cos  \left(k_x x+k_y y\right) \nonumber\\
	&&  +h_L S_{x,y,L} \cos \left(k_x \left(x+\Delta _x\right)+k_y \left(y+\Delta
	_y\right)\right)-s \sum_{z=1}^L S_{x,y,z}^2\Bigg\},
	\end{eqnarray}
which describes a three-dimensional lattice system with nearest-neighbor interactions and with (say, chemically) modulated bounding surfaces\footnote{Actually, critical Casimir measurements for continuously tunable boundary conditions have been already reported in the literature \cite{NHC2009}. There, a solid surface with a gradient in its adsorption preference for one of the two species forming a binary liquid mixture was considered. In this system  the interaction energy of a single colloidal particle suspended in a critical water-2,6-lutidine mixture above the solid surface was measured. The observed scaling functions are found to lie between those belonging to the limiting cases of $(--)$ and $(-+)$ boundary conditions. In the model considered here the chemical treatment is supposed to be more complicated, given by a wave-like pattern along the surfaces.} located at $z=1$ and $z=L$. In its present version the model allows one to consider both orthogonal (i.e., perpendicular to the surfaces) as well as longitudinal (i.e., parallel to the surfaces) Casimir forces (see Sec. \ref{sinusoidal-surface-fields}).  Here, $h_1=\beta H_1$ and $h_L=\beta H_L$ are the external fields acting only on the boundaries of the system. The parameter $s>0$ serves as to implement the constraint to have a value which ensures the existence of the partition function of the system for the given Hamiltonian. In \eq{eq:def_Ham_GM_def} one has 
\begin{equation}
	\label{eq:inte}
	K^{\|}=\beta J^{\|} \qquad \mbox{and} \qquad K^{\perp}=\beta J^{\perp},
\end{equation}
where $J^{\|}$ and $J^{\perp}$ are the strengths of the coupling constants along and perpendicular to the $L$ layers of the system, respectively. One can check that the relation $2K^{\|}+K^{\perp}-s \equiv \beta (2J^{\|}+J^{\perp})-s=0$ determines the inverse critical temperature $\beta_c$ of the bulk model:
\begin{equation}
	\beta_c=s/(2J^{\|}+J^{\perp}).
	\label{betac}
\end{equation}

In this specific example the modulation depends on the coordinates $x$ and $y$ in a wave-like way, specified by the applied surface fields $h_1\cos\left(k_x x+k_y y\right)\equiv h_1 \cos({\bf k}\cdot {\bf r})$ and $h_L \cos [k_x \left(x+\Delta_x\right)+k_y \left(y+\Delta_y\right)]\equiv h_L\cos({\bf k}\cdot ({\bf r}+{\bf \Delta}))$, so that their phases are shifted with respect to each other by $\Delta_x$ in $x$ direction and by $\Delta_y$ in $y$ direction. Here ${\bf r}=(x,y)$, ${\bf k}=(k_x,k_y)$, and ${\bf \Delta}=(\Delta_x,\Delta_y)$. Periodic boundary conditions are applied along the $x$ and $y$ axes, while missing neighbor (Dirichlet) boundary conditions are imposed in the $z$ direction: 
\begin{equation}
\label{bc_def_per}
S_{1,y,z}=S_{M+1,y,z}\,, \qquad S_{x,1,z}=S_{x,N+1,z} \qquad \text{and} \qquad S_{x,y,0}=0, \qquad S_{x,y,L+1}=0.
\end{equation}
Since in the following  we  envisage the case $M,N\gg 1$, we can always consider the directions $k_x$ and $k_y$ to coincide with $(2\pi p)/M $ and $(2\pi q)/N$ for some $p=1,\cdots, M$ and $q=1,\cdots,N$, respectively.

In Gaussian approximation the spatial fluctuations
change the exponent $\alpha$ from its mean field value $\alpha=0$ to
$\alpha=2-d/2$, $d<4$ \cite{BDT2000,Ma_S_K}. All other critical exponents characterizing the bulk behavior preserve their mean field values.

\subsubsection{Classical spherical model}
\label{sec:spherical_model}

We shall discuss separately the classical and the quantum version of the model.

We consider a  model which is constructed  on a $d$-dimensional hypercubic
lattice ${\cal L} \subset \mathbb{Z}^d$, where ${\cal L}=L_1\times L_2\times
\cdots L_d$ with $L_i=N_i a_i, i=1,\cdots,d$. $N_i$ is the
number of spins and $a_i$ is the lattice constant along the axis
$i$ with ${\bm e}_i$ as the unit vector along that axis, i.e., ${\bm e}_i\cdot {\bm e}_j=\delta_{ij}$.
With each lattice site $\bm{r}$ one associates a spin
variable $S_{\bm r}\in \mathbb{R}$, which obeys the constraint
\begin{equation}
\label{constraint}
\langle S_{\bm r}^2
\rangle=1, \quad \text{for all} \; {\bm r} \in {\cal L}.
\end{equation}
The average in \eq{constraint} is taken with respect to the following 
Hamiltonian of the model: 
\begin{equation}
\label{Hamsm}
\beta\,{\cal H}=-\frac{1}{2}\beta\sum_{{\bm r}, {\bm
		r}'}S_{\bm r}\,J({\bm r}, {\bm r}')S_{{\bm r}'}-\sum_{{\bm
		r}}h_{\bm r}\,S_{\bm r}+\sum_{{\bm
		r}}\lambda_{\bm r}\left(S_{{\bm r}}^2-1\right),
\end{equation}
where the Lagrange multipliers $\lambda_{\bm r}$, called spherical fields, are determined by the requirement that \eq{constraint} is fulfilled for all ${\bm r} \in {\cal L}$. Equations  \eqref{constraint} and \eqref{Hamsm} represent the most general definition of the so-called mean spherical model \cite{BK52,LW52,K73}. It deviates from the standard Ising model in that \eq{constraint} is fulfilled only \textit{on average} and not for any state of the system. Obviously, for a system with translational invariance one only needs a single spherical field\footnote{Actually, in 1952 Berlin and Kac \cite{BK52} introduced such a model, with a single spherical field, in which $\sum_{{\bm r}\in {\cal L} } S_{\bm r}^2={\cal N}$, where ${\cal N}=N_1 N_2 \cdots N_d$ is the total number of spins in the
system. This model is known as the spherical model. Slightly later, Lewis  and Wannier \cite{LW52} defined a model in which the above constraint is fulfilled only on average, i.e., $\sum_{{\bm r}\in {\cal L} } \langle S_{\bm r}^2
\rangle={\cal N}$ and, thus, introduced the {mean-spherical} model. The most general version of the model, as presented above, has been introduced by Knops \cite{K73}.}, i.e., $\lambda_{\bm r}=\lambda$ for all ${\bm r} \in {\cal L}$. Both for finite systems as well as in the thermodynamic limit, this model is exactly solvable for any spatial dimension $d$,  even in the presence of a spatially constant external magnetic field $h$. In the case of short-ranged interactions and for $2<d<4$ the bulk critical exponents of the model are given by \cite{J72,BDT2000,B82}
\begin{equation}
\label{crit_exp_SM}
\nu=1/(d-2),\qquad \delta= \frac{d+2}{d-2}, \qquad \beta=1/2, \qquad \eta=0, \qquad \gamma=2/(d-2). 
\end{equation} 
For $d=3$ for a system with an isotropic short-ranged interaction $J$, the critical coupling $K_c$, where $K=\beta J$, is also exactly known:  
\begin{equation}
\label{bulk_constraint}
K_c=\dfrac{1}{2}\dfrac{1}{(2\pi)^3}\int_{-\pi}^{\,\pi}\int_{-\pi}^{\,\pi}\int_{-\pi}^{\,\pi} \frac{d^3q}{3-\cos q_x-\cos q_y-\cos q_z}.
\end{equation}
In Ref. \cite{JZ2001} it has been shown that 
\begin{equation}
\label{JCart}
K_c=\frac{\left(\sqrt{3}-1\right) \left[\Gamma \left(1/24\right)\right]^2 \left[\Gamma
	\left(11/24\right)\right]^2}{192 \pi  ^3}\simeq 0.252731.
\end{equation}
According to Ref. \cite{W39}, this expression is equivalent to
\begin{equation}
\label{W39Rc}
K_c = \frac{4}{\pi ^2} \left(18+12 \sqrt{2}-10 \sqrt{3}-7 \sqrt{6}\right) 
\left\{K\left[\left(2-\sqrt{3}\right)
\left(\sqrt{3}-\sqrt{2}\right)\right]\right\}^2,
\end{equation}
where $K(x)$ is the complete elliptic integral of the first kind \cite{NIST2010}. 
The amplitude $\xi_0^+$, which is needed to form the scaling variable $x_\tau$, is also exactly known \cite{BD91,SP86,D96}. It can be expressed in terms of $K_c$:
\begin{equation}
\label{xi_ampl}
\xi_0^+=(4\pi K_c)^{-1}.
\end{equation}

One direction is singled out, say the one associated with the extension $L_d$, which will be kept finite, and will be called, for simplicity, $z$, so that ${\bm r}=({\bm r}_{\|},z)$.  At each of the $N_d$ sites along this finite extension there is a $(d-1)$-dimensional transverse layer containing a total of $A=N_1 N_2 \cdots N_{d-1}$ spins, where $A$ is large and  later will be taken to infinity. Periodic boundary conditions hold for each of the $N_d$ layers of the system while boundary conditions $(\tau)$ are imposed in the $z$ direction. We consider periodic $(p)$, antiperiodic $(a)$, and free (i.e., missing layers), or, equivalently, Dirichlet boundary conditions $(O)$.   Formally, the latter type of boundary conditions can be realized by placing a layer of spins with zero length on the top and the bottom of the film (i.e., at $z=0$ and $z=L_d+a_d$). We consider the following cases:

{\it (i)} nearest-neighbor interactions, i.e., $J({\bm r}, {\bm
	r}')=J(|{\bm r}-{\bm r}'|)=\ J_i$, if ${\bm r}-{\bm r}'=\pm a_i {\bm
	e}_i$, $a_i\in \mathbb{N} $, $i=1,\cdots,d$, and $J({\bm r}, {\bm r}')=0$ otherwise:
\begin{equation}
\label{eq:NN_SM}
J({\bm r}, {\bm
	r}')=\sum_{i=1}^d J_i[\delta({\bm r}-{\bm r}'-a_i {\bm
	e}_i)+\delta({\bm r}-{\bm r}'+a_i{\bm
	e}_i)];
\end{equation}

{\it (ii)} algebraic decay of the interaction: 
\begin{equation}
\label{eq:PL_SM}
J({\bm r}, {\bm
	r}')=J(|{\bm r}-{\bm r}'|=\hat{r})=\hat{r}^{-(d+\sigma)},
\end{equation}
where $\sigma>0$ controls the rate of the decay of the interactions. 

For the  envisaged boundary conditions all spins in a given layer are equivalent.  Thus, in order to fix their mean length one needs only $N_d$ Lagrange multipliers --- one for each layer, i.e., $\lambda({\bm r}_\|,z)=\lambda(z)\equiv\lambda_z, z=1,\cdots,N_d$. For the aforementioned  boundary conditions one has

{\it (i)} $S_{{\bm r}_\|,\, z}=S_{{\bm r}_\|,\, z+L_d}$ for periodic boundary conditions, 

{\it (ii)} $S_{{\bm r}_\|, \, z}=-S_{{\bm r}_\|, \,z+L_d}$ for antiperiodic boundary conditions, 

{\it (iii)} $S_{{\bm r}_\|, \,z=0}=S_{{\bm r}_\|, \, z=a_d+L_d}=0$ for free, or Dirichlet boundary conditions. 

In the following we shall measure the lengths along a given axis $i$ in units of $a_i$, for $i=1,\cdots,d$.

The spherical model  turns out to be equivalent \cite{S68,BDT2000,KT71,S88,KKPS92,K73} to the  limit $n\to\infty$ of the $O(n)$ models (see the definition below).  The bulk spherical model is exactly solvable in any dimension $d$, also in the presence of an external magnetic field \cite{BK52,LW52,J72,BDT2000}. If the boundary conditions and interactions are such that translational invariance is preserved the same is also true for finite systems ---  see, e.g., Refs. \cite{BDT2000,CD2004,DDG2006} for periodic boundary conditions and Ref. \cite{DG2009} for antiperiodic ones. For other boundary conditions a hybrid analytical plus numerical approach is needed if one insists on keeping the equivalence with  the limit $n\to \infty$ of the $O(n)$ models \cite{DGHHRS2012,DBR2014}. If such a requirement is abandoned, analytical results are available - see, e.g., Refs. \cite{Ba74,BJSW74,DBA97a,Cham2008}.  For certain results concerning semi-infinite systems see, e.g., Refs. \cite{Cos75,Cos76,BM77}. Concerning  the spherical model, there is a huge amount of literature and the above set of references is simply illustrative. For reviews the reader can consult, e.g., Refs. \cite{J72,KKPS92,BDT2000}.

\subsubsection{Quantum spherical model}
\label{sec:QSM_def}

The model is given by the Hamiltonian~\cite{V96,CDT2000}
\begin{equation}
{\cal H}=\frac 12g\sum_\ell {\cal P}_\ell ^2-\frac 12\sum_{\ell,\, \ell
	^{\prime }}{J}_{\ell \ell ^{\prime }}{\cal S}_\ell {\cal S}_{\ell^{
		\prime}}+\frac 12\mu \sum_\ell{\cal S}_\ell^2-H\sum_\ell{\cal S}_\ell ,
\label{eq:Ham_QSM}
\end{equation}
where ${\cal S}_\ell $ is an operator at site $\ell $. The
operators ${\cal P}_\ell $ play the role of ``conjugated'' momenta, i.e., 
\begin{equation}\label{eq:Ham_QSM_operators}
[{\cal S}_\ell ,{\cal S}_{\ell ^{\prime }}]=0, \quad [{\cal P}_\ell
,{\cal P}_{\ell^{\prime }}]=0, \quad  \mbox{and} \quad  [{\cal P}_\ell ,{\cal S}_{\ell
	^{\prime }}]=i\,\delta_{\ell \ell ^{\prime }}, \quad  \mbox{with} \quad \hbar =1.
\end{equation}
The coupling constant $g$ measures the strength of the quantum
fluctuations (in the remainder it will be called quantum parameter), $H$ is an
ordering magnetic field, and the spherical field $\mu $ is introduced as to ensure the constraint
\begin{equation}
\sum_\ell \left\langle {\cal S}_\ell ^2\right\rangle =N.  \label{eq5}
\end{equation}
Here $N$ is the total number of quantum spins located at sites $\ell
$ of a finite hypercubical lattice $\Lambda $ of size $L_1\times
L_2\times\cdots \times L_d=N$,  and $\left\langle \cdots
\right\rangle $ denotes the standard statistical average taken
with the Hamiltonian ${\cal H}$. In Ref. \cite{V96} it is demonstrated that the model defined in \eq{eq:Ham_QSM} and the quantum ${\cal O}(n)$
nonlinear sigma model are equivalent in the  limit of large $n$. 

We note that in the last few years there has been an increasing interest in the spherical approximation (or large $n$ limit), providing 
tractable models for quantum critical phenomena~\cite {V96,TW94,Nu95,NR98,CPT98,CDT2000,OCR2005,ORC2006,BS2012}. There are
various possibilities for the quantization of the spherical constraint. 
In general they lead to different universality classes associated with the
quantum critical point~\cite{V96,TW94,Nu95,NR98}. The
commutation relations for the operators ${\cal S}_\ell$ and ${\cal
	P}_\ell$ together with the kinetic energy term in the Hamiltonian (\eq{eq:Ham_QSM})
do not describe quantum Heisenberg-Dirac spins but quantum rotors~\cite{V96}. We recall that the quantum rotors model has
been widely used in the context of high-temperature superconductivity
(see, e.g., Ref. \cite{Sa2011} and references therein). 

\subsubsection{Ideal non-relativistic Bose gas}

The ideal Bose gas describes a 
many-body system exhibiting a phase transition from a normal phase to a condensed phase called Bose-Einstein condensate (BEC). This transition to BEC occurs below a critical temperature, which for a uniform three-dimensional gas consisting of non-interacting particles is given by
\begin{equation}
\label{eq:BEC_ct}
T_c= \frac{2\pi \hbar^2}{m k_B}\left[\frac{n}{\zeta(3/2)}\right]^{2/3},
\end{equation}
where here $n$ is the number density of particles and $m$ is the mass of the boson. For general $d>2$ the corresponding expression is \cite{GLB2007}
\begin{equation}
\label{eq:BEC_ct_gen_d}
T_c= \frac{2\pi \hbar^2}{m k_B}\left[\frac{n}{\zeta(d/2)}\right]^{2/d}.
\end{equation}

For $d=3$ the grand canonical
	potential of the ideal Bose gas in the box $L \times L \times L_\perp$ has the form \cite{H87,PB2011,MZ2006,PS2016}
\begin{equation}\label{eq:g-c-potent_Bose}
\Omega(T,\mu\,|\, d=3,{L,L_\perp}) = \beta^{-1} \sum_{\textbf{k}} \ln \left\{ 1 -
e^{-\beta \left[\varepsilon(\textbf{k})- \mu\right]}\right\}\,, \ \ \, \ \
\varepsilon(\textbf{k})= \frac{\hbar^2}{2 m} \left(k^{2}_{x} + k^{2}_{y}
+ k^{2}_{z}\right),
\end{equation}
where $\beta = (k_B T)^{-1}$, $m$ is the mass of the Bose particles, and $\mu$ is the chemical potential. 
In Eq. (\ref{eq:g-c-potent_Bose}) the sum runs over the set defined by the boundary
conditions for the Laplacian operator in the box $L \times L
\times L_\perp$. The dispersion relation in  \eq{eq:g-c-potent_Bose} is valid for non-relativistic particles.  The expression for the number of particles is consistent with \eq{eq:g-c-potent_Bose}:
\begin{equation}
\label{eq:av_N}
\left\langle N \right\rangle = \sum_{{\mathbf k}}
\frac{1}{{e^{\,\beta\left[\varepsilon(\mathbf{k})- \mu\right]}}- 1},
\end{equation}
and the mean occupation number $\left\langle n(\varepsilon) \right\rangle$ of the energy level $\varepsilon$ equals 
\begin{equation}
\label{eq:av_oc_number}
\left\langle n(\varepsilon) \right\rangle = 
\frac{1}{{e^{\,\beta\left[\varepsilon(\mathbf{k})- \mu\right]}}- 1}.
\end{equation}
In fact, if $\mu$ becomes equal to the lowest
value of $\varepsilon(\mathbf{k})$, i.e., here in the case  $\mu=0$, the occupancy of that particular level in the bulk limit becomes infinitely high, which
leads to the phenomenon of the Bose-Einstein condensation  (see below). 

In the limit of the film geometry (i.e., $L
 \rightarrow \infty$), the grand canonical thermodynamic
potential per area turns into 
\begin{eqnarray}
\Omega(T,\mu\,|\,d=3,L_\perp)&=& \lim_{L \rightarrow \infty}\frac{1}{L^2} \,\, \Omega(T,\mu\,|\,d=3,{L,L_\perp} )
= \frac{1}{(2\pi)^{2}\beta}\int_{\mathbb{R}^{2}}d^2 q\,\,
\sum_{k_z} \ln \left\{1 -
e^{-\beta\left[\varepsilon(\mathbf{q})+ \varepsilon(k_{z})
	- \mu\right]} \right\}
\nonumber\\ &=& -\frac{1}{(2\pi)^{2}}\int_{\mathbb{R}^{2}}d^2 q\,\,\sum_{k_z}
\frac{\varepsilon(\mathbf{q})}{{e^{\,\beta\left[\varepsilon(\mathbf{q})+
			\varepsilon(k_{z}) - \mu\right]}}- 1}\,,
\label{eq:pot-unit-area-Bose}
\end{eqnarray}
where $\mathbf{q}=(q_{x},q_{y})$ is a two-dimensional wave vector in the $(x,y)$ plane,
$\varepsilon(\mathbf{q})= \hbar^2 \left(q^{2}_x + q^{2}_y\right)/(2m)$, and
$\varepsilon\left(k_{z}\right)= \hbar^2 k^{2}_z/(2m)$. The last equality results from an integration by
parts with respect to $q=|\mathbf{q}|$. Equation  (\ref{eq:pot-unit-area-Bose}) allows one to generalize it straightforwardly to the $d$-dimensional ideal Bose gas in film geometry:
\begin{equation}
\Omega (T,\mu\,|\,d, L_\perp)= -\frac{1}{(2\pi)^{d}}\int_{\mathbb{R}^{d-1}}d^{d-1} q\,\,\sum_{k_z}
\frac{\varepsilon(\mathbf{q})}{{e^{\,\beta\left[\varepsilon(\mathbf{q})+
			\varepsilon(k_{z}) - \mu\right]}}- 1}. 
\label{eq:pot-unit-area-Bose-d-dim}
\end{equation}
The generalization of the expressions in Eqs. \eqref{eq:av_N} and \eqref{eq:av_oc_number} can be implemented directly. It is evident that in the bulk limit (i.e., $L_\perp \rightarrow \infty$), the Bose-Einstein condensation
of the ideal Bose gas occurs only for $\mu = 0$. Accordingly,
one distinguishes two regimes: $\mu < 0$
(normal phase) and $\mu = 0$ (condensed phase). For the  ideal Bose gas  in the bulk one has \cite{GB68})
\begin{equation}\label{eq:varphib}
\beta\Omega_{\text{bulk}}(T,\mu<0\,|\,d)=-\lambda^{-d}\, {\rm Li}_{1+d/2}\big(\rme^{\beta\mu}\big)=-\lambda^{-d}\, {\rm Li}_{1+d/2}\bigg[\rme^{-\left(\lambda/\xi_\mu\right)^2/\left(4\pi\right)}\bigg],
\end{equation}
where
\begin{equation}
\label{eq:de-Broglie-wavelength}
\lambda=\hbar \sqrt{2\pi \beta /m}\equiv h / \sqrt{2\pi m k_B T}
\end{equation}
is  the thermal de-Broglie wavelength, i.e., the mean thermal wavelength of the particles, 
and $\xi_\mu$ is the bulk correlation length which measures the thermodynamic distance from the  Bose-Einstein condensation point at $\mu=0$. In the case of the ideal Bose~gas  it is given by
\begin{equation}\label{eq:xiid}
\xi_\mu=\frac{\hbar}{\sqrt{2m(-\mu)}}.
\end{equation}
In \eq{eq:varphib} $\mathrm{Li}_n(z)$ is the polylogarithm function, also known as the Jonquière's function:
\begin{equation}
\label{eq:Li-def}
\mathrm{Li}_n(z)=\sum_{k=1}^\infty z^{\,k}/k^{\,n}.
\end{equation}
The functions $\mathrm{Li}_n(z)$ are directly related to the so-called Bose-Einstein functions \cite{PB2011}
\begin{equation}
\label{eq:BE-function}
g_\nu(z)=\frac{1}{\Gamma(\nu)}\int_{0}^{\infty}
\dfrac{x^{\nu-1}dx}{z^{-1}e^x-1}.
\end{equation}
One finds  \cite{PB2011} that $\mathrm{Li}_\nu(z)=g_\nu(z)$ for $0\le z<1$. (We note that occasionally the functions  $\mathrm{Li}_n(z)$ are denoted as $F(z,n)$ or $F_n(z)$ \cite{GB68}.) Due to the above relations one can encounter results for the Bose gas formulated in terms of different but otherwise equivalent functions. Here and in the following we shall use formulations in terms of the  polylogarithm functions  $\mathrm{Li}_n(z)$. 

Finally, we note that, for the 
ideal Bose gas  with no number density constraint of the particles, its grand canonical potential is equivalent to that of the two-component Gaussian model \cite{GD2006,BBZ2000,ZJ2002,DiRu2017}. However, if the gas is subject to a number density constraint, its phase behavior is actually equivalent to that of the two-component spherical model \cite{GB68,Ba83,PB2011}.

\subsubsection{Ideal relativistic Bose gas}

Formally, the energy spectrum of such a gas is expected to be of the form
\begin{equation}
\label{eq:rel_spectrum}
\varepsilon({\bf{k}})= c\hbar \sqrt{k^{2}_{x} + k^{2}_{y}
+ k^{2}_{z}}.
\end{equation}
Gunton and Buckingham \cite{GB68} generalized the study of the 
Bose-Einstein condensation to the single-particle energy spectrum $\varepsilon({\bf{k}})= c |\bk|^\sigma$, with $0\le\sigma\le2$. They found that the phase transition occurs at a
nonzero temperature $T_c$ for all $d>\sigma$, with the critical exponents for the spherical model with leading long-ranged interactions (for
$\sigma < d < 2\sigma$, c.f. \eq{critexplongranged})
\be
\label{eq:cr_exp_Bose}
\alpha=\frac{d-2\sigma}{d-\sigma}, \quad  \beta=\frac{1}{2}, \quad \gamma=\frac{\sigma}{d-\sigma}, \quad  \delta=\frac{d+\sigma}{d-\sigma}, \quad  \nu=\frac{1}{d-\sigma}, \quad \eta=2-\sigma. 
\ee
 Thus, in the model envisaged in Ref. \cite{GB68} the
Bose gas in its extreme relativistic state $\sigma=1$ is in a different universality class than the one in the non-relativistic case with $\sigma=2$.
 However, it was shown by Singh and Pandita \cite{SP83} that
if one employs the  energy spectrum 
\begin{equation}
\label{eq:rel_spectrum_improved}
\varepsilon({\bf{k}})= c \sqrt{m_0^2 c^2+\hbar^2\left(k^{2}_{x} + k^{2}_{y}
	+ k^{2}_{z}\right)}
\end{equation} 
and, at the same
time, allows for the possibility of particle-antiparticle pair production in the system --- as
suggested earlier by Haber and Weldon \cite{HW81,HW82}, the relativistic Bose
gas falls into the \textit{same} universality class as the non-relativistic one\footnote{This is due to the fact that if the rest mass $m_0$ is nonzero one can expand the dispersion relation for small values of $k^2=k^{2}_{x} + k^{2}_{y}
	+ k^{2}_{z}$. It will then become proportional to $k^2$, as in the non-relativistic classical case.} with critical exponents equal to those of the spherical model \cite{PB2011}. In Refs. \cite{SP85b,SP84a,SP84} the finite-size behavior of such a model  is investigated with periodic boundary conditions.

\subsubsection{Imperfect non-relativistic Bose gas}

The grand canonical partition function for a weakly interacting Bose gas in dimensions $d>2$
can be expressed, close to the transition point, as a functional integral with the statistical weight $e^{-{\cal H}_b[\mathbf{\Phi}]}$ (see, \eg\
Refs.~\cite{BBZ2000,ZJ2002,GD2006})  where ${\bf\Phi}(x)$ is a two-component real field, ${\cal H}_b$ is the $O(2)$ Landau-Ginzburg  Hamiltonian given by, c.f., \eq{Hb}, with ${\bf h}=0$, $\tau=-2m\mu/\hbar^2$  with  $\mu\le 0$, and $g=8\pi a\hbar^4/\lambda^2$, where $a$ is
the scattering length. 

For the ideal gas one has $g=0$, and ${\cal H}_b$ reduces to the
so-called Gaussian model (see, c.f., \eq{eq:Z_G}).

In the following we consider further microscopic details of a model for an interacting Bose gas. We shall deal  only with such models in which the repulsive pair interaction between identical bosons is described by associating
with each pair of particles a certain mean energy $a/V$ (with the parameter $a>0$ reflecting the strength of the interaction), where $V$ denotes the volume occupied by the system. The Hamiltonian of such an imperfect Bose gas \cite{Da72}, composed of $N$ particles, is defined as 
\begin{equation}
\label{HMF}
H  = H_{0} + H_{mf},
\end{equation}
which is the  sum of the kinetic energy 
\begin{equation}
\label{HPG}
H_{0} = \sum_{\bf{k}}\frac{\hbar^2 {\bf k}^{2}}{2m}{\hat n}_{\bf{k}},
\end{equation}
and of the term representing the mean-field approximation of the interparticle interaction\footnote{This interaction, being distance independent, is the same for each pair of bosons, and changes the critical behavior of the system from that of a weakly interacting Bose system, which belongs to the $O(2)$ model universality class \cite{ZJ2002}, to that of the spherical model. }: 
\begin{equation}
\label{mf}
H_{mf} = \frac{a }{V}\frac{N^2}{2 }. 
\end{equation}
The expression $\{\hat{n}_{\bf{k}}\}$  denotes the particle number operator, and the summation runs over the one-particle states $\{\bf{k}\}$. Here we follow  the generally adopted definition of the imperfect
	Bose gas for which the exact number $N(N-1)/2$ of pairs  is replaced in
$H_{mf}$ by $N^2/2$ (see the corresponding comment in  Ref. ~\cite{ZB2001}). 
The usual criticism of this model is, that it is unphysical, because each of the $N(N-1)/2=N^2/2+{\cal O}(N)$ pairs of bosons gives the same contribution to $H_{mf}$, independent of the distance between the two bosons forming each pair. $H_{mf}$ can be obtained by taking the limit $\gamma\to 0$ of a repulsive integrable Kac-type pair potential, such as $u_\gamma(x)=\gamma^d \,\rme^{-\gamma x}$, the  strength and inverse range of which are both controlled by the same parameter $\gamma>0$ \cite{JN2013,NJN2013,NP2011}. This  is analogous to the rigorous derivation of the van der Waals theory for a classical gas of particles interacting through an attractive Kac pair potential and a repulsive hard core \cite{KUH63}. The discussion above reveals the mean-field character of the model of an imperfect Bose gas as defined above. This has also been  demonstrated in Ref. \cite{DiRu2017} where the equivalence was shown of the imperfect Bose gas introduced above with an interacting Bose gas with $n$ internal degrees of freedom in the limit ${n\to\infty}$, i.e., in the limit of the spherical model.   We note that the latter model involves only a physically justified, short-ranged pair potential. 

\subsubsection{$O(n)$ lattice models} 

Models with $O(n)$ symmetry (so-called $n$-vector models) are defined by associating with each site ${\mbf r}$ on a lattice ${\mathcal L} \subset {\mathbb Z}^d$
an $n$-dimensional spin vector ${\mbf {s}_{\mbf r}}$ of unit length. The Hamiltonian for $O(n)$ models has the form
\begin{equation}
\label{eq:Ham_O_n}
{\cal H} = - J \sum_{\langle{{\mbf r},{\mbf r'}}\rangle} {\mbf{s}_{\mbf r}}\cdot{\mbf{s}_{\mbf{r'}}}= - J \sum_{\langle{{\mbf r},{\mbf r'}}\rangle}\sum_{\alpha=1}^d {s_{\mbf r}^\alpha}{s_{\mbf r'}^\alpha}. 
\end{equation}
For example, the $O(2)$, or $XY$, model can be formulated in terms of the angle associated with each site by writing ${\bf s_r}=(\cos \vartheta_{\mbf r}, \sin \vartheta_{\mbf r})$:
\begin{equation}
\label{eq:Ham_O_2}
{\cal H} = - J \sum_{\langle{{\mbf r}, {\mbf r'}}\rangle} \cos\left( \vartheta_{\mbf r}-\vartheta_{\mbf r'}\right).
\end{equation}
It is easy to see that by relaxing the constraint $ {\mbf s}_{\mbf r}^2=1$ to $\langle{\mbf s}_{\mbf r}^2\rangle=1$ (see \eq{constraint}) where the average has to be performed with respect to the corresponding Hamiltonian $\cal H$, leads to the $d$-dimensional spherical model \cite{BK52,LW52} defined above.

\subsection{Continuum models}

The field-theoretical approach allows one to use various techniques, such as the apparatus of partial differential equations, the variational calculus, the renormalization group, etc.   Within the mean field theory, exact results can be obtained for $d>d_u$, with the upper critical  dimension $d_u=4$ for the Ising bulk universality class or, more generally, for $O(n)$ systems with short-ranged interactions.

Renormalization group techniques 
offer a deeper 
understanding of critical phenomena, the derivation of certain general results, and in particular 
the development of an enhanced perturbation theory in  terms of
$\varepsilon=d_u-d$
\cite{WK74,Fi1974,B77,PT77,A84,D86,G92,F98,PV2002,ZJ2002}. Inter alia, the renormalization group theory succeeded in predicting the critical exponents and their dependence on $d$ and $n$. However,  
systems with symmetry-breaking boundaries  are, for technical reasons, up to now
beyond the reach of renormalization group methods. 
Therefore, in order to study spatially inhomogeneous three-dimensional systems, 
semi-empirical local functionals have been constructed 
\cite{FW69,FG78,FAY80,FU90b,BU98,OO2012}  
which require critical  bulk properties as an input. 
In $d=2$ there is an exact variational principle \cite{MF93,MF94}.
Within its so-called parametric representation, the critical bulk equation 
of state can be cast into a form which upon construction fulfills several 
analyticity requirements \cite{S69,J69,F71}. 
Concerning classical near-critical fluids, a hierarchy of integro-differential equations 
in momenta space has been proposed, accompanied by suitable approximations
\cite{PR85}. 

\subsubsection{Ginzburg-Landau $O(n)$ ${\mathbf \Phi}^4$ model}

The standard continuum field-theoretic description of bulk  $O(n)$ symmetric 
systems near a second-order phase transition is based on the  ${\mathbf \Phi}^4$ 
Landau-Ginzburg-Wilson (LGW) Hamiltonian ${\cal H}$. A $d$-dimensional critical system, confined by parallel plates, is described by the fixed point Hamiltonian ${\cal H}\equiv {\cal H}_b+ {\cal H}_s$, where ${\cal H}_b$ and ${\cal H}_s$ \cite{Bb83,D86} have the following form: 
\begin{eqnarray}
\label{Hb}
{\cal H}_b\left[{\mathbf \Phi};\tau,\mathbf{h},L\right]= \int d^{d-1}{ x}\int_0^L dz \left[\frac{1}{2}  {\left(\nabla{\mathbf\Phi}\right)}^2 +
\frac{1}{2}\tau\mathbf{\Phi}^2+\frac{1}{4}g\mathbf{\Phi}^4-\mathbf{h} \cdot  \mathbf{\Phi}\right],
\end{eqnarray}
and 
\begin{eqnarray}
\label{Hs}
{\cal H}_s\left[{\mathbf \Phi};c_1,c_2,{\mathbf h}_1,\mathbf{h}_2,L\right]=\int d^{d-1}{ x}\left[\frac{1}{2}c_1\mathbf{\Phi}^2\left({\mathbf x}, 0\right) + \frac{1}{2}c_2\mathbf{\Phi}^2\left({\mathbf x}, L\right)-\mathbf{h}_1 \cdot \mathbf{\Phi}({\mathbf x},0) -{\mathbf h}_2 \cdot \mathbf{\Phi}({\mathbf x},L)\right].
\end{eqnarray}
In Eq. (\ref{Hb}) $L$ is the film thickness, ${\mathbf \Phi}({\mathbf x},z)\in {\mathbb R}^n$ is the order parameter at the position $({\mathbf x},z)\in {\mathbb R}^d$ with 
$0 < z < L$ and ${\mathbf x}\in{\mathbb R}^{d-1}$,  $\tau \propto t= (T-T_c)/T_c$ is the bare reduced temperature with $\tau=0$ defining the bulk critical temperature, and $g > 0$
is the bare coupling constant which stabilizes the statistical weight below $T_c$. In \eq{Hs} $c_1$ and $c_2$ are
the so-called surface enhancements which characterize the surface
universality classes \cite{Bb83,D86} (see, c.f., Sect. \ref{sec:universality_classes}). The surface enhancements  $c_1$ and $c_2$ couple to the square of the order parameter at the boundaries, while the surface fields ${\mathbf h}_1$ and ${\mathbf h}_2$ are coupled linearly to it. In the
language of a lattice spin model $c$ takes into account that a surface spin has less neighbors than a bulk one and that the coupling between the surface spins in general differs from its bulk value. The surface fields  ${\mathbf h}_1$ and ${\mathbf h}_2$
explicitly break the symmetry  of the model with respect to the mapping ${\mathbf \Phi}(z) \to -{\mathbf\Phi}(z)$. 
In the case of such a broken symmetry at the surface, in principle also a surface term cubic in  ${\mathbf\Phi}$  needs to
be considered \cite{DC91}. However, concerning the investigation of the
{\em leading} critical behavior in the presence of a nonzero linear term $\mathbf{h}_1 \cdot \mathbf{\Phi}({\mathbf x},0)$ the cubic surface term can be neglected \cite{DC91}.

The partition function is obtained by performing the functional integral over ${\mathbf \Phi}({\mathbf x},z)$: 
\begin{equation}
\label{eq:Z_GL}
\mathcal{Z}_{{\rm GLW}} = \int \mathcal{D} {\mathbf \Phi}({\mathbf x},z) \exp\left(-{\cal H}[{\mathbf \Phi}({\mathbf x},z)]\right).
\end{equation}

\subsubsection{Continuous  mean field Landau theory and continuous Gaussian model}

If one drops the first term on the right-hand side of Eq. (\ref{Hb}), one obtains the mean field Landau theory; if, however, one considers 
$g =0$, $\tau \ge 0$, one is left with the so-called continuous Gaussian
	model. The latter is of basic interest, because the corresponding functional integral in the partition function,
\begin{equation}
\label{eq:Z_G}
\mathcal{Z}_{{G}} = \int \mathcal{D} {\mathbf \Phi}({\mathbf x},z)\, \exp\left\{-\int d^{d-1}{\mathbf x}\int_0^L dz \left[\frac{1}{2}  {\left(\nabla{\mathbf\Phi}\right)}^2 +
\frac{1}{2}\,\tau\mathbf{\Phi}^2-\mathbf{h} \cdot  \mathbf{\Phi}\right]\right\},
\end{equation}
can be
carried out exactly. In the  case $g 
\neq 0$  the utmost analytical result is to express $Z_{\rm GLW}$ in terms of an infinite series of
Gaussian functional integrals. The calculation of the partition function for the continuous Gaussian model in the bulk in the presence of an external field, can be found in many sources, e.g., Refs. \cite{BDFN92,G92,ZJ2002}.

Exact results concerning the Casimir amplitude and the scaling function of the Casimir force for the $O(n)$ continuous Gaussian model with periodic, antiperiodic, Dirichlet-Dirichlet, Neumann-Neumann, and Neumann-Dirichlet boundary conditions have been derived in Refs. \cite{S81,KD91,KD91b,KD92a}.

From \eq{Hb} one can directly read off the definition of the so-called mean-field Ginzburg-Landau Hamiltonian of a near critical system of the Ising type in the film geometry with thickness $L$ in $z$ direction:
\begin{eqnarray}
\label{HGL}
{\cal H}_{\rm GL}\left[ \phi;\tau,h,L\right]=\int_0^L dz \left[\frac{1}{2}  {\phi'}^2 +
\frac{1}{2}\tau\phi^2+\frac{1}{4}g\phi^4-h \phi\right],
\end{eqnarray} 
with $\phi'=d\phi/dz$.  If $\phi$ is dimensionless, in order to have $\tau$ also dimensionless one needs to have all lengths, i.e., $z$ and $L$, transferred too to dimensionless quantities by measuring them in, say, units of $\xi_0^+$ (how this can be achieved is shown in the footnote \ref{rem:rescaling}).  

For this model\footnote{\label{footnote-MF}
	\label{rem:rescaling} One can consider a more general formulation of the model in the form \cite{G92}	
\begin{eqnarray}
		\label{HGL-general}
		{\cal H}_{\rm GL}\left[ \phi;\tau,h,L\right]=\int_0^L dz \left[\frac{1}{2} \gamma {\phi'}^2 +
		a\tau\phi^2+\frac{1}{4}g\phi^4-h \phi\right].
\end{eqnarray} 
Following Refs. \cite{G92,CL95,PHA91,SHD2003} one can show that the bulk correlation length $\xi_b$ is given by 
\begin{equation}
\label{eq:MF-bulk-corr-length}
\xi_b^{-2}=\gamma^{-1}(2a\tau +3g\eta_b^2),
\end{equation}
where $\eta_b$ is the solution of the bulk equation of state
\begin{equation}
\label{eq:MF-bulk-eq-state}
2a\tau\eta_b+g\eta_b^3-h=0. 
\end{equation}
This  leads to 
\begin{equation}
	\label{eq:corr-lengths-t-h}
\xi_b(\tau\to 0^+,h=0)=\sqrt{\frac{\gamma}{2 a}} \tau^{-1/2}, \quad \xi_b(\tau\to 0^-,h=0)=\sqrt{\frac{\gamma}{4 a}} (-\tau)^{-1/2}, \quad \xi_b(\tau=0,h\to 0)=\sqrt{\frac{\gamma}{3}} \left(\sqrt{g} h \right)^{-1/3},
\end{equation}
which is consistent with \eq{eq:mean-field-critical-exponents}, for $h=\sqrt{g}h$.  For $\gamma=1$ and $a=1$, ${\cal H}_{\rm GL}$ in \eq{HGL-general} reduces to the standard Hamiltonian \ref{HGL}.}
one has \cite{G92,CL95,PHA91,SHD2003}
\begin{equation}
\label{eq:mean-field-critical-exponents}
\alpha=0,\quad \beta=1/2, \quad \gamma=1, \quad \delta=3, \quad \nu=1/2, \quad \eta=0, \quad \xi_0^+/\xi_0^-=\sqrt{2},\quad \xi_{0,\,\mu}/\xi_0^+=1/\sqrt{3},
\end{equation}
where $\xi_h$ is the correlation length $\xi(\tau=0, h)$
and $\xi_{0,\,h}$ is the corresponding correlation length amplitude $\xi_\mu(h \to
0)=\xi_{0,\,h} |h|^{-\nu/\Delta}$. 

Within this model exact results associated with the Casimir effect  have been derived in Refs. \cite{K97,GaD2006,DVD2016,DDV2017,DVD2018,DVD2020,DVD2020b}. 

\subsubsection{Model $\Psi$ for $^4$He}
\label{sec:psi-model}

Originally, the so-called $\Psi$-theory has been introduced by  Ginzburg and Sobyanin \cite{GS76,GS82,GS87,GS87b}. This theory has
been used to describe a variety of phenomena observed in helium films and it represents a portion of 
research on helium for which Ginzburg was awarded the Nobel prize in physics in 2003. The $\Psi$-theory has been successfully used to predict the size effects and the influence of external fields on the behavior of helium films, the contribution of ions and impurities to the thermodynamic functions of helium, the dependence of thermodynamic quantities and of the superfluid density on the superfluid velocity, etc. For a review of the available results the reader is referred to Refs. \cite{GS76,GS82,GS87b,GKMD2008}. In various respects, in studies of helium it plays a role similar to that of mean field theory in Ising-like systems. Since the critical exponents $\alpha=0, \nu=2/3$, and $\beta=1/3$ are numerically quite close to those of $^4$He, the $\Psi$-theory is, in fact, an effective theory which represents an approximation to the known scaling properties of $^4$He in $d=3$. 

Recently, in Ref. \cite{DRVD2019} three versions of this theory have been considered. There the authors calculated the critical Casimir force of a helium film  which is taken to be governed by short-ranged
interactions and to be in thermal equilibrium with its vapor. 

First, we  recall some basic expressions for the usual description of the phase behavior of $^4$He films. 

To this end we consider a film of thickness $L$ of liquid $^4$He which is in thermal equilibrium with its vapor. We suppose the two film interfaces to be parallel to the $(x,y)$ plane and the film thickness to be $L$  along the $z$ axis. Due to the coexistence of normal and superfluid fractions of $^4$He, one needs two order parameters: a one-component order parameter $\rho_n$ describing the \textit{normal} part of the fluid, and a parameter $\Psi_s=\eta \exp(i \varphi)$ representing the \textit{superfluid} part of it. In the film geometry $\eta=\eta(z)$ and $\varphi=\varphi(z)$ are real valued functions so that $|\Psi_s|=\eta$ with the identification of the superfluid density
\begin{equation}
\label{eq:rhos_eta}
\rho_s:=m |\Psi_s|^2=m\eta^2, 
\end{equation}
where $m$ is the mass of the helium atom.  A spatial gradient of the phase of the  function $\Psi_s$ gives rise to the superfluid velocity via the relation ${\mathbf{v}}_s=(\hbar/m) \nabla \varphi$. 
In the present review we consider only fluids at rest. Accordingly, one can take $\Psi_s$ to be a real function characterized solely by its amplitude $\eta$.  For temperatures well below the liquid-vapor critical point one usually takes the total density $\rho(z)=\rho_n(z)+\rho_s(z)$ to be constant within the film, i.e., $\rho$ is independent of $z$. This implies that near the $\lambda$ point of the normal-superfluid transition at liquid-vapor coexistence one can treat helium as an incompressible liquid (see Refs. \cite{GS76} and \cite{So73}).  

For the grand canonical potential ${\omega}(\mu,T; [\Psi_s, \rho]) $ per  area $A$ one has the functional
\begin{equation}
\label{eq:funct}
{\omega}(\mu,T; [\Psi_s, \rho]) =  \int_{0}^{L} \left\{\omega(z,\mu,T,\Psi_s,\rho, \dot{\Psi}_s) - \mu \rho \right\} dz,
\end{equation}
where $\omega(z,\mu,T,\Psi_s,\rho, \dot{\Psi}_s)$ is the local density of this potential per area $A$ and $\dot{\Psi}_s=d\Psi_s/dz$. Here $\omega=\omega_I(\mu,T, \rho)+\omega_{II}(\mu,T,\Psi_s,\dot{\Psi}_s)$, where $\omega_I$ is the local potential density of the normal fluid and $\omega_{II}$ is that one of the superfluid. Since $\mu, T$, and $\rho$ are constant throughout the thickness of the film, one concludes that upon integration the terms $\omega_I$ and $\mu \rho$ generate only bulk-like contributions. Therefore, in the present context one is interested only in contributions stemming from $\omega_{II}(\mu,T,\Psi_s,\dot{\Psi}_s)$.   Following Ref.  \cite{GS82}, one can write 
\begin{equation}
\label{eq:omegaII}
\omega_{II}=\omega_{II,0}+\frac{1}{2 m}|-i\hbar\dot{\Psi}_s|^2,
\end{equation}
where $\omega_{II,0}=\omega_{II,0}(\mu,T,|\Psi_s|^2)$ captures the corresponding bulk potential density of the macroscopically large system. The minimum of ${\omega}(\mu,T; [\Psi_s, \rho])$, considered as a functional of $\Psi_s$  and $\rho$, follows from solving the corresponding Euler-Lagrange equations. For the function $\Psi_s$ this leads to
\begin{equation}
\label{eq:psi_eq}
\frac{\hbar^2}{2m} \ddot{\Psi}_s=\Psi_s \frac{\partial \omega_{II,0}}{\partial |\Psi_s|^2}.
\end{equation}
For $\rho_s$ --- and thus for $\Psi_s$ --- the boundary conditions at both the substrate-fluid and the fluid-vapor interfaces turn out as 
\begin{equation}
\label{eq:bcpsis}
\rho_s(0)= \rho_s(L)=0 \Leftrightarrow \Psi_s(0)=\Psi_s(L)=0.
\end{equation}
In Ref. \cite{DRVD2019} it has been shown that the most general effective\footnote{Effective in the sense that the potential is given in terms of a polynomial of the order parameter.} field theory --- with the values of the critical exponents $\alpha=0, \nu=2/3$, and $\beta=1/3$ for $d=3$ being close to the experimentally measured ones for $^4$He --- is consistent with the following expression for $\omega_{II,0}$: 
\begin{equation}
\label{eq:omega_phi_expansion}
\beta\,\omega_{II,0} \simeq L^{-3} \xi_0^3x_\tau^3
A \left(-\sign(\tau)\left|\phi\right|^2+\frac{1-M}{2}\; \left|\phi\right|^4+\frac{M}{3}\; \left|\phi\right|^6\right),
\end{equation}
where $A$ and $M$ are parameters of the theory\footnote{Here the parameter $A$ here should not be mistaken with the surface area $A$.} to be determined from the experimental data (see below), and
\begin{equation}
\label{eq:psi_mod_def}
x_\tau=\frac{L}{\xi_\tau}, \quad \xi_\tau=\xi_0|\tau|^{-\nu}, \quad \xi_0\simeq\frac{\hbar \Psi_{s,e0} }{\sqrt{2m A k_B T_\lambda}}, \quad  \phi=\left|\frac{\Psi_s}{|\tau|^{\beta}\Psi_{s,e0}}\right|, \quad \tau=\frac{T_\lambda-T}{T_\lambda}.
\end{equation}
Here, $T_\lambda$ is the $\lambda$-transition temperature at liquid-vapor coexistence, $T_\lambda(\rho_\lambda) = 2.172$ K, with the bulk mass density $\rho_\lambda = 0.146$ g cm$^{-3}$ at the $\lambda$ point. The constants $\Psi_{s,e0}$ and $\xi_0$ are determined by further experimental data (see below). From there the value of the parameter $A$ in \eq{eq:omega_phi_expansion} follows from the above relations in \eq{eq:psi_mod_def}. We point out that here $\tau>0$ implies $T<T_\lambda$.

In the original formulation of the theory \cite{GS76,GS82,GS87,GS87b} one adopted $\xi_0\equiv \xi_0^+=1.63$ $\AA$. Currently, the best available experimental value is $\xi_0^+=1.432$ $\AA$, as reported in Ref. \cite{TA85}.

  In \eq{eq:psi_mod_def} $\Psi_{s,e0}$ is the amplitude of the temperature dependence of
the equilibrium value of $\Psi_s$ in bulk helium:
\begin{equation}
\label{eq:psi_eq_value}
\Psi_{s,e}(\tau) =\Psi_{s,e0}\, \tau^\beta = 0.23 \times 
10^{12} \, \tau^{1/3} \, {\rm cm}^{-3/2}.
\end{equation}
The spatial distribution of $\phi$ follows from \eq{eq:psi_eq}:
\begin{equation}
\label{eq:phi_spacial_eq}
\frac{\hbar^2}{2m} \Psi_{s,e0}^2 \,|\tau|^{-4/3}\,  \ddot{\phi}=(A/\beta)\, \phi \left[-\sign(\tau) + (1-M) \left|\phi\right|^2 + M\left|\phi\right|^4 \right],
\end{equation}
where the derivatives are taken with respect to $\zeta_\tau=z/\xi_\tau$. The condition  $0\le M \le 1$ ensures that $\phi=0$ is the only real solution of \eq{eq:phi_spacial_eq} for $T>T_\lambda$ (i.e., $\tau<0$).  

Attempts have been made to determine independently the value of the parameter $M$ of the theory. If one only considers second order phase transitions it is clear that  $0\le M < 1$. The value
$M = 1$ corresponds to a tricritical point. Distinct experiments render different "best" values of $M$. Accordingly, $M$ should be considered only as a fitting parameter, i.e., $M$ is not a universal quantity. Thus, universality is not borne out by the $\Psi$ theory.

\subsection{Bulk and surface universality classes  and boundary conditions}
\label{sec:universality_classes}

The notion of a bulk universality class, the corresponding universality hypothesis, as well as the gross features of bulk systems, on which such universality classes depend on, have been already discussed in Section \ref{sec:scaling}. There it has been stated that the bulk universality class is uniquely determined by the dimensionality $d$ of the system, by the symmetry of the disordered state, which is (spontaneously) broken in the ordered phase, and --- if the underlying interaction is of a leading long-ranged type (see Section \ref{sec:interactions}) --- by the parameter $\sigma$ which characterizes the spatial decay of this interaction.
Having in mind the definitions of the models presented in Sect. \ref{sec:models}, it is obvious that the bulk universality classes are named after the corresponding model. Thus the Ising universality class is characterized by the breaking of the $\mathbb{Z}_2$ symmetry of the corresponding effective Hamiltonian for a scalar order parameter; the $XY$ universality class corresponds to a two-component order parameter and a disordered phase with $O(2)$ symmetry; and the Heisenberg universality class is characterized by a vectorial order parameter with $O(3)$ symmetry. All these relevant cases are encompassed within the class of models with an $n$-component order parameter and a disordered phase with $O(n)$ symmetry. For any value of $n$, the models can be studied for various dimensionalities $d$, i.e., one can consider the Ising model in $d=2$, $d=3$, etc., and analogously the $XY$ and the  Heisenberg models. 

Before the thermodynamic limit is carried out, the finite extent of the systems becomes manifest due to the presence of surfaces. There, depending on the local interactions, the order parameter can be enhanced or, on the contrary, reduced.  Thus, in a system with surfaces its overall phase behavior is substantially enriched. The critical behavior of the system, including the behavior of the Casimir force, crucially   depends on the type of boundary condition which the surfaces impose on the order parameter. 
It is useful and common practice  
to describe the boundary conditions, imposed by the presence of surfaces of a certain type on the phase behavior of the system, by specifying the type of surface phase transition, which the semi-infinite system with such a surface would undergo.
Following Refs. \cite{D86,D97} we consider systems with only short-ranged or with subleading long-ranged interactions present. In this case, for the Ising bulk universality class in $3d$ it turns out that there are only three distinct symmetry preserving 
surface universality classes.  This divides the possible surface phase transitions into three groups \cite{Bb83,D86,D97}.
In particular, the surface may enhance the order parameter such that the system undergoes a second-order phase transition in the presence of an already ordered surface.  This is called the extraordinary transition ({\it E})  related to this surface. A surface may also suppress the order parameter with the result that bulk and surface simultaneously order at the bulk critical temperature with the surface being less ordered than the bulk.  This is known as the ordinary transition ({\it O}). Finally, there is a multi-critical point at which, within a mean field picture, the order at the surface equals the one in the bulk.  This transition is called  the  special, or the surface-bulk transition ({\it SB}). Obviously, in $d=2$ only the \textit{(O)} surface universality class is possible. For the $XY$, Heisenberg, and, more generally, $O(n\ge 2)$ models, all surface transitions mentioned above are possible if the dimension $(d-1)$ of the surface is larger than the lower critical dimension $d_l$ of the corresponding bulk universality class. If this condition is not obeyed, only the ordinary transition exists. In Ref. \cite{P90} it is noted, that for models other than Ising, a different classification scheme of the boundary conditions may be appropriate. In this context, we mention that recently \cite{Tol2021}  for the three-dimensional Heisenberg universality class the occurrence  of a special
phase transition was observed, exhibiting  unusual exponents, and a extraordinary phase with logarithmically decaying
correlations\footnote{This finding has been further supported by Ref. \cite{Hu2021} via Monte Carlo simulations of the three-dimensional $XY$ model. The authors report a large distance plateau of the \textit{two-point correlations}, decaying logarithmically with $L$. If confirmed, this would imply a very slow spatial decay of the concomitant  \textit{Casimir force}, which normally reflects the large distance behavior of the two-point correlations. For further details on this topic see Refs. \cite{PKMGM2021,TM2021,Tol2021b}.}.

If the spatial dimension $d$ of the system is high enough,
there are two possibilities for the occurrence of surface order above the bulk critical temperature $T_c$: {\it (i)} the surface may order spontaneously at a certain critical temperature $T_{c,s}>T_b$, or {\it (ii)} the
surface may be ordered externally by the presence of a surface field $h_s$. The bulk transition in the presence of an externally ordered surface is called normal transition. It has been shown rigorously that the normal and the extraordinary transitions only differ with respect to  corrections to scaling, so that both belong to the extraordinary surface universality
class \cite{BD94}. We stress, however, that {\it (i)} externally imposed boundary fields generate a specific preference for one of the possible ordered states of the system below its critical point, and that {\it (ii)} in systems with $O(n\ge 2)$ symmetry of the disordered state it is possible to impose external fields at the boundaries such that there is a richer set of boundary conditions than in the Ising case ($n=1$). For example one can impose the vectorial surface fields ${\bf h}_1$ and ${\bf h}_2$ such that they  form an angle $\alpha$ with respect to each other, which leads to the so-called twisted boundary conditions.

We clarify the above considerations by referring  to the example of the Ginzburg-Landau $O(n)$ $\mathbf{\Phi}^4$ model (see Eqs. \eqref{Hb} and \eqref{Hs}). Within mean field theory and the
dimensional regularization scheme for the field-theoretic
renormalization group,  $c_i > 0, i=1, 2$, defines the ordinary $(O)$
surface universality class and $c_i < 0$ defines the 
extraordinary $(E)$ surface universality class. 
The leading critical
behavior of a semi-infinite system with an $O$ or an $E$ surface is
described by the two stable renormalization group fixed
point values $c = +\infty$ and $c = -\infty$, respectively.
Finite positive or negative values of $c_i$ only yield corrections to
the leading behavior. Within this setting, $c = 0$ is an unstable
fixed point, so that $(\tau,c) = (0,0)$ has the meaning of a 
multicritical point at which both the bulk and the surface of a
semi-infinite system {\em simultaneously} undergo a second order phase
transition \cite{D86}. This multicritical point defines a surface
universality class in its own right which is commonly denoted as the surface-bulk $(SB)$ or  special universality class. 

Generically, a wall in contact with a binary liquid mixture exhibits a certain preferential affinity for one of the two  components so that the composition profile varies as a function of the perpendicular coordinate $z$. This situation can be represented by setting $c_1 \geq 0$ and $c_2 \geq 0$  or $c_2 \leq 0$ in Eq.~(\ref{Hs}), and by prescribing finite values for the surface fields ${h}_1$ and ${h}_2$. The phase transition in the bulk in presence of nonzero surface fields is
called the normal transition \cite{BD94}.
As already stated above, as far as the leading critical behavior is concerned, the normal transition is
equivalent to the extraordinary transition \cite{D86,BD94},
which can be represented by setting ${h}_i=0$ and
assigning $c_i < 0$ at the corresponding boundary. In the following for the extraordinary
transition we shall use the surface field picture  imposing $h_i\ne 0$.
In this situation, concerning the leading critical behavior of the two sub-cases of a system with a film geometry formed by  extraordinary surfaces with {\it (i)} $h_1h_2>0$ or {\it (ii)} $h_1h_2 < 0$, it is
sufficient to investigate the limits $h_1, h_2 \to \pm \infty$
\cite{D86}, i.e., the two principle sub-cases are $h_1 =
h_2 \to +\infty$, and $h_1 = -h_2 \to +\infty$. In order to simplify the
notation we refer to the case $h_1h_2>0$  as $(+,+)$ boundary
conditions and to the case $h_1h_2 < 0$  as $(+,-)$ boundary conditions. Any of these two sub-cases  
represents an example of the $E$ surface
universality class.  Note that $h_i\to \pm\infty$ leads to $|\mathbf{\Phi}(\mathbf{x},z_i)|\to \pm\infty$, where $z_i,i=1,2,$ is the $z$-coordinate of the boundaries. The last relation can be used in its own right, which is often done so as a definition of the $"+"$ or of the $"-"$ boundary conditions. Finally, we remark  that one can
combine a symmetry breaking $E$ surface 1 with a symmetry conserving
$O$ or $SB$ surface 2. In some cases, like in mean field theory \cite{K97},
the information needed for the combinations $(O,E)$ and $(SB,E)$ can be extracted from the analysis of the cases $(+,-)$ and $(+,+)$, respectively. Within our terminology the widely used notion of  Dirichlet boundary conditions,  under which the order parameter at the surface is suppressed to zero,  provides a representation of the ordinary surface universality class.

In order to exhaust the set of boundary conditions to be discussed in the present review, we mention  the  periodic [e.g., $\mathbf{\Phi}(\mathbf{x},z)=\mathbf{\Phi}(\mathbf{x},z+L)$] and antiperiodic [e.g., $\mathbf{\Phi}(\mathbf{x},z)=-\mathbf{\Phi}(\mathbf{x},z+L)$] boundary conditions under which the system does not exhibit manifest boundaries, i.e., the spatial translational invariance in $z$ direction is unbroken.  These boundary conditions are, however, widely used in numerical simulations and in analytical calculations because they give rise to the most easily accessible finite-size effects.

\subsection{Basic hypotheses of finite-size scaling theories}
\label{sec:FSS}

The basic concept of the phenomenological finite-size scaling theory
at criticality has been proposed by Fisher \cite{F71} 
and by Fisher and Barber \cite{FB72}. There are several reviews on this subject \cite{Ba83,P90,Ped90,BDT2000}, with Ref.  \cite{Ped90} being a set of review articles by various authors on finite-size scaling theory as well as on related topics such as numerical simulations of the finite-size behavior of systems with at least one finite dimension. The renormalization-group perspective on that theory is discussed in Refs. \cite{ZJ2002} and \cite{Ba83}. A set of basic papers on finite-size scaling theory is reprinted in Ref. \cite{C88}.  

In the thermodynamic limit of a macroscopic  system its  thermodynamic response functions, such as the susceptibility (see \eq{susscf2}) and the specific heat, diverge at the critical point $\{\tau = 0, h = 0\}$. This behavior is modified in finite systems. According to the phenomenological
theory, rounding and shifting of the divergences in the
thermodynamic functions, which characterize the finite systems, occur when the increasing bulk correlation length
$\xi$ becomes comparable to the smallest linear
size $L$ of the system. More specifically, it is predicted that
asymptotically, i.e., both $L$ and $\xi$ being large on microscopic scales, finite-size effects are controlled by the ratio $L/\xi$ only. 
In fact, upon approaching a continuous phase transition, $\xi$ and $L$ are the relevant length scales which control the collective behavior of the system and therefore one expects to observe scaling behavior in terms of them.  This expectation is encoded within the finite-size scaling theory, which we recall here concerning its most prominent features. 

Since the Casimir effect is directly related to the behavior of the free energy of a finite system, we shall consider mainly this quantity in order to introduce that theory. We shall be mainly interested in the film geometry $\infty^{d-1}\times L$.  But since in numerical calculations it is not possible to directly study a system with infinite lateral extent we  consider the more general geometry $L_\parallel^{d-1}\times L_\perp$ with $L_\perp \ll L_\parallel$. 
Such a finite system does not undergo a phase transition of it own --- in fact, depending on the dimensionality $d$, a critical behavior 
is developed only if either $L_\parallel\to \infty$ or if both $L_\perp$, and $L_\parallel\to \infty$. For reasons of simplicity, we first consider the case that there is no bulk ordering field, i.e., $h=0$.  
In this case, instead of diverging at $T_c$ the, say susceptibility, exhibits a maximum at $T=T_{c,L_\perp}^{(\zeta)}$, where $(\zeta)$ is a short-hand notation for the dependence on the boundary conditions imposed on the system, and
\begin{equation}\label{fracshift}
\varepsilon_{L_\perp}^{(\zeta)} =
\left(T_c-T_{c,L_\perp}^{(\zeta)}\right)/T_c= b^{(\zeta)}
\left(L_\perp/\xi_0^+\right)^{-\lambda}
\end{equation}
defines the so-called fractional
shift \cite{FB72}, with $\lim_{L_\perp \to\infty} T_{c,L_\perp}^{(\zeta)}=T_c$, (typically)  $\varepsilon_{L_\perp}^{(\zeta)}>0$, and a universal amplitude $b^{(\zeta)}$. It characterizes the shift of the critical temperature of the finite-size system with its  asymptotic behavior for $L_\perp\gg a$ given by the shift exponent $\lambda$. Here,  $a$ and $\xi_0^+$ are microscopic length scales pertinent to the system\footnote{For lattice models $a$ is typically taken to be the lattice spacing, while for continuum models this is usually the mean distance between the constituents of the system; usually it is specified as the position of the minimum of the corresponding interaction potential between them; $\xi_0$ is proportional to the second moment of the pair potential.}. Furthermore,  $T_{*,L_\perp}^{(\zeta)}$ denotes the temperature at
which, upon approaching $T_c$, a certain finite-size quantity, say the susceptibility, 
first shows a significant (i.e., of the relative order of unity)
deviation
from its bulk limit. This allows one to introduce the 
fractional rounding \cite{FB72}
\begin{equation}\label{deltatau}
\delta_{L_\perp}^{(\zeta)}=\left(T_{*,L_\perp}^{(\zeta)}-T_c \right)/T_c \simeq
c^{(\zeta)}\left(L_\perp/\xi_0^+\right)^{-\theta}, \ L_\perp\gg a.
\end{equation}
The basic assertions of the 
phenomenological finite-size scaling theory are as follows:

{\it (i)} Close to $T_c$, the only relevant variable, the properties of a finite-size system depend on,  is $L_\perp/\xi^+(\tau) =\left(L_\perp/\xi_0^+\right) \tau^\nu$. 

{\it(ii)} The rounding occurs when $\xi^+(\tau)\simeq L_\perp$.

Assumption {\it (ii)} leads directly to the
conclusion $\theta=1/\nu$. The assertion, that $\xi(T) \simeq L_\perp$,  is the
only criterion, which determines the finite-size scaling effects in the
critical region, and leads to the
equalities $ \lambda=\theta=1/\nu$. This result follows from the
renormalization group derivation of finite-size scaling
\cite{Ba83,Br82,Br83,ZJ2002} (see also Refs.  \cite{BM78,C88})\footnote{Except in certain
special cases (such as the ideal Bose gas and the spherical model for the film
geometry with Dirichlet-Dirichlet \cite{BF73,B74,D93,CD2003} or Neumann-Neumann boundary conditions \cite{DBA97a}, for which there is a 
logarithmic shift of the type $\pm (\ln L)/L$ in 
$d=3$ and $\lambda=1$ in {\it all} other dimensions $d>2$), the
relation $ \lambda=\theta=1/\nu$ seems to be valid quite generally.}. We emphasize that the relationship $\lambda=1/\nu$ is {\it not}
\cite{FB72} a necessary condition for finite-size scaling to
hold in general.\footnote{The formulation of the finite-size scaling theory for systems with a large shift of $T_{c,L_\perp}^{(\zeta)}$, i.e., for $\lambda < 1/\nu$, is presented in Ref. \cite{DB2003}, as well as a discussion of the size dependence of the quantities of interest in this peculiar case.}

The validity of the above hypotheses has been extensively tested and supported, e.g., by films of binary liquid mixtures with boundary conditions, which enforce the preference for  the same or the other coexisting phase near the two boundaries \cite{FN81,NF82,NF83,C88,Ped90,PE90,K94,BLF95,BLF95b,BDT2000,BLM2003,SZHHB2007,BHVV2008,HHGDB2008,GMHNHBD2009,NHC2009,TKGHD2009,AB2009,NDHCNVB2011,TZGVHBD2011,ZAB2011,DVD2015,DVD2018,DVD2018_matec}.  It turns out that a fluid, confined between walls which exert \textit{different} preferences, e.g., such that the fluid wets one wall and dries the
other,  may exhibit a phase behavior, which differs strongly from the one described above \cite{PE90}.  
Macroscopic arguments \cite{PE90,PEN91,PE92},
explicit mean-field analyses \cite{PE92}, exact results for systems in two spatial dimensions \cite{NN2008,NN2009}, and numerical results from Monte Carlo simulations \cite{BELF96,BLF95}  predict that, for a finite film thickness $L_\perp$ and above a certain critical wetting transition temperature $T_w$,  coexistence of two phases can only
occur for $T<T_{c,L_\perp}$, where the critical temperature $T_{c,L_\perp}$ lies below $T_w$ with  $T_w-T_{c,L_\perp} \propto L_\perp^{-1/\beta_s}$; the exponent $\beta_s$  describes the growth of the
wetting layer. For such special boundary conditions involving critical wetting, the location of the critical point of the $(d-1)$-dimensional system is determined by the (critical) wetting properties of the confining walls
rather than by the bulk critical properties.

We now explicitly state the consequences of the above phenomenological finite-size scaling postulates for the behavior of the singular part of the finite-size free energy $f^{(\zeta)}$ per area and $k_B T$.

\subsection{Finite-size scaling hypotheses for the free energy}
\label{sec:FSS_free_energy}

Similar to the procedure carried out for a bulk system (see \eq{singreg}),  the total free energy density $f^{\bc}(T,h,L_\perp,L_\parallel)$ per $k_B T$ of the finite system with the geometry $L_\parallel^{d-1}\times L_\perp$ and under boundary conditions\footnote{In the following, in order to simplify the notations, we shall omit the superscript $\bc$. Furthermore, if not stated otherwise, we shall suppose that periodic boundary conditions are imposed along the $(d-1)$ lateral   directions $L_\parallel$, i.e., $\bc$ will solely refer to the boundary conditions imposed along $L_\perp$. We note that such a rectangular system exhibits from all its possible geometric features only surfaces, whereas edges and corners are absent. The specific features of the finite-size behavior of systems possessing corners, such as the occurrence of "resonances" which lead to logarithmic contributions in $L$ towards their free energy, are discussed in Refs. \cite{BDT2000,K94}.} $\bc$ can be split into a {\it singular} 
($f^{\rm (s)}$) and \textit{nonsingular} (regular) part ($f^{\rm (ns)}$):
\begin{equation} \label{singregfinite}
f(T,h,L_\perp,L_\parallel)=f^{\rm (s)}(T,h,L_\perp,L_\parallel)+f^{\rm (ns)}(T,h,L_\perp,L_\parallel),
\end{equation}
where the singular part is responsible for the critical behavior in the thermodynamic limit.  As noted in Ref. \cite{PF84}, the analytic background term $f^{\rm (ns)}$ can be identified unambiguously only in the bulk system \footnote{It can be identified, e.g., as that  part of the bulk free energy $f_b(\tau,h)$, which is analytic in $\tau$ and $h$. }.  In fact, for $L_\perp$ and $L_\parallel$ finite, the "singular" part $f^{\rm (s)}(T,h,L_\perp,L_\parallel)$ is actually a real-analytic 
function of its variables $T$ and $h$ . For this reason in Ref. \cite{PF84} it has been suggested\footnote{See  Eq. (3.1) therein. Naturally, as for any such suggestion, this one has to be checked against  any model in which the corresponding quantities can be studied in all necessary detail.  } to define  $f^{\rm (ns)}(T,h,L_\perp,L_\parallel)$ as 
\begin{equation}
\label{free_energy_decomposition}
f^{\rm (s)}(T,h,L_\perp,L_\parallel)\equiv f(T,h,L_\perp,L_\parallel)-f^{\rm (ns)}_b(T,h). 
\end{equation}
Such a definition works well for periodic or antiperiodic boundary conditions, when the system does not have surfaces bounding it. In the case of, say, free boundary conditions, via \eq{free_energy_decomposition}  nonsingular contributions stemming from the surface free energies will be added to $f^{\rm (s)}$. Therefore, for such boundary conditions the above definition has to be modified\footnote{Practically, in Monte Carlo studies one often makes a finite-size scaling {\it Ansatz} and fits the parameters there, considering even $L$ as an effective parameter the value of which is to be determined from the best fit of the data.} . 
Independent of the chosen definition, however, the thermodynamic limit of $f^{\rm (s)}(T,h,L_\perp,L_\parallel)$ should  reproduce
the singular part of the bulk free energy:
\begin{equation}
\lim_{L_\perp,L_\parallel\to\infty}f^{\rm (s)}(T,h,L_\perp,L_\parallel)=f^{\rm (s)}_b(T,h). 
\end{equation}
The last relation ensures that\footnote{\label{footnote-nonsingular-h} As noted in Ref. \cite{P90} and commented on in Ref. \cite{BDT2000}, it can be shown that close to $T_c$ the size dependent part of $f^{\rm (ns)}(T,h,L_\perp,L_\parallel)$ is of the order smaller that $O(L^{-d})$. Thus, the   dependence of $f^{\rm (ns)}(T,h,L_\perp,L_\parallel)$ on $h$ can be neglected in the finite-size scaling region where $f^{\rm (s)}(T,h,L_\perp,L_\parallel)$ scales $\propto L^{-d}$ (see \eq{PFhypot}).  } 
\begin{equation}
\lim_{L_\perp,L_\parallel\to\infty}f^{\rm (ns)}(T,h,L_\perp,L_\parallel)=f^{\rm (ns)}_b(T,h). 
\end{equation}
If the boundary conditions are such that the system is bounded by actual surfaces, one has contributions, stemming from these surfaces, to the free energy of the system. In order to keep the discussion simple\footnote{For three-dimensional systems, e.g., the corrections due to the nonzero ratio $L_\perp/L_\parallel$ are
		known to scale approximately as  $(L_\perp/L_\parallel)^2$ \cite{MN87}. Next, $L_\perp$ and $L_\parallel$ are definition dependent quantities. A natural choice for, say, $L_\perp$
		would be to take the distance between the planes defined by
		the positions of the nuclei of the top and the  bottom layers.
		However, there are certainly other possible definitions, which
		differ by a microscopic length. This implies that a quantitative comparison 
		 between experimental  and theoretical data is
		only possible,  if they are accompanied by a precise definition
		of what $L_\perp$ and $L_\parallel$ are.}, here we consider a system in film geometry $\infty^{d-1}\times L_\perp$, in which  it  possesses only two surfaces orthogonal to $L_\perp$. For such a system one has \cite{P90}
\begin{equation}
	\label{eq:nonsingular-energy-decomposition}
f^{\rm (ns)}(T,h,L_\perp)=f^{\rm (ns)}_b(T,h)+\frac{1}{L_\perp} f^{\rm (ns)}_{\rm surface}(T,h)+\cdots.
\end{equation}
The accumulated, available results tell that for $T>T_c$ and for systems with short-ranged interaction one has \cite{P90}
\begin{equation}
	\label{eq:total-energy-decomposition}
	f(T,h,L_\perp)=f_b(T,h)+\frac{1}{L_\perp} f_{\rm surface}(T,h)+O(e^{-L_\perp/\xi}),
\end{equation} 
where both $f_b(T,h)$ and $f_{\rm surface}(T,h)$ are nonsingular functions of $T$ and $h$. 
From this and by taking into account that near $T_c$ the nonsingular contributions are slowly varying functions of their variables, we conclude that the ellipses in  \eq{eq:nonsingular-energy-decomposition} represent, in the envisaged case, exponentially small corrections; only then Eqs.  \eqref{eq:nonsingular-energy-decomposition} and \eqref{eq:total-energy-decomposition} are in agreement with each other\footnote{In Ref. \cite{P90} it has been suggested that the correction term is of the order of $O(L_\perp^{-{(d+1)}})$.}.  Next, based on Eqs.  \eqref{eq:nonsingular-energy-decomposition} and \eqref{eq:total-energy-decomposition}, one can propose an operational definition of $f^{\rm (ns)}_{\rm surface}(T,h)$: one calculates 
\begin{equation}
	\label{eq:one-definition}
	\Delta f(T_0, h, L_\perp)=f(T_0,h,L_\perp)-f_b(T_0,h)
\end{equation}
for $T_0$ close to, but above $T_c$, and takes the corresponding Taylor expansion as the non-singular part $f^{\rm (ns)}_{\rm surface}(T,h)$ at $T$ near $T_c$.

The free energy $f(T,h,L_\perp)$ has a more complicated behavior than the one given by \eq{eq:total-energy-decomposition}, especially for $T<T_c$, for systems with soft modes, or systems governed by long-ranged interactions, or with boundary conditions leading to the occurrence  of an interface inside the fluid film. These issues, as far as they concern the behavior of the Casimir forces, are discussed in Sect.  \ref{more_on_Casimir}. We note that surface-like contributions to $f(T,h,L_\perp)$ are of no importance for the behavior of the Casimir force (see below). Therefore we shall not go in any further details concerning the surface free energy $f_{\rm surface}(T,h)$.

Since one normally defines the Casimir force as an appropriate derivative of the excess free energy, here we concentrate on its finite size behavior.  According to the finite-size scaling theory, the singular part of the excess free energy density \cite{PF84} near the critical point of the corresponding infinite system is 
\begin{equation}
\label{PFhypot}
\beta f^{\rm (s)}(T,h,L_\perp,L_\parallel)=L_\perp^{-d}X_f(a_\tau \tau L_\perp^{1/\nu},a_h h L_\perp^{\Delta/\nu},L_\perp/L_\parallel).
\end{equation}
Here it is supposed that $d<d_u$, which is the upper critical dimension of the corresponding bulk system above which the critical exponents of the theory are mean-field like\footnote{For $O(n\ge 1)$ models with short-ranged interactions one has $d_u=4$.}. In \eq{PFhypot}  $X_f$ is the universal scaling function of the free energy, while $a_\tau$ and $a_h$ are non-universal metric factors, which can be chosen to be equal to those in the bulk --- which, in turn, can be fixed by suitable normalization conditions.
 One normally chooses them from the behavior of the bulk correlation length $\xi(\tau,h)$, with $a_\tau=(\xi_0^+)^{-1/\nu}$ and $a_h=(\xi_{0,h})^{-\Delta/\nu}$, where $\xi(\tau\to 0^+,0)=\xi_0^+\tau^{-\nu}$ and $\xi(0,h\to 0)=\xi_{0,h}^+h^{-\nu/\Delta}$. 
Compared to bulk systems, universal finite-size scaling functions such as $X_f$ depend not only on the bulk universality class, but also on 
the surface universality classes characterizing the bounding surfaces (i.e., the boundary conditions imposed onto the order parameter close to the surfaces) and, in general, on the geometry of the system under consideration.
By noting that the
finite system undergoes no phase transition at $\tau = h = 0$, one concludes
that the scaling function is analytic at the origin as a function of\footnote{The dependence on the aspect ratio  $\rho=L_\perp/L_\parallel$ can be complicated: if, e.g., $d$ is sufficiently large, the finite system characterized by $\rho=0$ exhibits a phase transition of its own at a certain temperature $T_{c,L}$ (see below).} 
\begin{equation}
x_\tau=a_\tau \tau L_\perp^{1/\nu}, \qquad \mbox{and} \qquad x_h=a_h hL_\perp^{\Delta/\nu}.
\end{equation} 
This fact explains why, contrary to the case of the bulk system (see \eq{BFEScaling}), for the function in \eq{PFhypot} no distinct "$\pm$" functions are needed. The hypothesis in \eq{PFhypot} stated above represents a strong form of
finite-size scaling,  which incorporates the equality of
critical exponents defined for $\tau<0$ and $\tau>0$, 
hyperscaling exponent relations, 
and the two-scale factor universality (hyperuniversality) relations among the critical point amplitudes.  A review of the available results  for the universal critical point
amplitude ratios is given in Ref. \cite{PHA91}. Concerning finite-size systems one often uses the notion  "critical region" which replaces the notion "critical point" for macroscopically large systems. One defines it as the set of values of $\tau$, $h$, and $L$  such that $|x_\tau| \lesssim 1$ and $|x_h|\lesssim 1$. 
This is the domain of the thermodynamic parameters within which the most pronounced finite-size effects occur near the critical point of the corresponding macroscopic  system. 

In the above discussion the finite system exhibits the geometry $L_\parallel^{d-1}\times L_\perp$, i.e., as such it does not have any infinite extent. If, however, a  finite-size system is sufficiently high-dimensional, it can support 
its own critical point at $T=T_{c,L_\perp}$ --- as it is the case for $\infty^2\times L_\perp$ Ising "slabs" in $d=3$.  This will become manifest via a singularity of $X_f(x_\tau, x_h,L_\perp/L_\parallel=0)$ at $x_\tau=x_\tau^c\ne 0$ and\footnote{Here it is tacitly assumed that the plus $\Leftrightarrow$ minus symmetry 
of the Hamiltonian of the system, i.e., the $h \Leftrightarrow -h$ symmetry is not violated. The opposite case of being violated will be discussed later. Here we only mention that in this case $x_h^c\ne0$.} $x_h=x_h^c=0$, which is characterized by the critical exponents of the corresponding $(d-1)$-dimensional bulk  system. It is a very challenging task to analytically describe this dimensional crossover from a $d$-dimensional type of  behavior to the $(d-1)$-dimensional one. This situation is discussed phenomenologically  in more details in Refs. \cite{F71}, \cite{Ba83}, and \cite{BDT2000}. For such a case, analytical results for the excess free energy and for the Casimir force are presented in Ref.  \cite{CD2004}. 
For systems with $L_\parallel^{d-1}\times L_\perp$  or  $\infty\times L_\parallel^{d-2}\times L_\perp$ one can define an apparent shifted  critical point by the location of, e.g., the specific heat maximum. Accordingly, $x_\tau^{\rm max}$ is determined implicitly by the equation 
\begin{equation}
\frac{\partial^3}{\partial x_\tau^3} X_f(x_\tau,x_h,L_\perp/L_\parallel)_{{\big |}_{\mbox{\small $ x_\tau=x_\tau^{max},x_h=0$ }}}=0.
\end{equation}
In general $x_{\tau}^{\rm max}$ depends on the quantity used for its definition; e.g., the specific heat and the susceptibility attain maxima at two different values of the scaling variable $x_\tau$.

\subsection{Thermodynamic Casimir force: basic properties}
\label{sec:Cas_definition_simplest}
For a system in film geometry $\infty^{d-1}\times L$, $L\equiv L_\perp$, with boundary conditions $\bc$ imposed along the spatial direction of finite extent $L$, and with total free energy ${\cal F}_{ {\rm tot}}^{\bc}$, 
one can introduce a generalized force per area ($F_{\rm Cas}^{\bc}$) conjugated to $L$:   
\begin{equation}
\label{CasDef}
\beta F_\Cas^{\bc}(T,h,L)\equiv- \frac{\partial}{\partial L}f_{\rm ex}^{\bc}(T,h,L)
\end{equation}
where
\begin{equation}
\label{excess_free_energy_definition}
f_{\rm ex}^{\bc}(T,h,L) \equiv f^{\bc}(T,h,L)-L f_b(T,h)
\end{equation}
is the so-called excess (over the bulk) free energy per area and per $k_B T$. In terms of that one has 
\begin{equation}
	\label{eq:Cas-def-2}
\beta F_\Cas^{\bc}(T,h,L)=	- \frac{\partial}{\partial L}f^{\bc}(T,h,L)+f_b(T,h).
\end{equation}
In what follows all free energies are taken in units of $k_B T$, if not stated otherwise. Here   $f^{\bc}(T,h,L)\equiv \lim_{A\to\infty}{\cal F}_{ {\rm tot}}^{\bc}/A$  is the free energy per area $A$ of the system. $F_\Cas^{\bc}(T,h,L)$ has the meaning of an effective \emph{pressure}
acting on the boundaries of the system due to the fact that they are a finite distance $L$ apart from each other, separated by a fluid medium. The structural and thermal properties of this medium are encoded in the free energy per area $f^{\bc}(T,h,L)$ of this medium;  $f_{\rm ex}^{\bc}$ equals the difference between the free energy per area of the film and of a portion of the bulk system occupying the same region of space as the film does. It contains the contribution $f_{{\rm surface}}^{\bc}=f_{{\rm surf s}, 1}^{\bc}+f_{{\rm surf s}, 2}^{\bc}$ of the two surface tensions (surface energies)  $f_{{\rm surf s}, 1}^{\bc}$ and  $f_{{\rm surf s}, 2}^{\bc}$ of the film, which are independent of $L$,  and thus drop out from the differentiation with respect to $L$ in \eq{CasDef}. 

 Equation (\ref{CasDef}) defines the so-called {\it thermodynamic} or, more restrictively, {\it critical}  Casimir force\footnote{The notion ``critical" tells that this force is caused by  finite-size effects due to critical fluctuations in the system, if the system is thermodynamically close to its bulk critical point. It is possible that strong finite-size effects can occur within a given thermodynamic system which are not due to critical fluctuations but, e.g., due to Goldstone modes, capillary condensation effects, interface localization - delocalization transitions, wetting and drying phenomena, etc. In those cases, instead of the notion ``critical", we use the notion "thermodynamic Casimir force" .}  introduced initially by M. E. Fisher and P. G. de Gennes \cite{FG78}. Since $\lim_{L\to \infty}f^{\bc}(T,h,L)/L=f_b(T,h)$, independent of the applied boundary conditions, one has $\lim_{L\to \infty} F_\Cas^{\bc}(T,h,L)=0$. 
How fast $F_\Cas^{\bc}(T,h,L)$ decays for $L\to \infty$ depends on how rapidly the excess free energy approaches its limiting value. The finite-size effects in the system reflect the presence of the two surfaces confining the system, the possible appearance of lateral geometrical features of the boundary conditions, and the presence of an interface in the system, which, eventually, is induced by the boundary conditions.  The order parameter profile between the surfaces mirrors these features. 
If the direct interactions are short ranged, the correlation length $\xi$  sets the spatial scale at which the ordering degrees of freedom are influenced by the confinement. 
Therefore, if $\xi$ approaches $L$, the excess free energy is strongly influenced by the presence of the surfaces,  i.e., the Casimir force is important. This is the case near the critical point of the system, as well as for a system which, below the critical point, exhibits a continuous symmetry of the order parameter.  In the current section we focus on the case that the finite system is close to the bulk critical point of the macroscopic system. 
More complicated situations will be discussed in Sec. \ref{more_on_Casimir}.

Near the critical point of the bulk system, in accordance  with Eqs. \eqref{free_energy_decomposition}, \eqref{PFhypot}, and \eqref{BFEScaling}, for the singular part of the excess free energy density (in units of $k_B T$) one has
\begin{equation}
\label{excess_free_energy_scaling}
f_{\rm ex}^{\bc}(\tau,h,L)=L^{-(d-1)}X_{\rm ex}^{\bc}(x_\tau,x_h)+\mbox{corrections},
\end{equation}
which, for the Casimir force, leads to
\begin{equation}
\label{Casimir_scaling}
\beta F_\Cas^{\bc}(\tau,h,L)=L^{-d}X_\Cas^{\bc} (x_\tau,x_h)+\mbox{corrections},
\end{equation}
where the scaling functions $X_{\rm ex}^{\bc}$ and $X_\Cas^{\bc}$ are related \cite{BDT2000} according to
\begin{equation}
X_\Cas^{\bc}(x_\tau,x_h) =(d-1)X_{\rm{ex}}^{\bc}(x_\tau,x_h)-
\frac 1\nu x_\tau\frac{\partial}{\partial x_\tau}X_{\rm{ex}}^{\bc}(x_\tau,x_h)-\frac{\Delta}{\nu}x_h\frac{\partial}{\partial x_h}
X_{\rm{ex}}^{\bc}(x_\tau,x_h).
\label{rel_cas_excess}
\end{equation}
We emphasize that in the above expressions  $X_{\rm ex}$ and $X_\Cas$  are universal scaling functions which, in addition to the bulk universality class of the system, depend also on the realized effective boundary conditions $\bc$ as well as, in general, on the geometry of the system and on the spatial arrangement of the chemical composition of the surfaces.
At the critical point of the bulk system, one obtains from \eq{rel_cas_excess}
\begin{equation}
\beta_c F_\Cas^{\bc}(T=T_c,h=0,L)=(d-1)\Delta_\Cas^{\bc} L^{-d} \qquad \mbox{with} \qquad \beta_c=1/(k_B T_c), 
\end{equation}
where  
\begin{equation}
\label{DefDeltaCas}
\Delta_\Cas^{\bc}\equiv X_{\rm ex}^{\bc}(x_\tau=0,x_h=0)
\end{equation} 
defines the so-called Casimir amplitude $\Delta_\Cas^{\bc}$, which is a universal quantity.

In the above discussion we have focused on the simplest possible case, i.e., a system with film geometry the behavior of which depends  on only two relevant scaling variables, the temperature $T$ and the external field $h$. In Section \ref{more_on_Casimir} we shall consider generalizations of the above setting.

\subsection{Finite-size hypothesis for 
additional observables}
\label{sec:FSS_hyp_etc}

According to the finite-size scaling hypothesis due to Privman and Fisher, from \eq{PFhypot} one can derive the corresponding expressions for the finite-size behavior of other important thermodynamic quantities such as the magnetization, the susceptibility, and the specific heat. One obtains\footnote{We use the symbol "$\simeq$" instead of "$=$" in order to indicate that the corresponding equation is valid up to the corresponding  leading corrections.} for the 
\begin{itemize}
\item  magnetization\footnote{We note that here, and in all thermodynamic functions which follow from the free energy via differentiation with respect to the field $h$, we do not use the superscript $(s)$, because the regular part of the free energy is smaller than $O(L^{-d})$, i.e., smaller than the singular part, which in the critical region is $O(L^{-d})$ (see footnote \ref{footnote-nonsingular-h}).}
\begin{equation}\label{mfss}
m(\tau,h,L_\perp,L_\parallel)\equiv -\frac{\partial(\beta f^{(s)})}{\partial h}\simeq a_h L_\perp^{-\beta/\nu}X_m
\left(a_\tau \tau L_\perp^{1/\nu},a_h h L_\perp^{\Delta/\nu}, L_\perp/L_\parallel\right)
\end{equation}
with 
\begin{equation}
X_m\left(x_\tau ,x_h, \rho\right)\equiv -\frac{\partial }{\partial x_h}X_f\left(x_\tau ,x_h, \rho\right), \quad \rho=L_\perp/L_\parallel,
\end{equation}
and (see \eq{PFhypot})
\begin{equation}
\label{Xf-scaling-function}
X_f\left(x_\tau ,x_h, \rho\right)\equiv X_f(a_\tau \tau L_\perp^{1/\nu},a_h h L_\perp^{\Delta/\nu},L_\perp/L_\parallel).
\end{equation}
Within mean field-like theories one typically uses the behavior of the finite-size magnetization in order to calculate the excess free energy and, from that, the Casimir force, in accordance with \eq{CasDef}. In the next step one obtains the

\item  susceptibility
\begin{equation}\label{chifss}
k_BT \, \chi(\tau,h,L_\perp,L_\parallel)\equiv\frac{\partial m}{\partial h}\simeq a_h^2 L^{\gamma/\nu}
X_\chi\left(a_\tau \tau L_\perp^{1/\nu},a_h h L_\perp^{\Delta/\nu}, L_\perp/L_\parallel\right)
\end{equation}
with 
\begin{equation}
X_\chi\left(x_\tau ,x_h, \rho\right)=-\frac{\partial^2 }{\partial x_h^2}X_f\left(x_\tau ,x_h, \rho\right)=\frac{\partial }{\partial x_h}X_m\left(x_\tau ,x_h, \rho\right),
\end{equation}

\item  as well as the nonlinear susceptibility (i.e., the fourth partial derivative with respect to $h$):
\begin{eqnarray}\label{chi4fss}
k_BT \,\chi^{(4)}(T,h,L_\perp,L_\parallel)&\equiv&-\frac{\partial^4 (\beta f^{(s)})}{\partial h^4}\\
&\simeq &
a_h^4L^{(\gamma +2\Delta)/\nu}X_\chi^{(4)}\left(a_\tau \tau L_\perp^{1/\nu},a_h h L_\perp^{\Delta/\nu}, L_\perp/L_\parallel\right) \nonumber
\end{eqnarray}
with 
\begin{equation}
X_\chi^{(4)}\left(x_\tau ,x_h, \rho\right)=-\frac{\partial^4 }{\partial x_h^4}X_f\left(x_\tau ,x_h, \rho\right)=\frac{\partial^3 }{\partial x_h^3}X_m\left(x_\tau ,x_h, \rho\right).
\end{equation}

\item  Furthermore there is the singular part of the specific heat (per number of ordering degrees of freedom)
\begin{equation}\label{capfss}
C^{(s)}(T,h,L_\perp,L_\parallel)\equiv -k_B\frac{\partial^2  f^{(s)}}{\partial \tau^2}\simeq k_B a_\tau^2 L^{\alpha /\nu} X_C\left(a_\tau \tau L_\perp^{1/\nu},a_h h L_\perp^{\Delta/\nu}, L_\perp/L_\parallel\right),
\end{equation}
with
\begin{equation}
\label{relXcXf}
X_C\left(a_\tau \tau L_\perp^{1/\nu},a_h h L_\perp^{\Delta/\nu}, L_\perp/L_\parallel\right)={\partial^2 \over
\partial x_\tau^2 } X_f\left(a_\tau \tau L_\perp^{1/\nu},a_h h L_\perp^{\Delta/\nu}, L_\perp/L_\parallel\right),
\end{equation}

\item and the two-point correlation function
\begin{equation}
\label{BCLScaling_finite}
G(\mathbf{r};T,h,L_\perp,L_\parallel) \simeq D_1 r^{-(d-2+\eta)}X_G(\mathbf{r}/\xi;
D_2h|\tau|^{-\Delta},L_\perp/\xi,L_\perp/L_\parallel), \quad r=|\mathbf{r}|.
\end{equation}
The behavior of the correlation function is a helpful direct indicator of the type of finite-size corrections to the free energy --- 
and, thus, to the Casimir force --- within certain regions of the thermodynamic space.  We recall that, via the 
fluctuation-dissipation relation (see \eq{flucdiss}), the two-point correlation function $G(\mathbf{r};T,h,L_\perp,L_\parallel)$ is directly related to  $k_B T \,\chi(T,h,L_\perp,L_\parallel)$, 
and thus to the behavior of the free energy. If the correlations decay algebraically upon increasing the distance between the two points, the corresponding finite-size behavior is also expected to be given by a power law in $L=L_\perp$ in the limit  $L_\parallel\to \infty$; on the other hand, if it varies  exponentially,  one expects an exponential decay of the Casimir force \cite{D2001}. 
\end{itemize}

In the above expressions $X_m$, $X_\chi$, $X_\chi^{(4)}$, 
$X_C$, and $X_G$ are universal scaling functions\footnote{They depend, however,   on the boundary conditions applied to the system, as usual. Accordingly, one has to distinguish  $X_m^{\bc}$, $X_\chi^{\bc}$, $X_\chi^{(4),\bc}$, 
	$X_C^{\bc}$, and $X_G^{\bc}$. In order to keep the notation simple, we suspended  this dependence.}.  By using Eqs. \eqref{chifss} -\eqref{capfss} one can construct also universal
ratios of finite-size quantities evaluated at
the bulk critical point. In particular, for a system with cubic geometry, i.e., for $L_\perp=L_\parallel=L$, 
the ratio
$\beta\chi^{(4)}(0,0,L,L)/\left[\big(\beta\chi\left(0,0,L,L\right)\big)^2L^d\right]\simeq U_1$
is expected to be universal. Quantities like $U_1$ can be straightforwardly 
determined by Monte Carlo simulations or by transfer matrix methods. One important quantity of this type is the so-called Binder cumulant ratio $U_B$. If $\psi$ denotes the fluctuating order parameter density, so that $\langle\psi\rangle=m$, one has  \cite{B90}
\begin{equation}
\label{Binder_cumulant}
U_B=1-\frac{\langle\psi^4\rangle}{3 (\langle\psi^2\rangle)^2}. 
\end{equation} 
This ratio is very useful for inferring  the bulk critical temperature $T_c$ from data for  systems of finite-size. For three-dimensional Ising systems with  periodic boundary conditions, upon increasing $T$, $U_B$ tends to 0 above $T_c$, and upon decreasing $T$ it tends to $2/3$ below $T_c$, and at $T_c$ it assumes the universal  "{fixed point}" value $U^*\simeq 0.47$ \cite{FL91}. 

A review concerning universal critical amplitude ratios is provided in Ref. \cite{PHA91}.

\subsection{Further finite-size properties of the free energy related to the Casimir force}
\label{more_on_Casimir}

In Sects. \ref{sec:FSS_free_energy} and  \ref{sec:Cas_definition_simplest}
we have considered systems for which only two thermodynamic parameters, i.e., temperature $T$ and the external field $h$, control the leading behavior of the free energy and therefore of the critical Casimir force, such that the effects of the boundaries are captured by effective boundary conditions imposed on the order parameter. 

In general, however, additional parameters play a relevant role. In fact, the confining surfaces give rise to modifications of the interaction potentials near the surfaces, which can be either localized at the boundaries or can decay algebraically upon increasing the distance from the boundary.  In particular,  the interaction potentials near the surfaces generically generate  a  term {\it linear} in the order parameter $\Phi$, which represents the relevant degree of freedom for the description of the system. This term  adds locally to the effective Hamiltonian of the system and mimics the presence of symmetry-breaking effects at the surface. Similarly, missing neighbors at the surface effectively give rise to a term  {\it quadratic} in $\Phi$,  which is localized at the surface, as indicated in Eq.~\reff{Hs}.  
If the system is sufficiently close to the critical point, these interaction potentials at the surface are responsible for the emergence of effective boundary conditions. One might need to account for them explicitly when discussing properties beyond the leading order, such as for crossover phenomena.

Additional features arise if an interface is present within the system: depending on the symmetry of the disordered phase of the system, this interface might be sharp --- such as in the cases represented by the $O(n=1)$ model --- or diffuse as for $n\ge 2$. In addition we shall briefly describe the changes in the finite-size behavior of the system caused by the occurrence  of capillary condensation or wetting, and by the related phenomena of pinning - depinning phase transitions of interfaces located within the finite system.  
Furthermore, certain features of the system, which for sufficiently large values of $L$ are unimportant, can become relevant for the behavior of the excess free energy and, thus, of the Casimir force for smaller  values of $L$. Since typically this force is experimentally accessible also for such values of $L$, there is a need to address also these  contributions to the finite-size quantities. This will be discussed in the current section below. 

\smallskip 

\subsubsection{Size dependence of the singular part of the excess free energy} 

First, we recall some results for continuous phase transitions as obtained  from renormalization group theory. According to Sect. \ref{sec:scaling}, the singular part $f^{(s)}$ of the free energy density for a sample with a characteristic linear size $L$ exhibits the scaling form  \cite{Ba83,PF84,P90,BLH95,BDT2000,PV2002}
\begin{equation}
\label{generlaRGfinite}
f^{\rm (s)}(u_\tau,u_h,\{u_n\}, L)=b^{-d}f^{\rm (s)}(b^{y_\tau}u_\tau,b^{y_h}u_h,\{b^{y_n}u_n\}, L/b),
\end{equation}
where $u_1\equiv u_\tau$ and $u_2\equiv u_h$ are the relevant bulk scaling fields related to the temperature and the magnetic field, respectively  (see \eq{defscalingfields}), while $\{u_n\}$, $n\ge 3$, are the scaling fields associated with irrelevant operators. Choosing $b=L$, one obtains 
\begin{equation}
\label{generlaRGfiniteL}
f^{\rm (s)}(u_\tau,u_h,\{u_n\}, L)=L^{-d}f^{\rm (s)}(L^{y_\tau}u_\tau,L^{y_h}u_h,\{L^{y_n}u_n\}, 1), 
\end{equation}
from which, by taking into account the definitions given in \eq{defscalingfields} and by performing the appropriate differentiations, one can derive the results given in Sects.  \ref{sec:FSS_free_energy} and  \ref{sec:Cas_definition_simplest} 
The lists of arguments in Eqs. \reff{generlaRGfinite} and \reff{generlaRGfiniteL} indicate that finally $u_{n\ge 3}=0$ is taken to be zero so that $u_\tau$  and $u_h$ are the only
relevant scaling fields and that corrections of the order $L^{y_3}$  with  $y_3=-\theta/\nu\equiv-\omega$ are neglected\footnote{The notation $\theta$ is one of the standard ones used in the literature for discussing  corrections to scaling. But it differs from, and should not be mistaken with the one used in \eq{deltatau} for finite-size scaling of rounding.}, 
where $\omega$ is the leading correction to scaling exponent in the bulk. 
This is indeed the case for finite systems with periodic or antiperiodic boundary conditions, under which the finite system is not endowed with actual surfaces. The presence of actual surfaces might give rise to \textit{surface} related new relevant scaling fields.
Nonetheless, the basic expressions presented in Sects. \ref{sec:FSS_free_energy} and \ref{sec:Cas_definition_simplest} remain valid, provided they are understood as statements about the leading finite-size behavior of a system with boundary conditions $(\zeta)$ provided by the corresponding surface universality classes, i.e., $(\zeta)$ coincides with one of the combinations $(a,b)$ where $a,b \in (E,O,SB)$.
The corresponding scaling function $X_f^{(\zeta)}$ for the free energy density reflects this dependence.  The situation becomes significantly more complicated if one focuses on crossover phenomena. To this end we  consider the most complicated  case of a surface-bulk ($SB$) phase transition 
taking place at both surfaces which are, however distinct.  Any of the surfaces $(1)$ and $(2)$ are characterized by certain surface enhancements $c_i$ and  local surface fields $h_i$, $i=1,2$ ( see  Eqs. \eqref{Hb} and \eqref{Hs} in Sec. \ref{sec:models}). For such a system the singular part of the finite-size free energy density $f^{\rm (s)}$ exhibits the scaling form 
\begin{equation}
\label{surface_bulk}
f^{\rm (s)}(\tau,h,c_1,c_2,h_1,h_2, L)= L^{-d}X_f^{(SB)}(a_\tau\tau L^{y_{1/\nu}}, a_h h L^{\Delta/\nu},\{c_i L^{\phi/\nu}\},\{h_i L^{\Delta_1^{SB}/\nu}\}, u_3 L^{-\omega}), 
\end{equation}
where $\phi$ is the crossover surface critical exponent, $\Delta_1^{SB}$ is the corresponding surface critical exponent for the special universality class; we have retained also the leading corrections to scaling in the bulk.  In the case of $O$ and $E$ surfaces the corresponding surface enhancements do not contribute to the list of relevant scaling fields; in fact, under renormalization-group transformations $c_i$ flows towards fixed-point values, which are $+\infty$ for the $O$ transition and $-\infty$ for the $E$ transition.  
If one considers the crossover behavior from the $O$ to the $E$ universality class,  $f^{\rm (s)}$ has the scaling form 
\begin{equation}
\label{ordinary-extraordinary}
f^{\rm (s)}(\tau,h,h_1,h_2, L)=L^{-d}X_f(a_\tau\tau L^{{1/\nu}}, a_h h L^{\Delta/\nu},\{h_i L^{\Delta_1/\nu}\}, u_3 L^{-\omega}),
\end{equation}
where $\Delta_1$ is the gap surface critical exponent for the ordinary universality class. For the characterization of the surface-bulk universality class, two additional surface critical exponents ($\Delta_1^{SB}$ and $\phi$) are needed, which are independent of the bulk ones, while for the ordinary universality class one additional independent surface critical exponent ($\Delta_1$) exists. 
For the $E$ surface universality
class all surface exponents can be expressed in terms of bulk ones \cite{Bb83,D86}.
We note that the combination $(a,b) = (E,E)$ is not unique. For the Ising universality class ($n=1$) and a suitable choice of the surface fields, one can enforce either of the two competing "$+$" and "$-$" phases to be formed at the bounding surfaces. Thus, one has to distinguish between the $(+,+)$ (and the  equivalent $(-,-)$ boundary condition) and the $(+,-)$ boundary conditions. 
Obviously, the latter boundary condition induces the appearance of an interface within the system. In systems with a continuous order parameter (like for the $XY$ (i.e., $n=2$), or for the  Heisenberg universality class ($n=3$)), the application of  boundary fields at a given angle with respect to each other (i.e., "twisted" boundary conditions) generate helical magnetization profiles within the system \cite{FBJ73,BDR2011,BKJZ93}.

\smallskip 

\subsubsection{Asymptotic behavior of the excess free energy} 

At the bulk critical point $T_c$ (and in the absence of a bulk field), the total free energy per transverse area $A$, and in units of $k_B T$, of a
$d$-dimensional critical film of thickness $L$ and with boundary conditions $(a,b)$ 
at the parallel surfaces, takes the
asymptotic form
\begin{equation} \label{Catcrittemp}
f^{(a,b)}(T_c,L)= Lf_{b}(T_c)+f_{\mathrm{surf}}^{(a)}
(T_c)+f_{\mathrm{surf}}^{(b)}(T_c)+L^{-(d-1)}\Delta_{\rm Cas}^{(a,b)} + \cdots
\end{equation}
in the limits $A\to \infty $ and $L\gg \tilde{a}$ where $\tilde{a}$ is a characteristic microscopic length. 
Here $f_{\mathrm{surf}}$ is the surface free
energy (per area $A $ and in units of $k_BT$)  contribution\footnote{In a geometry which involves other geometrical features such as edges or corners, the free energy is expected to acquire  contributions associated with them. This might lead to interesting dependences on $L$ in the total free energy. For example, the presence of corners leads to so-called "resonant" corrections which are logarithmic in $L$ \cite{P90,K94}. 
} and $\Delta_{\rm Cas}^{(a,b)}$ is the critical 
Casimir amplitude. 
The power law dependence of the finite size Casimir term in \eq{Catcrittemp} follows from the scale invariance of the free
energy and has been initially derived in Refs.~\cite{FG78,S81}.  As already stated above, the amplitude $\Delta_{\rm Cas}^{(a,b)}$ is
\textit{universal}, depending on the bulk and the surface universality classes, i.e.,  $\Delta_{\rm Cas}^{(a,b)}=\Delta_{\rm Cas}^{(a,b)}(d,n)$, where $d$ is the dimensionality of the bulk system, and $n$ captures the symmetry of the disordered phase \cite{K94,KD92a,Bb83,Ba83,P90,BDT2000}, such as the $O(n)$ symmetry which characterizes Eq.~\reff{Hb}. 
In general, a decomposition of the type given in \eq{Catcrittemp} can be performed at any temperature $T$. Any one of the terms of this decomposition, i.e.,  $f^{(a,b)}(T,L)$, $f_b$, $f_{\mathrm{surf}}^{(a)}$ and $f_{\mathrm{surf}}^{(b)}$, can, in turn, be decomposed into a nonsingular and a singular part, where only the latter one exhibits a scaling behavior near the bulk critical point. Furthermore, we note that for $O(n)$-symmetric model systems ($n\geq 1$), depending on the boundary conditions $(a,b)$ and on 
$n$, the excess free energy
$f^{(a,b)}_{\mathrm{ex}}(T, L)$ defined according to Eq.~\reff{excess_free_energy_definition} may, or may not, contain contributions
independent of $L$.
For systems belonging to the Ising universality class ($n=1$),  these can be the
surface free energies $f_{\mathrm{surf}}^{(a)}(T)$ and $f_{\mathrm{surf}}^{(b)}
(T)$, i.e.,  contributions which are associated with the presence of each single surface in an otherwise unbounded medium, and the interface free energy $\sigma^{(a,b)}(T)$ (for reasons of brevity we
consider only its temperature dependence). 
For systems characterized by the $O(n\ge 2)$ symmetry,
the only contributions independent of $L$ stem from the
surface free energies, because the term analogous to the interface free energy is actually the
helicity modulus $\Upsilon (T)$.  The corresponding contribution to the free energy is of the order $\Upsilon (T)/L$ \cite{FBJ73,BDT2000,BDR2011}, and therefore it does depend on $L$. 

We now consider the finite-size part of the excess free energy: 
\begin{equation}
\label{def_of_finite_size_part}
\Delta f^{(a,b)}(T,L)\equiv f^{(a,b)}(T,L)-Lf_{b}(T)-
f_{\mathrm{surf}}^{(a)}(T)-f_{\mathrm{surf}}^{(b)}(T),
\end{equation}
which is obtained by subtracting from the excess free energy in Eq.~\reff{excess_free_energy_definition} the contributions of the two separate surfaces,  which are independent of $L$. After this subtraction, $\Delta f^{(a,b)}(T,L)$ vanishes as the film thickness $L$ increases. The last property reflects the fact that  $\Delta f^{(a,b)}(T,L)$ is the bona fide finite-size part of the total free energy of the system.

In line with \eq{excess_free_energy_definition},  for the singular part of $\Delta f^{(a,b)}(T,L)$ one has for $h=0$ \cite{K94,P90,BDT2000}
\begin{equation}
\Delta f^{(a,b)}_{\rm sing}(\tau,L)=L^{-(d-1)} \Delta X^{(a,b)}_{\mathrm{ex}}
(a_\tau  \tau L^{1/\nu };a_\omega L^{-\omega}),
\label{CasimirScaling General}
\end{equation}
where $\Delta X^{(a,b)}_{\mathrm{ex}}$ is a universal scaling function with 
\begin{equation}
\label{Cas_ampl_dletaf}
\Delta X^{(a,b)}_{\mathrm{ex}}(0;0)\equiv
\Delta^{(a,b)}_{\rm Cas}.
\end{equation}
In \eq{CasimirScaling General} we have retained the leading correction to scaling. It is straightforward to extend the above equations  to the case of the presence of an external ordering field $h$. Let us note that, in agreement with 	\eq{rel_cas_excess}, and since $f_{\mathrm{surf}}^{(a)}(T)$ and $f_{\mathrm{surf}}^{(b)}(T)$ do not depend on $L$, one has
\begin{equation}
	\label{eq:deltaXex-XCasimir}
 	X_\Cas^{(a,b)}(x_\tau,x_h)=(d-1)\Delta X_{\rm{ex}}^{(a,b)}(x_\tau,x_h)-
	\frac 1\nu x_\tau\frac{\partial}{\partial x_\tau}\Delta X_{\rm{ex}}^{(a,b)}(x_\tau,x_h)-\frac{\Delta}{\nu}x_h\frac{\partial}{\partial x_h}
	\Delta X_{\rm{ex}}^{(a,b)}(x_\tau,x_h).
\end{equation}

In view of the comparison with experiments, it is important to certify the behavior of $\Delta f^{(a,b)}$ 
and not only of its singular part $\Delta f^{(a,b)}_{\rm sing}$. It is the dependence of $\Delta f^{(a,b)}$ on $L$, which determines the \emph{total} force acting on the confining boundaries. The universal critical Casimir force associated with $\Delta f^{(a,b)}_{\rm sing}$ contributes to this total force. 
In addition to the corrections to scaling for $\Delta f^{(a,b)}_{\rm sing}$ due to various irrelevant fields (see the discussion above), relevant contributions have to be expected from the nonsingular terms $\Delta f^{(a,b)}_{\rm ns}$ contained in $\Delta f^{(a,b)}(T,L)$.
As will be discussed below, there are cases in which the thickness dependence of the critical Casimir force and of the contribution to the total force stemming from $\Delta f^{(a,b)}_{\rm ns}$ is the same; however, due to their distinct  dependence on temperature $T$, they can be disentangled nonetheless.                    

In Ref.~\cite{P90} (see also Ref. \cite{KD92b}) it has been argued that in systems with short-ranged interactions the nonsingular contribution $\Delta f^{(a,b)}_{\rm ns}(T,L)$  at $T=T_c$ is proportional to $L^{-d}$.
Accordingly, near $T=T_c$ one has 
\begin{equation}
\Delta f^{(a,b)}(T,L)=L^{-(d-1)}\Delta X^{(a,b)}_{\mathrm{ex}}(a_\tau \tau L^{1/\nu};a_\omega L^{-\omega})+
O(L^{-d}), \quad a_\tau \tau L^{1/\nu}=O(1).
\label{finalCasimir}
\end{equation}
However, there are arguments,  that the nonsingular corrections in \eq{finalCasimir} are, in fact, exponentially small \cite{TD2010} (see also the discussion in Sect. \ref{sec:FSS_free_energy}).  If the system leaves the finite-size scaling region towards high
temperatures, i.e., if $a_\tau \tau L^{1/\nu}\gg 1$, the accumulated analytical and numerical evidences suggest that the 
bulk 
limit  is actually approached exponentially in $L$ \cite{Ped90,K94,BDT2000},
such 
that \cite{P90}
\begin{equation}
\Delta f^{(a,b)}(T,L)=O\left(\exp (-\mathrm{const.}\: L/\xi )\right), \quad  L/\xi\gg 1. 
\label{excorrectionsun}
\end{equation}
It has been shown, however, that 
these leading exponential corrections in \eq{finalCasimir} can 
stem not only from the singular contribution $\Delta f^{(a,b)}_{\rm sing}$ contained in  $\Delta f^{(a,b)}$ --- in which case they have to be of the form given by  Eq.~\reff{excorrectionsun} --- but also from the nonsingular
part $\Delta f^{(a,b)}_{\rm ns}$ \cite{CD99}, which both adopt the nonuniversal form
\begin{equation}
\Delta f^{(a,b)}(T,L)=O\left(\exp [-c(T)L] \right),
\label{eq:fab-exp-nsing}
\end{equation}
in which $c(T)$ is a nonsingular function of $T$. In certain cases \cite{BGD2012} these two mechanisms, related to the singular and nonsingular contributions of $\Delta f^{(a,b)}(T,L)$, can  even be in mutual competition, i.e., they interchange their roles in determining the leading behavior of the force as function of $T$. Examples of systems with such kind of corrections are those  in which critical Casimir forces  and electrostatic interactions compete, like in electrolytes \cite{CM2010,NDHCNVB2011,Bier2011,BGD2012,PCM2012,MKMD2014,Pousaneh2014}.

For systems with short-ranged interactions the overall size corrections relative to the thermodynamic bulk  behavior have to vary exponentially in $L$ for $T>T_c$. Because near $T_c$  the regular part of the free energy is a weakly varying function of $T$ and $L$, the finite-size part of the nonsingular free energy  decays exponentially for $T>T_c$. According to this understanding of the total finite-size part of the excess free energy one has
\begin{equation}
\Delta f^{(a,b)}(T,L)=L^{-(d-1)}\left[\Delta X^{(a,b)}_{\mathrm{ex}}(a_\tau \tau L^{1/\nu})+a_\omega L^{-\omega} \Delta X^{(a,b)}_{\mathrm{ex}, \omega}(a_\tau \tau L^{1/\nu})\right]+\cdots,
\label{finalCasimirExpansion}
\end{equation}
where an expansion with respect to the correction to scaling argument is performed. The ellipses in Eq.~\eqref{finalCasimirExpansion} refer to corrections of higher order than the ones we retained; $\Delta X^{(a,b)}_{\mathrm{ex}}(x_\tau)$ and $\Delta X^{(a,b)}_{\mathrm{ex},\omega}(x_\tau)$ decay exponentially for $x_\tau\gg 1$. 

The situation is more complicated if $a_\tau \tau L^{1/\nu }\ll -1 $. In
systems with $O(n\ge 2)$ symmetry, due to the spin wave excitations below $T_c$, the
correlation length remains infinite for $T<T_c$ (in the absence of an external field).
It is generally assumed 
that if there is no diffuse interface in the system, 
$\Delta f^{(a,b)}(T,L)=O(L^{-(d-1)})$ is of the same
order as in the critical finite-size scaling region \cite{D96,D98,BDT2000}. 
If, however, an interface is formed within the system due to the boundary
conditions, one has  $\Delta f^{(a,b)}(T,L)=\Upsilon (T)
L^{-1}+O(L^{-(d-1)})$ \cite{BDT2000,DG2009,BDR2011}, where $\Upsilon (T)$ is the helicity modulus. 
Finally, in the case of an Ising system below $T_c$ the corrections are expected to be exponentially small upon increasing $L$, because away from $T_c$ the 
corresponding correlation length is finite.
However,  this depends also on whether there is an interface within the film and possibly on its properties. In turn, the presence of such an interface depends on the effective boundary conditions which are realized at the surfaces of the film. In addition, this interface can be smooth or rough corresponding to $T<T_r$ and $T>T_r$ respectively; for $d=2$ the roughening temperature $T_r$ is zero  \cite{Priv92}.  If there is no interface present, the free energy approaches its bulk value exponentially upon increasing $L$, and therefore for $T<T_c$  the corresponding Casimir force vanishes analogously \cite{ES94,K97}. 
If in the two-dimensional Ising model the presence of an interface is enforced by the boundary conditions, the ensuing excess free energy decays as $L^{-2}$ for $T_w<T<T_c$ 
\cite{ES94,NN2008,NN2009,RZSA2010,AM2010} and exponentially below $T_w$, where $T_w$ is the wetting transition temperature\footnote{The value of $T_w$ for the $2d$ Ising model is reported in Sect. \ref{sec:models}.} \cite{Di88}. 
In the more general case that there is 
an interface, heuristic arguments \cite{PEN91,PE92}
concerning the form of the {\it singular} contributions to the excess free energy
due to the interface wandering suggest the following behavior: if $d<3$ and $T_w <T \ll T_c$  one has $\Delta f^{(a,b)}(T,L)\sim L^{-\tau}$, where $\tau=2(d-1)/(3-d)$ is the
so-called exponent for thermal wandering\footnote{This exponent has been introduced in Refs.  \cite{LF86,LF86b} in order to account phenomenologically
for fluctuation effects at wetting transitions. It shall not be confused with the reduced temperature notation $\tau$. }; if $d\geq 3$, one expects again an exponential decay of
$\Delta f^{(a,b)}(T,L)$, as in the case without an interface in the
system. The explicit mean field results \cite{PE92,K97} render  $\Delta f^{(a,b)}
(T,L)\sim \exp(-L/(2\xi))$. 
It remains to be seen if the {\it leading} order
finite-size contributions are of universal or nonuniversal character. For the singular part of $\Delta f^{(a,b)}(T \lesssim T_w,L)$ one has \cite{PEN91}
\begin{equation}
\Delta f^{(a,b)}_{\rm sing}(T,L)=L^{-\tau} \Delta X^{(a,b)}_{\mathrm{ex},w}
(a_\tau^{(w)}  \tau_w L^{1/\beta_s },h \tau_w^{-\Delta}),
\label{wetting}
\end{equation}
where $\tau_w=(T_w-T)/T_w$; $\beta_s$ is the critical
exponent which governs the growth of the wetting film at critical wetting  with 
thickness $l\propto |\tau_w|^{-\beta_s}$. We note that $\beta_s=0$ for short-ranged interactions and $\beta_s=1$ for systems with dispersion forces \cite{PE90,PE92,PEN91}. 

\smallskip 

\subsubsection{Excess free energy and Casimir force in systems with a weak anisotropy} 
 
Since we consider the film geometry, it is natural to allow for
an anisotropy of the interactions in the system which reflects
this geometry. To this end one chooses the lateral interaction constant $J_\|$
along the surface to be different
from  $J_\perp$ which holds perpendicular to the film. Since this kind of anisotropy, known as weak anisotropy, does not change
the universality class of the bulk system, one might surmise  that the scaling functions of the finite system will be the same as for the isotropic system. However, it has been argued  \cite{ChD2004,Do2008} that this is not the case and that one should expect these functions to be actually {\em nonuniversal} and to depend on the ratio $J_\perp/J_\parallel$. It has been shown  \cite{ChD2004,Do2008} that the main reason for this state of affairs is the need to generalize  the standard hyperuniversality hypothesis \cite{SFW72,A74,G75,HAHS76,W76,PHA91} (see Eq. \ref{HScalingH}). As it has been argued in Ref. \cite{DG2009}, on general grounds and explicitly demonstrated for the spherical model under periodic and anti-periodic   boundary conditions, and later confirmed in Ref. \cite{DC2009} on the basis of renormalization group considerations,  it is 
possible to relate the scaling functions of the normalized free
energy densities of the anisotropic system to those of the isotropic one. The corresponding relation between the scaling functions of the free energy is given by
\begin{equation}\label{relXf}
X_{f}^{\left({\zeta}\right)}(x_\tau|J_\perp,J_\parallel)=
\left[\frac{\xi_\perp(T)}{\xi_\parallel(T)}\right]^{d-1} X_{f}^{\left({\zeta}\right)}(x_\tau|J_\perp=J_\parallel),
\end{equation}
where $\xi_\parallel(\tau\to 0^+)\simeq \xi_{\|,0}^+\;\tau^{-\nu}$ and $\xi_\perp(\tau\to 0^+)\simeq \xi_{\perp,0}^+\;\tau^{-\nu}$ are the correlation lengths in the anisotropic system while $X_{f}^{\bc}(x_\tau)$ is the universal scaling function of the isotropic one. Equation  (\ref{relXf}) implies a similar relation for the Casimir force scaling functions:
\begin{equation}\label{relaniso}
    X_{\rm Cas}^{(\zeta)}(x_\tau|J_\perp,J_\parallel)=\left(\xi_\perp/\xi_\parallel\right)^{d-1}X_{\rm Cas}^{(\zeta)}(x_\tau|J_\perp=J_\parallel),
 \end{equation}
including a relation between the Casimir amplitudes in the anisotropic and isotropic system:
\begin{equation}\label{relDeltaGen}
\Delta_{\rm Cas}^{(\zeta)}(d|J_\perp,J_\parallel)=\left(\xi_\perp/\xi_\parallel\right)^{d-1} \Delta_{\rm Cas}^{(\zeta)}\left(d|J_\perp=J_\parallel\right).
\end{equation}
According to Refs. \cite{DG2009} and \cite{DC2009},  in the above expressions one can replace $\xi_\perp/\xi_\parallel$ by  $\sqrt{J_\perp/J_\|}$  (see, c.f., Eq. \eqref{relxi}). In Ref. \cite{KD2010} it is shown that \eq{relaniso} is also valid for the Gaussian model with $2<d<4$ for antiperiodic, Neumann-Neumann, Dirichlet-Dirichlet, and Dirichlet-Neumann boundary conditions. 

If the weak anisotropy exhibits more diverse features such as distinct  interactions along the diagonals of the unit cells of the lattice, or, more generally, between next-nearest neighbors, the overall picture becomes even more complicated \cite{ChD2004,DW2021,DWKS2021}.  For example, for a system with cubic geometry, represented  on a cubic lattice, at the bulk critical point and for periodic boundary conditions it has been suggested that the singular part $f^{(s)}$ of the total free energy ${\cal F}_{\rm tot}$ per $k_B T$,
\begin{equation}
	\label{eq:Dohm}
	{\cal F}_{\rm tot}(T=T_c,h=0,L)=L^d f^{(s)}(T=T_c,h=0,L),
\end{equation} 
depends on $d(d+1)/2-1$ nonuniversal parameters giving rise to the so-called multiparameter universality. For the two-dimensional Ising model on a square lattice with periodic boundary conditions, these predictions have been verified by high-precision Monte Carlo simulations  (see Ref. \cite{DWKS2021}). There the considered interactions are isotropic ferromagnetic couplings between nearest
neighbors and an anisotropic coupling between next-nearest neighbors
along one diagonal.
For the more general case of a $d$-dimensional cube the  hypothesis of multiparameter universality \cite{Do2008,D2018} states:  
\begin{equation}
	\label{eq:Dohm_gen_d}
	f^{(s)}(\tau,h,L)=L^{-d} X_{f,\,{\rm cube}}(\hat{a}_\tau \tau \hat{L}^{1/\nu},\hat{a}_h \hat{h}\hat{L}^{\Delta/\nu}; \bar{\mathbf{A}}),
\end{equation}
where $\hat{L}=L\left(\det \mathbf{A}\right)^{-1/(2d)}$, $\hat{h}=h \left(\det \mathbf{A}\right)^{1/4}$, $\bar{\mathbf{A}}=\mathbf{A}/\left(\det \mathbf{A}\right)^{1/d}$, and $\det \mathbf{A}>0$. For the free energy density $\hat{f}^{(s)}=f^{(s)}\left(\det \mathbf{A}\right)^{1/2}$ of a parallelepiped with volume $\hat{V}=V \left(\det \mathbf{A}\right)^{-1/2}$ and with anisotropy matrix $\hat{\mathbf{A}}/\left(\det \hat{\mathbf{A}}\right)=\mathbf{1}$ (i.e., after transferring the anisotropic system to the isotropic one) one obtains  
\begin{equation}
	\label{eq:Dohm_isotropic}
	\hat{f}^{(s)}(\tau, \hat{h},\hat{L})=\hat{L}^{-d}X_{f,\,{\rm cube}}(\hat{a}_\tau \tau \hat{L}^{1/\nu},\hat{a}_h \hat{h}\hat{L}^{\Delta/\nu}; \bar{\mathbf{A}}),
\end{equation}
Here the dimensionless $d\times d$ matrix $\mathbf{A}$ is defined as 
\begin{equation}
	\label{eq:A-matrix}
	\mathbf{A}\equiv (A_{\alpha,\,\beta}), \quad A_{\alpha,\, \beta}=\lim_{N\to\infty}\frac{1}{N}\sum_{i,j} (x_{i,\alpha}-x_{j,\alpha})(x_{i,\,\beta}-x_{j,\,\beta})K_{i,j},
\end{equation}
where $K_{i,j}=\beta J_{i,j}$ describes the interaction between the spins on the lattice sites $i$ and $j$,  where $\mathbf{x}_i\equiv (x_{i1}, x_{i2}, \cdots, x_{id})$ are the lattice sites in units of the lattice constants. 

Equation \reff{eq:Dohm_isotropic}  relates the expression for the anisotropic system to the one for the isotropic system. Indeed, 
the above equation has the structure of \eq{PFhypot}, due to Privman and Fisher, for the isotropic system (with $L_\perp=L_\|$). But the dependence on $\bar{\mathbf{A}}$ amounts to the additional dependence on $d(d+1)/2-1$ nonuniversal parameters. The two additional "standard" nonuniversal constants $\hat{a}_\tau$ and $\hat{a}_h$ can be expressed, as usual, in terms of the  amplitudes $\hat{\xi}_0^+$  and $\hat{\xi}_h$ of the asymptotic behavior of the bulk correlation lengths for ($T>T_c, \hat{h}=0$) and for ($T=T_c, \hat{h}\to 0$), respectively, of the transformed isotropic system. 

Therefore, by
means of an appropriate rescaling of lengths, a transformation
from a weakly anisotropic to an asymptotically isotropic system is always possible,
provided that the anisotropy matrix $\mathbf{A}$ is positive definite
and that the rescaling is performed along the $d$
nonuniversal directions of its principal axes \cite{Dohm_2006}. In general, these axes 
differ from the symmetry axes of the system. 
This rescaling distorts, e.g., the shape and the boundary conditions in a
nonuniversal way from, say, a cube to a parallelepiped,  and from periodic b.c. in rectangular directions to ones in nonrectangular directions. This nonuniversality is captured by the dependence of
the scaling function $X_f$ on the anisotropy matrix $\mathbf{A}$, in addition to the dependences on $\hat{a}_\tau$ and $\hat{a}_h$.

The above results simplify substantially in the limit $n\to\infty$  of $O(n)$ models characterized  by the matrix $\mathbf{A}$  in the case of a film geometry with periodic boundary conditions.  As shown in Ref. \cite{ChD2004},  for such a set-up one has the following relation between the Casimir amplitudes $\Delta_{\rm Cas, \,aniso}$ of the anisotropic system and $\Delta_{\rm Cas, \,iso}$ for the isotropic one: 
\begin{equation}
	\label{eq:Casimir_amplitudes_aniso_iso}
	\Delta_{\rm Cas, \,aniso}= \left[(\bar{\mathbf{A}}^{-1} )_{d,d}\right]^{-d/2}\,\Delta_{\rm Cas, \,iso},
\end{equation}
where $(\bar{\mathbf{A}}^{-1} )_{d,d}$ is the $d$-th diagonal element
of the inverse of matrix $\bar{\mathbf{A}}$. For the scaling behavior of the singular part of the free energy of the anisotropic system one finds
\begin{eqnarray}
	\label{eq:free-energy-film-aniso}
	f^{(s)}(\tau,h,L)=L^{-d} \left[(\bar{\mathbf{A}}^{-1} )_{d,d}\right]^{-d/2}\, X_{f,\, {\rm film, \, iso}}\left( \left[ L \left((\mathbf{A}^{-1} )_{d,d}\right)^{1/2}\bigg/\xi'\right]^{1/\nu}\right),
\end{eqnarray}
where $X_{f, \, {\rm film, \, iso}}$ is the scaling function of the isotropic film system, $\xi'$ is its bulk correlation length, and $(\mathbf{A}^{-1} )_{d,d}$ is the $d$-th diagonal element
of the inverse of the matrix $\mathbf{A}$.

In the present context, we recall that the Casimir force is observable if
the ordering degrees of freedom can enter and leave the system. Therefore,
this force can be realized only for "fluids". However, most fluids are
isotropic. Exceptions are ordered liquid crystals and, potentially, the
"fluid" of Cooper pairs when a superconducting thin film is connected to
a bulk sample of the same material \cite{W2004}. The latter hypothesis \cite{remDohm} requires and calls for a thorough
theoretical exploration.

\smallskip 

\subsubsection{Size dependence of the excess free energy in systems with long-ranged interactions}

The study of the Casimir effect in systems with
long-ranged interactions, such as dispersion forces \cite{P2006,I2011,KL2003,MN76}, which fall off at large distances $\propto r^{-d-\sigma}$ with $\sigma > 0$, reveals certain peculiarities compared with the case of
short-ranged interactions. Specifically, one considers the pair potentials of the interacting atoms or molecules\footnote{The situation is even more complicated for binary liquid mixtures of $A$ and $B$ particles, where one has to distinguish the $A-A$, $B-B$, and $A-B$ interactions.} to decay at large distances as $r^{-3-\sigma}$, where
$\sigma = 3$ represents the standard and $\sigma = 4$ the retarded 
van der Waals interactions. We assume that retardation
sets in for distances $r >\xi_{\rm ret}$ , where the retardation length\footnote{The retardation length corrects the van der Waals energy with respect to the finite speed of light $c$, i.e., it describes the crossover from the nonretarded (standard) van der Waals force characterized by $\sigma=3$ to a retarded, Casimir one, characterized by $\sigma=4$. One can also consider an interaction which smoothly interpolates between the standard and the retarded (i.e., Casimir) van der Waals
interactions, depending on the value of $\xi_{\rm ret}$. For $d=3$, this can be achieved by the expression  $F_{\rm tot}(T,h,L)=-\left[A_{\rm Ham}/(6\pi)\right] L^{-3}\left[ 1+5L/(3\xi_{\rm ret})\right]^{-1}$ \cite{DV2012}. 
\\ \hspace*{0.3cm}	
As stated, $\xi_{\rm ret}$ is a length specific for the corresponding medium, depending, inter alia, on the speed of light in that medium and the electronic structure of its constituents. For $^4$He one has $\xi_{\rm ret}=193\, \AA$ \cite{GC99}. }  $\xi_{\rm ret}$ is a length specific for the corresponding medium. In the most general case
one can consider $\sigma > 2$ as a real continuous variable governing
the decay of the pair potential. 

In view of the long-ranged character of this interaction, there is 
a direct attraction between the surfaces bounding the system, which is not mediated by the fluctuations of the medium between them. If $T<T_c$ 
one can readily argue that the  $L$-dependent part of
the excess free energy, due to the direct inter-particle interaction, is
of the order of $L^{-\sigma+1}$. Furthermore, the bounding surfaces exert a substrate potential which decays  $\propto z^{-\sigma}$ as function of the distance $z$ from them, i.e., any particle of the fluid is influenced by an external, algebraically decaying force field. 
In the critical region there are still remnants of the direct long-ranged interaction.  Due to the large   compressibility of the critical fluid (infinite at $T=T_c$), richly structured density profiles emerge. Against this background, the 
\textit{fluctuations} of the order parameter give rise to an additional force, i.e., the critical Casimir force,  which acts on the confining surfaces. The discussion above leads to the necessity to define the Casimir force as given by \eq{CasDef}, which is usually employed for systems with short-ranged interaction, so that one can extract from the total force, which acts on the confining plates of the system, that contribution, which is due to the fluctuations. Below we explain how this procedure can be implemented.  The total force acting between the bounding surfaces is
\begin{equation}
\label{totDef}
\beta F_{\rm tot}^{\bc}(T,h,L)\equiv- \frac{\partial}{\partial L}f_{\rm ex}^{\bc}(T,h,L). 
\end{equation}
For $\sigma>2$ and $T\ne T_c$ this force has the form \cite{VD2015} 
\begin{equation}
\label{way}F_{\mathrm{tot}}(T,h,L) \simeq (\sigma-1) H_A(T,h) L^{-\sigma}\xi_{\mathrm{ret}}^{\sigma-d}.
\end{equation}
One typically considers the case $d=\sigma$ (which corresponds to the  standard case of van der Waals interactions) and omits the apparent dependence on the so-called retardation length $\xi_{\mathrm{ret}}$ \cite{CC88,GC99,DV2012,VD2015,VD2017}. Here $H_{A}$ is the Hamaker term\footnote{ $H_{A} <0$ corresponds to attraction between the plates, $H_{A} >0$ to repulsion.}, the dependence of which on $T$ and $h$ is given by the so-called Hamaker constant \cite{P2006,I2011} 
\begin{equation}\label{HamConstant}
A_{\rm Ham}(T,h)=-12\pi H_{A}(T,h).
\end{equation}
The Hamaker constant is a constant only in the sense that it does not depend on  $L$.  It depends smoothly on $T, h$, and the material properties of the fluid medium and the walls. The factor $12\pi$ in \eq{HamConstant} is there due to historical reasons, according to which the interaction energy between two substrates at a finite separation $L$ in the case of standard van der Waals interactions (i.e., $d=\sigma=3$), and away from any phase transition, is \cite{P2006,I2011}
\begin{equation}\label{deltaomegaincomprfluid}
f_{\rm{ex}}(T,h,L)=-\dfrac{1}{12\pi}A_{\rm Ham}(T,h) L^{-2}=H_{A}(T,h)L^{-2}.
\end{equation}
Near the critical temperature $T_c$ of the bulk system, \eq{way} is no longer valid because the critical fluctuations of the order parameter lead to additional  contributions to the total force. For such a system, following Refs. \cite{DSD2007,VD2015}, near the bulk critical point \eq{totDef} can be written as
\begin{equation}
\label{Casimir_dispersion}
\beta F_{\rm tot}^{\bc}(T,h,L)=L^{-d} X^{\bc}_{\rm crit}(x_\tau,x_h;x_\omega,x_b,x_{s,1},x_{s,2})+\beta(\sigma-1)H_A(T,h)L^{-\sigma} \xi_{\mathrm{ret}}^{\sigma-d},
\end{equation}
with
\begin{equation}
\label{Cas_nonpolar}
F_\Cas(T,h,L)=F_{\rm tot}^{\rm (sing)}(T,h,L),
\end{equation}
where 
$x_\omega=a_\omega L^{-\omega}$ is the usual correction-to-scaling variable. The exponent $\omega$ is the same as for systems with short-ranged interactions, and 
\begin{equation}
\label{dispersion_scalin_variables}
x_b= b L^{-\omega_l}, \qquad x_{s,i}= s_i L^{-\omega_{s}}, i=1,2,
\end{equation}
with 
\begin{equation}
\label{omega_def_dispersion}
\omega_l=\sigma-(2-\eta) \quad \mbox{and}\quad \omega_{s}= \sigma-(d+2-\eta)/2. 
\end{equation}
In \eq{dispersion_scalin_variables} the parameter $b$ measures the strength of the long-ranged part of the fluid-fluid interaction while the parameters  $s_i$, $i=1,2$, describe the contrast between  the physical properties of the fluid and of the substrates limiting the system;  in \eq{omega_def_dispersion} $\eta$ is the standard bulk critical exponent characterizing the decay of the two-point correlation function at $T_c$ in a system with purely short-ranged interactions.
The form of \eq{Casimir_dispersion} describes the most general case in which the constituents of the $d$-dimensional slab are mutually interacting with a strength $\propto b r^{-d-\sigma}$ (at large distances between them), while the boundaries are exerting substrate potentials onto these constituents $\propto s_i z_i^{-\sigma}$, where $z_1$ and $z_2$ are the distances between a given position within the slab and the boundaries $i=1$ and $i=2$. 

Accordingly, one expects
a crossover from a regime governed by the critical Casimir force (the magnitude of which is
of the order of $L^{-d}$, see Eq.~(\ref{Casimir_scaling})), to the one governed by the direct attraction with the magnitude of the order of $L^{-\sigma}$ (see \eq{way}). However, for van der Waals type of  interactions one has  $d=\sigma=3$, i.e., these two effects are of the same order, and each of them will  dominate in different temperature regions. Wetting films provide a relevant physical realization of this case. To this end  one considers a wetting layer near bulk criticality which intrudes between two noncritical phases, and takes into account the effects of long-ranged correlations and the long-ranged van der Waals forces\footnote{In this case both finite-size and van der Waals forces give rise to a leading contribution to the free energy of the wetting layer  $\propto L^{-2}$ in  $d=3$.} \cite{NI85}.

Further details concerning the definitions of the above parameters, as well as concerning  the finite-size behavior of systems governed by dispersion interactions, are provided in Refs.~\cite{DKD2003,DDG2006,DRB2007,DRB2009,D2001,DR2001,VD2015,VD2017}.  Here we  only mention that  there is a critical thickness  $L_{\rm crit}$ of the system such that for $L<L_{\rm crit}$ the long-ranged tails of the interaction turn out to produce leading order contributions to the free energy of the system even at $T_c$ and thus can influence even the sign of  both the Casimir and the total force \cite{VD2015,VD2017}.

\bigskip 

\subsubsection{Finite-size scaling within the mean-field regime}

If the dimension $d$ of a system is above the upper critical\footnote{We recall that for $O(n)$ systems with short-ranged interactions one has $d_u=4$.} dimension, i.e., if $d>d_u$, 
the critical exponents of the system are independent of $d$ and attain their mean-field values. Moreover, hyperscaling, as well as the related two-factor universality (see above), are no longer valid \cite{B82,P90,BDT2000,ZJ2002,PV2002}. In addition to studying systems with fixed dimension $d>d_u$, one can also construct effective theories, starting, e.g., from the $({\mathbf \Phi}^2)^2$ model (see Eqs.~\eqref{Hb} and \eqref{Hs}), by replacing ${\mathbf \Phi}({\mathbf x}, z)$ with its statistical mean value ${\mathbf M}(z)$ which is determined by minimizing ${\cal H}={\cal H}_b+{\cal H}_s$ with respect to ${\mathbf M}(z)$. Within the film geometry, the corresponding model becomes effectively one-dimensional and its parameters carry the information about the actual dimension $d$ of the system. Such a model is characterized by the mean field critical exponents 
\begin{equation}
\label{eq:MF_crit_exponents}
\beta_{MF}=1/2,\qquad \alpha_{MF}=0, \qquad \mbox{and} \qquad \nu_{MF}=1/2.
\end{equation}
All the other critical exponents can be obtained from these values 
by using the scaling relations. Formally the mean field critical exponents satisfy the scaling relations for systems with short-ranged interactions and $d=4$. 

An extensive list of relations between the variety of possible surface critical exponents in the mean-field case is compiled in Ref.~\cite{Bb83}. Here we report only the values of those exponents which were mentioned in the previous discussion and which are relevant in the present context: 
\begin{equation}
\label{eq:surface_MF_crit_exponents}
\Delta_{1,MF}^{O}\equiv \Delta_{1,MF}=1/2,\qquad \Delta_{1,MF}^{SB}=1, \qquad \mbox{and} \qquad \phi_{MF}=1/2.
\end{equation} 
Within the  mean-field ${\mathbf \Phi}^4$ theory the  Eqs.~\eqref{surface_bulk} and \eqref{ordinary-extraordinary} --- applied to the singular part of the finite-size free energy density --- remain valid with the corresponding critical exponents replaced by the mean-field values reported in Eqs.~\eqref{eq:MF_crit_exponents} and \eqref{eq:surface_MF_crit_exponents}.  
However, one should  bear in mind that the corresponding scaling function $X_{f}^{MF}$ 
contains a nonuniversal pre-factor which can be expressed in terms of the inverse of  the fourth-order coupling constant $g$ in ${\cal H}_b$ (see \eq{Hb} \cite{K97,K99,VMD2011,DSD2007,GaD2006}).

The theoretical analysis becomes significantly more challenging  for a system with dimension $d>4$ if one aims at studying  its behavior without resorting to the replacement of the order parameter by its mean value, which is obtained by minimizing the Hamiltonian considered as a functional of the mean value of the profile of the order parameter profile. In order to understand why this approach is much more challenging, we recall the renormalization-group prediction for the singular part of the free energy \cite{P90} (compare \eq{generlaRGfiniteL}):
\begin{equation}
\label{men_field_d}
f^{\rm (s)}(\tau,h,L)=L^{-d}X_f^{MF}(a_\tau\tau L^{y_\tau},a_h h L^{y_h},u L^{y_u}),
\end{equation}
where 
\begin{equation}
\label{mean_field_RG_exponents}
y_\tau=2, \qquad y_h=\frac{d+2}{2}, \quad \mbox{and} \quad y_u=4-d
\end{equation}
are the renormalization-group eigenexponents at the Gaussian fixed point, and $u$ is proportion to the coefficient $g$ in \eq{Hb} of the fourth-order term in the Ginzburg-Landau Hamiltonian.  
The delicate point is that the scaling function 
$X_f^{MF}(\omega_\tau,\omega_h,w)$ is a singular function of its last variable $w=u L^{y_u}$ as $w\to 0$ and therefore it cannot be expanded in terms of it: $w$ provides an example of the so-called dangerous irrelevant variable \cite{P90,F74}. 
Furthermore, it turns out that the actual nature of this singularity depends on the number of spatial directions along which the film is effectively infinite (e.g., two in a  three-dimensional film). Accordingly, the topic of finite-size scaling above the upper critical dimension is still an active research topic \cite{LM2014,KB2014,WY2014,KB2015,FlBKW2015,FBKW2016,KB2017,GEMZGD2017,GGT2021}.

 With this we finish our brief account of finite-size scaling theory. In the following chapter we apply the knowledge presented in the current one in order to study the Casimir force in thermodynamic systems undergoing phase transitions. Strong emphasis will be given to systems exhibiting continuous  (i.e., critical) phase transitions. 

Thus, in the following, we present the available exact results for the thermodynamic Casimir effect. 

%

\section{Exact results for the thermodynamic Casimir effect in chains}

In the current section we consider chains with nearest neighbor interaction $J$ on one-dimensional lattices. All lengths are measured in units of the lattice constant $a$. 

Before reporting the available exact results for several models, we note that recently one-dimensional and quasi one-dimensional systems have been objects of serious experimental interest, see, e.g., Ref. \cite{Balandin2022a} and references therein. Some of these systems, like ${\rm TaSe_3}$, are
quasi one-dimensional in the sense that they have strong covalent bonds in one
direction, i.e., along the atomic chains, and weaker bonds in the perpendicular
plane \cite{Stolyarov2016}. Other systems are
more properly considered to be truly one-dimensional materials, in that they have covalent bonds only
along the atomic chains and only much weaker van der Waals interactions
in the perpendicular directions \cite{Balandin2022}. These one-dimensional "van der Waals materials" have emerged as an entirely new research field, which encompasses interdisciplinary studies by physicists, chemists, materials
scientists, and engineers \cite{Balandin2022a}. The chains considered here can be seen as the simplest possible examples of such one-dimensional materials.

\subsection{One-dimensional Ising model}

We start by discussing systems with discrete symmetry. Among them, the simplest case is the one-dimensional Ising chain of $N$ spins. We  recall that, for Ising chains with short-ranged interactions,  $T_c=0$ is an essential critical point, and that in terms of $\tau=\exp(-2K)\to 0$, $K=\beta J$, and $h$ one can identify the usual power laws with the exponents \cite{B82}
\be
\label{eq:crit_exponents}
\alpha=\gamma=\nu=\eta=1, \beta=0, \delta=\infty, \; \mbox{but such that}\; \beta \delta=1. 
\ee

The scaling functions for the Casimir force in the one-dimensional Ising model have been determined in Ref. \cite{RZSA2010} for periodic, antiperiodic, free, and fixed boundary conditions. We note that the correlation length for the macroscopically long Ising chain  is 
\begin{equation}
\label{eq:corr_length_1d_Ising} 
\xi_\tau=-1/\ln \tanh(K), \quad \xi_\tau \simeq (2\tau)^{-1} \quad  \mbox{for} \quad K\gg 1.  
\end{equation}
If $N$ is the total number of spins in the system, the results for the scaling functions, in terms of the scaling variable $x_\tau=L/\xi_\tau$,  are as follows: 

a) for {\it periodic} boundary conditions
\begin{equation}
\label{eq:Xcas_per_1d_Ising}
X_{\rm Cas}^{(p)}(x_\tau)=-\frac{x_\tau}{\exp(x_\tau)+1}; 
\end{equation}

b) for {\it antiperiodic} boundary conditions
\begin{equation}
\label{eq:Xcas_aper_1d_Ising}
X_{\rm Cas}^{(ap)}(x_\tau)=\frac{x_\tau}{\exp(x_\tau)-1}.
\end{equation}

\begin{figure}[!htb]
	\includegraphics[width=0.95\columnwidth]{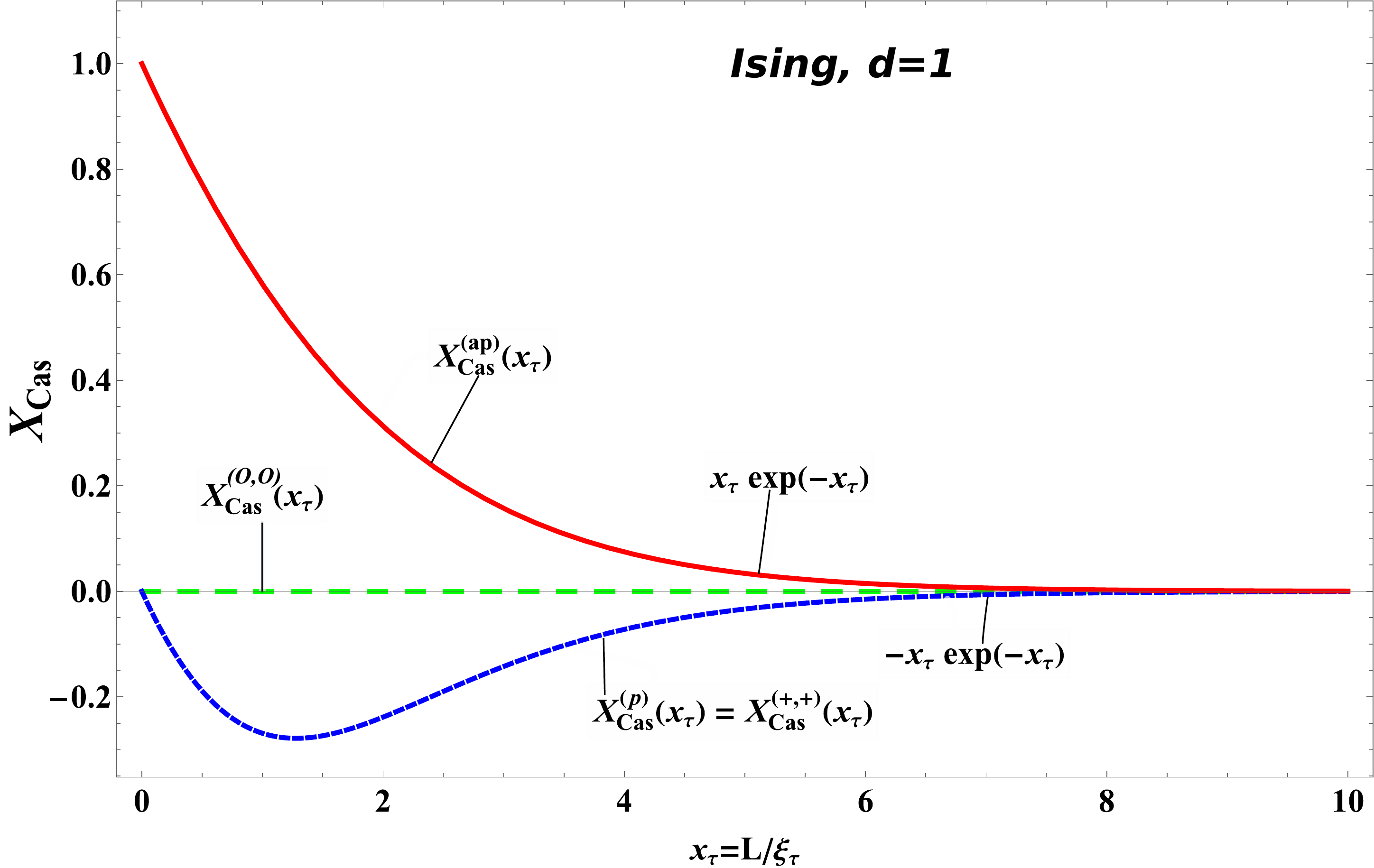}
	\caption{The scaling functions $X_{\rm Cas}$ for the critical Casimir force of the one-dimensional Ising model as a function of the scaling variable $x_\tau$ (see \eq{eq:corr_length_1d_Ising}). For  $x_\tau \gg 1$ one has $X_{\rm Cas}^{(p)}(x_\tau)\simeq -x_\tau \exp (-x_\tau)$ and $X_{\rm Cas}^{(ap)}(x_\tau)\simeq x_\tau \exp (-x_\tau)$ --- see Eqs. \eqref{eq:Xcas_per_1d_Ising} and \eqref{eq:Xcas_aper_1d_Ising}, respectively. For antiperiodic boundary conditions the critical Casimir force is repulsive, whereas for the other boundary conditions it is attractive. }
	\label{Fig:1d_Ising_Casimir}
\end{figure}

Fox {\it fixed} boundary conditions (i.e., the spins at both ends of the chain are fixed to the same value), one has $X_{\rm Cas}^{(+,+)}=X_{\rm Cas}^{(p)}$, while for {\it free}  boundary conditions (i.e., missing neighbors) one finds $X_{\rm Cas}^{(O,O)}=0$. The scaling functions $X_{\rm Cas}^{(p)}(x_\tau)$ and $X_{\rm Cas}^{(ap)}(x_\tau)$ are shown in Fig. 	\ref{Fig:1d_Ising_Casimir}.

\subsection{One-dimensional models with continuous  symmetry in the presence of boundary fields} 
\label{sec:continuum}
\label{1d_systems}
We consider two one-dimensional models ($d=1$) with continuous $O(n)$ spin symmetry: the XY ($n=2$) and the  Heisenberg ($n=3$) chain. The definitions of these models are given in Sect. \ref{sec:models}.

As shown in Ref. \cite{DaR2017}, if the boundary fields are non-zero,  the Casimir force $F_{\rm Cas}$ of these systems displays a very rich and interesting behavior. It turns out that near $T=0$ this force exhibits scaling and --- depending on the angle between the vectorial boundary fields $\mathbf{H}_1$ and $\mathbf{H}_N$, and on the value of the temperature scaling variable  $x\propto N k_B T/J$ --- the force can be attractive or repulsive. Explicitly, one finds the following properties: 
\begin{enumerate} 
	\item[(i)] At  low temperatures, for which $x={\cal O}(1)$, and for 
	\begin{equation}
	\label{eq:constraint}
	N\gg J\left(\frac{1}{H_1}+\frac{1}{H_N}\right),
	\end{equation}
	the leading behavior of the Casimir force can be written in the form
	\begin{equation}
	\label{eq:1dCas_gen}
	\beta F_{\rm Cas}(T,N,{\bf H}_1,{\bf H}_N)=N^{-1}X(\psi,x),
	\end{equation}
	where $x$ is  the scaling variable reported above,  $\psi=\psi_N-\psi_1$ (see \eq{eq:system_angles}), and $X$ is a universal scaling function. Equation (\ref{eq:1dCas_gen}) implies that, under the constraint formulated in \eq{eq:constraint}, $X_{\rm Cas}$ depends only on the scaling variable $x$ and the angle $\psi$. The latter parameter effectively describes the boundary conditions of the system. Unlike the Ising model, here the boundary conditions depend {\it continuously} on a single parameter---in our notation $\psi$. 
	
	\item[(ii)] When $x\to 0^+$ the scaling function of the Casimir force becomes positive, i.e., the force turns repulsive, provided that $\psi \ne 0$. In that latter case $X_{\rm Cas}\propto  x^{-1}$ so that the overall dependence of the force on $N$ is of the order of $N^{-2}$ (see \eq{eq:1dCas_gen}).
	
	\item[(iii)] If $x\gtrsim 1$, the sign of the scaling function is opposite to the sign of $\cos(\psi)$: for $0<|\psi|<\pi/2$, $\cos(\psi)>0$ and $X<0$ so that the force is attractive, while for $\pi/2<|\psi|<\pi$,  $\cos(\psi)<0$ and $X>0$ so that the force  is repulsive. For $x\gg 1$ the force vanishes exponentially. 
	
	\item[iv)] For any $\psi$ with  $0<|\psi|<\pi/2$, the Casimir force changes from attractive to repulsive upon lowering  the  temperature from a moderate value to zero for a fixed system size $N$. 
	
	\item[v)] If $\psi=0$, the force is attractive for {\it any} value of the scaling variable $x$. 
	
\end{enumerate} 

These one-dimensional models have been studied analytically for free (frequently called ``open'' or Dirichlet boundary conditions) and for periodic boundary conditions \cite{F64,J67,J67b,S68,PB2011}. However, the properties discussed above manifest themselves  only via  the presence of boundary fields. 

Below we shall present results specific for the XY and for the Heisenberg models, respectively.

\subsubsection{The one-dimensional XY model}
\label{sec:1dXY_model}

The definition of this model and the explanation of the corresponding  notations are provided in Sect. \ref{sec:models}.  The Casimir force in this system is given {\it exactly} by the following expression \cite{DaR2017}: 
	\begin{equation}
	\label{FCas}
	\beta  F_{\text{Cas}}=
	\frac{ 2\sum _{k=1}^{\infty } 
		\cos \left[k (\psi_1-\psi_N)\right] \log \left[\frac{I_k(K)}{I_0(K)}\right]\frac{I_k\left(h_1\right)}{I_0\left(h_1\right)} \left(\frac{I_k(K)}{I_0(K)}\right)^{N-1} \frac{I_k\left(h_N\right)}{I_0\left(h_N\right)}
	}{1+2 \sum _{k=1}^{\infty } \cos \left[k (\psi_1-\psi_N)\right] \frac{I_k\left(h_1\right)}{I_0\left(h_1\right)} \left(\frac{I_k(K)}{I_0(K)}\right)^{N-1} \frac{I_k\left(h_N\right)}{I_0\left(h_N\right)}},
	\end{equation}
where $I_k(z)$ is the modified Bessel function of the first kind, and 
\begin{equation}
\label{eq:def_parameters}
\psi \equiv \psi_1-\psi_N, K\equiv \beta J,\, h_1\equiv \beta H_1, h_N\equiv\beta H_N. 
\end{equation}
We note that the force depends only on the difference $\psi_1-\psi_N$,  and not on $\psi_1$ and $\psi_N$ separately.

In the limit $\beta J\gg 1$, i.e., for $T\to 0$, one obtains
\begin{equation}
\label{eq:F_Cas_scaling}
\beta F_{\rm Cas}(x)=\frac{1}{N_{\rm eff}}X_{\rm Cas}(\psi,x,h_{\rm eff})
\end{equation}
where 
\begin{equation}
\label{eq:F_Cas_scaling_function}
X_{\rm Cas}=-x\frac{ \sum _{k=1}^{\infty } k^2
	\cos \left(k \psi\right) \exp
	\left[-\frac{1}{2}
	k^2\left(h_{\rm eff}^{-1}+x\right)\right]}{
	1+2 \sum _{k=1}^{\infty } \cos \left(k \psi\right) \exp \left[-\frac{1}{2}
	k^2\left(h_{\rm eff}^{-1}+x\right)\right]}
\end{equation}
and
\begin{equation}
\label{eq:scaling_variables}
x\equiv \frac{N_{\rm eff}}{K}, \qquad h_{\rm eff}^{-1}=h_1^{-1}+h_N^{-1}, \qquad N_{\rm eff}=N-1.
\end{equation}
Here the ratio $x$ serves as a temperature  dependent scaling variable, which in systems with a non-zero transition temperature takes the form $x=\tau^{\nu}(L/\xi_0^+)=L/\xi_\tau^+$, with  $\tau$ as the reduced temperature, i.e., $\tau= (T-T_c)/T_c$, $L$ as the characteristic linear size of the finite system, and $\nu$ as the correlation length exponent ($\xi^+=\xi_0^+ \tau^{-\nu}$).  With an effective transition temperature  $T=0$ and $K \propto 1/T$, the definition in \eq{eq:scaling_variables} is consistent with $x=L/\xi_\tau^+$ under the assumption $\nu=1$, because  $x\propto K^{-1}N= (k_B T/J)N\equiv L/\xi_\tau^+$, with $L=N a$ and $\xi_\tau^+=a(k_B T/J)^{\nu}$, where $a$ is the lattice constant of the chain. 

Obviously,  if the relation in \eq{eq:constraint} is fulfilled,  
one has $x\gg h_{\rm eff}^{-1}$, and in \eq{eq:F_Cas_scaling} one can safely ignore $h_{\rm eff}$. In this case the behavior of the force is the one stated in \eq{eq:1dCas_gen}.

The representation of $X_{\rm Cas}$ given by \eq{eq:F_Cas_scaling} is suitable for all values of $x$ except the limit $x \ll 1$. In that limit one obtains 
\begin{eqnarray}
\label{eq:F_Cas_scaling_small_x}
\lefteqn{X_{\rm Cas}(\psi,x,h_{\rm eff})=-\frac{x}{2
		\left(x+h_{\rm eff}^{-1}\right)}}\\
&&+\frac{x}{2\left(x+h_{\rm eff}^{-1}\right)^2}\frac{ \sum _{n=-\infty}^{\infty} \left(2 n \pi
	+\psi\right)^2
	\exp{\left[-\frac{\left(2 n \pi +\psi\right)^2}{2
			\left(x+h_{\rm eff}^{-1}\right)}\right]}}{
	\sum _{n=-\infty}^{\infty}
	\exp{\left[-\frac{\left(2 n \pi +\psi\right)^2}{2
			\left(x+h_{\rm eff}^{-1}\right)}\right]}}. \nonumber
\end{eqnarray}
Provided the constraint in \eq{eq:constraint} is fulfilled and that the asymptotic behavior of $X_{\rm Cas}$ is given by Eqs. \eqref{eq:F_Cas_scaling} and \eqref{eq:F_Cas_scaling_small_x}, one obtains
\begin{equation}
\label{X_cas_ass}
X_{\rm Cas}(\psi,x)=\left\{ \begin{array}{ll}
\frac{1}{2x} \psi^2-\frac{1}{2}+{\mathcal O}\left\{\frac{2 \pi}{x}\exp{\left[-\frac{2 \pi^2}{x}\right]}\max\left\{ (\pi -\psi ) \exp{\left[\frac{2 \pi\, \psi }{x}\right]}, (\pi +\psi ) \exp{\left[-\frac{2 \pi\, \psi }{x}\right]}\right\}\right\}, & x\to 0^+, \\
-x \cos(\psi) \exp(-x/2), & x \gg 1. 
\end{array}
\right.
\end{equation}
\begin{figure}[h]
	\includegraphics[width=\columnwidth]{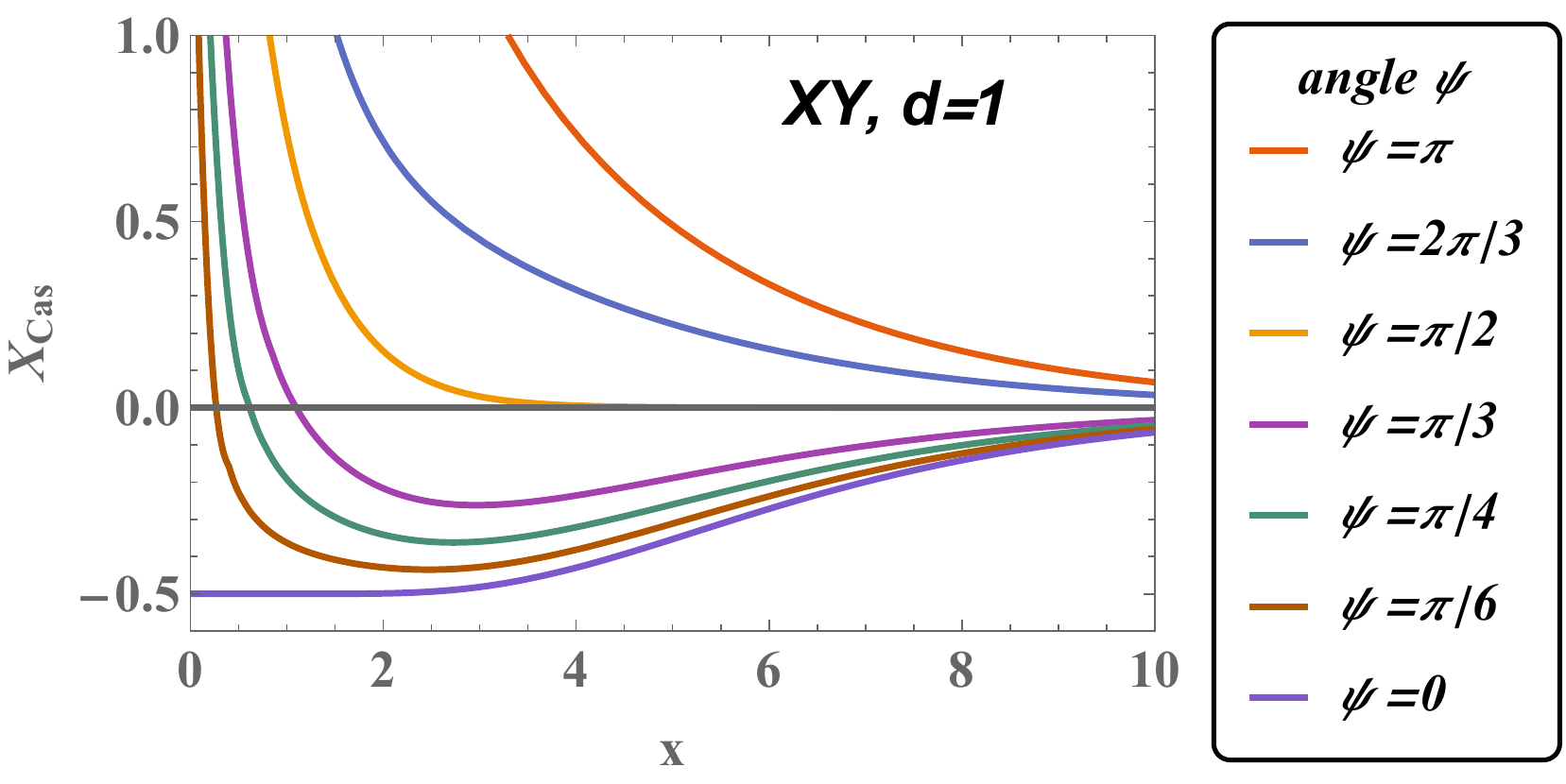}
	\caption{The scaling function $X_{\rm Cas}$ for the critical Casimir force of the XY model in $d=1$ as a function of the scaling variable $x=(N-1)/K$ (see \eq{eq:scaling_variables}) for seven values of the phase difference $\psi$. For $x \gg 1$, $X_{\rm Cas} \propto -x \cos(\psi) \exp(-x/2)$,  and for $x \to 0^+$, $X_{\rm Cas} \propto -1/2 +\psi^2/(2x)$.}
	\label{Fig:1d_X_Cas}
\end{figure}
\begin{figure}[h]
	\includegraphics[width=\columnwidth]{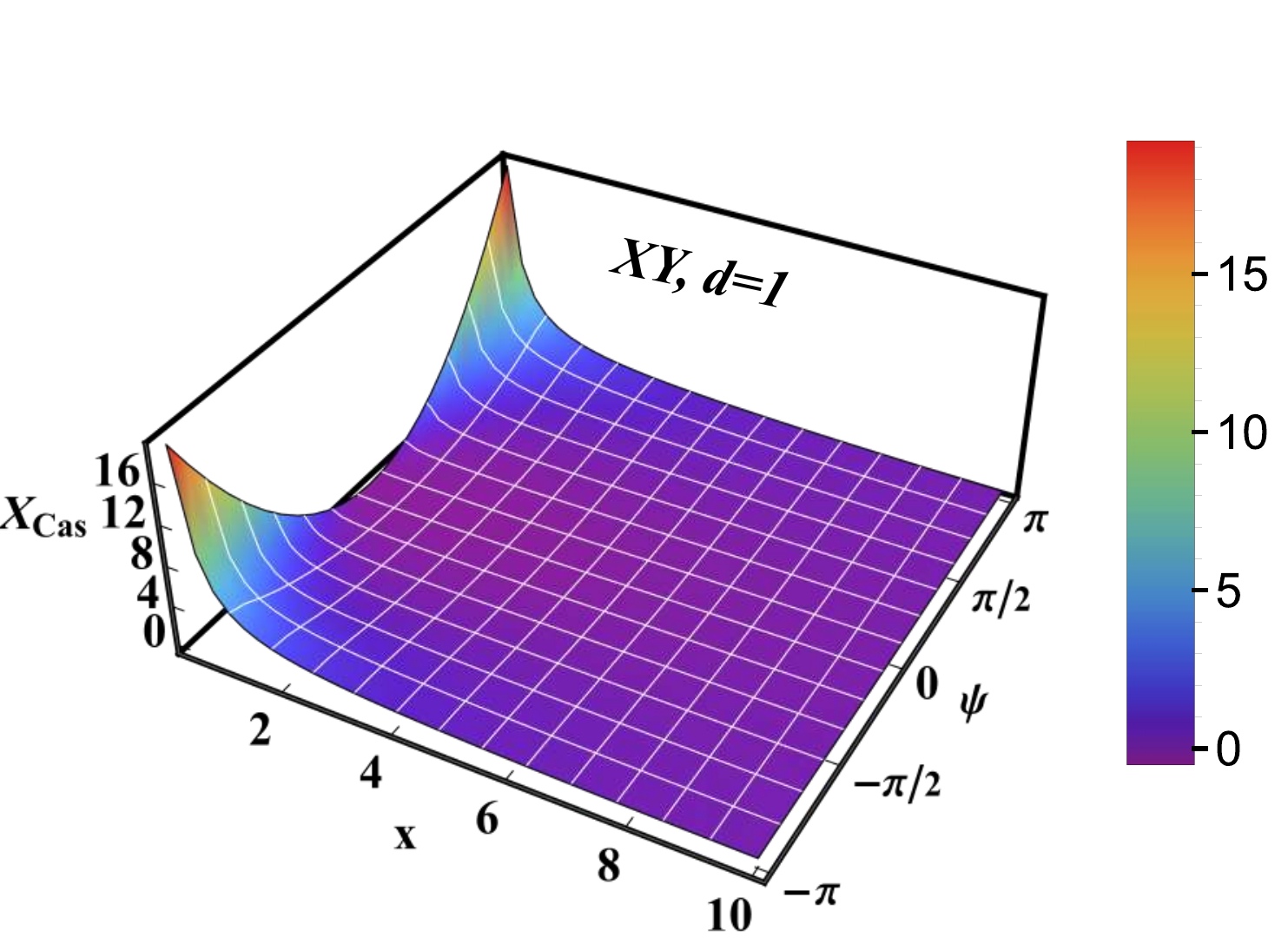}
	\caption{The surface of the scaling function $X_{\rm Cas}(\psi,x)$ for the critical Casimir force of the one-dimensional XY model as a function of the scaling variables $x=(N-1)/K$ (see \eq{eq:scaling_variables}) and phase difference $\psi$. The white lines correspond to $x=const$ and $\psi=const$, respectively. The horizontal plane ($x\gg 1$) marks the value $X_{\rm Cas}=0$.}
	\label{Fig:1d_X_Cas_Surface}
\end{figure}

From \eq{eq:F_Cas_scaling_small_x} one can also derive an expression for the behavior of the system at low temperatures which  retains the dependence on $H_1$ and $H_N$:
\begin{eqnarray}
\label{low_T_behavior}
\beta F_{\rm Cas}&=&-\frac{1}{2}\frac{1}{
	\left(J/H_1 + J/H_N+N-1\right)}\nonumber \\ && +\frac{1}{2}K\frac{\left(\psi _1-\psi
	_N\right)^2 }{
	\left(J/H_1+J/H_N+N-1\right)^2}.
\end{eqnarray}
This result can be also  derived directly by realizing that the ground state of the system is a spin wave such that the end spins are twisted with respect to each other at the angle $\psi=\psi_1-\psi_N$.

Equations \eqref{eq:F_Cas_scaling}, \eqref{eq:F_Cas_scaling_small_x},  \eqref{X_cas_ass}, and \eqref{low_T_behavior} confirm the validity of the statements (i)-(iv) in the first part of the present section. For example, \eq{eq:F_Cas_scaling} reveals that, if $\psi=0$, the force is attractive for {\it any} value of the scaling variable $x$; \eq{X_cas_ass} confirms this behavior for small and large values of the scaling variable $x$.

The behavior of the scaling function $X_{\rm Cas}(\psi,x)$ for a variety of values of $\psi$ as a function of the scaling variable $x$ is shown in Fig. \ref{Fig:1d_X_Cas}. Figure \ref{Fig:1d_X_Cas_Surface} shows a three-dimensional plot of this function for $x\in[0,10]$ and $\psi\in[-\pi,\pi]$.

\subsubsection{The one-dimensional Heisenberg model}
\label{sec:1dH_model}

The definition of this model together with an explanation of the corresponding notations are given in Sect. \ref{sec:models}.

\begin{figure}[h]
	\includegraphics[width=\columnwidth]{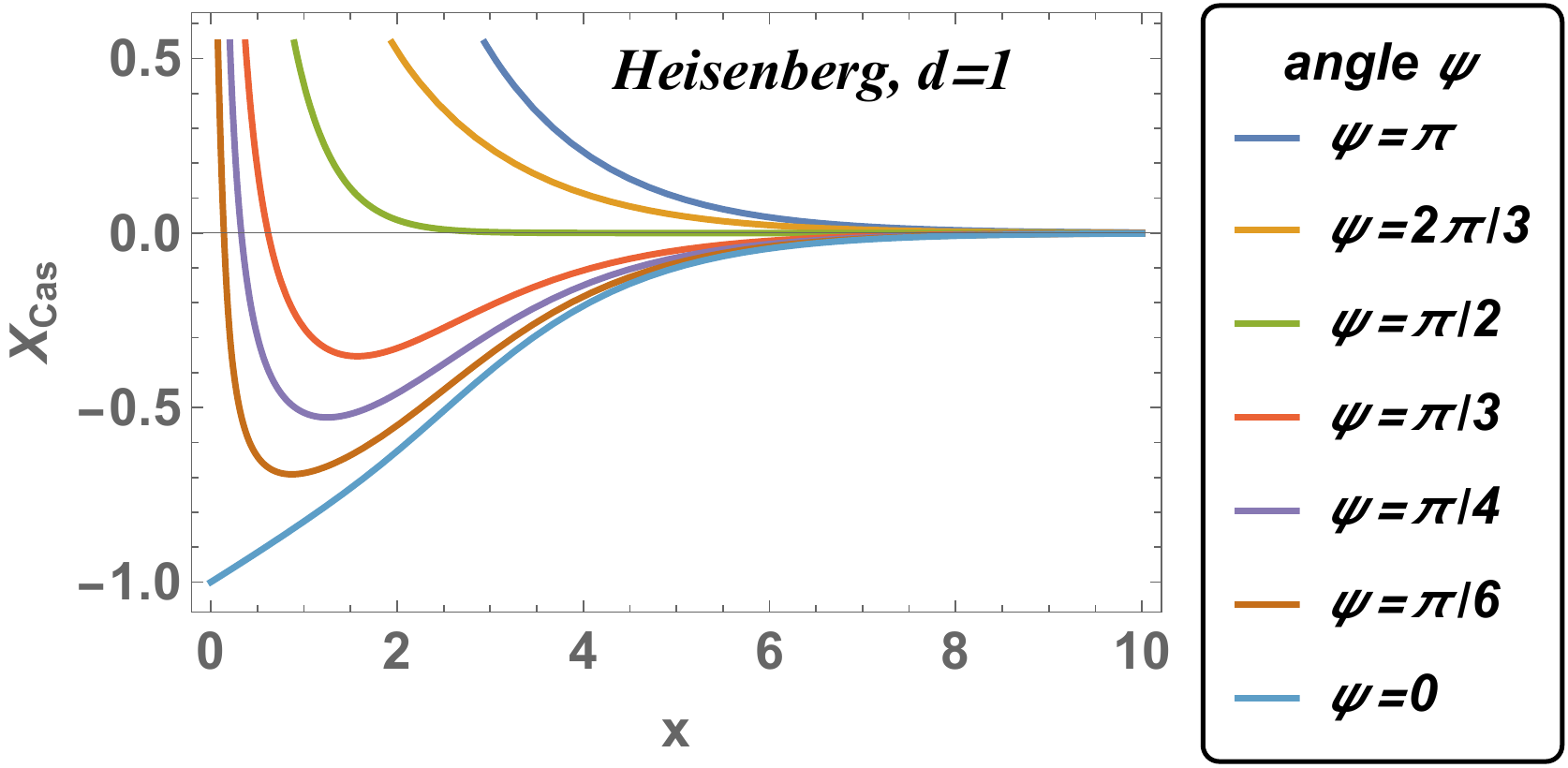}
	\caption{The scaling function $X_{\rm Cas}$ for the critical Casimir force of the one-dimensional Heisenberg model as a function of the scaling variable $x=(N-1)/K$  (see \eq{eq:scaling_variables}), for seven  values of the phase difference $\psi=\psi_h$.}
	\label{Fig:1d_X_H_Cas}
\end{figure}

The {\it exact} expression for the Casimir force within the one-dimensional Heisenberg model \cite{DaR2017} is
	\begin{equation}
	\label{eq:def_FCas_Heis}
	\beta  F_{\text{Cas}}=\frac{ \sum_{n=1}^{\infty} (2n+1)P_n \left(\cos\psi_h\right) \ln\left[\frac{I_{n+1/2}(K)}{I_{1/2}(K)}\right] \frac{I_{n+1/2}(h_1)}{I_{1/2}(h_1)} \frac{I_{n+1/2}(h_N)}{I_{1/2}(h_N)} \left[\frac{I_{n+1/2}(K)}{I_{1/2}(K)} \right]^{N-1}}{1+ \sum_{n=1}^{\infty} (2n+1)P_n \left(\cos\psi_h\right) \frac{I_{n+1/2}(h_1)}{I_{1/2}(h_1)} \frac{I_{n+1/2}(h_N)}{I_{1/2}(h_N)} \left[\frac{I_{n+1/2}(K)}{I_{1/2}(K)}\right]^{N-1}},
	\end{equation}
	where $\psi_h$ is the angle between the vectors ${\mathbf H}_1$ and ${\mathbf H}_N$,  where  $I_{n+1/2}(x)$ is the modified Bessel function of the first kind with  half-integer index so that $I_{1/2}(x)=\sqrt{2/(\pi x)}\sinh(x)$; $P_n(x)$ is the Legendre polynomial of degree $n$, and $K$, $h_1$, and $h_N$ are defined in \eq{eq:def_parameters}. 
In the limit $T\to 0$, so that $h_1\gg 1$, $h_N\gg 1$, and $K\gg 1$, one obtains \eq{eq:F_Cas_scaling}, with $\psi_h$ replacing $\psi$,
where the scaling variable $x$, as well as $h_{\rm eff}$, are  defined as in \eq{eq:scaling_variables} while the scaling function $X_{\rm Cas}$ is 
	\begin{equation}
	\label{eq:def_FCas_Heis_scaling}
	X_{\rm Cas}(\psi_h,x,h_{\rm eff})=-\frac{x}{2}\frac{ \sum_{n=1}^{\infty} n(n+1)(2n+1)P_n \left(\cos\psi_h\right)  \exp\left[-\frac{1}{2} n(n+1)\left(x+h_{\rm eff}^{-1}\right)\right]}{1+ \sum_{n=1}^{\infty}(2n+1)P_n \left(\cos\psi_h\right)  \exp\left[-\frac{1}{2} n(n+1)\left(x+h_{\rm eff}^{-1}\right)\right]}.
	\end{equation}
As in the case of the $XY$ model, if the relation in \eq{eq:constraint} is fulfilled, in  \eq{eq:def_FCas_Heis_scaling} one can neglect $h_{\rm eff}$. In this case the scaling function  $X_{\rm Cas}$ depends only on the scaling variable $x$ and the angle $\psi_h$ which parametrizes the boundary conditions of the system  (compare \eq{eq:1dCas_gen}). The behavior of  $X_{\rm Cas}$ in this regime is shown in Fig. \ref{Fig:1d_X_H_Cas}. Since $P_1(\cos \psi_h)=\cos\psi_h$, and in view of the fast decay of the terms in the sums in \eq{eq:def_FCas_Heis_scaling}, for moderate and large values of $x$ the sign of the force is determined by the sign of $\cos \psi_h$.

The expression for $X_{\rm Cas}$ given by \eq{eq:def_FCas_Heis_scaling} is reliable for all values of $x$,  except in the limit $x \ll 1$. In this limit $x\ll 1$, the leading behavior of the Casimir force is  
	\begin{equation}
	\label{eq:def_FCas_Heis_ass}
	X_{\rm Cas}(\psi_h,x,h_{\rm eff})=-1+
	\frac{h_{\rm eff}^{-1}}{h_{\rm eff}^{-1}+x}
	+
	\frac{x (1-\cos\psi_h)}{\left(h_{\rm eff}^{-1}+x\right)^2}
	+x\frac{
		\coth 
		\left(\frac{1}{h_{\rm eff}^{-1}+x}\right)
		-1}{\left(h_{\rm eff}^{-1}+x\right)^2}.
	\end{equation}
One can derive the first three terms in that expansion also by considering the dependence on the Heisenberg chain length $N$  of the ground state energy of the one-dimensional Heisenberg model, assuming that it corresponds to a spin wave configuration. Explicitly, concerning the behavior of the Casimir force in the limit $T\to 0$, from \eq{eq:def_FCas_Heis_ass} one obtains 
\begin{eqnarray}
\label{low_T_behavior_Heis}
\beta F_{\rm Cas}&=&-\frac{1}{
	J/H_1 + J/H_N+N-1}\nonumber \\ && +K \frac{1-\cos\psi_h}{
	\left(J/H_1+J/H_N+N-1\right)^2},
\end{eqnarray}
with $\cos \psi_h=(\mathbf{H}_1\cdot \mathbf{H}_N)/(H_1H_N)$.

\begin{figure}[h]
	\includegraphics[width=\columnwidth]{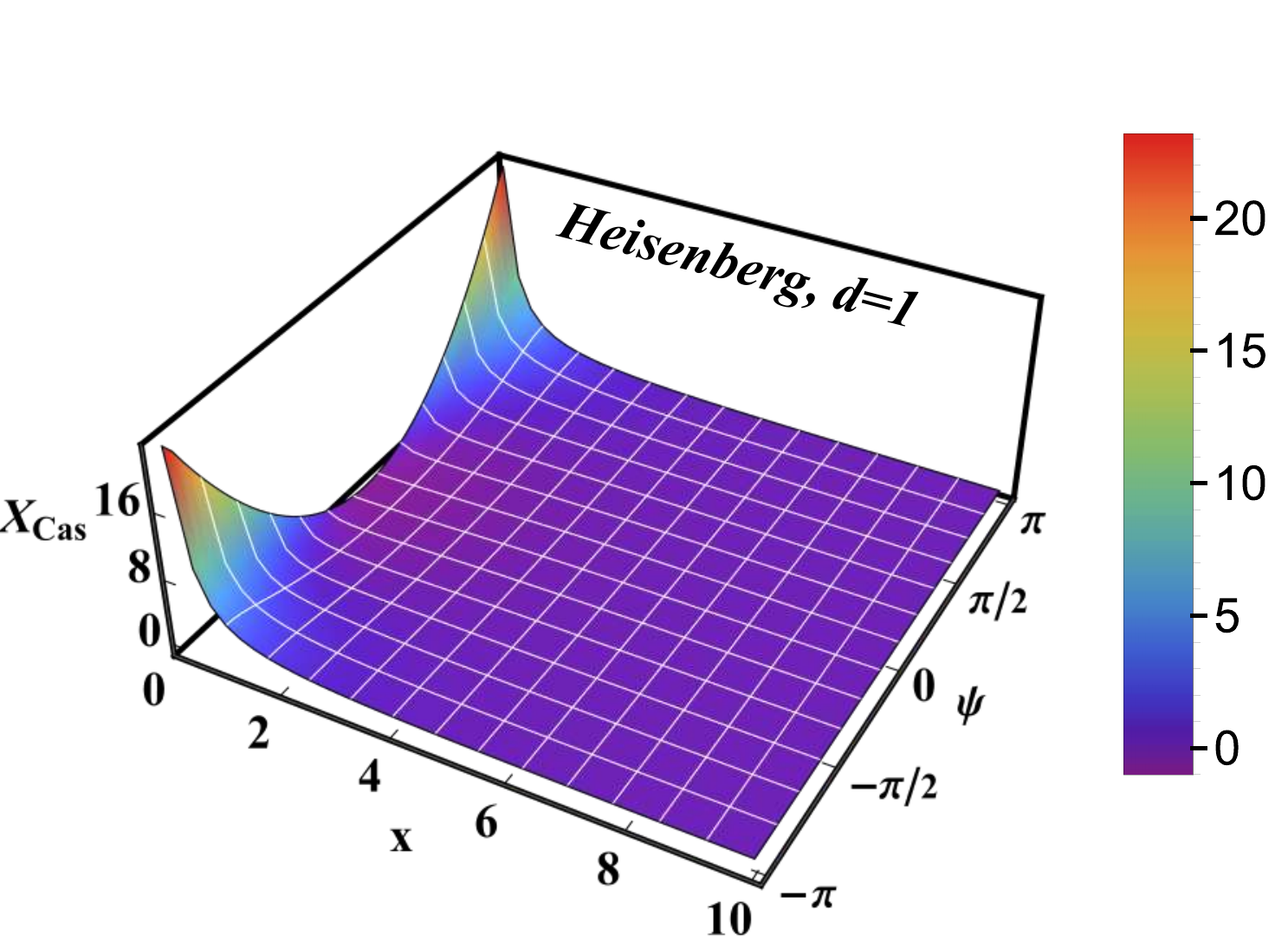}
	\caption{The surface of the scaling function $X_{\rm Cas}(\psi,x)$ for the critical Casimir force of the Heisenberg model as a function of the scaling variables $x=(N-1)/K$ (see \eq{eq:scaling_variables}) and $\psi$. The white lines correspond to $x=const$ and $\psi=const$, respectively. The horizontal plane ($x\gg 1$) marks the value $X_{\rm Cas}=0$. } 
	\label{Fig:1d_X_H_Cas_Surface}
\end{figure}

The behavior of the scaling function $X_{\rm Cas}(\psi,x,h_{\rm eff}=0)$, i.e., when \eq{eq:constraint} is valid, for seven values of $\psi$ as a function of the scaling variable $x$ is shown in Fig. \ref{Fig:1d_X_H_Cas}, while Fig. \ref{Fig:1d_X_H_Cas_Surface} shows a three-dimensional plot of this function for $x\in[0,10]$ and $\psi\in[-\pi,\pi]$.  
Thus, for the overall behavior of the Casimir force as a function of $\psi_h$, one arrives at the same set of conclusions for the  Heisenberg model as for the $XY$ model considered as a function of $\psi$ and as summarized in the statements (i)-(v) above. 

If $h_1=\beta H_1\to 0$ and $h_N=\beta H_N\to 0$, the system under consideration turns into one with Dirichlet boundary conditions, a case which was studied by M. E. Fisher in Ref. \cite{F64}. In this case $\beta f^{(O,O)}(T,N)=-(N-1)\left(\ln \sinh K-\ln K\right)$, which leads to $F_{\rm Cas}^{(O,O)}=0$ for these boundary conditions, similar to the one-dimensional Ising chain.

\section{Exact results for the thermodynamic Casimir effect in two-dimensional strips}
\label{Ising_UC}

\subsection{Exact results for $d=2$ obtained within conformal invariance}
\label{conformal_invariance}

Conformal invariance and the corresponding conformal field theory \cite{FQS84,BCN86,C86,C89,C96,CH93,ketov1995conformal,Gaberdiel2000,FMS2012conformal} are very useful and successful in determining finite-size properties of two-dimensional classical statistical systems
or of finite-temperature properties in one-dimensional
quantum systems at their corresponding critical point. These two phenomena are related, because a
$(1+1)$-dimensional quantum field theory at temperature
$T$ is given by a Euclidean-space functional integral
on a strip of width $L_\tau\sim T$ in the imaginary time
direction (see Sect. \ref{QuantumCritCasEffect} for more details). More specifically, in two dimensions the Casimir amplitudes for systems with strip geometry can
be obtained by using conformal invariance. According to this theory, in $d=2$ the universality
class of a critical system is characterized by the
value of a single scalar parameter $c$ called central charge or
conformal anomaly number. The value of $c$ cannot be determined
from general conformal invariance considerations alone. However, the
requirement that the \textit{critical} statistical mechanical system, which
one can describe in terms of a conformal field theory, has a unitary transfer
matrix, leads to a ``quantization'' of the possible values of $c<1$.
According to Ref. \cite{FQS84}, one has
\[
c=1-\frac 6{m(m+1)},
\]
where $m\in \left\{ 3,4,5,\cdots \right\}$. Table \ref{table:2dIsingPotts} provides the identifications
of critical and tricritical models with given values of $m$
\cite{FQS84}. The universality class of these two-dimensional models is uniquely determined by the
value of $c$ (or $m$).
\begin{table}[h!]
\begin{tabular}{lll}
\hline\hline
$m$ & model & $c$ \\ \hline\hline
$3$ & critical Ising & $1/2$ \\
$4$ & tricritical Ising & $7/10$ \\
$5$ & critical 3-state Potts & $4/5$ \\
$6$ & tricritical 3-state Potts & $6/7$ \\
$\infty $ & critical Gaussian & $1$
\end{tabular}
\caption{Identification of critical and tricritical models with the corresponding central charge $c=1-6/[m(m+1)]$.}
\label{table:2dIsingPotts}
\end{table}

The following general results for the Casimir amplitudes are known \cite{C86,BCN86,A86}:

\textit{(a) periodic boundary conditions:}
\begin{equation}
\label{CasAmPerGen}
\Delta^{(p)}=-\frac \pi 6c
\end{equation}

\textit{(b) ordinary and} $(+,+)$ \textit{boundary conditions:}
\begin{equation}
\label{CasAmPPOOGen}
\Delta ^{(O,O)}=\Delta ^{(+,+)}=-\frac \pi {24}c.
\end{equation}
For other boundary conditions, the amplitudes are also known
\cite{C86} for the critical Ising (see below) and the 3-state Potts model (see the second column in Table 2 in Ref. \cite{C86}). For the
3-state Potts model one can consider $(s,s)$ boundary conditions, which
means that equal surface Potts fields are applied, such that the Potts spins
at the boundaries have the same value $s$. In this case  $\Delta^{(O,O)}=\Delta^{(s,s)}$
for any Potts state $s$.

\subsection{Exact results for the critical Casimir amplitudes in $d=2$ stemming from the analytic solution of the corresponding model}
The known critical Casimir amplitudes for the two-dimensional Ising model are summarized in Table
\ref{table:2dIsing}. Exact analytical results
are available for periodic $(p)$,
antiperiodic $(a)$, ordinary $(O)$, $(+,+)$ and $(+,-)$ boundary conditions.
The results for periodic, $(+,+)$, and $(+,-)$ boundary conditions are
obtained via exact solutions, whereas those for antiperiodic and ordinary boundary conditions are derived by using conformal invariance methods. We note that in the two-dimensional Ising model the  extraordinary and the special (surface-bulk) surface phase transitions (and with them the corresponding boundary conditions) do not exist.

\begin{table}[ht]
\begin{tabular}{llllll}
\hline\hline  \vspace*{-0.3 cm}\\
$\DS{\Delta^{(O,O)}_\Cas}$ & $\DS{\Delta^{(+,+)}_\Cas}$ & $\DS{\Delta^{(p)}_\Cas}$ & $\DS{\Delta^{(+,-)}_\Cas}$ &
$\DS{\Delta^{(a)}_\Cas}$ & $\DS{\Delta^{(O,+)}_\Cas}$ \vspace*{0.1 cm}\\ \hline \\
 $\DS{-\frac \pi{48}} $ & $\DS{-\frac \pi{48} }$ & $\DS{-\frac \pi{12}} $ & $\DS{\frac{23\pi}{48}
} $ & $\DS{\frac \pi6} $ & $\DS{\frac \pi{24}} $ \\ \\ \hline\hline
\end{tabular}
\caption{Exact Casimir amplitudes for the two-dimensional Ising model.}
\label{table:2dIsing}
\end{table}

In Refs. \cite{INW86} and \cite{KD2010}, for the anisotropic Ising model with coupling $J_\|$ parallel to the strip and $J_\perp$ perpendicular to it, one has 
\begin{equation}
	\label{eq:ampl-relation-Ising}
	\Delta_{\rm anisotropic}^{(\tau)}=(\xi_{0,\perp}/\xi_{0,\|}) \Delta_{\rm isotropic}^{(\tau)}.
\end{equation}
This has been proven in Ref. \cite{KD2010} for periodic and  antiperiodic boundary conditions, and in Ref. \cite{INW86} for ordinary boundary conditions. In \eq{eq:ampl-relation-Ising},  $\xi_{0,\perp}$ and $\xi_{0,\|}$ are the  correlation length amplitudes of the so-called true correlation lengths (i.e., determined from the exponential decay of the correlation functions) orthogonal and along the Ising strip, respectively. One has (see \eq{HIsingdef})
\begin{equation}
	\label{eq:correlation-length-ratio}
\frac{\xi_{0,\perp}}{\xi_{0,\|}}=\frac{\sinh(2\beta_c J_\|)}{\sinh(2\beta_c J_\perp)}. 
\end{equation}
The precise definition of the two-dimensional Ising model and the notations used in the current section have been introduced in Sect. \ref{sec:models}.

In Ref. \cite{DSE2016} (see also Ref. \cite{DSE2015})  the authors have introduced the so-called alternating boundary conditions and determined the corresponding critical Casimir amplitudes for them. In this case  two types of configurations are considered:

(i) a strip of width $L$ with periodically alternating spin boundary conditions fixed to $+1$ or $-1$ on both edges, with periodicity $a$ and lateral shift $\delta \ge 0$;

(ii) a strip of width $L$ with homogeneous 
lower boundary consisting of $+1$ spins and alternating domains of $-1$ and $+1$  spins of distinct lengths $a$ and $b$, respectively, on the upper boundary. 

In the limit $L\ll a, b, \delta$ one obtains
\begin{equation}
(i) \; \Delta^{(I)}=-\frac{\pi}{48}+\frac{\pi}{2}\frac{\delta}{a}, \qquad \mbox{configuration (i)},
\end{equation}
and
\begin{equation}
(ii) \; \Delta^{(II)}=-\frac{\pi}{48}\frac{b-23 a}{a+b}, \qquad \mbox{configuration (ii)}.
\end{equation}
In configuration $(ii)$ the amplitude changes sign depending on the
relative number of $+1$ and $-1$ spins on the boundaries, and it vanishes for $a = b/23$. 

In the opposite limit $L\gg a, b, \delta$, as expected,  one obtains the critical Casimir amplitudes pertinent for the system with homogeneous boundary conditions: ordinary for $a=b$, fixed $+1$ spin for $b>a$ and fixed $-1$ spin for $a>b$. 
	
We remark that boundary conditions of type \textit{(i)} have been utilized in Ref. \cite{NN2014} to study, among other quantities, the lateral critical Casimir force in the two-dimensional Ising model. This has been accomplished numerically by using the exact diagonalization of the transfer matrix. The authors have studied the force as function of temperature, the strength of the surface magnetic field, and the geometric parameters $L, a$, and $\delta$. Furthermore, in Ref. \cite{TZGVHBD2011} the behavior of the critical Casimir force acting on a  colloidal particle 
immersed in a binary liquid mixture of water and 2,6-lutidine film and close to a substrate --- which is chemically
patterned with periodically alternating stripes of antagonistic adsorption preferences --- has been studied both theoretically and experimentally.

\subsection{Exact results for the scaling functions of the critical Casimir force in $d=2$}

Below we present explicit results for the behavior of the excess free energy and the critical Casimir force for the two-dimensional Ising model with various boundary conditions. To this end 
one can use the transfer matrix  method which renders the partition function of the two-dimensional Ising model. This has proved  to be an effective approach to obtain exact results for critical Casimir scaling functions. The mapping onto a theory of  free fermions \cite{SML64} allows one to diagonalize the transfer matrix on the  square lattice  with various boundary conditions.
For macroscopically long strips the free energy is given by the largest eigenvalue of the transfer matrix. By considering the lattice to be wrapped on a cylinder with circumference $M$ and height $L$   with  suitable
boundary conditions  at its ends (at $l=1$ and $l=L$; see Fig. \ref{Fig:square}), and  by taking  a transfer matrix along the axis of the cylinder, it is possible to obtain all expressions for  Casimir scaling functions in integral form. If the transfer is taken along the axis of the cylinder (i.e., along $l$), all  information about the surface properties is encoded in the
initial and final states. In this case the transfer matrix  gives
the Boltzmann factor of the interaction energy of a given and two neighboring periodic rows. Due to translational symmetry, the transfer matrix can be easily diagonalized and the whole difficulty in solving the problem
reduces to the determination of the  initial and final states in terms of the diagonalization basis. If the transfer is taken in the direction
perpendicular to the cylinder axis (i.e., along $m$) the surface states enter into the elements of the transfer matrix
and hence  into its spectrum. In that case,   diagonalization of the transfer matrix  becomes much more difficult and represents a  major challenge for the  calculations. Of course, rigorous analysis shows that the expressions for the critical Casimir force, as obtained from both approaches, are equivalent \cite{DM2013}.

As it will be discussed, the  critical Casimir force obtained by using the transfer
along the cylinder axis has  an integral representation from which one can derive the scaling limit and the exact asymptotic behavior. Below we shall present the corresponding results for the scaling functions of the critical Casimir force, which demonstrate explicitly their dependence on the imposed boundary conditions. For the temperature dependence of the force 
we shall use the scaling variable
\begin{equation}\label{Eq:defoftau}
	x_\tau=\tau\,(L/\xop)^{1/\nu},
\end{equation}
(with $\nu=1$ for the two-dimensional Ising model). Since, however, in Sects. \ref{sec:plus-plus-ordinary-ordinary},  \ref{sec:plus-minus}, \ref{sec:periodic}, \ref{sec:antiperiodic}, and \ref{sec:ordinary-plus}  only the temperature dependence of the Casimir force scaling function is considered, for reasons of simplicity of the notation the subscript $\tau$ in the scaling variable $x_\tau$ will be omitted.

\subsubsection{Excess free energy and critical Casimir force for  $(+,+)$ and $(O,O)$ boundary conditions}
\label{sec:plus-plus-ordinary-ordinary}

\begin{figure}[htb]
	\includegraphics[angle=0,width=\columnwidth]{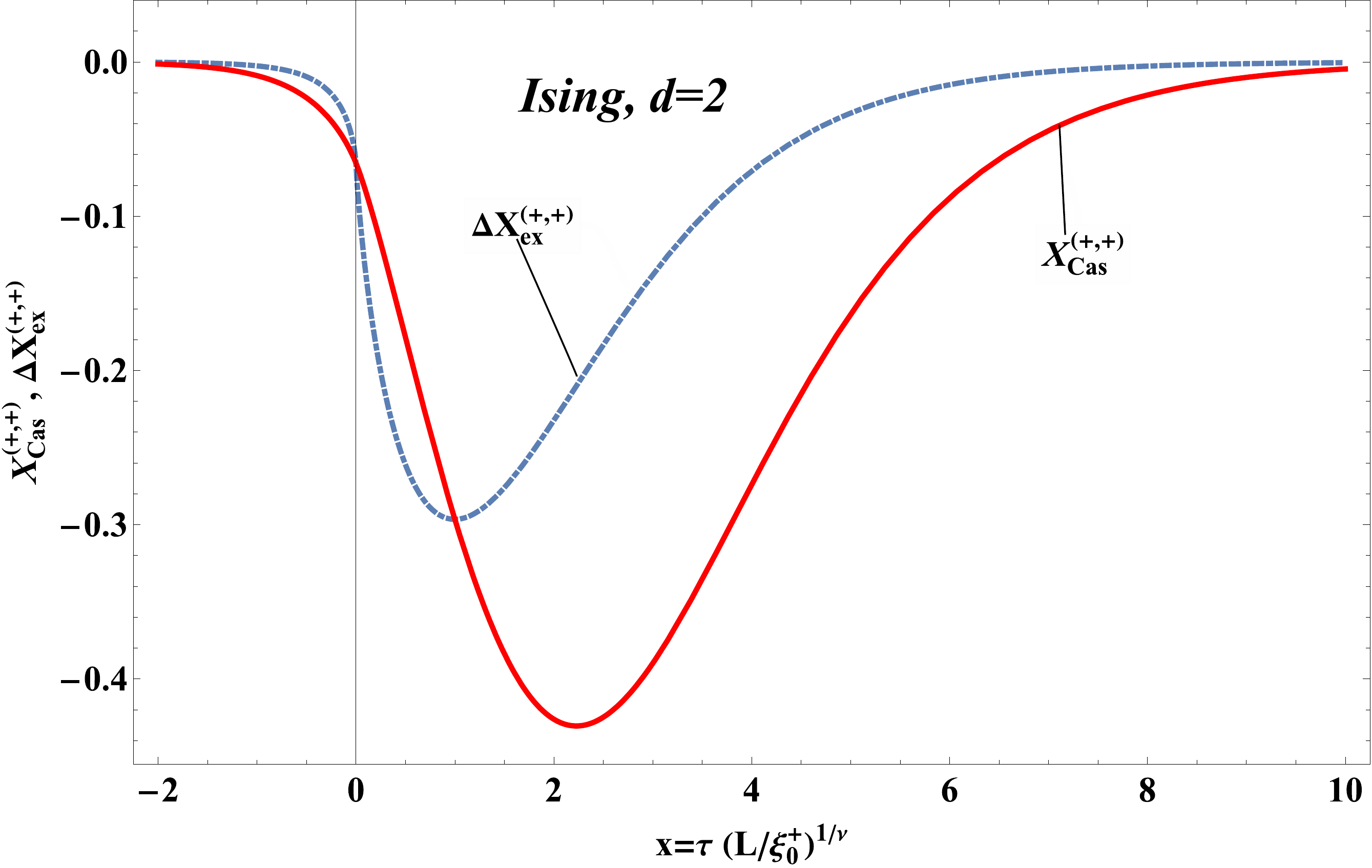}
	\caption{The finite-size scaling functions of the finite-size part of the excess free energy
		$\Delta X_{\rm ex}^{(+,+)}$ (see \eq{CasimirScaling General}) and of the Casimir force $X^{(+,+)}_{\rm Cas}$ (see \eq{Casimir_scaling}) for
		$(+,+)$ boundary conditions. These functions are  negative everywhere, 
		and both have a minimum \textit{above} $T_c$: the former attains its minimum
		$\Delta X_{\rm ex}^{(+,+)}\simeq -0.2965$  at $x \simeq 1.00$, while the latter
		attains its minimum $X^{(+,+)}_{\rm Cas} \simeq -0.4305$ at $x\simeq 2.23$. Due to \eq{eq:DeltaXex-Casimir-d2} the curves intersect at $x=0$ and at the minimum of 	$\Delta X_{\rm ex}^{(+,+)}$. }
	\label{pluscomparison}
\end{figure}

In the limit $h_1\to\infty$ with $h_L=h_1$ one obtains the so-called $(+,+)$ boundary conditions. Within the Ising model this corresponds to "frozen" spins within the boundary layers, the values of which are fixed to $+1$ if $h_1\to+\infty$. Accordingly, via the coupling $J$, an effective magnetic field of strength $J$ acts on the two layers adjacent to the boundary layers. Thus, one can consider a system with boundary layers which are exposed to a magnetic field $J$. Such a system is equivalent to one with $(+,+)$ boundary conditions. In the following, concerning both situations, i.e., $h_1\to \infty$ or $h_1=J$, we shall refer to as $(+,+)$ boundary conditions. If $h_1=h_L=0$ one has a system with so-called "ordinary" boundary conditions. Equivalently, one also uses the notions "Dirichlet" or "free"  boundary conditions.

In the case of $(+,+)$ boundary conditions, from the results obtained in Ref. \cite{ES94} (see also Refs. \cite{YF80}, \cite{BDT2000}, and \cite{K94}) one can
extract that the scaling function $\Delta X^{(+,+)}_{\rm ex}(x)$ of the finite-size part of the excess \textit{free energy}  (see \eq{CasimirScaling General}) is
\begin{equation}
\label{xppexscaling}
\Delta X^{(+,+)}_{\rm ex}=
                       \DS{-\frac 1{4\pi}\int_{2|x|}^\infty{\rm d}y \frac{y}{\sqrt{y^2-(2x)^2}}\;\ln \left[1+\frac{y+2x}{y-2x}\exp \left(-y\right)\right]
,}
\end{equation}
where
\begin{equation}
\label{IsingInfo}
\xop=\frac{1}{4K_c}, \quad K_c=\frac{\ln (1+\sqrt{2})}{2}, \quad \text{with} \quad \frac{\xop}{\xom}=2.
\end{equation}
For $(O,O)$  \BC\,  the corresponding result follows \cite{ES94} from the relation 
\begin{equation}\label{Eq:ppoo}
\Delta X^{(O,O)}_{\rm ex}(x)=\Delta X^{(+,+)}_{\rm ex}(-x).
\end{equation}
\begin{figure}[htb]
	\includegraphics[angle=0,width=\columnwidth]{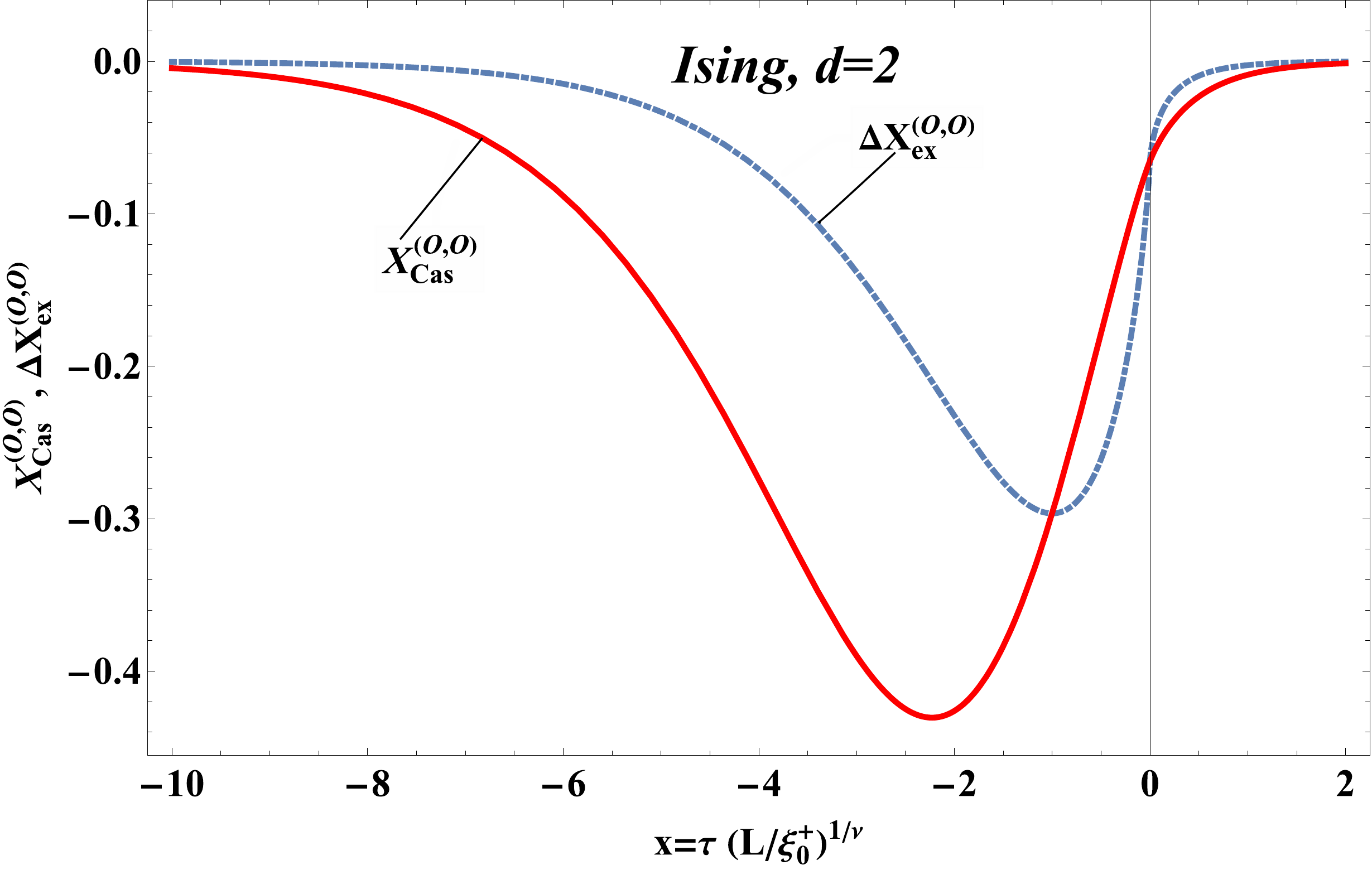}
	\caption{The finite-size scaling functions of the finite-size part $\Delta X_{\rm ex}^{(O,O)}$ of the excess free energy
	and of the critical  Casimir force $X^{(O,O)}_{\rm Cas}$ for 
		$(O,O)$ boundary conditions. These functions are  negative everywhere, 
		and both have their minimum \textit{below} $T_c$: the former attains its minimum
		$\Delta X_{\rm ex}^{(O,O)}\simeq -0.305$  at $x \simeq -1.00$, while the latter
		attains its minimum $X^{(O,O)}_{\rm Cas}\simeq -0.43$ at $x\simeq -2.23$. The curves here are mirror images of those in Fig. \ref{pluscomparison}. Due to \eq{eq:DeltaXex-Casimir-d2} the curves intersect at $x=0$ and at the minimum of 	$\Delta X_{\rm ex}^{(O,O)}$. }
	\label{excessordinary}
\end{figure}
From Eq. (\ref{xppexscaling}), at the critical point $x=0$  one obtains the following  critical Casimir amplitudes for the \textit{force}:
\begin{eqnarray}
\Delta X^{(+,+)}_{\rm ex}(0) &=&\Delta X^{(O,O)}_{\rm ex}(0)=\Capp =\Caoo \nonumber
\\ &=&-\frac 1{4\pi}\int_0^\infty {\rm d}y\ln \left[1+\exp (-y) \right]
=-\frac{\pi}{48}.
\end{eqnarray}
The behavior of $\Delta X_{\rm ex}^{(+,+)}$  and of  $\Delta X_{\rm ex}^{(O,O)}$ is illustrated in Figs.~\ref{pluscomparison} and \ref{excessordinary}, respectively. 
Having in mind that according to \eq{eq:deltaXex-XCasimir} one has, in $d=2$ with $\nu=1$, 	
\begin{equation}
	\label{eq:DeltaXex-Casimir-d2}
	X_{\rm Cas}^{(+,+)}(x)=\Delta X^{(+,+)}_{\rm ex}(x)-x\frac d{dx} \Delta X^{(+,+)}_{\rm ex}(x),
\end{equation}
which renders \cite{ES94}  the finite-size scaling functions of the critical Casimir \textit{force}
$X_{\rm Cas}^{(+,+)}(x)$ and $X_{\rm Cas}^{(O,O)}(x)$: 

\begin{equation}\label{XcasppIsing2d}
X_{\rm Cas}^{(+,+)}(x)=
                      \DS{-\frac{1}{4\pi} \int_{2|x|}^\infty {\rm d}y  \frac{y^2}{1+
\frac{y-2x}{y+2x}\; e^{y}}\frac{1}{\sqrt{y^2-(2x)^2}},}
\end{equation}
with
\begin{equation}\label{Eq:ppooCas}
X_{\rm Cas}^{(+,+)}(x)=X_{\rm Cas}^{(O,O)}(-x).
\end{equation}
The behavior of $X_{\rm Cas}^{(+,+)}$ is illustrated in Fig.~\ref{pluscomparison}, while the one of $X_{\rm Cas}^{(O,O)}$ is shown in Fig.~\ref{excessordinary}. $X_{\rm Cas}^{(+,+)}$ attains its minimum $X_{\rm Cas}^{(+,+)}\simeq
-0.4305$ at $x\simeq 2.2286$ (see the full line in Fig.~\ref{pluscomparison}).
The finite-size scaling functions of the Casimir force
$X^{(+,+)}_{\rm Cas}$ and of the finite-size part $\Delta X_{\rm ex}^{(+,+)}$ of the excess free energy attain their minimum at rather different values of the scaling argument, although on
the same side of $T_c$. Furthermore, because $X_{\rm Cas}^{(+,+)}(x)=X_{\rm Cas}^{(O,O)}(-x)$,  
$X_{\rm Cas}^{(+,+)}$ has a
minimum \textit{above} $T_c$, while $X_{\rm Cas}^{(O,O)}$ attains its minimum
\textit{below} $T_c$. 

The asymptotic behavior of $X_{\rm Cas}^{(+,+)}(x)$ and of 
$X_{\rm Cas}^{(O,O)}(x)$ follows from Eq. (\ref{XcasppIsing2d}).
Due to the absence of a genuine phase transition in the two-dimensional Ising strip,
the only singularity in the behavior of $X^{(+,+)}_{\rm Cas}(x)$ stems from the contribution of the bulk free energy $f_b(x\,|\,d=2)$ near $x=0$. Accordingly, in
the limits $x\rightarrow 0$ and $x\rightarrow \pm \infty $ one has
\begin{equation}
X^{(+,+)}_{\rm Cas}(x)=\left\{
\begin{array}{lrl}
\DS{-\frac 1{16} \sqrt{\frac{|x|}{\pi}}\exp (-2|x|)}, & x\rightarrow -\infty, & \tau <0,  \\
& \\
\DS{-\frac \pi {48}-\frac{1}{2\pi} x-\frac 1{4\pi } x^{2-\alpha} \ln |x|+O(x^2)}, &
x\rightarrow 0, & \tau \to 0,
\\
& \\
\DS{-x^2 \,\exp \left( -x\right)} , & x\rightarrow +\infty, & \tau > 0.
\end{array} \right.   \label{XCasasplus}
\end{equation}
The expansion around $x=0$ is consistent with the logarithmic singularity at $T=T_c$ of the specific heat in the two-dimensional Ising model. This is in line with the corresponding bulk critical exponent $\alpha=0$.

Equation (\ref{XCasasplus}) is consistent with the following asymptotic expansion of the scaling function $\Delta X^{(+,+)}_{\rm ex}(x)$ of the finite-size part of the excess free energy:
\begin{equation}
\Delta X^{(+,+)}_{\rm ex}=\left\{
\begin{array}{l}
\DS{-\frac 1{32 \sqrt{\pi |x|}}\exp (-2|x|)},    \quad x\rightarrow -\infty, \;\tau <0,  \\
 \\
\DS{-\frac \pi {48}+\frac{1}{2\pi} x^{2-\alpha_s} \ln |x|+2 c \, x -\frac 1{4\pi } x^{2-\alpha} \ln |x|+O(x^2)}, 
 \quad x\rightarrow 0,  \tau \to 0,
\\
 \\
\DS{-x \,\exp \left( -x\right)}, \qquad  x\rightarrow +\infty, \tau > 0,
\end{array} \right.   \label{Xexasplus}
\end{equation}
where $\alpha=0$, $\alpha_s=\alpha+\nu=1$, and $c$ is a number.  $\Delta X^{(+,+)}_{\rm ex}(x)$ has an infinite slope at $x=0$ while for $X^{(+,+)}_{\rm Cas}(x)$ this slope is finite. This is due to the contribution $\propto x \ln |x|$, which stems from the surface free energy contribution to the scaling function of the excess free energy --- see Eqs. (4.44) and (4.45) in Ref. \cite{MW73} and recall that in the current systems there are two surface free energies,  because the systems are finite in one (i.e., the normal) direction, while in Ref. \cite{MW73} the case of semi-infinite systems is considered. The number $c$ is (part of) the value of the surface free energy at the critical point of the two-dimensional infinite system. 

\subsubsection{Excess free energy and critical Casimir force for  $(+,-)$ boundary conditions}
\label{sec:plus-minus}

Here we consider a $d$-dimensional Ising strip with $(+,-)$
boundary conditions, so that for $T<T_c$ an interface develops parallel to the planes bounding the strip. Introducing \cite{P90} the finite-size interface free energy $\sigma^{(+,-)}(T,L)$  via the definition 
\begin{equation}
f_{\rm ex}^{(+,-)}(T,L)=f_{\rm ex}^{(+,+)}(T,L)+\sigma^{(+,-)}(T,L),
\label{interfacedefinition}
\end{equation}
one concludes that the corresponding scaling functions of $f_{\rm ex}^{(+,-)}$, $f_{\rm ex}^{(+,+)}$, and $\sigma^{(+,-)}$ are related by the
equation
\begin{equation}
X^{(+,-)}_{\rm ex}(x) = X^{(+,+)}_{\rm ex}(x)+X_\sigma^{(+,-)}(x),
\label{XpmexXsigma}
\end{equation}
where
\begin{equation}
 f_{\rm ex}^{(+,-)} =  L^{-(d-1)}  X^{(+,-)}_{\rm ex}(x),\quad f_{\rm ex}^{(+,+)} = L^{-(d-1)} X_\sigma^{(+,+)}(x),  \quad \sigma^{(+,-)} = L^{-(d-1)} X_\sigma^{(+,-)}(x). 
\label{Xsigma-def}
\end{equation}
Here $X_\sigma^{(+,-)}$ is the finite-size scaling function of the
\textit{finite-size} interface free energy $\sigma^{(+,-)}(T,L)$. The corresponding relation between the Casimir \textit{forces} follows as 
\begin{equation}\label{Caspppmcon}
F_{\rm Cas}^{(+,-)}(T,L)=F_{\rm Cas}^{(+,+)}(T,L)-\frac{\partial}{\partial
L}\,\sigma^{(+,-)}(T,L).
\end{equation}
Hence one obtains
\begin{equation}
X_{\rm Cas}^{(+,-)}(x)=X_{\rm Cas}^{(+,+)}(x)+\delta X_{\rm Cas}^{(+,-)}(x),
\label{relationplusminus}
\end{equation}
where
\begin{equation}\label{DeltapmRelation}
\delta X_{\rm Cas}^{(+,-)}(x)=(d-1)X_\sigma^{(+,-)}(x)-\frac{x}{\nu}
\frac d{dx}X_\sigma^{(+,-)}(x).
\end{equation}
For any $d$, one has $\sigma^{(+,-)}(T,L\rightarrow \infty) =\sigma(T)$, where the interface free energy $\sigma(T)$
is independent of the specific choice of the boundary conditions applied in order to create the interface.

For the two-dimensional Ising model the following exact result  is known
\cite{O44}:
\begin{equation}\label{OnsagerSgma}
\beta\,\sigma(T)=\left\{\begin{array}{lll}
2K-\mathrm{arcsinh}\left[1/\sinh (2K)\right], & &  T< T_c,
\\	
0, && T\ge T_c.
\end{array}\right.
\end{equation}
It can be shown that $\sigma^{(+,-)}(T,L)>\sigma(T)\geq 0$ \cite{ES94}.
At the critical temperature, $\sigma(T_c)=0$ and \cite{ES94,S93}
\begin{equation}
\beta_c \,\sigma^{(+,-)}(T_c,L)=\frac \pi 2L^{-1}-\sqrt{2}\pi L^{-2}+\frac{\pi}{4}
\left(1-\frac{1}{96}\pi^2\right) L^{-3}+\cdots ,
\label{intffe}
\end{equation}
hence
\begin{equation}\label{Deltapm}
\Delta^{(+,-)}_{\rm Cas} =\Delta^{(+,+)}_{\rm Cas}+\frac{\pi}{2} =\frac{23}{48}\pi \qquad (d=2).
\end{equation}
The next-to-leading order terms in Eq. (\ref{intffe}) yield
corrections to scaling.

The low temperature behavior of $\sigma^{(+,-)}(T,L)$ is given by  \cite{ES94,S93,AS86,MS96}
\begin{equation}\label{sigmaexpansion}
\beta\sigma^{(+,-)}(T,L)=\beta\sigma(T)+\frac{\pi^2}{2\sinh [\beta\sigma(T)]}L^{-2}+
O(L^{-3}),
\end{equation}
which implies 
\begin{equation}
F_{\rm Cas}^{(+,-)}(T,L)=F_{\rm Cas}^{(+,+)}(T,L)+\frac{\pi^2}{\sinh
[\beta\sigma(T)]}L^{-3}+O(L^{-4}), \, T\ll T_c.  \label{fcpmb}
\end{equation}
\begin{figure}[htb]
\center\includegraphics[angle=0,width=0.85\columnwidth]{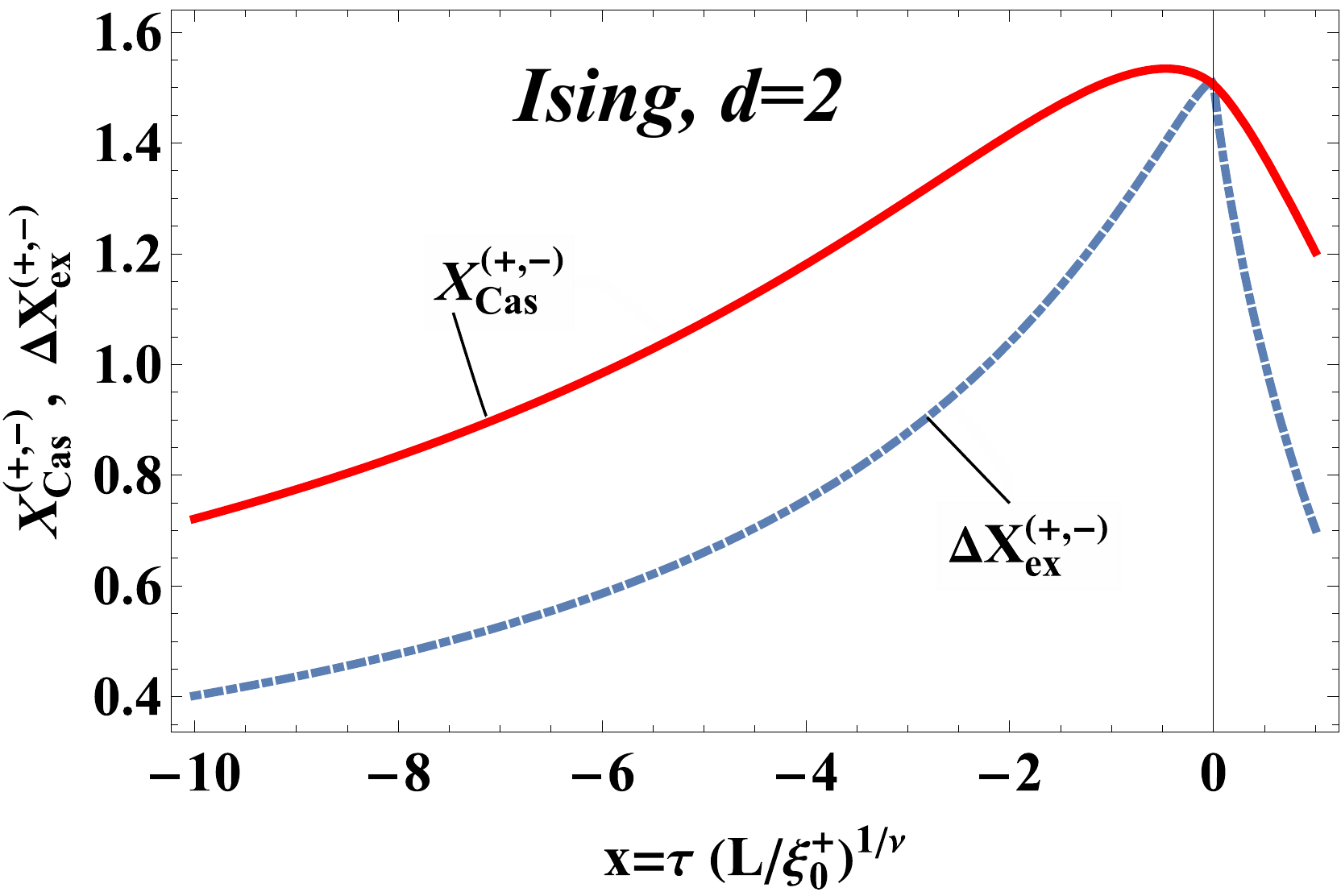}
\caption{The finite-size scaling function $X_{\rm Cas}^{(+,-)}(x)$ of the Casimir force
 and of the finite-size part of the excess free energy $\Delta X_{\rm ex}^{(+,-)}(x)$ for a system with $(+,-)$ boundary conditions. For such boundary conditions, as expected, the force is {\it positive},
i.e., it corresponds to a {\it repulsion} of the surfaces bounding the system.
It attains its maximum $X_{\rm Cas}^{(+,-)}\simeq 1.5331$ at $x=-0.4821$, i.e., 
{\it below} $T_c$. There the behavior of the finite size part of the excess free energy and of the Casimir force is dominated by the contributions which are due to the formation of a free interface inside the system. In the limit $x\to-\infty$, due to the presence of this kind of fluctuating interface,  the scaling function $X_{\rm Cas}^{(+,-)}$ decays to zero not exponentially, as $X_{\rm Cas}^{(+,+)}$ does, but algebraically,  $\propto -\pi^2/x$  (see Eq. (\ref{asXCaspm}). Above $T_c$, i.e., for $x\ge 0$ this kind of interface is not present and $X_{\rm Cas}^{(+,-)}(x)$, similar to $X_{\rm Cas}^{(+,+)}$,  decays to zero exponentially for $x\to +\infty$.}
\label{plusminuscasimir}
\end{figure}

The scaling function $X_\sigma^{(+,-)}(x)$ of $\sigma^{(+,-)}(T,L)$ is
also known \cite{ES94,AS86,S93}. It is given by the implicit
equation
\begin{equation}\label{Xpms}
X_\sigma^{(+,-)}(x)=y(x)/\sin[y(x)],
\end{equation}
where $y=y(x)$ follows implicitly from the equation
\begin{equation}\label{pmXs}
    y(x)\cot y(x)=x
    \end{equation}
 with $y\in [0,\pi]$, i.e., $x\in (-\infty , 1]$. At higher temperatures, i.e., for $x \gg 1$, one finds
an exponential decay of $\sigma^{(+,-)}(T,L)$ \cite{ES94}:
\begin{equation}\label{Xsas}
X_\sigma ^{(+,-)}(x)\simeq 2\, x^2 \exp (-x).
\end{equation}
Similar to Eq. (\ref{Xpms}), one can  derive also the behavior of the scaling function (see \eq{relationplusminus})
\begin{equation}\label{dXpmCas}
    \delta X_{\rm Cas}^{(+,-)}(x)=\frac{y^2 \sin(y)}{y-\frac{1}{2}\sin(2y)}, \quad y\in [0,\pi],
\end{equation}
for the force, where $y=y(x)$ is the solution of  \eq{pmXs}.
Close to $T_c$, i.e., for $|x|\ll 1$, one can obtain an explicit expansion \cite{ES94} of this function:
\begin{equation}\label{dXpmas}
\delta X_{\rm Cas}^{(+,-)}(x)=
                              {\DS \frac \pi 2-\frac{\pi^2-8}{\pi^3}x^2-
\frac{8}{3}\frac{5\pi^2-48}{\pi^5}x^3+O(x^4)},    \qquad x\to 0.
\end{equation}
Equation \eqref{dXpmas} shows that $\delta X_{\rm Cas}^{(+,-)}(x)$ is positive and attains its maximum value $\pi/2$ at $T=T_c$. Since $F_{\rm Cas}^{(+,-)}
(T,L)$ and $F_{\rm Cas}^{(+,+)}(T,L)$ are related via the relation
(\ref{Caspppmcon}), the former function turns out to have a maximum
$X_{\rm Cas}^{(+,-)}\simeq 1.5331$ below $T_c$ at $x\simeq -0.4621$  (see Fig. \ref{plusminuscasimir}). In the limit $x\rightarrow -\infty$, one finds 
\begin{equation}\label{asXCaspm}
\delta X_{\rm Cas}^{(+,-)}(x)\simeq -\pi^2/x+O(x^{-2}),
\end{equation}
which leads to a low-temperature asymptotic form of $F_{\rm Cas}^{(+,-)}(T,L)$
of the type given by Eq. (\ref{fcpmb}).
We remark that close to $T_c$ both $F_{\rm Cas}^{(+,-)}(T,L)$ and
$F_{\rm Cas}^{(+,+)}(T,L)$ carry nonuniversal corrections $\propto 1/L$  to their leading-order scaling behavior \cite{ES94}.

\subsubsection{Excess free energy and Casimir force for periodic boundary conditions}
\label{sec:periodic}

Certain results concerning the excess free energy for periodic boundary conditions can
be inferred from Ref. \cite{FF69} and have been presented in Refs.  \cite{K94,BDT2000} and \cite{DK2004}.

\begin{figure}[h!]
	\includegraphics[angle=0,width=0.95\columnwidth]{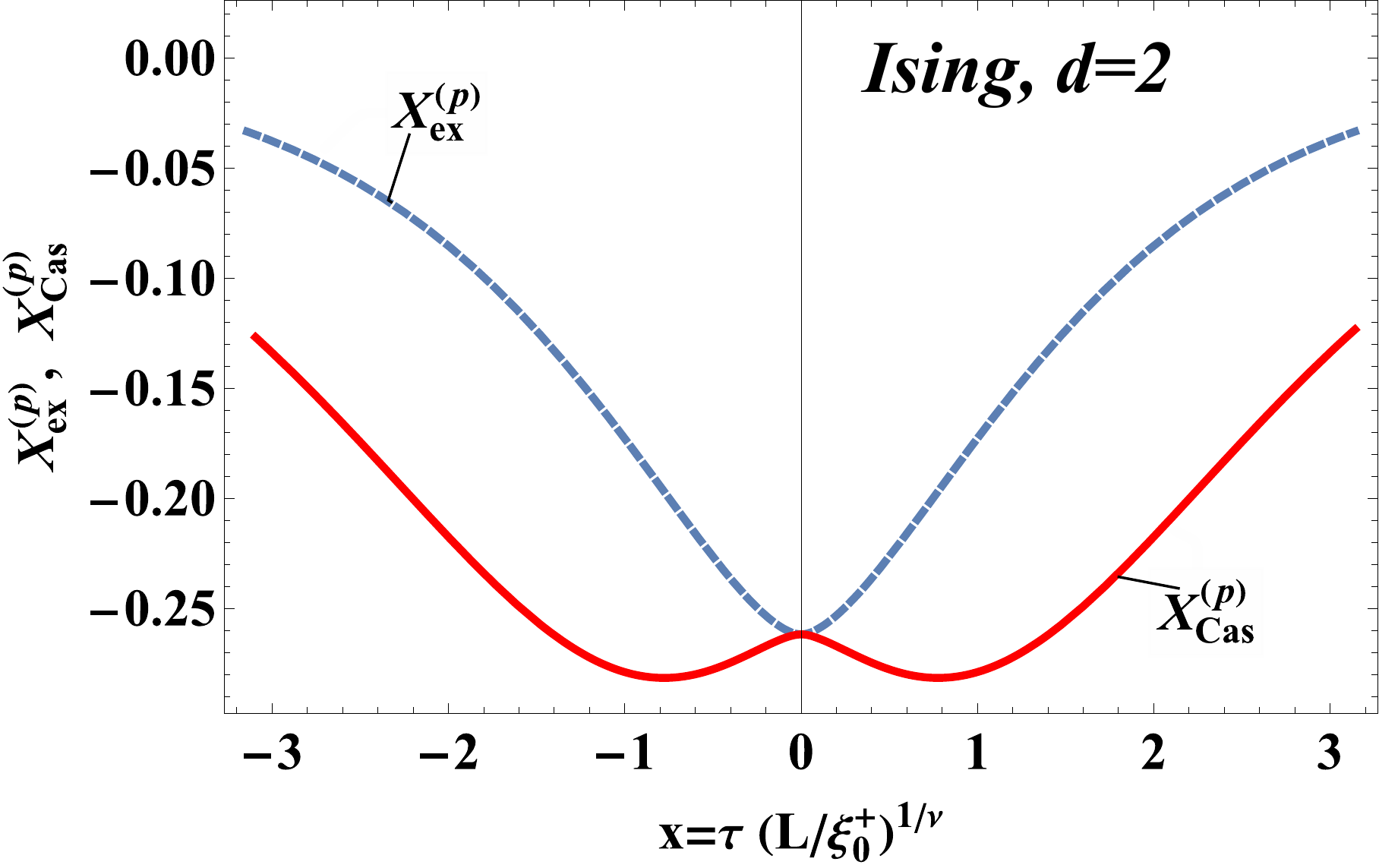}
	\caption{The finite-size scaling functions of the excess free energy
		$X_{\rm ex}^{(p)}$ and of the Casimir force $X^{(p)}_{\rm Cas}$ for periodic
		$(p)$ boundary conditions. These functions are  negative everywhere and symmetric around $x=0$, i.e., $T=T_c$. The excess
		free energy scaling function attains its minimum $-\pi/12$ at $T=T_c.$ The Casimir force, however, exhibits two symmetric and equally deep minima  below \textit{and} above $T_c$, at $x=\pm 0.78$, separated by a local maximum of the Casimir force at $T=T_c$ given by the corresponding critical Casimir amplitude $\Delta^{(p)}=-\pi/12$  (see Eq. (\ref{CasAmPerGen})). 
		At the two minima one has $X^{(p)}_{\rm Cas}\simeq -0.282$.}
	\label{excessperiodic}
\end{figure}
At the critical point one obtains
\begin{equation}\label{Xexper2d}
f^{(p)}_{\rm ex}(T_c,L) =\frac{\pi}{12}L^{-1}+O(L^{-3} \ln^3 L).
\end{equation}
Within the interval $-\pi < x < \pi$, the scaling function of the excess free energy can be represented as a series expansion \cite{rem1}:
\begin{equation}\label{efeIsing2d}
X_{{\rm ex}}^{(p)}(x)=-\frac{\pi}{12}-\frac{1}{4\pi} x^2 \ln \left | \frac{x
}{x_0}\right| -\pi \sum_{i=2}^{\infty} \binom{1/2}{i}
 \left(
\frac{x}{\pi}\right)^{2i}(1-2^{-2i+1})\zeta(2i-1),
\end{equation}
where $x_0=\pi\exp(1/2-\gamma)$ with Euler's constant $\gamma\simeq 0.577216$
and the Riemann zeta function $\zeta$ \cite{HD}.  Concerning the Casimir force scaling function $X_{{\rm Cas}}^{(p)}$ for periodic boundary conditions  \cite{rem1} one finds from  \eq{efeIsing2d} 
\begin{eqnarray}\label{efeI}
X_{{\rm Cas}}^{(p)}(x)=-\frac{\pi}{12}+\frac{1}{4\pi}\left(1+ \ln \left
| \frac{x}{x_0}\right | \right) x^2-\pi \sum_{i=2}^{\infty}
\binom{1/2}{i}
\left(
\frac{x}{\pi}\right)^{2i}(1-2i)(1-2^{-2i+1})\zeta(2i-1). 
\end{eqnarray}
Both the excess free energy as well as the Casimir force are symmetric functions  of $x$, i.e.,  one has
\begin{equation}\label{perrel}
X_{{\rm ex}}^{(p)}(x)=X_{{\rm ex}}^{(p)}(-x), \quad X_{{\rm Cas}}^{(p)}(x)=X_{{\rm Cas}}^{(p)}(-x), \quad x \in [0,\pi].
\end{equation}
The behaviors of $X_{{\rm ex}}^{(p)}(x)$ and of $X_{{\rm Cas}}^{(p)}(x)$ are shown in Fig. \ref{excessperiodic}.
Since for periodic boundary conditions there are no surface or interface free energies, one has $X_{{\rm ex}}^{(p)}(x)\equiv \Delta X_{{\rm ex}}^{(p)}(x)$ (see \eq{CasimirScaling General}).

Starting from
 the formula for the partition function of the  square
Ising lattice wrapped on a torus,  provided by Ferdinand and Fisher \cite{FF69},
a   simple  integral representation for the Casimir force scaling function for periodic boundary conditions
was  derived in Ref.~\cite{RZSA2010}:
\begin{equation}\label{period}
X_{{\rm Cas}}^{(p)}(x)=\frac{1}{\pi}\int_{-\infty}^{\infty}dy \sqrt{x^2+y^2}\left[\tanh\sqrt{x^2+y^2}-1\right].
\end{equation}

\subsubsection{Casimir force for antiperiodic boundary conditions}
\label{sec:antiperiodic}

\begin{figure}[h!]
	\includegraphics[angle=0,width=0.95\columnwidth]{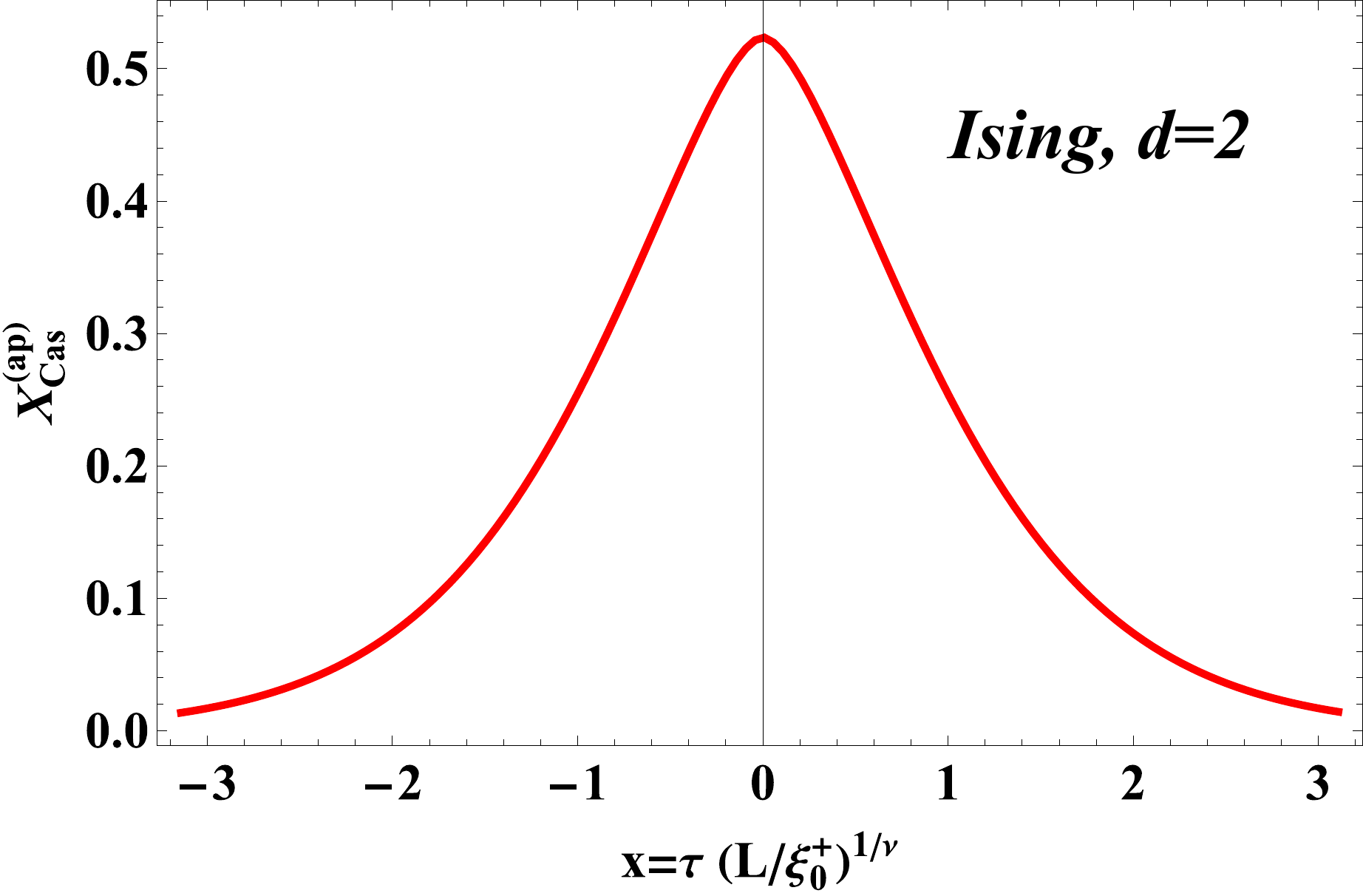}
	\caption{The finite-size scaling function of the  Casimir force $X^{({\rm ap})}_{\rm Cas}$ for the  antiperiodic
		$({\rm ap})$ boundary condition. For the antiperiodic boundary condition there is a floating and fluctuating interface, enforced by the imposed boundary conditions. Accordingly, the corresponding scaling function is positive everywhere. 
		Both for the periodic and the antiperiodic boundary conditions the Casimir force is  symmetric about $T_c$.}
	\label{excessaperiodic}
\end{figure}
For the antiperiodic boundary condition
 the exact expression of the partition function, calculated by using the Grassman path integral (see Ref. \cite{WH2002}), was used in order to derive the scaling function of the Casimir force \cite{RZSA2010}. Its
integral form is  similar to the one of the periodic boundary
condition (see Eq.~(\ref{period})):
\begin{equation}\label{antiper}
X_{{\rm Cas}}^{({\rm ap})}(x)=\frac{1}{\pi}\int_{-\infty}^{\infty}dy\sqrt{x^2+y^2}\left[\coth\sqrt{x^2+y^2}-1\right].
\end{equation}
The behavior of $X_{{\rm Cas}}^{({\rm ap})}(x)$  is shown
in Fig. \ref{excessaperiodic}.
As expected, for the antiperiodic boundary condition the critical Casimir force is repulsive.

\subsubsection{Casimir force for $(O,+)$ boundary conditions}
\label{sec:ordinary-plus}

\begin{figure}[h!]
	\includegraphics[angle=0,width=0.95\columnwidth]{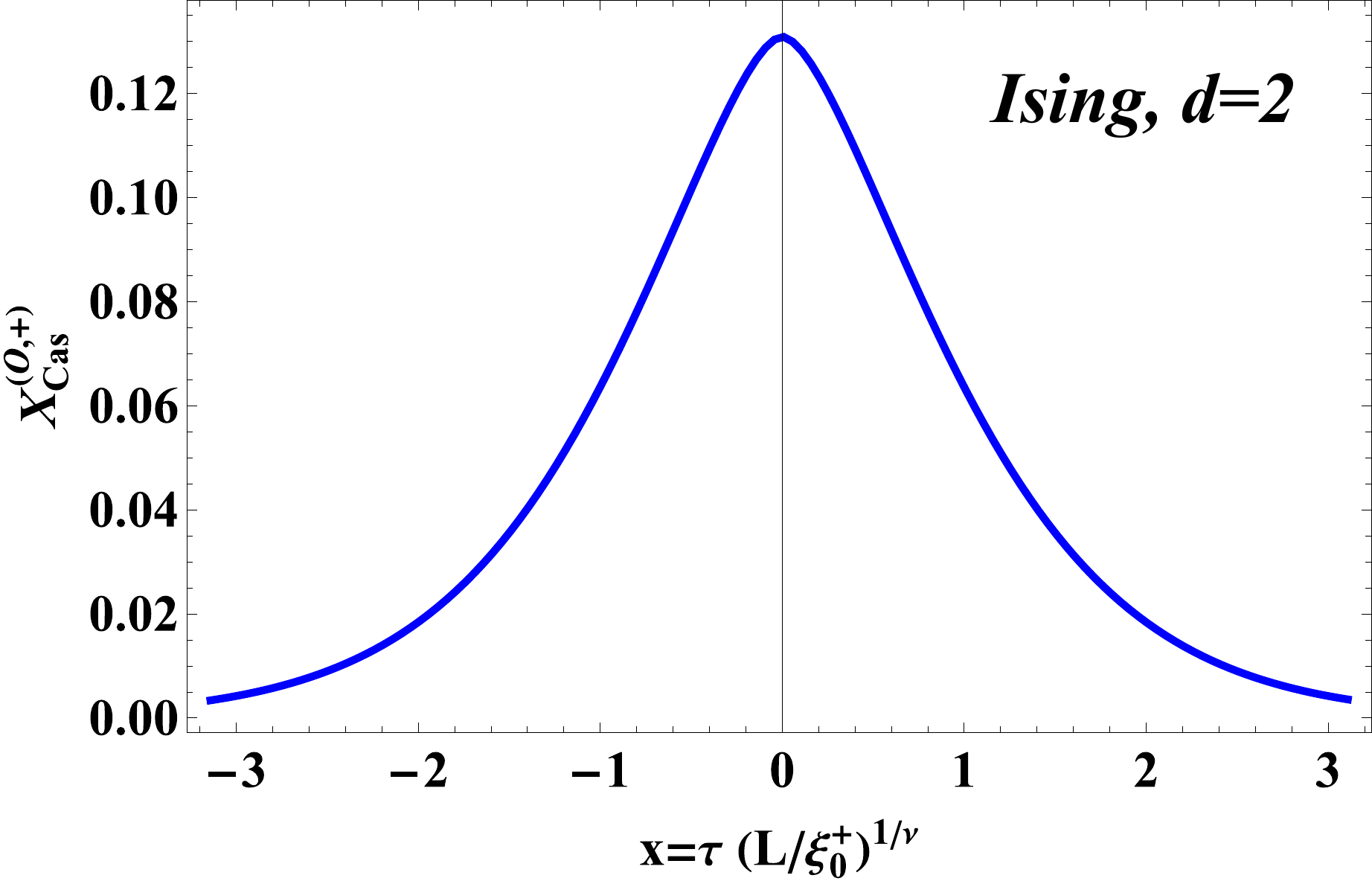}
	\caption{The finite-size scaling function $X^{(O,+)}_{\rm Cas}$ of the  Casimir force  for $(O,+)$ boundary conditions. As for the antiperiodic boundary condition, the scaling function is  positive everywhere. For $(O,+)$ and for $({\rm ap})$ boundary conditions the scaling functions of the Casimir force are symmetric about $T_c$.}
	\label{fig:xCas_ord_plus}
\end{figure}

Starting from the result derived in Ref. \cite{AM2010} for Ising strips with boundary fields, one can obtain the corresponding result for  $(O,+)$ boundary conditions by taking the corresponding appropriate limit there (see below). The final result is \cite{rem1}
\begin{equation}
\label{eq:XCas_ord_plus}
X_{\rm Cas}^{(O,+)}(x)=\frac{1}{2 \pi }\int_{0}^{\infty}\sqrt{y^2+x^2} \left[\coth \sqrt{y^2+x^2}-1\right] dy.
\end{equation}
Its behavior is shown in Fig. \ref{fig:xCas_ord_plus}. 
We note that 
\begin{equation}
\label{eq:XCas_ord_plus_antiper}
X_{\rm Cas}^{(O,+)}(x)=\frac{1}{4} X_{{\rm Cas}}^{(ap)}(x). 
\end{equation}

\subsubsection{Casimir force in the presence of boundary fields }

We use $(h_1,\pm h_L)$ as short-hand notations of these boundary conditions, depending on whether $h_1 h_L>0$, or $h_1 h_L<0$. In this case the Casimir force  has been studied in Refs.  \cite{NN2008,NN2009,AM2010,DM2013}. Below we provide a short summary of these results concerning the behavior of the scaling functions of the Casimir force.

\begin{itemize}
	\item \textit{subcase}  $h_1h_L \ge 0$
\end{itemize}
 For this subcase the most general result has been derived in Ref. \cite{AM2010}. One uses the transfer matrix  method for calculating the partition function of the two-dimensional Ising model. This has proved itself to be an effective approach in order to obtain exact results for critical Casimir scaling functions. Mapping this procedure to a theory of  free fermions \cite{SML64} allows one to diagonalize the transfer matrix on the  square lattice  with various boundary conditions. For macroscopically  long strips the free energy is given by the largest eigenvalue of the transfer matrix. By considering the lattice wrapped on a cylinder with circumference $M$ and height $L$   with  suitable
boundary conditions  at its ends, and  by taking  the  transfer matrix along the axis of the cylinder  (as in  Ref. \cite{RZSA2010}), it is possible to obtain most of the expressions for the Casimir scaling functions which have been presented above.

\begin{figure}[h!]
	\includegraphics[angle=0,width=0.95\columnwidth]{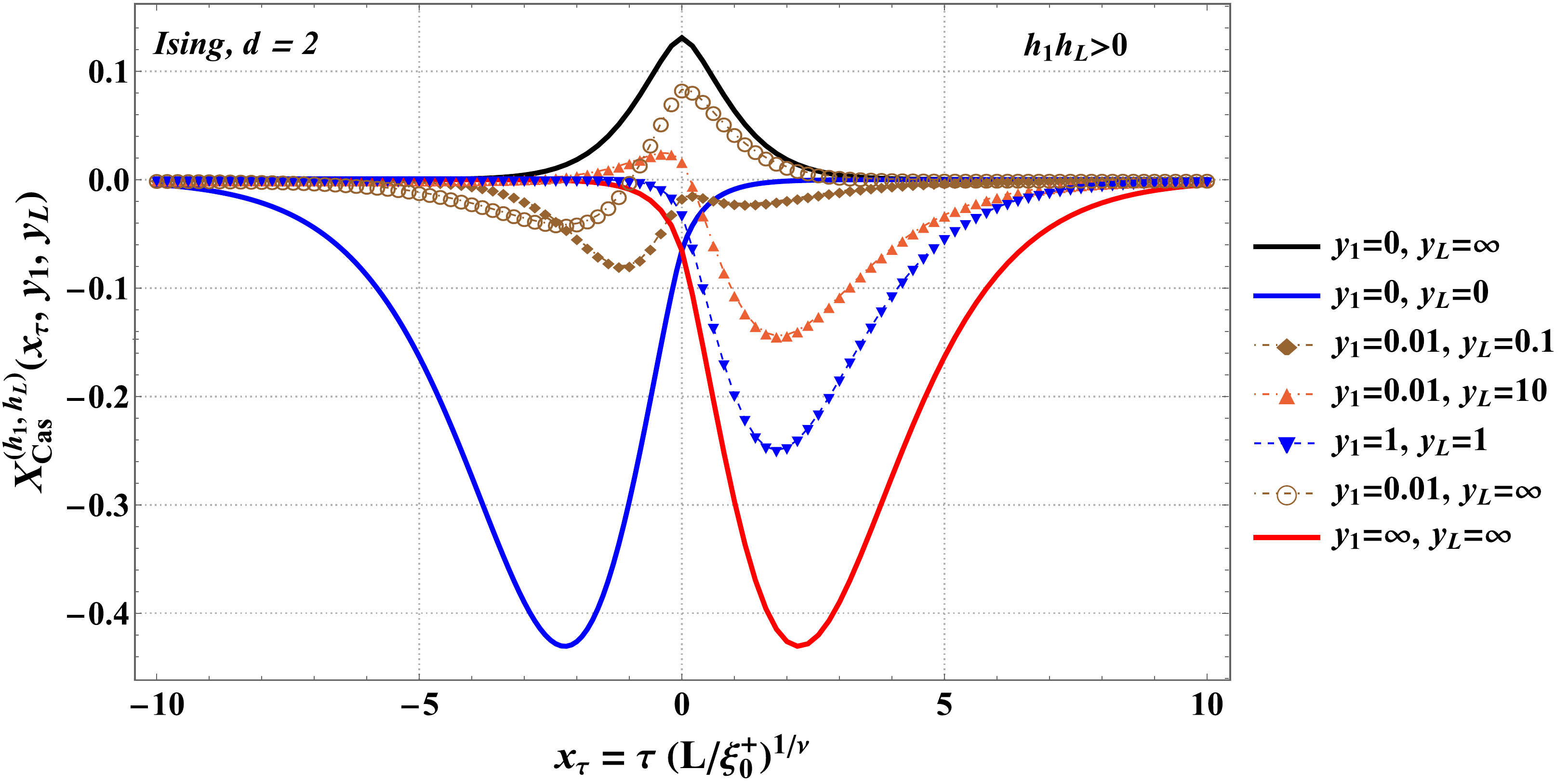}
	\caption{The finite-size scaling functions $X^{(h_1,h_L)}_{\rm Cas}$ of the  Casimir force  for  $(h_1,h_L)$ boundary conditions. We note that certain scaling functions can have both positive and negative parts. In general, they are {\it not} symmetric about $T_c$. The full lines represent universal scaling functions which correspond to the limiting values of $y_1$ and $y_L$. (The symbols serve to better identify the corresponding lines.)}
	\label{fig:xCas_plus_plus_finite}
\end{figure}

We consider the case that $h_1$ and $h_L$ have finite values. We shall denote such boundary conditions as $(h_1,h_L)$. For them, the Casimir force scaling function $X^{(h_1,h_L)}_{\rm Cas}$ depends on the temperature scaling variable $x_\tau$ and the variables characterizing the two scaling fields (see Eqs. \eqref{CasDef} and \eqref{ordinary-extraordinary}):
\begin{equation}
\label{eq:Ising_scaled_fields}
y_1=|\tilde{h}_1|^{\nu/\Delta_1} L,\quad  y_L=|\tilde{h}_L|^{\nu/\Delta_1} L,\qquad \mbox{with} \quad \tilde{h}_j=\beta h_j, j=1,L, \quad \mbox{and} \quad \Delta_1=1/2,  \quad \nu=1.
\end{equation}
In the scaling limit one has\footnote{This expression follows from Eq. (3) in Ref. \cite{AM2010} but there the sign of $x_\tau$ has been mixed up; this sign is misprinted in the corresponding expression in Ref.  \cite{AM2010}.  } \cite{rem1}
\begin{eqnarray}
\label{eq:Ising_finite_pos_fields}
\lefteqn{ X_{{\rm Cas}}^{(h_1,h_L)}(x_\tau,y_1,y_L)=-\frac{1}{\pi}\int_0^{\infty} \sqrt{x_\tau^2+u^2} }\\
&& \times \left  \{\frac{\left[\sqrt{x_\tau^2+u^2}-x_\tau\right]}{\left[\sqrt{x_\tau^2+u^2}+x_\tau\right]}\frac{\left[\sqrt{x_\tau^2+u^2}+x_\tau+2e^{2K_c}y_1\right] \left[\sqrt{x_\tau^2+u^2}+x_\tau+2e^{2K_c}y_L\right]}{\left[\sqrt{x_\tau^2+u^2}-x_\tau-2e^{2K_c}y_1\right] \left[\sqrt{x_\tau^2+u^2}-x_\tau-2e^{2K_c}y_L\right]}\exp{\left(2\sqrt{x_\tau^2+u^2}\right)}+1\right \}^{-1} du.\nonumber
\end{eqnarray}
From this expression it is clear that if $y_1={\cal O}(1)$ or  $y_L={\cal O}(1)$, the corresponding scaling function depends on the critical coupling $K_c$ of the system, which is a {\it non-universal} quantity. Thus, one can obtain universal scaling functions only if both $y_1$ and $y_L$ tend either to $0$ or to  $\pm \infty$. {\it (i)}  If $y_1, y_L\to 0$, one obtains the case of $(O,O)$ boundary conditions; {\it (ii)} if $y_1, y_L \to \infty$, the case of $(+,+)$ boundary conditions is encountered, and {\it (iii)}  if one of the surface scaling variables $y_1$ or $y_L$ tends to $\infty$ while the other one tends to zero, one obtains the case of $(O,+)$ boundary conditions. In the case $x_\tau=0$,   \eq{eq:Ising_finite_pos_fields} renders the corresponding critical amplitudes (see Table \ref{table:2dIsing}). The case $h_1=h_L$ has been investigated in detail in Refs.  \cite{MCD99,NN2008,NN2009}. For this case numerical results  have been also reported in Refs.  \cite{MCD99,MDB2004} which follow from applying the quasi-exact density-matrix renormalization-group (DMRG) method. The current result in  \eq{eq:Ising_finite_pos_fields} allows one to study  the force for any combination of the magnitudes of $y_1$ and $y_L$, provided that $y_1 y_L\ge 0$. By performing the numerics one observes that $X_{{\rm Cas}}^{(h_1,h_L)}(x_\tau,y_1,y_L)$ can change sign upon varying the temperature scaling variable $x_\tau$, provided that $y_1$ and $y_L$  differ significantly in magnitude \cite{AM2010}. The scaling function  $X_{{\rm Cas}}^{(h_1,h_L)}(x_\tau,y_1,y_L)$ is depicted in Fig. 	\ref{fig:xCas_plus_plus_finite} for various combinations of $y_1$ and $y_L$.
The crossover from one universality class to another is described by finite and nonzero values of the scaling fields $y_1$ and $y_L$. 

\begin{itemize}
 \item \textit{subcase}  $h_1h_L \le 0$
\end{itemize}

We again consider the case that $h_1$ and $h_L$ have finite values. We denote such boundary conditions, with opposite signs of the surface fields, as $(h_1,-h_L)$ boundary conditions.

\begin{figure}[h!]
	\includegraphics[angle=0,width=0.95\columnwidth]{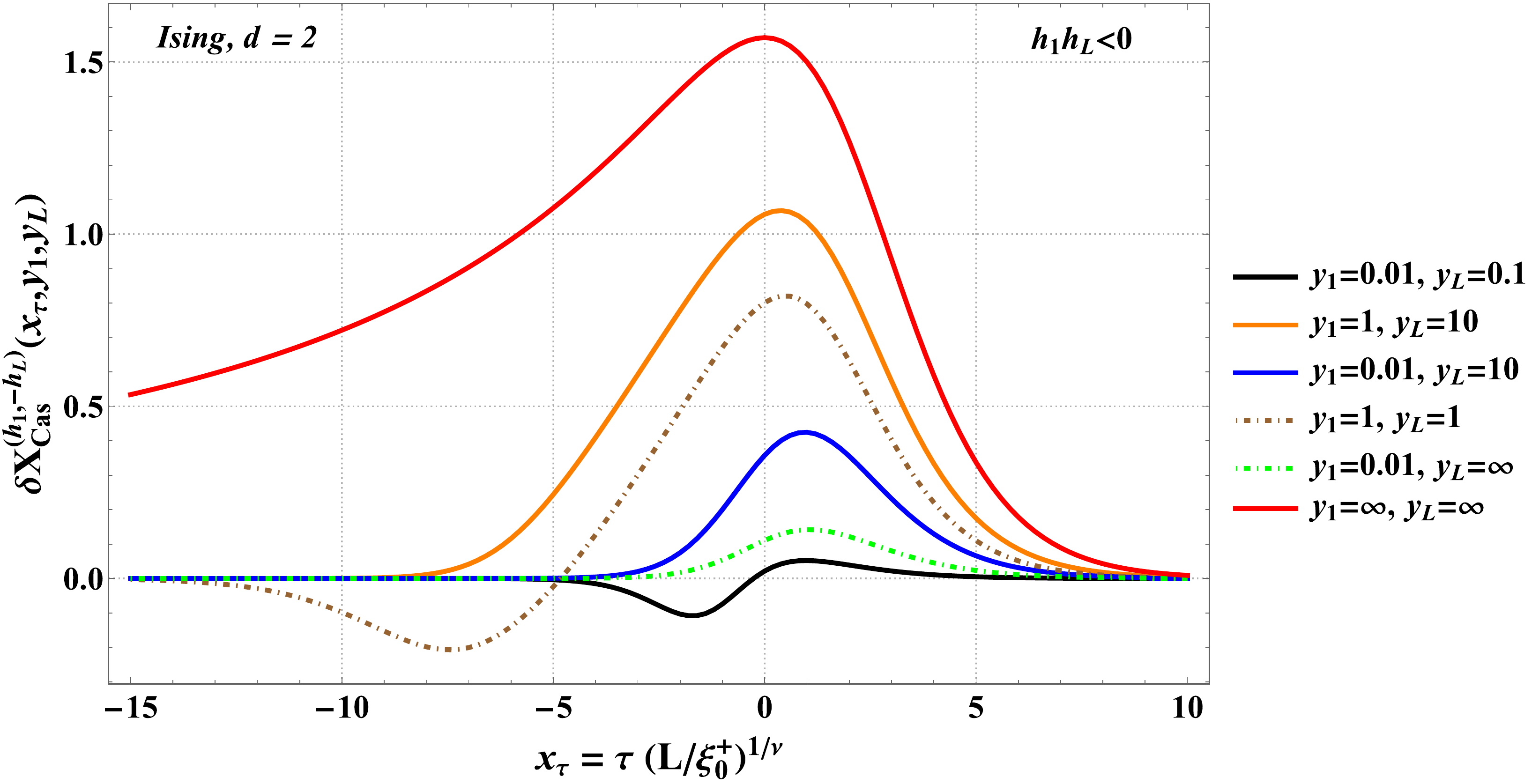}
	\caption{The finite-size scaling functions $\delta X^{(h_1,-h_L)}_{\rm Cas}$ of the excess Casimir force  for $(h_1,h_L)$, $h_1h_L<0$, boundary conditions. Note that the scaling functions can be both positive and negative. In general, they are {\it not} symmetric about $T_c$. The full red line represents a universal scaling function which corresponds to the limiting values  $y_1\to \infty$ and $y_L\to\infty$.}
	\label{fig:deltaxCas_plus_minus_fields}
\end{figure}

\begin{figure}[h!]
	\includegraphics[angle=0,width=0.95\columnwidth]{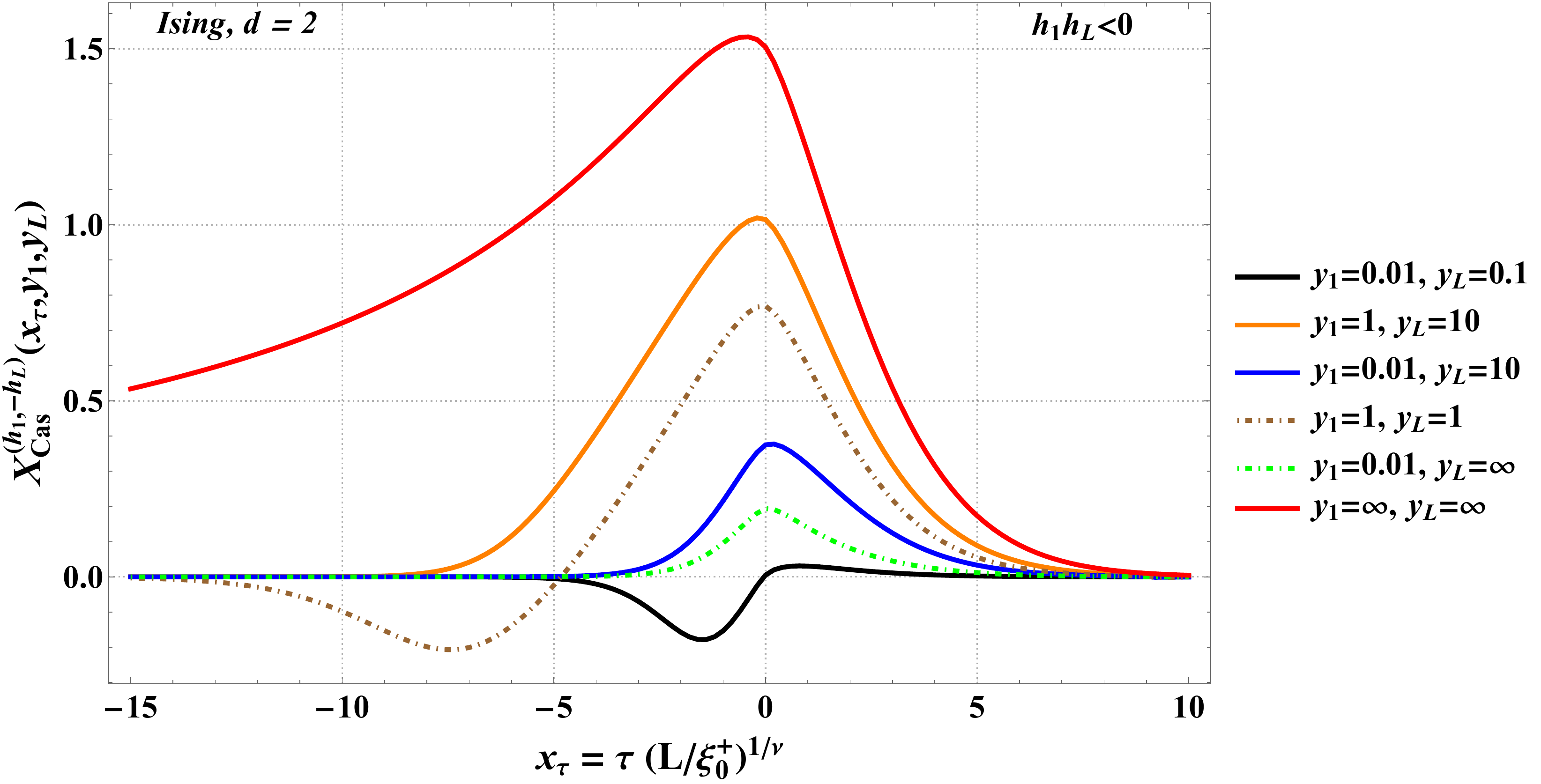}
	\caption{The finite-size scaling functions of the Casimir force $X^{(h_1,-h_L)}_{\rm Cas}$ for boundary conditions $(h_1,h_L)$ with $h_1h_L<0$. We note that certain scaling functions can have both positive and negative parts. In general, they are {\it not} symmetric about $T_c$. We note that the Casimir force is non-monotonic and can have both a negative minimum and a positive maximum. The comparison with Fig. \ref{fig:deltaxCas_plus_minus_fields} shows that the dominant contribution to the force is stemming from the finite-size effects related to the occurrence of a fluctuating interface inside the system. They change completely the behavior of the scaling functions from  $X^{(h_1,h_L)}_{\rm Cas}$ to $X^{(h_1,-h_L)}_{\rm Cas}$ (see \eq{eq:X_Cas_plus_minus_finite}).}
	\label{fig:xCas_plus_minus_fields}
\end{figure}

The critical Casimir scaling function  is
\begin{equation}
\label{eq:X_Cas_plus_minus_finite}
X_{{\rm Cas}}^{(h_1,- h_L)}(x_\tau,y_1,y_L) = X_{{\rm Cas}}^{(h_1,h_L)}(x_\tau,y_1,y_L)+\delta X_{{\rm Cas}}^{(h_1,-h_L)}(x_\tau,y_1,y_L), 
\end{equation}
where $\delta X_{{\rm Cas}}^{(h_1,-h_L)}$ stems from the presence  of an interface inside the system  (see \eq{relationplusminus} for the case of $(+,-)$ boundary conditions). We note that, by definition, the scaling fields $y_1$ and $y_L$  are positive, i.e., $y_1>0$ and $y_L>0$, independent of the sign of $h_1$ and $h_L$  (see \eq{eq:Ising_scaled_fields}).  We introduce the notation $\mathbf{r}=(x_\tau,y_1,y_L)$. With this,  $\delta X_{{\rm Cas}}^{(h_1,-h_L)}(x_\tau,y_1,y_L)$ can be written compactly \cite{AM2010} as \footnote{The factor $N^2$ in the denominator of Eq. (6) in Ref. \cite{AM2010} is a misprint. The correct equation  is the one given here.}
\begin{equation}
\label{eq:delta_X_plus_minus_finite_fields}
 \delta X_{{\rm Cas}}^{(h_1,-h_L)}(\mathbf{r})=\frac{u_0^2-u_0\; \mathbf{r}\cdot { \mathbf{\nabla}} u_0}{\sqrt{x_\tau^2+u_0^2}},
\end{equation}
where  $u_0(\mathbf{r})$ is the  solution\footnote{Note that in Ref. \cite {AM2010} the sign of $x_\tau$ is mixed up. The correct equation is the one given here. } of the so-called "quantization condition" \cite{AM2010}
\begin{equation}
\label{eq:quant_condition}
e^{2iu}=-\frac{\left[i u+x_\tau\right]}{\left[i u-x_\tau\right]}\frac{\left[i u-x_\tau-2e^{2K_c}y_1\right] \left[i u-x_\tau-2e^{2K_c}y_L\right]}{\left[i u+x_\tau+2e^{2K_c}y_1\right] \left[i u +x_\tau+2e^{2K_c}y_L\right]}. 
\end{equation}
In \eq{eq:delta_X_plus_minus_finite_fields} the derivatives of $u_0$  are to  be calculated by using  \eq{eq:quant_condition}. After performing these cumbersome calculations one arrives at
\begin{equation}
\label{eq:delta_X_plus_minus_finite_fields_final}
\delta X_{{\rm Cas}}^{(h_1,-h_L)}(x_\tau, y_1, y_L)=
\frac{u_0^2 \sqrt{u_0^2+x^2_\tau}}{u_0^2+x_\tau (x_\tau+1)+a y_1\frac{ u_0^2-x_\tau (a y_1+x_\tau)}{u_0^2+(a y_1+x_\tau)^2}+a y_L\frac{
		u_0^2-x_\tau (a y_L+x_\tau)}{u_0^2+(a y_L+x_\tau)^2}},
\end{equation}
where $a=2\exp(2K_c)$. In certain intervals of $x_\tau$, \eq{eq:quant_condition} has more than one solution. One looks  either for real or for purely imaginary solutions. If the solution is real we take the smallest nonzero $u$; if the solutions are imaginary we take the one with the largest absolute value\footnote{These rules are following from the requirement of a minimal value of Onsager's $\gamma$ function \cite{O44}. Here we spare the reader the details of the corresponding derivation. In a more explicit form, the rules for choosing  the proper solution can be inferred from Ref. \cite{MS96}  (see also Table 1 in Ref. \cite{NN2009}). As shown there, one has to pay attention to four distinct intervals of $x_\tau$.}.   The result of the evaluation of  $\delta X_{{\rm Cas}}^{(h_1,-h_L)}(x_\tau, y_1, y_L)$ is shown in Fig. \ref{fig:deltaxCas_plus_minus_fields}, while for such boundary conditions the scaling function of the full Casimir force  $X_{{\rm Cas}}^{(h_1,-h_L)}(x_\tau, y_1, y_L)$  is shown in Fig.	\ref{fig:xCas_plus_minus_fields}.
There $\delta X_{{\rm Cas}}^{(h_1,-h_L)}(x_\tau, y_1, y_L)$ and $X_{{\rm Cas}}^{(h_1,-h_L)}(x_\tau, y_1, y_L)$
are plotted as function of  $x_\tau$ for $K_\|=K_\perp$ (see Sect. \ref{eq:two-dimensional-Ising-model-definition}) and several choices of the scaling variables $y_1$ and $y_L$. The change of sign of the Casimir scaling function  occurs for cases in which the scaling variables $y_1 = y_L$
are associated with the  localization-delocalization transition \cite{A80,PE90}.
This feature persists for a weakly broken symmetry, i.e., for $y_1\approx y_L$ and for $y_{i}\ll 1, i={1,L}$. 
 For strongly asymmetric strips the excess scaling function of the critical Casimir force is always positive.

The behavior of the Casimir force for finite values of $h_1$ and $h_L$ can be easily understood in terms of the framework of the general theory. According to \eq{ordinary-extraordinary}, if the scaling fields $x_{h,i}, i=1, L$, associated with the surface fields $h_{i}, i=1, L$, are finite, one expects that the singular part of the free energy, and thus of the Casimir force, depends on the combinations  $x_{h,i}=h_i  L^{\Delta_1/\nu}, i=1, L$, (see \eq{ordinary-extraordinary}). Having in mind that for the two-dimensional Ising model $\Delta_1=1/2$ and $\nu=1$, one has  $y_i=x_{h,i}^{\nu/\Delta_1}=(h_i)^{\nu/\Delta_1}L$. This implies that Eqs. \eqref{eq:Ising_finite_pos_fields}  and  \eqref{eq:X_Cas_plus_minus_finite} are in full agreement with the functional form given by \eq{ordinary-extraordinary}. If $y_i, i=1,L$, increase one observes a crossover from the corresponding ordinary surface universality class to the normal one with $+$ or $-$ boundary conditions, depending on the sign of the corresponding fields $h_i, i=1,L$. 

Because of its relative simplicity the two-dimensional Ising model offers the possibility to study the behavior of the fluctuation induced force not only near the critical point $T_c$ of the infinite system but also below $T_c$. As explained in Sec. \ref{more_on_Casimir}, if the occurrence of an interface within the system is enforced by the boundary conditions, the excess free energy of the two-dimensional Ising model decays  $\propto L^{-2}$ for $T_w<T<T_c$ 
\cite{ES94,NN2008,NN2009,RZSA2010,AM2010} and, as a function of $L$, exponentially below $T_w$, where $T_w$ is the wetting transition temperature. The behavior of the force near $T_w$ of the system is of special interest because in the vicinity of $T_w$  the force changes sign (see below). We shall comment briefly   on the analytical results for the two-dimensional Ising model near $T_w$ without covering them exhaustedly. 
\begin{figure}[htb]
	\includegraphics[angle=0,width=0.85\columnwidth]{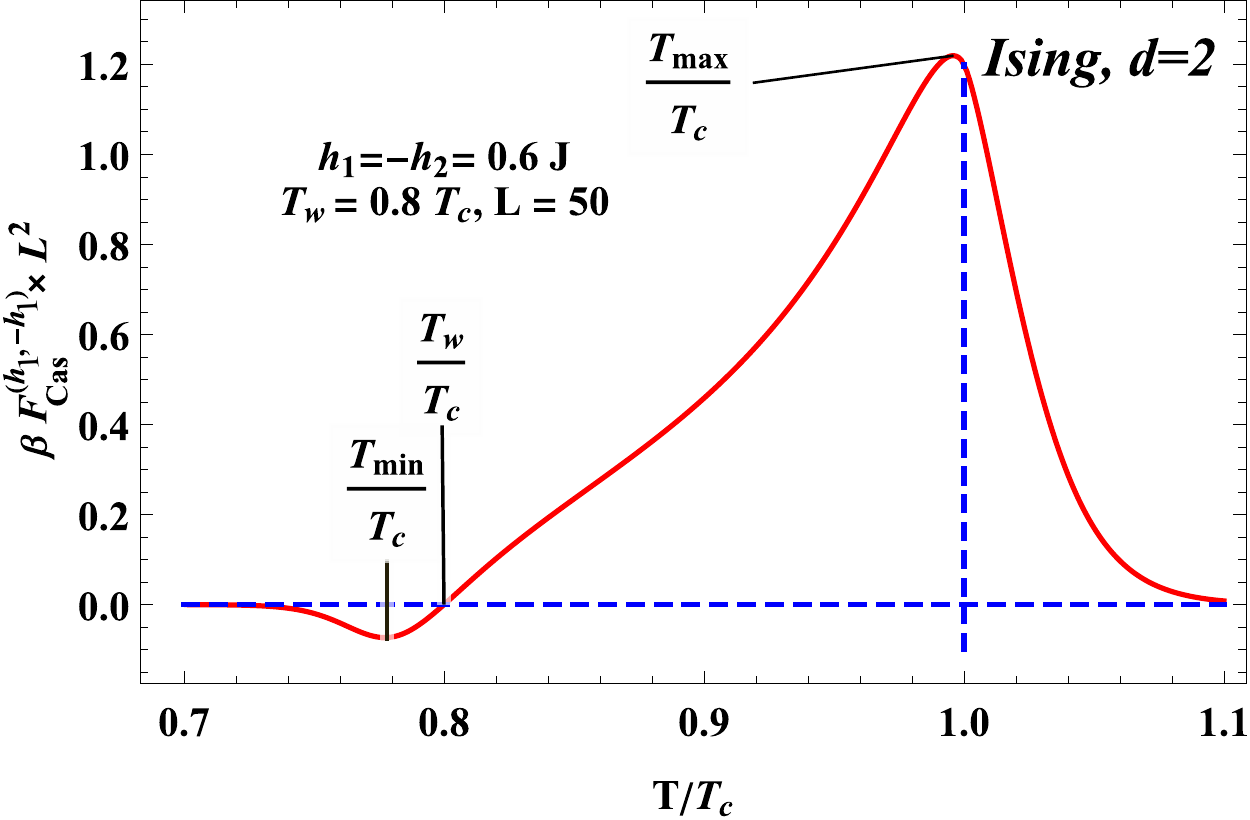}
	\caption{The finite-size behavior of the rescaled Casimir force \cite{remMarek} per area $\beta F^{(h_1,-h_1)}_{\rm Cas} L^2$ as obtained numerically via transfer matrix calculations for the force with $L=50$ and $h_1=-h_2=0.6J$ for boundary conditions $(h_1,-h_1)$. Near $T_c$ this equals the scaling function $X^{(h_1,-h_1)}_{\rm Cas}$ of the  Casimir force for boundary conditions with finite surface fields of opposite signs (see the red curve in Fig. \ref{fig:xCas_plus_minus_fields}).
    For the surface field $h_1=J$ the boundary conditions here are equivalent to the $(+,-)$ ones.  At a finite distance in temperature below $T_c$ the force (in this region often called  the solvation force) attains its minimum below the wetting transition temperature $T_w$, becomes zero at $T^*$ slightly above $T_w$, and reaches its maximum slightly below $T_c$.  The difference between the zero of the Casimir
    force $T^*$ and the wetting temperature $T_w$ is not visible on the present scale.}
	\label{caspmfiniteh1}
\end{figure}
In this respect, the case $h_1=-h_L$ and $h_1<J$, which we denote  as $(h,-h)$ boundary conditions,
has been analyzed by applying the transfer matrix approach parallel to the surfaces ~\cite{NN2008,NN2009}.
Special attention has been paid to the  behavior of the  Casimir force near the wetting temperature of a single wall. In this case 
 the typical variation  of the Casimir force as a function of temperature \cite{remMarek} is shown 
 in Fig. \ref{caspmfiniteh1} for $h_1=0.6\, J$, $L=50$  (compare with Fig. 2 in Ref. \cite{NN2008}). In the isotropic case (see \eq{eq:two-dimensional-Ising-model-definition}) one has $K_\|=K_\perp=K$ \cite{NN2008} (compare the curve corresponding to $y_2=1$ in
 Fig.~5 of Ref. \cite{AM2010}).
For this system the wetting transition temperature is $T_w=0.8 T_c$.
A comparison between the exact analytical curve and the numeric result \cite{remAnia}  for $L=90$ and $(+,-)$ boundary conditions, which for $h_1=J$ renders $T_w=0$,
is presented in Fig. \ref{caspmdifffiniteh1}. We observe in Fig.  \ref{caspmfiniteh1} that, if $h_1<J$, for
low temperatures the Casimir force is negative (attractive) and has a minimum at $T_\mathrm{min}<T_w$. 
The force is still negative at the wetting temperature $T_w$ and
has a zero at $T=T^*$, where $T^*>T_w$. Above $T^*$ the force is positive
(i.e., repulsive) and has a maximum at $T_\mathrm{max}<T_c$ (for $L$ sufficiently large). This is
in contrast to the case $T_w=0$  in which the Casimir force is positive for all
temperatures (see the red curve in Fig. \ref{fig:xCas_plus_minus_fields}). For fixed $h_1> 0$ and
$L\to\infty$, $T^*$ approaches $T_w$ exponentially \cite{NN2008}:
\begin{equation}
\label{tast}\frac{T^*-T_w}{T_c}=A\left(h_1\right) {\rm e}^{-B\left(h_1\right)L},\quad L\to\infty,
\end{equation}
where $A\left(h_1\right)$ and $B\left(h_1\right)$ are positive functions of the surface field $h_1$, such that
for $h_1\to 0$, i.e., $T_w \to T_c$, one has
\begin{equation}
\lim_{h_1\to 0} A\left( h_1\right)=0.
\end{equation}
This result differs from the corresponding one obtained within
mean field theory, according to which, as function of $L$, $T^*$ is exponentially shifted below $T_w$. It also differs from the corresponding result for the
restricted
solid--on--solid (RSOS) model, within which $T^*$ is equal
to $T_w$ \cite{PE92}.
\begin{figure}[htb]
\includegraphics[angle=0,width=\columnwidth]{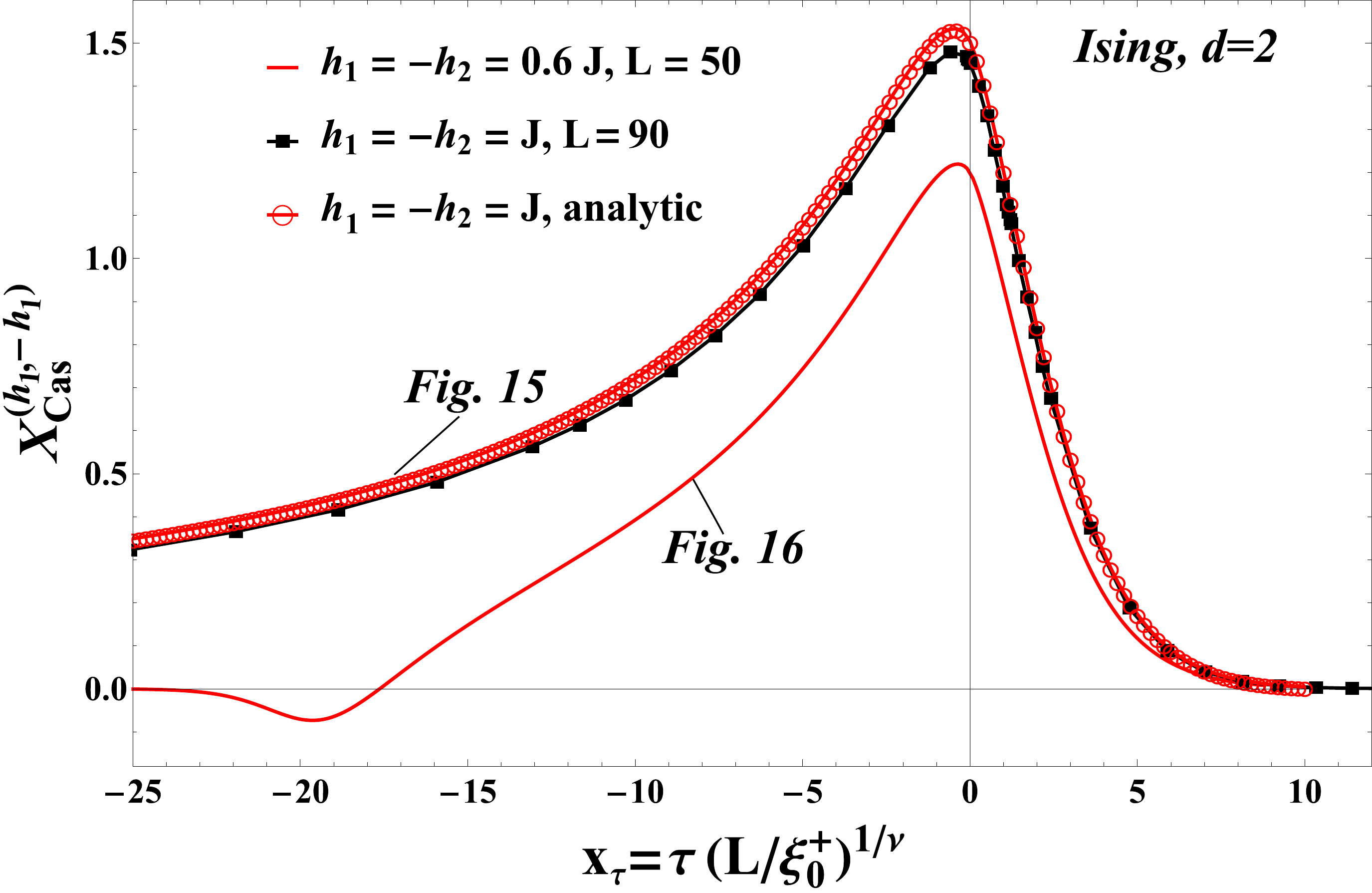}
\caption{Comparison between the finite-size scaling functions $X^{(h_1,-h_1)}_{\rm Cas}$ of the  Casimir force $F^{(h_1,-h_1)}_{\rm Cas}$ for boundary conditions with finite surface fields and  opposite signs for moderate values of $L$ (i.e., $L=50$ and  $L=90$) with the corresponding analytical result valid for $L\gg 1$. If the surface field is $h_1=J$, the boundary conditions are equivalent to the $(+,-)$ ones. The red curve marked with empty circles is the same as in Fig. 	\ref{fig:xCas_plus_minus_fields}, while the full red curve is the same as the corresponding one in Fig. 	\ref{caspmfiniteh1}. The position of the wetting temperature, $T_w/T_c=0.8$ for $h_1=-h_2=0.6 \,J$ and $L=50$, is now at $x\equiv x_{\rm w}=-17.6275$, while the minimum value is attained at $x_{\rm min}=-19.6$. (The symbols serve as to better identify the lines.) }
\label{caspmdifffiniteh1}
\end{figure}

Within the two-dimensional Ising model, by changing the strength of the surface field $h_1$, one can vary the wetting temperature in the range $0\le T_w\le T_c$.
If $T_w$ is sufficiently far below the critical temperature $T_c$, one can consider the scaling limit  $L\to \infty, T\to T_w$, with
the variable $x'=L\ln W(T,h_1)\sim (T_w-T)L>0$ fixed. Here $W(T,h_1)=(\cosh 2K^*+1)\left(\cosh 2K- \cosh 2\beta h_1 \right)$, with $K=\beta J$ and $K^*$ related via $\sinh 2K \sinh 2K^*=1$. 
In this scaling limit, for large $L$ the corresponding force exhibits in  leading order a decay  $\propto L^{-3}$:
\begin{equation}\label{Eq:pmfiniteh1wet}
F^{(h,-h)}_{\rm Cas}(T,h_1,L)=-L^{-3} G^{(h,-h)}_{\rm Cas}(x',h_1),
\end{equation}
where the scaling function $G^{(h,-h)}_{\rm Cas}$ is given by 
\begin{equation}\label{Eq:pmfiniteh1wetsf}
 G^{(h,-h)}_{\rm Cas}(x',h_1)=\frac{x'^3H^2(x')(H^2(x')-1)}{\left[\sinh v(h_1)\right]\left[2+x'(H^2(x')-1)\right]}
\end{equation}
with $v(h_1)=2(K-K^*)|_{T=T_w(h_1)}$. The function $H(x')$ is  given implicitly as the negative solution of the
equation
\begin{equation}\label{Eq:pmfiniteh1wetqc}
 e^{-x'H(x')}=\frac{H(x')-1}{H(x')+1}.
\end{equation}
The scaling function $G^{(h,-h)}_{\rm Cas}$ is shown in Fig.~6 of Ref. \cite{NN2009}, while $H(x')$ is shown here in Fig. \ref{Hofxprime}.
\tcr{\begin{figure}
\includegraphics[angle=0,width=0.8\columnwidth]{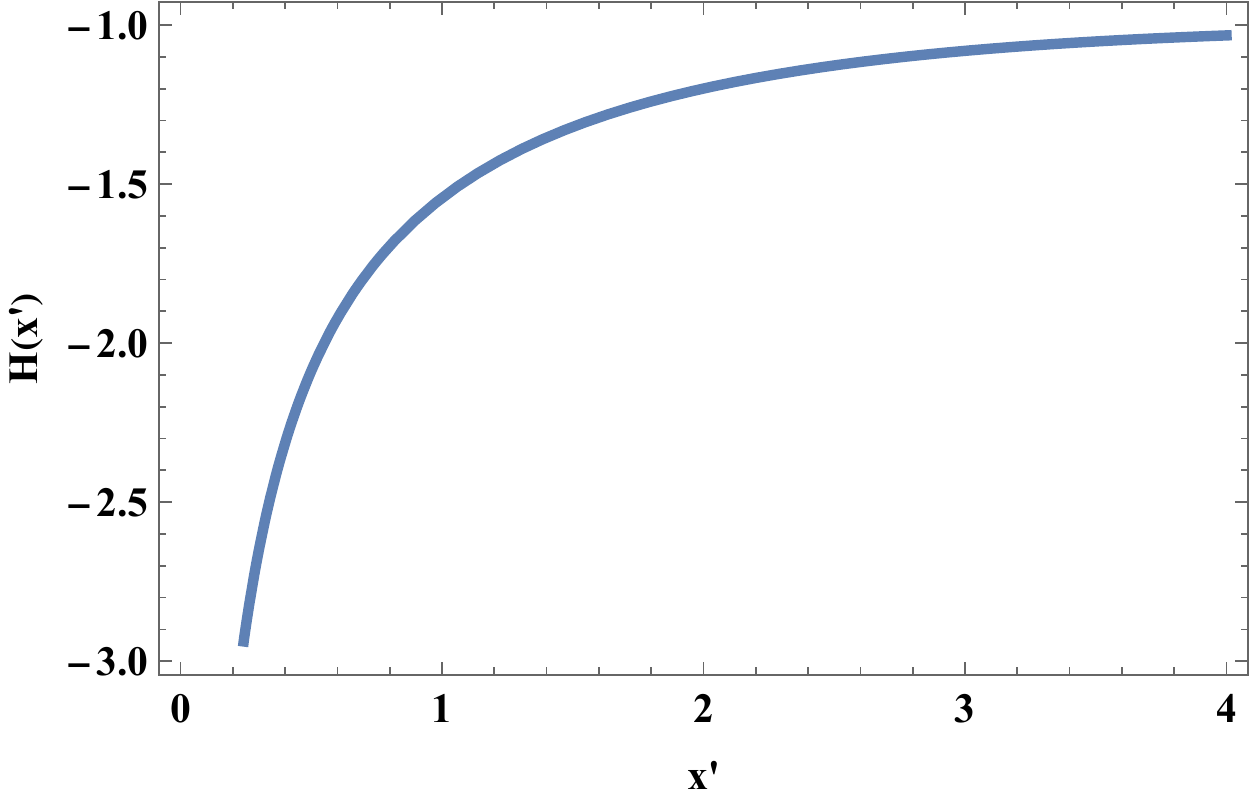}
\caption{$H(x')$ as defined in \eq{Eq:pmfiniteh1wetqc}.
\label{Hofxprime}}
\end{figure}}
It
is positive,
increases linearly from zero, exhibits
 a maximum at $x'\approx 3.22$, and decays exponentially for large $x'$.
Introducing the scaling variable
\begin{equation}
\label{eq:y-variable-wetting}
y=\frac{A_0}{k_B T_c} \frac{h_1}{\tau^{\Delta_1}}
\end{equation}
and considering the limits  $(L\to \infty, T\to T_c)$ and $h_1\to 0$, allows one to study the scaling limit for $T_w\to T_c$.
The corresponding
 scaling function is $X^{(h,-h)}_{\rm Cas}(x,y)$, i.e., 
\begin{equation}\label{Eq:pmfiniteh1}
F^{(h,-h)}_{\rm Cas}(T,h_1,L)=L^{-2} X^{(h,-h)}_{\rm Cas}(x,y),
\end{equation}
where $x$ is further given by Eq. (\ref{Eq:defoftau})
with the gap exponent $\Delta_1=1/2$ for the two-dimensional Ising model. In Ref.  \cite{NN2008} the constant $A_0$ (see \eq{eq:y-variable-wetting}) has been chosen as
\begin{equation}\label{Eq:A0def}
A_0=\left[\left(1+\sqrt{2}\right)/\ln\left(1+\sqrt{2}\right)\right]^{1/2}\simeq 1.655,
\end{equation}
and $y$ is given by \eq{eq:y-variable-wetting}, so that $y=1$ corresponds to $T=T_w$. Thus, for $y < 1$ the function $X^{(h,-h)}_{\rm Cas}(x,y)$ gives the Casimir force $F^{(h,-h)}_{\rm Cas}(T,h_1,L)$ for $T$ below the wetting transition temperature $T_w$, while for $y>1$ it characterizes its behavior above $T_w$. The scaling functions $X^{(h,-h)}_{\rm Cas}(x,y)$ and $X^{(+,-)}_{\rm Cas}(x)$ are related according to 
\begin{equation}\label{Eq:relfiniteinfiniteh1}
X^{(+,-)}_{\rm Cas}(x)=\lim_{y\to\infty} X^{(h,-h)}_{\rm Cas}(x,y).
\end{equation}
For selected values of $y$, the dependence of the  scaling function $X^{(h,-h)}_{\rm Cas}(x,y)$ of the Casimir force is shown in Fig. 8 of Ref. \cite{NN2009}.  
Upon increasing  $y$,  the scaling function approaches
the one for $(+,-)$ boundary conditions (see Eqs. (\ref{XcasppIsing2d}), (\ref{relationplusminus}), (\ref{Xpms}), (\ref{pmXs}),
and (\ref{dXpmCas})), while for $y\to 0$ it approaches the one pertinent to the $(O,O)$ surface universality class  (see Eqs. (\ref{XcasppIsing2d})
and (\ref{Eq:ppooCas})).

\section{Exact results for the thermodynamic Casimir effect in the thin film geometry ($d>2$)}

Naturally, the Casimir effect has been most thoroughly studied for systems which belong to the Ising universality class. Physical realizations of such systems are simple, nonpolar fluids, binary liquid mixtures close to their demixing point, binary alloys, etc.  Since the bulk correlation length $\xi$ is finite both above
and below the critical point ($T=T_c$, $\mu=\mu_c$), in view of the absence of additional structures within the finite systems, such as interfaces, the Casimir force is long-ranged only
in the finite-size critical region defined by $L/\xt=O(1)$ and $L/\xm=
O(1)$. Outside of this region the Casimir force is supposed to decay  upon increasing $L$, which is the same as the expected behavior of the finite-size corrections. The current review is focused on systems with film geometry. We note that results concerning the behavior of Casimir-like forces in fully finite systems are also available \cite{H2016,H2017,HH2017,VED2013,EB2016,NMD2016}. We shall briefly comment on them in Sec. \ref{some_other_results}. We mention that mean field results concerning the Casimir force for a canonical ensemble with fixed  magnetization \cite{GVGD2016,GGD2017,RSVG2019} are also available. 

\subsection{Exact mean field results for the Ising universality class}

The definition of the Ising type mean-field
approximation is given in \eq{HGL}. The finite-size scaling functions for the Casimir force, $X_{\rm Cas}^{(a,b)}(x_\tau)$ with $x_\tau \equiv\tau (L/\xop)^{1/\nu}=\tau (L/\xop)^{2}=\sign(\tau)(L/\xi)^{1/\nu}$, with $\nu=1/2$, have been reported 
in Refs.  \cite{K97,GD2006,ZSRKC2007,DVD2016,DVD2018,DVD2018_matec,VDD2019b,DVD2019c,DVD2020,DVD2020b,DVD2020b,Dan2021} for a variety of boundary conditions.  We recall that within mean-field theory the critical behavior is characterized by the critical bulk exponents $\nu=1/2$ and $\alpha=0$. Formally, one can consider the scaling laws between the critical exponents to be still valid, provided the dimension of the system is set to  $d=4$ in any relation in which the dimension explicitly enters. In this way one straightforwardly obtains all the remaining critical exponents. 

Below we consider ordinary (Dirichlet), "$+$", "$-$", and surface-bulk (SB) boundary conditions applied to any of the two surfaces of the film. In terms of the order parameter $\phi^{(a, b)}(z)$ at, say, the surface "\textit{a}" (at $z=0$), within mean field theory they can be specified as follows:
	\\
(i) ordinary (or Dirichlet boundary condition):
\begin{equation}
	\phi^{(O, b)}(z=0)=0;
	\label{eq:Dirichlet_bc_MF}
\end{equation}
(ii) "$+$"  boundary condition:
\begin{equation}
	\lim_{z\to 0}\phi^{(+, b)}(z)=\infty;
	\label{eq:plus_bc_MF}
\end{equation}
(iii) "$-$"  boundary condition:
\begin{equation}
	\lim_{z\to 0}\phi^{(-, b)}(z)=-\infty;
	\label{eq:minus_bc_MF}
\end{equation}
(iv) special (or "SB" boundary condition):
\begin{equation}
	\left(d \phi^{({\rm SB}, b)}\big / d z\right)_{|z=0}=0.
	\label{eq:SB_bc_MF}
\end{equation}

It is convenient to introduce the scaling variables
\begin{equation}\label{isvarxt}
	l_t\equiv {\rm sign}(\tau) \, L/\xi_\tau^+ ={\rm sign}(\tau)\, \left(L/\xi_0^+ \right)\sqrt{|\tau|},
\end{equation}
and
\begin{equation}\label{isvarxh}
	l_h\equiv {\rm sign}(h)\, L/\xi_h={\rm sign}(h)\,\left(L/\xi_0^+\right)\sqrt{3}\left(\sqrt{g}\, |h|\right)^{1/3}, 
\end{equation}
which are the temperature and the field scaling variables, respectively. Here it is taken into account that for the current mean field model $\xi_{0,h}/\xi_0^+=1/\sqrt{3}$ \cite{SHD2003}, $\nu=1/2$, and $\Delta=3/2$.

For the Casimir force scaling functions  $X_{\Cas}^{\bc}$ the following results are known:

\subsubsection{For $(+,+)$ boundary conditions with}

\hspace*{2cm} (I) {\textit  {zero  external bulk field}}

it has been shown \cite{K97} that

{\it (i)} if $x_\tau \geq 0$,  one has
\begin{equation}
	\label{eq:Xcas-pp-mf}
X_{\rm Cas}^{(+,+)}(x_\tau)=-[2K(k)]^4 k^2(1-k^2),
\end{equation}
where $k=k(x_\tau)$, $1/\sqrt{2}\leq k <1$, solves the parametric equation
\begin{equation}\label{pekpos}
    x_\tau=[2K(k)]^2(2k^2-1);
\end{equation}

{\it (ii)} if $-\pi^2 \leq x_\tau \leq 0$,  one has 
\begin{equation}
X_{\rm Cas}^{(+,+)}(x_\tau)=-4[K(k)]^4,
\end{equation}
where $k=k(x_\tau)$, $0\leq k \leq 1/\sqrt{2}$ again solves the parametric equation (\ref{pekpos});

\begin{figure}[h!]
\includegraphics[angle=0,width=\columnwidth]{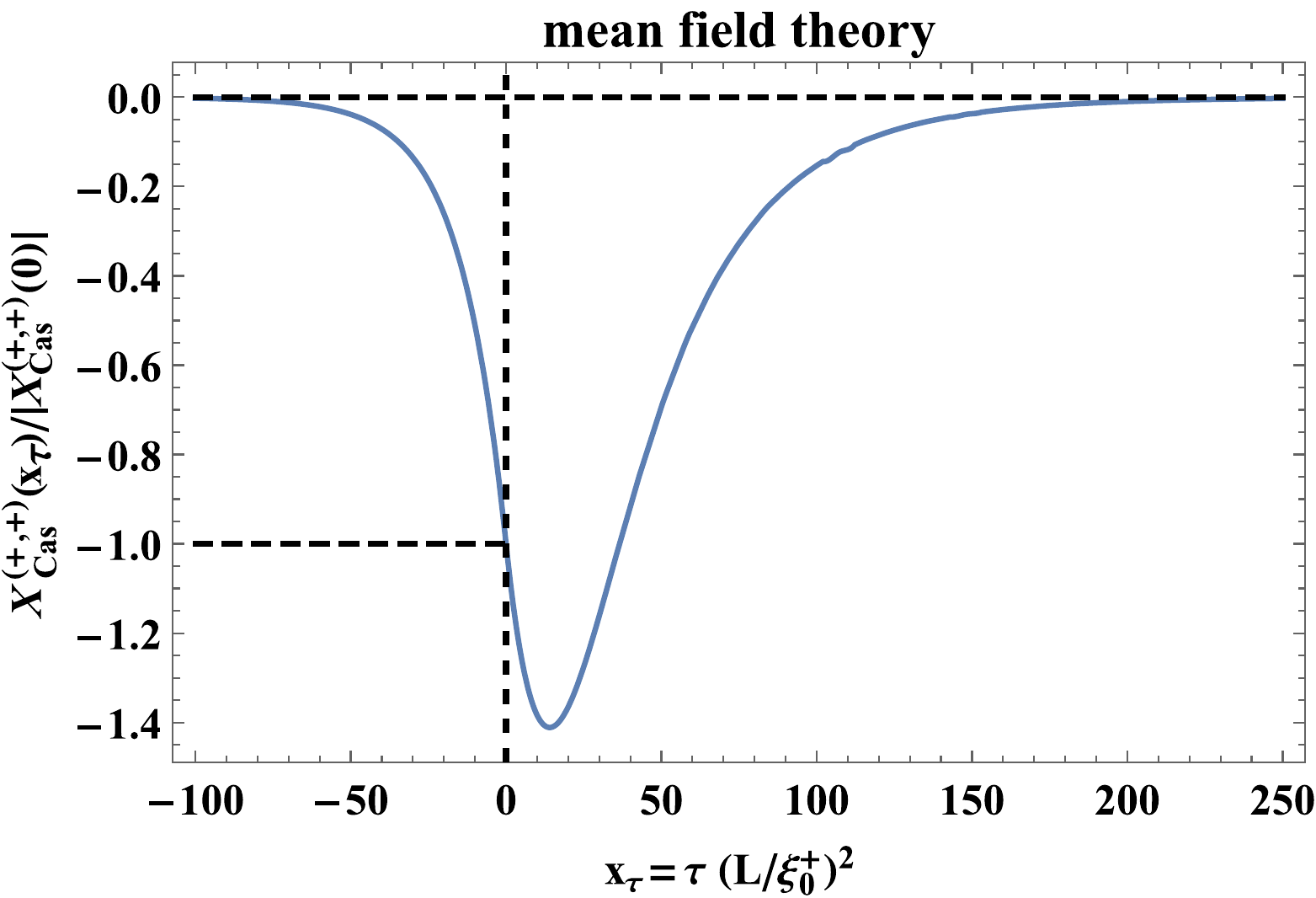}
\caption{The normalized zero-field finite-size scaling function $X_{\rm Cas}^{(+,+)}$ of
the Casimir force as a function of the scaling variable $x_\tau=\tau (L/\xi_0^+)^2$, obtained within 
mean field theory for the film geometry with $(+,+)$ boundary
conditions. The function is normalized by its value at the bulk critical point. It is negative
and has a minimum $X_{\rm Cas}^{(+,+)}(x_{\rm min})/|X_{\rm Cas}^{(+,+)}(0)|\simeq -1.411$ {\it  above} $T_c$ at $x_{\rm min}\simeq 14.055$, as in the case of the two-dimensional Ising model.}
\label{CasPlusMField}
\end{figure}

{\it (iii)} if $x_\tau \leq -\pi^2$, one has 
\begin{equation}
X_{\rm Cas}^{(+,+)}(x_\tau)=-4[K(k)]^4(1-k^2)^2,
\label{eq:plus-plus}
\end{equation}
where $k=k(x_\tau)$, $0\leq k <1$, solves the parametric equation
\begin{equation}\label{pekneg}
    x_\tau=-[2K(k)]^2 (1+k^2).
\end{equation}
In the above equations $K\equiv K(k)$ is the complete elliptic integral of the
first kind. The behavior of $X_{\rm Cas}^{(+,+)}(x_\tau)$ is shown in Fig.
\ref{CasPlusMField}. The function is negative and has a minimum above $T_c$. As follows from \eq{eq:plus-plus}, it is attained at $k=k_{\rm min}$, which solves the equation
\begin{equation}
	\label{eq:kmin-plus-plus}
	K\left(k\right)-2 E\left(k\right)=0,
\end{equation}
with $E(k)$ the complete elliptic integral of the second kind. This equation renders $k_{\rm min}\simeq 0.909$. 	According to Eqs.  \eqref{eq:Xcas-pp-mf} and \eqref{pekpos} this leads to the minimum $X_{\rm Cas}^{(+,+)}(x_{\rm min})/|X_{\rm Cas}^{(+,+)}(0)|\simeq -1.411$ at $x_{\rm min}\simeq 14.055$. 

\hspace*{2cm} (II) {\textit  { non-zero external bulk field}}

\begin{figure}[h!]
	\centering
	\includegraphics[width=\columnwidth]{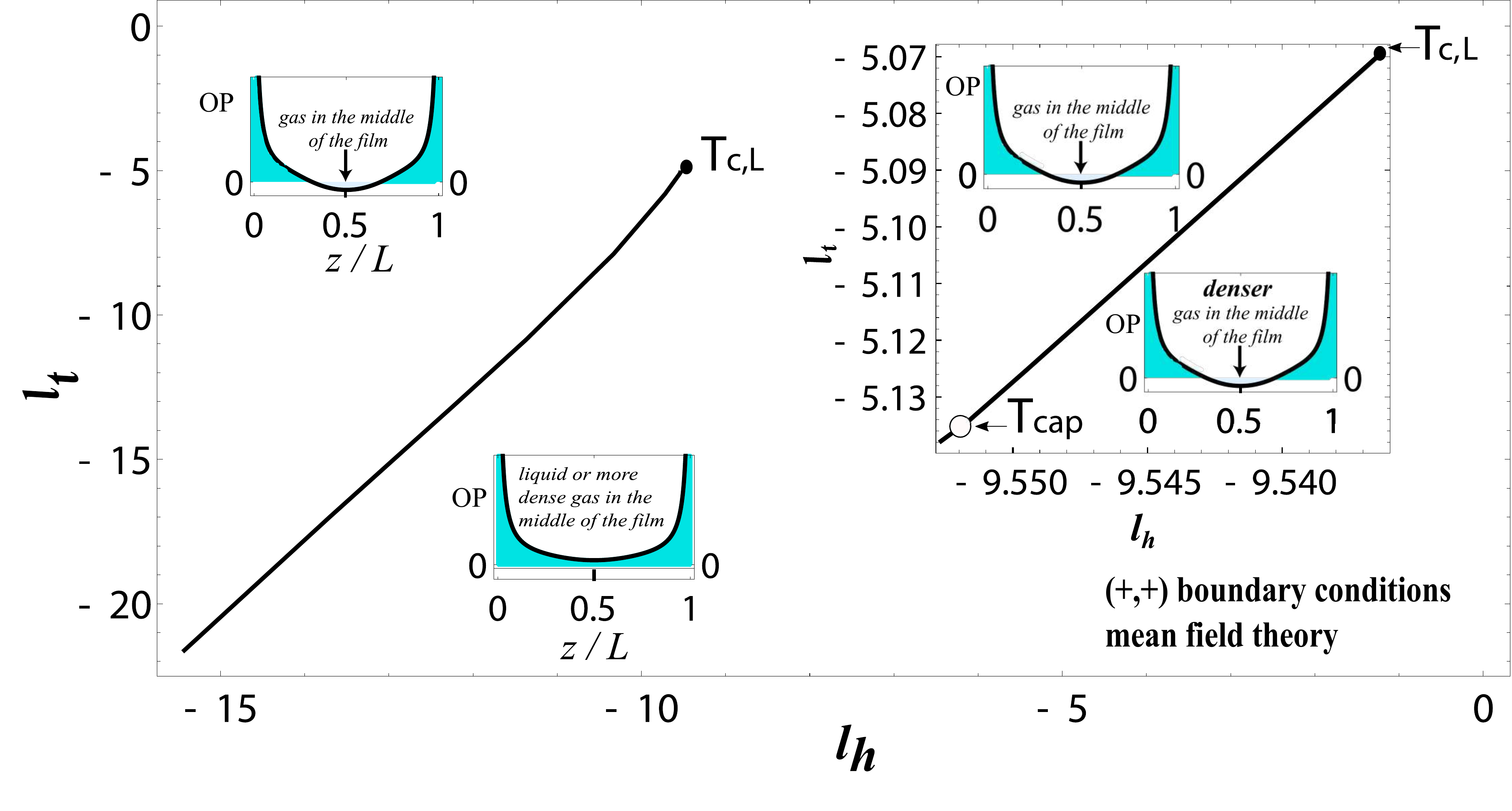}
	\caption{Phase diagram of a capillary of thickness $L$ and with $(+,+)$ boundary conditions in terms of the scaling variables $l_h$ and $l_t$ (Eqs. \eqref{isvarxt} and \eqref{isvarxh}).  Provided the "+" boundary conditions are interpreted as a strong preference of the walls for the "liquid" phase, this phase is always present near the boundaries and gives rise to the increase of the order parameter at the walls. Accordingly, capillary condensation occurs between a gas-like phase in the middle of the system with a liquid-like phase  everywhere else in the system. The spatial regions, in which these phases are stable, are depicted by the two order parameter (OP) profiles sketched in the insets where the darker colored regions represent the "liquid" phase (positive adsorption) while the dimly colored regions  represent the "gas" phase (negative adsorption). The phase coexistence line of the first-order phase transition terminates at
	the critical point of the capillary at $T_{c,L} = (l_t^{(c)},l_h^{(c)})=(-5.06935, -9.53633)$, i.e., at this point the film exhibits a bona fide  phase transition. At such a point the susceptibility of the film diverges; this is how the position of $T_{c,L}$ has been determined. The $(l_h,l_t)$ inset on the right shows the pre-capillary-condensation curve, as determined in Ref.  \cite{DVD2015}, where above $ T_{\rm cap}=(-5.13834, -9.55252)$ and below $ T_{c,L}$ the jump of the order parameter in the middle of the system is from a less dense gas to a more dense gas. However, for $T\le T_{\rm cap}$ the system jumps from a ''gas'' to a ''liquid'' state. Note that in the $(l_h,l_t)$ plane the two points $T_{c,L}$ and $T_{\rm cap}$ are very close to each other.} 
	\label{fig:PD}
\end{figure}

The results for this case have been derived in Ref. \cite{DVD2016}.
The phase diagram of this model has been studied in detail in  Ref. \cite{DVD2015}.  It is presented in figure \ref{fig:PD} in terms of the scaling variables $l_t$ (\eq{isvarxt}) and $l_{h}$ (\eq{isvarxh}). 

The scaling function of the Casimir force $X_{\rm Cas}^{(+,+)}$ in this case takes the form
\begin{equation} \label{Casimir_final}
X_{\rm Cas}^{(+,+)}(l_t,l_h)= X_{b}^4-x_m^4+ {\rm sign}(l_t) \, l_t^2 \left(X_{b}^2-x_m^2\right) -\frac{2}{3 \sqrt{6}}l_h^{3} \left(X_{b}-x_m\right).
\end{equation}

Below we briefly explain the meaning of the quantities appearing in this equation. 
The Casimir force has been determined by using \eq{eq:Casimir}. Therein, for the pressure in the film system one has \begin{equation} \label{FIX}
P_L(\tau,h)=\frac{1}{g L^4} p\left( l_t,l_h \right),
\end{equation}
with\footnote{The expression for $P_L$, and thus of $p$, can be obtained in several ways: as a functional derivative of the free energy functional \cite{DVD2016}; basically the same but re-deriving the details of the functional calculus for the example of a mean field model \cite{INW86}; via the stress tensor formalism \cite{SHD2003,EiS94,VD2013}.   It can be shown that $P_L$ is the first integral of the corresponding differential equation governing the behavior of the order parameter; it is also equal to the pressure acting on the bounding surface at $z=L$; as well as equal to the corresponding component ($T_{zz}$) of the stress tensor (see Ref.  \cite{DVD2016} for details).}
\begin{equation} \label{FIc}
p\left( l_t,l_h \right)={X'}^2
-X^4-{\rm sign}(l_t) \, l_t^2 X^2 +\frac{2}{3 \sqrt{6}}l_h^{3} X.
\end{equation}
Here 
\begin{equation}\label{isvar2}
X(\zeta|l_t,l_h)=\sqrt{\frac{g}{2}}L^{\beta/\nu} \phi(z|\tau,h,L) 
\end{equation}
is the scaling function of the order parameter $\phi$ with  $\beta=\nu=1/2$. Hereafter the prime in \eq{FIc} stands for differentiation with respect to the variable $\zeta=z/L, \zeta \in [0,1]$. Similarly, for the bulk system one has 
\begin{equation} \label{pbX}
P_b(\tau,h)=  \frac{1}{g L^4}p_b(l_t,l_h),
\end{equation}
where
\begin{equation} \label{pbX1}
p_b(l_t,l_h)= -X_{b}^4-{\rm sign}(l_t) \, l_t^2 X_{b}^2 +\frac{2}{3 \sqrt{6}}l_h^{3} X_{b}.
\end{equation}
From Eqs. (\ref{FIX}) and (\ref{pbX}), for the Casimir force (\eq{eq:Casimir}) one obtains 
\begin{equation} \label{CasimirX}
F_{\rm Cas}^{(+,+)}(\tau,h,L)= \frac{1}{g L^4} X_{\rm Cas}^{(+,+)}(l_t,l_h).
\end{equation}
Its scaling function $X_{\rm Cas}^{(+,+)}$ is
\begin{equation} \label{CasimirX1}
X_{\rm Cas}^{(+,+)}(l_t,l_h)= p\left( l_t,l_h \right)-p_b(l_t,l_h).
\end{equation}
\begin{figure}[h!]
	\centering
	\includegraphics[width=3.0in]{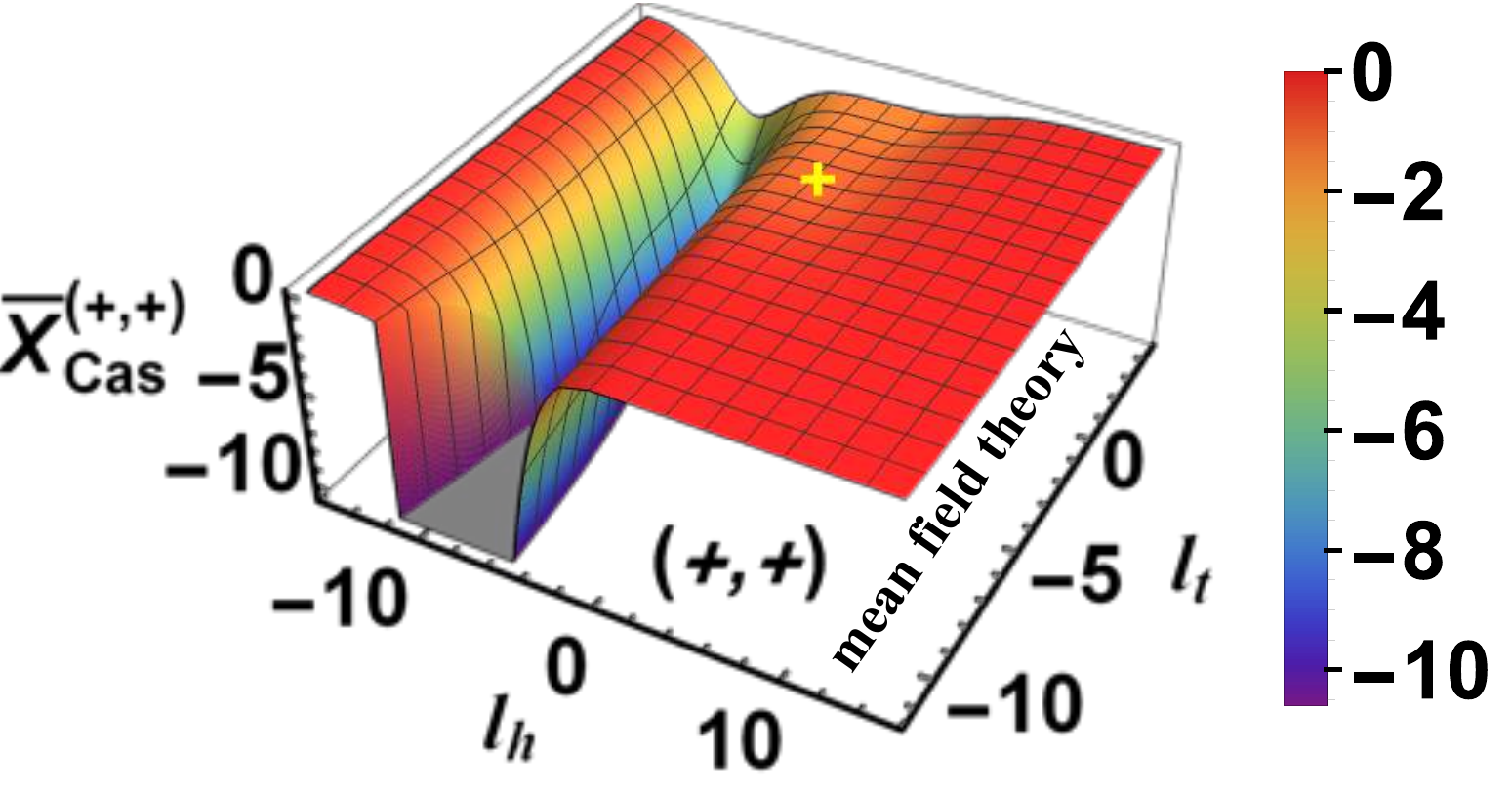} \quad
	\includegraphics[width=3.0in]{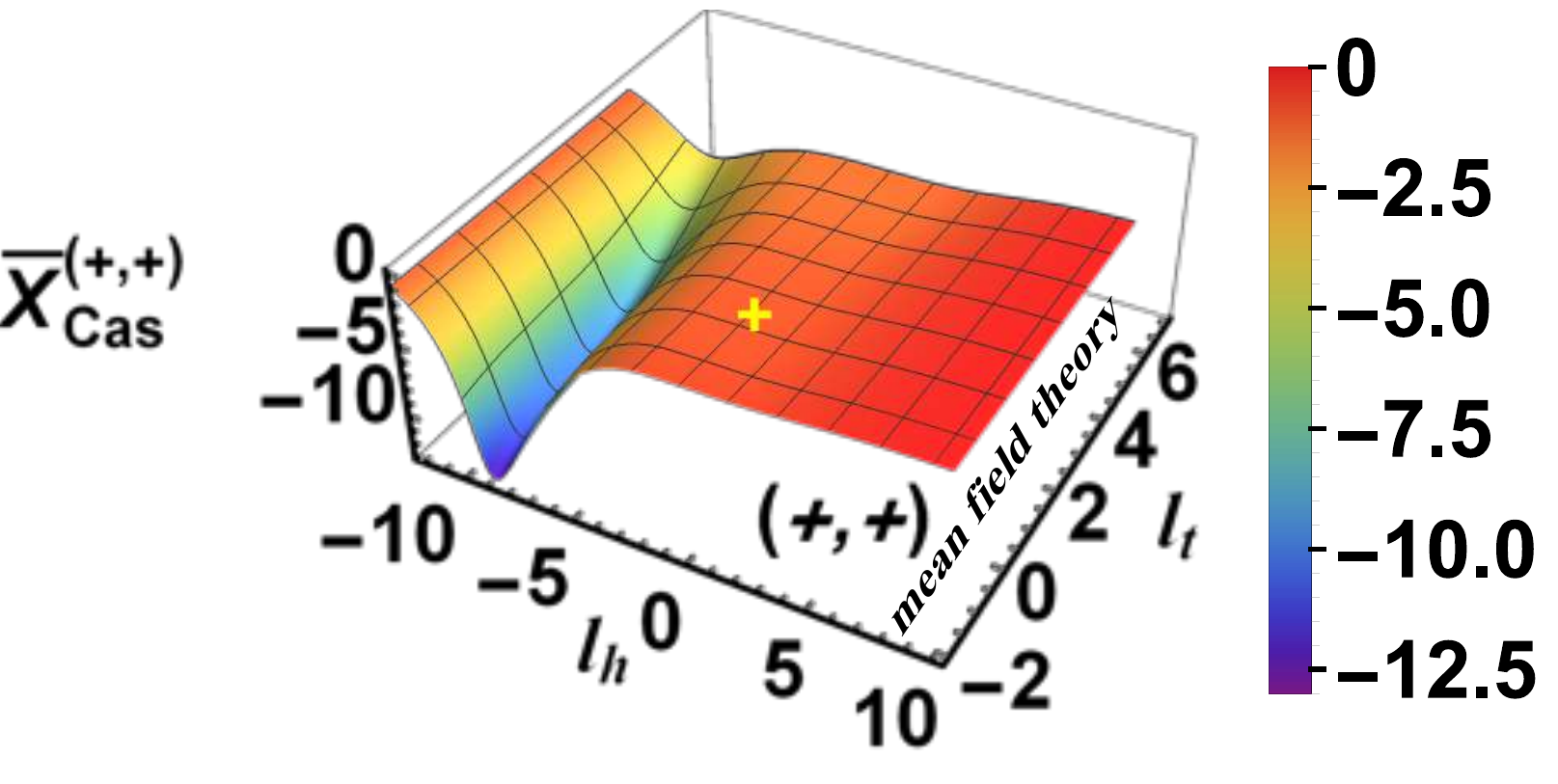}
	\caption{Relief plots of the normalized Casimir force as a function of both $l_t$ and $l_h$. The black lines correspond to $l_h=const.$ or $l_t=const.$ The left panel shows the behavior of $\bar{X}_{\rm Cas}^{(+,+)}$ for a larger temperature interval. The gray "surface of the lake in the valley" is the limit bounding the values shown in the plot from below. The right panel shows in more detail the temperature behavior of the force near the bulk critical point. The yellow cross marks the value of the Casimir force at  $(l_t=0,l_h=0)$.}
	\label{fig:DP}
\end{figure}
The latter relation leads to  \eq{Casimir_final} with $x_m$ as the value of the scaling function of the order parameter profile in the middle of the system, i.e., $X^{(+,+)}(\zeta=1/2|l_t,l_h)=x_m(l_t,l_h)$.  As shown in  Ref.  \cite{DVD2015}, $x_m$ is determined by 
\begin{equation}  \label{TrEq}
12 \wp \left(\frac{1}{2};g_{2},g_{3}\right) - {\rm sign}(l_t)l_t^2 -6 x_m^{2}=0
\end{equation}
so that $x_m$ gives rise to a continuous order parameter profile in the interval $(0,1)$, and satisfies the relation 
\begin{equation}  \label{eq:condition}
6 \sqrt{3}\, x_m \left({\rm sign}(l_t)l_t^2 +2 x_m^{2}\right)- \sqrt{2} \, l_{h}^{3}>0.
\end{equation}
In \eq{TrEq}, $\wp \left(\xi;g_{2},g_{3}\right)$ is the Weierstrass elliptic function \cite{AS,NIST2010}, the invariants  $g_{2}$ and $g_{3}$ of which are given by the expressions
\begin{equation} \label{g1} 
g_2=\frac{1}{12} l_t^{\,4}+p\left( l_t,l_h \right), \quad \mbox{and} \quad 
g_3=-\frac{1}{216} \left[l_{h}^{\,6}+ l_t^{\,6}-36  \, p\left( l_t,l_h \right) l_t^{\,2}\right].
\end{equation}
\begin{figure}[h!]
	\centering
	\includegraphics[width=3.in]{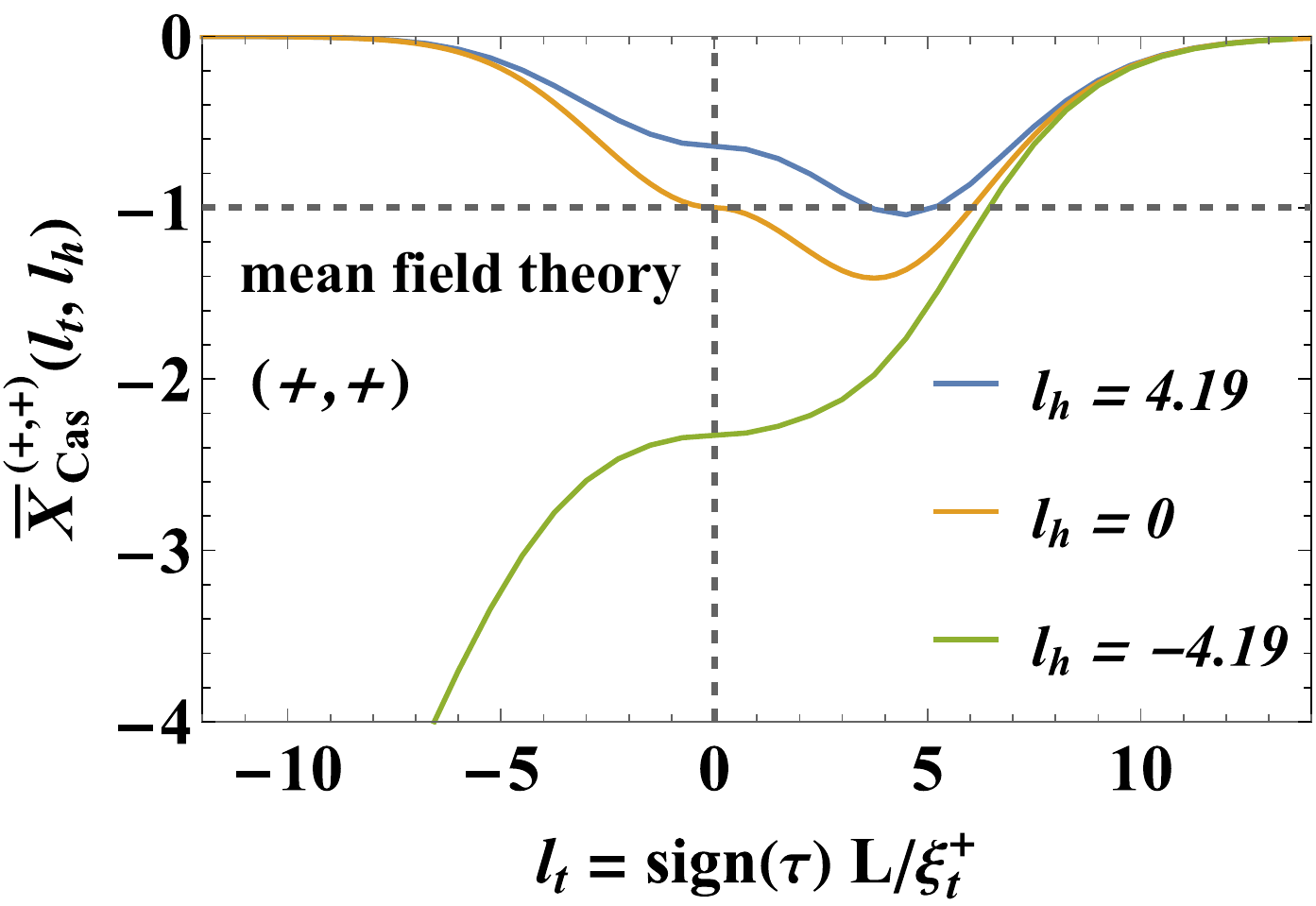} \quad
	\includegraphics[width=3.3in]{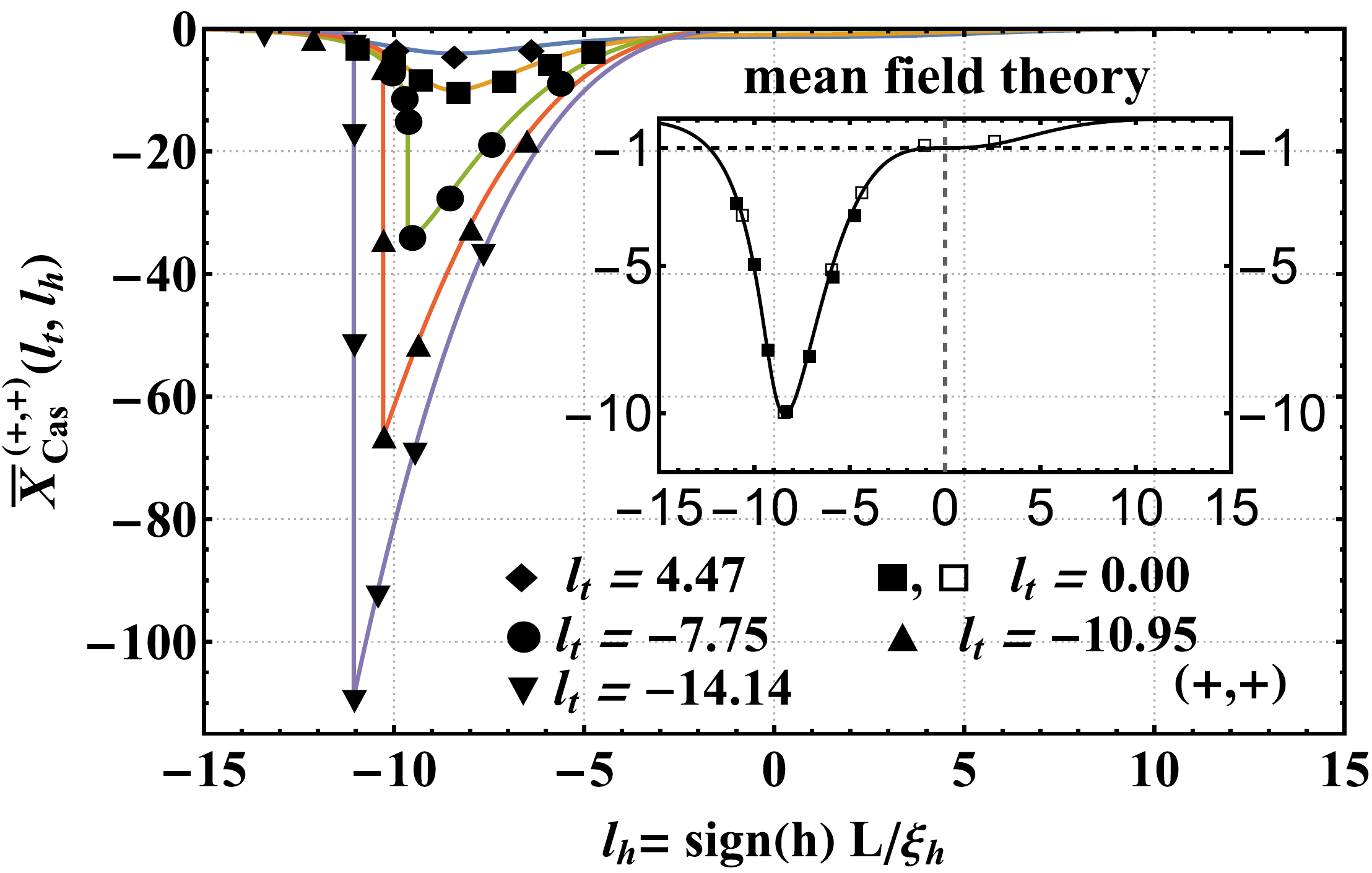}
	\caption{The left panel shows the dependence of the normalized finite-size scaling function $\overline{X}_{\rm Cas}^{(+,+)} (l_t,l_h) $ (such that $\overline{X}_{\rm Cas}^{(+,+)} (0,0) =-1$) on the temperature scaling variable $l_t$ for three values of the field scaling variable $l_h$: $l_h=0$, $l_h=\pm 4.19$.
		The right panel shows  the dependence of the normalized thermodynamic Casimir force $\overline{X}_{\rm Cas}^{(+,+)} (l_t,l_h)$ on the field scaling variable $l_h$ for five values of the thermal scaling variable $l_t$. The markers on the curves, including the inset curve
		representing the blow-up in the case $l_t = 0$, show an excellent agreement between the numerical
		results, obtained in Ref. \cite{SHD2003} (filled markers) and in Ref. \cite{VD2013} (empty squares), and the analytic
		results (solid lines) presented here.}
	\label{fig:CFPP}
\end{figure}

If there is more than one such solution $x_ m$, one has to take that one which leads to an order parameter profile which corresponds to the minimum of the free energy functional.
The precise mathematical procedure how this can be implemented, despite the divergence of the free energy, caused by the boundary conditions for the order parameter, which takes infinite values at the top and the bottom surfaces of the system, is explained in details in   Ref. \cite{DVD2015} (see Eq. (3.27) therein and the text around it). 

The behavior of the normalized finite-size scaling function $\overline{X}_{\rm Cas}^{(+,+)}(l_t,l_h) \equiv X_{\rm Cas}^{(+,+)}(l_t,l_h)/|X_{\rm Cas}^{(+,+)}(0, 0)|$  of the Casimir force is shown in Fig. \ref{fig:CFPP}. 

The relief map of the Casimir force, as function of both $l_t$ and $l_h$, is shown in Fig. \ref{fig:DP} where panel (a) presents the force on a larger scale, and panel (b) provides an enlarged view of the region close to the bulk critical point $(l_t=0,l_h=0)$.

\subsubsection{$(+,-)$ boundary conditions}

For such boundary conditions the two phases, i.e., the "+" one ("liquid") and the "-" one ("gas") occur near the corresponding boundary for any values of $T$ and $h$, i.e., they always coexist. Thus there is no phase transition from one phase to the other inside the film for any finite values of the temperature and the external field.

As before we start by discussing first the simpler case 

\hspace*{2cm} (I) {\textit { zero  external bulk field.}}

\begin{figure}[h!]
	\includegraphics[angle=0,width=\columnwidth]{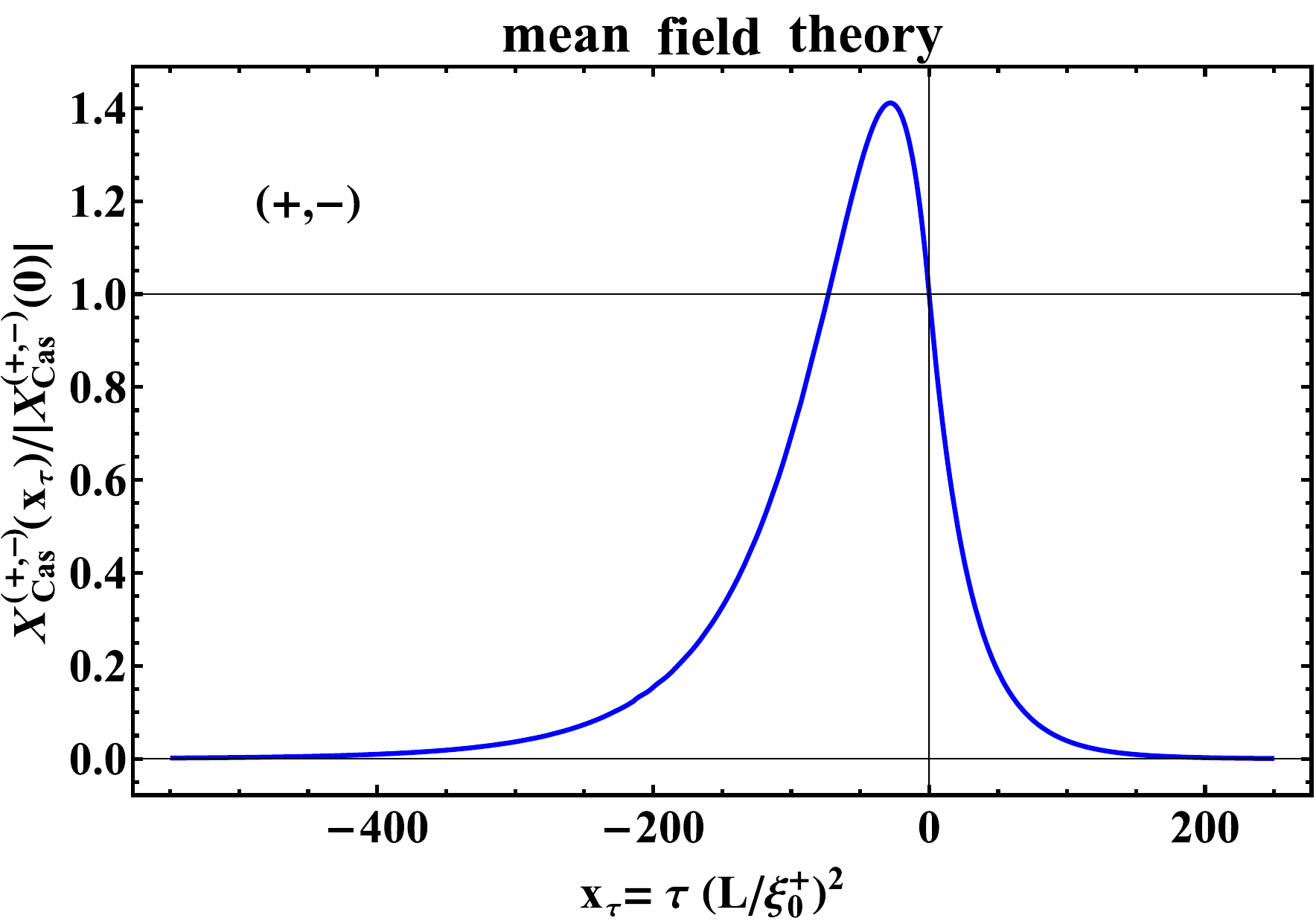}
	\caption{The normalized zero-field finite-size scaling function $X_{\rm Cas}^{(+,-)}$ of
		the Casimir force as a function of the scaling variable $x_\tau=\tau (L/\xi_0^{+})^2$, obtained within 
		mean field theory for the film geometry with  $(+,-)$ boundary
		conditions. The function is normalized by its value at the bulk critical point.
		It is positive and has a maximum {\it  below} $T_c$ (at $x_{\rm max}\simeq -28.110$, and  $X_{\rm Cas}^{(+,-)}(x_{\rm max})/|X_{\rm Cas}^{(+,-)}(0)|\simeq 1.411$), as in the case of the two-dimensional Ising model.}
	\label{CasPlusMinusMF}
\end{figure}

{\it (i)} If $x_\tau\geq 2\pi^2$ with $x_\tau=\sign(\tau)(L/\xi)^{1/\nu}$, one has
\begin{equation}
\label{eq:pm_MF_Cas}
X_{\rm Cas}^{(+,-)}(x_\tau)=\left[2K(k)\right]^4(1-k^2)^2,
\end{equation}
where $k=k(x_\tau)$, $0\leq k <1$, solves  the parametric equation
\begin{equation}\label{xdefpm}
x_\tau=2\left[2K(k)\right]^2(k^2+1).
\end{equation}

{\it (ii)} If $0\leq x_\tau\leq 2\pi^2$, one has 
\begin{equation}
X_{\rm Cas}^{(+,-)}(x_\tau)=\left[2K(k)\right]^4,
\end{equation}
where $k=k(x_\tau)$, $0\leq k \leq 1/\sqrt{2}$, is to be determined from the equation 
\begin{equation}\label{xdefpmneg}
x_\tau=-2\left[2K\right]^2(2k^2-1).
\end{equation}

{\it (iii)} If $x_\tau\leq 0$, one has 
\begin{equation}
\label{eq:pm_MF_Cas_m}
X_{\rm Cas}^{(+,-)}(x_\tau)=64\; k^2 (1-k^2)\left[K(k)\right]^4,
\end{equation}
where $k=k(x_\tau)$, $1/\sqrt{2}\leq k <1$,  solves the parametric equation (\ref{xdefpmneg}).

The behavior of
$X_{\rm Cas}^{(+,-)}(x_\tau)$ is shown in Fig. \ref{CasPlusMinusMF}. The scaling function is positive and has a maximum \textit{below} $T_c$. The maximum is attained for $k=k_{\rm max}$, which solves the equation \reff{eq:kmin-plus-plus}
	and renders $k_{\rm max}\simeq 0.909$. 	According to Eqs. \eqref{xdefpmneg} and \eqref{eq:pm_MF_Cas_m}  this leads to the maximum $X_{\rm Cas}^{(+,-)}(x_{\rm max})/|X_{\rm Cas}^{(+,-)}(0)|\simeq 1.411$ at $x_{\rm max}\simeq -28.110$.

One can check that the mean-field scaling functions $X_{\rm Cas}^{(+,+)}(x_\tau)$ and $X_{\rm Cas}^{(+,-)}(x_\tau)$ are related as \cite{VGMD2009}:
\begin{equation}\label{relscfunctionspppm}
X_{\rm Cas}^{(+,+)}(x_\tau)=-\frac{1}{4} X_{\rm Cas}^{(+,-)}(-x_\tau/2).
\end{equation}
Thus, for the corresponding mean-field Casimir amplitudes one has
\begin{equation}\label{deltarelpppm}
\frac{\Delta_{\rm Cas}^{(+,+)}}{ \Delta_{\rm Cas}^{(+,-)}}=-\frac{1}{4}.
\end{equation}

\hspace*{2cm} (II) {\textit { {non-zero external bulk field}}}

\begin{figure}[h!]
	\includegraphics[width=8cm]{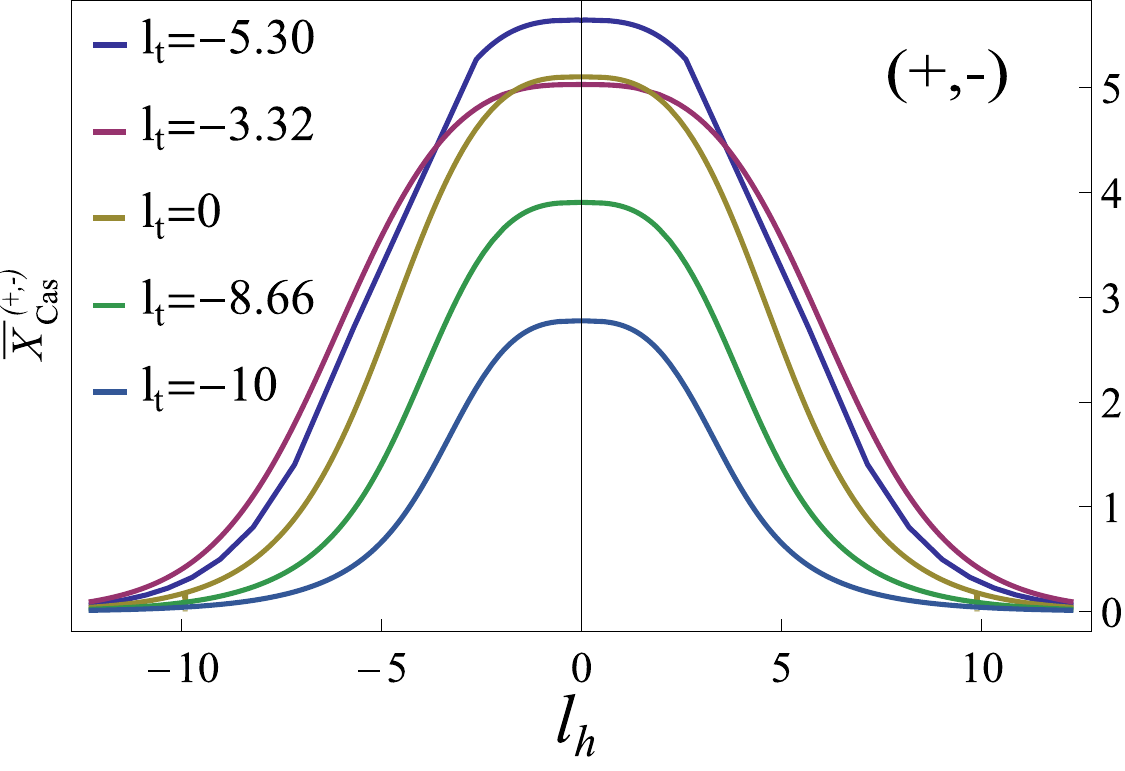}\quad
	\includegraphics[width=8cm]{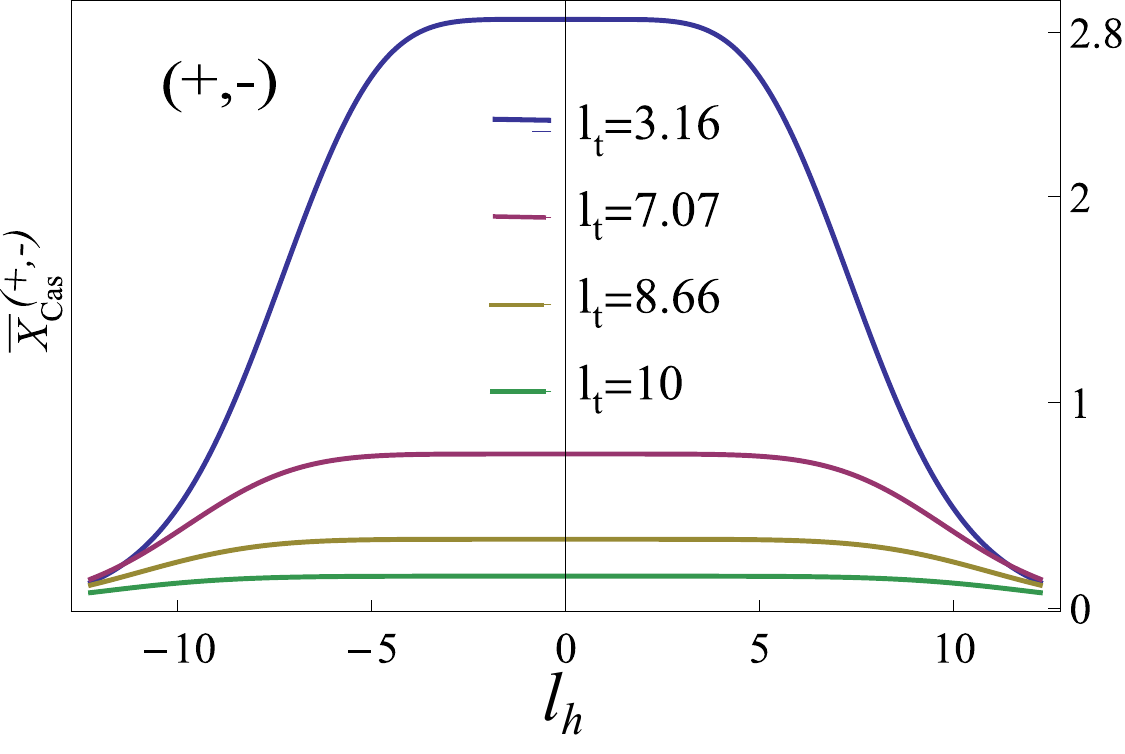}
	\caption{
		Normalized scaling function of the critical Casimir force for $(+,-)$ boundary conditions as function of $l_h$ for nine values of $l_t$.  } 
	\label{fig:dif_lt_as_lh1}
\end{figure}

\begin{figure}[h!]
	\includegraphics[width=8cm]{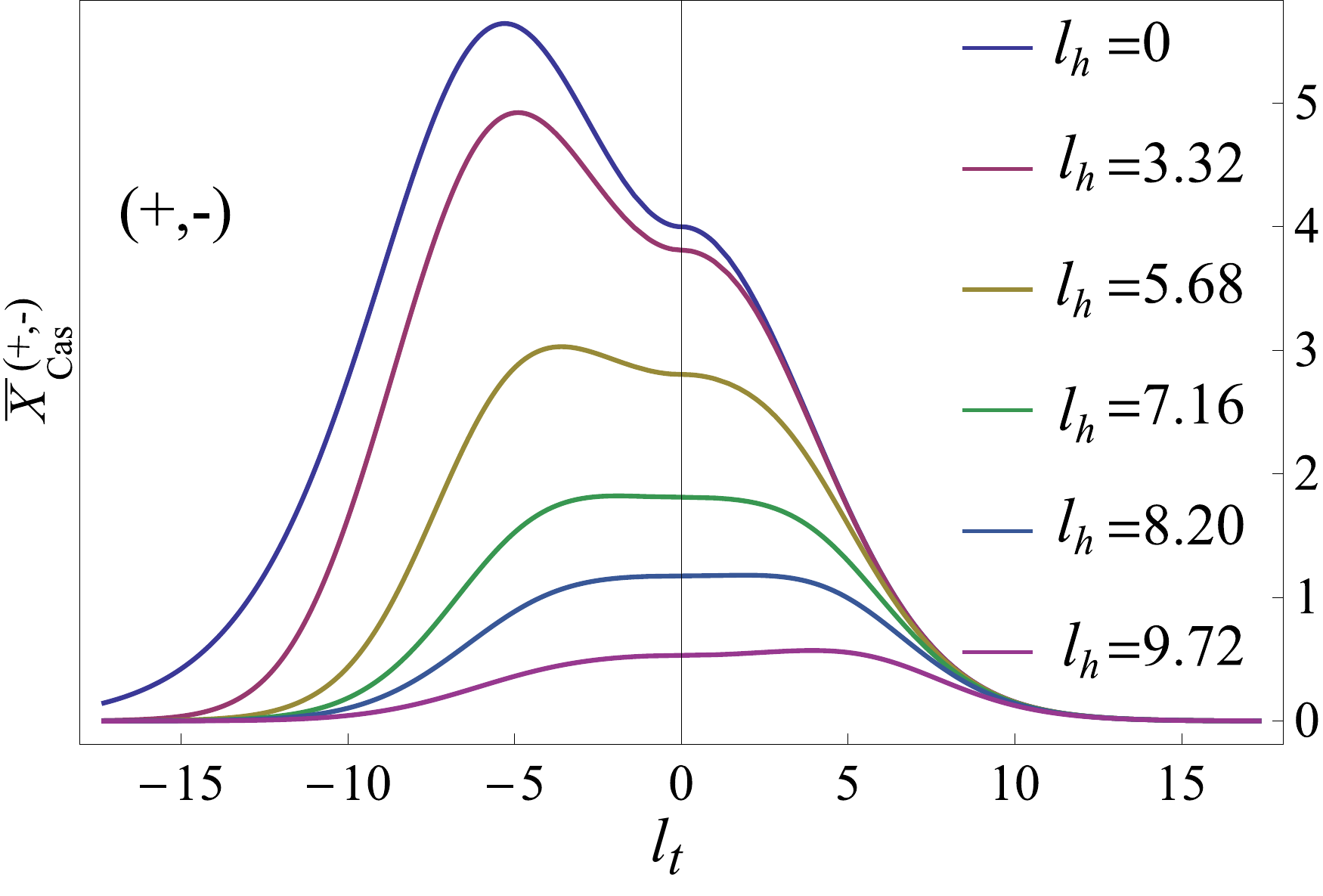}
	\includegraphics[width=8.35cm]{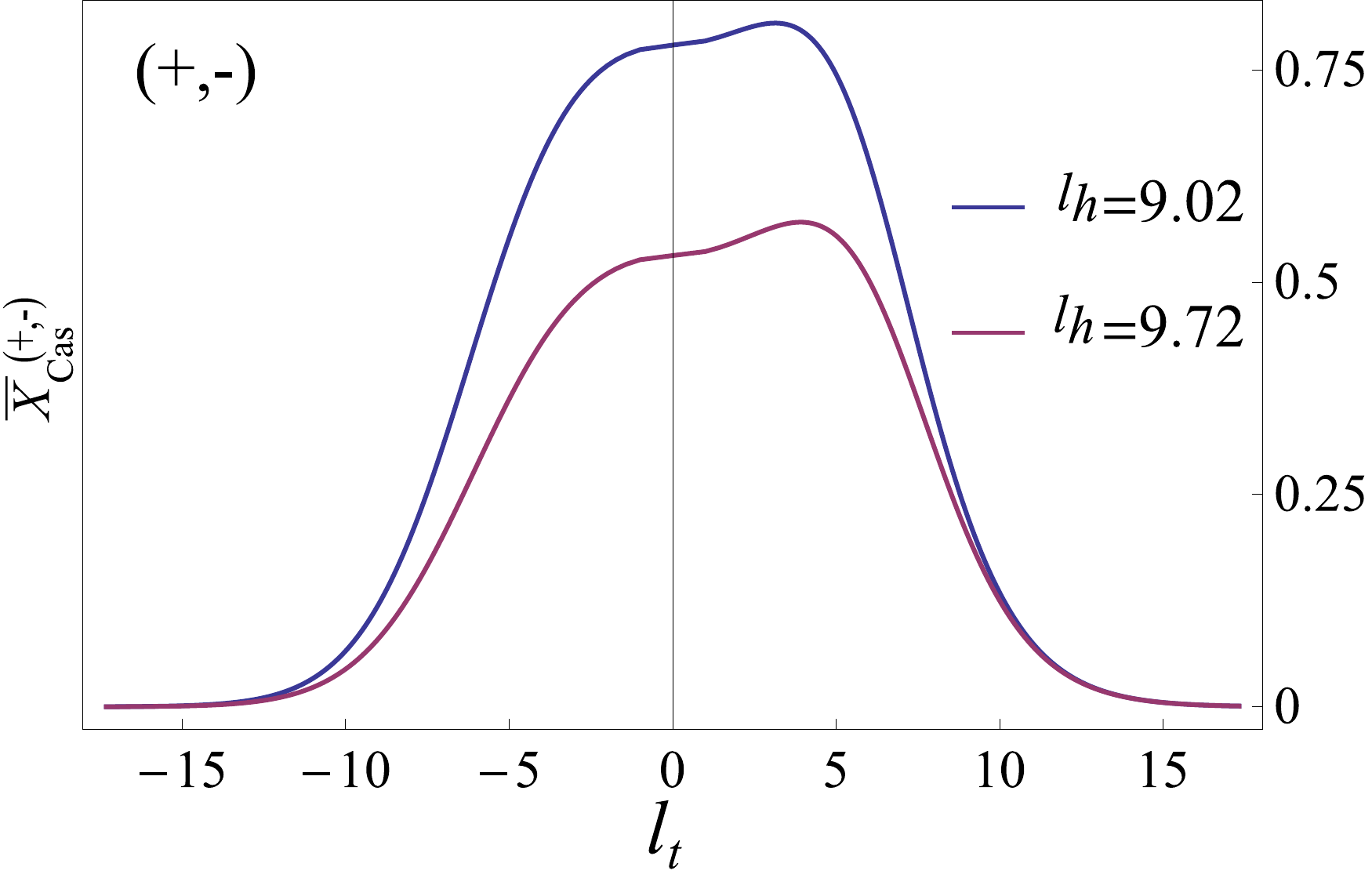}
	\caption{
		Normalized scaling function of the critical Casimir force for $(+,-)$ boundary conditions as function of $l_t$ for seven values of $l_h$. The comparison between the panels demonstrates that the maximum of the force moves from below $T_c$ (i.e., for $l_t<0$) for relatively small values of $l_h$ (see the left panel) to above $T_c$ (i.e., for $l_t>0$) for larger values of $l_h>0$ (see the right panel).} 
	\label{fig:dif_lh_as_lt}
\end{figure}

The results for the behavior of the Casimir force in the case $h\ne 0$ have been derived in Ref. \cite{DVD2018}. In this case Eqs. \reff{FIX} - \reff{CasimirX1} are still valid, but the Casimir force is not given by \eq{Casimir_final} because, for $(+,-)$ boundary conditions,  $X'\ne 0$ in the midplane of the system. Moreover, there is no point inside the system where $X'=0$. In this case, the scaling function has to be inferred from \eq{CasimirX1} but with the pressure $p$ in the film to be determined implicitly from the equation 
\begin{equation}
\label{eq:main_eq}
1=f(l_t,l_h,p)
\end{equation}
where
\begin{equation}
\label{eq:main_eq_2}
f(l_t,l_h,p)= \int_{-\infty}^{\infty} \frac{dX}{\sqrt{P(X)}}, \qquad \mbox{with} \quad P(X)= X^4+{\rm sign}(l_t) \, l_t^2 X^2 -\frac{2}{3 \sqrt{6}}l_h^{3} X+p > 0.
\end{equation} 
Figure \ref{fig:dif_lt_as_lh1} shows the scaling function of the Casimir force as a function of $l_h$ for nine values of $l_t$. These data are normalized by the critical  Casimir amplitude $\Delta_{\rm Cas}^{(+,+)}$. The normalized  scaling function of the force is denoted by $\bar{X}_{\rm Cas}^{(+,-)}(l_t,l_h)$.
Figure \ref{fig:dif_lh_as_lt} shows the scaling function of the critical Casimir force as function of $l_t$ for seven values of $l_h$.

Using the exact expressions for the scaling function of the Casimir force one can derive the corresponding asymptotes for large values of $l_t$ or $l_h$: 
%
%
\begin{equation}
\label{eq:as_h0}
X_{\rm Cas}^{(+,-)}(l_t,l_h=0)\simeq \left\{ \begin{array}{cc} 
16\; l_t^{\,4} \; \exp(-l_t), & l_t \gg 1 \\
16 \; l_t^{\,4} \; \exp(l_t/\sqrt{2}), & l_t \ll -1.
\end{array}\right .
\end{equation}
The above expressions are equivalent to the corresponding results derived in Ref. \cite{K97} in terms of $x_t$. The result for $T<T_c$ contains the same exponential decay  as the one predicted in Ref.  \cite{PE92}. 
The asymptotes $X_{\rm Cas}^{(+,-)}(l_t=0,l_h)$ for $|l_h|\gg 1$ are 
\begin{equation}
\label{eq:as_XCas_h}
p-p_b\equiv X_{\rm Cas}^{(+,-)}\simeq 48\; l_h^4\;\exp(- |l_h|),  \qquad l_t=0, |l_h|\gg 1.
\end{equation}
Thus, the Casimir force decays exponentially not only for large values of $|l_t|$, but also for large values of $|l_h|$.
%
%

\begin{figure}[h!]
	\includegraphics[width=\columnwidth]{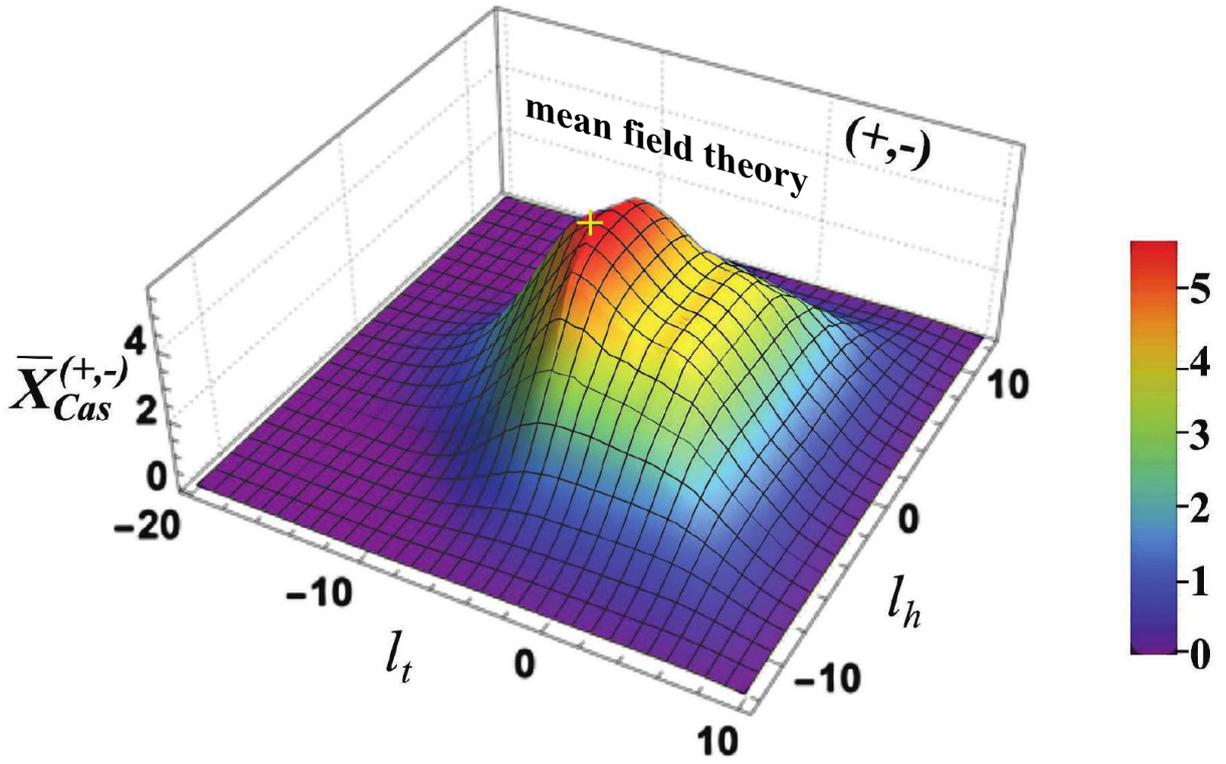}
	\caption{
		Normalized scaling function of the critical Casimir force for $(+,-)$ boundary conditions as function of temperature  and field scaling variables $l_t$ and $l_h$, respectively. The relief map is symmetric about the plane $l_h=0$. The maximum occurs at $(l_t=-5.30, l_h=0)$ and is marked by a cross.}
	\label{fig:CF_all}
\end{figure}

Evaluating $P(X)$ at the bulk value of the order parameter, i.e., $X=X_b$, and by taking into account \eq{pbX1} one obtains the equation $P(X_b)=p(l_t, l_h)-p_b(l_t, l_h)>0$. Taking into account \eq{CasimirX1}, one concludes that $X_{\rm Cas}^{(+,-)}(l_t,l_h)>0$, i.e., the Casimir force is {\it repulsive} for $(+,-)$ boundary conditions for any values of $l_t$ and $l_h$. Using the symmetry of the above equation under the change of  variables $(X\to -X, l_h\to -l_h)$, one  obtains the relation $p(l_t,l_h)=p(l_t,-l_h)$. 
One can check that this is valid also for $p_b$. Thus, one concludes that 
\begin{equation}
\label{eq:symmetry}
X_{\rm Cas}^{(+,-)}(l_t,l_h)=X_{\rm Cas}^{(+,-)}(l_t,-l_h).
\end{equation}
This, and the fact that the derivative of $p$ with respect to $l_h$ can be zero only for $l_h=0$, lead to the conclusion that $X_{\rm Cas}^{(+,-)}(l_t,l_h)$ exhibits an  extremum for $l_h=0$. As explained above, it is attained for $k_{\rm max}^2=0.826115$ determined as the root of the equation  $2=K(k)/E(k)$ (see \eq{eq:kmin-plus-plus}). In fact the extremum is  a maximum. For $k=k_{\rm max}$ the force scaling function normalized with $\Delta_{\rm Cas}^{(+,+)}$ reaches the value $5.64$, which is attained at $l_t=-5.30$. Thus, in the $(l_t,l_h)$ plane the force has a {\it single global maximum}, which upon the normalization by $\Delta_{\rm Cas}^{(+,+)}$ is
\begin{equation}
\label{eq:max_CF}
\max_{\left(l_t, l_h\right)}\bar{X}_{\rm Cas}^{(+,-)}(l_t,l_h)=\frac{X_{\rm Cas}^{(+,-)}(l_t=-5.30,l_h=0)}{\Delta_{\rm Cas}^{(+,+)}}=5.64. 
\end{equation} 
The overall (temperature-field) relief map of the force is shown in Fig. \ref{fig:CF_all}. The global maximum of the scaling function of the force, which is attained at $(l_t=-5.30, l_h=0)$, is clearly visible and is marked by a cross. (In order to keep the number of normalization constants at a minimum we are using $\Delta_{\rm Cas}^{(+,+)}$ both for $X_{\rm Cas}^{(+,-)}$ and $X_{\rm Cas}^{(+,+)}$.)

As expected, the external field strongly influences the behavior of the force. The strongest force occurs at $l_h=0$ (see Fig. 	\ref{fig:dif_lt_as_lh1}). For moderate strengths of the field the maximal value of the force is encountered at {\it negative} values of $l_t$ (see the left panel of Fig. \ref{fig:dif_lh_as_lt}), while for stronger fields the maximum value is encountered at {\it positive} values of $l_t$ (see the right panel of Fig. \ref{fig:dif_lh_as_lt}). This occurs due to the competing effects of the temperature $T$ and the field $h$ on the fluctuations of the system in distinct regions of the $(T,h)$ plane.

\begin{figure}[htb!]
	\centering
	\includegraphics[width=\columnwidth]{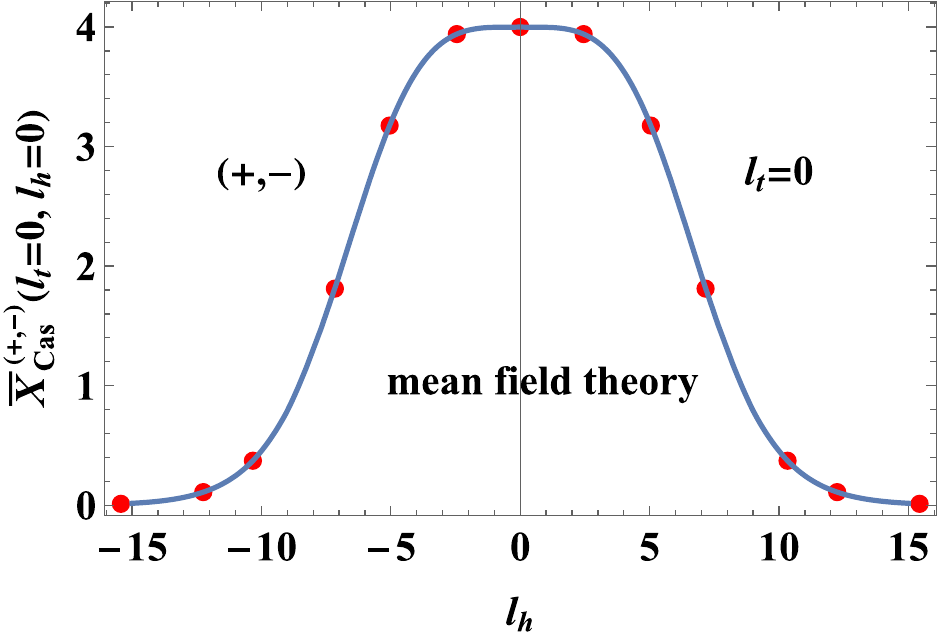}
	\caption{
		Comparison of the behavior of the scaling function of the Casimir force obtained analytically (solid line) with that one obtained numerically in Ref. \cite{VD2013} (dots). Following Ref. \cite{VD2013}, the scaling function $X_{\rm Cas}^{(+,-)}(l_t,l_h)$ is normalized by $\Delta_{\rm Cas}^{(+,+)}$ , i.e., the figure shows the behavior of $\bar{X}_{\rm Cas}^{(+,-)}(l_t,l_h)=X_{\rm Cas}^{(+,-)}(l_t,l_h)/\Delta_{\rm Cas}^{(+,+)}$ as function of $l_h$ for $l_t=0$. } 
	\label{fig:comparison}
\end{figure}

The comparison of the analytical results concerning the field dependence of the force with those  obtained numerically in Ref. \cite{VD2013} for the case $T=T_c$ is presented in Fig. \ref{fig:comparison}. There is excellent agreement. 

\subsubsection{$(O,O)$ boundary conditions}

\begin{figure}[htb]
	\centering
	\includegraphics[width=0.45\textwidth]{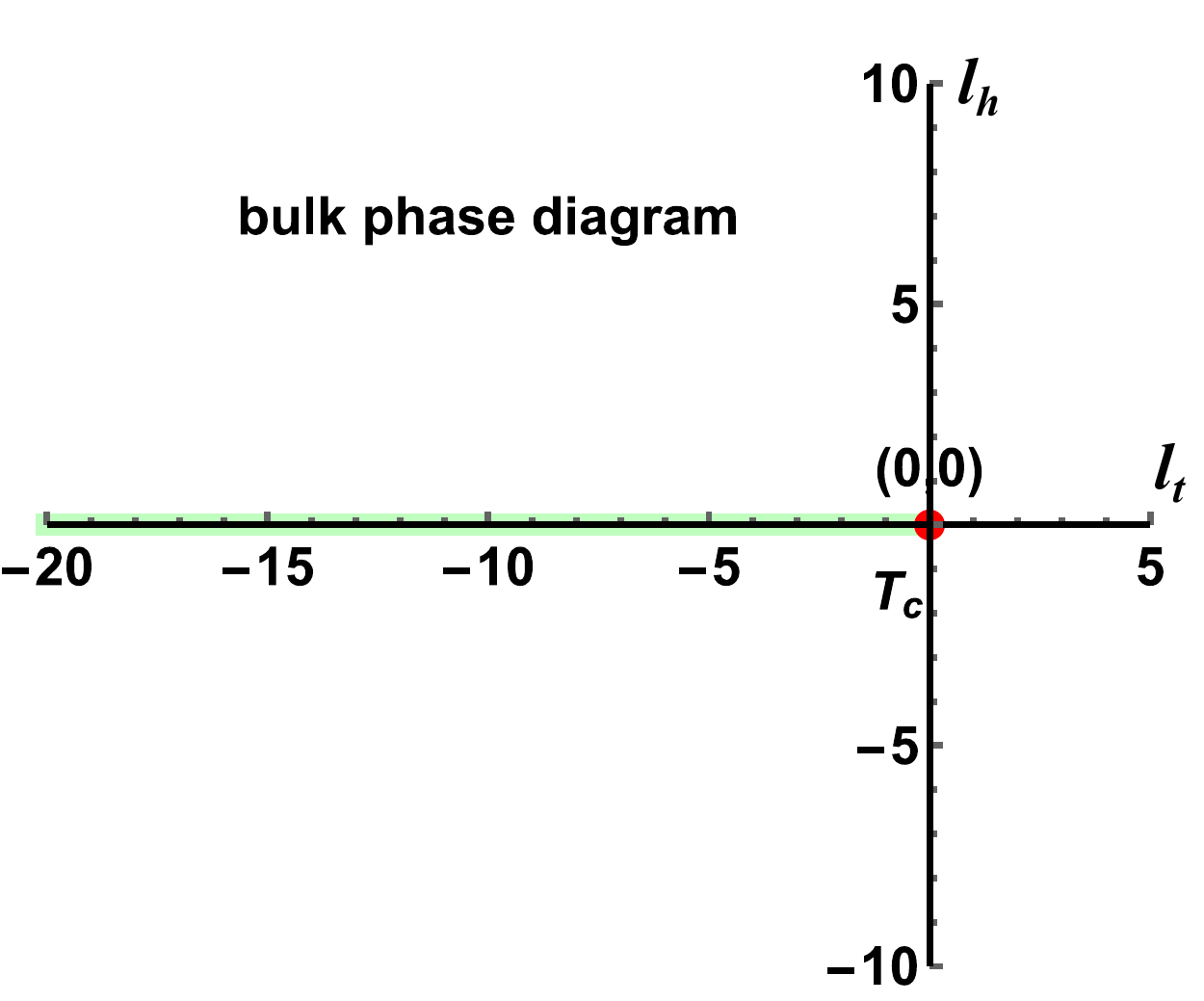}
	\hspace{10mm}
	\hfill
	\includegraphics[width=0.45\textwidth]{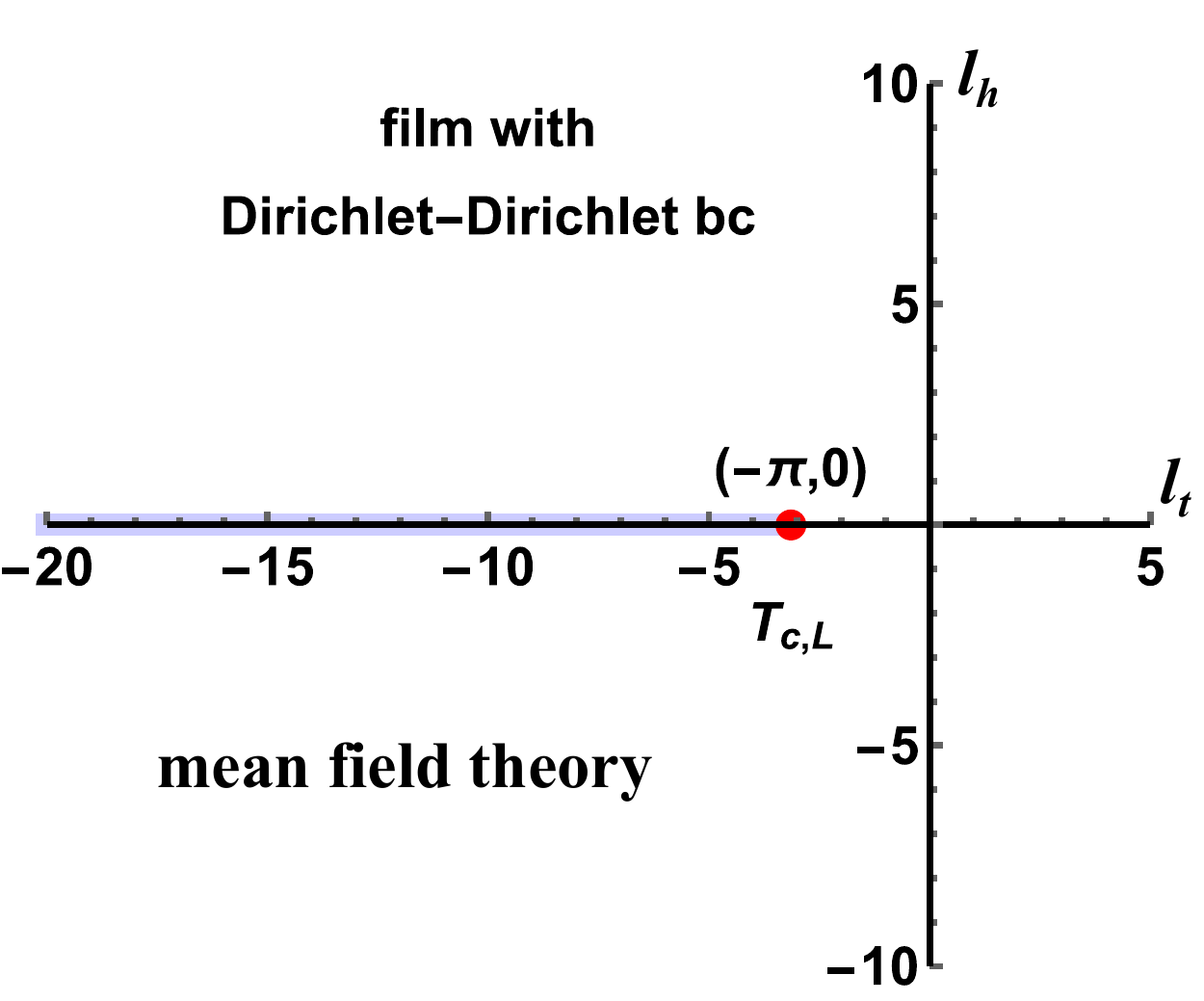}
	\caption{Phase diagram in the $(l_t, l_h)$ plane of the bulk (left panel) and of films with  Dirichlet-Dirichlet boundary conditions (right panel, mean field theory). In the bulk system first-order phase transitions occur upon crossing the phase coexistence line at $l_h=0$  for temperatures $T\in(0,T=T_c)$. At $T=T_c$ the bulk system exhibits a second-order phase transition. For films the coexistence line is at $l_h=0$ and at temperatures  $T\in(0,T=T_{c,L})$. The second-order phase transition occurs at $T=T_{c,L}$, i.e., at $(l_t,l_h) = (-\pi,0)$, or, due to $x_\tau=l_t^{1/\nu}$, equivalently at $x_\tau=-\pi^2$.}
	\label{fig:pd-bulk}
\end{figure}

The phase diagram, in terms of $l_t$ and $l_h$, of a \textit{film} under Dirichlet-Dirichlet boundary conditions  is shown on the right panel of Fig. \ref{fig:pd-bulk}; the left panel shows the \textit{bulk} phase diagram in terms of these variables. The film system exhibits a line of first-order phase transitions $\left\{T<T_{c,L},l_h=0\right\}$ which occur upon crossing it by varying the field variable $l_h$. In terms of $l_t$ and $l_h$, the coordinates of the critical point $T_{c,L}$ are $\left(l_t=-\pi,l_h=0\right)$. If $T<T_{c,L}$, the equilibrium order parameter profile with minimal free energy is positive for $h>0$  and negative  for $h<0$. More details can be found in Refs.  \cite{NF82,Bb83,VDD2019b,DVD2019c,Dan2021}.

In the case of Dirichlet-Dirichlet boundary conditions the  results for the corresponding scaling function of the Casimir force have been derived in Refs. \cite{GaD2006,ZSRKC2007,VDD2019b,DVD2020b,DVD2020,Dan2021} for 

\hspace*{2cm} (I) {\textit {zero external bulk  field.}} 

In the  case of zero bulk field the corresponding results for the scaling function of the Casimir force have been obtained\footnote{The functional dependences reported below seems to differ in form from those given in Ref. \cite{ZSRKC2007}. However, they become equivalent, if one takes into account the different definitions of the argument of the  function $K$ tacitly used in Ref.  \cite{ZSRKC2007}. In terms of the conventional notation the argument in their Eq. (5) should be the square root of it.} in Refs. \cite{GaD2006,ZSRKC2007}. They read as follows: 

{\it (i)} if $x_\tau\equiv \tau (L/\xi_0^+)^{1/\nu}\leq-\pi^2$ one has 
\begin{eqnarray}
\label{eq:MF_DD_Cas}
X_{\rm Cas}^{(O,O)}(x_\tau)&=&-\frac{1}{4}x_\tau^2 \left(\frac{1-k^2}{1+k^2}\right)^2= -4 (1-k^2)^2 \left[K(k)\right]^4, 
\end{eqnarray}
where $k(x_\tau)$, $0\leq k <1$, is given implicitly by
\begin{equation}\label{xdefDD}
x_\tau=-4 (1+k^2)\left[K(k)\right]^2.
\end{equation}
(The two expressions for $X_{\rm Cas}^{(O,O)}(x_\tau)$ reported in \eq{eq:MF_DD_Cas} are equivalent due to the relation given in \eq{xdefDD});

{\it (ii)} if $-\pi^2\leq x_\tau\leq 0$, one has
\begin{equation}
	\label{eq:DD-Cas-x-negative}
X_{\rm Cas}^{(O,O)}(x_\tau)= -\frac{1}{4}x_\tau^2;
\end{equation}

{\it (iii)} if $x_\tau> 0$, one has
\begin{equation}
	\label{eq:DD-Cas-x-positive}
	X_{\rm Cas}^{(O,O)}(x_\tau)=0.
\end{equation}

\begin{figure}[tbp]
	\includegraphics[angle=0,width=\columnwidth]{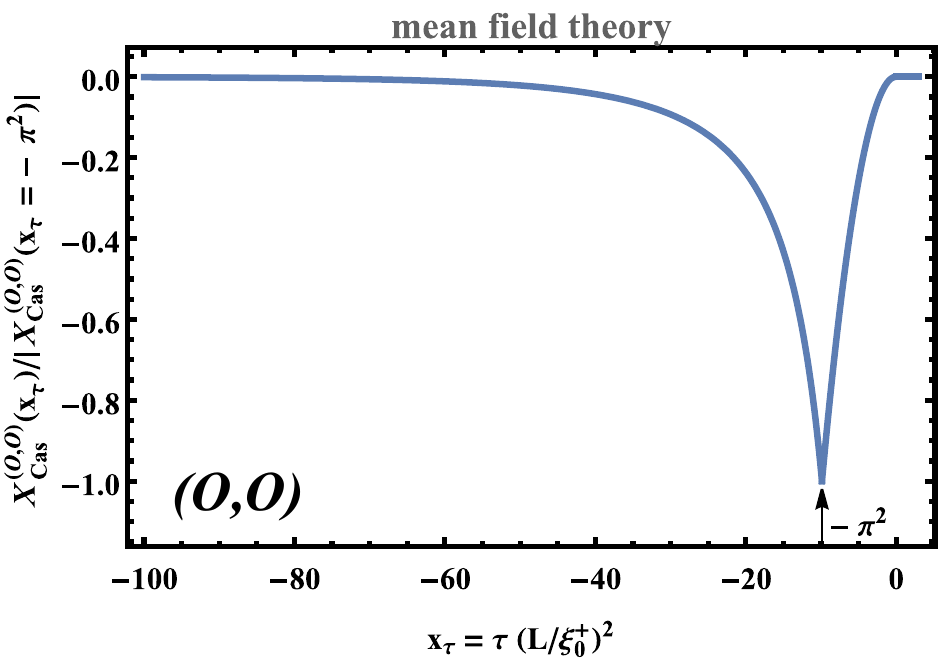}
	\caption{The zero-field finite-size scaling function $X_{\rm Cas}^{(O,O)}$ of
		the Casimir force as a function of the scaling variable $x_\tau=\tau (L/\xi_0^+)^2$, obtained by
		mean field theory  for films with Dirichlet-Dirichlet boundary
		conditions. The function is normalized by its value at the film critical point $x_\tau=-\pi^2$.
		It is negative and has a minimum {\it  below} $T_c$ at $x_\tau=-\pi^2$.}
	\label{CasPlusDirDirMF}
\end{figure}

The behavior of
$X_{\rm Cas}^{(O,O)}(x_\tau)$ is shown in Fig. \ref{CasPlusDirDirMF}.
The cusp-like singularity of the force stems from the fact that for films a nonzero solution for the order parameter profile appears only for $\tau < -\lambda_0$, where $\lambda_0=\pi^2/L^2$ is the smallest eigenvalue of the Laplacian with Dirichlet boundary conditions, while for the bulk system a nonzero solution exists already for $\tau<0$. 

\hspace*{2cm} (II) {\textit  { non-zero external bulk field}}

\begin{figure}
\centerline{\includegraphics[width=0.475\textwidth]{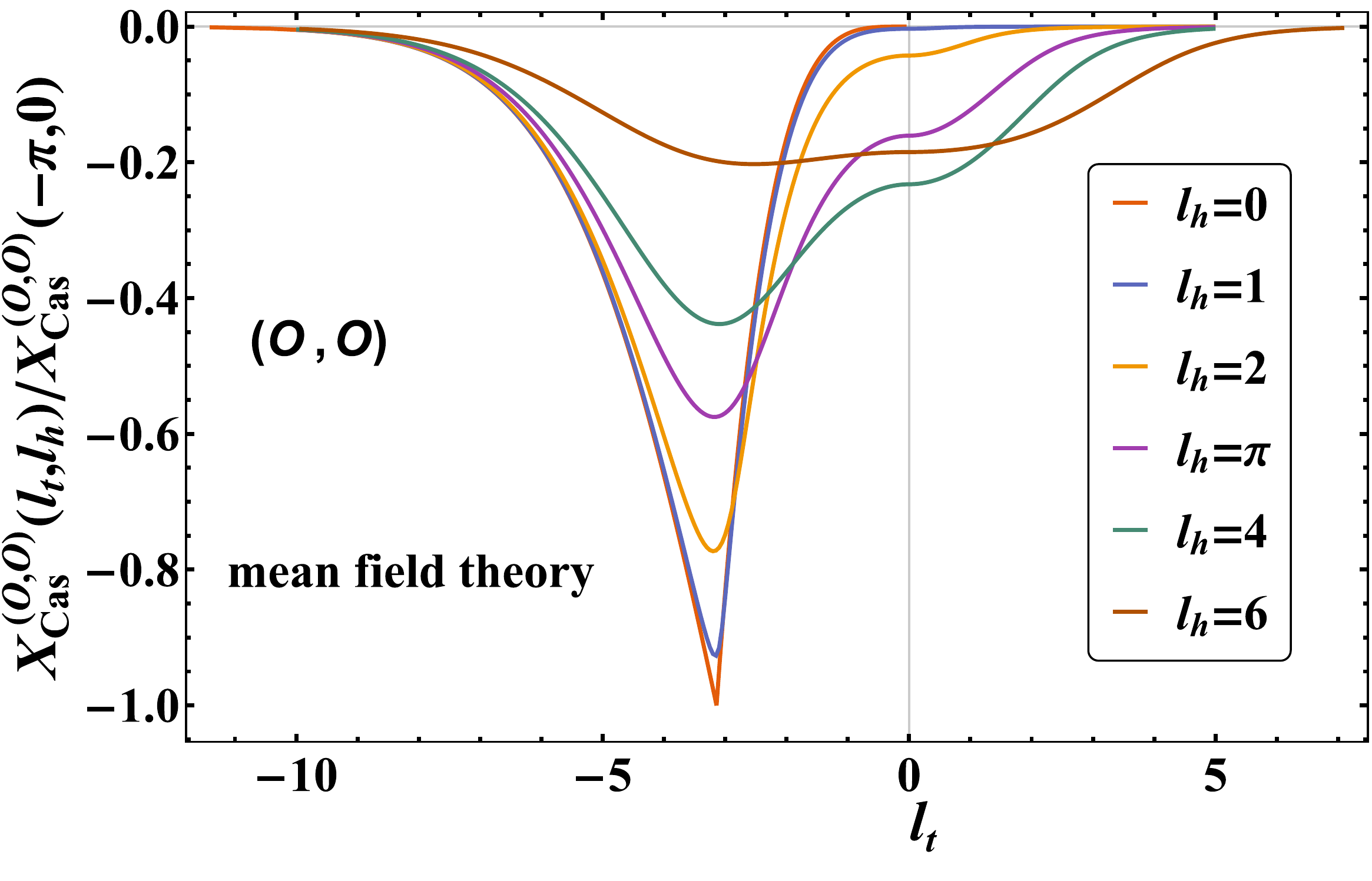}\qquad \includegraphics[width=0.475\textwidth]{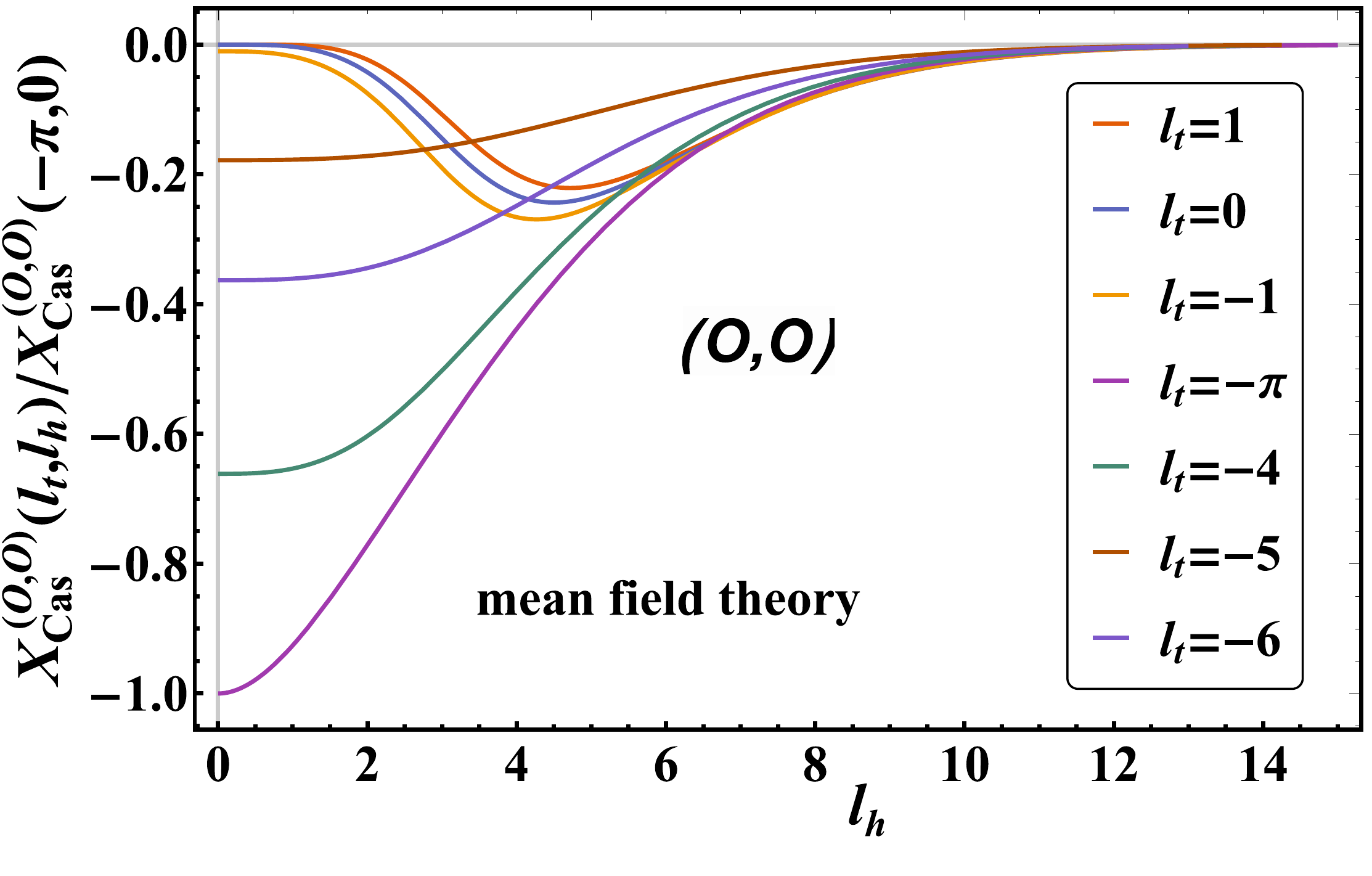}}
\caption{Scaling function of the Casimir force as a function of $l_t$ for six fixed values of $l_h$ (left panel) and as a function of $l_h$ for seven fixed values of $l_t$ (right panel). The function is normalized by its value at the critical point of the film. The force is attractive and non-monotonic.} 
\label{figCasimirltlh}
\end{figure}

\begin{figure}[tbh!]
	\centerline{ \includegraphics[width=\columnwidth]{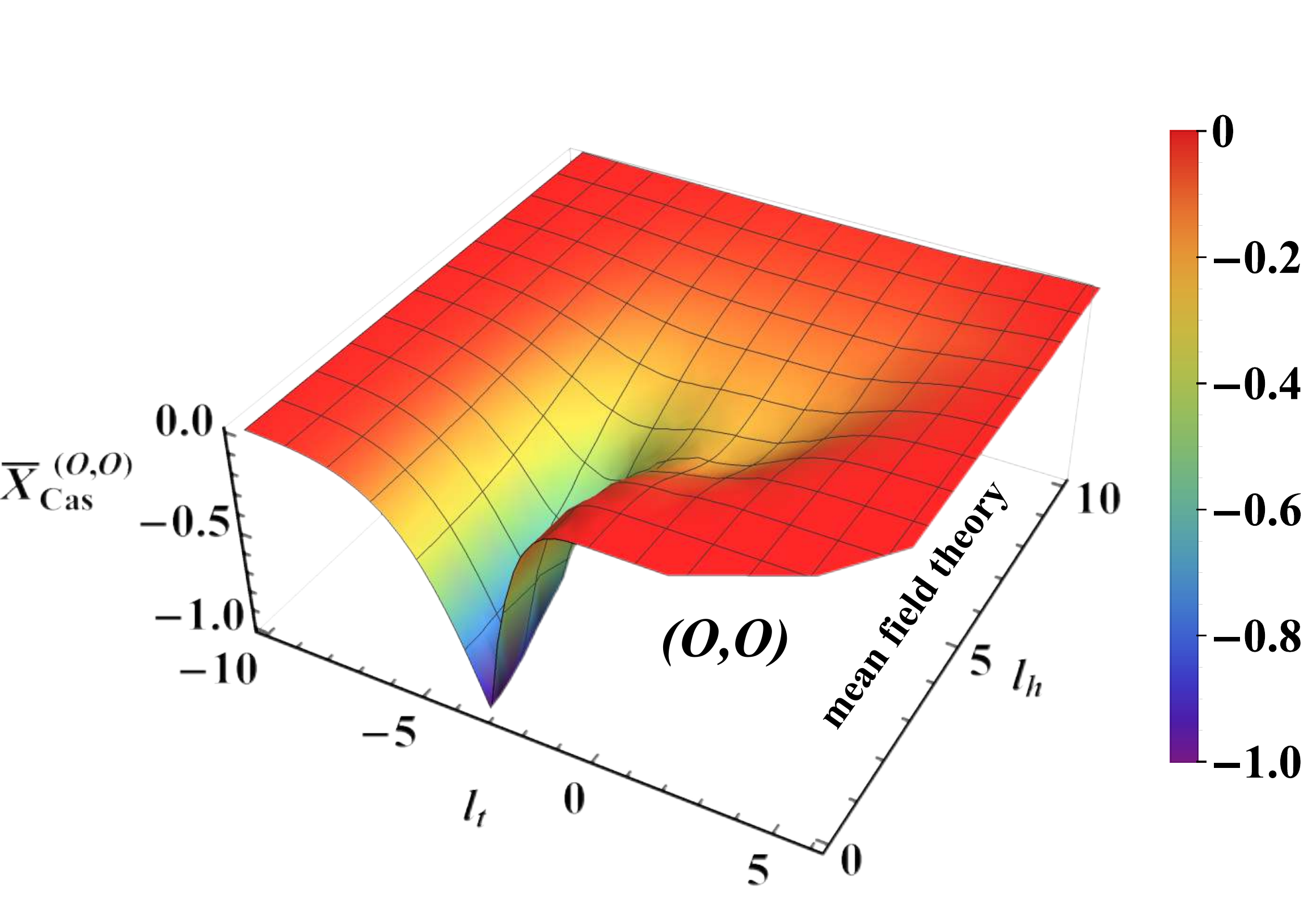}}
	\caption{Scaling function of the Casimir force as a function of both $l_t$ and $l_h$. The function is normalized by its value at the critical point of the film, i.e., the figure shows the behavior of $\bar{X}_{\rm Cas}^{(O,O)}=X_{\rm Cas}^{(O,O)}(l_t,l_h)/X_{\rm Cas}^{(O,O)}(-\pi,0)$. The force is attractive and non-monotonic.} 
	\label{figCasimirltlh3D}
\end{figure}

Results for the critical Casimir force in the presence of an external ordering field have been obtained in Refs. \cite{VDD2019b,DVD2020b,DVD2020,Dan2021}. Therein the results are derived in terms of the scaling variables $l_t$ and $l_h$ (see \eq{isvarxt} for $l_t$ and \eq{isvarxh} for $l_h$). This case is somewhat similar to the one of $(+,+)$ boundary conditions in that the order parameter has a zero derivative in the middle of the film and varies monotonically from the boundary up to this point.  The Eqs.  \eqref{Casimir_final} - \eqref{TrEq}, as well as \eqref{g1}, are valid  also for $(O,O)$ boundary conditions. The "only" difference concerns \eq{eq:condition}, which turns into 
\begin{equation}  \label{eq:condition_ordinary}
	6 \sqrt{3}\, x_m \left({\rm sign}(l_t)l_t^2 +2 x_m^{2}\right)- \sqrt{2} \, l_{h}^{3}<0,
\end{equation}
supposing that the scaling function $X^{(O,O)}$ of the order parameter profile  increases in the interval $\zeta\in [0,1/2], \zeta=z/L$, starting from $X^{(O,O)}(\zeta=0|l_t,l_h)=0$. Performing the corresponding numerical evaluations one obtains the results displayed in Fig.~\ref{figCasimirltlh} (left panel),  showing the behavior of the scaling function of the Casimir force as function of the temperature scaling variable $l_t$ for six fixed values of the field scaling variable $l_h$, while the right panel in Fig.~\ref{figCasimirltlh} demonstrates this behavior as a function of $l_h$ for seven fixed  values of $l_t$.
The overall dependence of the Casimir force on {\it both} the temperature, i.e., $l_t$ and the field, i.e., $l_h$ scaling variables, which provide the temperature-field relief map of the force, is shown in Fig. \ref{figCasimirltlh3D}. 

\subsubsection{$(O,+)$ boundary conditions}
From the symmetry of the order parameter profile $\phi^{(+,-)}(z)$ for
$(+,-)$ boundary conditions it follows that, within
mean field theory, the case of $(O,+)$ boundary conditions
can be obtained from the corresponding results for the order parameter profile $\phi^{+,-}(z)$ by the simple transformation
$L \to 2L$: it is
then given by $\phi^{(O,+)}(z)=\phi^{(+,-)}(z+L)$ for
$0\le z \le L$. Indeed, then one would have $\phi^{(O,+)}(z=0)=\phi^{(+,-)}(L)=0$. Thus, for the Casimir force it follows that, with

\hspace*{2cm} (I) {\textit { {zero external bulk field,}}}

\begin{equation}
X_{\rm Cas}^{(O,+)}(x_\tau)=\frac{1}{2^4} X_{\rm Cas}^{(+,-)}(2^2 x_\tau)=\frac{1}{16} X_{\rm Cas}^{(+,-)}(4x_\tau).
\end{equation}
This property of $X_{\rm Cas}^{(O,+)}(x)$ has been stated in Ref. \cite{K97}.
Accordingly, $X_{\rm Cas}^{(O,+)}(x_\tau)>0$ corresponds to a repulsive Casimir force. 
Here both $X_{\rm Cas}^{(+,-)}$ and
$X_{\rm Cas}^{(O,+)}$ exhibit a maximum {\it below} $T_c$.

\hspace*{2cm} (II) {\textit {{non-zero external bulk field}}}

In this case one has \cite{rem1}
	\begin{equation}
	X_{\rm Cas}^{(O,+)}(x_\tau, x_\mu)=\frac{1}{2^4} X_{\rm Cas}^{(+,-)}(2^2x_\tau,2^3x_\mu)=\frac{1}{16} X_{\rm Cas}^{(+,-)}(4x_\tau,8x_\mu),
\end{equation}
where we have used that within mean field theory $d=4$, $1/\nu=2$, and $\Delta/\nu=3$. 

\subsubsection{$(S\!B,+)$ boundary conditions}

Due to the symmetry of the order parameter profile $\phi^{(+,+)}(z)$ for
$(+,+)$ boundary conditions, one finds that, within
mean field theory the case of $(S\!B,+)$ boundary conditions
follows from the corresponding results for the order parameter profile upon the simple transformation
$L \to 2L$: it is
given by $\phi^{(S\!B,+)}(z)=\phi^{(+,+)}(z+L)$ for
$0\le z \le L$. Indeed, this leads to  $\dot{\phi}^{(S\!B,+)}(z=0)=\dot{\phi}^{(+,+)}(z=L)=0$, where the dot denotes differentiation with respect to $z$. Thus, for the Casimir force it follows that, with 

\hspace*{2cm} (I) {\textit { {zero external bulk field,}}}

\begin{equation}
X_{\rm Cas}^{(S\!B,+)}(x_\tau)=\frac{1}{16} X_{\rm Cas}^{(+,+)}(4x_\tau).
\end{equation}
This result has been established in Ref. \cite{K97}.

Both $X_{\rm Cas}^{(+,+)}$ and $X_{\rm Cas}^{(S\!B,+)}<0$ are negative
and have a minimum {\it above} $T_c$. In the case of

\hspace*{2cm} (II) {\textit {{non-zero  external bulk field}}}

one has \cite{rem1}
\begin{equation}
	X_{\rm Cas}^{(S\!B,+)}(x_\tau,x_\mu)=\frac{1}{16} X_{\rm Cas}^{(+,+)}(4x_\tau,8x_\mu).
\end{equation}

\subsubsection{$(S\!B,O)$ boundary conditions}
\label{sec:SB_O_bc}

Due to the symmetry of the order parameter profile $\phi^{(O,O)}(z)$ for
	$(O,O)$ boundary conditions one finds that within
	mean field theory $\phi^{(S\!B,O)}$ 
	can be obtained from the corresponding results for the order parameter profile $\phi^{(O,O)}(z)$ by the simple transformation
	$L \to 2L$: it is
	given by $\phi^{(S\!B,O)}(z)=\phi^{(O,O)}(z+L)$ for
	$0\le z \le L$. Indeed, this leads to  $\dot{\phi}^{(S\!B,O)}(z=0)=\dot{\phi}^{(O,O)}(z=L)=0$, where the dot denotes differentiation with respect to $z$. Thus, for the Casimir force, with 

\hspace*{2cm} (I) {\textit { {zero external bulk field,}}}

one obtains \cite{rem1}

\begin{equation}
\label{eq:MF_O_SB}
	X_{\rm Cas}^{(S\!B,O)}(x_\tau)=\frac{1}{16} X_{\rm Cas}^{(O,O)}(4x_\tau).
\end{equation}

Since, within mean field theory,  $X_{\rm Cas}^{(O,O)}$ is negative, this holds for $X_{\rm Cas}^{(S\!B,O)}<0$, too, i.e., the Casimir force for $(S\!B,O)$ boundary conditions is \textit{attractive} within mean field theory. This contradicts the expectation that the force is repulsive if the boundary conditions at the bounding surfaces differ. If, however, the fluctuations are fully taken  into account, like in the Gaussian model with equal boundary conditions, the force becomes \textit{repulsive} (see \eq{eq:Gaussian_O_SB} below). 
Equation \eqref{eq:MF_O_SB} shows also, that $X_{\rm Cas}^{(S\!B,O)}$ exhibits a minimum {\it above} $T_c$,  similar to the case of ${(O,O)}$ boundary conditions. 

In the case of

\hspace*{2cm} (II) {\textit { {non-zero  external bulk field}}}

one has \cite{rem1}
\begin{equation}
	X_{\rm Cas}^{(S\!B,O)}(x_\tau,x_\mu)=\frac{1}{16} X_{\rm Cas}^{(O,O)}(4x_\tau,8x_\mu).
\end{equation}

All mean field scaling functions of the Casimir force carry an unspecified amplitude, which can be fixed by a suitable normalization, so that a universal scaling function emerges which keeps the functional form of $X_{\rm Cas}^{(a,b)}$.

\subsection{$XY$ universality class}
\label{parallel plates}

The critical exponents\footnote{The quoted values are obtained analytically via renormalization group methods. As far as Monte Carlo results are concerned, one finds, e.g., $\alpha = - 0.0146 ( 8 )$, $\nu = 0.671 55 ( 27 )$, and $\eta = 0.0380 ( 4 ) $\cite{CHPRV2001}.} of systems, which belong to the $O(n=2)$, or $XY$, universality class of the order parameter, are \cite{KS2001,ZJ2002,PV2002}
\begin{equation}
\label{eq:crit_exp}
\alpha = -0.011\pm 0.004  ,\qquad  \nu= 0.6703 \pm 0.0013, \qquad \eta =  0.0354\pm 0.0025.
\end{equation}
Since hyperscaling holds, all critical bulk exponents  can be determined from, say, $\nu$ and $\eta$ by applying suitable  scaling relations. 

\subsubsection{Exact mean field results for the $XY$ model with twisted boundary conditions}
\label{sec:XY_model_twisted}

The finite-size scaling functions  $X_{\rm Cas}^{(a,b)}(x_\tau)$ for the Casimir force in the XY model within mean field
theory, with $x_\tau \equiv\tau (L/\xop)^{1/\nu}=\tau L^2$, have been analyzed
in Ref. \cite{BDR2011}.  We recall that for this model the critical behavior is characterized by the classical critical exponents $\nu=1/2$,  $\alpha=0$, and $\eta=0$. Formally, one can consider the scaling laws between the mean field critical exponents to be still valid, provided the dimensionality of the system is set to  $d=4$ in any relation in which the spatial dimension enters explicitly. This way one straightforwardly obtains all remaining mean field critical exponents. Finally, we  recall that within the standard continuum approach one can relate $\xop$ to the parameters of the mean field model (see  footnote \ref{footnote-MF} on page  \pageref{footnote-MF}).

The authors of Ref. \cite{BDR2011} have studied a system in film geometry $\infty^2\times L$ fixed on a cubic lattice.  The two-component  vector order parameter of this system is taken to lie in the horizontal planes of the system, which contain the   $x$ and $y$ directions, and occupies the vertices of the cubic lattice. At the two opposite sides of the film the  orientations of the vectors are fixed with a prescribed angle $\alpha$ between them, with $0 \le  \alpha \le \pi$, which comprise the essence of the so-called twisted boundary conditions. Within mean field theory the authors have analyzed the Casimir force as a function of temperature $T$, the angle $\alpha$, and the film thickness $L$. For the lattice version of the model the calculations have been carried out numerically, while for its continuum version, i.e., the Ginzburg-Landau mean field theory of the three-dimensional $XY$ model, exact analytical results have been derived. It has been shown that the force depends  continuously on both the angle $\alpha$ and temperature, and it can be attractive or repulsive. By varying $\alpha$ and $T$ one controls {\it both} the sign {\it and} the magnitude of the Casimir force in a reversible way. Furthermore, in the case $\alpha=\pi$, an additional phase transition have been found, which  occurs only in the finite system ($L<\infty$) and is associated with the spontaneous symmetry breaking concerning the direction of the rotation of the vectors through the body of the system, i.e.,  in top-down view clockwise or anticlockwise. 

 The behavior of the system is found analytically by minimizing the free energy functional (per area)
\begin{equation}
{\cal F}\left[ {\bf m};\tau,L\right]=\int_{-L/2}^{L/2} dz\,\left[\frac{b}{2}\left|\frac{d{\bf m}}{dz}\right|^2+\frac{1}{2}a\tau\left|\textbf{m}\right|^2+\frac{1}{4}g\left|{\bf m}\right|^4\right],
\label{LGenergyfunctional}
\end{equation}
with respect to ${\bf m}$, where ${\bf m}(z)$ is the magnetization profile of the order parameter of the system in $z$-direction, $-L/2<z<L/2$, which is normal to the film surfaces. By switching to polar coordinates, i.e.,
\begin{equation}
{\bf m}(z)=\Phi(z)\left(\cos\varphi(z),\sin\varphi(z)\right),
\end{equation}
the free energy functional can be rewritten as
\begin{equation}
{\cal F}\left[\Phi,\varphi;\tau,L\right]=\int_{-L/2}^{L/2} dz\,\left[\frac{b}{2}\left(\frac{d\Phi}{dz}\right)^2+\frac{b}{2}\Phi^2\left(\frac{d\varphi}{dz}\right)^2+\frac{1}{2}a\tau\Phi^2+\frac{1}{4}g\Phi^4\right]. \label{LGfreeenergy}
\end{equation}
The twisted boundary conditions are specified via
\begin{equation}
\varphi(z=\pm L/2)=\pm \alpha/2,\qquad \mbox{and} \qquad \Phi(z=\pm L/2) = \infty.
\label{boundaryconditions}
\end{equation}
Minimization with respect to $\varphi(z)$ leads to
\begin{equation}
\Phi^2(z) \left(\frac{d\varphi}{dz}\right)=P_\varphi
\label{Pvarphidef}
\end{equation}
with an integration constant $P_\varphi$ independent of $z$, which indicates the degree of twist in the system. The condition, which follows from minimizing with respect to $\Phi(z)$, is 
\begin{equation}
b\frac{d^2\Phi}{dz^2}=b\frac{P^2_\varphi}{\Phi^3}+a\tau\Phi+g\Phi^3.
\label{Phieom}
\end{equation}
Note that, due to reflection symmetry in Eq. (\ref{Phieom}) about $z=0$, and the boundary conditions imposed on $\Phi$ (\eq{boundaryconditions}), one has $\Phi(z)=\Phi(-z)$ and, thus, $\Phi'(z)=-\Phi'(-z)$, whence $\Phi'(0)=0$. Then, from the symmetry of Eq. (\ref{Pvarphidef}) under the constraints imposed on $\varphi$  in \eq{boundaryconditions} one infers  $\varphi(z)=-\varphi(-z)$, which leads to $\varphi(0)=0$.

For the first integral of the above system of equations one finds 
\begin{equation}\label{solPhiprime}
{P}_\Phi=-\frac{1}{2}b \left[\frac{P_\varphi^2}{\Phi^2}+\left(\frac{d\Phi}{dz}\right)^2 \right]  +  \frac{1}{2} a\, \tau\, \Phi^2+\frac{1}{4} g\, \Phi^4,
\end{equation}
where $P_\Phi$ is another integration constant independent of $z$. Taking into account that $\Phi'(0)=0$ one can conveniently express $P_\Phi$ as
\begin{equation}\label{PhiPhinode}
	{P}_\Phi=-\frac{1}{2}b\frac{P_\varphi^2}{\Phi_0^2}  +  \frac{1}{2} a\, \tau\, \Phi_0^2+\frac{1}{4} g\, \Phi_0^4,
\end{equation}
where $\Phi_0=\Phi(z=0)$, from which it follows that
\begin{equation}\label{Phifirstintegral}
	\left(\frac{d\Phi}{dz}\right)^2=P^2_\varphi\left(\frac{1}{\Phi_0^2}-\frac{1}{\Phi^2}\right)+\hat{a} \tau \left(\Phi^2-\Phi_0^2\right)
	+\frac{\hat{g}}{2}\left(\Phi^4-\Phi_0^4\right),
\end{equation}
where 
\begin{equation}\label{hatvar}
	\hat{a}=\frac{a}{b}, \qquad \hat{g}=\frac{g}{b}.
\end{equation}
The last result allows one to express the boundary conditions as 
\begin{equation}
\frac{L}{2}=
\int_{\Phi_0}^{\infty}d\Phi\,\frac{1}{\sqrt{P^2_\varphi\left(\Phi_0^{-2}-\Phi^{-2}\right)+\hat{a}\tau\left(\Phi^2-\Phi_0^2\right)+\frac{\hat{g}}{2}\left(\Phi^4-\Phi_0^4\right)}} \label{lengthcondition} 
\end{equation}
and
\begin{equation}
\frac{\alpha}{2}=P_\varphi\int_{\Phi_0}^\infty \frac{d\Phi}{\Phi^2}\frac{1}{\sqrt{P^2_\varphi\left(\Phi_0^{-2}-\Phi^{-2}\right)+\hat{a}\tau\left(\Phi^2-\Phi_0^2\right)+\frac{\hat{g}}{2}\left(\Phi^4-\Phi_0^4\right)}}. \label{twistcondition}
\end{equation}
Equations \reff{lengthcondition}  and \reff{twistcondition} relate $\Phi_0$ and the integration constant $P_\varphi$ to the external parameters $L$ and $\alpha$ of the system. 

Within this model the Casimir force is
\begin{equation}\label{Casimir-XY}
F_{\rm Cas}^{(\alpha)}(\tau,L)=\frac{1}{2}b\frac{P_\varphi^2}{\Phi_0^2}  -  \frac{1}{2} a\, \tau\, \Phi_0^2 -\frac{1}{4} g\, \Phi_0^4 -\frac{1}{4g}(at)^2 \theta(-\tau),
\end{equation}
where $\theta(x)$ is the Heaviside step function; the dependences on $L$ and $\alpha$ are hidden in the dependences of $P_\varphi$ and $\Phi_0$ on them. This takes into account that the bulk free energy density $f_b$ for the system is $f_b(\tau<0)=-(a \tau)^2/(4g)$ while $f_b(\tau>0)=0$. The Casimir force is sensitive to the boundary conditions via $P_\varphi$ and $\Phi_0$.

$F_{\rm Cas}^{(\alpha)}(\tau,L)$  exhibits the expected scaling. In terms of the variables
\begin{equation}
 \Phi_0=\sqrt{\frac{2}{\hat{g}}}X_0 L^{-1},
P_\varphi=\frac{2}{\hat{g}} X_\varphi L^{-3}, \quad \mbox{and} \quad  \hat{a}\tau= x_\tau L^{-2}
\label{scalingvariables}
\end{equation}
the Casimir force reads
\begin{equation}\label{cas}
F_{\rm Cas}^{(\alpha)}(\tau,L)=\frac{b}{\hat{g}}L^{-4}X_{\rm Cas}^{(\alpha)}(x_\tau),
\end{equation}
where
\begin{equation}\label{casscalingfunction}
X_{\rm Cas}^{(\alpha)}(x_\tau)=\left\{ \begin{array}{cc}
                             X_\varphi^2 / X_0^2- X_0^2\left(x_\tau+X_0^2 \right),  & x_
                             \tau \ge 0, \\
                              X_\varphi^2 / X_0^2- \left(\frac{1}{2} x_\tau+X_0^2\right)^2, & x_\tau \le 0.
                            \end{array} \right. 
\end{equation}
Taking into account that mean field theories for systems with short-ranged interactions  are effective theories for $d=4$, one can conclude, that Eq. (\ref{cas}) is in full agreement with the expected scaling behavior of the Casimir force as described in \eq{Casimir_scaling}.  
In order to obtain $X_{\rm Cas}^{(\alpha)}(x_\tau)$, in \eq{casscalingfunction} one has, for  $x_\tau$ and $\alpha$ fixed,  to determine $\Phi_0$ and $P_\phi$ by solving Eqs. \reff{lengthcondition} and \reff{twistcondition}. Using the definitions in \eq{scalingvariables} leads to the amplitudes $X_0$ and $X_\varphi$. This has been carried out in Ref. \cite{BDR2011}.   

\begin{figure}[ht]
\includegraphics[angle=0,width=\columnwidth]{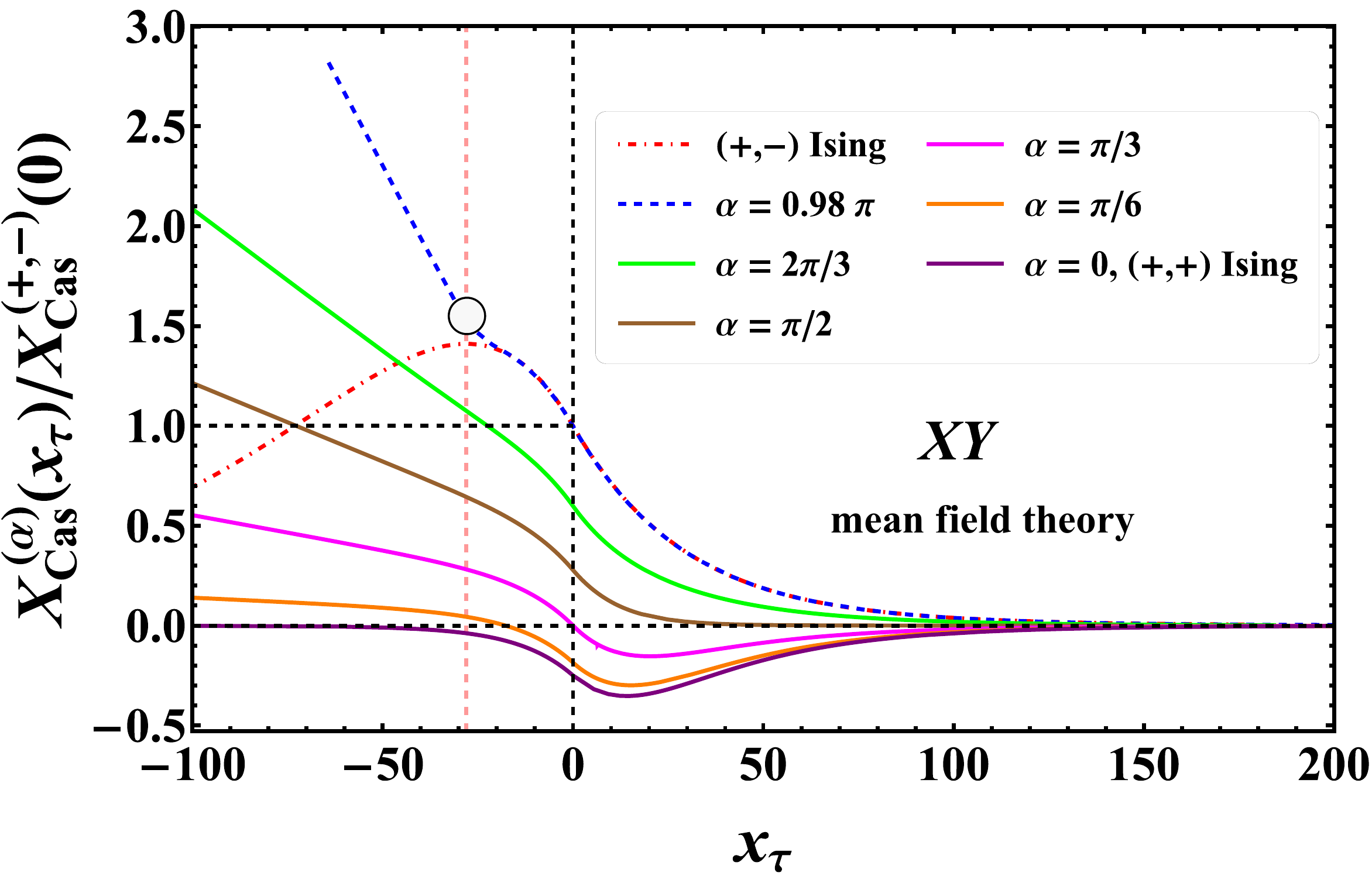}
\caption{Scaling function $X_\textrm{Cas}^{(\alpha)}$ for the Casimir force for twisted boundary conditions with twist angle $\alpha$,  normalized by the corresponding critical value of $X_{\rm Cas}^{(+,-)}$ for the Ising model with $(+,-)$ boundary conditions. The plot shows the behavior of $X_\textrm{Cas}^{(\alpha)}$ as function of $x_\tau=\tau (L/\xi_0^+)^{1/\nu}$  for six values of
$\alpha$. The value of the force at the critical point $x_\tau=0$ depends on $\alpha$. The dash-dotted red curve is the Casimir scaling function for the Ising 
case with $(+,-)$ boundary conditions. For $x_\tau$ sufficiently large this curve coincides with the one for the present model with a twist  angle $\alpha\approx\pi$. The force has a kink-like singularity  at $x_\tau \simeq-28.11$ (its position is marked by the dashed vertical line) for $\alpha=\pi$. This is due to the spontaneous  symmetry breaking of the direction of rotation of the order parameter through the body of the film (clockwise or counterclockwise). For $0<\alpha<\pi/2$ the curves can change sign upon varying the temperature, i.e., $x_\tau$. The change of sign  can be also obtained by varying $\alpha$ for fixed,  moderate values of $x_\tau$. The circle indicates the region where the kink-like singularity in the behavior of the force occurs (see the main text).}
\label{plot:Casimir}
\end{figure}

In terms of the variables
\begin{equation}\label{newvarscaling}
q=x_\tau/X_0^2 \qquad \mbox{and} \qquad p=X_\varphi/X_0^3
\end{equation}
the scaling function of the Casimir force turns into 
\begin{equation}\label{casscalingfunctionpandtau}
X_{\rm Cas}^{(\alpha)}(x_\tau)=\left\{ \begin{array}{cc}
                             X_0^4[p^2- \left(1+q \right)],  & x_\tau \ge 0 \\
                              X_0^4[p^2- \left(1+q/2 \right)^2], & x_\tau \le 0
                            \end{array} \right.,
\end{equation}
where $X_0=X_0(x_\tau,\alpha)$, $p=p(x_\tau,\alpha)$ and $q=q(x_\tau,\alpha)$.
The scaling functions for the Casimir force as a function of $x_\tau$ can be inferred numerically from Eq. (\ref{casscalingfunctionpandtau}). These scaling functions are plotted in Fig. \ref{plot:Casimir}.

%

\begin{figure*}[ht]
	\begin{center}
		\includegraphics[angle=0,width=0.47\columnwidth]{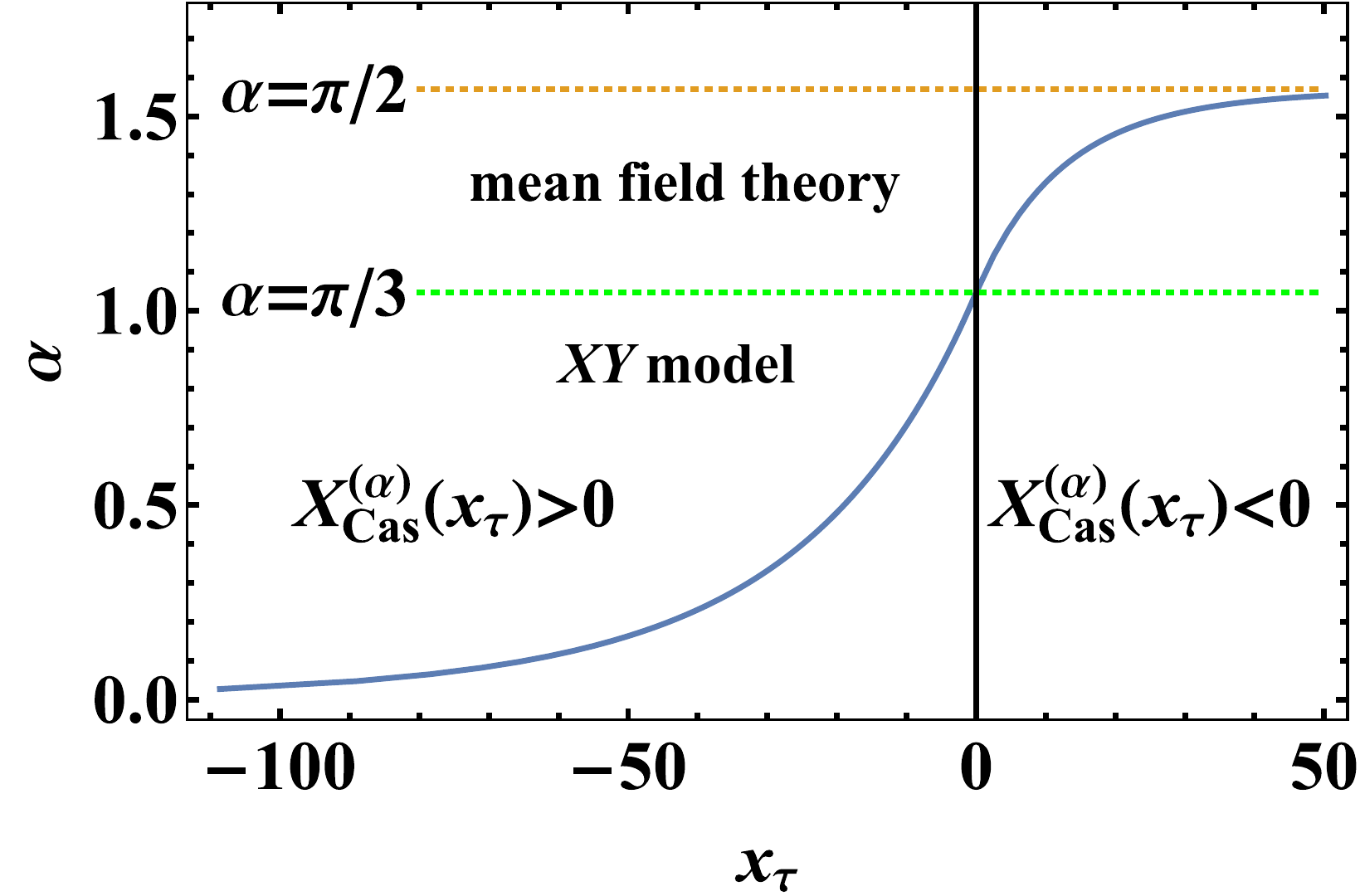}
		\includegraphics[angle=0,width=0.48\columnwidth]{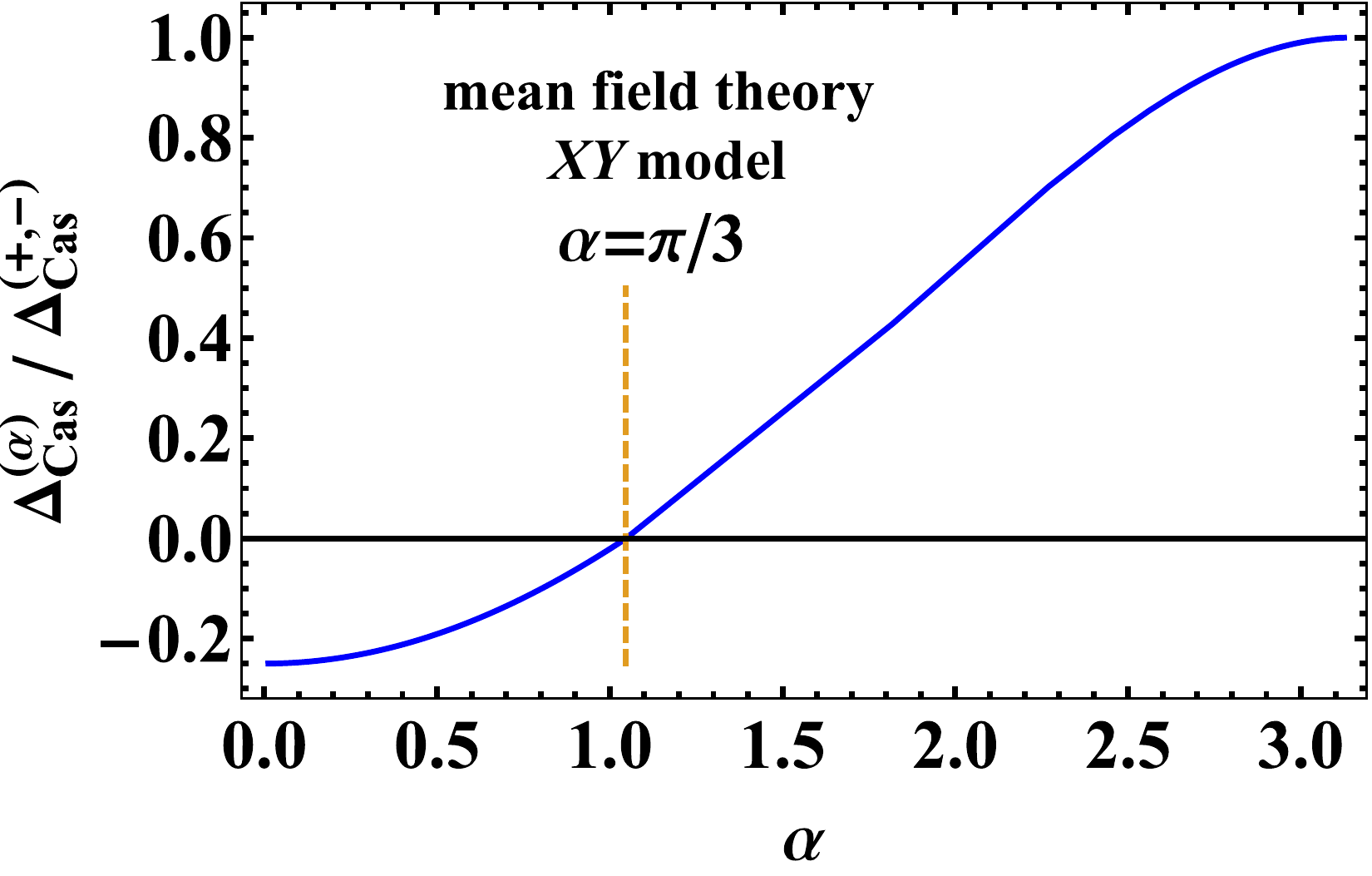}
	\end{center}
	\caption{ (left panel) The positions $x_{\tau,0}^{(\alpha)}$ of the zeros of the Casimir
		force in the $(x_\tau,\alpha)$-plane (full blue line). (right panel) The  Casimir amplitude $\Delta_{\rm
			Cas}^{(\alpha)}$ normalized by $\Delta_{\rm Cas}^{(+,-)}$ as a function of the twist angle $\alpha$ (full blue line). This curve has been
		initially reported in Ref. \cite{K97} based on different representations for the amplitudes derived therein. The Casimir amplitude
		changes sign at $\alpha=\pi/3$. In the left panel the dotted lines indicate paths along which the temperature varies at constant $\alpha$.}
	\label{plot:Casimir_zeros_and_amplitudes}
\end{figure*}

From Eq. (\ref{casscalingfunctionpandtau}) one can conclude certain general
properties of the Casimir force. For example, taking into account that $p$ is a function of $x_\tau$ and $\alpha$, i.e.,
$p=p(\tau|\alpha)$, allows one to determine the coordinates $x_{\tau,0}^\alpha$ of
the zeros of the Casimir force for a given angle $\alpha$. According to Eq.
(\ref{casscalingfunctionpandtau}) one has $X_{\rm Cas}^{(\alpha)}=0$ for
$p(\tau|\alpha)=\sqrt{1+\tau}$, with $\tau\ge 0$, and for $p(\tau|\alpha)=
1+\tau/2$ if $-2\le\tau\le 0$. A plot of the positions of these zeros in the
$(x_\tau,\alpha)$-plane  is presented in Fig.
\ref{plot:Casimir_zeros_and_amplitudes}.

Figures \ref{plot:Casimir} and  \ref{plot:Casimir_zeros_and_amplitudes} indicate how, by changing, e.g., the twist angle $\alpha$, one
can, at a given temperature $\tau$, render the Casimir force either repulsive or
attractive. For $0<\alpha<\pi/2$, this can also be achieved by changing the
temperature, i.e., the scaling variable $x_\tau$, at a given fixed value of
$\alpha$. We also conclude that, upon $\alpha \to 0$ the position of the zero of the Casimir force tends to $-\infty$. This implies that for
$\alpha=0$ the Casimir force is attractive for all temperatures. 
For $\alpha=0$ one has $X_{\rm Cas}^{(\alpha=0)}(x_\tau)\equiv X_{\rm Cas}^{(+,+)}(x_\tau)$, i.e., as expected it coincides with the known result
for the Ising model \cite{K97}. In Fig. \ref{plot:Casimir} the behavior of $X_{\rm Cas}^{(+,+)}(x_\tau)$
is shown  as the dashed line at the bottom.

\begin{figure}[!h]
	\includegraphics[width=\columnwidth]{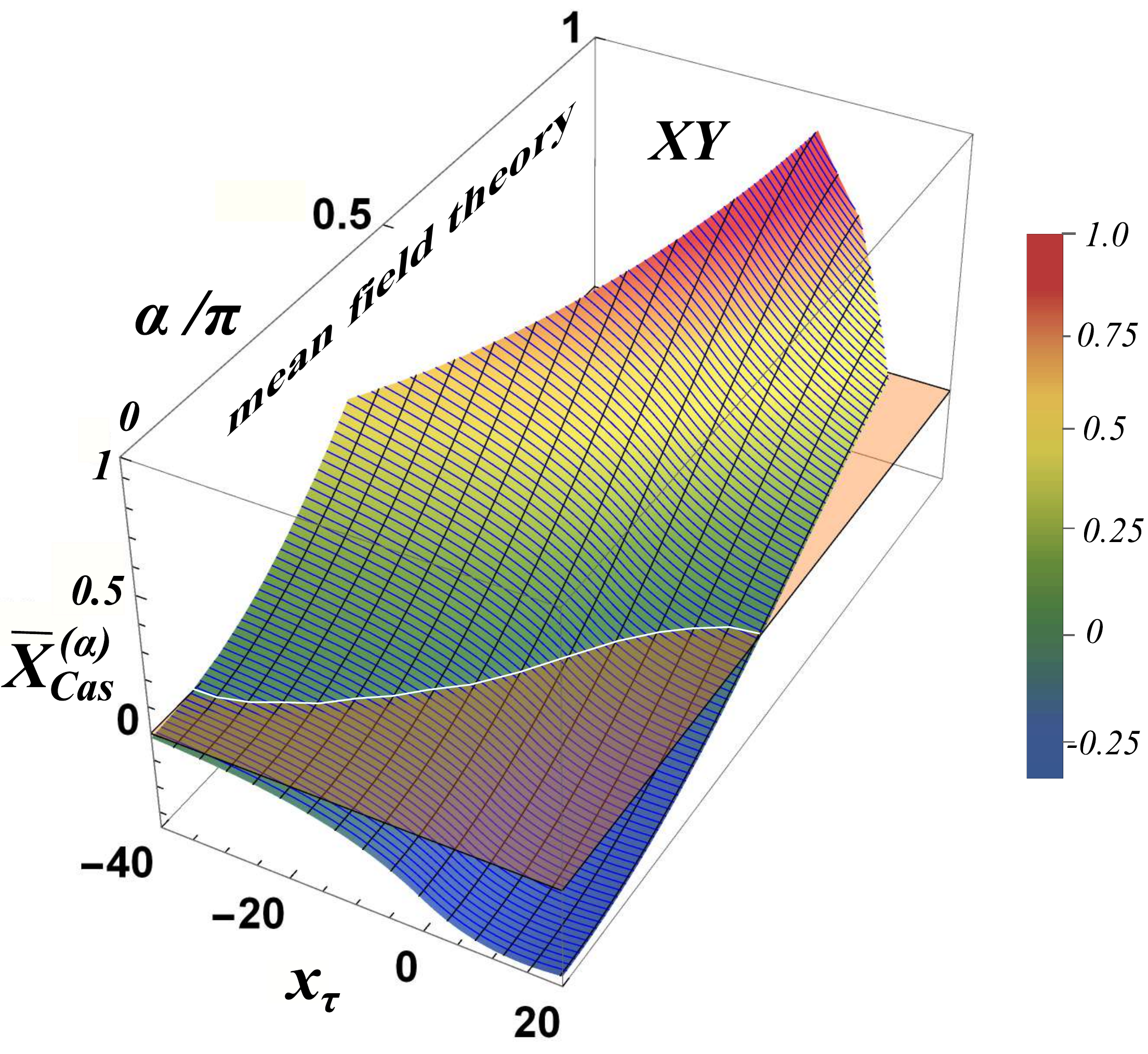}
	\caption{The normalized scaling function $\overline{X}^{(\alpha)}_{\rm Cas}(x_\tau)=X^{(\alpha)}_{\rm Cas}(x_\tau)/X^{(+,-)}_{\rm Cas}(0)$  of the $XY$ model for twisted boundary conditions as function of $x_\tau$ and $\alpha$ with the bulk field $h=0$. The colored surface intersects the plane $\overline{X}^{(\alpha)}_{\rm Cas}(x_\tau)=0$ along the white line (see the left panel in Fig. \ref{plot:Casimir_zeros_and_amplitudes}). $\overline{X}^{(\alpha)}_{\rm Cas}(x_\tau)$ is positive above the ocher part of the plane, i.e., the force is repulsive.  Below the brownish  part of the plane $\overline{X}^{(\alpha)}_{\rm Cas}(x_\tau)$ is negative, i.e., the force is attractive. The lines on the surface $\overline{X}^{(\alpha)}_{\rm Cas}(x_\tau)$ correspond to its intersections with the planes $x_\tau={\rm const}$ or $\alpha={\rm const}$.  }
	\label{Fig:3d_Casimir_Force_XY}
\end{figure}

For $x_\tau \to-\infty$ the following  asymptotic expression for $X_{\rm Cas}^{(\alpha)}(x_\tau)$ has been derived in Ref.  \cite{BDR2011}:
\begin{equation}\label{casscalingfunctionasympXY}
X_{\rm Cas}^{(\alpha)}(x_\tau\to -\infty)\simeq \frac{1}{2}  \alpha^2 \left[|x_\tau|+4
 \sqrt{2|x_\tau|}+\frac{1}{2} \left(48-3 \alpha ^2\right)\right].
\end{equation}
According to \eq{cas}, \eq{casscalingfunctionasympXY} implies that in this asymptotic regime one has 
\begin{equation}
\label{casMFas}
\beta F_{\rm Cas}^{(\alpha)}(\tau,L)\simeq \frac{1}{2} \alpha^2 \frac{b}{\hat{g}}   |x_\tau|L^{-4} = \frac{1}{2}\frac{a b}{g} \alpha^2  |\tau| L^{-2},
\end{equation}
i.e., due to the presence of helicity in the system, the leading behavior of $\beta F_{\rm Cas}^{(\alpha)}(\tau,L)$  in this regime is of the order of $L^{-2}$.  The scaling function $X^{(\alpha)}_{\rm Cas}(x_\tau)$ of the $XY$ model with twisted boundary conditions  is shown in Fig. \ref{Fig:3d_Casimir_Force_XY} as a function of $x_\tau$ and $\alpha$.

\subsubsection{Exact solution of the $\Psi$ model for $^4$He}

The $\Psi$ model (or $\Psi$-theory) has been briefly introduced in Sect. \ref{sec:psi-model}. The order parameter, which governs the behavior of the system,  minimizes a certain free energy functional (see Eqs.  \eqref{eq:omega_phi_expansion} and \eqref{eq:phi_spacial_eq}).  This is conceptually similar to  mean field theory.  However, the $\Psi$-theory differs from mean field theory in that the $\Psi$-theory is an effective  theory for $d=3$, in contrast to mean field theory, which holds  in $d=4$. The corresponding critical exponents are distinct: say, $\nu=2/3$ in the $\Psi$-theory, while $\nu=1/2$ in mean field theory.

 In Ref. \cite{DRVD2019} the Casimir force,  acting on $^4$He wetting films in thermal equilibrium with their vapor phase, has been analyzed within  the $\Psi$ model. This leads to
\begin{equation}\label{eq:Xcas_general}
\beta F_{\rm Cas}(T,L;M,Q)=Q x_\tau^3 \left[p(\phi_0(x_\tau,M), M)-\frac{3+M}{6}\right] L^{-3}. 
\end{equation}
Here
\begin{equation}
\label{eq:p}
p=\sign(\tau) \phi_0^2 - \frac{1}{2}(1-M) \phi_0^4 - \frac{1}{3}M\phi_0^6
\end{equation}
and  $\phi_0=\phi_0(x_\tau,M)$, which follows from 
 \begin{equation}
 \label{eq:phi_not_y}
 x_\tau \equiv \frac{L}{\xi_\tau}= \int_{0}^{1} \frac{dy}{\sqrt{y\left(1-y\right)\left[1-\frac{1}{2}(1-M)\phi_0^2 \left(1+y\right)- \frac{1}{3}M \phi_0^4\left(1+y^2+y \right)\right]}}=\frac{4 \sqrt{3}}{b} K(k),
 \end{equation}
is the value of the order parameter in the middle of the system.  In order to keep notations simple, we have introduced  the variables
 \begin{equation}
k=\sqrt{ 2 \sqrt{3} a}\; \frac{\phi_0 }{b}, \quad a =\sqrt{(3+M+2M\phi_0^2) (1+M(3-2\phi_0^2)) },\quad 
 b =\sqrt{12-6 M\phi_0^4+\phi_0^2 ( \sqrt{3} a-9(1-M))}\,.
 \end{equation}
In \eq{eq:Xcas_general} the constant $Q>0$ and the parameter $M$ of the $\Psi$ theory have to be determined experimentally (see below).  We note that, according to \eq{eq:psi_mod_def}, $x_\tau>0$ for $T<T_\lambda$. 

For $0\le M\le 1$, the rhs of \eq{eq:phi_not_y} is a monotonically increasing function of $0\le\phi_0^2\le 1$. Therefore, if it exists, there is a single solution of \eq{eq:phi_not_y}. Thus, this equation can be inverted uniquely, which renders $\phi_0(x_\tau,M)$ for $0\le M\le 1$. Since $\phi_0\ge 0$, the minimal value of $x_\tau$, for which such a solution exists, is given by the rhs of \eq{eq:phi_not_y} upon setting $\phi_0=0$ in it. This leads to the relation $x_\tau(\phi_0=0,M)=\pi$, which implies that for $x_\tau>\pi$ only (see below) there is a non-zero solution for $\phi_0$ and, therefore, for the order parameter profile $\phi(\zeta_\tau,M)$   (see \eq{eq:psi_mod_def} concerning the definition of the scaled coordinate $\zeta_\tau$ and of the order parameter $\phi$).  This statement is valid independently of the value of $M$. Finally, using the above arguments, one can determine the position of the minimal value of $p$ as function of $x_\tau$. For $\tau>0$, $p$ is an increasing function of $\phi_0$ and thus of $x_\tau$ for $0\le \phi_0\le 1$. The lowest value of $p$ is attained at the minimal value of $\phi_0$, i.e., at $x_\tau=\pi$, which is also the position at which the Casimir force attains its largest negative value.  Thus, within the class of effective theories considered above, one can conclude that the position of the largest negative value of the Casimir force does not depend on $M$. Experimental investigations \cite{GSGC2006} yield for the position of the minimum $x_\tau=3.2\pm 0.18$, which \textit{de facto} coincides with the  result obtained above. The value of the minimum and the shape of the scaling function of the Casimir force depend on $M$. Since the minimum of $p$ is located at $x_\tau=\pi$, and since  $\phi_0(x_\tau=\pi,M)=0$, one infers from  \eq{eq:Xcas_general} that the minimum value of the scaling function of the Casimir force is 
\begin{equation}\label{eq:XCas_min}
X_{\rm Cas}^{\rm min} \equiv \min_{\mbox{\normalsize $x_\tau$}} X_{\rm Cas}(x_\tau,M)=X_{\rm Cas}(x_\tau=\pi,M)=-\frac{3+M}{6} \pi^3 Q.
\end{equation}
The minimum deepens as $M$ increases. In obtaining \eq{eq:XCas_min} we have taken into account that the scaling function of the Casimir force is given by
\begin{equation}\label{eq:XCas}
X_{\rm Cas}(x_\tau,M)=Q x_\tau^3 \left[p(\phi_0(x_t,M), M)-\frac{3+M}{6}\right], \quad Q>0.
\end{equation}
We note that the term $(3+M)/6$ is the value of $p$  (see \eq{eq:p}) for $\tau>0$ and $\phi_0=1$. Since $p$ is a  monotonically increasing function of $0\le \phi_0\le 1$, this implies that $p<(3+M)/6$ for \textit{any} values of $x_\tau$ and $M$, i.e., 
\begin{equation}
\label{eq:Xcas_neg}
X_{\rm Cas}(x_\tau,M)\le 0.
\end{equation}
Accordingly, within the $\Psi$ model the Casimir force is \textit{attractive}. 

We now briefly comment on how, within the $\Psi$ model, one can determine the value of the constant $Q$. In Ref. \cite{DRVD2019} three different approaches for achieving this goal have been presented. The standard approach is based on the  original papers by Ginzburg and Sobyanin \cite{GS76,GS82}. It provides $Q=Q^{\rm GS}$, where 
\begin{equation}
\label{eq:QGS}
Q^{\rm GS}=\frac{1}{2}\beta T_{\lambda } \Delta C_{\mu } \xi _0^3\;\sqrt{\frac{3+M}{3}}=\sqrt{\frac{3+M}{3}} \times \left \{\begin{array}{cccl}
0.119, & {\rm for} &\xi_0=1.63\, \AA{} & \mbox{ (Refs. \cite{GS76,GS82})}\\
0.081, &  {\rm for} &\xi_0=1.432\, \AA& \mbox{(Ref. \cite{TA85})}
\end{array} \right. .
\end{equation}
This is consistent with the value of the constant $A$ (see Eqs. \reff{eq:omega_phi_expansion}
and 
\reff{eq:psi_mod_def}), which is equal to  $A=[(3+M)/3]^{1/2}\beta T_{\lambda } \Delta C_{\mu }$. Here $\Delta C_\mu $ is the ``specific heat jump''
at the $\lambda$ point: $\Delta C_\mu = \Delta C_p = 0.76 \times 10^7$ erg cm$^{-3}$ K$^{-1}$ \cite{GS76,GS82}. Another approach \cite{DRVD2019}, is based on the currently best known values of $\xi_0$ and $\Psi_{s,e0}$ (see \eq{eq:psi_eq_value}), which leads to 
\begin{equation}
\label{eq:QPsi_xi}
Q=Q^{(\xi,\Psi)}=\frac{1}{2}\frac{(\hbar \Psi_{s,e0})^2}{2m k_B T_\lambda}\xi_0=0.106.
\end{equation}
Here the value of $\Psi_{s,e0}$ is taken from Refs. \cite{GS76,GS82}, while the value of $\xi_0$ stems from Ref.  \cite{TA85}.

\begin{figure}[h!]
	\includegraphics[width=\columnwidth]{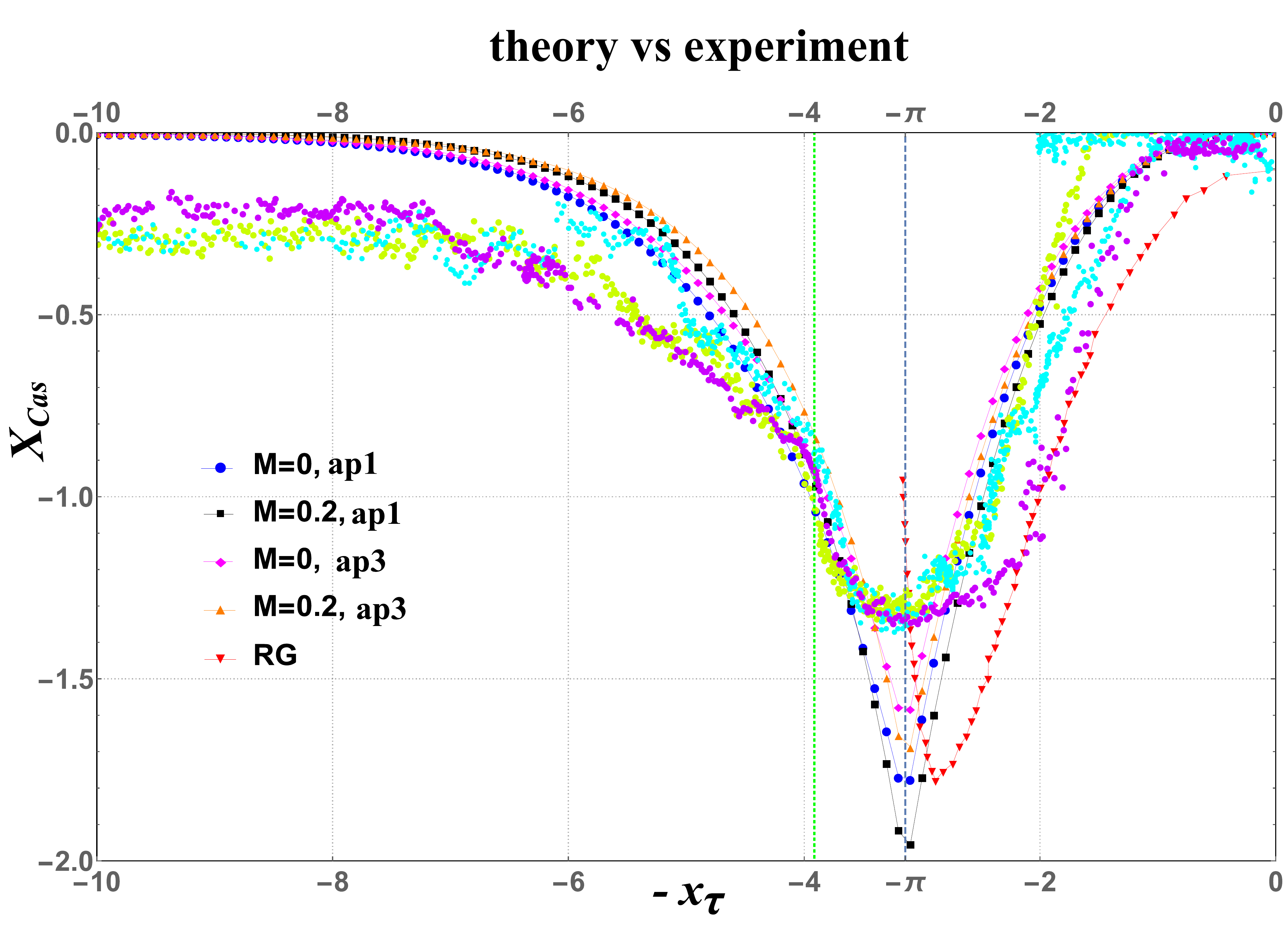}
	\caption{
		Comparison of experimental data \cite{GSGC2006} for the Casimir force acting on $^4$He films (scattered data points) with the predictions of the $\Psi$ theory for $M=0$ and $M=0.2$, and for two types of approaches (solid curves with symbols). The prediction of renormalization group theory (RG) \cite{D2014} is presented as the red curve with inverted triangles, the wiggles are numerical artifacts. (The solid lines are decorated by symbols  in order to be able to identify more easily the various lines.)  The green dotted vertical line at $-x_\tau=-3.91$ (in the current section $x_\tau>0$ means $T<T_\lambda$) marks the Monte Carlo prediction for the occurrence of the  Kosterlitz-Thouless phase transition in the film \cite{H2009}. The blue dashed vertical line indicates $x_\tau=\pi$. Due to the scatter of the experimental data it is not possible to identify that one of the four analytical $\Psi$-theory curves, which matches the experimental data best. We note that $X_{\rm Cas}$ is nonzero above $T_\lambda$, i.e., for $x_\tau<0$. For that part of the scaling function one finds agreement between the experimental data and the RG analysis in Refs. \cite{KD92a,KD92b}. } 
	\label{fig:PsiSRCF_comp}
\end{figure}

In the following we compare these theoretical results with the available experimental ones. The results are presented in Fig. \ref{fig:PsiSRCF_comp}. For convenience, the following short-hand notations are introduced:  \textit{(i)} \textit{ap1} corresponds to the $\Psi$-theory as formulated in Refs. \cite{GS76,GS82};  \textit{(ii)} \textit{(ap2)} corresponds to the same theory but with the value of the  amplitude $\xi_0$ as determined in Ref. \cite{TA85};  \textit{(iii)} \textit{ap3} corresponds to that theory, in which the constant $Q$, determined via $\Psi_{s,e0}$ and $\xi_0$, is taken from Ref. \cite{TA85}. The experimental results pertain to the Casimir force acting on horizontally oriented liquid $^4$He wetting films,  supported by a stack of substrates such that the films are in thermal equilibrium with their vapor phase. The experimental signal consists of the thickness of the $^4$He wetting films from which the strength of the critical Casimir force can be  inferred \cite{KD92b}. This is carried out for temperatures at, as well as close to, the critical end point of the $\lambda$-transition of $^4$He from its normal to its superfluid state. This continuous phase transition of $^4$He, referred to as the $\lambda$-transition because of the shape of the temperature dependence of the specific heat curve,  occurs  at the temperature  $T_\lambda=2.172$ $^\circ$K \cite{GS76} and at a saturated-vapor pressure $p_\lambda =0.05$ atm with mass density  $\rho_\lambda=0.1459$ g/cm$^3$ \cite{W67,GS82}.  (We note that while the density varies  continuously through the transition, its temperature gradient varies discontinuously \cite{W67,DB98}.)
To be precise, we stress that upon expressing  the experimental data in Ref. \cite{GSGC2006} (given as function of $(T/T_\lambda-1)L^{1/\nu}$)  in terms of the variable $L/(\xi_0 |\tau|^{1/\nu})$, the value of $\nu$ as given in \eq{eq:crit_exp}, and the data for $\xi_0=1.432$ $\AA$, as reported in Ref. \cite{TA85},  have been used.  Figure~\ref{fig:PsiSRCF_comp} also contains some results obtained from various theoretical approaches and Monte Carlo simulations. As shown, the minimum is positioned at $x_\tau=\pi$, independent of the value of $M$, while the experiment yields  $x_\tau=3.2\pm 0.18$, which is \textit{de facto}  consistent with our results.  Within \textit{ap1}, the  maximal absolute value of the (negative, i.e., attractive) force is reached for $M=0$; the extremum is $-1.848$ while the experimental value is $-1.30$. Inspection of the plot shows that the closest agreement between theory and experiment is obtained for $M=0$. The next best theoretical curve is the one corresponding to $M=0.2$ for which the minimal, negative value of the force does not differ too much from the one corresponding to the  case $M=0$. One has $X_{\rm Cas}^{\rm min}(x_\tau=\pi,M=0.2)=-2.036$. 	The figure also	presents  a comparison of the experimental data for the Casimir force  with the predictions of the  approach \textit{ap3} with  $M=0$ or $M= 0.2$, for which one  observes again relatively good agreement. In that case one has $X_{\rm Cas}^{\rm min}(x_\tau=\pi,M=0)=-1.643$ and $X_{\rm Cas}^{\rm min}(x_\tau=\pi,M=0.2)=-1.753$.

Below, we briefly comment on other available analytical approaches towards the behavior of the Casimir force acting on  $^4$He films. We start by discussing mean field theory. 

As already explained, the $\Psi$-theory is conceptually similar to mean field theory in that an effective free energy is minimized, without taking the fluctuations fully into account. Within the $\Psi$-theory, the fluctuations are partially taken into account by assigning the value $2/3$ to the critical exponent $\nu$. Furthermore, the   $\Psi$-theory is an effective theory for $d=3$, while standard mean field theory characteristically applies to  $d\ge 4$. However, the latter theory contains a single, non-universal parameter,  the value of which is determined by information from outside this theory in order to be able to  connect with experimental data. In Ref. \cite{ZSRKC2007}  renormalization group arguments are utilized in order to determine this nonuniversal parameter.  In this mean-field case the minimum of the force occurs at $x_{{\rm min, MF}}=\tau(L/\xi_0 )^{1/\nu_{{\rm MF}}} =\pi^2$, where $\nu_{{\rm  MF}}=1/2$. If renormalization group input is used in order  to enhance mean field theory  (see Refs. \cite{ZSRKC2007,MGD2007}), the minimum of the scaling function of the force is $X_{\rm Cas, min, MF}=-6.92$
at $x_{{\rm min, MF}}=\pi^2$. We note, however, that the mean field scaling variable can be redefined as $\tau^{\nu_{\rm MF}} (L/\xi_0 )$, with $\nu_{\rm MF}=1/2$. In that case the minimum of the force is at   $\hat{x}_{{\rm \tau,min}}^{{\rm MF}}=\sqrt{\tau} (L/\xi_0 )=\pi$. Within the $\Psi$-theory, one also has the minimum at $x_{{\rm \tau,min}}=\tau^{2/3} (L/\xi_0 )=\pi$. Thus, if one formulates the scaling variables in both theories in terms of $L$ divided by the corresponding correlation length, i.e., in terms of $L/\xi_{{\rm MF}}$ or $L/\xi_\Psi$,  both theories have a minimum at $\pi$. The agreement with the experimental data is, however, better for the $\Psi$-theory, because the correct value of $\nu$ (see \eq{eq:crit_exp}), is quite close to the value $\nu=2/3$ of the $\Psi$-theory. 

The fully fledged renormalization group approach is, apart from MC simulations, the only one which does not require any external input, as long as one insists on the full validity of the universality hypothesis.  The first attempt in that direction has been made in Refs. \cite{KD91,KD91b,KD92a,KD92b}, where the authors studied within the $\varepsilon$ expansion the behavior of the critical Casimir force in $^4$He films above $T_\lambda$. Recent progress below $T_\lambda$ has been made in Ref. \cite{D2014};   in Fig. \ref{fig:PsiSRCF_comp} see  the red curve marked with inverted triangles. This proposed theory holds only for temperatures above a certain temperature $T_{\rm film}^*$, close to but below $T_\lambda$, and breaks down below $T_{\rm film}^*$. Nevertheless, this temperature interval $[T_{\rm film}^*, T_\lambda]$ encompasses the position of the minimum of the force. It is reported to be at $x_{\rm min, RG}=\tau_{\rm min, RG}(L/\xi_0 )^{1/\nu}=-4.73$ (with $\nu=0.671$). If one expresses this in terms of $x_\tau$, one obtains  $\hat{x}_{{\rm \tau,min}}^{{\rm RG}}=2.84$. In Ref. \cite{D2014} the value of the scaling function of the Casimir force at the minimum  is not reported, but from the presented plot (see Fig. $1(b)$ therein), it can be estimated as $X_{{\rm Cas, min, RG}} \simeq -1.8$, which is very close to the above results of the $\Psi$ theory within approach 1 with $M=0$. 

In view of the criticisms of the $\Psi$-theory summarized above, it is useful to recall also the main advantages offered by it: it is relatively simple to apply, and it provides analytical results reasonably close to the corresponding experimental data (see Fig. \ref{fig:PsiSRCF_comp}). Therefore, it can be a useful tool to obtain approximate expression for the experimental behavior of such physically rich systems as helium films. In view of the issues described above, this is especially the case,  if one keeps in mind the limited state of the art of more advanced theories,  such as  renormalization group theory for temperatures below the critical one (see Fig. \ref{fig:PsiSRCF_comp}).

\subsection{Gaussian model}

 Within the Gaussian model the Casimir force has been studied in Refs.  \cite{S81,KD92a,DK2004,MZ2006,GD2006,DiRu2017,BDT2000,DaR2017,KD2010}. We point out right at the beginning, that for the Gaussian model the scaling functions of the excess free energy and of the Casimir force  coincide, up to a factor of two, with the corresponding scaling functions of the ideal Bose gas in film geometry within the grand canonical ensemble. The factor of two takes into account that the order parameter in the case of a Bose gas is complex and has, therefore, twice as many components  than the corresponding classical one, i.e., the $O(2n)$ classical model is equivalent to the corresponding $O(n)$ ideal Bose gas \cite{DiRu2017}. 
 
For the $d$-dimensional ($2<d<4$) Gaussian $O(n)$ model in film geometry, Refs. \cite{KD91,KD92a} provide explicit results for the Casimir amplitudes, as well as the scaling functions of the excess free energy and of the Casimir force. The scaling functions for the excess free energy and for the Casimir force are related via the simple relation (see \eq{rel_cas_excess} for $h=0$)
\begin{equation}
\label{eq:Gaus_ex_Cas_sf}
X_{\rm Cas}^{{\rm (\zeta)}}(x_\tau|d)=(d-1)X_{\rm ex}^{{\rm (\zeta)}}(x_\tau|d)-2x_\tau \frac{\partial }{\partial x_\tau}X_{\rm ex}^{{\rm (\zeta)}}(x_\tau|d).
\end{equation}
Here $\zeta$ stands for the corresponding boundary conditions, and we have taken into account that for the Gaussian model $\nu=1/2$, and $x_\tau=\tau L^{1/\nu}$. For the various boundary conditions the following results have been derived:

\subsubsection{periodic boundary conditions}
The corresponding Casimir amplitude is 
\begin{equation}
\label{eq:Gaus_per_Cas}
\Delta_{\rm Cas}^{{\rm (p)}}(d, n)=-n\; \pi^{-d/2} \Gamma(d/2)\zeta(d)<0
\end{equation}
with $\zeta(z)$ the Riemann's $\zeta$ function, while the corresponding scaling function for the excess free energy is 
\begin{equation}
\label{eq:Gaus_per_ex_sf}
X_{\rm ex}^{{\rm (p)}}(x_\tau|d, n)=-n \;2^{-d} \pi^{-d/2} \dfrac{2\sqrt{\pi} }{\Gamma\left(\frac{d+1}{2}\right)}x_\tau^{\,d/2}\int_{1}^{\infty} dx \dfrac{(x^2-1)^{(d-1)/2}}{\exp\left(x \sqrt{x_\tau}\right)-1} \;< 0.
\end{equation}
It is possible to express the scaling function of the excess free energy in terms of higher functions \cite{BDT2000,DiRu2017}:
\begin{equation}
\label{eq:excess_per_equiv}
X_{\rm ex}^{{\rm (p)}}(x_\tau|d, n)=-n\; 2^{1-d/2} \pi ^{-d/2} x_\tau^{\,d/4} \sum _{k=1}^{\infty }
\frac{K_{d/2}\left(k \sqrt{x_\tau}\right)}{k^{d/2}}.
\end{equation}
The asymptotic behavior for $x_\tau\to\infty$ is
\begin{equation}
\label{eq:excess_p_as}
X_{\rm ex}^{{\rm (p)}}(x_\tau|d, n) \simeq -n\;  (2 \pi )^{-(d-1)/2} x_\tau^{ (d-1)/4}e^{-\sqrt{x_\tau}}, \qquad x_\tau\to\infty.
\end{equation}
From  \eq{eq:excess_per_equiv} one obtains for the excess free energy scaling function in $d=3$ 
\begin{equation}
\label{eq:excess_p_equiv_d3}
X_{\rm ex}^{{\rm (p)}}(x_\tau|d=3, n)=-\frac{n}{2 \pi } \left[\text{Li}_3\left(e^{-\sqrt{x_\tau}}\right)+\sqrt{x_\tau}\,
	\text{Li}_2\left(e^{-\sqrt{x_\tau}}\right)\right]
\end{equation}
 and 
\begin{equation}
\label{eq:Cas_p_equiv_d3}
X_{\rm Cas}^{{\rm (p)}}(x_\tau|d=3, n)=-\frac{n}{\pi } \left[\text{Li}_3\left(e^{-\sqrt{x_\tau}}\right)+
 \sqrt{x_\tau}\, \text{Li}_2\left(e^{-\sqrt{x_\tau}}\right)
	-\frac{1}{2}x_\tau \ln
	\left(1-e^{-\sqrt{x_\tau}}\right)\right] < 0
\end{equation}
for the scaling function of the attractive Casimir force.

\subsubsection{antiperiodic boundary conditions}
The corresponding Casimir amplitude is 
\begin{equation}
\label{eq:Gaus_aper_Cas}
\Delta_{\rm Cas}^{{\rm (ap)}}(d,n)=n\; (1-2^{-d+1})\pi^{-d/2} \Gamma(d/2)\zeta(d)>0,
\end{equation}
while the corresponding scaling function is 
\begin{equation}
\label{eq:Gaus_aper_ex_sf}
X_{\rm ex}^{{\rm (ap)}}(x_\tau|d,n)=n \;2^{-d} \pi^{-d/2} \dfrac{2\sqrt{\pi} }{\Gamma\left(\frac{d+1}{2}\right)}x_\tau^{\,d/2}\int_{1}^{\infty} dx \dfrac{(x^2-1)^{(d-1)/2}}{\exp\left(x \sqrt{x_\tau}\right)+1} \; >0,
\end{equation}
which can be expressed as 
\begin{equation}
\label{eq:excess_aper_equiv}
X_{\rm ex}^{{\rm (ap)}}(x_\tau|d,n)=n\; 2^{1-d/2} \pi ^{-d/2} x_\tau^{d/4} \sum _{k=1}^{\infty }(-1)^{k+1}
\frac{K_{d/2}\left(k \sqrt{x_\tau}\right)}{k^{d/2}}
\end{equation}
with the asymptotic behavior, for $x_\tau\to\infty$,  
\begin{equation}
\label{eq:excess_ap_as}
X_{\rm ex}^{{\rm (ap)}}(x_\tau|d,n) \simeq n\;  (2 \pi )^{-(d-1)/2} x_\tau^{ (d-1)/4}e^{-\sqrt{x_\tau}}, \qquad x_\tau\to\infty.
\end{equation}
From  \eq{eq:excess_aper_equiv} one obtains for the excess free energy scaling function in $d=3$ 
\begin{equation}
\label{eq:excess_ap_equiv_d3}
X_{\rm ex}^{{\rm (ap)}}(x_\tau|d=3,n)=-\frac{n}{2 \pi } \left[\text{Li}_3\left(-e^{-\sqrt{x_\tau}}\right)+\sqrt{x_\tau}\,
\text{Li}_2\left(-e^{-\sqrt{x_\tau}}\right)\right]
\end{equation}
 and 
\begin{equation}
\label{eq:Cas_ap_equiv_d3}
X_{\rm Cas}^{{\rm (ap)}}(x_\tau|d=3,n)=-\frac{n}{\pi } \left[\text{Li}_3\left(-e^{-\sqrt{x_\tau}}\right)+
\sqrt{x_\tau}\, \text{Li}_2\left(-e^{-\sqrt{x_\tau}}\right)
-\frac{1}{2}x_\tau \ln
\left(1+e^{-\sqrt{x_\tau}}\right)\right] \; >0
\end{equation}
for the scaling function of the repulsive Casimir force.

\subsubsection{Dirichlet or ordinary boundary conditions}
The corresponding Casimir amplitude is 
\begin{equation}
\label{eq:Gaus_DD_Cas}
\Delta_{\rm Cas}^{{\rm (O,O)}}(d,n)=-n\; 2^{-d} \pi^{-d/2} \Gamma(d/2)\zeta(d) \;<0
\end{equation}
while the corresponding scaling function for the excess free energy is 
\begin{equation}
\label{eq:Gaus_DD_ex_sf}
X_{\rm ex}^{{\rm (O,O)}}(x_\tau|d,n)=-n \;2^{-d} \pi^{-d/2} \dfrac{2\sqrt{\pi} }{\Gamma\left(\frac{d+1}{2}\right)}x_\tau^{\,d/2}\int_{1}^{\infty}dx \dfrac{(x^2-1)^{(d-1)/2}}{\exp\left(2x \sqrt{x_\tau}\right)-1} \; <0,
\end{equation}
which equals 
\begin{equation}
\label{eq:excess_DD_equiv}
X_{\rm ex}^{{\rm (O,O)}}(x_\tau|d,n)=-n\; 2^{1-d} \pi ^{-d/2} x_\tau^{\,d/4} \sum _{k=1}^{\infty }
\frac{K_{d/2}\left(2 k \sqrt{x_\tau}\right)}{k^{d/2}}.
\end{equation}
Its asymptotic behavior for $x_\tau\to\infty$ is
\begin{equation}
\label{eq:excess_DD_as}
X_{\rm ex}^{{\rm (O,O)}}(x_\tau|d,n) \simeq -n \; 2^{-d} \pi ^{-(d-1)/2} x_\tau^{ (d-1)/4} e^{-2 \sqrt{x_\tau}}, \qquad x_\tau\to\infty.
\end{equation}
From \eq{eq:excess_DD_equiv}  one has for $d=3$
\begin{equation}
\label{eq:excess_DD_equiv_d3}
X_{\rm ex}^{{\rm (O,O)}}(x_\tau|d=3,n)=-\frac{n}{16 \pi } \left[\text{Li}_3\left(e^{-2 \sqrt{x_\tau}}\right)+2 \sqrt{x_\tau}\,
	\text{Li}_2\left(e^{-2 \sqrt{x_\tau}}\right)\right]
\end{equation}
for the excess free energy scaling function and 
\begin{equation}
\label{eq:Cas_DD_equiv_d3}
X_{\rm Cas}^{{\rm (O,O)}}(x_\tau|d=3,n)=-\frac{n}{8 \pi} \left[\text{Li}_3\left(e^{-2 \sqrt{x_\tau}}\right)+2 \sqrt{x_\tau}\,
	\text{Li}_2\left(e^{-2 \sqrt{x_\tau}}\right)-2 x_\tau \ln \left(1-e^{-2
		\sqrt{x_\tau}}\right)\right] \; <0
\end{equation}
for the scaling function of the attractive Casimir force. 

\subsubsection{Neumann or surface-bulk or special boundary conditions}
In this case the Casimir amplitude is 
\begin{equation}
\label{eq:Gaus_SB_Cas}
\Delta_{\rm Cas}^{{\rm (SB,SB)}}(d,n)=-n\; 2^{-d} \pi^{-d/2} \Gamma(d/2)\zeta(d) \; <0,
\end{equation}
while the corresponding scaling functions for the excess free energy and for the attractive Casimir force are 
\begin{equation}
\label{eq:Gauss_SB_SB_OO}
X_{\rm ex}^{{\rm (SB,SB)}}(x_\tau|d,n)=X_{\rm ex}^{{\rm (O,O)}}(x_\tau|d,n) \quad \mbox{and} \quad X_{\rm Cas}^{{\rm (SB,SB)}}(x_\tau|d,n)=X_{\rm Cas}^{{\rm (O,O)}}(x_\tau|d,n),
\end{equation}
respectively. 

\subsubsection{Mixed or ordinary-special boundary conditions}
The corresponding Casimir amplitude is 
\begin{equation}
\label{eq:Gaus_O,SB_Cas}
\Delta_{\rm Cas}^{{\rm (O,SB)}}(d,n)= n (1-2^{-d+1}) 2^{-d} \pi^{-d/2} \Gamma(d/2)\zeta(d)\; >0
\end{equation}
while the corresponding scaling function for the excess free energy is 
\begin{equation}
\label{eq:Gaus_O_SB_ex_sf}
X_{\rm ex}^{{\rm (O,SB)}}(x_\tau|d,n)=n \;2^{-d} \pi^{-d/2} \dfrac{2\sqrt{\pi} }{\Gamma\left(\frac{d+1}{2}\right)}x_\tau^{\,d/2}\int_{1}^{\infty} dx \dfrac{(x^2-1)^{(d-1)/2}}{\exp\left(2x \sqrt{x_\tau}\right)+1}
\end{equation}
so that 
\begin{equation}
\label{eq:excess_o_sb__equiv}
X_{\rm ex}^{{\rm (O,SB)}}(x_\tau|d,n)=n\; 2^{1-d/2} \pi ^{-d/2} x_\tau^{d/4} \sum _{k=1}^{\infty }(-1)^{k+1}
\frac{K_{d/2}\left(k \sqrt{x_\tau}\right)}{k^{d/2}}.
\end{equation}
The asymptotic behavior for $x_\tau\to\infty$ is
\begin{equation}
\label{eq:excess_o_sb_as}
X_{\rm ex}^{{\rm (O,SB)}}(x_\tau|d,n) \simeq n \; 2^{-d}\pi ^{-(d-1)/2} x_\tau^{ (d-1)/4} e^{-2\sqrt{x_\tau}}, \qquad x_\tau\to\infty.
\end{equation}
From \eq{eq:excess_o_sb__equiv}  one obtains for the excess free energy scaling function in $d=3$
\begin{equation}
\label{eq:excess_o_sb_equiv_d3}
X_{\rm ex}^{{\rm (O,SB)}}(x_\tau|d=3,n)=-\frac{n}{16 \pi } \left[\text{Li}_3\left(-e^{-2 \sqrt{x_\tau}}\right)+ 2\sqrt{x_\tau}\,
\text{Li}_2\left(-e^{- 2\sqrt{x_\tau}}\right)\right]
\end{equation}
 and 
\begin{equation}
\label{eq:Cas_o_sb_equiv_d3}
X_{\rm Cas}^{{\rm (O,SB)}}(x_\tau|d=3,n)=-\frac{n}{8\pi} \left[\text{Li}_3\left(-e^{-2\sqrt{x_\tau}}\right)+ 2\sqrt{x_\tau}\,
\text{Li}_2\left(-e^{-2 \sqrt{x_\tau}}\right)-2 x_\tau \ln \left(1+e^{-2
	\sqrt{x_\tau}}\right)\right] \; >0
\end{equation}
for the scaling function of the repulsive Casimir force.

\subsubsection{Sinusoidal surface fields}
\label{sinusoidal-surface-fields}

Here we consider a Gaussian model in which the boundary conditions are represented by surface fields which can be considered to form a wave-like pattern on the bounding surfaces. The corresponding waves can exhibit a phase shift relative to each other.  

 Following Ref. \cite{DaR2017}, we discuss a discrete  Gaussian model, which consists of $L\in \mathbb{N}$ two-dimensional layers with the Hamiltonian 
	\begin{eqnarray}
	\label{eq:def_Ham_GM}
	-\beta \mathcal{H}&=&\sum _{x=1}^M \sum _{y=1}^N \Bigg\{K^\| \sum _{z=1}^L S_{x,y,z} \left(S_{x+1,y,z}+S_{x,y+1,z}\right)+K^\perp\sum _{z=1}^{L-1} 
	S_{x,y,z} S_{x,y,z+1}+h_1 S_{x,y,1} \cos  \left(k_x x+k_y y\right) \nonumber\\
	&&  +h_L S_{x,y,L} \cos \left(k_x \left(x+\Delta _x\right)+k_y \left(y+\Delta
	_y\right)\right)-s \sum_{z=1}^L S_{x,y,z}^2\Bigg\}, \quad \mbox{where} \quad S_{x,y,z}\in \mathbb{R}.
	\end{eqnarray}
The Hamiltonian corresponds to a system with nearest-neighbor interactions $K^\|=\beta J^\|$ in lateral ($\|$) and $K^\perp=\beta J^\perp$ in orthogonal ($\perp$) directions with chemically modulated bounding surfaces located at $z=1$ and $z=L$.  Here, $h_1=\beta H_1$ and $h_L=\beta H_L$ are the external fields acting only on the boundaries of the system. In the example considered above, the modulation depends simultaneously on the lateral coordinates $x$ and $y$  specified by the applied surface fields $h_1\cos\left(k_x x+k_y y\right)\equiv h_1 \cos({\bf k}\cdot{\bf r})$ and $h_L \cos [k_x \left(x+\Delta_x\right)+k_y \left(y+\Delta_y\right)]\equiv h_L\cos({\bf k} \cdot ({\bf r}+{\bf \Delta}))$. The phases of surface fields are shifted with respect to each other by $\Delta_x$ in $x$ direction and by $\Delta_y$ in $y$ direction. Here, the two-component vectors are ${\bf r}=(x,y)$, ${\bf k}=(k_x,k_y)$, and ${\bf \Delta}=(\Delta_x,\Delta_y)$. Periodic boundary conditions are applied along the $x$ and $y$ axes. In the $z$ direction the model exhibits missing neighbors (Dirichlet) boundary conditions with surface fields present. These boundary conditions are formulated as follows:
\begin{equation}
\label{bc_Gaussian-model}
S_{1,y,z}=S_{M+1,y,z}, \qquad S_{x,1,z}=S_{x,N+1,z} \quad \mbox{and} \quad 
S_{x,y,0}=0, \qquad S_{x,y,L+1}=0.
\end{equation}
Given these boundary conditions, the Hamiltonian  in Eq. (\ref{eq:def_Ham_GM}) can be rewritten as 
\begin{equation}
	\label{eq:def_Ham_GM_final}
	-\beta \mathcal{H}=\sum _{x=1}^M \sum _{y=1}^N \sum _{z=1}^L S_{x,y,z} \Bigg\{K^\| \left(S_{x+1, y, z}+S_{x,y+1,z}\right) + K^\perp
	S_{x,y,z+1} +\delta_{1, \,z} h_1 \cos  \left[{\bf k}\cdot{\bf r} \right] + \delta_{L,\, z} h_L \cos \left[{\bf k}\cdot({\bf r}+{\bf \Delta}) \right]-s\; S_{x,y,z}\Bigg\}.
\end{equation}
Since we shall consider the case $M,N\gg 1$,  we can choose the wave vector components $k_x$ and $k_y$ to coincide with $(2\pi p)/M $ and $(2\pi q)/N$ with  $p\in \{1,\cdots, M\}$ and $q\in \{1,\cdots,N\}$, respectively. Taking the limits $M,N\to \infty$ one keeps $k_x$ and $k_y$ fixed, which formally corresponds to fixed  $p/M$ and $q/N$. 
The parameter $s>0$ on the right hand side of \eq{eq:def_Ham_GM_final} ensures the occurrence of a nonzero inverse critical temperature\footnote{The Hamiltonian in  \eq{eq:def_Ham_GM_final}  can be diagonalized in a standard way via Fourier transformation (for details see Appendix C in Ref. \cite{DaR2017}). 	Mathematically, $s>0$ ensures that the coefficients in front of the quadratic terms in the Fourier transform of the Hamiltonian are negative for $\beta < \beta_c$; the model is well defined only for $\beta < \beta_c$. The partition function does not exist for  $\beta > \beta_c$.}  $\beta_c$ of the system; in the bulk models satisfies $2K^{\|}+K^{\perp}-s \equiv \beta (2J^{\|}+J^{\perp})-s=0$, i.e., one has
\begin{equation}
	\beta_c=s/(2J^{\|}+J^{\perp}).
	\label{betacGM}
\end{equation}

Since the boundary fields at the top and at the bottom of the system are shifted with respect to each other, the Casimir force, which acts on the bounding planes at $z=1$ and $z=L$, has both an orthogonal ($\beta F^{(\perp)}_{\rm Cas}$) component and lateral ones ($\beta F^{(\|,\alpha)}_{\rm Cas}$, $\alpha=x$ or $\alpha=y$).  They can be written in the form 
\begin{equation}
\label{eq:gen_force}
\beta F^{(\cdots)}_{\rm Cas}=L^{-3}\left(\frac{J^\perp}{J^\|}\right) X^{(\cdots)}_{\rm Cas}(x_\tau,x_k,x_1,x_L),
\end{equation}
where the ellipses $(\cdots)$ stand either for $\perp$ or $(\|,\alpha)$, with $\alpha=x$ or $\alpha=y$. Here,
\begin{equation}
\label{eq:field_scaling_def}
x_{1}=\sqrt{L K^\|}\frac{h_1}{K^\perp}, \quad \mbox{and}\qquad x_{L}=\sqrt{L K^\|}\frac{h_L}{K^\perp}
\end{equation}
are the scaling  variables associated with the strengths of the surface fields, $x_\tau$  is the temperature scaling variable 
\begin{equation}
\label{eq:xt_and_xk}
x_\tau=L\sqrt{2\left(\frac{\beta_c}{\beta}-1\right)\left[ 2\frac{J^{\|}}{J^\perp}+1\right]}, \quad \mbox{and} \quad  x_k=\sqrt{\dfrac{J^{\|}}{J^{\perp}}}\; L k, 
\end{equation}
with $k=\sqrt{k_x^2+k_y^2}$, is the scaling variable related to the wave vector of the surface modulation. 
The  partition function of the bulk system exists only for $\beta<\beta_c$, i.e., $T>T_c$. The partition function of the finite system exists under the less demanding constraint
\begin{equation}
\label{eq:T_finite}
\left(\beta_c/\beta-1\right)\left[ 2J^{\|}+J^\perp\right] +J^{\perp} \left[1-  \cos\left(\frac{\pi 
}{L+1}\right)\right]>0.
\end{equation}
Concerning the Casimir force,  the properties of both the bulk  and of the finite system enter.  Thus, we will always assume $\beta<\beta_c$ in all expressions concerning the Casimir force within the Gaussian model. 

If $h_1={\cal O}(1)$ and $h_L={\cal O}(1)$, the Casimir force  $F^{(\cdots)}_{\rm Cas}$ contains a field dependent contribution which provides the {\it leading} contribution to the force being of the order of $L^{-2}$. Of course, there is a longitudinal ($\|$) component of the force only, if the amplitudes of neither of the two surface fields are zero. This implies that while the transverse component ($\perp$) of the force $F^{(\perp)}_{\rm Cas}$ has both field-dependent ($\Delta F^{(h,\perp)}_{\rm Cas}$) and field-independent contributions  ($\Delta F^{(0,\perp)}_{\rm Cas}$), i.e.,
\begin{equation}
\label{eq:ort_force}
F^{(\perp)}_{\rm Cas}\equiv \Delta F^{(0,\perp)}_{\rm Cas}+\Delta F^{(h,\perp)}_{\rm Cas},
\end{equation}
the longitudinal force ($\|$) contains only a  field-dependent contribution $\Delta F^{(h,\alpha)}_{\rm Cas}$. (The superscript $h$ stands for the pair $(h_1,h_L)$); $\Delta F^{(h,\perp)}_{\rm Cas}$ vanishes in the limits $h_1, h_L \to 0$.

{\it $\bullet$  transverse Casimir force}

\begin{figure}[h]
	\includegraphics[width=\columnwidth]{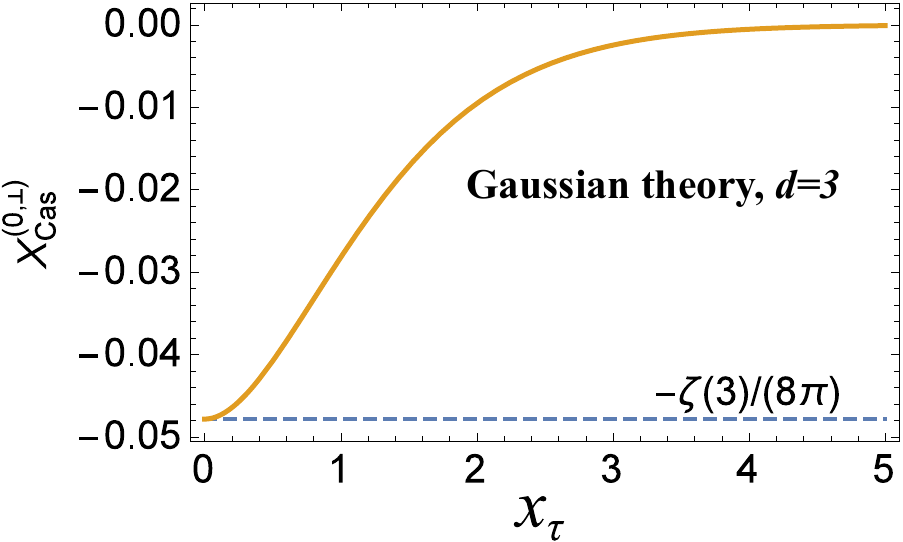}
	\caption{Gaussian scaling function $X^{(0,\perp)}_{\rm Cas}(x_\tau)$ of the field-independent contribution to the transverse Casimir force (see Eqs. \eqref{eq:Cas_no_field} and  \eqref{eq:X_Cas_no_field_sf}) as function of the temperature scaling variable $x_\tau$ (see \eq{eq:xt_and_xk}). The horizontal dashed line marks the Casimir amplitude $X^{(0,\perp)}_{\rm Cas}(0)=-\zeta(3)/(8\pi)$.}
	\label{Fig:3d_G_X_Cas_Zero_Field}
\end{figure}

 Concerning the  contribution of the field-independent term $\beta\Delta F^{(0,\perp)}_{\rm Cas}$ to the total transverse Casimir force, one finds 
 \begin{equation}
 	\label{eq:Cas_no_field}
 	\beta\Delta F^{(0,\perp)}_{\rm Cas}=-\frac{1}{2}\int _{-\pi }^{\pi
 	}\int _{-\pi }^{\pi }\frac{d\theta
 	_1d\theta _2}{ (2 \pi )^2}\;\delta
 	\left\{\coth [(1+L)\, \delta]-1\right\} ,
 \end{equation}
 where $\delta=\delta\left(\theta_1,\theta_2\big|\,\beta_c/\beta,J^\|/J^\perp\right)$ is defined by the expression
 \begin{equation}
 	\label{eq:def_delta}
 	\cosh\delta = 1+\left(\frac{\beta_c}{\beta}-1\right)\left(1+ 2\frac{J^{\|}}{J^\perp}\right) +\frac{J^\|}{J^\perp}\left(2-\cos \theta_1-\cos \theta_2\right).
 \end{equation}
 Within the present discrete Gaussian model the result in \eq{eq:Cas_no_field} is an {\it exact} expression for $\beta\Delta F^{(0,\perp)}_{\rm Cas}$. Since $\coth(x)>1$ for $x>0$ one immediately concludes that $\Delta F^{(0,\perp)}_{\rm Cas}<0$, i.e., it is an {\it attractive} force, for {\it all} values of $L$. In order to obtain scaling and, thus, the scaling form of $\Delta F^{(0,\perp)}_{\rm Cas}$. One has to consider the regime $L\gg 1$. 
For the scaling behavior of the force one finds 
\begin{equation}
\label{eq:F_Cas_no_field}
\beta\Delta F^{(0,\perp)}_{\rm Cas}=L^{-3}\left(\frac{J^\perp}{J^\|}\right) X^{(0,\perp)}_{\rm Cas}(x_\tau)
\end{equation}
where $X^{(0,\perp)}_{\rm Cas}(x_\tau)$ is the universal scaling function 
\begin{eqnarray}
\label{eq:X_Cas_no_field_sf}
X^{(0,\perp)}_{\rm Cas}(x_\tau)=-\frac{1}{8 \pi}\Bigg\{
\text{Li}_3\left(e^{-2
	x_\tau}\right)+2 x_\tau
\text{Li}_2\left(e^{-2
	x_\tau}\right) -2 x_\tau^2 \ln \left(1-e^{-2
	x_\tau}\right)\Bigg\};
\end{eqnarray}
the scaling variable $x_\tau$ is given by \eq{eq:xt_and_xk}.
$X^{(0,\perp)}_{\rm Cas}(x_\tau)$ is a monotonically increasing function of $x_\tau$. The behavior of $X^{(0,\perp)}_{\rm Cas}(x_\tau)$ is shown in Fig. 
\ref{Fig:3d_G_X_Cas_Zero_Field}. One can infer from this figure that this scaling function coincides, upon taking into account the difference in notations, with that one given in \eq{eq:Cas_DD_equiv_d3} for the three-dimensional Gaussian model with Dirichlet-Dirichlet boundary conditions. 

At the critical point, i.e., at $x_\tau=0$, one obtains the well known Casimir amplitude for the Gaussian model with Dirichlet-Dirichlet boundary conditions:
\begin{equation}
\label{eq:F_Cas_no_field_ampl}
X^{(0,\perp)}_{\rm Cas}(x_\tau=0)=-\frac{\zeta(3)}{8 \pi}.
\end{equation}

For $ x_\tau\gg 1$ and $x_\tau\to 0$ one finds the following asymptotic behaviors: 
\begin{equation}
\label{eq:asX0_Cas}
X^{(0,\perp)}_{\rm Cas}\simeq 
\left\{\begin{array}{lcr}
-\frac{1}{8\pi} \exp(-2 x_\tau) \left[1+2 x_\tau
\left(1+x_\tau\right)\right],& x_\tau\gg 1&\\
&& \\
-\frac{1}{8 \pi
}\zeta (3)+\frac{1}{48 \pi} x_\tau^2 \left(6-4
x_\tau+x_\tau^2\right),& x_\tau\to 0.
\end{array} \right.
\end{equation}

The field induced ($h \ne 0$) contribution to the transverse Casimir force is as follows.  

{\it (i)} If $0<k_x<2\pi$ or $0<k_y< 2\pi$:
\begin{eqnarray}
\label{eq:Casimir_tr_pq}
\lefteqn{\beta \Delta F^{(h,\perp)}_{\rm Cas}=\frac{\lambda\sinh (\lambda)}{32 K^{\perp}}}\\
&&\times\left\{  \left[h_1^2+h_L^2-2 h_L h_1
\cos(\mathbf{k\cdot\Delta})\right]^2 \text{csch}^2\left(\frac{1+L}{2} \lambda\right)-\left[h_1^2+h_L^2+2 h_L h_1
\cos(\mathbf{k\cdot\Delta})\right]^2 \text{sech}^2\left(\frac{1+L}{2} 
\lambda\right)\right\},
\nonumber
\end{eqnarray}
where $\mathbf{k}=(k_x,k_y)$ and  $\mathbf{\Delta}=(\Delta_x,\Delta_y)$. 

{\it (ii)} If $k_x=2\pi$ and $k_y=2\pi$ 
\begin{eqnarray}
\label{eq:Casimir_tr}
\lefteqn{ \beta \Delta F^{(h,\perp)}_{\rm Cas}=\frac{\lambda\sinh (\lambda)}{32 K^{\perp}} }\\
&&\times\left\{  \left[h_1-h_L
\cos \left(2 \pi  ( \Delta
_x+ \Delta
_y)\right)\right]^2 \text{csch}^2\left[\frac{1+L}{2} \lambda\right]-\left[h_1+h_L
\cos \left(2 \pi  ( \Delta
_x+ \Delta
_y)\right)\right]^2 \text{sech}^2\left[\frac{1+L}{2} 
\lambda\right]\right\}. \nonumber
\end{eqnarray}
In Eqs. \reff{eq:Casimir_tr_pq} and \reff{eq:Casimir_tr} $\lambda$ is defined so that $\lambda={\rm arcosh} (\Lambda)$ for $\Lambda\ge 1$, and $\lambda= \arccos(\Lambda)$ for $\Lambda\le 1$ with 
\begin{equation}
	\label{eq:Lambda_def}
	\Lambda = 1+\left(\frac{\beta_c}{\beta}-1\right)\left[ 2\frac{J^{\|}}{J^\perp}+1\right]+2\dfrac{J^{\|}}{J^{\perp}}\left[\sin^2 \frac{k_x}{2} +\sin^2 \frac{k_y}{2}\right].
\end{equation}
 We note that
\begin{itemize}
	\item[-]
	if $h_1={\cal O}(1)$, $h_L={\cal O}(1)$, and 
	\begin{equation}
	\label{eq:w_def}
	w=L\lambda/2
	\end{equation}
	is such that $w={\cal O}(1)$ too, the Casimir force is of the order of ${\cal O}(L^{-2})$, despite the fact that the system is at a temperature {\it above} the bulk critical temperature; 
	
	\item[-] if $h_1$ and $h_L$ are such that  the field-dependent scaling  variables $x_1={\cal O}(1)$ and $x_L={\cal O}(1)$ (see \eq{eq:field_scaling_def}), the Casimir force $\beta\Delta F^{(h,\perp)}_{\rm Cas}$ reads in terms of $w$ as
	\begin{equation}
	\label{eq:sc_funct_field}
	\beta\Delta F^{(h,\perp)}_{\rm Cas}=L^{-3}\left(\frac{J^\perp}{J^\|}\right) X^{(h,\perp)}_{\rm Cas}(w,x_1,x_L),
	\end{equation}
	where the scaling function $X^{(h,\perp)}_{\rm Cas}(w,x_1,x_L)$ is given by,
	
	{\it (i)} if $0<k_x<2\pi$ or $0<k_y< 2\pi$,
	\begin{eqnarray}
	\label{eq:scaling_h_pq}
	\lefteqn{X^{(h,\perp)}_{\rm Cas}(w,x_1,x_L)=\frac{1}{8} w^2}\\
	&& \times \left\{[x_1^2+x_L^2-2x_1x_L \cos\left(\mathbf{k\cdot\Delta}\right)] \text{csch}^2 w -  [x_1^2+x_L^2+2x_1 x_L \cos\left(\mathbf{k\cdot\Delta}\right)] \text{sech}^2 w \right\}, \nonumber
	\end{eqnarray}
	and,
	
	{\it (ii)} if $k_x=2\pi$ and $k_y=2\pi$,
	\begin{eqnarray}
	\label{eq:scaling_h}
	\lefteqn{X^{(h,\perp)}_{\rm Cas}(w,x_1,x_L) = \frac{1}{8} w^2}\\
	&& \times \left\{[x_1-x_L \cos 2 \pi  ( \Delta
	_x+ \Delta
	_y)]^2 \text{csch}^2 w -  [x_1+x_L \cos 2 \pi  ( \Delta
	_x+ \Delta
	_y)]^2 \text{sech}^2 w \right\}. \nonumber
	\end{eqnarray}
	The latter expressions imply  that in the regime considered here ($x_1={\cal O}(1), x_L={\cal O}(1), w={\cal O}(1)$) the field-dependent part of the force is of the order of $L^{-3}$, as it is the case for the  field-independent part of it. 
\end{itemize}

The asymptotic behavior of $\Delta F^{(h,\perp)}_{\rm Cas}$ for $w\gg 1$ is also known:
\begin{eqnarray}
\label{eq:scaling_h_as}
\beta\Delta F^{(h,\perp)}_{\rm Cas}{ _{\big|_{\mbox{\small{$w \gg 1$}}}}}\simeq -\dfrac{2w^2}{K^{\perp}L^2} e^{-2w}h_1 h_L 
\cos\left(\mathbf{k\cdot\Delta}\right).
\end{eqnarray}
This implies that in this limit the transverse component of the force is exponentially small as a function of $L$, and attractive  {\it or } repulsive depending on the products $h_1 h_L \cos[\mathbf{k\cdot\Delta}]$.  

\begin{figure}[!h]
	\begin{floatrow}
		\ffigbox{\includegraphics[scale = 0.5]{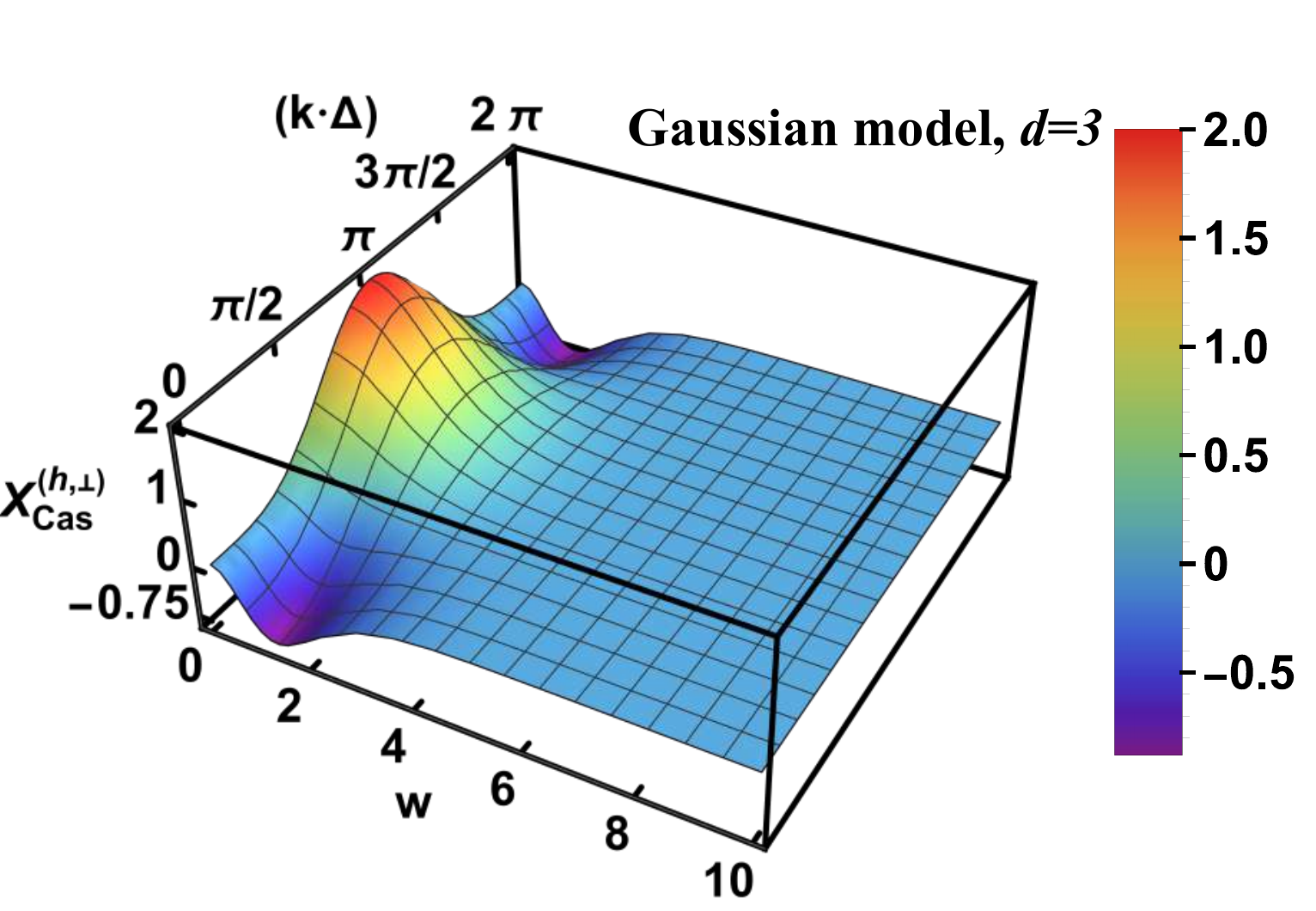}}{\caption{The scaling function $X^{(h,\perp)}_{\rm Cas}(w,x_1,x_L)$ (see Eqs. \eqref{eq:ort_force}, 	\eqref{eq:sc_funct_field}, and  \eqref{eq:scaling_h_pq}) of the field-dependent contribution to the transverse Casimir force of the Gaussian model as a function of $w\in (0,10]$  (see \eq{eq:w_def}) and $\mathbf{k\cdot \Delta}\in [0,2\pi]$ for $x_1=x_L=1$.  $X^{(h,\perp)}_{\rm Cas}(w,x_1,x_L)$ can be positive or negative. }\label{Fig:3D_G_h1_eq_hL_Legend}}
		\ffigbox{\includegraphics[scale = 0.5]{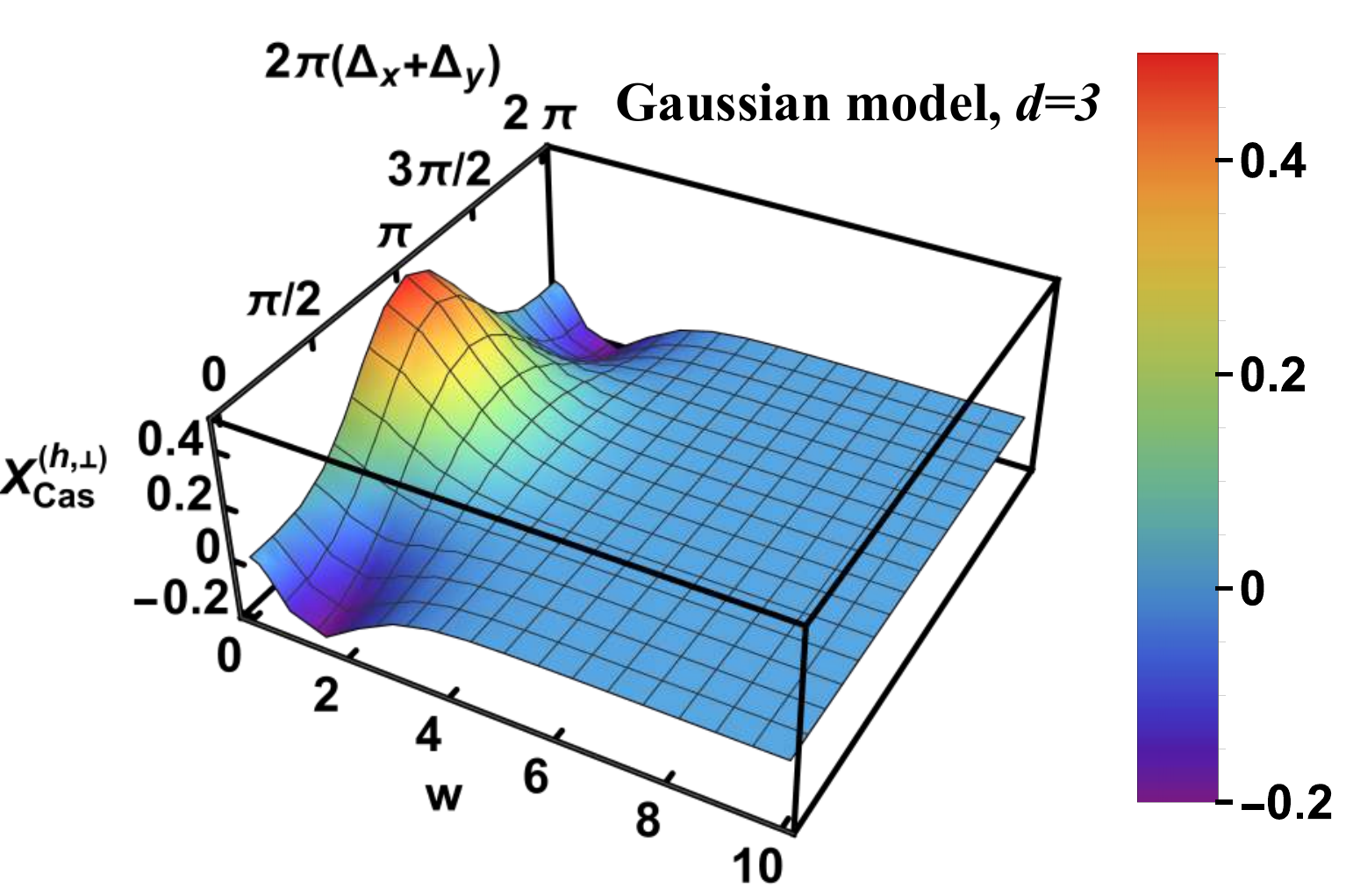}}{\caption{The scaling function $X^{(h,\perp)}_{\rm Cas}(w,x_1,x_L)$ (see \eq{eq:scaling_h}) as a function of $w\in(0,10]$ and $\Delta_x+ \Delta_y\in [0,1]$ for $x_1=x_L=1$. Also in this case  $X^{(h,\perp)}_{\rm Cas}(w,x_1,x_L)$ can be positive or negative.  The comparison between Figs. \ref{Fig:3D_G_h1_eq_hL_Legend} and \ref{Fig:3D_G_h1_eq_hL_Legend_MN} shows that the maximal values of these scaling functions are smaller in the present figure.}\label{Fig:3D_G_h1_eq_hL_Legend_MN}}
	\end{floatrow}
\end{figure}

The behavior of the scaling function $X^{(h,\perp)}_{\rm Cas}(w,x_1,x_L)$ is shown in Fig. \ref{Fig:3D_G_h1_eq_hL_Legend}  {\it (i)} if $0<k_x<2\pi$ or $0<k_y< 2\pi$ , and in Fig. \ref{Fig:3D_G_h1_eq_hL_Legend_MN}
 {\it (ii)}, if $k_x=2\pi$ and $k_y=2\pi$. 
The comparison between  Figs. \ref{Fig:3D_G_h1_eq_hL_Legend} and  \ref{Fig:3D_G_h1_eq_hL_Legend_MN} tells, that in Fig. \ref{Fig:3D_G_h1_eq_hL_Legend_MN} the maximal values of the function $X^{(h,\perp)}_{\rm Cas}(w,x_1,x_L)$  are  smaller than in Fig. \ref{Fig:3D_G_h1_eq_hL_Legend} . 

\begin{figure}[!th]
	\begin{floatrow}
		\ffigbox{\includegraphics[scale = 0.35]{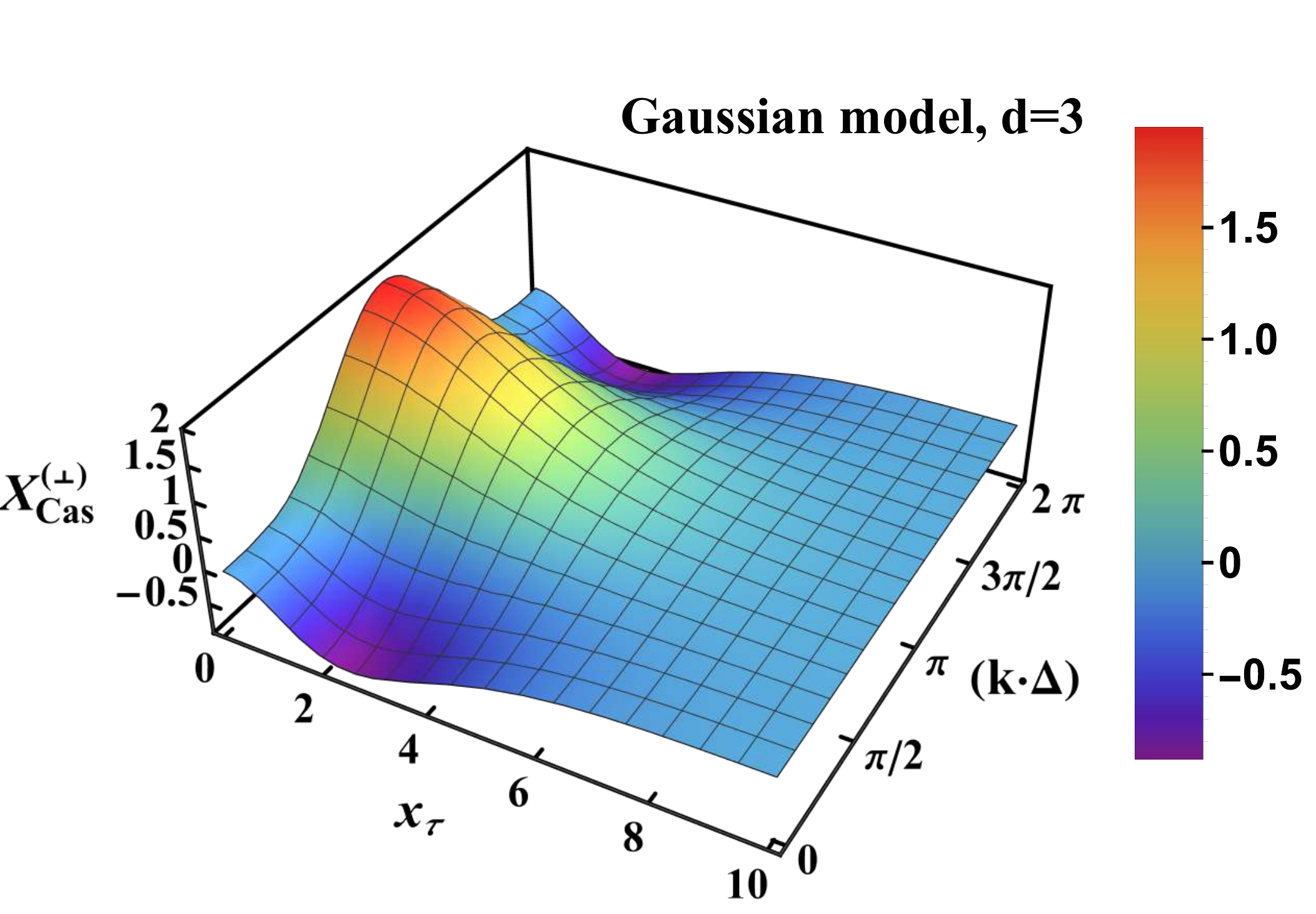}}{\caption{The scaling function 
				$X^{(\perp)}_{\rm Cas}(x_\tau,x_k,x_1,x_L)$ of the total transverse Casimir force (see Eqs. \eqref{eq:gen_force} and  \eqref{eq:ort_force})
				as a function of $x_\tau\in(0,10]$ and
			 $\mathbf{k\cdot\Delta}\in [0,2\pi]$ for $x_k=0.1$ and $x_1=x_L=1$. $X^{(\perp)}_{\rm Cas}$ can be positive or negative. }	\label{Fig:3D_G_h1_eq_hL_Legend_total_force}}
		\ffigbox{\includegraphics[scale = 0.325]{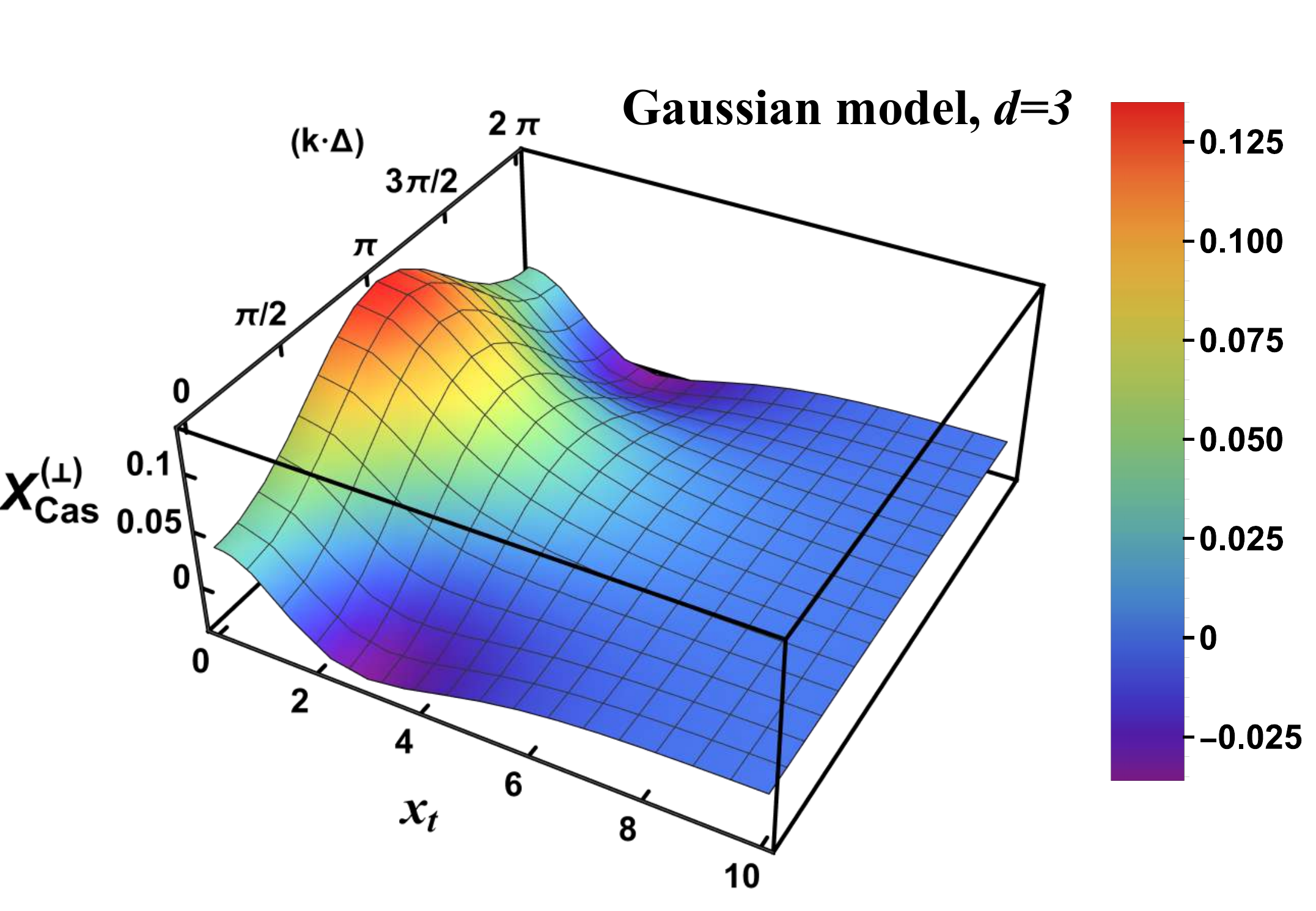}}{\caption{The same as Fig. \ref{Fig:3D_G_h1_eq_hL_Legend_total_force} but  $x_L=0.1$ so that $x_1=10 x_L$.  The scaling function is predominantly positive, i.e., repulsive. }\label{Fig:3D_G_h1_not_eq_hL_Legend_total_force_1}}
			\end{floatrow}
		\begin{floatrow}
			\ffigbox{\includegraphics[scale = 0.475]{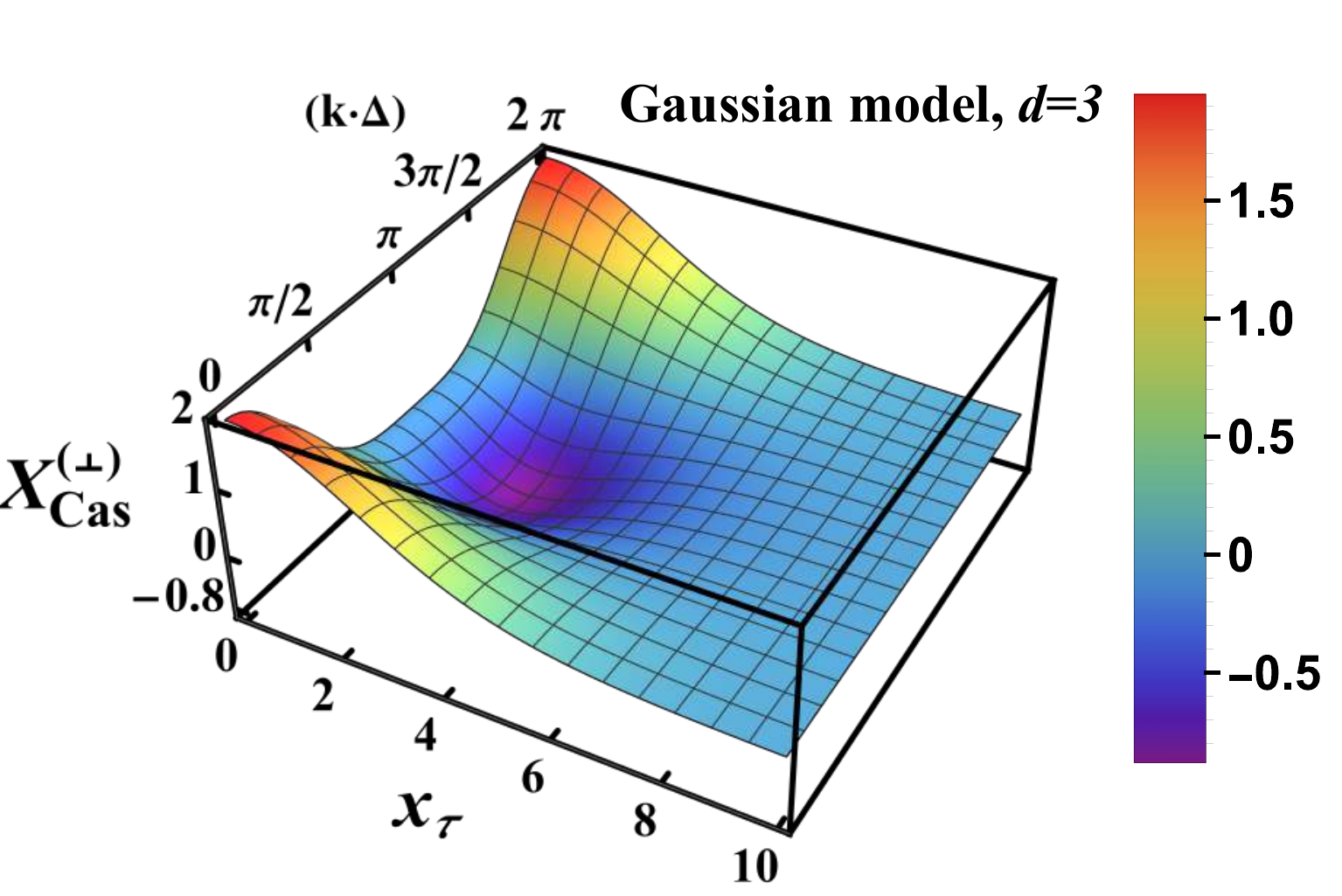}}{\caption{The scaling function 
					$X^{(\perp)}_{\rm Cas}(x_\tau,x_k,x_1,x_L)$
					as a function of  $x_\tau\in(0,10]$ and $\mathbf{k\cdot\Delta}\in [0,2\pi]$ for $x_k=0.1$, $ x_1=-x_L=1$. The scaling function can be positive or negative.}\label{Fig:3D_G_h1_not_eq_hL_Legend_total_force_2}}
		\ffigbox{\includegraphics[scale = 0.37]{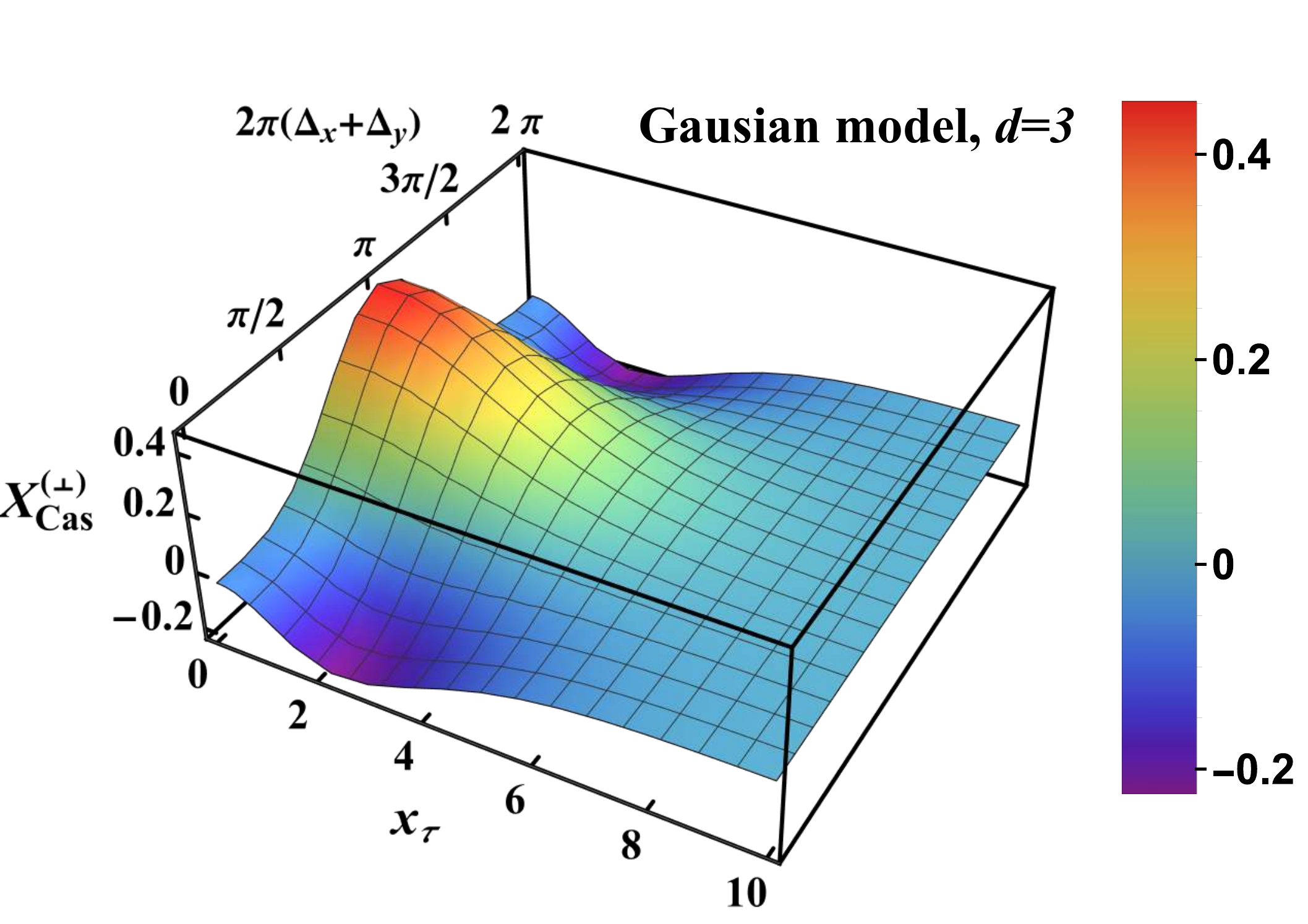}}{\caption{The scaling function 
				$X^{(\perp)}_{\rm Cas}(x_\tau,x_k=0,x_1,x_L)$
				as a function of $x_\tau\in (0,10]$ and $(\Delta_x+\Delta_y)\in [0,1]$ for $x_1=x_L=1$. $X^{(\perp)}_{\rm Cas}$ can be positive or negative. }\label{Fig:3D_G_h1_eq_hL_Legend_total_force_MN}}
	\end{floatrow}
\end{figure}

\begin{figure}[!h]
	\begin{floatrow}
		\ffigbox{\includegraphics[scale = 0.5]{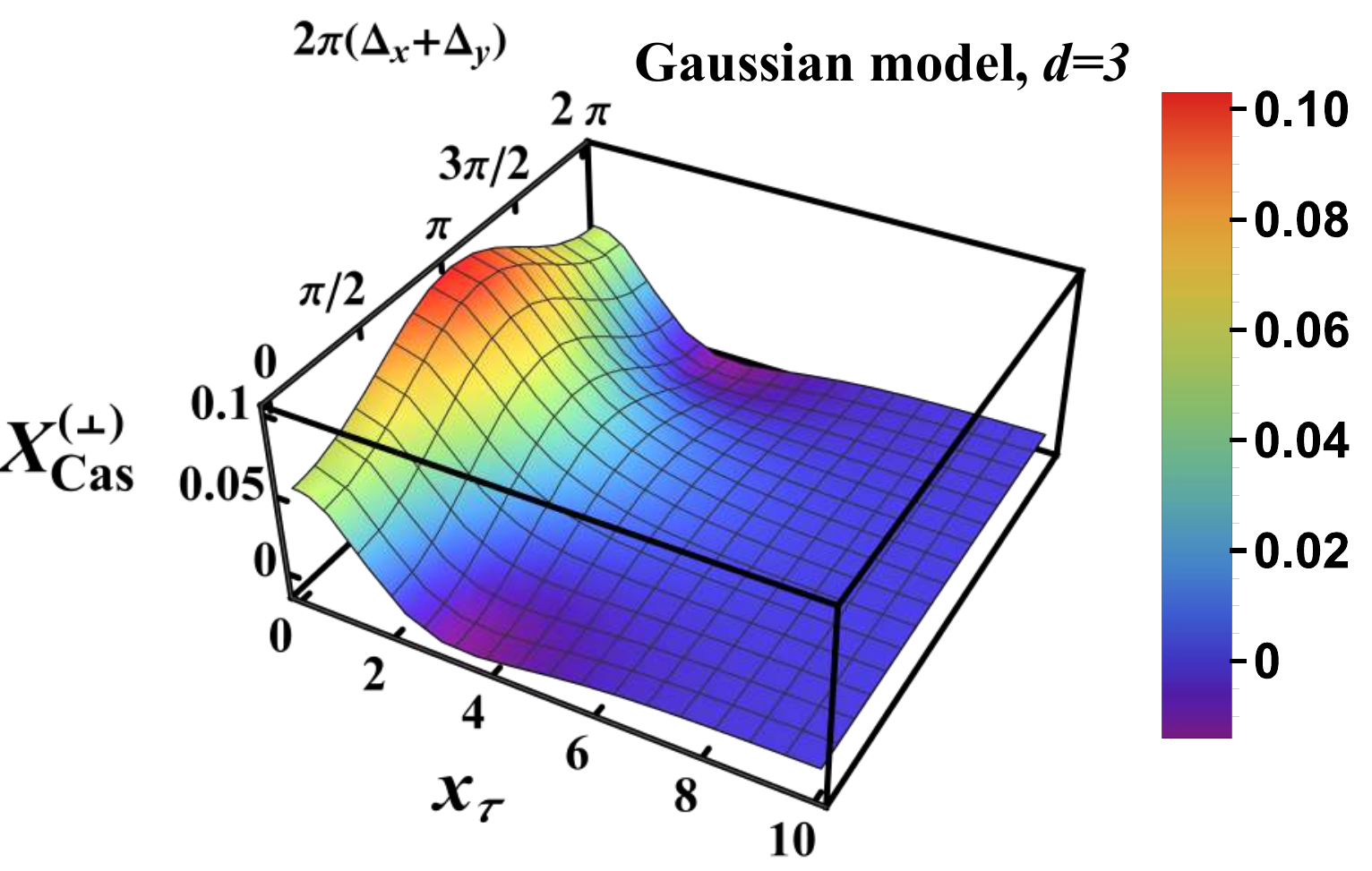}}{\caption{The scaling function 
				$X^{(\perp)}_{\rm Cas}(x_\tau,x_k=0,x_1,x_L)$
				as a function of $x_\tau \in (0,10]$ and $\Delta_x+\Delta_y\in [0,1]$  for $x_1=10 x_L=1$. The scaling function is predominantly positive.}\label{Fig:3D_G_h1_not_eq_hL_Legend_total_force_1_MN}}
		\ffigbox{\includegraphics[scale = 0.48]{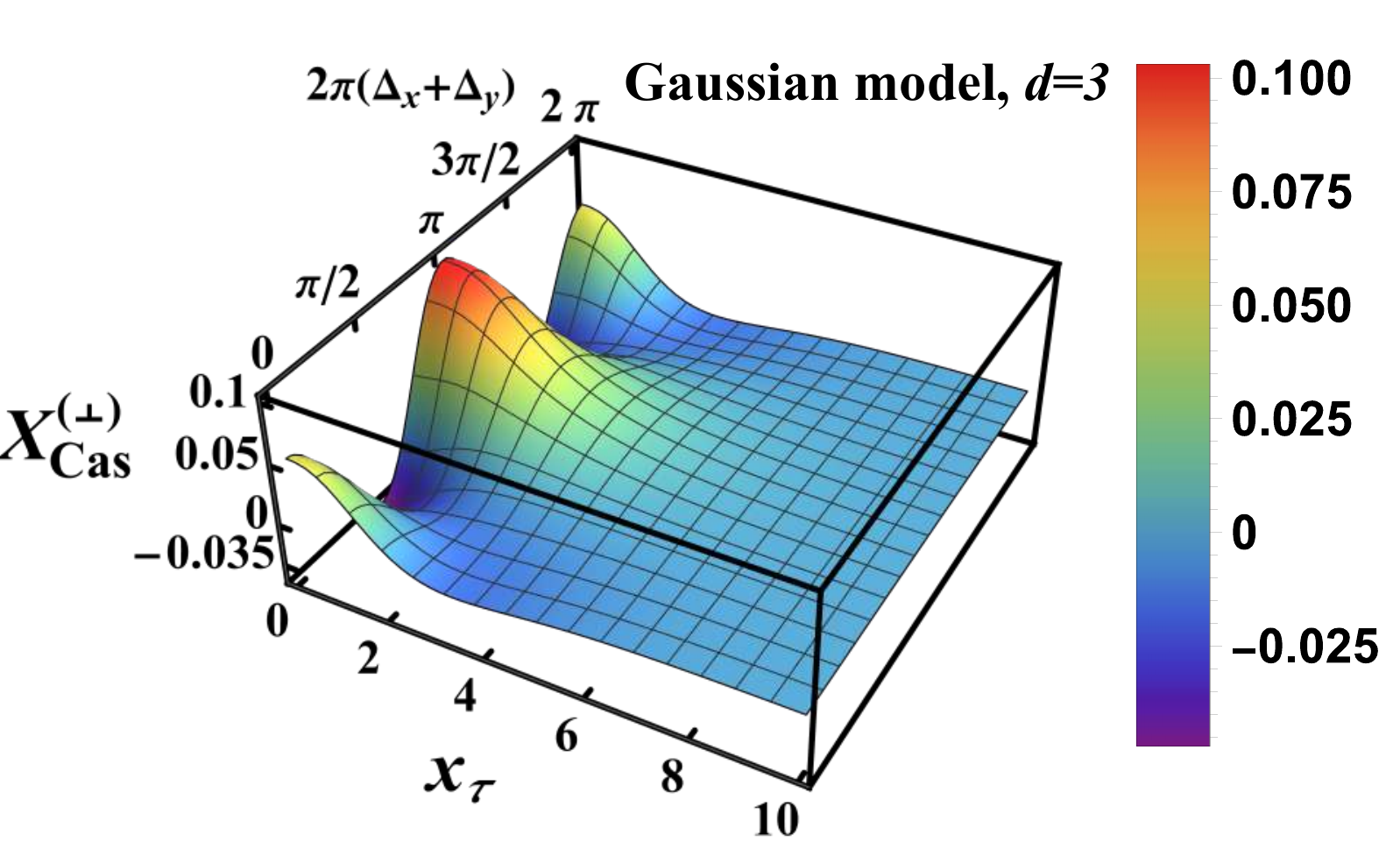}}{\caption{The scaling function 
				$X^{(\perp)}_{\rm Cas}(x_\tau,x_k=0,x_1,x_L)$
				as a function of $x_\tau\in (0,10]$ and $\Delta_x+\Delta_y\in [0,1]$  for $10 x_1= x_L=1$. The scaling function can be positive or negative. }\label{Fig:3D_G_h1_not_eq_hL_Legend_total_force_11_MN}}
	\end{floatrow}
\end{figure}

\begin{figure}[h]
	\includegraphics[width=\columnwidth]{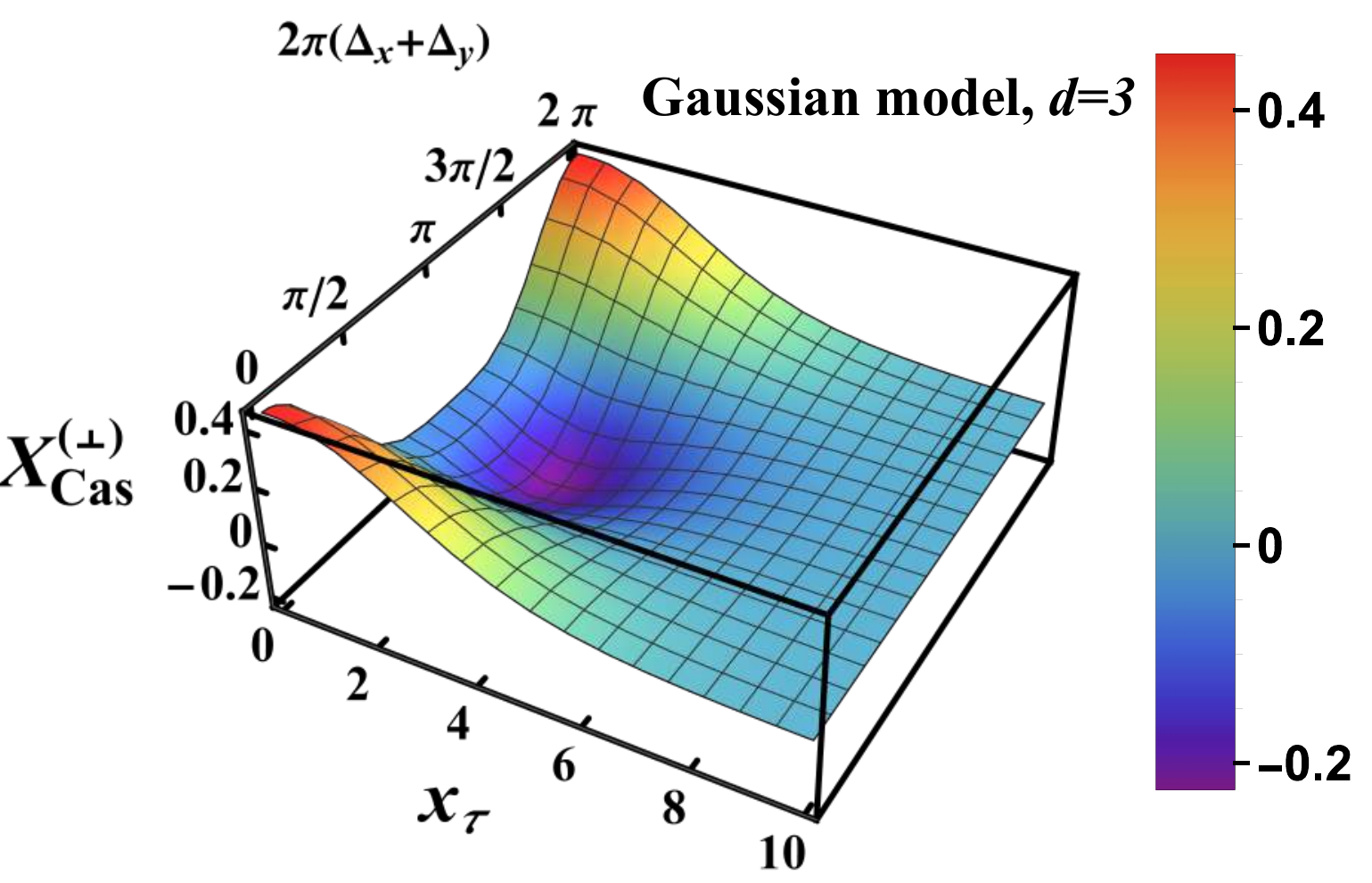}
	\caption{The scaling function 
		$X^{(\perp)}_{\rm Cas}(x_\tau,x_k=0,x_1,x_L)$ (see Eqs. \eqref{eq:ort_force}, \eqref{eq:X_Cas_no_field_sf}, and  \eqref{eq:Casimir_tr})
		as function of $x_\tau\in (0,10]$ and $\Delta_x+\Delta_y\in [0,1]$  for $x_1= -x_L=1$ or $x_1=-x_L=-1$. The scaling function can be positive and negative.}
	\label{Fig:3D_G_h1_not_eq_hL_Legend_total_force_2_MN}
\end{figure}

We now turn  to the behavior of the \textit{total} transverse Casimir force $F^{(\perp)}_{\rm Cas}$ (\eq{eq:gen_force} and \eq{eq:ort_force}). 
The behavior of its corresponding scaling function  $X^{(\perp)}_{\rm Cas}(x_t,x_k,x_1,x_L)$ is depicted in Figs. \ref{Fig:3D_G_h1_eq_hL_Legend_total_force} - \ref{Fig:3D_G_h1_not_eq_hL_Legend_total_force_2} for the case that {\it (i)} $0<k_x<2\pi$ or $0<k_y< 2\pi$ and in Figs. \ref{Fig:3D_G_h1_eq_hL_Legend_total_force_MN} - \ref{Fig:3D_G_h1_not_eq_hL_Legend_total_force_2_MN} for the case that {\it (ii)} $k_x=2\pi$ and $k_y=2\pi$ with $x_k=0$ (\eq{eq:xt_and_xk}). In case {\it (i)} the function $X^{(\perp)}_{\rm Cas}$ is symmetric about interchanging $x_1$ and $x_L$, while in case {\it (ii)} this is not so. The last property implies that, if $x_1\ne x_L$, in case {\it (ii)} one has to consider separately the sub-case $x_1\gg x_L$ and $x_1 \ll x_L$. 
Figures \ref{Fig:3D_G_h1_eq_hL_Legend_total_force}  and \ref{Fig:3D_G_h1_eq_hL_Legend_total_force_MN} show the behavior of the force for $x_1=x_L$. If $x_1\ne x_L$, this behavior is presented in Figs.  \ref{Fig:3D_G_h1_not_eq_hL_Legend_total_force_1} and \ref{Fig:3D_G_h1_not_eq_hL_Legend_total_force_2} for case {\it (i)} and in Figs. \ref{Fig:3D_G_h1_not_eq_hL_Legend_total_force_1_MN}, \ref{Fig:3D_G_h1_not_eq_hL_Legend_total_force_11_MN}, and \ref{Fig:3D_G_h1_not_eq_hL_Legend_total_force_2_MN} for case {\it (ii)}. The figures  \ref{Fig:3D_G_h1_not_eq_hL_Legend_total_force_1} and \ref{Fig:3D_G_h1_not_eq_hL_Legend_total_force_1_MN} represent the situation $x_1\gg x_L$, i.e., $x_1=10 x_L$, while Figs. \ref{Fig:3D_G_h1_not_eq_hL_Legend_total_force_2} and  \ref{Fig:3D_G_h1_not_eq_hL_Legend_total_force_2_MN} correspond to the case $x_1=-x_L=1$. 

The comparison of these figures with Figs. \ref{Fig:3D_G_h1_eq_hL_Legend} and \ref{Fig:3D_G_h1_eq_hL_Legend_MN} shows, that one might expect from the data presented in Fig. \ref{Fig:3d_G_X_Cas_Zero_Field}, that the  field-independent contribution $X^{(0,\perp)}_{\rm Cas}(x_t)$ to the overall behavior of the force is rather small, at least in the cases presented here.  

{\it $\bullet$ longitudinal Casimir force}

For the  field-dependent contribution $\Delta F^{(h,\alpha)}_{\rm Cas}(L)$ to the longitudinal component of the Casimir force along the $\alpha$ axis, $\alpha=x,y$, one has,

{\it (i)} if $0<k_x<2\pi$ or $0<k_y< 2\pi$,
\begin{equation}
\label{eq:Casimir_longit_pq}
\beta\Delta F^{(h,\alpha)}_{\rm Cas}(L)=-\frac{h_1 h_L }{4 K^{\perp}} k_\alpha \sin (\mathbf{k\cdot\Delta}) \frac{\sinh(\lambda)}{\sinh[\lambda(L+1)]},
\end{equation}
and,

{\it (ii)} if $k_x=2\pi$ and $k_y=2\pi$, 
\begin{equation}
\label{eq:Casimir_longit}
\beta\Delta F^{(h,\alpha)}_{\rm Cas}(L) = -\frac{ \pi \sin [2 \pi (\Delta_x+ \Delta_y)]}{2 K^{\perp}} h_L \times \Bigg\{ h_1 \frac{\sinh(\lambda)}{\sinh[(L+1)\lambda]} + h_L \cos[2 \pi (\Delta_x+ \Delta_y)]  \left[\Lambda-\frac{\sinh(\lambda)}{\tanh [(L+1)\lambda]}\right]\Bigg\}.
\end{equation}

In the limit $L\lambda\gg 1$, the above expressions reduce to, 

{\it (i)} if $0<k_x<2\pi$ or $0<k_y< 2\pi$,
\begin{equation}
\label{eq:scaling_h_as_x_pq}
\beta\Delta F^{(h,\alpha)}_{\rm Cas}(L) \simeq -\dfrac{k_\alpha}{2 K^{\perp}}  \sinh(\lambda) e^{-(L+1)\lambda} h_1 h_L \sin\left(\mathbf{k\cdot\Delta}\right),
\end{equation}
and, 

{\it (ii)} if $k_x=2\pi$ and $k_y=2\pi$, 
\begin{equation}
\label{eq:scaling_h_as_x}
\beta\Delta F^{(h,\alpha)}_{\rm Cas}(L) \simeq -\frac{ \pi h_L^2}{4 K^{\perp}} \sin [4 \pi (\Delta_x+ \Delta_y)] \left\lbrace \Lambda-\sinh[\lambda)\right\rbrace-\dfrac{\pi}{K^{\perp}}  \sinh(\lambda) e^{-(L+1)\lambda} h_1 h_L \sin [2 \pi (\Delta_x+ \Delta_y)].
\end{equation}
In the first sub-case (see \eq{eq:scaling_h_as_x_pq}), the  limit $L\gg 1$ of the lateral force is zero. In the second sub-case (see \eq{eq:scaling_h_as_x}), if the mean value of the external field on the upper surface is nonzero, the lateral force tends to a finite, well-defined limit, which is proportional to the surface area of the system. Obviously, this force has the meaning of a purely local surface force.

Subtracting from  $\Delta F^{(h,\alpha)}_{\rm Cas}$ its $L$-independent part, one obtains the lateral force $\delta F^{(h,\alpha)}_{\rm Cas}(L)$ which acts on the upper surface due to the presence of the lower one. In the case  $p=M$ and $q=N$  one obtains 
\begin{eqnarray}
\label{eq:F_long}
\lefteqn{\beta\delta F^{(h,\alpha)}_{\rm Cas}(L) \equiv  \beta\left[\Delta F^{(h,\alpha)}_{\rm Cas}(L)-\lim_{L\to\infty}\Delta F^{(h,\alpha)}_{\rm Cas}(L)\right]}\\
&& =-\frac{ \pi h_L }{2 K^{\perp}} \sin [2 \pi (\Delta_x+ \Delta_y)] \sinh(\lambda)  \Bigg\{ h_1/ \sinh[(L+1)\lambda] + h_L \cos[2 \pi (\Delta_x+ \Delta_y)] [1-\coth (L+1)\lambda]\Bigg\}. \nonumber
\end{eqnarray}
In the other sub-case,  i.e., if  $p\ne  M$ or $q\ne N$, one has $\beta\delta F^{(h,\alpha)}_{\rm Cas}(L)\equiv \beta\Delta F^{(h,\alpha)}_{\rm Cas}(L)$. 

In terms of the scaling variables, for $\beta\delta F^{(h,\alpha)}_{\rm Cas}(L)$ one has 
\begin{equation}
\label{eq:sc_funct_field_long}
\beta\delta F^{(h,\alpha)}_{\rm Cas}(L)=L^{-3}\left(\frac{J^\perp}{J^\|}\right) X^{(h,\alpha)}_{\rm Cas}(w,x_1,x_L),
\end{equation}
where, in the subcase 

{\it (i)}, if $0<k_x<2\pi$ or $0<k_y< 2\pi$,

one has 
\begin{equation}
\label{eq:Casimir_longit_pq_scaling}
X^{(h,\alpha)}_{\rm Cas}=-\pi x_1 x_L \, p_\alpha \sin (\mathbf{k\cdot\Delta}) \frac{w}{\sinh(2w)}. 
\end{equation}
Here $p_\alpha=p$ for $\alpha=x$, and $p_\alpha=q$ for $\alpha=y$. 

In the subcase

\begin{figure}[!h]
	\begin{floatrow}
		\ffigbox{\includegraphics[scale = 0.5]{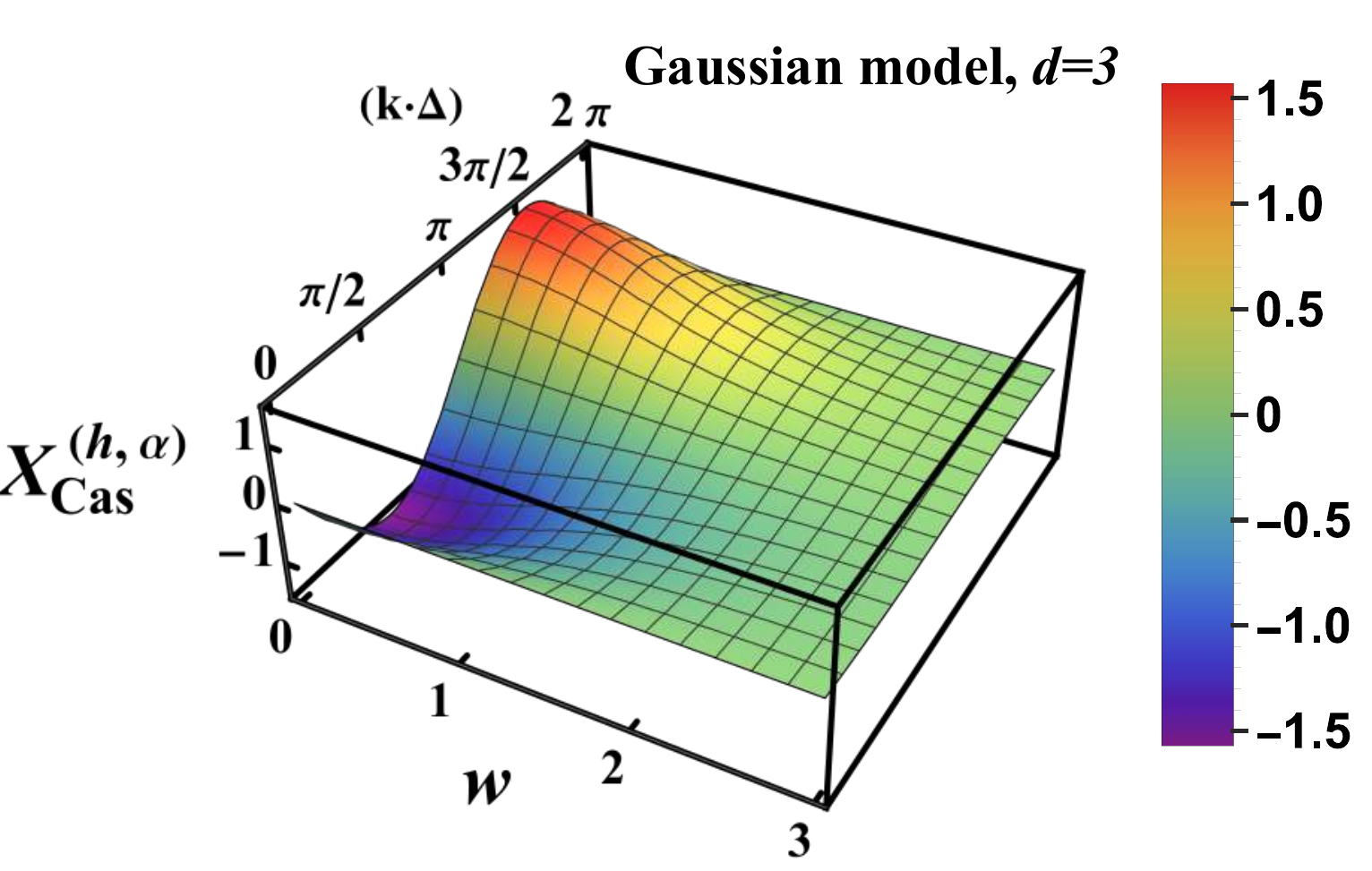}}{\caption{The Gaussian scaling function $X^{(h,\alpha)}_{\rm Cas}(w,x_1,x_L)$ (see Eqs. \eqref{eq:F_long} - \eqref{eq:Casimir_longit_pq_scaling}) as function of $w\in (0,3]$ and of  $\mathbf{k\cdot\Delta}\in [0,2\pi]$ for $x_1=x_L=1$.}\label{Fig:3D_G_h1_eq_hL_Legend_lat}}
		\ffigbox{\includegraphics[scale = 0.5]{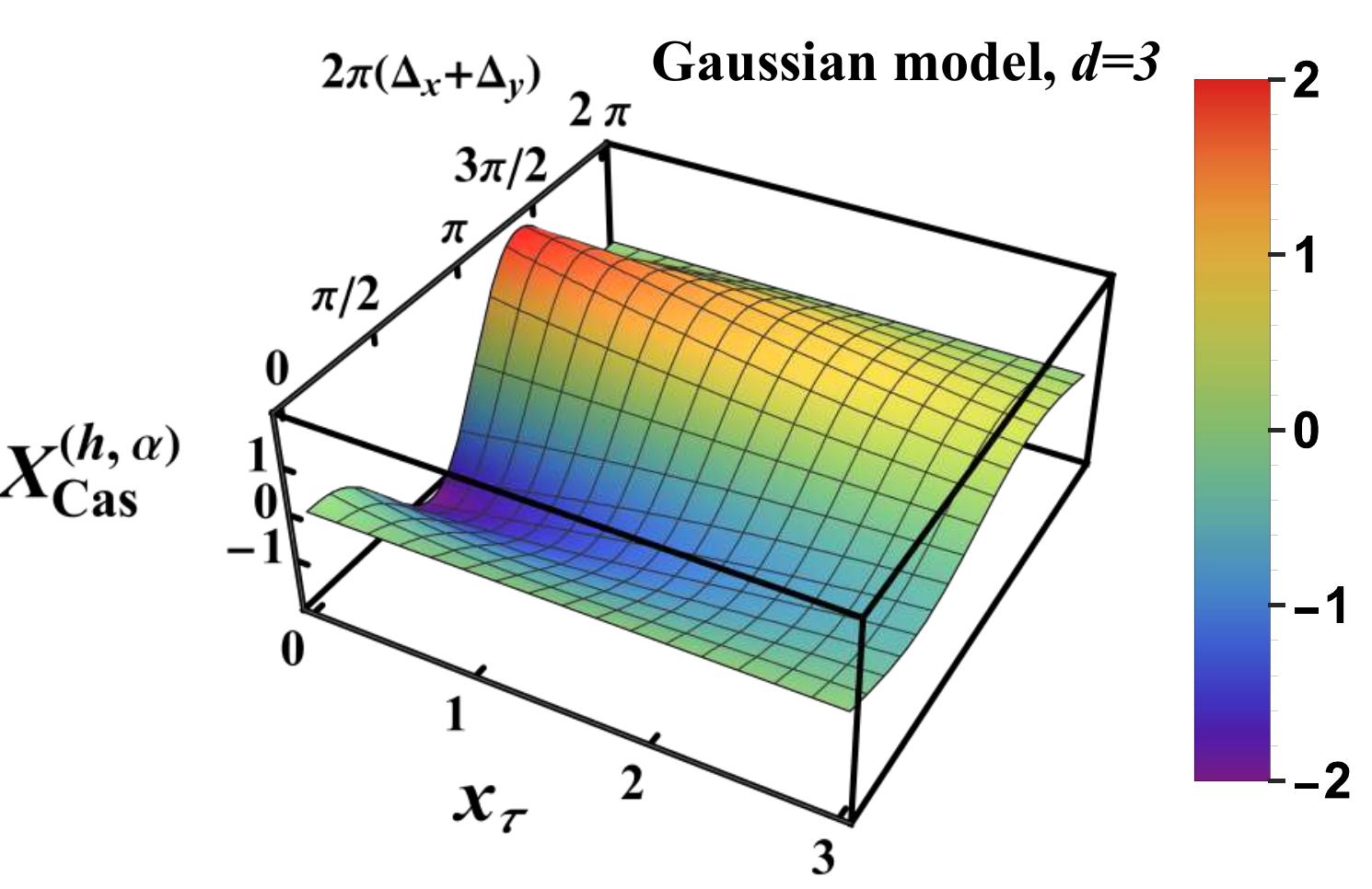}}{\caption{The scaling function $X^{(h,\alpha)}_{\rm Cas}(x_\tau,x_1,x_L)$ (see \eq{eq:F_long_scaling}) as function of $w\in(0,3]$ and of $\Delta_x+ \Delta_y\in [0,1]$ for $x_1=x_L=1$.   }\label{Fig:3D_G_h1_eq_hL_Legend_MN_lat}}
	\end{floatrow}
\end{figure}

{\it (ii)}, if $k_x=2\pi$ and $k_y=2\pi$,

one has 
\begin{equation}
\label{eq:F_long_scaling}
X^{(h,\alpha)}_{\rm Cas}=-\pi x_L w \sin [2 \pi (\Delta_x+ \Delta_y)]\; \Bigg\{ x_1/ \sinh(2w) + x_L \left(\cos[2 \pi (\Delta_x+ \Delta_y)]\right) [1-\coth (2w)]\Bigg \}.  
\end{equation}
Equation \reff{eq:sc_funct_field_long} implies that in the scaling regime the longitudinal Casimir force is of the \textit{same} order of magnitude as the transverse component of the force. Its behavior is visualized in Figs. \ref{Fig:3D_G_h1_eq_hL_Legend_lat} and \ref{Fig:3D_G_h1_eq_hL_Legend_MN_lat}. 
The  scaling function  $X^{(h,\alpha)}_{\rm Cas}$ can be positive or negative, independently of the values of $x_1$ or $x_L$.

We finish this section, which is concerned with the Casimir force within the Gaussian model for twisted boundary conditions, by clarifying the physical meaning of the regimes $w={\cal O}(1)$ and  $w \gg 1$ in terms of the temperature $T$. One has to consider two sub-cases:  

{\it (i)} $0<k_x<2\pi$ or $0<k_y< 2\pi$.

In this case, in order to have $\lambda$ to be small, it is necessary that $\beta/\beta_c\to 1$ and $k_\alpha\to 0$, $\alpha=x,y$. Under these conditions one has
\begin{equation}
\label{eq:lambda_eq}
\lambda \simeq \sqrt{2\left(\frac{\beta_c}{\beta}-1\right)\left[ 2\frac{J^{\|}}{J^\perp}+1\right]+\dfrac{J^{\|}}{J^{\perp}}\left[k_x^2 +k_y^2\right]}.
\end{equation}
This implies (see \eq{eq:w_def}) 
\begin{equation}
\label{eq:omega_eq_simple}
w=\frac{1}{2}\sqrt{x_\tau^2+x_k^2},
\end{equation}
where $x_\tau$ and $x_k$ are defined in \eq{eq:xt_and_xk}. From \eq{eq:omega_eq_simple} it follows that in order to have $w = {\cal O}(1)$ one needs to have simultaneously $x_\tau={\cal O}(1)$ and $x_k={\cal O}(1)$. Taking into account that $1/\nu=2$ for the Gaussian model and that  $x_\tau^2$ takes its expected form $a_\tau \tau L^{1/\nu}$ with $\tau=(T-T_c)/T_c$. The condition $x_k={\cal O}(1)$ implies that in order to encounter the regime $w = {\cal O}(1)$, one needs to have a modulation with wave vector $k\lesssim L^{-1}$, which includes, e.g., the  case $k=0$. If $x_k\gg 1$, one has, even at the critical point $\beta=\beta_c$, $w\gg 1$, and, according to \eq{eq:scaling_h_as},  the field contributions to the Casimir force are, in this regime, exponentially small. 

{\it (ii)} $k_x=2\pi$ and $k_y=2\pi$.

As it follows from \eq{eq:Lambda_def}, this sub-case reduces to the previously considered one with $k_x=k_y=0$. This implies that $w=x_\tau/2$ (see \eq{eq:omega_eq_simple}). 

If $w = {\cal O}(1)$, from Eqs. \eqref{eq:Casimir_tr_pq} and \eqref{eq:Casimir_tr} with $h_1 = {\cal O}(1)$ and $h_L = {\cal O}(1)$, respectively, one obtains  $\Delta F^{(h,\perp)}_{\rm Cas} = {\cal O}(L^{-2})$, i.e., in this case the transverse force  is one order of magnitude {\it larger} in $L$ than the common transverse Casimir force, which is of the order of ${\cal O}(L^{-3})$.  

\subsubsection{Relations between Casimir force scaling functions and amplitudes for various boundary conditions}

The following relations between certain  Casimir amplitudes (see \eq{DefDeltaCas}) within the Gaussian model are known \cite{KD92a,KD2010}:
\begin{equation}
\label{eq:rel_Cas_ampl_per_OO}
\Delta_{\rm Cas}^{{\rm (p)}}=2^d \Delta_{\rm Cas}^{\rm (O, O)},\quad 
\Delta_{\rm Cas}^{{\rm (ap)}}=2^d \Delta_{\rm Cas}^{\rm(O, {\rm SB})}, \quad 
\Delta_{\rm Cas}^{{\rm (O,O)}}=\Delta_{\rm Cas}^{({\rm SB}, {\rm SB})}, \quad 
\Delta_{\rm Cas}^{{\rm (O,{\rm SB})}}=-(1-2^{-d+1})\Delta_{\rm Cas}^{\rm(O, O)}.
\end{equation}

\begin{figure}[h]
	\includegraphics[width=\columnwidth]{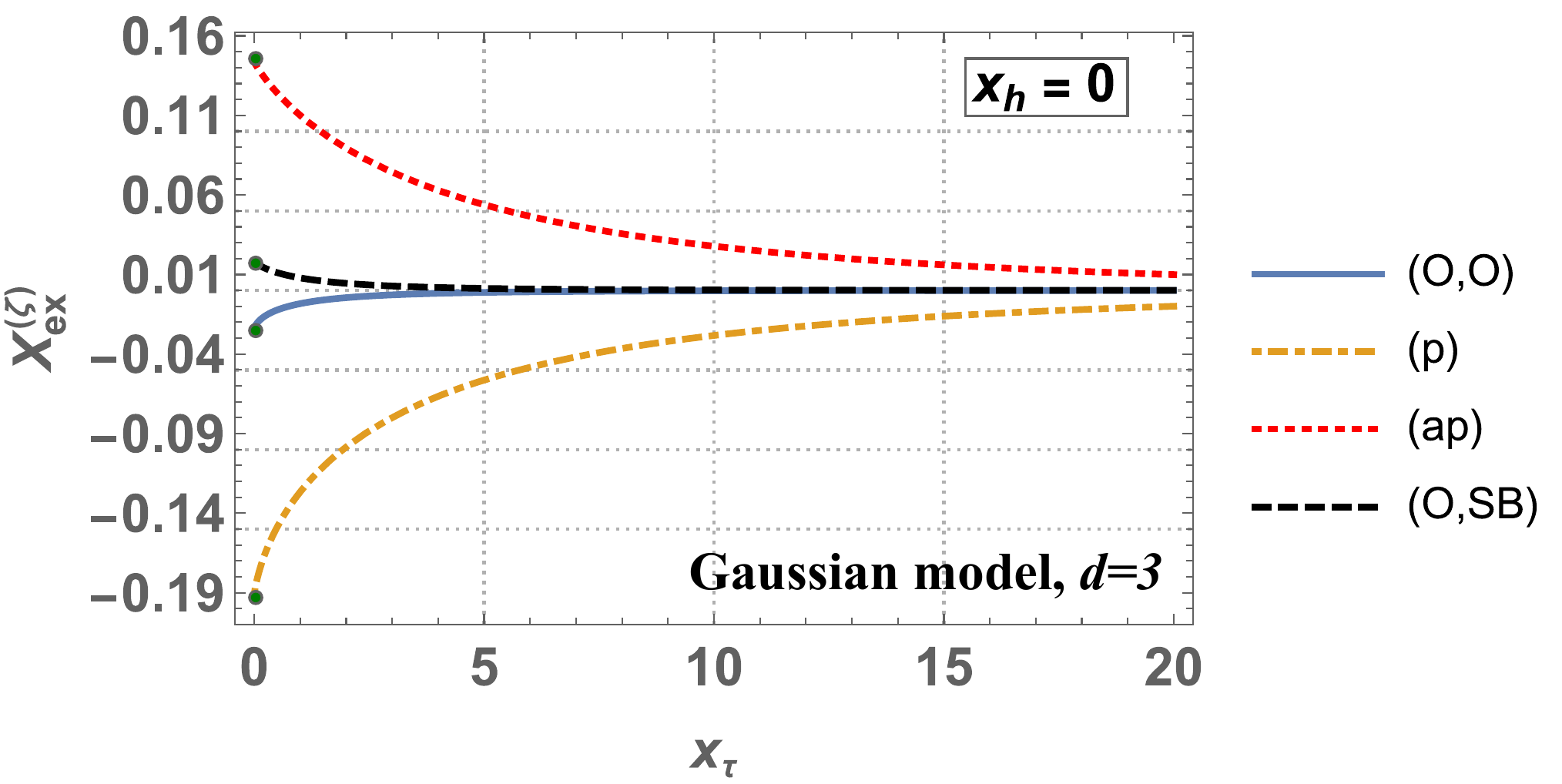}
	\caption{The scaling functions $X^{(\zeta)}_{\rm ex}(x_\tau)$ for the Gaussian model (see Eqs. \eqref{eq:excess_p_equiv_d3}, \eqref{eq:excess_ap_equiv_d3}, \eqref{eq:excess_DD_equiv_d3}, and  \eqref{eq:excess_o_sb_equiv_d3}) as functions of the temperature scaling variable $x_\tau$ for four combinations of boundary conditions in the case $(d,n)=(3,1)$. The dots correspond to the critical Casimir amplitudes $\Delta_{\rm Cas}^{(\zeta)}(d=3,n=1)$ for the free energy at $x_\tau=0$ (see Eqs. \eqref{eq:Gaus_per_Cas},
		\eqref{eq:Gaus_aper_Cas},
		\eqref{eq:Gaus_DD_Cas}, and
		\eqref{eq:Gaus_O,SB_Cas}). }
	\label{Fig:3d_Gauss_ex} 
\end{figure}

\begin{figure}[h]
	\includegraphics[width=\columnwidth]{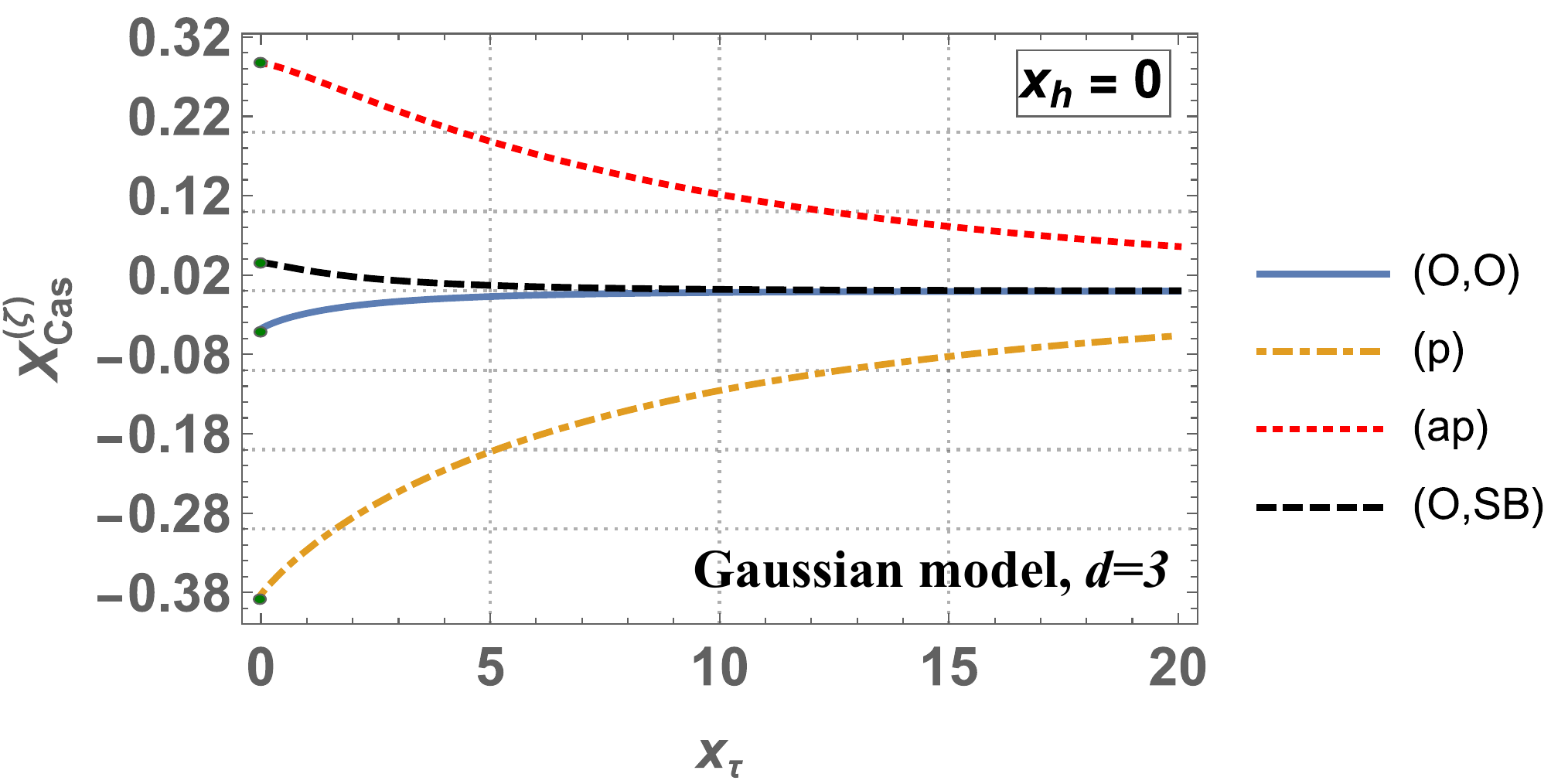}
	\caption{The scaling functions $X^{(\zeta)}_{\rm Cas}(x_\tau)$ for the Gaussian model (see Eqs. \eqref{eq:Cas_p_equiv_d3}, \eqref{eq:Cas_ap_equiv_d3}, \eqref{eq:Cas_DD_equiv_d3}, and  \eqref{eq:Cas_o_sb_equiv_d3}) as functions of the temperature scaling variable $x_\tau$ for four combinations of boundary conditions in the case $(d,n)=(3,1)$. The dots correspond to the critical Casimir amplitudes $2\Delta_{\rm Cas}^{(\zeta)}(d=3,n=1)$ for the Casimir force at $x_\tau=0$ (see Eqs. \eqref{eq:Gaus_per_Cas},
		\eqref{eq:Gaus_aper_Cas},
		\eqref{eq:Gaus_DD_Cas}, and 
		\eqref{eq:Gaus_O,SB_Cas}). }
	\label{Fig:3d_Gauss}
\end{figure}

The scaling functions of the excess free energy with periodic, anti-periodic, ordinary (i.e., Dirichlet), and mixed (i.e., (ordinary, surface-bulk)) boundary conditions are shown in Fig. \ref{Fig:3d_Gauss_ex}. This comparison is made for $d=3$ and $n=1$. 
The behavior of the scaling functions of the Casimir force with the same boundary conditions is shown in Fig. \ref{Fig:3d_Gauss}. One clearly sees that the scaling functions decay much slower for periodic and anti-periodic boundary conditions (see Eqs. \eqref{eq:excess_p_as} and \eqref{eq:excess_ap_as}) as compared to other boundary conditions, which are realized by actual,  physical boundaries in the system. It is somewhat surprising that periodic boundary conditions (see Eqs. \eqref{eq:excess_DD_as} and \eqref{eq:excess_o_sb_as}), which seem to be the least invasive way of implementing a finite size system, have the strongest finite-size impact on the Casimir force. The simulation community is encouraged to take note. 

Concerning the scaling functions \cite{KD2010}  for $n=1$, and $2<d<4$,  one has 
\begin{equation}
	\label{eq:rel_Cas_scaling_functions}
X_{\rm Cas}^{({\rm SB,SB})}(x_\tau)=X_{\rm Cas}^{\rm(O,O)}(x_\tau)=2^{-d}X_{\rm Cas}^{(p)}{(4x_\tau)},\quad X_{\rm Cas}^{\rm (SB, O)}=2^{-d}X_{\rm Cas}^{\rm(ap)}(4x_\tau), \qquad X_{\rm Cas}^{\rm (ap)}(x_\tau)=2^{1-d}X_{\rm Cas}^{(p)}(4x_\tau)-X_{\rm Cas}^{(p)}(x_\tau).
\end{equation}
These results hold also for $n>1$ \cite{rem1} (see also Ref. \cite{KD92a}). 

Obviously, the relations in  \eq{eq:rel_Cas_ampl_per_OO} follow from \eq{eq:rel_Cas_scaling_functions}. 

In Ref. \cite{KD2010} also  a Gaussian model has  considered with coupling constant $J_\perp$ across the film and $J_\|$ parallel to it. It has been shown that for all boundary conditions $(\zeta)$ mentioned above one finds 
\begin{equation}
	\label{eq:isi-aniso-Casimir-scaling-relation}
X_{\rm Cas,anisotropic}^{(\zeta)}(x_\tau)=(\xi_{0,\perp}/\xi_{0,\|})^{d-1}X_{\rm Cas, isotropic}^{(\zeta)}(x_\tau), \qquad \mbox{with} \qquad \xi_{0,\perp}/\xi_{0,\|}=(J_\perp/J_\|)^{1/2} . 
\end{equation}
Here, $\xi_{0,\perp}$ and $\xi_{0,\|}$ are the correlation length amplitudes in the direction perpendicular and along the film surfaces, respectively. 
In addition, in Ref. \cite{KD2010} it  has been shown that this relation is also valid for the  critical Casimir amplitudes of the two-dimensional Ising model for periodic and antiperiodic boundary conditions.

\subsection{Casimir effect in the  limit $n\to\infty$ (spherical model)}
\label{sec:spherical-model}

We recall that the unlimited, translationally invariant,  standard spherical model \cite{BK52} is equivalent\footnote{H. E. Stanley demonstrated \cite{S68} that in the thermodynamic limit the free energy density
of the classical $n$-vector model  converges to that one of the spherical model when the spin dimensionality
	$n$ tends to infinity. In addition, all correlation functions
	of the $n$-vector spin model converge, in the limit $n \to \infty$, to the corresponding
	correlation functions of the spherical model \cite{S88}. For a detailed
	review of subsequent analytic efforts in this direction, such as demonstrating the equivalence of the two models for systems
	with disorder, we refer the reader to Ref. \cite{KKPS92}.} to the limit $n\to\infty$ of the corresponding ensemble of $n$-component vectors \cite{S68,KT71,S88,BDT2000,KKPS92,K73}, i.e., to the $O(n\to\infty)$ model. However, for the spherical model with surfaces or, more generally, without translational symmetry, a careful analysis is necessary, which takes into account that the aforementioned  equivalence is preserved, only if one imposes so-called spherical constraints on the spins of the system,  such that the mean square value of {\em each} spin of the system is the same \cite{K73}. For quite some time this model has been considered  as to be analytically intractable \cite{BJSW74,BM77}. However, in Ref. \cite{DBR2014},  for the case of Dirichlet boundary conditions, via exact calculations this model has been analytically reduced  to a one-dimensional model, the properties of which have been studied  either numerically near the critical region, or in an exact, analytical manner in the low-temperature regime. The  Ginzburg-Landau-Wilson (GLW) version of the $O(n\to\infty)$ model has been studied, with the same boundary conditions, in Refs. \cite{DGHHRS2012,DR2014,DGHHRS2014}. 

The Casimir effect in the three-dimensional  spherical model has been studied in Refs. \cite{D96,D98,DG2009,DDG2006,CD2004,DGHHRS2012,DR2014,DBR2014}. The spherical model is the only nontrivial statistical-mechanical model which  can be solved exactly in its bulk limit for any dimension $d$, even in the presence of an external magnetic field $h$. The Casimir force for this model has been derived both for short-ranged \cite{D96,D98,DG2009,DGHHRS2012,DR2014,DBR2014} as well as, in the case of periodic boundary conditions, for long-ranged, algebraically decaying  interactions \cite{DDG2006,CD2004}. In the current section we briefly review the available results  for this model with a focus on those, which are directly relevant for the Casimir force.

Within this model, and by keeping translational invariance inspite of its finite system size, the Casimir force has been studied in Refs.  \cite{D96,D98,CD2004} for periodic and in Ref. \cite{DG2009} for antiperiodic boundary conditions. There, exact analytical results have been obtained for the corresponding scaling function and the Casimir amplitude in spatial dimensions $2<d<4$. This encompasses three-dimensional films, for which an exact expression for the Casimir amplitude in closed form has been derived. We remark that this amplitude is the only one, which is  known exactly for $d=3$, and which is nontrivial, i.e., non-Gaussian. 

Results for the quantum version of the spherical model, subject to periodic boundary conditions are also available \cite{CDT2000}. Various  quantizations of the classical model are possible \cite{BDT2000,O72,V96,VZ92,MVZ97,N95}.  Among them are certain versions of the Bose gas \cite{NJN2013,MVZ97,NP2011,SP85}. We also mention the large-$n$ limit of the so-called 2+1 Gross-Neveu model \cite{CT2011}, which represents a broader class of four fermionic models, which are mathematically very similar to the three-dimensional spherical model.  The  Casimir amplitude equals the negative value of the Casimir amplitude of the three-dimensional spherical model subject to antiperiodic boundary conditions \cite{DG2009}.  

\subsubsection{Casimir force for periodic boundary conditions}

We consider two main sub-cases: systems with short-ranged, and with algebraically decaying long-ranged  interactions of either the subleading (van der Waals) or of the leading type (see Sect. \ref{sec:interactions}). Since for periodic boundary conditions all spins in the system are equivalent, one needs only a single equation for determining the mean length of the spins, i.e., only one spherical field equation. 

{\bf (I)} {\it Casimir force within the spherical model with short-ranged interactions}

We shall separately consider the general case $2<d<4$ in its own right. Special attention will be paid to the case  $d=3$ for which  analytical expressions in closed form will be presented. Since the spherical model with periodic boundary conditions is equivalent to the  limit $n\to\infty$ of the corresponding $O(n)$ models, the results within this model for $2<d<4$ provide a direct check of the corresponding $\varepsilon$-expansion renormalization group results for $O(n)$  models upon carrying out the limit $n\to\infty$ therein. 

$\blacktriangleright$ \textit{{The case ${\boldmath d=3}$}} 

 The results in that case have been reported in Refs. \cite{D96} and \cite{D98}. For the scaling function of the excess free energy (\eq{excess_free_energy_scaling}) one obtains
 \begin{equation}
 X^{(p)}_{{\rm {ex}}}(x_\tau,x_h) =-\frac 1{2\pi }\Bigg[ \frac 16\left(
 y_L^{3/2}-y_\infty ^{3/2}\right) +\sqrt{y_L}\;{\rm {Li}_2}\left( \exp \left(
 -\sqrt{y_L}\right) \right)  +{\rm {Li}_3}\left( \exp \left( -\sqrt{y_L}%
 \right) \right) \Bigg ]  +\frac{1}{8\pi}\left[x_h^2\left( \frac 1{y_\infty }-\frac 1{y_L}\right) -
 x_\tau\left( y_\infty -y_L\right)\right] ,  \label{xecff} 
\end{equation}
 where ${\rm {Li_p}}(z)$ are the polylogarithm functions;
 \begin{equation}
  \label{sv_tau_SM}
  x_\tau\equiv \tau (L/\xi_0^+)^{1/\nu}=4\pi K_c \tau L, \quad 
  \mbox{and} \quad 
  x_h\equiv h (L/\xi_{0,h})^{\Delta/\nu}=\sqrt{\frac{4\pi}{K_c}}\;hL^{5/2}, \qquad \text{with} \qquad \xi_{0,h}=\left(\frac{K_c}{4\pi}\right)^{1/5},
  \end{equation}  
 are the temperature and the field scaling variables, respectively. $K_c$ is given by \eq{JCart} (see also \eq{W39Rc}), and $\xi_0^+$ by \eq{xi_ampl}. Here we have used the following expressions  \cite{BD91,SP86,D98,BDT2000} for the correlation length in the finite system:
 \begin{equation}
 \label{xiL}
 \xi_L(\tau,h)=L/\sqrt{y_L(\tau,h)}, 
 \end{equation} 
 with  $\xi_\infty(\tau,h)=\lim_{L\to\infty}\xi_L(\tau,h)$ in the cases in which the bulk limit $L\to\infty$ is finite\footnote{It is infinite for $T<T_c$ and $h=0$.}. In Eq. \eqref{xecff}, $y_L\equiv y_L(x_\tau,x_h)$ and $y_\infty\equiv y_\infty(x_\tau,x_h)$ are determined implicitly by the equation
 \begin{equation}
  -x_\tau=\frac{x_h^2}{y_L^2}-2\ln \left[ 2\sinh \left( \frac 12\sqrt{%
  y_L}\right) \right] ,  \label{eqffSM}
  \end{equation}
  for the finite system, and by the equation 
  \begin{equation}
  -x_\tau=\frac{x_h^2}{y_\infty ^2}-\sqrt{y_\infty } 
  \label{eqifSM}
  \end{equation} 
 for the bulk system if $x_h\ne 0$, or if $x_h=0$ but $x_\tau\ge 0$. If $x_h=0$ and $x_\tau<0$ one has $y_\infty=0$. 
 
 In terms of the scaling variables $x_\tau$ and $x_h$ for 
 the Casimir force \cite{D98} $\beta F^{(p)}_{{\rm {Cas}}}(t,h,L)$, one has
 \be 
 \label{eq:d3_Cas}
 \beta F^{(p)}_{{\rm {Cas}}}(\tau,h,L) = L^{-3}X^{(p)}_{{\rm {Cas}}}(x_\tau,x_h),
\ee 
with the scaling function 
 \begin{equation}
 \label{fcas}
  X^{(p)}_{{\rm {Cas}}}(x_\tau,x_h) = \frac{1}{\pi}\Bigg\{ \frac {3}{8}x_h^2\left(\frac{1}{y_L}-%
 \frac {1}{y_\infty }\right) +\frac{1}{8}x_\tau\left(y_L-y_\infty \right)   - \left[ \frac 16\left( y_L^{3/2}-y_\infty ^{3/2}\right) +%
 \sqrt{y_L}\;{\rm {Li}_2}\left( \exp \left( -\sqrt{y_L}\right) \right) +{\rm {%
 Li}_3}\left( \exp \left( -\sqrt{y_L}\right) \right) \right] \Bigg\}. 
 \end{equation}
 \begin{figure}[h]
 \includegraphics[angle=0,width=\columnwidth]{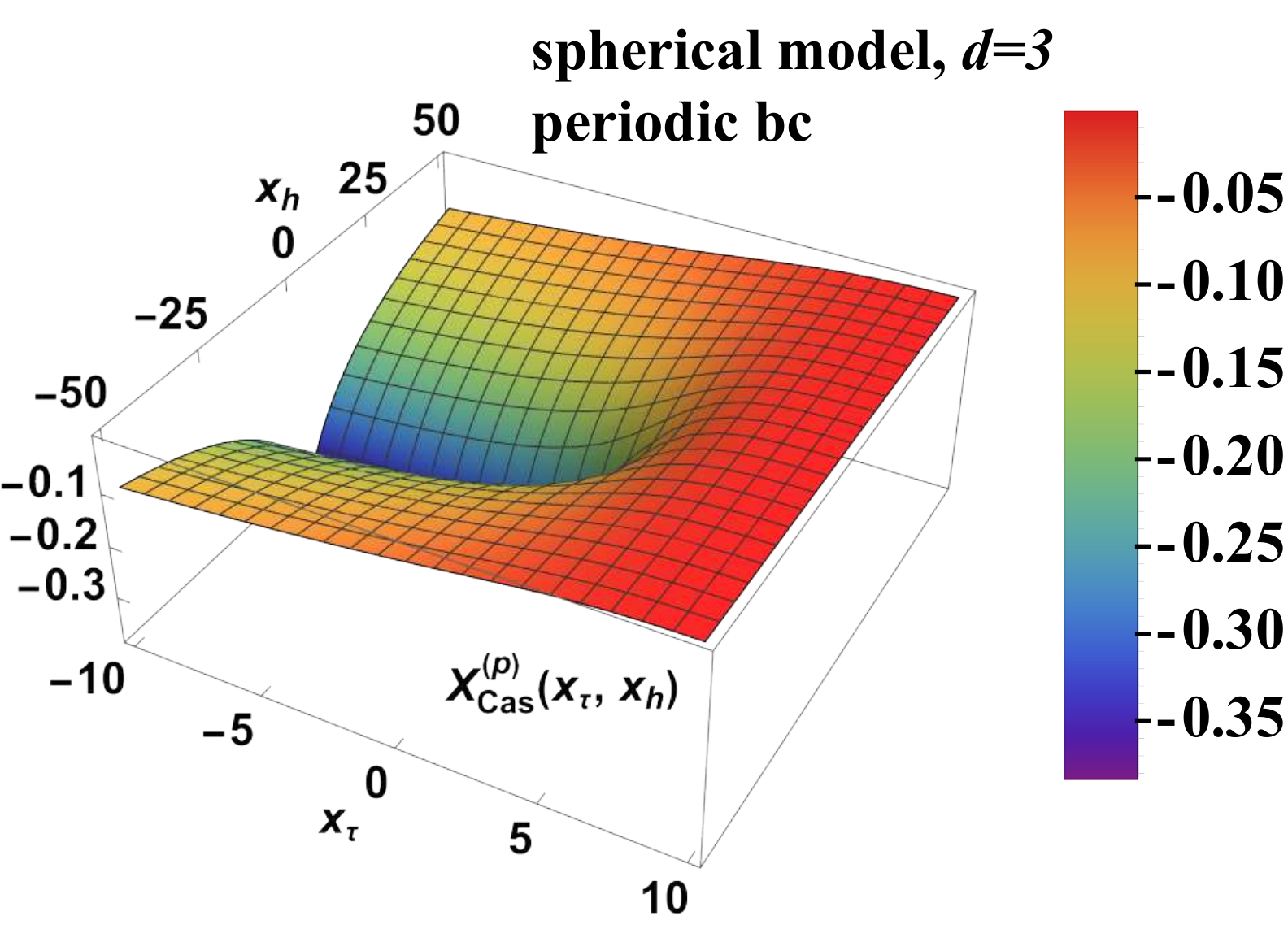}
 \caption{The scaling function $X^{(p)}_{{\rm {Cas}}}(x_\tau,x_h)$ (see \eq{fcas}) of the Casimir force  for the spherical model as a function of both $x_\tau$ and $x_h$. In the figure $x_\tau \in [-10,10]$ while $x_h\in[-50,50]$. Currently, the spherical model is the only three-dimensional model which, despite of exhibiting Gaussian types of fluctuations, is characterized by non-trivial, i.e., by non-Gaussian critical exponents, for which this manifold of the Casimir force is analytically available.}
 \label{CasForceSMper}
 \end{figure}
\begin{figure}
	\centering
		\includegraphics[width=0.475\linewidth]{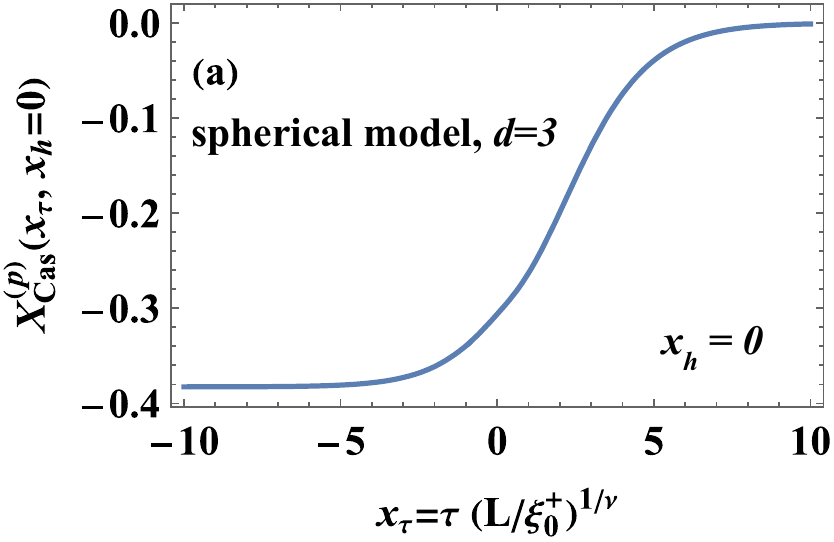}
		\includegraphics[width=0.475\linewidth]{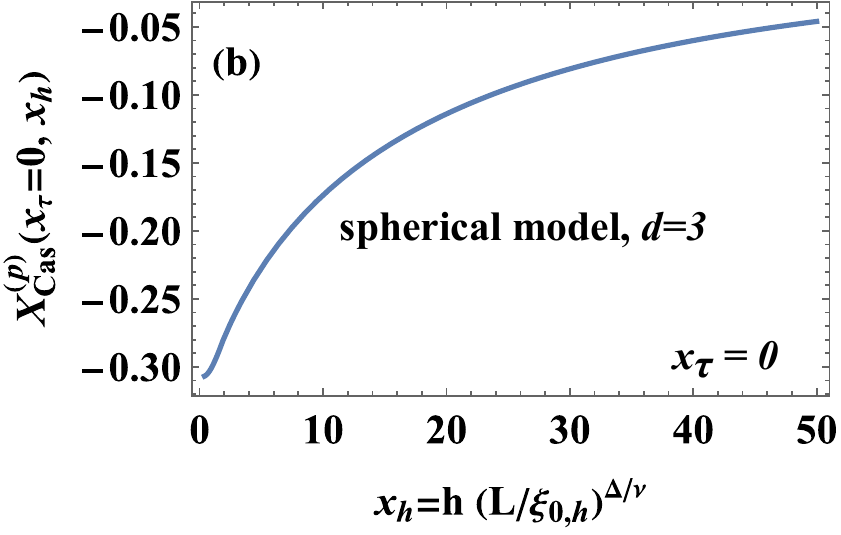}
	\caption{The left panel (a) depicts the Casimir force scaling function $X^{(p)}_{{\rm {Cas}}}$ as a function of the temperature scaling variable $x_\tau$ for $x_h=0$. The right panel (b)  shows $X^{(p)}_{{\rm {Cas}}}$ as a function of the field scaling variable $x_h$ for $x_\tau=0$.  Both curves are cross-sections of the Casimir force scaling function presented in Fig. \ref{CasForceSMper}.}
	\label{fig:Cas_per_cross_sections}
\end{figure}
Obviously, one has $X^{(p)}_{{\rm {Cas}}}(x_\tau,x_h)=X^{(p)}_{{\rm {Cas}}}(x_\tau,-x_h)$. The behaviors of the scaling functions of the Casimir \textit{force} are shown in Figs. \ref{CasForceSMper} and \ref{fig:Cas_per_cross_sections}.
 
 For the Casimir amplitude of the excess free energy \textit{density} (in units of $k_BT_c$) at criticality within the three-dimensional
 spherical model with periodic boundary conditions one has the exact expression \cite{D98}
 \begin{equation}
 X^{(p)}_{{\rm {ex}}}(x_\tau=0, x_h =0)=\Delta_{\rm Cas}^{(p)} =-\frac{2}{5\pi }\zeta (3)\simeq -0.153051.  \label{delta}
 \end{equation}
 The numerical value of this amplitude has  initially been reported in Ref.  \cite{D96}. Knowing the amplitude $\Delta_{\rm Cas}^{(p)} $ of the excess free energy, for the Casimir force at criticality one has 
 \begin{equation}
 \label{SM_crit_point}
 X^{(p)}_{{\rm {Cas}}}(x_\tau=0,x_h=0)=2\Delta_{\rm Cas}^{(p)}=-\frac{4}{5\pi}\zeta(3)\simeq -0.306101.
 \end{equation}
 The low-temperature behavior of the scaling function $X^{(p)}_{{\rm {Cas}}}$ of the Casimir force is also exactly known. In Ref. \cite{D96} it has been shown that 
 \begin{equation}
 \label{low_T_Casimir}
 X^{(p)}_{{\rm {Cas}}}(x_\tau\to-\infty,x_h=0)=-\frac{1}{\pi}\zeta(3)\simeq -0.382627.
 \end{equation}
If $x_\tau\to+\infty$, one infers from Eqs. \eqref{eqffSM} and \eqref{eqifSM} that $y_L, y_\infty \to \infty$ and, thus, from Eq. \eqref{fcas}, that 
\begin{equation}
X^{(p)}_{{\rm {Cas}}}[x_\tau\to \infty,x_h={\cal O}(1)]={\cal O}[\exp(-x_\tau)].
\end{equation}

$\blacktriangleright$ \textit{ {The case ${{2<d<4}}$}} 

The results for this case have been reported in Refs. \cite{DK2004} and \cite{DG2009}. The most general case studied so far concerns the film geometry, in which one allows for anisotropic interactions, which reflect the geometry, i.e., one takes the coupling constant in the Hamiltonian along the surface, i.e., $J_\parallel$ to be distinct from the one perpendicular to the film, i.e., $J_\perp$. 

In the film geometry $\infty^{d-1}\times L_\perp$ with periodic boundary conditions and $J_1=J_2=\cdots=J_{d-1}$, $J_\perp=J_d$, the corresponding result for the excess free energy density $f^{(p)}_{\rm ex}(\beta,L_\perp|d, J_\|, J_\perp)$ in units of $k_BT_c$ \cite{DK2004} is 
\be
\label{eq_Xes_SM}
f^{(p)}_{\rm ex}(\beta,L_\perp|d, J_\|, J_\perp)=L_{\perp}^{-(d-1)}\left(J_\perp/J_\parallel\right)^{(d-1)/2} 
X_{{\rm ex},{\rm SM}}(x_1|d),
\ee
where SM stands for the "spherical model" and 
\begin{equation}
\label{fex_standard_final_per}
X_{{\rm ex},{\rm SM}}(x_1|d)= \frac{1}{2} x_1(y_L-y_\infty)-
\frac{\Gamma(-d/2)}{2(4\pi)^{d/2}}\left(y_L^{d/2}-y_\infty^{d/2} \right)
   -y_L^{d/2} \frac{2}{(2\pi)^{d/2}} \sum_{n=1}^{\infty} \frac{K_{d/2}(n\sqrt{y_L})}{(n\sqrt{y_L})^{d/2}}.
\end{equation}
Here $y_L(x_1)$ is the solution of the equation
\begin{equation}\label{yfinal_per}
- x_1 =\frac{\Gamma(1-d/2)}{(4\pi)^{d/2}} y_L^{d/2-1}
   +y_L^{d/2-1} \frac{2}{(2\pi)^{d/2}} \sum_{n=1}^{\infty} \frac{K_{d/2-1}(n\sqrt{y_L})}{(n\sqrt{y_L})^{d/2-1}}
\end{equation}
for finite film thickness $L_\perp$, while $y_\infty(x_1)$ is the solution of the equation 
\begin{equation}\label{ybfinal}
- x_1=\frac{\Gamma(1-d/2)}{(4\pi)^{d/2}} y_\infty^{d/2-1},
\end{equation}
which holds for the bulk system if $x_1>0$;  for $x_1\le 0$ one has  $y_\infty=0$.
In Eqs. \eqref{fex_standard_final_per} - \eqref{fex_standard_final_per} one has  
\begin{equation}\label{xtdef}
    x_1=\frac{1}{2}b_\perp \left(\frac{b_\parallel}{b_\perp}\right)^{(d-1)/2}\left(K_c-K\right)\,
    L_\perp^{1/\nu}, \qquad \nu=1/(d-2),
\end{equation}
with 
\begin{equation}\label{bdef}
b_\perp=J_\perp/\sum_{i=1}^d J_i, \qquad b_\parallel=J_\parallel/\sum_{i=1}^d J_i, \qquad 
\mbox{and} \qquad
K_c \equiv 2\beta_c \sum_{j=1}^d J_j=
    \int_0^\infty dx \prod_{j=1}^d e^{-x b_j}I_0(x b_j).
\end{equation}

For the Casimir force one finds  \cite{DK2004} 
\begin{eqnarray}
\label{cassm}
 \lefteqn{\beta F_{{\rm Cas}}^{(p)}(\beta,L_\perp|d, J_\|,J_\perp) = L_\perp^{-d}\left(\frac{J_\perp}{J_\parallel}\right)^{(d-1)/2}  \left\{ \frac{1}{2}x_1 (y_L-y_\infty)\right.} \\
 && \left.-(d-1)\left[\frac{1}{2}\frac{\Gamma(-d/2)}{(4\pi)^{d/2}}
\left(y_L^{d/2}-y_\infty^{d/2}\right)\right. \left. + \frac{2}{(2\pi)^{d/2}}
y_L^{d/4}\sum_{q=1}^{\infty}\frac{K_{d/2}(q\sqrt{y_L})}{q^{d/2}}\right]
\right\}. \nonumber
\end{eqnarray}
Equations  (\ref{fex_standard_final_per}) - (\ref{cassm}) demonstrate, that the Casimir force with periodic boundary conditions in a system with anisotropic interactions can be written as 
\begin{equation}\label{Casscaling_ani}
\beta F^{(p)}_{\rm Cas}(\beta,L_\perp|d, J_\|, J_\perp) =L_\perp^{-d} \left(\frac{J_\perp}{J_\parallel}\right)^{(d-1)/2} X_{\rm Cas}^{(p)}(x_1|d),
\end{equation}
where $X_{\rm Cas}^{(p)}$ is a universal scaling function pertinent to the \textit{isotropic} system in which a suitable definition of the scaling variable is taken (see Eq. (\ref{xtdef})). The scaling variable $x_1$ is of the form $x_1=a_\tau({\bm b})\; \tau L_\perp^{1/\nu}$ where ${\bm b}=(b_\|, \cdots, b_\|,b_\perp)$ is a $d$-dimensional vector; explicitly, here  $a_\tau = K_c\, b_\perp \left(b_\parallel/b_\perp\right)^{(d-1)/2}/2$.  This implies,  that all effects of that type of anisotropy considered here can be incorporated into the prefactor $\left(b_\perp / b_\parallel \right)^{(d-1)/2}=\left(J_\perp / J_\parallel \right)^{(d-1)/2}$ of the scaling function on the r.h.s. of Eq. \eqref{fex_standard_final_per}, and into the nonuniversal factor $a_\tau$ which enters the definition of the temperature scaling variable $x_1$ (\eq{xtdef}). With regard to the Casimir amplitudes, Eq. (\ref{cassm}) leads to the following relation between the amplitudes in anisotropic and in isotropic systems:
\begin{equation}\label{CasAmplRel_aniso}
\Delta_{\rm Cas}^{(p)}(d| J_\perp,J_\parallel) = \left(\frac{J_\perp}{J_\parallel}\right)^{(d-1)/2} \Delta_{\rm Cas}^{(p)}(d|J_\perp=J_\parallel).
\end{equation}
We remark that, in line with the concept of universality, the value of the Casimir amplitude in the isotropic system does not depend on $J\equiv J_\perp=J_\parallel$. In order to achieve agreement  with the relations in Eqs.  \eqref{relaniso} and (\ref{relDeltaGen}), one only has to take into account that \cite{DK2004}
\begin{equation}\label{relxi}
\frac{\xi_\perp}{\xi_\parallel}=\sqrt{\frac{J_\perp}{J_\parallel}}.
\end{equation}

{\bf (II)} {\it Casimir force in the spherical model with subleading long-ranged interactions of the van der Waals type}

The results for this case have been reported in Ref. \cite{DDG2006} for a system with periodic boundary conditions in the  direction of finite extent. Specifically, the influence of
long-ranged pair interactions --- the  potentials of which decay with the distance $r$ between the interacting spins as $b\, r^{-(d+\sigma)}$ for  $r\to\infty$, with $2<\sigma<4$ and $2<d+\sigma\leq 6$ --- on the Casimir effect has been studied at and near the bulk critical temperature $T_{c}$ for $2<d<4$.  This type of 
  interactions decays sufficiently fast to leave bulk critical
  exponents and other universal bulk quantities unchanged relative to those of systems with short-ranged interactions, i.e., they
  are irrelevant in the renormalization group sense. Yet they
  entail important modifications of the standard scaling behavior of
  the excess free energy and of the Casimir force $F^{(p)}_{\rm Cas}(\beta,L|d,\sigma)$. 
  
  Concerning the most general case of systems confined to a $d$-dimensional slab of
    macroscopic lateral extent and finite thickness $L$, --- which in the limit $L\to\infty$ undergo 
    a continuous bulk phase transition  and can be 
    described in terms of an $O(n)$ symmetrical Hamiltonian ---  one can argue that $F^{(p)}_{\rm Cas}(\beta,L|d,\sigma)$  decomposes as follows: 
\begin{equation}
\label{Cas_subleding}
\beta F^{(p)}_{\rm Cas}(\beta,L|d,\sigma)\simeq L^{-d}\left[
  X_{\rm Cas}^{(p)}\left(\frac{L}{\xi_\infty}\Bigg\vert d\right)+ g_\omega\,
  L^{-\omega}X_\omega\left(\frac{L}{\xi_\infty}\Bigg\vert d\right)
 +g_\sigma\,
  L^{-\omega_\sigma}\, X_\sigma\left(\frac{L}{\xi_\infty}\Bigg\vert d,\sigma\right)\right].
\end{equation}
 Here
  $X_{\rm Cas}^{(p)}$, $X_\omega$, and $X_\sigma$ are universal scaling
  functions, with $X_{\rm Cas}^{(p)}$ being the scaling function pertinent to a system with purely short-ranged interactions (see the results discussed above); $g_\omega$ and $g_\sigma$ are the scaling fields associated
  with the leading corrections to scaling and with long-ranged 
  interactions, respectively; $\omega$ \cite{ZJ2002} (for the spherical model $\omega=4-d$ \cite{DDG2006}) and
  $\omega_\sigma=\sigma+\eta-2$ \cite{D86} are the corresponding
  correction-to-scaling exponents, where $\eta$ denotes the standard
  bulk exponent of the two-point correlation function at $T_c$ without long-ranged
  interactions; $\xi_\infty$ is the (second-moment) bulk correlation
  length (which itself carries corrections to scaling). For $T>T_c$, as function of $L$ the
  contribution $\propto g_\sigma$ decays 
  algebraically rather than exponentially. This is in contrast to the scaling functions $X_{\rm Cas}^{(p)}$ and $X_\omega$ which for large scaling variables decay exponentially, and hence the contributions stemming from $X_\sigma$ become
 {\it dominant} within a certain region of temperatures and $L$. It has been argued\footnote{See Eqs. (2.28) and  (2.31) in Ref. \cite{DDG2006} and the text connected with them. This is consistent with the result of  Iagolnitzer and Souillard \cite{IS77} who proved, by using the
 	Griffiths-Sherman-Kelly inequalities \cite{Griff67,KS68}, that the two-point net
 	correlation function of a ferromagnetic system,  the interactions of which 
 	decay algebraically as a function of the distance, cannot decay more rapidly than
 	the potential. Equation \eqref{as_SLRI} actually reinforces that the correlations decay at large distances exactly as  the interaction potential does. That the decay mode of the correlations  at large distances is mirrored by the type of $L$-dependence one observes in a given system, has been used in Refs. \cite{D2001,CD2004,DSD2007}. The prefactor $b$ of the interaction potential (see above) can be determined via the Fourier transform of the interaction $J({\mathbf k})=J({\mathbf 0})\left[1-v_2 k^2+v_\sigma k^\sigma-v_4 k^4+{\cal O}(k^6)\right]$ as to $b=v_\sigma/v_2$, providing the system is  fully isotropic \cite{D2001,DKD2003,DDG2006}. }   that \cite{DDG2006} 
 \begin{equation}
 \label{as_SLRI}
 \beta F^{(p)}_{\rm Cas}(\beta,L|d,\sigma)\propto b\,  L^{-(d+\sigma)}t^{-\gamma}.
 \end{equation}
Although the scaling function $X_\omega$  belongs to the  universality class of short-ranged interactions, the
scaling field $g_\omega$ incorporates, in general, also contributions due to the long-ranged tails of the interaction. We refer the interested reader to Ref. \cite{DDG2006} where explicit results about the aforementioned mixing of the corrections to scaling between  long-ranged forces and Wegner-type corrections have been
reported; the contributions which are due to the
corrections to scaling ruled by the Wegner exponent always  produce only corrections, i.e., they do never become dominant -- neither near the critical point,
where they are of the order of $L^{-\omega}$, nor apart form it, where they decay exponentially.

In Ref. \cite{DDG2006} the above conclusions are supported by 
deriving exact results for spherical and Gaussian models. For the case $d+\sigma =6$, which includes the one of 
nonretarded van der Waals interactions in $d=3$, it is shown that within the
spherical model the 
power laws of the corrections to scaling, which are proportional to $b$,  are modified by logarithms. Using
general renormalization group ideas it has been  shown, that these logarithmic singularities originate from the degeneracy $\omega\to \omega_\sigma$ with $\omega_\sigma=4-d$ which  specifically occurs for the spherical model if $d+\sigma=6$, in conjunction with the dependence of $g_\omega$ on $b$. The scaling function $X_{\rm Cas}^{(p)}$, as explained above, is known from Refs. \cite{D96,D98,DK2004}.  In Ref. \cite{DDG2006} explicit results are derived for the functions $X_{\omega}$
and $X_\sigma$. The corresponding expressions are, however, too cumbersome in order to make their presentation useful here.

{\bf (III)} {\it Casimir force within the spherical model with leading, algebraically decaying interactions}

The results concerning this case have been reported in Ref. \cite{CD2004}. Therein  interactions have been considered which decay with the distance $r\to \infty$ between the spins as $r^{-(d+\sigma)}$, with\footnote{For such systems $d_l=\sigma$ is the lower critical dimension, while $d_u=2\sigma$ is the upper critical one \cite{FMN72,J72,BDT2000}. The above relations imply that for $d<\sigma$ the system does not undergo a phase transition, while for $d>2\sigma$ the critical exponents acquire their mean-field values.} $0<\sigma<2$ and $\sigma<d<2\sigma$. According to the classification given in Section \ref{sec:interactions} these quantify as leading long-ranged interactions. In this case the universality class, i.e., the critical exponents, the scaling functions etc., do depend on the interaction parameter $\sigma$. For example, for the critical exponents \cite{BDT2000} one has 
\begin{equation}
\label{critexplongranged}
\nu =\frac{1}{d-\sigma}, \qquad \eta=2-\sigma ,\qquad \gamma= \frac{\sigma}{d-\sigma}, \qquad \beta=1/2, \qquad \delta=\frac{d+\sigma}{d-\sigma}.
\end{equation}
All the other critical exponents, which are not reported here (e.g., $\alpha$), can be obtained by applying the corresponding scaling relations. From Eqs. \eqref{crit_exp_SM} and \eqref{critexplongranged} one infers,  that the case of short-ranged interactions corresponds formally to the limit $\sigma\to 2$ taken in the above expressions. 

We start the presentation of the results pertinent to the case of leading-order long-ranged interactions by  noting first, that from Eqs. (3.33)-(3.35) and (3.50) in Ref. \cite{BDT2000} one can determine the nonuniversal amplitudes $\xi_0^+$ and $\xi_{0,h}$ of the bulk correlation lengths within this model. The corresponding results are
\begin{equation}
\label{xitodsigma_1}
\xi_0^+ (d,\sigma)= \left[|D_{d,\sigma}|/K_c\right]^{1/(d-\sigma)}
\end{equation}
and 
\begin{equation}
\label{xihodsigma_1}
\xi_{0,h}(d,\sigma)= 
\left[K_c \lvert D_{d,\sigma}\rvert \right]^{1/(d+\sigma)}
\end{equation}
where 
\begin{equation}\label{constD}
D_{d,\sigma}=2\pi\left[(4\pi)^{d/2}\Gamma\left(\frac d2\right)
\sigma\sin\left(\frac{\pi d}\sigma\right)\right]^{-1}.
\end{equation}
Explicitly, one has 
\begin{equation}
\label{xitodsigma}
\xi_0^+ (d,\sigma)= \left[2\sigma (4\pi)^{d/2-1}
     \Gamma
   \left(d/2\right) \left|\sin \left(d \pi/\sigma \right)\right|K_c \right]^{-1/(d-\sigma)}
\end{equation}
and
\begin{equation}
\label{xihodsigma}
\xi_{0,h}(d,\sigma)= 
\left[ 
\frac{K_c}{2\sigma (4\pi)^{d/2-1}\Gamma \left(d/2\right)\left|
   \sin \left(d \pi/\sigma\right)\right|}
   \right]^{1/(d+\sigma)}.
\end{equation}
Here all length are measured in units of the lattice constant. Thus, the proper scaling variables for films of thickness $L$, in the presence of long-ranged interactions considered above, are 
\begin{equation}
  \label{sv_tau_SM_LRI}
  x_\tau\equiv \tau (L/\xi_0^+)^{1/\nu}= \left[K_c/|D_{d,\sigma}| \right] \; \tau L^{d-\sigma}
  \end{equation}
  and 
\begin{equation}
  \label{sv_h_SM_LRI}
  x_h\equiv h (L/\xi_{0,h})^{\Delta/\nu}=\frac{1}{\sqrt{K_c |D_{d,\sigma}|}}\;hL^{(d+\sigma)/2}.
  \end{equation}  
If $\sigma<d<2\sigma$, the excess free energy exhibits the scaling form \cite{CD2004} 
\begin{equation}\label{excess}
\beta f^{(p)}_{\rm ex}(\beta,H,L|d,\sigma) = L^{-(d-1)}X^{(p)}_{{\rm {ex}}}(x_\tau,x_h|d,\sigma),
\end{equation}
with 
\begin{equation}\label{excessscaling}
	X^{(p)}_{{\rm {ex}}}(x_\tau,x_h|d,\sigma)=|D_{d,\sigma}|\left[-\frac12x_h^{ 2}\left(
	\frac1{y_L}-\frac1{y_\infty}\right)
	-\frac12x_\tau\left(y_L-y_\infty\right)-\frac\sigma{2d}\left(y_L^{d/\sigma}
	-y_\infty^{d/\sigma}\right) -\frac{1}{2}{\cal K}_{d,\sigma}(y_L)\right].
\end{equation}
In Eq. (\ref{excessscaling}) $y_L$ is the solution of the
spherical field equation for the finite system which is obtained by
minimizing the excess free energy with respect to $y_L$:
\begin{equation} \label{fses}
-x_\tau=\frac{x_h^2}{y_L^2}-y_L^{d/\sigma-1}
-\frac\partial{\partial y_L}{\cal K}_{d,\sigma}(y_L).
\end{equation}
For the bulk system the corresponding equation is
\begin{equation}\label{ises}
-x_\tau=\frac{x_h^2}{y_\infty^2}-
y_\infty^{d/\sigma-1}.
\end{equation}
The function ${\cal K}_{d,\sigma}(y)$ is defined as
\begin{eqnarray}\label{kfun}
{\cal K}_{d,\sigma}(y)=\frac{1}{|D_{d,\sigma}|}\frac{\sigma}{(4\pi)^{d/2}}\sum_{l=1}^{\infty}
\int_0^\infty dx \, x^{-d/2-1}\left[\exp\left(-\frac{l^2}{4x}\right) \right] E_{\sigma/2}\left(-x^{\sigma/2}y\right),
\end{eqnarray}
where $E_a(x)\equiv E_{a,1}(x)$, and
\begin{equation}\label{mittagt}
E_{\alpha,\,\beta}(z)=\sum_{k=0}^{\infty}\frac{z^k}{\Gamma(\alpha
k+\beta)}
\end{equation}
are the Mittag-Leffler functions. For a review of the properties
of $E_{\alpha,\,\beta}(z)$ and of other functions related to them see Refs.~\cite{GM97,B89}, as far as their applications in statistical and continuum
mechanics are considered.

The finite-size scaling
function of the Casimir force \cite{CD2004} for the system under consideration
is
\begin{equation}\label{Casimir-per}
X^{(p)}_{{\rm {Cas}}}(x_\tau,x_h|d,\sigma)=|D_{d,\sigma}|\left[\frac{\sigma+1}{2}x_h^{2}\left(
\frac1{y_L}-\frac1{y_\infty}\right)-\frac{\sigma-1}{2}x_\tau\left(y_L-y_\infty\right)
-\frac{\sigma(d-1)}{2d}  \left(y_L^{d/\sigma}-
y_\infty^{d/\sigma}\right)-\frac12(d-1){\cal K}_{d,\sigma}(y_L)\right].
\end{equation}
In the limit $\sigma\to 2^-$ Eqs.
(\ref{excess})-(\ref{Casimir-per}) reproduce  the corresponding 
expressions for the case of short-ranged interactions
\cite{D96,D98,DK2004}. In this limiting case the above
equations simplify substantially, due to  $E_{1,1}(z)=\exp(z)$, and because 
the function ${\cal K}_{d,\sigma}(y)$ defined in Eq. (\ref{kfun})
reduces to 
\begin{equation}
{\cal K}_{d,2}(y)=\frac{2^{d/2+2}}{|\Gamma(1-d/2)|}y^{d/4}\sum_{l=1}^\infty
l^{-d/2}K_{d/2}(l\sqrt{y}),
\end{equation}
where $K_{\nu}$ is the modified Bessel function.

\subsubsection{ Casimir force for antiperiodic boundary conditions}

The results for anti-periodic boundary conditions  have been reported in Ref. \cite{DG2009}. 

$\blacktriangleright$ \textit{{The case ${{2<d<4}}$}}

In the present context the most general case studied so far is again the film geometry, in which one allows for an anisotropy of the interactions in the system which reflects this geometry, i.e., one takes the interaction constant in the Hamiltonian along the surface ($J_\parallel$), to be distinct from the one perpendicular to the film ($J_\perp$). Accordingly, for the excess free energy and the Casimir force one has

(i) for the excess free energy 
\begin{equation}
\label{fex_standard_final}
  \beta f^{(a)}_{\rm ex}(\beta,L|d,J_\|,J_\perp) = L^{-(d-1)}\left(\frac{J_\perp}{J_\parallel}\right)^{(d-1)/2}  \Bigg \{ \frac{1}{2} x_1(\tilde{y}_L-y_\infty)-\frac{1}{2}\frac{\Gamma(-d/2)}{(4\pi)^{d/2}} \left(\tilde{y}_L^{d/2}-y_\infty^{d/2} \right)-\frac{2\tilde{y}_L^{d/2}}{(2\pi)^{d/2}} \sum_{n=1}^{\infty}(-1)^n \frac{K_{d/2}(n\sqrt{\tilde{y}_L})}{(n\sqrt{\tilde{y}_L})^{d/2}} \Bigg \};
\end{equation} 
(ii) for the Casimir force
\begin{eqnarray}
\label{fcas_standard_final}
  \beta F^{(a)}_{\rm Cas}(\beta,L|d, J_\|,J_\perp)&=& L^{-d}\left(\frac{J_\perp}{J_\parallel}\right)^{(d-1)/2}  \Bigg \{ \frac{1}{2} x_1(\tilde{y}_L-y_\infty)-(d-1)\times\\&&\times\Bigg[\frac{1}{2}\frac{\Gamma(-d/2)}{(4\pi)^{d/2}} \left(\tilde{y}_L^{d/2}-y_\infty^{d/2} \right) \nonumber
   +\tilde{y}_L^{d/2} \frac{2}{(2\pi)^{d/2}} \sum_{n=1}^{\infty}(-1)^n \frac{K_{d/2}(n\sqrt{\tilde{y}_L})}{(n\sqrt{\tilde{y}_L})^{d/2}}\Bigg] \Bigg \}.
\end{eqnarray}
Here $x_1$ is the temperature dependent scaling variable defined in Eq. \eqref{xtdef},
while $\tilde{y}_L(x_1)$ is the solution of the equation 
\begin{eqnarray}\label{yfinal}
- x_1&=&\frac{\Gamma(1-d/2)}{(4\pi)^{d/2}} \tilde{y}_L^{d/2-1}
   +\tilde{y}_L^{d/2-1} \frac{2}{(2\pi)^{d/2}} \sum_{n=1}^{\infty}(-1)^n \frac{K_{d/2-1}(n\sqrt{\tilde{y}_L})}{(n\sqrt{\tilde{y}_L})^{d/2-1}}
\end{eqnarray}
for a film of thickness $L<\infty$ with antiperiodic boundary conditions, while $y_\infty$ is the solution of Eq. \eqref{ybfinal}  (and, thus, it is independent of the boundary conditions).

Equations (\ref{fex_standard_final})-(\ref{yfinal}) reveal  that the Casimir force for antiperiodic boundary conditions in a system with anisotropic interactions can be written as
\begin{equation}\label{Casscaling}
\beta F^{(a)}_{\rm Cas}(\beta,L_\perp|d, J_\|, J_\perp) =L_\perp^{-d} \left(\frac{J_\perp}{J_\parallel}\right)^{(d-1)/2} X_{\rm Cas}^{(a)}(x_1|d),
\end{equation}
\begin{figure}[htb!]
\includegraphics[angle=0,width=\columnwidth]{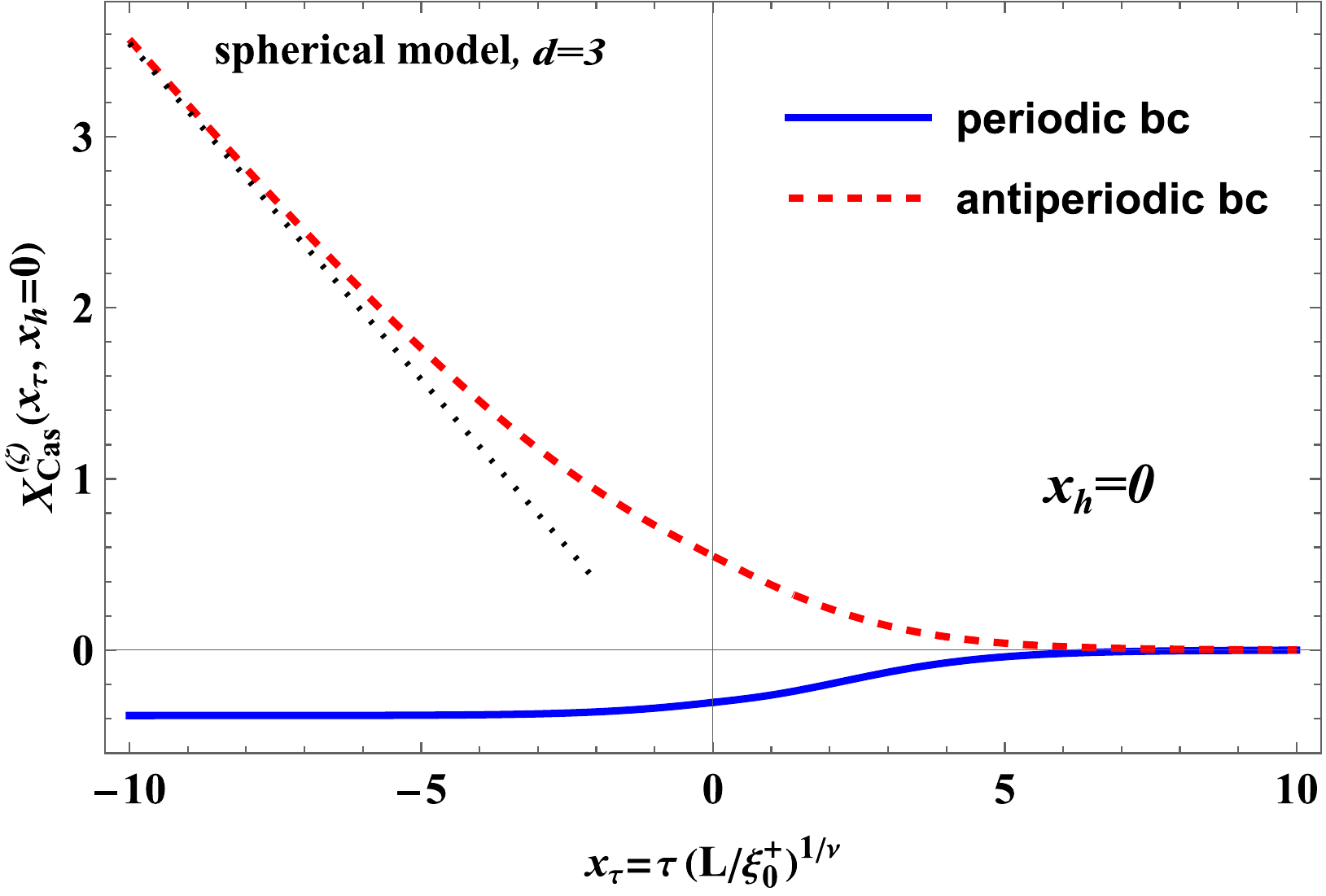}
\caption{The scaling functions $X_{\rm Cas}^{(\zeta)}(x_\tau,x_h=0)$ for the spherical model, for periodic ($\zeta=p$) or antiperiodic ($\zeta=a$) boundary conditions of the critical Casimir forces $F^{(a)}_{\rm Cas}$ and $F^{(p)}_{\rm Cas}$ for three-dimensional, isotropic systems.  The qualitative difference in the behavior of the two scaling functions is due to the contributions stemming from the helicity modulus (see Ref. \cite{DG2009}). This contribution is rather strong and dominates the behavior of the force below $T_c$ for antiperiodic boundary conditions. For all values of $x_\tau$ the force is attractive for periodic boundary condition and repulsive for antiperiodic boundary conditions. The dotted line represents the asymptote $X_{\rm Cas}^{(a)}(x_\tau) \mathop{\simeq}_{x_\tau\to-\infty}-\pi x_\tau/8-\zeta(3)/\pi$ (see Eq. \eqref{CasXas}) of $X^{(a)}_{\rm Cas}$ for large negative values of the temperature-dependent scaling variable. \label{Cas_plot_compare}}
\end{figure}
where $X_{\rm Cas}^{(a)}$ is a universal scaling function, provided a suitable definition of the scaling variable is used (see Eq. (\ref{xtdef})). The scaling variable $x_1$,  as in the case for periodic boundary conditions with short-ranged interactions, is of the form $x_1=a_\tau({\bm b})\, \tau L^{1/\nu}$. This means that, similar to the case of periodic boundary conditions, all effects of the anisotropy can again be incorporated into the prefactor $\left(b_\perp / b_\parallel \right)^{(d-1)/2}=\left(J_\perp / J_\parallel \right)^{(d-1)/2}$ of the scaling function on the r.h.s. of Eq. (\ref{Casscaling}) and into the nonuniversal factor $a_\tau$ which enters the definition of the temperature scaling variable $x_1$.  Concerning the amplitudes of the critical Casimir forces, Eq. (\ref{Casscaling}) leads to the following relation among the amplitudes for the anisotropic and for the isotropic systems:
\begin{equation}\label{CasAmplRel}
\Delta_{\rm Cas}^{(a)}(d| J_\perp,J_\parallel) = \left(\frac{J_\perp}{J_\parallel}\right)^{(d-1)/2} \Delta_{\rm Cas}^{(a)}(d).
\end{equation}
As in the case of periodic boundary conditions, in order to achieve agreement with the relations in Eqs.  \eqref{relaniso} and (\ref{relDeltaGen}) one only has to take into account Eq. \eqref{relxi}. 

$\blacktriangleright$ \textit{{The case ${ d=3}$}} 

In this particular but important case, concerning the scaling function of the critical Casimir force in a system with isotropic interactions,  one obtains the following explicit expression:
\begin{eqnarray}
\label{fcas_standard_final_d3}
  X_{\rm Cas}^{(a)}(x_\tau) &=& \frac{1}{8\pi} x_\tau(\tilde{y}_L-y_\infty)-\frac{1}{6 \pi }\left(\tilde{y}_L^{3/2}-y_\infty^{3/2}\right)
  -\frac{\sqrt{\tilde{y}_L}}{\pi}\text{Li}_2\left[-\exp{\left(-\sqrt{\tilde{y}_L}\,\right)}\right]-
  \frac{1}{\pi}\text{Li}_3\left[-\exp{\left(-\sqrt{\tilde{y}_L}\,\right)}\right],
\end{eqnarray}
where 
\begin{equation}\label{yfd3solsf}
\sqrt{\tilde{y}_L} =2\, \rm{arccosh}\left [ \frac{1}{2}\, \exp{ (x_\tau/2)} \right]
\end{equation}
and \cite{BDT2000} 
\begin{equation}\label{ybfinald3}
\sqrt{y_\infty}=\left \{\begin{array}{cc}
                     x_\tau,  & x_\tau\ge 0\\
                     0, & x_\tau \le 0
                   \end{array}
 \right..
\end{equation}
For the  amplitude of the critical Casimir force in the anisotropic case one finds the exact, closed form
\begin{equation}\label{CasAmAdditional}
\Delta_{\rm Cas}^{(a)}= \left(\frac{J_\perp}{J_\parallel}\right)\left[\frac{1}{3}\rm{Im} \left(
   \rm{Li}_2\left(\mathit{i}^{2/3}\right)\right)-\frac{\zeta (3)}{6 \pi }\right].
\end{equation}
Using the relation $\mathrm{Im}(\mathrm{Li}_2(e^{\mathit{i}\theta}))=\mathrm{Cl}_2(\theta)$ between the polylogarithm and the Clausen function \cite{CC2007}
\begin{equation}
\mathrm{Cl}_2(\theta)=\sum_{k=1}^{\infty}\frac{\sin(k\theta)}{k^2},
\end{equation}
the Casimir amplitude can be expressed as 
\begin{eqnarray}\label{CasAmfinal}
\Delta_{\rm Cas}^{(a)} = \left(\frac{J_\perp}{J_\parallel}\right) \left[\frac{1}{3} {\rm Cl}_2\left (\frac{\pi}{3} \right)-\frac{\zeta
   (3)}{6 \pi }\right] \simeq  0.274543 \left(\frac{J_\perp}{J_\parallel}\right).
\end{eqnarray}
In Fig. \ref{Cas_plot_compare} we present the scaling function $X_{\rm Cas}^{(a)}(x_\tau)$ of the critical Casimir force as a function of the temperature scaling variable $x_\tau$. We observe that $X_{\rm Cas}^{(a)}(x_\tau)>0$ for all $x_\tau$, i.e., the Casimir force for  antiperiodic boundary conditions is always a repulsive force. Furthermore, from Eqs. (\ref{fcas_standard_final_d3}), (\ref{yfd3solsf}), and (\ref{ybfinald3}) one can infer that $x_\tau\gg 1$ implies  $\tilde{y}_L, y_\infty\gg 1$, so that in this regime the scaling function $X_{\rm Cas}^{(a)}(x_\tau)$ decays exponentially, while for $x_\tau \ll -1$ one has
\begin{equation}\label{CasXas}
X_{\rm Cas}^{(a)}(x_\tau\to-\infty)\simeq -\pi x_\tau/8-\zeta(3)/\pi.
\end{equation}
We recall that for $d=3$, periodic boundary conditions,   and  $x_h=0$, the Casimir force is given  by the expression\cite {DK2004}
\begin{eqnarray}
\label{fcas_standard_per_d3}
\beta F^{(p)}_{\rm Cas}&=& L^{-3}
  \Bigg \{ \frac{1}{8\pi} x_\tau(y_L-y_\infty) -\frac{1}{6 \pi }\left(y_L^{3/2}-y_\infty^{3/2}\right)
  -\frac{\sqrt{y_L}}{\pi}\text{Li}_2\left(e^{-\sqrt{y_L}}\right)-
  \frac{1}{\pi}\text{Li}_3\left(e^{-\sqrt{y_L}}\,\right)\Bigg\},
\end{eqnarray}
where $y_L$, in accordance with Eq.  \eqref{eqffSM}, is given by 
\begin{equation}
\label{ylsol}
\sqrt{y_L}=2\, \rm{arcsinh}\left [ \frac{1}{2}\, \exp{(x_{\tau}/2)} \right].
\end{equation}
The comparison between the force for antiperiodic and periodic boundary conditions is shown in Fig. \ref{Cas_plot_compare}. One finds that the contribution of the helicity energy in a system with antiperiodic boundary conditions is so strong, that the Casimir force converts from being completely  attractive for periodic boundary conditions into being entirely repulsive for antiperiodic boundary conditions. This feature can eventually be used for practical purposes, if one applies a certain external ordering field, which  causes the spins, dipoles, etc. at the boundary to order ferromagnetically or antiferromagnetically.  By changing the extent of helicity, the force passes from being attractive, through zero, into being repulsive, as in the case of twisted boundary conditions for the $XY$ model - see Sect. \ref{sec:XY_model_twisted}.

\subsubsection{Casimir force for free boundary conditions}

For this kind of boundary conditions three different versions of the spherical model have been considered (Refs. \cite{DGHHRS2012,DGHHRS2014,DGHHRS2014c,DBR2014,DBR2015}; see Sect. \ref{sec:spherical_model} for the usual definition of the model).  In Ref.  \cite{DBR2014} the authors   investigated a mean spherical model on a simple cubic three-dimensional lattice with the microscopic Hamiltonian
\begin{equation}\label{eq:H}
\mathcal{H}_{\mathrm{M}}=-J\sum_{\langle (i,\mathbf{r}_\|),(i',\mathbf{r}'_\|)\rangle}s(i,\mathbf{r}_\|)s'(i',\mathbf{r}'_\|)+J\sum_i\Lambda_i\bigg(\sum_{\mathbf{r}_\|} s^2(i,\mathbf{r}_\|)-|A|\bigg).
\end{equation}
Here, the first summation is taken over nearest-neighbor spins, $s$ and $s'\in\mathbb{R}$, which lie either in the same layer or in adjacent layers, $i=1,\ldots,L$ labels the layers along the $z$-direction, $\mathbf{r}_\|\in A$ specifies the location of the spin $s(i,\mathbf{r}_\|)$ in each  layer $i$,  $J>0$ is a ferromagnetic interaction constant, and
 $\Lambda_i$ are Lagrange multipliers enforcing the mean spherical constraints $\langle\sum_{\mathbf{r}_\|} s^2(i,\mathbf{r}_\|)\rangle=|A|$, where $|A|$ is the number of spins in area $A$. Free and periodic boundary conditions are applied along the directions perpendicular and parallel to the layers $i$, respectively. The model corresponds to the limit $n\to\infty$ of an $n$-component vector model with  fixed  spin length and with nearest-neighbor coupling $J$. 
 
 In Refs. \cite{DGHHRS2012,DGHHRS2014} two families of $\phi^4$ models called model A and model B were considered\footnote{The notions of model A and model B should not be mixed up with the models A and B describing critical dynamics.}. 
 
 Model B is a lattice model with the reduced Hamiltonian
\begin{equation} \label{eq:Hl}
\mathcal{H}_{\mathrm{B}}=\sum_{\bm{x}={(i,\mathbf{r}_\|})}\bigg[
\frac{1}{2} \sum_{\alpha=1}^{3}  (\bm{\phi}_{\bm{x}+\bm{e}_\alpha}- \bm{\phi}_{\bm{x}})^2
+\frac{\tb}{2} \bm{\phi}_{\bm{x}}^2
+\frac{g}{4!\, n}
 |\bm{\phi}_{\bm{x}}|^4\bigg],
\end{equation}
where $\bm{\phi}_{\bm{x}}$ is a classical $n$-component vector spin at lattice site $\bm{x}=(i,\mathbf{r}_\|)$ with the spatial unit vectors $\bm{e}_\alpha, \alpha=1,2,3$.  We have adjusted the notation of Ref. \cite{DGHHRS2012} in order to facilitate comparison with Ref. \cite{DBR2014}. The boundary conditions are again free and periodic for the directions perpendicular and parallel to the layers $i$, respectively. 

Model A differs from model B in that the coordinates parallel to the layers are taken to be continuous. In Ref. \cite{DGHHRS2012,DGHHRS2014c} it has been shown that the $n\to\infty$ limit of model B is equivalent to $n$ copies of a constrained Gaussian model for a one-component field $\Phi_{\bm{j}}$ with the Hamiltonian
\begin{equation} \label{eq:HG}
\mathcal{H}_{\text{G}}=\frac{1}{2}\sum_{\bm{x}={(i,\mathbf{r}_\|})}\bigg[
\sum_{\alpha=1}^3  ({\Phi}_{{\bm{x}}+\bm{e}_\alpha}- {\Phi}_{{\bm{x}}})^2
+V_i\Phi_{\bm x}^2-\frac{3}{ g}\,(V_i-\tb  )^2
\bigg].
\end{equation}
By taking particular care of the parameters in \eq{eq:HG}, notably if  $n\to \infty$, $g\to \infty$, with the ratio $\tb/g$ fixed, and after carrying out the identifications $\tb/g=-\beta J/6$, $\Phi_{\bm x}\equiv\Phi_{(i,\mathbf{r}_\|})=\sqrt{\beta J}\,s{(i,\mathbf{r}_\|})$ and $V_i=2(\Lambda_i-3)$, one can transfer the 
Hamiltonian $\mathcal{H}_{\mathrm{B}}$ into the microscopic one, i.e.,  $\mathcal{H}_{\mathrm{M}}$.

Before passing to the presentation of some of the numerical results obtained in Refs. \cite{DBR2014,DGHHRS2012,DGHHRS2014}, we  summarize some of the analytical results derived there. First, in Ref. \cite{DBR2014} it has been shown that in the low-temperature regime
\begin{equation}
\label{eq:low_T_regime}
-x_\tau \ K/K_c \gg \ln L 
\end{equation}
the Casimir force $\beta F_{\rm Cas}^{(O,O)}$ is the sum of two terms: that of a leading order, temperature-independent term $\beta F_{\rm Cas,\tau=0}^{(O,O)}(L)$ the  behavior of which can be calculated \textit{exactly}, and of a term which reflects the leading temperature-dependent contribution $\beta F_{\rm Cas,\tau\ne 0}^{(O,O)}(T,L)$ for which the leading $L$-behavior can be found explicitly. For $\beta F_{\rm Cas,\tau=0}^{(O,O)}(L)$ one finds: 
\begin{eqnarray}
\label{F0}
\beta F_{\rm Cas,\tau=0}^{(O,O)}(L)= -\frac{1}{(2\pi)^2}\int_{-\pi}^\pi\int_{-\pi}^\pi dq_x dq_y \frac{\mathsf{v}(q_x,q_y)}{\exp[2L\,\mathsf{v}(q_x,q_y)]-1},  
\end{eqnarray}
where
\begin{equation}
\label{v}
\mathsf{v}(q_x,q_y)=\cosh^{-1}\left[3-\cos q_x-\cos q_y\right]
\end{equation}
\begin{figure*}[t!]
	\begin{center}
		\includegraphics[angle=0,width=\columnwidth]{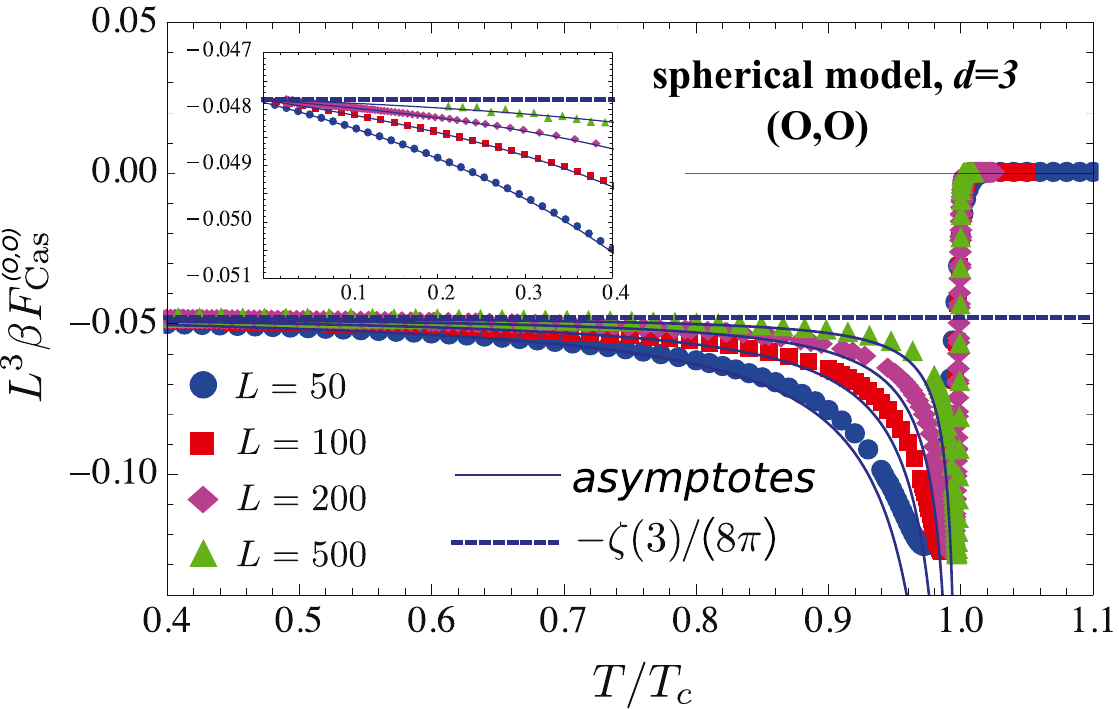}
	\end{center}
	\caption{Numerical results for the scaled Casimir force (symbols) compared with the analytic expressions for the low temperature asymptotes given by thin blue lines  (see Eqs. \eqref{d1} and \eqref{d2}, for $L=50, 100, 200$, and $500$; $L$ is measured in units of the lattice constant). The black thin horizontal line corresponds to zero, and the thick dashed line corresponds to the low temperature limit given by \eq{d1}. The inset focuses on the low temperature behavior. One observes excellent scaling for $T>T_c$.} 
	\label{Figure_AsymptoticComparisonPlot}
\end{figure*}
so that $\beta F_{\rm Cas,\tau=0}^{(O,O)}(L)<0$. Expanding  $\beta F_{\rm Cas,\tau=0}^{(O,O)}(L)$ into powers of $1/L$, one finds 
\begin{equation}
\label{d1}
\frac{\beta F_{\rm Cas,\tau=0}^{(O,O)}(L)}{A}=-\frac{1}{8\pi L^3}\Bigg[\zeta(3)+\frac{7}{8}\zeta(5)L^{-2}
+O\left(L^{-4}\right)\Bigg],
\end{equation}
while for $\beta F_{\rm Cas,\tau\ne 0}^{(O,O)}(T,L)$ one has 
\begin{equation}
\label{d2}
\frac{\beta F_{\rm Cas,\tau \ne 0}^{(O,O)}(T,L)}{A}=-\frac{1}{4(K-K_c)L^4}\Big[a+b\ln L+O\left(L^{-2}\right)\Big],
\end{equation}
with
\begin{equation}
\label{eq:a_and_b_constants}
a=\frac{\zeta'(-2)}{4}\left[2+3K\left(\frac{1}{2}\right)-21\ln 2+6\ln(2\pi)\right]\\
-\frac{3\zeta''(-2)}{4} \simeq 0.0224639,  \quad 
\mbox{and} \quad 
b=-\frac{3\zeta'(-2)}{2}\simeq 0.0456727, 
\end{equation}
where $\zeta$ is Riemann's $\zeta$-function, and $K(k)$ is the complete elliptic integral of the first kind with modulus $k$. We note that  Eq. \eqref{d2} is well defined and can be safely used even at very low temperatures: for $T\approx 0$ one has $K\gg K_c$, and thus $1/(K-K_c)\sim T$.  In addition, Eqs. \eqref{d2} and  \eqref{eq:a_and_b_constants} imply that  $\beta F_{\rm Cas,\tau\ne 0}^{(O,O)}(T,L)<0$.

In Fig. \ref{Figure_AsymptoticComparisonPlot} the range of accuracy of the above analytical expressions is checked against the full numerical results.
The asymptotic results turn out to be accurate for moderately low absolute temperatures, corresponding to  $T \lesssim 0.8 \,T_c$. The larger $L$ the better is the approximation provided by the asymptotes. This is expected, because the product $4\pi(K-K_c)L\gg \ln L$ is the variable which governs the observed behavior. As indicated by the inset, which tracks the scaled Casimir force down to $T=0$, the asymptotic forms, given by Eqs. \eqref{d1} and \eqref{d2}, are quite accurate at low temperatures,  for any $L\gg 1$. In Ref. \cite{DGHHRS2014}  a very precise value of the Casimir amplitude of the force has been  determined numerically\footnote{In Ref. \cite{DGHHRS2014} special care has been taken for the precision of the numerical result of the Casimir amplitude $\Delta_{\rm Cas}^{(O,O)}$. The authors report $\Delta_{\rm Cas}^{(O,O)}= -0.01077340685024782(1)$.}:
\begin{equation}
	\label{eq:Cas_ampl_D_D_SM}
\Delta_{\rm Cas}^{(O,O)}= -0.011(3).
\end{equation}

\begin{figure*}[h!]
 \begin{center}
 \includegraphics[angle=0,width=\columnwidth]{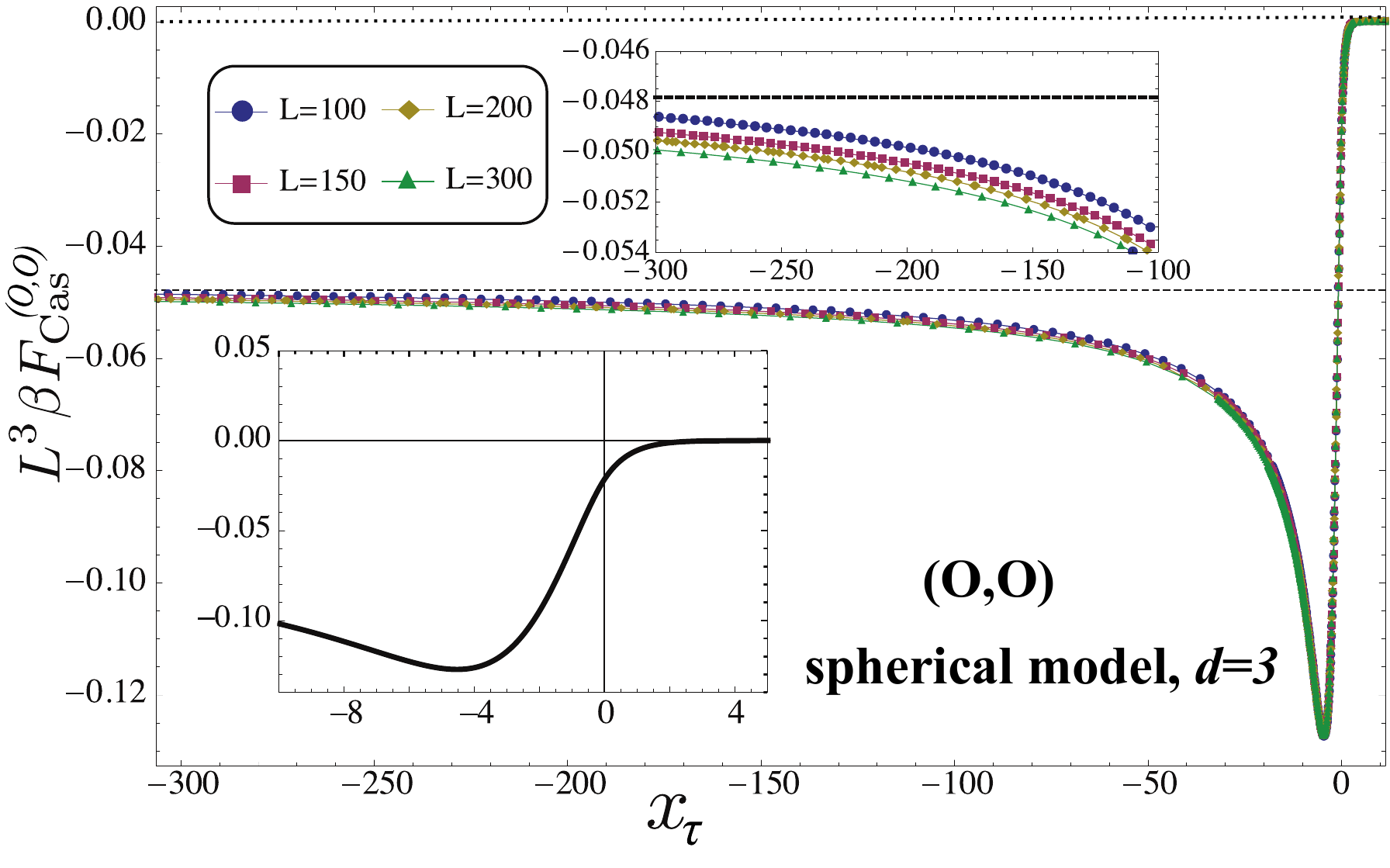}
 \end{center}
 \caption{Numerical results for the scaled Casimir force $L^3\beta F_{\textrm{Cas}}^{(O,O)}$ as a function of the scaling variable $x_\tau=\tau(L/\xi_0^+)^{1/\nu}$ with $\nu=1$ in the temperature region close to and well below the critical temperature $T_c$, allowing corrections to scaling to be only linear in $1/L$.  Data for $L=100,150,200$, and $L=300$ are presented ($L$ is measured in units of the lattice constant). The lower inset shows a blow-up of the region close to $T_c$ and demonstrates the excellent scaling prevailing there in that all curves for different $L$ coincide with a master curve. The upper inset shows a blow-up of the region $x\in(-300,-100)$ and depicts the spread of the scaled curves well below $T_c$. This incomplete scaling is due to the presence  of a term $\propto \ln L$  in that region (see Eq. \eqref{d2}). The zero-temperature limit $-\zeta(3)/(8\pi)$ is indicated as a horizontal dashed line.}
 \label{Figure_Casimir_scaled}
 \end{figure*}

The behavior of the critical Casimir force near $T_c$ is presented
in Fig.  \ref{Figure_Casimir_scaled}. The plotted results agree
with the expected behavior of the Casimir force in systems with broken continuous symmetry \cite{BDT2000,K94}. Specifically, the Casimir force scales as $L^{-3}$ both well
below and near $T_c$; the scaling function of the force tends to a
nonzero constant for $x_\tau \to -\infty$. The force is negative, i.e., it is attractive for all temperatures, as one expects if 
the boundary conditions are the same at both bounding surfaces.

\section{Critical Casimir effect in quantum systems}
\label{CritCasEffectQuantumSystems}

In the previous sections we considered mainly systems described by classical statistical mechanics. In the present context there are, however, also many-body systems which require to be described by quantum statistical mechanics. Among them, there are systems governed by Bose-Einstein or Fermi-Dirac statistics for which, nevertheless, the thermal fluctuations are still of primary importance. In addition, there are systems, in which quantum fluctuations of non-thermal nature are the ones which govern the phase behavior of the system. In Sec. \ref{Cas_Effect_Quantum_Thermal} we review the exact results available concerning the Casimir effect  associated with the first group of systems, and in Sec. \ref{QuantumCritCasEffect} the one concerned with systems belonging to the second group. 

\subsection{Critical Casimir effect in quantum systems driven by thermal fluctuations}
\label{Cas_Effect_Quantum_Thermal}

We start with discussing the results available for Bose gases.

On dimensional grounds (see \eq{CasimirScaling General}) the singular finite-size part of the excess grand potential of the Bose gas per area for boundary conditions $\zeta$ can be expressed as
\begin{equation}\label{eq:fresUp}
\Delta \Omega^{(\zeta)}(T,\mu|d,L)=L^{-(d-1)}\,\Delta X_{\rm ex}^{(\zeta)}(L/\lambda,L/\xi),
\end{equation}
where $\lambda$ is  the thermal de-Broglie wavelength  (\eq{eq:de-Broglie-wavelength})
and  the bulk correlation length $\xi$ diverges upon approaching the Bose-Einstein condensation point. In the case of the ideal Bose~gas, $\xi\equiv \xi_\mu$ is given by \eq{eq:xiid}. Using \eq{eq:fresUp}, one can determine the Casimir force, or 
the reduced Casimir pressure,
\begin{equation}
\beta{F}^{(\zeta)}_{\rm Cas}=-\frac{\partial}{\partial L}\Delta \Omega^{(\zeta)}(T,\mu\,|\,d,L).
\end{equation}
The force exhibits the scaling form
\begin{equation}\label{eq:betaFC}
\beta{F}^{(\zeta)}_{\rm Cas}(T,\mu,L)=L^{-d} X_{\rm C as}^{(\zeta)}(L/\lambda,L/\xi),
\end{equation}
where the scaling function $X_{\rm C as}^{(\zeta)}$ can be expressed in terms of $\Delta X_{\rm ex}^{(\zeta)}$ and its derivatives (compare with \eq{eq:deltaXex-XCasimir}) as
\begin{equation}\label{eq:Ydrel}
X_{\rm C as}^{(\zeta)}(x_\lambda,x_\xi)=\bigg[(d-1)-x_\lambda\frac{\partial}{\partial x_\lambda}-x_\xi\frac{\partial}{\partial x_\xi}\bigg] \Delta X_{\rm C as}^{(\zeta)}(x_\lambda,x_\xi),
\end{equation}
where 
\be
\label{eq:sc_var_Bose}
x_\lambda=L/\lambda, \quad \mbox{and} \quad x_\xi=L/\xi.
\ee 
In the limit  $L/\lambda\gg 1$ one expects that 
\be
\label{eq:limits_Bose}
\lim_{x_\lambda\to\infty}\Delta X_{\rm ex}^{(\zeta)}(x_\lambda,x_\xi)=\Delta X_{\rm ex}^{(\zeta)}(x_\xi)={\cal O}(1),\quad \mbox{and} \quad \lim_{x_\lambda\to\infty} X_{\rm C as}^{(\zeta)}(x_\lambda,x_\xi)=X_{\rm C as}^{(\zeta)}(x_\xi)={\cal O}(1),
\ee
i.e., the corresponding limits exist and are provided by finite functions.

\subsubsection{Ideal Bose gas}

The Casimir force in the ideal Bose gas has been studied in Refs.  \cite{MZ2006,GD2006,DiRu2017}. Below we present the corresponding results for periodic, antiperiodic, Dirichlet, Neumann-Neumann, and Dirichlet-Neumann boundary conditions, which are available in closed analytic form. In Ref. \cite{DiRu2017} results for the Robin boundary condition have been reported. For the latter, the eigenfunctions $e_f(z)$ of the  operator $-\partial_z^2$ are required to satisfy, along the normal $z$~direction of the film, the relations
\begin{equation}\label{eq:BCR}
(\partial_z-c_1)e_f|_{z=0}=0=(\partial_z +c_2)e_f|_{z=L}
\end{equation}
at the boundary planes ${z=0}$ and $z=L$, respectively,  where $c_1$ and $c_2$ are arbitrary non-negative, inverse microscopic lengths.  In this case, the results derived for the scaling functions involve implicit solutions and rather cumbersome expressions.  Therefore we refrain from presenting them here; instead, the interested reader is referred to Ref. \cite{DiRu2017}. With Robin boundary conditions these scaling functions depend on the two additional scaling variables $c_1L$ and $c_2L$. 

Before presenting the explicit expressions for all but Robin boundary conditions, we recall the following  remark made previously. As explained when the scaling functions for the Gaussian model are reported, the expressions for the ideal Bose gas,  say $X_{{\rm ex}, B}^{{\rm (\zeta)}}$, can be immediately obtained from those of the Gaussian model $X_{\rm ex, GM}^{{\rm (\zeta)}}$ via
\begin{equation}
\label{eq:GM_Bose}
X_{{\rm ex}, B}^{{\rm (\zeta)}}(x_\xi|d)=X_{\rm ex, GM}^{{\rm (\zeta)}}(x_\xi^2|d,n=2),
\end{equation}
i.e., the scaling function of the two-component Gaussian model, i.e., $n=2$, coincides, after the suitable re-definition $x_\tau\rightarrow x_\xi^2$ of the scaling variables, with that of the ideal Bose gas studied in the grand canonical ensemble. 

For the  convenience of the reader and in order to avoid misconceptions, for the ideal Bose gas we just and only provide the corresponding expressions for the scaling function of the excess free energy for $2<d<4$. To this end, we mainly use the expressions reported in Ref. \cite{DiRu2017}.

\subsubsection*{$\bullet$ periodic boundary conditions}
For periodic boundary conditions one has (\eq{eq:excess_per_equiv})
\begin{equation}
\label{eq:excess_per_equiv_BG}
X_{{\rm ex},B}^{{\rm (p)}}\left(x_\xi|d\right)=-4\; (2\pi)^{-d/2} x_\xi^{d/2} \sum _{k=1}^{\infty }
\frac{K_{d/2}\left(k x_\xi\right)}{k^{d/2}}.
\end{equation}
The amplitude of the Casimir force is (\eq{eq:Gaus_per_Cas})
\begin{equation}
\label{eq:Gaus_per_Cas_BG}
\Delta_{{\rm Cas},B}^{{\rm (p)}}(d)=-2\; \pi^{-d/2} \Gamma(d/2)\zeta(d)=2\Delta_{{\rm Cas, GM}}^{{\rm (p)}}(d,n=1)=\Delta_{{\rm Cas, GM}}^{{\rm (p)}}(d,n=2).
\end{equation}

\subsubsection*{$\bullet$ antiperiodic boundary conditions}
The amplitude of the Casimir force is  (\eq{eq:Gaus_aper_Cas})
\begin{equation}
\label{eq:Gaus_aper_Cas_BG}
\Delta_{{\rm Cas}, B}^{{\rm (ap)}}=2\; (1-2^{-d+1})\pi^{-d/2} \Gamma(d/2)\zeta(d)=2 \Delta_{{\rm Cas, GM}}^{{\rm (ap)}}(d,n=1)=\Delta_{{\rm Cas, GM}}^{{\rm (ap)}}(d,n=2)
\end{equation}
while the corresponding scaling function of the excess free energy is (\eq{eq:excess_aper_equiv})
\begin{equation}
\label{eq:excess_aper_equiv_BG}
X_{{\rm ex}, B}^{{\rm (ap)}}(x_\xi|d)=4\; (2\pi )^{-d/2} x_\xi^{d/2} \sum _{k=1}^{\infty }(-1)^{k+1}
\frac{K_{d/2}\left(k x_\xi\right)}{k^{d/2}}.
\end{equation}

\subsubsection*{$\bullet$ Dirichlet or ordinary boundary conditions}
The amplitude of the Casimir force in the case of Dirichlet-Dirichlet boundary conditions is  (\eq{eq:Gaus_DD_Cas})
\begin{equation}
\label{eq:Gaus_DD_Cas_BG}
\Delta_{{\rm Cas}, B}^{{\rm (O,O)}}(d)=-\; 2^{-d+1} \pi^{-d/2} \Gamma(d/2)\zeta(d)=2\Delta_{{\rm Cas, GM}}^{{\rm (O,O)}}(d, n=1)=\Delta_{{\rm Cas, GM}}^{{\rm (O,O)}}(d, n=2)
\end{equation}
and the corresponding scaling function of the excess free energy is  (\eq{eq:excess_DD_equiv})
\begin{equation}
\label{eq:excess_DD_equiv_BG}
X_{{\rm ex}, B}^{{\rm (O,O)}}(x_\xi|d)=-4\; ( 4\pi) ^{-d/2} x_\xi^{d/2} \sum _{k=1}^{\infty }
\frac{K_{d/2}\left(2 k x_\xi\right)}{k^{d/2}}.
\end{equation}

\subsubsection*{$\bullet$ Neumann, surface-bulk, or special boundary conditions}
The amplitude of the Casimir force in the case of Neumann-Neumann boundary conditions is 
\begin{equation}
\label{eq:Gaus_SB_Cas_BG}
\Delta_{{\rm Cas},B}^{{\rm (SB,SB)}}(d)=-\; 2^{-d+1} \pi^{-d/2} \Gamma(d/2)\zeta(d)=2\Delta_{{\rm Cas, GM}}^{{\rm (SB,SB)}}(d,n=1)=\Delta_{{\rm Cas, GM}}^{{\rm (SB,SB)}}(d,n=2)
\end{equation}
while for the corresponding scaling function of the excess free energy one has
\begin{equation}
\label{eq:Gauss_SB_SB_OO_BG}
X_{{\rm ex}, B}^{{\rm (SB,SB)}}(x_\xi|d)=X_{{\rm ex}, B}^{{\rm (O,O)}}(x_\xi|d).
\end{equation}

\subsubsection*{$\bullet$ Mixed, or ordinary-special boundary conditions}
The amplitude of the Casimir force in the case of Dirichlet - Neumann boundary conditions is  (\eq{eq:Gaus_O,SB_Cas})
\begin{equation}
\label{eq:Gaus_O,SB_Cas_BG}
\Delta_{{\rm Cas}, B}^{{\rm (O,SB)}}(d)=-2 (1-2^{-d+1}) 2^{-d} \pi^{-d/2} \Gamma(d/2)\zeta(d)=2\Delta_{\rm Cas, GM}^{{\rm (O,SB)}}(d, n=1)=\Delta_{\rm Cas, GM}^{{\rm (O,SB)}}(d, n=2)
\end{equation}
while the corresponding scaling function  of the excess free energy is (\eq{eq:excess_o_sb__equiv})
\begin{equation}
\label{eq:excess_o_sb__equiv_BG}
X_{{\rm ex},B}^{{\rm (O,SB)}}(x_\tau|d)=4\; (2\pi) ^{-d/2} x_\xi^{d/2} \sum _{k=1}^{\infty }(-1)^{k+1}
\frac{K_{d/2}\left(k  x_\xi\right)}{k^{d/2}}.
\end{equation}

\subsubsection{Imperfect Bose gas}

The Casimir force for the imperfect Bose gas has been studied in Refs. \cite{NJN2013,JN2013,NP2011,DiRu2017}. In $d=3$ it has been studied with periodic, Dirichlet, and Neumann boundary conditions \cite{NP2011}. The more general case of a $d$-dimensional system ($2<d<4$)  with periodic boundary conditions has been analyzed in Refs. \cite{NJN2013} and \cite{DiRu2017}. In Ref. \cite{DiRu2017} an $O(n)$ model of an interacting Bose gas with $n$ internal degrees of freedom  has been considered in the limit $n\to\infty$.  For this model the case of Dirichlet-Dirichlet boundary conditions has been studied. This analysis reveals that numerical and analytically exact results for the scaling function in $d=3$ with such boundary conditions follow from those of the
$O(2n)$ $\Phi^4$ model in the limit $n\to\infty$, i.e., from the spherical model with a two-component order parameter. We recall that the thermodynamics of the \textit{ideal} Bose gas is defined only for $\mu<0$ and the condensate forms at $\mu=0$ for $T<T_c$. The thermodynamics of the \textit{imperfect} Bose gas is defined for arbitrary
values of $\mu$, and the region of the $\mu-T$  phase diagram, where the condensate occurs, corresponds to $\mu>\mu_c(T)>0$. In Ref. \cite{NJN2013} it is shown that above the bulk
condensation temperature
the Casimir force decays exponentially as
function of $L$, with the bulk correlation length providing the relevant length
scale. For $T=T_c$ and for $T<T_c$
it decays algebraically. As function of $L/\xi$ the scaling function varies monotonically in any $d\in(2,4)$. These properties are in a full agreement with the corresponding ones for the spherical model \cite{D96,D98}. 

Below we present some explicit expressions for the model of the imperfect Bose gas as reported in Ref. \cite{NJN2013}. 
For $d\in (2,4)$ and below the bulk condensation temperature, i.e., $T<T_c(\mu)$ and for $\hat{\mu}\ge 0$ the scaling function $X_{{\rm ex}, {\rm IB}}(x|d)$ takes the form 
\begin{equation}
\label{eq:ex_f_e_B_int}
-X_{{\rm ex},\, {\rm IB}}(x|d)= \frac{\zeta(d/2)}{4 \pi} x \left[\sigma(x)\right]^2 + \frac{\Gamma(-d/2)}{2^d \pi^{d/2}}  \left[\sigma(x)\right]^d + 
\frac{2^{2-d/2}}{\pi^{d/2}} \sum_{n=1}^{\infty} \, \left[\frac{\sigma(x)}{n}\right]^{d/2} K_{d/2}\left[n\,\sigma(x)\right],
\end{equation}
with $\sigma(x)$ as a solution of
\begin{equation}
\label{eq:sigma_eq}
x \,\zeta\left(d/2\right)\pi^{d/2-1}-\frac{\Gamma(1-d/2)}{2^{d-2}}\left[\sigma(x)\right]^{d-2}=
2^{3-d/2}\left[\sigma(x)\right]^{d/2-1}\sum_{n=1}^{\infty}n^{-(d/2-1)}K_{d/2-1}[n\sigma(x)]\;.
\end{equation}
The scaling variables are 
\be
\label{eq:def_var_Imp_Bose}
x=\hat{\mu}\left(L/\lambda\right)^{d-2}, \quad \hat{\mu}=(\mu-\mu_c)/\mu_c,
\ee
with
\begin{equation}
\mu_c(T)={\rm Li}_{d/2}(1)\left[a/\lambda^d\right] =\zeta\left(d/2\right)\left[a/\lambda^d\right].
\end{equation}
and $a$ as defined in \eq{mf}.
Introducing the identifications (see \eq{yfinal_per})
\be
\label{eq:identif_Imp_B_SM}
\sigma \to \sqrt{y_L}, \quad \mbox{and} \quad -x_1\to x\, \zeta \left(d/2\right),
\ee
one obtains (see \eq{fex_standard_final_per} with $J_\|=J_\perp$ and $y_\infty=0$), that 
\be
\label{eq:rel_Imp_Bose_SM}
X_{{\rm ex},\, {\rm IB}}=2 X_{{\rm ex},{\rm SM}}(x_1|d).
\ee
Equation \eqref{eq:rel_Imp_Bose_SM} states that the scaling function of the imperfect Bose gas is two times the corresponding one for the spherical model, similar to the connection between the scaling functions of the ideal Bose gas and the Gaussian model. The last implies, and it indeed follows, that 
\begin{equation}
\label{x_limit}
\lim_{x\to\infty} X_{{\rm ex},\, {\rm IB}}(x|d)= -2\,\frac{\Gamma(d/2)}{\pi^{d/2}}\,\zeta(d).
\end{equation}
Above the condensation temperature, i.e., $T>T_c(\mu)$, the corresponding expression for the scaling function is reported in Ref. \cite{NJN2013}, but only for the case $x\ll -1$. The result is  
\begin{equation}
\label{eq:ex_f_e_B_int_large_min_x}
X_{{\rm ex},\, {\rm IB}}(x|d)= -
\frac{2^{2-d/2}}{\pi^{d/2}} \sum_{n=1}^{\infty} \, \left[\frac{\sigma(x)}{n}\right]^{d/2} K_{d/2}\left[n\,\sigma(x)\right],
\end{equation}
which follows from \eq{fex_standard_final_per} if one takes into account that, in the considered regime, $y_L- y_\infty \to 0$. In \eq{eq:ex_f_e_B_int_large_min_x} $\sigma(x)$ has to be determined from \eq{eq:sigma_eq}.

\subsubsection{Relativistic Bose gas}

Bose-Einstein condensation can occur only if the particle number is conserved \cite{H87}. Thus, any discussion about the Bose-Einstein condensation for a relativistic Bose gas must also take the  anti-particles into account \cite{HW81,HW82,SP83,SP84,SP84a,SP85c}. Below we shall formulate the corresponding model in the way as used in Ref. \cite{SP85c}. We do this,  because this produces  expressions concerning the free energy in the film geometry, once one  obtains the corresponding results for the Casimir force. 

In Ref. \cite{SP85c} the authors consider an ideal Bose gas composed of $N_1$ particles
and $N_2$ antiparticles, each of mass $m$, confined to a
three-dimensional cuboidal cavity $L_1 \times L_2 \times L_3$ with periodic boundary conditions. Since particles and antiparticles are 
created only in pairs, the system is governed by the conservation
of the number $Q=N_1-N_2$, which
may be considered as a kind of generalized ``charge''. In
equilibrium the chemical potentials of the two species have the same absolute value but the opposite sign, i.e., $\mu_1=-\mu_2=\mu$. With respect to the occupation numbers $N_1$ and $N_2$ this implies 
\be
\label{eq:N1_N2_Bose}
N_1=\sum_{\bk}\left[e^{\;\beta(\varepsilon(\bk)-\mu)}-1\right]^{-1},\,\, N_2=\sum_{\bk}\left[e^{\;\beta(\varepsilon(\bk)+\mu)}-1\right]^{-1},
\ee
where $\varepsilon(\bk)=\sqrt{{\bk}^2+m^2}$. Here we are using the units $\hbar=c=k_B=1$, so that  in the above expression $\beta=1/(k_B T)$.
We stress that here  both $\varepsilon$ and $\mu$ include the rest energy $m$ of the particle, or of the antiparticle.
The condition $|\mu|\le m$ ensures that the mean occupation numbers in the various
states are positive definite.   
There are two case, which are symmetric relative to each other: $\mu>0$ and $\mu<0$. If, for definiteness, one assumes $\mu>0$, it follows that $Q>0$, i.e., $N_1>N_2$. In view of the conservation of $Q$, $\mu$ keeps its sign.  Thus, in what follows we consider $\mu>0$. 

	For periodic boundary conditions, the eigenvalues $k_i, i= 1,2, 3$ of the wave vector $\mathbf{k}$ are given by
$k_i =(2\pi/L_i)\,n_i$, where $n_i =0, \pm 1, \pm 2, . . . $. The pressure $P$ in the grand canonical ensemble \cite{SP85c,PS2016} is 
\begin{equation}
	\label{eq:gran_pot_P}
	P=-\frac{1}{\beta V} \sum_{{\mathbf n}} \left[\ln\left(1-e^{\beta(\varepsilon({\mathbf n})-\mu)}\right)\right]+\left[\ln\left(1-e^{\beta(\varepsilon({\mathbf n})+\mu)}\right)\right], 
\end{equation}
where ${\mathbf n}=\{n_1,n_2,n_3\}$, and $V=L_1 L_2 L_3$. For the film geometry, i.e., in the limits $L_1,L_2\to\infty$ after setting $L_3=L$, the pressure $P$ takes the form
\begin{equation}
	\label{eq:scaling-P-rel-Bose}
	P=\frac{m^4}{2\pi^2}X+L^{-3}\frac{1}{\pi  \beta }\left[2 y_L\,
	{\rm Li}_2\left(e^{-2 y_L}\right)+{\rm Li}_3\left(e^{-2
		y_L}\right)\right],
\end{equation}
where 
\beq
\label{eq:y_L_def-Bose}
y_L=\frac{1}{2}\sqrt{m^2-\mu^2}L, \quad \mbox{and} \quad 
	X(\beta,\mu)=2\sum_{j=1}^\infty \cosh(j\beta\mu) \frac{K_2(j\beta m)}{(j \beta m)^2}.
\eeq
In \eq{eq:y_L_def-Bose} $K_2(z)$ is the modified Bessel function of the second kind,  while  $\mathrm{Li}_n(z)$ in \eq{eq:scaling-P-rel-Bose} is the polylogarithm function, also known as the Jonquière's function - see \eq{eq:Li-def}.
The functions $\mathrm{Li}_n(z)$ are directly related to the  Bose-Einstein functions \cite{PB2011} - see \eq{eq:BE-function}. 
One finds \cite{PB2011} that $\mathrm{Li}_\nu(z)=g_\nu(z)$ for $0\le z<1$. We note that sometimes $\mathrm{Li}_n(z)$ is denoted as $F(z,n)$ or $F_n(z)$ \cite{GB68}. Due to the aforementioned diversity of notations one can encounter results for the Bose gas formulated in terms of distinct, but otherwise equivalent functions. Here, following Ref. \cite{D2020}, we use expressions in terms of the polylogarithm functions  $\mathrm{Li}_n(z)$. As we shall see later, technically this is an important aspect because the available identities for these functions lead to closed form, explicit expressions for the  amplitude of the Casimir force within this model.

In accordance with standard thermodynamic relations, for the "charge" density one has
\begin{equation}
	\label{eq:cgarge_density}
	\rho\equiv \frac{Q}{V}=\left(\frac{\partial P}{\partial \mu}\right)_{T}.
\end{equation}
The bulk critical point $\beta_c$ (i.e., for $L\to\infty$) is determined implicitly by the condition
\begin{equation}
	\label{eq:crit_point}
	\rho=\frac{m^3}{2\pi^2} W(\beta_c,m),
\end{equation}
where 
\begin{equation}
	\label{eq:W_function}
	W(\beta,\mu) = m \left(\frac{\partial X}{\partial \mu} \right)_\beta=2\sum_{j=1}^\infty \sinh(j\beta\mu) \frac{K_2(j\beta m)}{j \beta m}.
\end{equation}

One can show that in the case $\tau L=O(1)$, the scaling function of the excess free energy is
\cite{D2020}
\be
\label{eq:Xex_Bose}
X_{{\rm ex}}(x_\tau)=\frac{1}{\pi}\left[\frac{2 }{3  }\left(y_L^3-y_\infty^3\right)+2 y_L^2\ln \left(1-e^{-2 y_L}\right) -2 y_L
\text{Li}_2\left(e^{-2 y_L}\right)-\text{Li}_3\left(e^{-2
	y_L}\right)\right],
\ee
where $y_L$ is given by 
\begin{equation}\label{eq:y_L_via_x_tau}
	y_L(x_\tau)={\rm arcsinh}\left[\frac{1}{2}\exp\left(\frac{x_\tau}{2\pi} \right)\right],
\end{equation}
while 
\be
\label{eq:yi_Bose}
y_\infty=\left\{\begin{array}{ll}
	x_\tau/(2 \pi), & x_\tau>0 \\
	0, & x_\tau<0.
\end{array}\right. 
\ee
Here the temperature scaling variable is  
\begin{equation}\label{eq:x_tau}
	x_\tau \equiv \beta_c m^2 L
	\left[  
	W(\beta,m)-W(\beta_c,m)
	\right] \\
	\simeq\left( \beta_c^2 m^2\left|\frac{\partial W}{\partial \beta}\right|_{\beta=\beta_c}\right)\, L\tau,\; \tau=\frac{T-T_c}{T_c}. 
\end{equation}

\begin{figure}[h!]
	\includegraphics[width=\columnwidth]{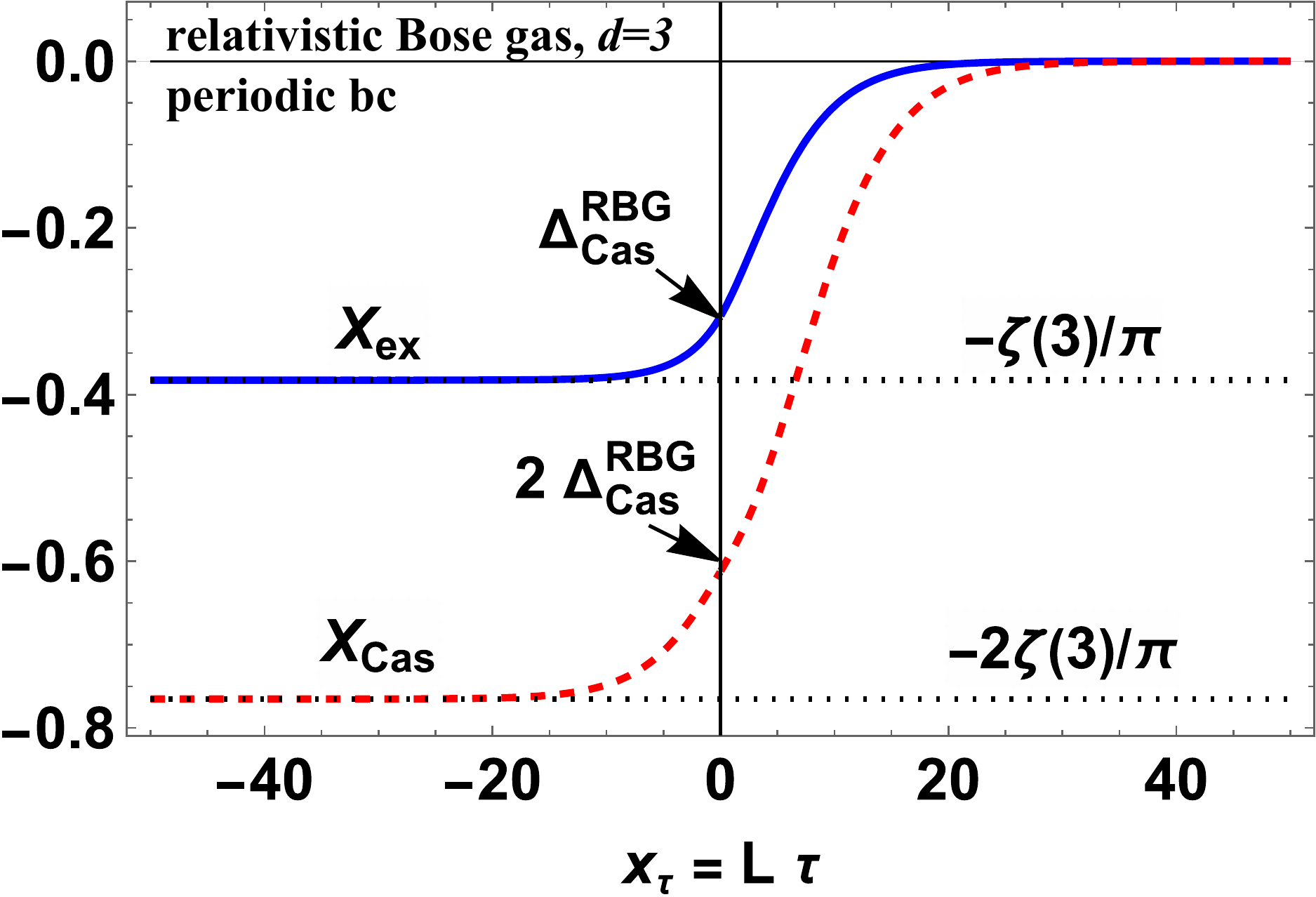}
	\caption{ The behavior of the scaling functions $X_{\rm ex}$ of the excess free energy (see \eq{eq:Xex_Bose}), and $X_{\rm Cas}$ of the Casimir force (see \eq{eq:XCas_final}),  respectively, for the relativistic Bose gas (RBG). Both scaling functions are negative, monotonic, and attain universal negative constants for $x_\tau\to-\infty$ (see \eq{eq:low-temperature-Rel_Bose-gas}) at low temperatures. This is in full agreement with \eq{eq:rel_SM_RBG} and with the results reported before for the spherical model. The amplitude of the Casimir force is $\Delta_{\rm Cas}^{\rm RBG} =(4/5)\times [-\zeta(3)/\pi]$ (see \eq{eq:Cas_ampl_RBG}) where $-\zeta(3)/\pi$ is the asymptotic value of $X_{\rm ex}$ for $x_\tau\to-\infty$ (see \eq{eq:low-temperature-Rel_Bose-gas}).} 
	\label{fig:PsiSRPR}
\end{figure}

It is interesting to compare the above expression for $X_{{\rm ex}}(x_\tau)$ (\eq{eq:Xex_Bose}) with the corresponding one for the spherical model reported in \eq{xecff}. There, expressing  $x_\tau$ in terms of $y_L^{(\rm SM)}$ and $y_\infty^{(\rm SM)}$, by using  \eq{eqffSM}
for the finite-size system, and by using \eq{eqifSM}
for the bulk system, one can show, after implementing the identifications   
\be
\label{eq:transformation}
\dfrac{1}{2}\sqrt{y_L^{\left(\rm SM\right)}}\to y_L^{\left(\rm Bose \right)}, \quad \dfrac{1}{2}\sqrt{y_\infty^{\left(\rm SM\right)}}\to y_\infty ^{\left(\rm Bose\right)},
\ee
that \cite{D2020}
\be
2X_{{\rm ex}}^{\left(\rm SM \right)}=X_{{\rm ex}}^{\left(\rm Bose\right)}.
\label{eq:rel_SM_RBG}
\ee
Equation \eqref{eq:rel_SM_RBG} means that the relativistic Bose gas (RBG) is mathematically equivalent to the two component spherical model. We recall that the same holds also for the imperfect Bose gas \cite{NJN2013,DiRu2017}. For the scaling function of the corresponding Casimir force one obtains the following explicit expression:
\begin{equation}\label{eq:XCas_final}
X_{\rm Cas}^{\rm RBG}(x_\tau)=-\frac{2}{\pi} \bigg [\frac{1}{3}
\left(y_L^3-y_\infty^3\right)+2 y_L \text{Li}_2\left(e^{-2
	y_L}\right) +\text{Li}_3\left(e^{-2 y_L}\right)-y_L^2 \ln
\left(1-e^{-2 y_L}\right)\bigg ].
\end{equation}
One finds $y_L\ge y_\infty$ so that all terms inside the square brackets are positive, i.e., $X_{\rm Cas}(x_\tau)\le 0$. Therefore  the Casimir force for the relativistic Bose gas is always attractive.  The behavior of $X_{\rm Cas}(x_\tau)$ is shown in Fig. 	\ref{fig:PsiSRPR}.
The  corresponding Casimir amplitude is
\begin{equation}\label{eq:Cas_ampl_RBG}
\Delta_{\rm Cas}^{\rm RBG} =-\frac{4\zeta(3)}{5\pi}.
\end{equation}
For $x_\tau\to-\infty$ one has $y_L\to 0$ (see \eq{eq:y_L_via_x_tau}), while $y_\infty=0$ (see \eq{eq:yi_Bose}) and from \eq{eq:Xex_Bose}, and \eq{eq:XCas_final} one obtains  
\begin{equation}
	\label{eq:low-temperature-Rel_Bose-gas} 
X_{{\rm ex}}(x_\tau\to-\infty)= -\frac{\zeta (3)}{\pi } \quad \mbox{and} \quad	X_{\rm Cas}^{\rm RBG}(x_\tau\to-\infty)= -\frac{2\zeta (3)}{\pi }.
\end{equation}

\subsection{Critical Casimir effect in quantum systems driven by quantum fluctuations }
\label{QuantumCritCasEffect}
 Concerning the QED Casimir effect, the corresponding Casimir force is determined by the leading spatial dependence of the interaction between material bodies, caused by changes of the allowed fluctuations of the {\it electromagnetic field} between them. Instead, the thermodynamic Casimir effect describes the corresponding leading distance dependence of the force between such bodies, which is due to changes of the {\it critical fluctuations} of the order parameter of the solvent, induced by the boundary conditions imposed on them. In the first case, the fluctuations are \textit{Gaussian}, whereas in the second case they are \textit{non-Gaussian}. In the latter case this reflects the occurrence  of a critical point in the thermal phase behavior of  the fluctuating medium.   In addition, there is a sort of "mixed" case. In it, one faces circumstances in which the fluctuations are again of quantal character, as in the quantum Casimir effect, but are related to a critical point $g=g_c$ of a certain {\it quantum parameter}\footnote{Any 
 non-analyticity at $g=g_c$ of the ground state energy of the bulk system as function of a certain 
quantal parameter $g$ defines a quantum phase transition
 with respect to $g$. This phase transition is typically related to a 
 qualitative change in the character of the correlations in the ground state.} $g$. Also in this case  critical fluctuations play an important role, which makes the system similar to the thermodynamic Casimir effect.  However these fluctuations are not related to a temperature driven critical point, but to the quantal parameter $g$, with respect to which the system exhibits a critical point. This phenomenon is called {\it quantum critical Casimir effect} (QCC).

Usually, quantum phase transitions involve fluctuations which arise from short-term
changes of the energy as described by Heisenberg's uncertainty principle. The parameter $g$ mentioned above reflects the role of these fluctuations although its definition in specific models can be quite involved and difficult to relate directly  to energy fluctuations.  Nevertheless, due to their quantum nature it is clear quite generally that 
quantum phase transitions tend to occur at $T=0$ or at very low temperatures\footnote{\label{footnote:very-low} This means that $k_B T \ll h \nu$, where $\nu$ is a characteristic frequency of the excitations in the quantum system (see also below).}, at which
quantum fluctuations dominate over thermal fluctuations.  As just stated, because of the temperature induced fluctuations, which are always present at nonzero temperature, a quantum critical point is important only at $T=0$ or for $T$ sufficiently small (see footnote \ref{footnote:very-low}). 

Below we present some general theoretical remarks and definitions concerning the QCC effect.

For a system close to its quantum critical point $g_c$, one can again consider the distance dependence of the interaction between objects immersed in the corresponding  fluctuating quantum field. Naturally,  this leads to the familiar type of the Casimir force described by the corresponding Casimir amplitudes and scaling functions.   In addition one can  study also the leading temperature dependence of the energy of the system, departing from its  ground state at $T=0$ upon increasing the temperature. That leads to {\it temporal} Casimir amplitudes and scaling functions. Finally,  one can consider how a system, close to $g_c$ and of finite extent in at least one spatial dimension, will behave if its temperature is close to $T=0$. In this case, one encounters a combination of the two  effects  mentioned above. Thus, the study of the Casimir effect in a critical quantum system exhibits peculiarities as compared with the properties of classical systems.  This is due to the interplay between the fluctuation induced
forces related to finite temporal or spatial extents. This stems from the presence of a  phase transition at $T=0$, which reveals a particularly rich critical behavior in various regimes at low temperatures.
 
Currently, the study of quantum phase transitions form a very rich and active research topic, both theoretically and experimentally.  In the present review we do not intend in any way to fully cover the richness of this topic.  The interested reader can find the corresponding information in books \cite{NO98,C2001,Sa2011,D2011,Con2017}, and in specialized reviews \cite{H95,SGCS97,KB99,S2000,V2000,V2003,CS2005,S2006,SK2011,S2008,D2011}. Here, we summarize only certain basic concepts and facts, which are needed in order to be able to present results related to the Casimir effect  in such systems \cite{DT99,BDT2000,CDT2000,CTD2000b,V2003,PCC2009,CT2011}. 

Before passing to theoretical details, we point out  that the research activities concerning quantum critical phase transitions  are not only of theoretical interest.  The quantum paraelectric-ferroelectric transition near its quantum critical point serves as an example for the experimental relevance of this topic. More specifically, the dielectric susceptibility of the quantum paraelectric   ${\rm SrTiO}_3$ near its quantum ferroelectric critical point exhibits a temperature dependence $\propto T^{-2}$  for $T\to 0$ \cite{VSIW2004,C2006}. This experimental result can be explained by following the line of arguments presented in, e.g., Ref. \cite{PCC2009} (see below). However, as far as we are aware of, there are no experiments yet aimed at verifying the predictions pertaining to the quantum critical Casimir force. 

In quantum systems \cite{H76} static and
dynamic fluctuations are not independent, because the Hamiltonian $\hat{H}$ determines
not only the partition function, but also the time evolution of any observable $\hat{A}$ via
the Heisenberg equation of motion:
\begin{equation}\label{eq:HEM}
i\hbar \frac{d \hat{A}}{dt}=\left[\hat{A},\hat{H}\right].
\end{equation}
Therefore, in quantum systems the energy $E_c$, associated with the correlation (lifetime or equilibration) time $t_c$, is also the typical fluctuation energy for static fluctuations. If one considers time-dependent phenomena in the vicinity of a critical point $T_c$, one observes that temporal correlations of the
order parameter decay slower upon approaching the critical point $T_c$. This phenomenon is called \textit{critical slowing down}. The typical lifetime of temporal
order-parameter fluctuations is called correlation time $t_c$. One usually observes that
in the vicinity of a critical point, it diverges as a power law:
\begin{equation}\label{eq:corr_time}
t_c\propto \xi_\tau^z\propto |\tau|^{-\nu z},
\end{equation}
where $z$ is the so-called \textit{dynamical critical exponent} \cite{HH77}. Accordingly,  the typical energy scale $E_c$ associated with
temporal fluctuations is 
\begin{equation}\label{eq:typ_energy}
E_c=\hbar /t_c \propto |\tau|^{\nu z}\propto (T-T_c)^{\nu z}.
\end{equation}
 Quantum effects are unimportant and classical fluctuations are dominant
as long as $E_c \ll k_B T_c$.  Naturally,
$E_c \ll k_B T_c$ always holds  sufficiently close to $T_c>0$,
i.e., the classical critical behavior dominates in the temperature region
close to $T_c$, provided $T_c>0$.  However, the width of this region
shrinks to zero upon lowering $T_c$. In other words, if one has at
ones disposal a parameter which can be used to lower $T_c$, one can attain
a regime in which the quantum critical behavior dominates. If a
zero-temperature phase transition takes place,
the quantum critical regime occurs close to $T=0$. These so-called quantum phase transitions can be associated
with  nonanalyticities  of ground state properties of the system at $g = g_c$. The corresponding point in the relevant parameter space is called a quantum critical point. Among the interesting phenomena in condensed matter systems, 
related to quantum critical points, are: \textit{(i)} Anderson localization; \textit{(ii)} quantum Hall effect; \textit{(iii)} paramagnetic–ferromagnetic transition of uniaxial ferromagnets in a transverse magnetic field; \textit{(iv)} Mott–Hubbard transition; \textit{(v)} Fermi-liquid spin-density wave transition. On any of these topics there is a  plethora of research reports. For this reason we simply refer the interested reader to the reviews and books on the quantum critical phenomena and to the references therein \cite{H95,SGCS97,S2000,V2000,C2001,V2003,CS2005,S2006,S2008,Sa2011,D2011,SK2011,Con2017}. We stress that all these phase transitions are  realized by tuning some of the coupling constants in the Hamiltonian instead of temperature.

Near a quantum critical point (QCP) temperature provides a low energy  cutoff for quantum fluctuations. On general grounds, the associated finite time scale can be seen as appearing through the uncertainty relation  $\Delta t\sim \hbar/(k_B T)$. This "cutoff" manifests itself as "finite size"; the clearest way to see  this is via Feynman's path integrals. Within this formalism, at a quantum critical point \cite{C96,C2001,SGCS97,Sa2011,CS2005} temperature plays the role of a finite-size effect in time. Using the pathintegral description of the \textit{quantum} statistical mechanics of such systems, considerable insight is gained concerning the understanding of their behavior by reducing the corresponding mathematics to that one of \textit{classical} statistical mechanics of a system, in which time appears as an extra dimension. In particular, this allows one to deduce scaling properties for the nonzero-temperature behavior, which is then formulated in terms of the theory of finite-size scaling. Indeed, the expression for the partition function of quantum systems as a Feynman path integral in
imaginary time $\tau ={\rm i}\,t$ (see, e.g., Refs. \cite{P88,C96,BDT2000}) has the form 
\begin{equation}\label{partition11}
{\cal Z}=\int{\cal D}[p]{\cal D}[q]\exp\left[-{1\over \hbar}\int_0^{\beta\hbar}
{\rm d}\tau {\cal L}\{p,q \}\right],
\end{equation}
where ${\cal L}$ is the Lagrangian, and the functional integral is over the degrees of freedom $\{p\}$ and $\{q\}$. The boundary condition in the imaginary time direction is periodic for bosonic and antiperiodic for fermionic degrees of freedom. The Lagrangian itself is a lattice sum (or integral) over a local Lagrangian density.  Formally, it can be regarded as the partition function for a $(d+1)$-dimensional classical system in slab geometry of thickness 
\begin{equation}
	\label{eq:L-tau}
	L_\tau=\beta\hbar.
\end{equation}
In the limit $\beta \hbar \to \infty$ one obtains a ``purely classical", $(d+1)$-dimensional system and, hence, the deviation of the temperature from the quantum critical point at $T=0$ can
be considered as a finite-size effect with $L_\tau$ being large, but finite. Taking into account that  in the vicinity of a quantum critical point, characterized by the dynamic exponent
$z$, time scales as (length)$^z$ (see \eq{eq:corr_time}), one concludes that the scaling properties of the quantum system are
identical with those of an effective classical system in spatial dimension $d_{\rm eff}=d + z$ (see below). This crossover from $d$ to $d+z$ spatial dimensions is called dimensional crossover rule. Since typically $z>0$ one has $d_{\rm eff}>d$. Thus, it can happen that a quantum critical system with dimension $d$ exhibits Gaussian behavior with mean-field critical exponents at $T=0$ if $d_{\rm eff}>d_u$, where $d_u$ is the upper critical dimension of the corresponding classical system.  
\begin{figure}[h!]
	\includegraphics[width=\columnwidth]{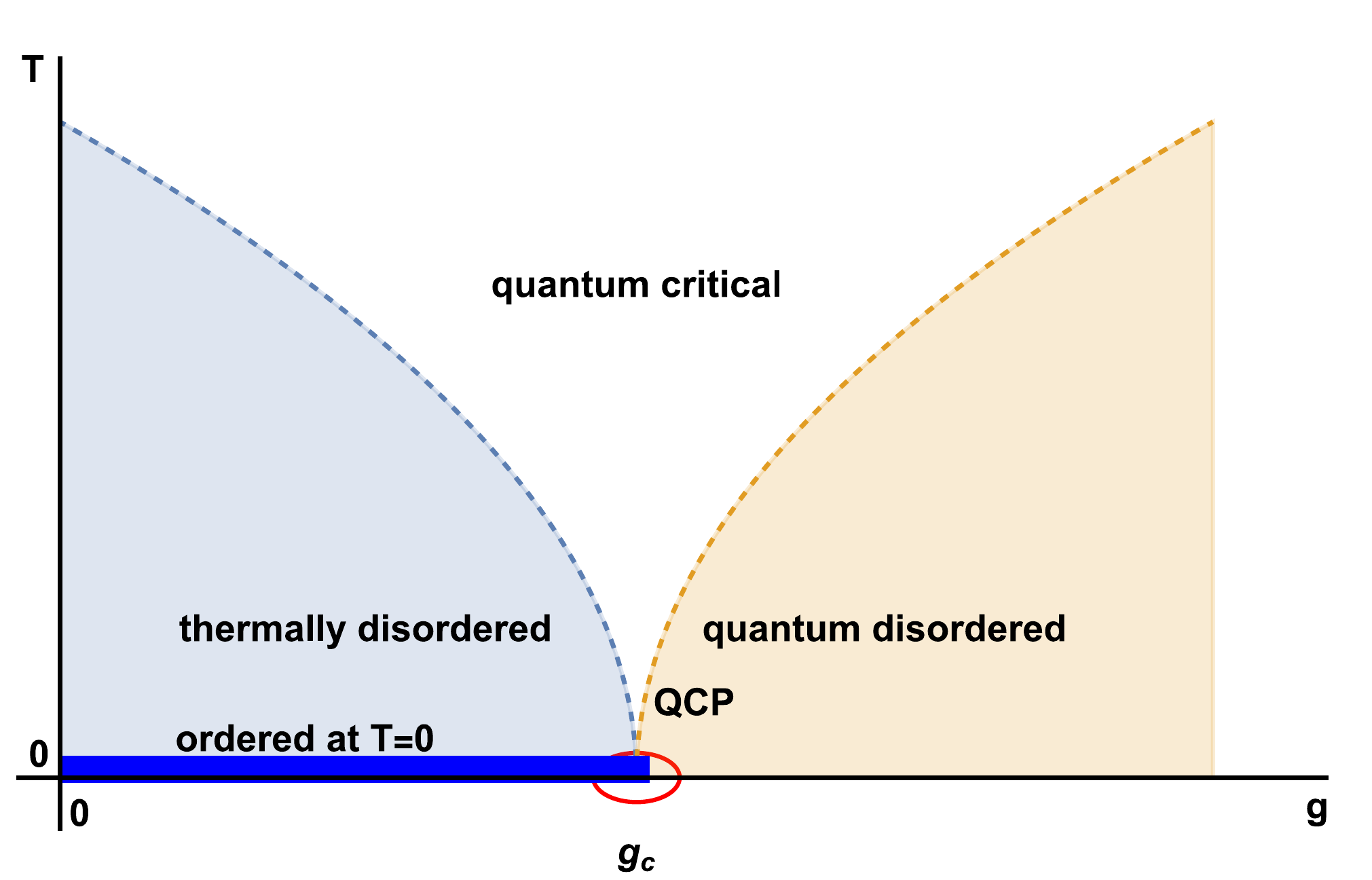}
	\caption{ Schematic phase diagram in the vicinity of a quantum critical point (QCP). The horizontal axis represents the control parameter $g$ used to tune the system through the quantum phase transition, and the vertical axis is the temperature $T$. The  long-ranged order is  present only at $T=0$ for $g<g_c$. The dashed crossover lines indicate the boundaries $k_B T\propto |g-g_c|^{\nu z}$ of the quantum critical region in which the leading quantum critical singularities are observed. The case shown corresponds to $\nu z<1$.  }
	\label{fig:QPDT0}
\end{figure}

\begin{figure}
	\includegraphics[width=0.88\columnwidth]{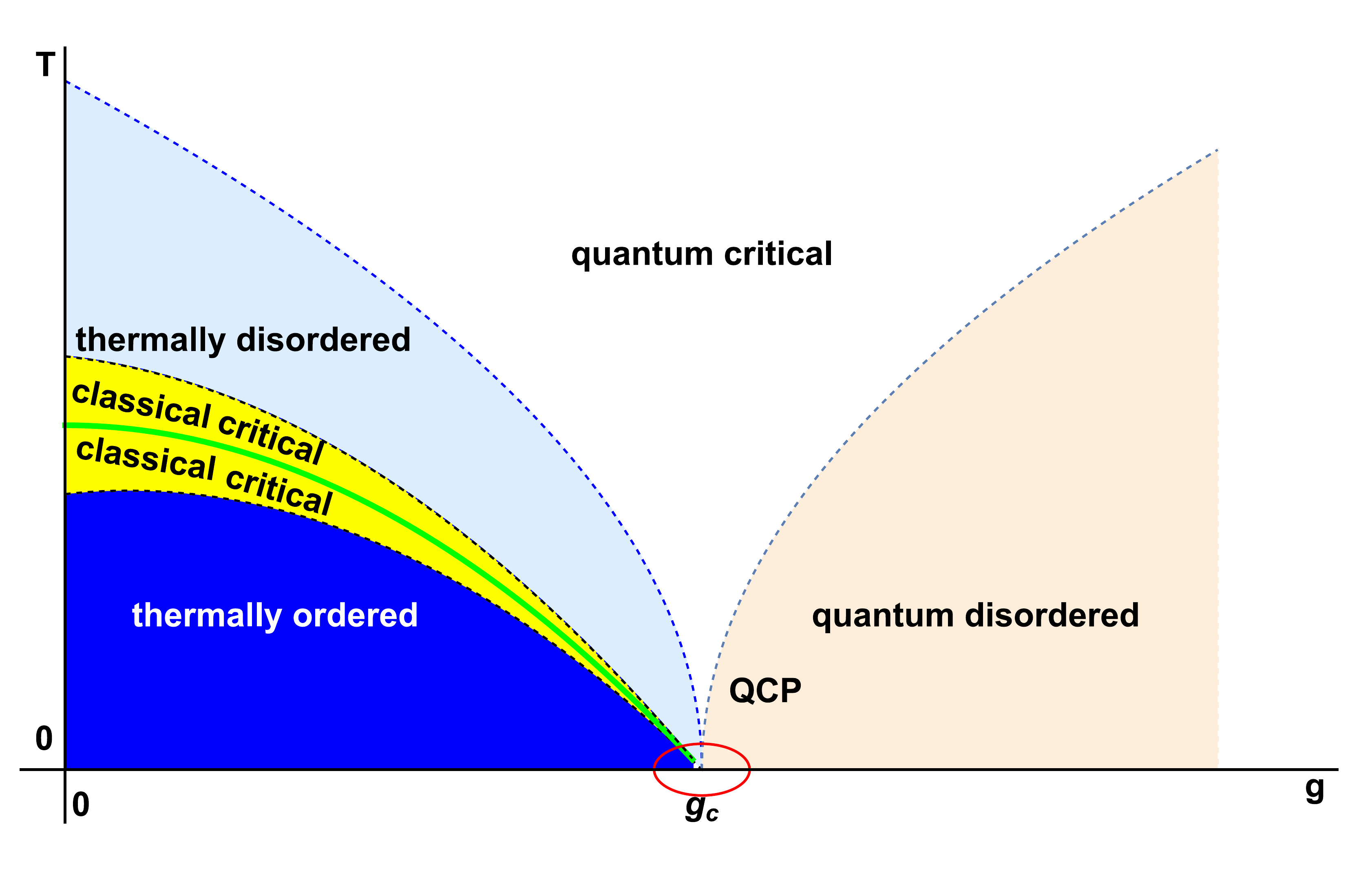}
	\caption{ The phase diagram corresponds to the case in which order can prevail also at \textit{nonzero} temperatures. The thin solid line $T_c(g)$ marks the nonzero-temperature boundary between the thermally ordered and the thermally disordered phase. Close to this line of singularities, the critical behavior is the \textit{classical} (i.e., governed by the temperature and not by the quantum parameter $g$) one. Here, the notion "classical critical behavior" should not be mistaken for the Gaussian, or a mean-field like, behavior.  
	}
	\label{fig:QPDnonT}
\end{figure}

The schematic phase diagrams for systems with a quantum critical point, which order at $T=0$ or at $T\ne 0$, are shown in Figs. 	\ref{fig:QPDT0} and \ref{fig:QPDnonT}, respectively. The plots are in accordance with Refs. \cite{Sa2011,V96,V2000,V2003,KBS2010}. From the quantum critical point two lines emanate, which delineate the so-called quantum critical region. There the physical properties are dominated by the \textit{thermal} excitations of the quantum critical ground state. This region is
bounded by a thermally disordered domain and a quantum disordered domain, respectively. The case of a system, which exhibits long-ranged order only at zero temperature, is shown in Fig. \ref{fig:QPDT0}, while the case, in which long-ranged order at nonzero temperature can also be seen, is shown in Fig. \ref{fig:QPDnonT}. There, the corresponding phase diagram contains,  in addition, a regime where \textit{classical} critical behavior can be observed.

\subsubsection{Finite-size behavior of the excess free energy, Casimir force, and Casimir amplitudes}

In this section we focus on certain more concrete predictions concerning the behavior of the excess free energy, the Casimir force, and the Casimir amplitudes, thereby following the above line of arguments. We consider quantum systems in the film geometry $L\times
\infty ^{d-1}\times L_\tau$, where $L_\tau=\hbar /(k_BT)$ (see \eq{eq:L-tau}) is the finite extent in the temporal (imaginary time) direction.  We suppose that various boundary conditions are imposed across the finite spatial extent such as periodic ones.  

For simplicity, in the remainder we set $\hbar =k_B=1$.
The free energy  per area of the finite system in units of $k_B T$ is denoted as  $f(T,g,H|L,d)$;  $f_b(T,g,H|d)$ is the corresponding bulk expression (per volume), and $H$ is an external ordering field. The parameter $g$  governs its phase transition at $T=0$, and $g=g_c$ denotes its quantum critical point. In accordance with the dimensional crossover rule explained above, the
Privman - Fisher hypothesis  for finite classical systems \cite{PF84} (see \eq{PFhypot} as well as Sects. \ref{sec:FSS_free_energy}  and  \ref{sec:Cas_definition_simplest}),
can be applied to the quantum case. This leads to \cite{C94,CDT2000,CTD2000b,CPV2014,Con2017}
\begin{equation}
L^{-1} f_{\rm ex}(T,g,H,L|d)=(TL_\tau)L^{-(d+z)}X_{{\rm ex}}
(x_1,x_2,\rho|d) \label{hypot1}
\end{equation}
with the finite-size scaling variables
\begin{equation}
x_1=L^{1/\nu} \delta g,\quad x_2=hL^{\Delta/\nu}, \quad \rho = L^z/L_\tau .
\end{equation}
Here (see \eq{excess_free_energy_definition})
\begin{equation}
\label{eq:Xex-standard-quantum}
f_{\rm ex}(T,g,H,L|d)=f(T,g,H|L,d)-L f_b(T,g,H|d),
\end{equation}
$\Delta$ and $\nu$ are the
usual bulk critical exponents, $h\propto H$ is a suitably normalized, 
external (magnetic) field $H$, $\delta g \propto g-g_c$, and $ X_{{\rm ex}}$ is the
universal scaling function of the excess free energy. According to the
definition of the Casimir force (\eq{CasDef}), one obtains
\begin{equation}
F_{L,{\rm Cas}}(T,g ,H|d)=(TL_\tau)L^{-(d+z)}X_{{\rm Cas}}(x_1,x_2,\rho|d),
\label{defCasimir}
\end{equation}
where the {\it universal} finite-size scaling function $X_{\rm Cas}(x_1,x_2,
\rho|d)$ of the Casimir force can be expressed in terms of the excess free energy
$X_{\rm ex}\equiv X_{\rm ex}(x_1,x_2,\rho|d)$: 
\begin{equation}
X_{\rm Cas}(x_1,x_2,\rho|d)=-(d+z)X_{{\rm ex}}-\frac 1\nu x_1\frac{\partial
X_{\rm ex}}{\partial x_1} +\frac\Delta \nu
x_2\frac{\partial X_{\rm ex}}{\partial x_2}+z\,\rho
\frac{\partial X_{\rm ex}}{\partial \rho}.
\label{DefXCasimir}
\end{equation}
It follows from Eq. (\ref{defCasimir}) that one can consider the general case
of the  amplitudes of the Casimir force, which are functions of  the scaling variable $\rho $:
\begin{equation}
\Delta_{\rm Cas}(\rho|d) = X_{{\rm ex}}(0,0,\rho|d). \label{DeltaDef}
\end{equation}
The classical amplitudes $\Delta_{\rm Cas}^{(a,b)}$ introduced by Eq.
(\ref{Catcrittemp}) for $(a,b)\equiv (p)$ (periodic boundary conditions)
are special cases of $\Delta_{{\rm Cas}}(\rho|d)$ for  $\rho =0$, i.e., at
$T=0$. According to the dimensional crossover rule,
with respect to its scaling behavior the quantum system is formally equivalent to a $(d+z)$-dimensional classical one. More precisely, the dimensional crossover rule asserts that the critical singularities
with respect to $g$ of a $d$-dimensional quantum system\footnote{A direct consequence of the equivalence between a
quantum system and a classical system with an extra “temporal” dimension is, that along with the common spatial correlation
length $\xi_g$, which diverges upon approaching the phase transition as $\xi_g\sim (\Delta g)^{-\nu}$, where $\Delta g\equiv (g-g_c)/g_c$ is the deviation of the control parameter
from its critical value, there appears a distinct correlation “length”, $\xi_\tau \sim \xi_g^z$ in the time direction; $\nu$ and $z$ are
the correlation length exponent and the dynamic scaling exponent, respectively. In such systems the dispersion relation is $\omega=c\, q^z$ \cite{PCC2009}; the value $z=1$ of the dynamical critical exponent  refers to a linear dispersion relation $\omega \simeq c\,q$ with momentum $q$. Formally, starting from the standard finite-size scaling theory (see Chapter \ref{FSS_theory}), \cite{PCC2009,VRV2015,SGCS97} one adopts the substitutions  $L\to L_\tau, (T-T_c)/T_c\to \Delta g\equiv (g-g_c)/g_c, \xi\sim[(T-T_c)/T_c]^{-\nu}\to \xi_g\propto (\Delta g)^{-\nu}$, and $\xi_\tau=\xi_g^z\propto \Delta g^{-z\nu}$ and  obtains that, near $T = 0$  and about
$g_c$, the $d$-dimensional quantum system with respect to scaling is formally equivalent to a classical system with dimensionality $d+z$, i.e., with $d$ finite spatial and $z$ finite temporal dimensions, and with a critical temperature $T_c > 0$.} at $T = 0$  around
$g_c$ are formally equivalent to those of a classical system with dimensionality $d+z$  and a critical temperature $T_c > 0$. This facilitates to investigate low-temperature effects (by considering an effective system with $d$ infinite spatial and $z$ finite temporal dimensions) within the framework of  finite-size scaling theory. 

In addition to the above statements concerning the standard excess free energy and the amplitudes of the Casimir force,
one can introduce analogously the ``temporal excess
free energy density'' $f^{\,\rm t}_{\rm ex}$
\begin{equation}
f^{\,\rm t}_{\rm ex}(T,g,H|d)=f_b(T,g,H|d)-f_b(0,g,H|d) \label{deffext}
\end{equation}
and the ``temporal Casimir amplitudes''
\begin{equation}
f^{\,\rm t}_{\rm ex}(T,g_c,0|d) =TL_\tau^{-d/z} \Delta_{\rm Cas}^{\rm t}(d).
\label{Deltatdef}
\end{equation}
Whereas the common amplitudes characterize the leading 
corrections in $L$ to the bulk free energy density at the critical point, the
``temporal amplitudes'' determine the leading temperature-dependent
corrections of the ground state energy for an {\it infinite} system at
its quantum critical point $g_c$. 

If the quantum parameter $g$ is close to $g_c$, one expects
\begin{equation}
f^{\,\rm t}_{\rm ex}(T,g,H|d) =TL_\tau^{-d/z}X_{\rm ex}^{\rm t}(x_1^{\rm t},
x_2^{\rm t}|d),
\label{defxextgen}
\end{equation}
i.e., one has a scaling function
$X_{\rm ex}^{\rm t}(x_1^{\rm t},x_2^{\rm t}|d)$, which is
the analogue of $X_{\rm ex}(x_1,x_2,\rho)$. Here, the scaling variables  are 
\begin{equation}
x_1^{\rm t}=L_\tau^{1/(\nu\, z)} \Delta g \quad \mbox{and}\qquad x_2^{\rm t}=hL_\tau^{\Delta /(\nu\, z)},
\end{equation}
so that 
\begin{equation}
\Delta_{\rm Cas}^{\rm t}(d)= X_{\rm ex}^{\rm t}(0,0|d). \label{defCastgen}
\end{equation}
We finally note that if $z=1$, the temporal excess free energy
coincides, up to a (negative) normalization factor, with the one proposed by Neto and Fradkin \cite{NF93} concerning the
nonzero temperature generalization of the $C$-function of Zamolodchikov 
(see also Ref. \cite{DT99}).

The predictions in Eqs. \eqref{hypot1} -- \eqref{defCastgen} can be checked in detail by using the $d$-dimensional quantum spherical model as a paradigm (see below) \cite{BDT2000,CDT2000,CTD2000b}. Detailed results for the Casimir force are also available for a few variants of the Bose gas \cite{MZ2006,GD2006,NJN2013,JN2013,NP2011,DiRu2017}. Some related results are also available for certain  topological phase transitions \cite{GC2018}. We stress, however, that the two lengths $L$ and $L_\tau$ have different physical meaning: $L$ is a spatial length, while $L_\tau$ is an effective one which is proportional to the inverse temperature (see 	\eq{eq:L-tau}). Thus, the derivative of $f_{\rm ex}$ with respect to $L$ and of $f_{\rm ex}^t$ with respect to $L_\tau$ have also a very distinct meaning: the first corresponds to a force conjugate to $L$  (the Casimir force), while the second one is related to the behavior of the internal energy (such as the temperature corrections to the ground state of the system). Thus, the measurements related to these quantities require very different approaches. Since the occurrence of quantum phase transitions is a quite involved topic, it is not yet obvious what is the best actual physical system representing them and  what is the way in which such corresponding measurements can be performed. Future studies are expected to address such issues. Currently, $^3$He films seem to be good candidates for such type of measurements (see, e.g. Ref. \cite{SIK2016}). 

\subsubsection{Casimir amplitudes for the  $d$-dimensional quantum spherical model}
\label{sec:Cas_ampl_long_ranged}

We present some results concerning the amplitudes defined above, which are derived within 
the $d$-dimensional quantum spherical model (see Sec. \ref{sec:QSM_def}) and within the $O(n)$ quantum $\varphi^4$ model in
the limit $n \to \infty$. These two models belong to the same universality class and offer the possibility to investigate the interplay of quantum and classical
fluctuations in an exact manner \cite{BDT2000,CDT2000,CTD2000b}. We stipulate that  these models are governed by algebraically decaying interactions $\propto r^{-(d+\sigma)}$, where $0<\sigma \leq 2$ controls how rapid the  interaction\footnote{The  case $\sigma=2$ provides formally the results pertinent to the case of short-ranged interactions.} decreases as function of  the distance $r$. In this case the dynamic critical exponent is $z=\sigma/2$ \cite{V96,CDT2000} (see \eq{eq:corr_time}), which provides the opportunity analytically to study systems with $z=1$, i.e., with linear dispersion relation, as well as those with $z\ne 1$, i.e., with nonlinear one. 

For these models one obtains the following results:

\textit{(i)} amplitudes of the Casimir forces:

\textit{(a)} if $d=\sigma=2$,
\begin{equation}
\Delta _{{\rm Cas}}\left(\rho= 0|d=2\right) =-\frac{2\zeta (3)}{
5\pi }\approx -0.153051. \label{qsmCA}
\end{equation}
Here, $\zeta (x)$ is Riemann's zeta function. This result coincides with that one of the classical spherical model \cite{D98} with periodic boundary conditions and $d=3$ (see \eq{delta}). The limit of large $n$ of the so-called
2+1 Gross-Neveu model (GN) \cite{CT2011}, which represents a broader
class of four fermionic models, is mathematically  
very similar to that of the three-dimensional spherical model. 
The corresponding Casimir amplitude $\Delta_{\rm Cas}^{GN}$ which is exactly equal in absolute values but opposite in sign 
to the Casimir amplitude of the three-dimensional spherical
model, subject to antiperiodic boundary conditions \cite{DG2009} (see \eq{CasAmfinal}). One has 
\begin{eqnarray}\label{GN}
\Delta_{\rm Cas}^{GN}&=&-\left[\frac{1}{3} {\rm Cl}_2\left (\frac{\pi}{3} \right)-\frac{\zeta
   (3)}{6 \pi }\right] =-0.274543.
\end{eqnarray}
\textit{(b)} if $d=\sigma=1$,
\begin{equation}
\Delta_{\rm Cas}(\rho=0|d=1)=-0.3157.
\label{deltasigma1}
\end{equation}

\textit{(ii)} "temporal" Casimir amplitudes

From  \eq{Deltatdef} one obtains a general expression for the
temporal Casimir amplitudes of the present model, for a
system with the  geometry $\infty ^d\times L_\tau$ and at the quantum critical point $(\lambda=\lambda_c, h=0)$:
\begin{eqnarray}
\Delta _{{\rm Casimir}}^{{\rm t}}\left(d,\sigma \right) =
-\frac{k_d}{4\sqrt{\pi }\sigma }\Gamma \left(
\frac d\sigma \right) \Gamma
\left( -\frac 2d\sigma -\frac12\right)y_0^{d/z+1} -\frac{k_d}{\sigma \sqrt{\pi }}\Gamma \left( \frac
d\sigma
\right)\sum_{m=1}^\infty\frac{\left( 2y_0^2\right) ^{d/\sigma+1/2}K_{d/\sigma + 1/2}\left( m y_0\right) } {\left( m
	y_0\right) ^{d/\sigma +1/2}},
\label{xexter}
\end{eqnarray}
where $k_d^{-1}=(4\pi)^{d/2}\Gamma(d/2)/2$ and  $K_\nu(x)$ is the
Bessel function of the second kind (also known as MacDonald's function).
The scaling variable $y_0$ is the solution of the equation which can be obtained by requiring  the partial derivative of the r.h.s. of the above expression with respect to $y_0$ to be equal to zero\footnote{For details see Eqs. (16), (17), and (26) in Ref. \cite{CDT2000}.}. In the general case, this solution can be found only numerically. However, in the
particular case $d/\sigma=1$, Eq.~(\ref{xexter}) simplifies considerably;  for $0<\sigma \leq 2$ one obtains 
\begin{equation}
\Delta _{{\rm Cas}}^{{\rm t}}(d=\sigma)=-\frac{16}{5\sigma }
\frac{\zeta (3)}{(4\pi )^{\sigma /2}}\frac 1{\Gamma (\sigma /2)}, \quad 0<\sigma \leq 2.
\label{delta_eq_sigma}
\end{equation}
If $d=\sigma=2$, the ''temporal" Casimir amplitude $ \Delta
_{{\rm Cas}}^{{\rm t}}$ reduces to the ''normal'' Casimir amplitude $\Delta _{{\rm
Cas}}\left( \rho=0\right) $, given by Eq.~(\ref{qsmCA}). This reflects the existence of a special symmetry in that case between the ''temporal'' and the spatial  dimensions of the system (they are then both characterized by the same spectrum $\propto k^2$).

One can derive the following relation between the temporal amplitudes:
\begin{equation}\label{dec}
\frac{\Delta _{{\rm Cas}}^{{\rm t}}(d=\sigma )}{\Delta _{{\rm
Cas}}^{{\rm t}}(d=\sigma=2)}=\frac{8\pi}{\sigma(4\pi)^{\sigma/2}
\Gamma(\sigma/2)}.
\end{equation}
The r.h.s. of Eq. (\ref{dec}) is a
decreasing function of $\sigma$.

\subsubsection{Scaling function of the Casimir force for the $d$-dimensional quantum spherical model}

$\bigstar$ \textit{{The case of short-ranged interaction}}

The temporal excess free energy density, with $L=\infty$ and $H=0$, has been investigated in Ref. \cite{DT99} for $d=1$, $d=2$, and $d=4$. Using these results one can obtain the corresponding expressions for the scaling function of the Casimir force and for the Casimir amplitudes. The critical behavior and certain finite-size properties of this model have been studied in Refs. \cite{CDPT97,CPT98} for $1 < d < 3$. One expects that all results can be cast into the form
\begin{equation}
f^{\,\rm t}_{\rm ex}(T,g|d) =TL_\tau^{-d/z}X_{\rm ex}^{\,\rm t}(x_1^{\rm t}|d).
\label{eq:defxextgen_H0}
\end{equation}

In order to proceed we introduce the following notations: the normalized quantum parameter\footnote{The notation of $\lambda$ for the normalized quantum parameter should not be mixed up with the thermal wavelength for which the same symbol is used.} $\lambda =\sqrt{g/J}$, the normalized temperature $t=T/J$, $b=(2\pi t)/\lambda$, and the shifted spherical field $\phi =\mu/J-2d$. The results of Ref. \cite{DT99} are formulated in terms of Zamolodchikov's C-function\footnote{Zamolodchikov's C-theorem concerns the zero-temperature behavior of quantum systems.
	It establishes the existence of a dimensionless function C of the coupling constants with
	monotonic properties along the renormalization group trajectories \cite{Za86,Za87}. Since
	the basic assumptions underlying the C-theorem are not specific for two spatial dimensions, there is 
	considerable interest  in generalizations of  Zamolodchikov's results for $d\ne 2$ as well as for nonzero temperatures.} $C(t,\lambda|d)$ \cite{Za86,Za87}, as defined by Neto and Fradkin \cite{NF93}. Due to the relation
\begin{equation}\label{eq:relation_C_feex_t}
	f^{\,\rm t}_{\rm ex}(t,\lambda|d)=-\frac{n(d)}{\upsilon^d\beta^{d+1}} C(t,\lambda|d)=-n(d)T L_\tau^{-d} C(t,\lambda|d),
\end{equation}
their transformation to  $f^{\,\rm t}_{\rm ex}(T,\lambda|d)$ is, however, straightforward, 
where $n(d)$ is a positive real number (which depends only on the spatial dimension $d$ of the system) and $\upsilon$ is the characteristic velocity (e.g., the velocity of quasi-particles) in the system with $\upsilon \beta=L_\tau$. For bosons
\begin{equation}\label{eq:n_on_d_bosons}
n(d)=\frac{\Gamma\left(\frac{d+1}{2}\right)\zeta(d+1)}{\pi^{(d+1)/2}},
\end{equation}
and, for the considered model, $\upsilon=\sqrt{gJ}$, with $n(1)=\pi/6, \; n(2)=\zeta(3)/(2\pi)$, and $ n(4)=3\zeta(5)/(4\pi^2)$.

For the temporal excess free energy density one has the following results:

\begin{itemize}
	\item if $d=1$:
	
	\begin{equation}\label{eq:d1_tem_efe}
	f_{\rm ex}^t(t,\lambda|d=1 )=-T^2\frac{\pi^{3/2}}{36\sqrt{2gJ}}y_0^{1/4}\exp \left( -\sqrt{y_0}\right),
	\end{equation}
where 	
\begin{equation}
y_0=\left(\frac{\lambda}{t}\right)^2\phi_0=\left(\frac{L_\tau}{\xi_\lambda}\right)^2, \quad \mbox{ with} \quad \phi _0=64\exp \left( -4\pi /\lambda \right), \quad L_\tau\equiv \frac{\lambda}{t}=\frac{\upsilon}{T} \quad \mbox{and} \quad \xi_\lambda\equiv\phi_0^{-1/2}. \label{eq:sfed1}
\end{equation}
As \eq{eq:sfed1} shows, the solution $\phi_0$ of the corresponding spherical field equation for the zero-temperature system has an essential singularity at $\lambda=0$ (see also Ref. \cite{S93}). Such a type of solution is known from diverse problems, such as one-dimensional anharmonic crystals \cite{PT86} and the quantum
nonlinear $O(n)$ sigma model in the limit of large $n$ \cite{Sen93,JG94}.

With the identifications associated with \eq{eq:sfed1}, the temporal excess free energy density turns into 
	\begin{equation}\label{eq:d1_tem_efe_scaling}
f_{\rm ex}^t(t,\lambda|d=1)=T L_\tau ^{-1}Y\left( L_\tau/\xi_\lambda\big|d=1\right) \quad \mbox{with} \quad Y(x|d=1)= -\frac{\pi^{3/2}}{36\sqrt{2}}x^{1/2}\exp \left( -x\right).
\end{equation}
This result is fully in line with \eq{eq:defxextgen_H0}. From the above expressions, one infers that $Y$ is a negative, monotonically increasing function of $L_\tau\propto 1/T$; obviously, $Y(0|d=1)=0$, i.e., $\Delta_{\rm Cas}^{\rm t}(d=1)=0$. At  zero temperature the system has an essential singularity at $\lambda=0$. 

	\item if $d=2$:

In this case the critical point  is located at $\lambda =\lambda _c=1/%
{\cal W}_2(0)$ $\approx 3.1114$ and at $T=0$; here 
\begin{equation}
{\cal W}_d(\phi )=\frac 1{2(2\pi )^d}\int_{-\pi }^\pi dq_1\ldots \int_{-\pi
}^\pi dq_d\left( \phi +2\sum_{i=1}^d\left( 1-\cos q_i\right) \right) ^{-1/2}.
\label{ew:Watson_d}
\end{equation}
For the temporal excess free energy density one has
\begin{equation}
f_{\rm ex}^t(t,\lambda|d=2)=T L_\tau ^{-2}Y\left( L_\tau/\xi_\lambda\big| d=2\right)
 \label{f_ex_t_d2}
 \end{equation}	
with
\begin{equation}
Y(x|d=2)=\frac{1}{2\pi} \left[ x(y-y_0)+\frac 16\left(
y^{3/2}-y_0^{3/2}\right) +\sqrt{y}{\rm Li}_2\left( \exp \left( -\sqrt{y}%
\right) \right) +{\rm Li}_3\left( \exp \left( -\sqrt{y}\right) \right)
\right]
\label{eq:d2_quantum_sc_function}
\end{equation}
and 
\begin{equation}
x=\pi \left( 1/\lambda -1/\lambda _c\right) \lambda /t,\quad \sqrt{y}=2{\rm arcsh}\left( \frac 12\exp \left( -2x\right) \right) \quad \mbox{and} \quad 
\sqrt{y_0}=\left\{ 
\begin{tabular}{lll}
$-4x$, & $\lambda >\lambda _c$ \\ 
$0$,  & $\lambda \leq \lambda _c$%
\end{tabular}
\right. .  \label{eq:solution}
\end{equation}	
The  amplitude of the Casimir force is 
 $\Delta_{\rm Cas}^{\rm t}(d=2)\equiv Y(x=0|d=2)=2\zeta(3)/(5\pi)$. 
 \begin{figure}
 	\includegraphics[width=0.88\columnwidth]{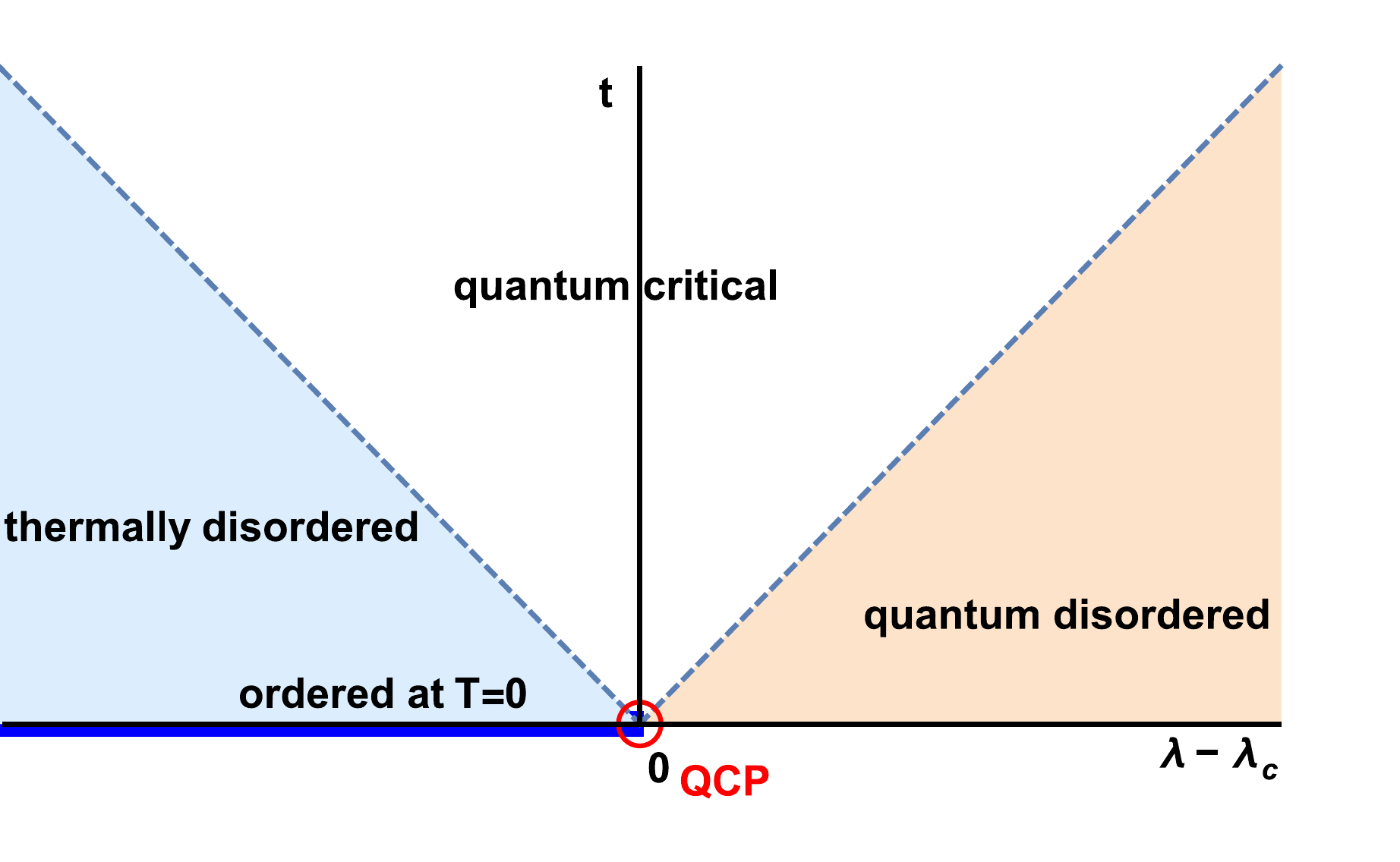}
 	\caption{Schematic phase diagram of the quantum spherical model with short-ranged interactions for $d=2$. Long-ranged order is present only at $T=0$ for $\lambda<\lambda_c$. The phase diagram should be compared with that one given in Fig. \ref{fig:QPDT0} for $\nu z=1$. For the model under consideration, $\nu=1$ and $z=1$. The phase diagram for the present model coincides with the
 		one of the quantum Ising model in  $d=1$, as well as with the nonlinear $O(n)$ sigma
 		model in the limit $n\to\infty$  (see, e.g., Ref. \cite{Sa2011}).
 	}
 	\label{fig:QSPherModT0}
 \end{figure}
 
 In view of the crossover rule (see Sect. \ref{QuantumCritCasEffect}), one expects that the scaling function of the  quantum model considered here with $d=2$ and $z=1$ is the same as the classical model with dimension $d_{\rm classical}=d_{\rm quantum}+z=3$. We stress that the meaning of the scaling variables (see \eq{eq:solution}), as well as the corresponding phase diagram (see Fig. \ref{fig:QSPherModT0}), are different for the classical and the quantum model. Comparing Eqs.  \eqref{eq:d2_quantum_sc_function} and \eqref{eq:solution} with the corresponding ones for the classical spherical model, one concludes that the statement of the equivalence for the scaling functions is indeed correct for the models under consideration (see Eqs. \eqref{fcas_standard_final_d3} -- \eqref{ybfinald3}). 
 
The phase diagram of the quantum spherical model,  with $d=2$ and $z=1$,  is shown in Fig. \ref{fig:QSPherModT0}. From the
above equations one can verify the various behaviors of $y$ as function of $x$ in three regions: \textit{(i)}  the so-called renormalized classical (i.e., thermally disordered) region, where $y$ tends to zero exponentially as a function of $x$ ($x\gg 1)$; \textit{(ii)} the quantum critical region, where $y=O(1)$ (for $x=O(1)$); \textit{(iii)} the quantum disordered region, where $y$ diverges as $4x^2$
for $x\ll -1$. 
	
	\item if $d=4$:
	
	In this case there is a line in the $(T,g)$ plane of temperature driven second-order phase transitions which ends at a quantum critical point $(T=0,g=g_c)$. The phase diagram is similar to the one shown in Fig. \ref{fig:QPDnonT}. Since in this case $d_{\rm classical}=d_{\rm quantum}+z=5>d_u=4$, where $d_u$ is the upper critical dimension for systems with short-ranged interactions, hyperscaling is no longer valid. In the most
	general case, in principle it might be even possible that the function $f_{\rm ex}^t(t,\lambda|d=4)$ cannot be cast into a finite-size scaling form. However, it turns out that due to the presence of a dangerous irrelevant variable \cite{F83,P90}, a modified finite-size scaling function emerges  exponentially close (on the scale of  $L_\tau$)  to the line of nonzero-temperature phase transitions. But outside of this region, the standard scaling turns out to be restored. We refrain from reporting further details of the case $d=4$ and refer the interested reader to Ref.  \cite{DT99}. 
\end{itemize}

$\bigstar$ \textit{{long-ranged interactions}}
\label{sec:LRI_QSM_scaling_functions}

In the current section we are presenting results for the scaling functions of the excess free energy and of the Casimir force for the quantum spherical model with long-ranged interactions (see Sect. \ref{sec:QSM_def}), and for the $O(n)$ quantum $\varphi^4$ model in
the limit $n \to \infty$. These two models belong to the same universality class and offer the possibility to investigate the interplay of quantum and classical
fluctuations in an exact manner \cite{BDT2000,CDT2000,CTD2000b}. In Sect.   \ref{sec:Cas_ampl_long_ranged} we have presented the corresponding results for the Casimir amplitudes.  We consider  algebraically decaying interactions $\propto r^{-(d+\sigma)}$, where $0<\sigma \leq 2$ controls how rapidly this interaction decreases as function of the distance $r$. In this case the dynamic critical exponent is $z=\sigma/2$ \cite{V96,CDT2000} (see \eq{eq:corr_time}). 

For a system with film geometry $L\times \infty ^{d-1}\times L_\tau$ and for $\frac 12\sigma <d<\frac 32\sigma $, the excess free
energy exhibits, in full accordance with Eq. (\ref{hypot1}),
the finite size scaling form
\begin{equation}
\frac1L f_{d,\sigma }^{{\rm ex}}(t,\lambda ,h;L)=\left( TL_\tau
\right) L^{-(d+z)}X_{{\rm ex}}(x_1,x_2,|d,\sigma ),
\label{eq10}
\end{equation}
where the universal scaling function $X_{{\rm ex}}(x_1,x_2,
\rho|d,\sigma
)$ of the excess free energy has the form
\begin{eqnarray}
\label{Xscaling}
\lefteqn{X_{{\rm ex}}(x_1,x_2,\rho|d,\sigma ) =\frac 12x_1\left(
y_0-y_\infty \right) +\frac 12x_2^2\left( \frac 1{y_0}-\frac
1{y_\infty }\right)-\frac{k_d}{4\sqrt{\pi }\sigma }\Gamma \left(
\frac d\sigma \right) \Gamma
\left( -\frac d\sigma -\frac 12\right) \left( y_0^{\frac d\sigma +\frac
	12}-y_\infty ^{\frac d\sigma +\frac 12}\right)} \nonumber \\ &&-\frac{k_d}{\sigma \sqrt{\pi }}\Gamma \left( \frac
d\sigma \right)
\sum_{m=1}^\infty \left[ \frac{\left( 2y_0\right) ^{\frac d\sigma +\frac
		12}K_{\frac d\sigma +\frac 12}\left( m\frac{\sqrt{y_0}}\rho\right) }
{\left( m\frac{\sqrt{y_0}}\rho\right) ^{\left( \frac d\sigma +\frac
		12\right) }}-\frac{\left(2y_\infty\right) ^{\frac d\sigma +\frac 12}
	K_{\frac d\sigma +\frac12}\left( m\frac{\sqrt{y_\infty}}\rho\right)
}{\left( m\frac{\sqrt{y_\infty}}\rho\right)
	^{\left( \frac d\sigma +\frac 12\right) }}\right] \;\;\;\;\;\;\;\;\;\; \\
&& -\frac 14\frac \sigma {\left( 4\pi \right)
	^{\frac d2}}\sum_{n=1}^\infty\int_0^\infty x^{-\frac\sigma4-\frac d2-1}
\exp \left( -\frac{n^2}{4x}\right)
G_{\frac\sigma 2,1-\frac \sigma 4}
\left( -x^{\frac\sigma2}y_0\right) dx \nonumber \\ 
&&-\frac{\sqrt{2}}{(2\pi
	)^{\frac{d+1}2}}\!\sum_{n=1}^\infty
\!\sum_{m=1}^\infty \int\limits_0^\infty \frac{dz}{m^dz^{\frac32}}\!{\cal F}
_{\frac d2-1,\sigma }\left(\frac z{m^\sigma }\right)\exp
\left[ -zy_0-\frac{n^2}{4z\,\rho^2}\right],\nonumber 
\end{eqnarray}
where
$k_d^{-1}=\frac 12(4\pi )^{\frac d2}\Gamma (d/2)$,
\begin{equation}
G_{\alpha ,\,\beta }\left( t\right) =\frac 1{\sqrt{\pi }}\sum_{k=0}^\infty
\frac{\Gamma \left( k+1/2\right) }{\Gamma \left( \alpha k+\beta \right) }
\frac{t^k}{k!}, \quad \mbox{and} \quad {\cal F}_{\nu ,\sigma }\left( y\right) =\int_0^\infty x^{\nu +1}J_\nu \left(
x\right) \exp \left( -yx^\sigma \right) dx. 
\label{eq:G_and_F}
\end{equation}
The function $G_{\alpha ,\,\beta }\left( t\right)$ has been introduced in Ref.  \cite{Cha94}, and $K_\nu (x)$ and $J_\nu (x)$ are the MacDonald and Bessel
functions, respectively. In Eq.~(\ref{Xscaling}) $y_0$ is the solution of the corresponding
spherical field equation (see \eq{eq5}) which follows by requiring the partial
derivative of the r.h.s. of Eq.~(\ref{Xscaling}) with respect to $y_0$ to
be zero\footnote{For details see Eqs. (16), (17), and (24) in Ref. \cite{CDT2000}.}. We denote the solution of the corresponding bulk
spherical field equation as $y_\infty $. The scaling variables in \eq{Xscaling} are
\begin{equation}\label{eq:scaling_var_QLRISM}
x_1=L^{-1/\nu }\left(\frac1\lambda
-\frac1\lambda
_c\right), \quad x_2=hL^{\Delta /\nu }, \quad \rho=L^z/L_\tau, \quad \mbox{with}\quad  L_\tau =\lambda /t.
\end{equation}
The critical value $\lambda_c$  of $\lambda$ is
\begin{equation}
\lambda _c^{-1}=\frac 12(2\pi )^{-d}\int d^d{q}(U({q})/{J})^{-\frac 12},
\label{eq9}
\end{equation}
where the integration runs over the first Brillouin zone. Here $U(q)$ is the Fourier transform of
the interaction, with the energy scale
fixed such that $U(0) = 0$: $U({q})=2{J}\sum_{i=1}^d\left( 1-\cos q_i\right) $ for nearest
neighbor interactions and it has the functional form $U({q})\simeq 
{J}\rho _\sigma |{q} |^\sigma $, $0<\sigma <2$, for long-ranged
interactions ($\rho _\sigma >0$ is taken equal to
one here, i.e., its value is incorporated in $J$).  We recall that $\nu ^{-1}=d-\frac 12\sigma $, $\Delta /\nu =\frac 12\left(d+\frac 32\sigma \right)$, and $z=\frac 12\sigma $ are the critical
exponents of the present model~\cite{V96}. 

The Casimir forces in the considered system scale as 
\begin{equation}
F_{{\rm Cas}}^{d,\sigma }(T,\lambda ,h;L)=\left( TL_\tau \right)
L^{-(d+z)}X_{{\rm Cas}}(x_1,x_2,\rho|d,\sigma ).  \label{eq12}
\end{equation}
According to Eq. (\ref{DefXCasimir}), the {\it universal} scaling function $X_{{\rm Cas}}(x_1,x_2,\rho|d,\sigma )$ of the Casimir force
 can be expressed in terms of the excess free energy, i.e., $X_{{\rm ex}}\equiv X_{{\rm
		ex}}(x_1,x_2,\rho|d,\sigma )$ (see \eq{DefXCasimir}).
	
If the quantum parameter $\lambda $ is close to $\lambda _c$, one
finds, in accordance with Eqs. (\ref{deffext}) and (\ref{defxextgen}),
a scaling
function $X_{{\rm ex}}^{{\rm t}}\left( x_1^t,x_2^t|d,\sigma \right) $, i.e.,
\begin{equation}
f_{d,\sigma }(t,\lambda ,0;\infty )-f_{d,\sigma }(0,\lambda ,0;\infty
)=TL_\tau ^{-d/z}X_{{\rm ex}}^{{\rm t}}\left(
x_1^{\,t},x_2^{\,t}|d,\sigma
\right). \label{defxext}
\end{equation}
For the current model \cite{CDT2000} the most general expression for the temporal excess free energy is 
\begin{eqnarray}
&&X_{{\rm ex}}^{{\rm t}}(x_1,x_2|d,\sigma )=\frac 12x_1\left(
y_t^2-y_\infty^2\right)+\frac 12x_2^2\left( \frac 1{y_t^2}-\frac
1{y_\infty^2 }\right)-\frac{k_d}{4\sqrt{\pi }\sigma }\Gamma \left(
\frac d\sigma \right) \Gamma
\left( -\frac d\sigma -\frac 12\right) \left(y_t^{\frac dz+1}-
y_\infty ^{\frac dz+1}\right)\nonumber\\
&& -\frac{k_d}{\sigma \sqrt{\pi }}\Gamma \left( \frac d\sigma
\right)\sum_{m=1}^\infty\frac{\left( 2y_t^2\right) ^{\frac d\sigma+\frac12}
	K_{\frac d\sigma +\frac 12}\left( m y_t\right) } {\left( m
	y_t\right) ^{\frac d\sigma +\frac 12}}.
\label{xext}
\end{eqnarray}
Here  the scaling  variables are 
\begin{equation}\label{eq:scalin_temporal_QSMLRI}
x_1=L_\tau
^{-1/(\nu z)}\left(\frac1\lambda-\frac1\lambda_c\right), \quad x_2=hL_\tau^{\Delta/(z\nu) },\quad \mbox{and} \quad y_t=L_\tau\phi_0^{1/2}.
\end{equation}
In Eq.~(\ref{xext}) $y_t$ is the solution of the corresponding
spherical field equation (see \eq{eq5}) which follows by requiring the partial
derivative of the r.h.s. of Eq.~(\ref{xext}) with respect to $y_t$ to be zero.

\subsubsection{Further results for the Casimir amplitudes in quantum critical systems}

For $n=1, 2$, and $n=3$, quantum two-dimensional $O(n)$ models have been  studied recently \cite{RHRLDHR2016} 
via quantum world-line Monte Carlo (QMC) simulations of the transverse field Ising model and by  non-perturbative renormalization group theory. The interaction in these models is taken to be short-ranged and $z=1$ for the case considered there. 
It has been shown that the corresponding results are in very good agreement with the ones known from studies of the classical systems with one extra dimension (see Refs. \cite{VGMD2009} for $n=1$, \cite{HGS2011} for $n=2$, and \cite{DK2004} for $n=3$). 
For example,  via a non-perturbative renormalization group approach, the Casimir amplitudes have been obtained as 
$\Delta_{\rm Cas}(d=3,n=1)=-0.1527$,  $\Delta_{\rm Cas}(d=3,n=2)=-0.3006$, and $\Delta_{\rm Cas}(d=3,n=3)=-0.4472$. They agree very  well with the previously known classical Monte Carlo results for $n=1$ and $n=2$ reported in Ref. \cite{VGMD2009}: $\Delta_{\rm Cas}(d=3,n=1)=-0.1520(2)$ and $\Delta_{\rm Cas}(d=3,n=2)=-0.2993(7)$. 

We note that within a reasonably good approximation one has $\Delta_{\rm Cas}(d=3,n)\simeq n \; \Delta_{\rm Cas}(d=3,n=1)$.

We conclude this section by mentioning that, according to our knowledge, there are only few results available concerning the Casimir effect in quantum critical systems which are governed by the fluctuations related to certain quantum parameters. This situation is expected to improve due to the growing interest in quantum phase transitions. 

\section{Casimir effect for curved geometries}

In this section we focus on the sphere-sphere \cite{BE95,ER95} and the  sphere-plate \cite{ER95,HSED98} geometries. For the following reasons these geometries are of particular interest. 
\begin{itemize}
\item{These geometries are easy to control experimentally. With the sphere radii fixed, the only dynamic variable is the distance of closest approach between two spheres  and between a sphere and a plane. In contrast, experimentally it is very demanding to keep two plates parallel to each other with sufficiently high accuracy ($10^{-5}$ rad for $1$ cm diameter plane disks \cite{L97}).} 

\item{These geometries are directly relevant for experiments with colloidal suspensions (see, e.g., Refs.  \cite{BeE1985,NDNS2016,MD2018}) and references therein.}

\item{The sphere-plane geometry is particularly important for interpreting direct force measurements carried out by atomic force microscopy \cite{MR98,RLM99,HCM2000,KMM2000,CMKM2005,2005b,GSM2018}, total internal reflection microscopy \cite{SZHHB2007,HHGDB2008,NHC2009,ZAB2011,NDHCNVB2011,TZGVHBD2011,HKTAGB2021}, or optical tweezers \cite{PCTBDGV2016,MCSGDV2019,MCSGDV2019b,CMGV2021}.}
\end{itemize}

\subsection{Sphere-sphere geometry}

Most of the available results for the two-sphere geometry have been established by using the method of conformal invariance \cite{C87,C89}, which is a valuable and powerful tool for investigations \textit{at} the critical point of the system. 

We denote $f^{(a,b)}(T_c,r,R_1,R_2)$ as  the free
energy density (divided by $k_BT$) of a bulk critical fluid in which two
spheres are immersed with radii $R_1$ and $R_2$ and at center-to-center distance $r$. The boundary conditions $(a)$ and
$(b)$ are imposed at the surfaces of the spheres. As before, we introduce the excess potential $f^{(a,b)}_{\rm ex}(T_c,r,R_1,R_2)$ as the difference of the free energy densities
of the fluid system with and without the spheres immersed
into it:
\begin{equation}
f^{(a,b)}_{\rm ex}(T_c,r,R_1,R_2)=f^{(a,b)}(T_c,r,R_1,R_2)-
f^{(a,b)}(T_c,\infty ,R_1,R_2).
\label{eq:def_excess_energy_spheres}
\end{equation}
The critical system is characterized by three length scales: $r,R_1$, and $R_2$. Scale invariance in $d=2$ requires \cite{C87,BE95}
$f^{(a,b)}_{\rm ex}$ to be a function of two independent, scale-invariant
combinations of these lengths such as $r^2/(R_1R_2)$ and $R_1/R_2$. Conformal invariance in
\textit{general} dimension $d$ leads to the result that
$f^{(a,b)}_{\rm ex}$ depends on a \textit{single}
variable $\kappa =(2R_1R_2)^{-1}|r^2-R_1^2-R_2^2|$:
\begin{equation}
f^{(a,b)}_{\rm ex}(T_c,r,R_1,R_2)=X^{(a,b)}_{\rm ex}(\kappa),\quad
1<\kappa <\infty .
\label{sphere}
\end{equation}
It turns out to be possible to make specific predictions, on general grounds, concerning the behavior of $X^{(a,b)}_{\rm ex}(\kappa)$ in the
limits $\kappa \rightarrow 1$ and $\kappa \rightarrow \infty $. First, we note that the case of \textit{two} spheres with radii $R_1$ and $R_2$ (with, say,  $R_2>R_1$) in a bulk critical fluid is
conformally equivalent to that of a \textit{single} sphere of radius $R_1$ immersed
in a critical fluid of radius $R_2$. The last finding implies that these two
cases can be described by the same universal scaling function
$X^{(a,b)}_{\rm ex}(\kappa)$. Moreover, it is evident that the asymptotic form
of the scaling function $X^{(a,b)}_{\rm ex}(\kappa \to 1^{+} )$ is determined by
the Casimir amplitude $\Delta^{(a,b)}_{\rm Cas}$ for the parallel plates geometry.
Indeed, by considering the case of concentric spheres ($r=0$) in the limit
$R_1,R_2\rightarrow \infty $ at fixed $L=R_2-R_1$ one obtains \cite{BE95}
\begin{equation}
X^{(a,b)}_{\rm ex}(\kappa)\simeq S_d\Delta^{(a,b)}_{\rm Cas}
[2(\kappa -1)]^{-(d-1)/2}, \quad \kappa \to 1^+, \label{sphereclose}
\end{equation}
where $S_d=2\pi^{d/2}/\Gamma(d/2)$ is the surface area of the unit sphere in $\mathbb{R}^d$. This expression determines the Casimir
interaction of two spheres, which nearly touch each other, and are immersed in an unbounded
critical fluid. With $D=r-R_1-R_2$ as the distance between the closest
points of the two spheres, and for $D\ll R_1,R_2$, one has $\kappa \simeq
1+D(1/R_1+1/R_2)$ and
\begin{equation}
	\label{eq:fex-two-spheres}
f^{(a,b)}_{\rm ex}(T_c,r,R_1,R_2)\simeq S_d \Delta^{(a,b)}_{\rm Cas}
\left\{ 2\left[1+D\left(\frac 1{R_1}+\frac 1{R_2}\right)\right]
\right\}^{-(d-1)/2}.
\end{equation}
In the opposite limit $r\gg R_1,R_2$, by using the so-called operator product expansion \cite{BE95}, and by performing the ``small-sphere expansion'' one
finds 
\begin{equation}
f^{(a,b)}_{\rm ex}(T_c,r,R_1,R_2)=-B^{(a,b)}\left(\frac{R_1 R_2}{r^2}
\right)^{x_\psi},  \label{sphereaway}
\end{equation}
where $\psi$ is the order parameter operator $\varphi $, if the two boundary
conditions $(a=b=e)$ correspond to the extraordinary surface universality
class. In this case the
scaling exponent is $x_\varphi =\beta/\nu$ (for the $d=3$ Ising model $x_\varphi
\simeq 0.518$, and for the $d=3$ XY model $x_\varphi \simeq 0.519$). If at least
one of the two spheres does not exhibit a symmetry-breaking boundary
condition, the
operator $\psi$ corresponds to the local energy density operator
$\varepsilon$, and the corresponding scaling exponent is $x_\varepsilon
=d-1/\nu=(1-\alpha)/\nu$ (for the $d=3$ Ising model $x_\varepsilon \simeq 1.41$, and for the
$d=3$ XY model $x_\varepsilon \simeq 1.51$). The amplitude $B^{(a,b)}$ factorizes
\begin{equation}
	\label{eq:B-two-spheres}
	B^{(a,b)} =A^{(a)}_\psi A^{(b)}_\psi /B_\psi ,
\end{equation}
where $A^{(a)}_\psi$ and $A^{(b)}_\psi$ are the amplitudes of the critical
profiles of the operator $\psi$ in a semi-infinite system bounded by a planer
surface of type $(\zeta = a,b)$ in the normal $z$ direction, i.e., $\langle \psi (z)
\rangle_{\mathrm{half\ space}}^{(\zeta)} =A^{(\zeta)}_\psi (2z)^{-x_\psi}$. The
factor $B_\psi $ is the amplitude of
the bulk correlation function: $\langle \psi (\mathbf{R}_1)\psi (\mathbf{R}_2)
\rangle_{\mathrm{bulk}}=B_\psi r^{-2x_\psi}$. Although neither
$A^{(\tau)}_\psi$ nor $B_\psi$ are
universal, their combination $B^{(a,b)}$ does so. Its value for the
$d=2$ Ising model is  known exactly (with $x_\varphi =1/8$ and $x_\varepsilon =1)$:
$B^{(+,+)}=\sqrt{2}$ and $B^{(a,b)}=1$ for all other
boundary conditions. In $d=4-\varepsilon $ dimensions,  renormalization group
calculations yield \cite{BE95}:
\begin{eqnarray}
	\label{eq:RG-two-spheres}
B^{(+,+)} &=&45\varepsilon ^{-1}\left[ 1-\frac{62}{27}\varepsilon +O\left( \varepsilon
^2\right) \right] , \\
B^{(O,O)} &=&\frac 12\left[1+O(\varepsilon^2) \right]
\end{eqnarray} 
and
\begin{equation}
\label{eq:RG-two-spheres-SB}
B^{(SB,SB)} =\frac 12\left[1+\frac 23\varepsilon +O(\varepsilon^2) \right].
\end{equation}

We  now turn to the full functional form of $X^{(a,b)}_{\rm ex}(\kappa)$. 
The corresponding results are quite scarse and are of mean-field character. In Ref. 
\cite{ER95} the properties of $X^{(a,b)}_{\rm ex}(\kappa )$ at $d \lesssim d_u=4$ have been derived. For  boundary conditions
with broken order parameter symmetry, i.e., $(SB,+)$, $(O,+)$, $(+,+)$,
and $(+,-)$, the leading behavior in the $\varepsilon$-expansion follows from mean-field theory, while for symmetry
	preserving boundary conditions, i.e., $(O, O)$, $(O, SB)$, and $(SB, SB)$,
the leading behavior is that of the Gaussian model. Since
the Gaussian model is conformally invariant for all $d>2$, one can determine the corresponding critical behavior for any dimension $d$. The results
obtained in Ref. \cite{ER95} are given below.

For boundary conditions with broken symmetry of the order parameter and concentric geometry (then $r=0$, $R_1<R_2$) one has
\begin{equation}
X^{(a,b)}_{\rm ex}\left(\kappa[\rho(x)]\right)=\frac{9 S_d}{2^{d+1}\pi^2 \varepsilon}
\int_0^x {\rm d}y \frac{{\rm d} \ln(\rho(y))}{{\rm d}y}y,
\end{equation}
where $\kappa(\rho)=\frac{1}{2}(\rho+\rho^{-1})$, $\rho=R_1/R_2$, and $\varepsilon =
4-d$; the function $\rho(x)$ is given below:\\
{\it (i) For $(+,+)$ boundary conditions:}
\begin{equation}
\rho(x)=\exp[-\psi (x)]
\end{equation}
with
\begin{equation}
\psi (x)=2 \omega ^{-1/2} K\left([(1+1/\omega )/2]^{1/2}\right).
\end{equation}
Here $K(\kappa)$ is the complete elliptic integral of the first kind \cite{NIST2010} and
$\omega =\sqrt{1-x}$.

{\it (ii) For $(SB,+)$ boundary conditions:}

the function $\psi (x)$ is given by
\begin{equation}
\psi (x)=-\omega ^{-1/2} K([(1+1/\omega )/2]^{1/2}).
\end{equation}
{\it (iii) For $(+,-)$ boundary conditions:}
\begin{equation}
\rho(x)=\exp[-\varphi (x)]
\end{equation}
where
\begin{equation}
\varphi (x)=2^{3/2}x^{-1/4}\frac{2}{\delta }K\left(\frac{2}{\delta }-1\right)
\end{equation}
with
\begin{equation}
\delta =1+[(1+x^{-1/2})/2]^{1/2}.
\end{equation}
{\it (iv) For $(+,O)$ boundary conditions:}

One has
\begin{equation}
\rho(x)=\exp\left\{-2^{1/2}x^{-1/4}\frac{2}{\delta}K\left(\frac{2}{\delta}-1\right)\right\}.
\end{equation}
From the above expressions it follows that  for $(+,+)$
and $(+,SB)$ boundary conditions $X^{(a,b)}_{\rm ex}<0$, i.e., the Casimir force is
{\it attractive}. For $(+,-)$ and $(O,+)$ boundary conditions one has
$X^{(a,b)}_{\rm ex}>0$, i.e., the force is {\it repulsive}.

The boundary conditions $(O,O)$, $(O,SB)$, and $(SB,SB)$ belong to the group of  symmetry preserving  boundary conditions. For these boundary conditions the $O(n)$
Gaussian model yields \cite{ER95}
\begin{equation}
	\label{eq:Gaussian_O_SB}
X^{(a,b)}_{\rm ex}(\kappa)=\frac{n}{d-2}\sum_{l=0}^{\infty }
\ln\left[1-\delta^{(a,b)}\, \rho(\kappa)^{2\lambda}\right]\, \times \frac{1}{2\lambda} \binom{d-3+l}{l} 
\left[ 2l(d-2+l)+\frac{1}{2} (d-2)^2
\right],
\end{equation}
where $\lambda =l+(d-2)/2$, $\delta^{(a,b)}=(1,-1,1)$ for $(a,b)\in \{(O,O), (O,SB), (SB,SB)\}$, $\binom{d-3+l}{l}$ is the Newton's binomial, and $\kappa(\rho)=\frac{1}{2}(\rho+\rho^{-1})$. One observes that for all values of $\kappa$ (or
$\rho$) the interaction is attractive if $(a)=(b)$ and repulsive if $(a)\ne
(b)$. This is \textit{not} true for $(O,SB)$ boundary conditions within mean-field theory, for which the interaction is attractive (see Sect. \ref{sec:SB_O_bc} above). 

Although some of the results are available only within mean-field theory,
interpolating them smoothly, as a function of $d$, and by taking into account the exact results for the Ising model in $d=2$ yields
reliable estimates of $X^{(a,b)}_{\rm ex}(\kappa)$ within the Ising
universality class and for the boundary
conditions $(+,+)$, $(+,-)$, $(O,+)$, and $(O,O)$.

\subsection{Sphere-plate geometry}

The Casimir interaction of a single sphere with a planar boundary follows
from the results for the  interaction between two-spheres, with radii $R_1$ and $R_2$, in the limit $R_2\rightarrow
\infty$. Re-denoting for simplicity the radius $R_1$ of the remaining finite sphere by $R$, and taking the  distance
from its center to the planar boundary to be $r+L$ (so that $L$ is the surface-to-surface distance), one obtains $\kappa =1+L/r$
and from Eqs. (\ref{sphereclose}) and (\ref{sphereaway}) it follows that
\begin{equation}
f^{(a,b)}_{\rm ex}(T_c,D,r)=\left\{\begin{array}{c}
S_d\Delta^{(a,b)}_{\rm Cas}\left(R/2L\right)^{(d-1)/2},\quad L\ll R,
\\ -B^{(a,b)} \left(R/2L\right)^{x_{\scalebox{0.7}{$\psi$}}},\quad L\gg R.
\end{array} \right.  \label{spherewall}
\end{equation}
Here $\psi=\varphi$ if both boundaries have  symmetry-breaking boundary conditions, and $\psi=\varepsilon$ otherwise. For the two-dimensional Ising model one has $B^{(+,+)}=\sqrt{2}$, $B^{(a,b)}=1$ for all other
	boundary conditions, $x_\varphi=1/8$, and $x_\varepsilon=1$. 

From Eqs. (\ref{sphereclose}), (\ref{sphereaway}), and (\ref{spherewall}) one
concludes that the Casimir interaction between two spheres, as well as between a
sphere and a wall, is very long-ranged. Indeed, the extraordinary surface
universality class   ($x_\varphi(d=3)\simeq 0.518$ implies  the 
decay $\propto  r^{2 x_\varphi} = r^{-1.04}$ (see \eq{sphereaway}) of the Casimir potential energy of two widely separated colloidal
particles (two spheres) in a one-component fluid at its  liquid-vapor critical point, or at
the consolute point of a binary liquid mixture. This holds for the $d=3$ Ising
bulk universality class and $(+,+)$ boundary conditions. Accordingly, the
Casimir potential energy decays almost as slowly as the Coulomb interaction
or Newtonian gravitation. (The thermodynamics of systems controlled by
gravitational forces is an extensively studied, but still controversial, topic \cite{LRT2001,CDR2009,LPRTB2014,MM2017,Robertson2019}.)
In this case the volume integral $\int {\rm d}^3r\ f^{(a,b)}_{\rm ex}(T_c,r,R_1,R_2)$
diverges, i.e., at criticality the total potential
energy of a homogeneous configuration of colloidal particles is 
super-extensive. For colloidal particles dissolved in helium at its lambda transition\footnote{This Gedankenexperiment ignores the actual behavior of these solutes to flocculate in the course of freezing out.} (which corresponds to
the XY model with $(O,O)$ boundary conditions), due to $x_\psi=x_\varepsilon=d-1/\nu\simeq 1.51$ the Casimir potential decays as
$r^{-3.02}$. For both boundary conditions [$(+,+)$ or (+,-)] the Casimir force decays much slower ($\propto r^{-2.04}$ and $\propto r^{-4.02}$, respectively)
than the van der Waals force ($\propto r^{-7}$), and the Casimir force is stronger than the van der Waals force for all $r$
larger or close to $R_1+R_2$. This implies that --- even if the omnipresent van der Waals
force alone
is not strong enough to produce a liquid-like/gas-like phase transition of colloidal particles --- 
the solutes are expected to form a condensed phase at the critical point of the solvent due to
the strong and long-ranged Casimir force. (In this condensed phase the presence of many-body Casimir forces plays an important role \cite{MHD2012,MHD2015,PCTBDGV2016}.) Since $\xi(T\rightarrow T_c) \rightarrow
\infty $, one expects aggregation or flocculation to occur due
to the Casimir force in a temperature range encompassing $T_c$.
Flocculation of colloidal particles in nearly critical fluids has been observed
experimentally and discussed theoretically \cite{MD2018}.

We now consider the generic case \cite{HSED98} of a spherical particle with radius $R$ immersed into a binary liquid mixture, at a
distance $L$ of closest approach from a planar wall, which gives rise to $(+,+)$ boundary conditions, i.e., of the two coexisting bulk phases the same one is enriched near the wall and near the spherical surface. The particle may be considered as a freely diffusing  colloidal particle but also as a sphere attached to the tip of an atomic force
microscope.

Close to $T_c$ the \textit{singular} contribution to the Casimir force (divided by $k_B T$) exhibits the scaling form
\begin{equation}
F_{\rm Cas}^{(+,+)}=R^{-1}X_{\rm Cas}^{(+,+)}(L/\xi, L/R) \qquad (T>T_c).
\end{equation}
The universal scaling function $X_{\rm Cas}^{(+,+)}$ corresponds to attraction and
has (as in the case of the parallel plates geometry) a maximum as function of temperature $T$,
with $L$ and $R$ fixed, at $T_{max}(L,R)$ {\it above} $T_c$.
The solvent is described by the standard $\varphi^4$ Hamiltonian (see also
Sect. 1.6.1)
\begin{equation}
H\{\varphi \}=\int_V {\rm d}V \left\{\frac{1}{2}(\nabla \varphi)^2+
\frac{\tau}{2}\varphi^2+\frac{u}{24}\varphi^4 -h \varphi \right\},
\label{HamSphere}
\end{equation}
for a scalar order parameter $\varphi({\bf r})$ in cylindrical coordinates
${\bf r}=({{\boldsymbol \rho}}, z)\in \mathbb{R}^d$ associated with the rotational symmetry axis normal to the wall surface. The boundary conditions are $\varphi =\infty$
at the wall and at the sphere surface, which  correspond to the critical adsorption fixed
point. The region $V$ is the half-space $z\ge 0$,  excluding the volume occupied
by the sphere. The field $h$ is conjugate to the deviation of the concentration
from its critical composition (if the solvent is a binary liquid mixture). The scaling functions are obtained from
the evaluation of the stress tensor by using the mean-field
order parameter profile. This profile diverges at the wall and at the
 surface of the sphere; it is obtained by numerically solving the Euler-Lagrange
equation which determines the minimum of $H\{\varphi \}$. 

If $R\gg L$, one can employ the so-called Derjaguin approximation \cite{D34}, in which the
presence of the sphere is replaced by  a stack of parallel discs of thickness d$\rho $ and radius $\rho$, at a distance $L(\rho)=
L+\rho^2/(2R)$ from the wall. In the limit  $L/R\rightarrow 0$ and at $T>T_c$,  the
scaling function $X_{\rm Cas}^{(+,+)}(L/\xi,L/R)$ of the Casimir force {\it on
the sphere} can be expressed in terms of $X_{\rm Cas}^{(+,+)}(L/\xi)$ which describes  the
force between two {\it parallel plates} at a distance $L$. For $d=3$ and $d=4$
this procedure leads to \cite{HSED98}
\begin{equation}
X_{\rm Cas}^{(+,+)}(x=L/\xi,y=L/R)=
\omega(d)\left(\frac{L}{R}\right)^{-(d+1)/2}
\int_0^\infty {\rm d}\alpha \,{ \alpha^{d-2}\over (1+\alpha^2/2)^d}
X_{\rm Cas}^{(+,+)}\left((L/\xi)(1+\alpha^2/2)\right),
\label{Derjaguin}
\end{equation}
where $\omega(3)=4\pi$ and $\omega(4)=12\pi$. In order to obtain a reliable
approximation for $d=3$, one can interpolate the exact results
available for $X_{\rm Cas}^{(+,+)}(L/\xi)$, $T>T_c$, for $d=2$ and $d=4$, and evaluate this interpolation for $d=3$. 

If $R\ll L, \xi $ one can apply the so-called small sphere expansion
\cite{BE95,ER95}. The expressions of the corresponding results for $d=4-\varepsilon$
are
quite cumbersome so that we refer the reader to Refs. \cite{HSED98,DS93a,DS93b}. Here we only quote the expression for $d=3$ \cite{HSED98}:
\begin{equation}
 X_{\rm Cas}^{(+,+)}(x,y)=
-\frac{B^{(+,+)}}{c_+}\frac{x^{1+(\beta /\nu)}}{2^{(\beta /\nu)}} P_+(x)
y^{-1-(\beta /\nu)} +\left[{B^{(+,+)}\over c_+^2}\right]^2
\frac{x^{1+2(\beta/\nu)}}{2^{2(\beta /\nu)}}
P_+(x)P'_{+}(x) y^{-1-2(\beta /\nu)}+ O\left(y^{-d-1+1/\nu}\right),
\label{ssphereexpd3}
\end{equation}
where $P_+(x)$ is the universal Ising scaling function $
P_+(z/\xi_+)=\langle \varphi(z)\rangle^{+}_{{\rm half} \: {\rm space},\, t \;> 0}/
\langle \varphi\rangle_{{\rm bulk},\, t\; > 0}$.
The universal amplitude $c_+$ belongs to the order parameter profile (for
$T>T_c$) for critical adsorption on a planar substrate: $P_+(x_+\to 0)=c_+ x_+^{-\beta/\nu}$ and $P_+(x_+\to \infty) \sim \exp{(-x_+)}$. In three dimensions
$B^{(+,+)} \simeq 7.73$ and $c_+\simeq 0.717$ \cite{HSED98} (see also \cite{FD95}). Since $2\beta /\nu=1.036$ is smaller than $-d+1/\nu= 1.41$, \eq{ssphereexpd3}
includes the two leading contributions ($\propto y^{-1.518}$ and $\propto y^{-2.026}$, respectively, with the correction term $\propto y^{-2.41}$). From these results one
infers that for values being typical for atomic force microscopy ($R\simeq
10^{-6}$~m, $L\simeq 10^{-8}$~m, and $T_c\simeq 300$~K), the singular part of
the Casimir
force should be of the order of $10^{-10}$~N, while the corresponding van der
Waals force $F_{\rm vdW}=2 A/[3 R\, y^2(2+y)^2]$ is of the order of $10^{-11}$~N for $A$ with a typical value of $A \simeq 10^{-20}$ J \cite{HSED98}. Therefore, near $T_c$
the critical Casimir force {\it dominates} the background dispersion forces.
Finally, it is worth mentioning that, within the small sphere expansion, the
Casimir force between two spherical particles turns out to be nonsymmetric with
respect to the deviations from the critical concentration of a critical binary $A-B$
liquid mixture --- in such a way, that the Casimir force is {\it enhanced}
if the concentration $c_A$ of the component $A$ preferably adsorbed by the (colloidal)
particles is {\it reduced}. This asymmetry is consistent with the
shape of experimentally observed flocculation diagrams \cite{BE85,GM92,JK97,MD2018}.

The Casimir force between concentric spheres has
been considered also at tricritical points \cite{RG97}. Up to now, exact results at $T=T_c$
are still not available.

\section{Further exact results}
\label{some_other_results}

The current section provides a brief review of certain additional, exact results concerning the thermodynamic Casimir effect, which are not directly related to the geometries of basic interest as the ones discussed in previous parts of the text.  

\subsection{Conformal invariant results}

The study of critical phenomena reveals that two-dimensional systems at their critical point represent a very special case. As already stated in Sect.  \ref{conformal_invariance}, \textit{at} criticality such systems can be described by conformal field theory (CFT) \cite{FQS84,C89}.  As discussed there, Casimir forces in a strip are related to the central charge of the CFT. In addition to the results already presented, there are also results for the Casimir amplitudes for a section of a circle ("pie slice") with opening angle $\phi$, as well as for a helical staircase of finite angular (and radial) extent \cite{KP1997}, for rectangular domains \cite{KV91}, for interactions between circles \cite{MVS2012}, for needles \cite{VED2013}, and for Janus particles \cite{SME2020}. Reference \cite{BEK2013} describes results between particles of any compact shape. In Ref. \cite{BEK2015} one considers the interaction between two wedges, or an array (strip) of wedges. More specifically, the following special cases have been considered: (i) the force between two wedges, (ii) between strips with triangular corrugations, (iii) between two truncated wedges with lateral shift, and (iv) between two strips with truncated corrugations and lateral shifts. The authors show that such fluctuation induced forces can be attractive or repulsive depending on the angle of the wedge, and that stable equilibrium can be obtained with truncated wedges and arrays of them. 

\subsection{Ising model}

As already stated in Sect. \ref{Ising_UC}, in the case of the Ising model there are also certain exact results which do not belong to the geometries of main interest discussed before. Here we briefly mention them in order to provide at least their most relevant aspects. 

We start by mentioning the series of studies on the Ising model in fully finite geometries \cite{H2016,H2017,HH2017}.  In such systems one has more than one option to define an effective force between their opposite sides. For any further details in that respect we refer the reader to these aforementioned papers.  Reference   \cite{H2016} reports results for the partition function of the square lattice Ising model defined on a rectangle, with Ref. \cite{H2017} describing the corresponding scaling limit. Reference  \cite{HH2017} provides information about the saling functions of the critical Casimir force within the two-dimensional Ising model for finite aspect ratios with  various boundary conditions.  

The studies in Refs. \cite{VED2013,EB2016} discuss the effective interaction of prolongated objects, i.e., needles and rod-like particles, when they are embedded in an Ising type medium at its critical point. 

\textit{Lateral} critical Casimir force has been studied in Refs. \cite{THD2008,TKGHD2009,TKGHD2010,TZGVHBD2011,TTD2013,MKMD2014,THD2015,TTD2015,NN2016}. For example, Ref. \cite{NN2016} analyzes the lateral force acting between two planar, chemically inhomogeneous
walls confining an infinite Ising strip of width $L$ and with surface fields $h_1$. The chemical inhomogeneity of each of the walls, which is described by the presence of segments exposed to  surface fields $-h_1$, is taken to have segments of 
size $N$ being shifted by a distance $M$ along the strip, with respect to the inhomogeneity on the opposite boundary. The authors use the exact diagonalization of the
transfer matrix, calculate numerically the lateral critical Casimir force, and discuss its properties. For example, they find
\begin{equation}
\label{eq:2dIsing_lateral}
f_{\rm Cas}^{\parallel}(T,L,M,N,h_1)=\frac{1}{L}X_{\rm Cas}^{\parallel}(x_\tau,M/L,N/L)+{\cal O}(L^{-2}),
\end{equation}
where $h_1>0$ is a finite surface field. We note that {\it (i)} the lateral force decays slower, i.e., $\propto L^{-1}$, instead of $L^{-2}$ as for the perpendicular force and that {\it (ii)} its leading behavior does not depend on $h_1$, provided $h_1>0$ at a certain fixed, finite value.

Finally, Ref. \cite{NMD2016} presents an exact derivation of the critical Casimir interactions between two defects in a planar lattice-gas Ising model. Each defect consists of a finite group of nearest-neighbor spins which interact through modified coupling constants. 

\section{Concluding remarks}

In the current report we have presented a review on the available exact results concerning the thermodynamic Casimir effect. As it is in general the case, exact results are useful for a detailed understanding of the corresponding phenomena and of the role of the parameters involved, for the clarification of the issue which parameters are to which extent essential, etc.  These results and properties can be used for testing the wide spectrum of possible approximate methods --- numerical ones like Monte Carlo simulations and molecular dynamics or analytic ones such density functional theory. Naturally, for technical reasons the scope of exact results is somewhat limited. Experience tells that innocent sounding extensions, like having a new nonzero parameter (even if spatially constant), might convert the task to one for which exact results are out of reach. A paradigm in that direction and in the present context is the two-dimensional Ising model for which the solution for zero external field is known since Onsager's work in 1944 \cite{O44}, for which despite numerous efforts the solution for a nonzero external field is not available even today. 

Up to now most of the exact results available belong to classical systems in the grand canonical ensemble. It is expected that in the future there will be attempts to extend them to dynamical systems, to quantum systems --- including systems with different types of quenches, to systems described by other, say, canonical ensembles, to systems exhibiting disorder, and topological phase transitions --- as well as to a combination of them. One can also consider ensemble dependent fluctuation induced forces - see, e.g., Ref. \cite{DR2022}, for which all the issues studied for the Casimir forces will be objects of investigation.

The critical Casimir effect is not only a topic of interest for academic investigations. Similar to the QED Casimir effect, for which the first practical applications are under discussion  (see, e.g., Refs. \cite{IMCB2014,FAKA2014,FMRA2014,WDTRRP2016,YHZM2018,PSS2020,MCKBS2021,GCMSM2021,XuGBJL2022,Schmidt2022} and the references  cited therein), also certain applications of the critical Casimir effect have been already studied  (see, e.g., Refs.  \cite{IMC2005,TZGVHBD2011,DLMP2016,NNVKBS2017,GSL2018,MBCHLKS2019,MCSGDV2019,VMKSK2021,SRJRGSBS2021,XiLSL2021,VGCL2021}).

\section*{Acknowledgments}

Stimulating and fruitful discussions over the years on various occasions with the late M. E. Fisher, as well as with C. Bechinger, K. Binder, R. Evans,  A. Gambassi, M. Kardar,  M. Krech, and A. Macio{\l}ek  are gratefully acknowledged. 

The authors benefited immensely from co-authorships on the subject under review by   M. Barmatz,  M. Bier, the late J. Brankov, H. Chamati, H. W. Diehl, P. Djondjorov, E. Eisenriegler, N. Farahmand Bafi,  M. Gross, F. K. P. Haddadan, A. Hanke, L. Harnau, S. Kondrat, M. Labb{\'e}-Laurent, T. F. Mohry,     M. Napi\'{o}rkowski, P. Nowakowski,      F. Parisen Toldin,             J. Rudnick,  C. Rohwer, F. Schlesener, M. Sprenger, H. Tanaka, M. Tr\"ondle, N. Tonchev, G. Valchev, V. Vassilev, O. Vasilyev, and G. Volpe.

\newpage
\section{List of main notations}
{\setstretch{1.5}
\begin{itemize}[\label{}]
\item $T$: temperature
\item $T_c$: bulk critical temperature
\item $T_{c,L}$: critical temperature of a system with one finite extension $L$
\item $\tau=(T-T_c)/T_c$: reduced temperature
\item $\mu$: chemical potential
\item $\bar{\mu}= \mu/(\kB T)$: reduced chemical potential
\item $\mu_c$: chemical potential of the critical point
\item $\Delta \mu=\mu-\mu_c$: undersaturation
\item $\dmb=\Delta \mu/(\kB T)$: reduced undersaturation
\item $\hat{\mu}=\left(\mu-\mu_c\right)/\mu_c$: dimensionless undersaturation
\item $h=H/(\kB T)$: reduced external ordering field $H$
\item $L$: film thickness or the characteristic size of a finite-size system
\item $a$: microscopic length scale; lattice constant
\item $\kB$: Boltzmann  constant, $\kB=1.380 6504(24)\times 10^{-23}$ $\rm{J/K}$
\item $c$: speed of light in vacuum, $c=299 792 458$ m/s
\item $\hbar$: reduced Planck constant $\hbar\equiv h/(2\pi)=1.054571628(53)\times 10^{-34}$ $\rm J s$
\item $A$: surface area
\item ${\cal F}_{ {\rm tot}}^{\bc}$: the total free energy per $k_B T$ of a system with boundary conditions $\zeta$
\item $f^{\bc}$: free energy per area and $k_BT$ for  boundary conditions (bc) $\zeta$
\item $f_b$: bulk free energy density in units of $k_BT$
\item $f_{\ex}^{\bc}\equiv f^{\bc}-L f_b$: excess free energy per area and $k_BT$  for bc  $\zeta$
\item $f_{\rm surf}^{\bc}$: surface free energy per area and $k_BT$  for bc  $\zeta$
\item $\Delta f_{\ex}^{\bc}\equiv f_{\ex}^{\bc} -\lim_{L\to\infty} f_{\ex}^{\bc}$: finite size contribution to the excess free energy per area and $k_BT$  for bc  $\zeta$
\item $f^{(s)}$: singular part  of the free energy density per $k_B T$
\item $f^{\rm (ns)}$: nonsingular part  of the free energy density per $k_B T$
\item $\Omega^{\bc}$: grand canonical potential for bc $\zeta$
\item $\Omega_{\ex}^{\bc}$: excess grand canonical potential for bc $\zeta$
\item $\omega^{\bc}$: grand canonical potential per area for bc $\zeta$
\item $\omega_b$: bulk grand canonical potential per volume
\item $\omega_{\ex}^{\bc}$: excess grand canonical potential per area for bc $\zeta$
\item $X_{\ex}^{\bc}$: scaling function of the excess free energy $f_{\ex}^{\bc}$ or of the excess grand canonical potential per area $\omega_{\ex}^{\bc}$ for bc $\zeta$
\item $\Delta \omega_{\ex}^{\bc}\equiv \omega_{\ex}^{\bc} -\lim_{L\to\infty} \omega_{\ex}^{\bc}$: finite size contribution to the  excess grand canonical potential per area for bc  $\zeta$
\item $\Delta X_{\ex}^{\bc}$: scaling function of the finite-size contribution to the excess free energy $\Delta f_{\ex}^{\bc}$ per area, or of the finite size contribution to the excess grand canonical potential per area $\Delta \omega_{\ex}^{\bc}$ for bc $\zeta$
\item $\omega_{s}^{\bc}$: surface grand canonical potential per area for bc $\zeta$
\item $F_{\Cas}^{\bc}$: Casimir force per area (Casimir pressure) for bc $\zeta$
\item $\Delta_{\Cas}^{\bc}$: Casimir amplitude for bc $\zeta$
\item $X_{\Cas}^{\bc}$: Casimir scaling function for bc $\zeta$
\item $F_{\rm Cas}^{\parallel}(L)$: Casimir force per area, i.e., Casimir pressure between two planar surfaces at a distance L
\item $\varphi_{\rm Cas}^{\parallel}(L)$: excess energy per area between two planar surfaces at a distance L

\item $\xi (\tau,\dmb)$: bulk correlation length
\item $\xi_\tau$: correlation length $\xi(\tau, \dmb=0)$
\item $\xi_\tau^\pm$: correlation length $\xi(\tau \gtrless 0, \dmb=0)$; sometimes for brevity the upper index is dropped 
\item $\xi_\mu$: correlation length $\xi(\tau=0, \dmb)$
\item $\xi_0^\pm$ : correlation length amplitude: $\xi_\tau(\tau \to 0^\pm)=\xi_0^\pm |\tau|^{-\nu}$
\item $\xi_{0,\,\mu}$ : correlation length amplitude $\xi_\mu(\mu \to
\mu_c)=\xi_{0,\,\mu} |\dmb|^{-\nu/\Delta}$
\item $\xi_l$: Matsubara frequencies $\xi_l=2\pi l k_B T/\hbar$, with 
$l\in \mathbb{N}^+$
\item $x=L/\xi$
\item $x_\tau \equiv \tau \; (L/\xi_0^+)^{1/\nu} $ : temperature scaling variable
\item $x_\mu \equiv \dmb \; (L/\xi_{0,\,\mu})^{\Delta/\nu}$: field scaling variable corresponding to reduced undersaturation

\item $\sigma$: surface tension
\item $\sigma^{\bc}$: tension of an interface generated by imposing bc $\zeta$ on a finite system
\item $X_\sigma^{\bc}$: scaling function of $\sigma^{\bc}$
\item $\cal H$: the Hamiltonian of a system or the corresponding free energy functional 
\item $\chi(T,h)$: the susceptibility (compressibility) of the system
\item $g$: a parameter governing a quantum phase transition
\item $g_c$: critical value of the parameter $g$ for a quantum phase transition
\item{$z$: dynamical critical exponent or one of the coordinates of the Cartesian coordinate system}
\item $L_\tau=\beta\hbar$: effective thickness of a quantum system at temperature $ T =1/(k_B \;\beta)$

\end{itemize} 
}




\begin{thebibliography}{100}
\bibitem{remMarek}
The data for the case {$L=50$} are by courtesy of {M}arek {N}api{\'o}rkowski.
\bibitem{remAnia}
The data for {$L=90$} are by courtesy of {Ania Maci{\`o}{\l}ek}.
\end{thebibliography}
\end{document}